\documentclass[12pt,oneside]{book}
\usepackage{epsfig}
\usepackage{graphics}
\usepackage{comment}
\usepackage[letterpaper,dvips,hmargin={1.65in,1.15in},vmargin={1in,1.15in}]{geometry}
\usepackage{setspace}
\usepackage{cite} 
\usepackage{fancyhdr}

\usepackage{dectab}
\usepackage{sigpage}
\usepackage{copyright}
\usepackage{rotating}
\usepackage{colordvi} 

\newcommand{\bgen}{{\tt Bgenerator}}
\newcommand{\evtgen}{{\tt EvtGen}}
\newcommand{\pythia}{{\tt PYTHIA}}
\newcommand{\bcharm}{{\tt B$\_$CHARM~Scenario~A}} 
\newcommand{\br}{\ensuremath{{\cal BR}}}
\newcommand{\pbarn} {\ensuremath{\mathrm{pb^{-1}}}}
\newcommand{\fbarn} {\ensuremath{\mathrm{fb^{-1}}}}
\newcommand{\ppbar} {\ensuremath{p\overline{p}}}
\newcommand{\degs}  {\ensuremath{^{\circ}}}
\newcommand{\lqcd}{\ensuremath{\Lambda_\mathrm{QCD}}}
\newcommand{\um}  {\ensuremath{\mu}m}

\newcommand{\bra}{\langle}
\newcommand{\ket}{\rangle}

\newcommand{\etal}{{\em et~al.}}
\newcommand{\brlbsemic}{\ensuremath{6.26}}
\newcommand{\brlbsemie}{\ensuremath{0.21}}

\newcommand{\brlbsemidatac}{\ensuremath{8.1}}
\newcommand{\brlbsemidatae}{\ensuremath{1.2}}
\newcommand{\brlbsemidatasyste}{\ensuremath{{+1.1 \atop -1.6}}}
\newcommand{\brlbsemidatabre}{\ensuremath{4.3}}

\newcommand{\brlbhadc}{\ensuremath{0.41}}
\newcommand{\brlbhadpte}{\ensuremath{{+0.06 \atop -0.08}}}
\newcommand{\brlbhade}{\ensuremath{0.19}}
\newcommand{\brlbhad}{\ensuremath{\brlbhadc \pm \brlbhade}}

\newcommand{\brlcstarc}{\ensuremath{0.30}}
\newcommand{\brlcstare}{\ensuremath{0.13}}
\newcommand{\brlcsstarc}{\ensuremath{0.49}}
\newcommand{\brlcsstare}{\ensuremath{0.17}}
\newcommand{\brsigcc}{\ensuremath{0.26}}
\newcommand{\brsigce}{\ensuremath{0.13}}
\newcommand{\brfzero}{\ensuremath{0.25}}
\newcommand{\brlcpipi}{\ensuremath{0.50}}
\newcommand{\brlcsemitau}{\ensuremath{1.74}}

\newcommand{\ndstarhadc}{\ensuremath{106}}
\newcommand{\ndstarhade}{\ensuremath{11}}

\newcommand{\ndhadc}{\ensuremath{579}}
\newcommand{\ndhade}{\ensuremath{30}}

\newcommand{\nlbhadc}{\ensuremath{179}}
\newcommand{\nlbhade}{\ensuremath{19}}

\newcommand{\nlbhad}{\ensuremath{\nlbhadc\ \pm \nlbhade}}
\newcommand{\ndhad}{\ensuremath{\ndhadc\ \pm \ndhade}}
\newcommand{\ndstarhad}{\ensuremath{\ndstarhadc\ \pm  \ndstarhade}}

\newcommand{\nlbsemic}{\ensuremath{1237}}
\newcommand{\nlbsemie}{\ensuremath{97}}

\newcommand{\ndsemic}{\ensuremath{4720}}
\newcommand{\ndsemie}{\ensuremath{100}}

\newcommand{\ndstarsemic}{\ensuremath{1059}}
\newcommand{\ndstarsemie}{\ensuremath{33}}

\newcommand{\nlbsemi}{\ensuremath{\nlbsemic\ \pm \nlbsemie}}
\newcommand{\ndsemi}{\ensuremath{\ndsemic\ \pm \ndsemie}}
\newcommand{\ndstarsemi}{\ensuremath{\ndstarsemic\ \pm \ndstarsemie}}

\newcommand{\rlbc}{\ensuremath{20.0}}
\newcommand{\rlbe}{\ensuremath{3.0}}
\newcommand{\rlbsyste}{\ensuremath{1.2}}
\newcommand{\rlbbre}{\ensuremath{{+0.7 \atop -2.1}}}
\newcommand{\rlbubre}{\ensuremath{0.5}}

\newcommand{\rdc}{\ensuremath{9.8}}
\newcommand{\rde}{\ensuremath{1.0}}
\newcommand{\rdsyste}{\ensuremath{0.6}}
\newcommand{\rdbre}{\ensuremath{0.8}}
\newcommand{\rdubre}{\ensuremath{0.9}}

\newcommand{\rdstarc}{\ensuremath{17.7}}
\newcommand{\rdstare}{\ensuremath{2.3}}
\newcommand{\rdstarsyste}{\ensuremath{0.6}}
\newcommand{\rdstarbre}{\ensuremath{0.4}}
\newcommand{\rdstarubre}{\ensuremath{1.1}}

\newcommand{\rlb}{\ensuremath{\rlbc\ \pm \rlbe}}
\newcommand{\rd}{\ensuremath{\rdc\ \pm \rde}}
\newcommand{\rdstar}{\ensuremath{\rdstarc\ \pm \rdstare}}

\newcommand{\rphi}{\ensuremath{r}-\ensuremath{\phi}}
\newcommand{\lxy}{\ensuremath{L_{xy}}}
\newcommand{\pt}{\ensuremath{P_T}}
\newcommand{\ctau}{\ensuremath{c\tau}}
\newcommand{\gevc}{\ensuremath{\rm GeV/c}}
\newcommand{\gevcsq}{\ensuremath{{\rm GeV}/{\rm c}^{2}}}
\newcommand{\mevcsq}{\ensuremath{{\rm MeV}/{\rm c}^{2}}}
\newcommand{\mevc}{\ensuremath{{\rm MeV}/{\rm c}}}
\newcommand{\Lb}{\ensuremath{\Lambda_b}}
\newcommand{\Lc}{\ensuremath{\Lambda_c^+}}
\newcommand{\Lamc}{\ensuremath{\Lambda_c}}
\newcommand{\Bd}{\ensuremath{B^0}}
\newcommand{\Bu}{\ensuremath{B^+}}
\newcommand{\Bs}{\ensuremath{B_s}}
\newcommand{\B}{\ensuremath{B}}
\newcommand{\Dstar}{\ensuremath{D^{*+}}}
\newcommand{\D}{\ensuremath{D^+}}
\newcommand{\Ds}{\ensuremath{D_s^+}}
\newcommand{\Dzero}{\ensuremath{D^0}}
\newcommand{\seqds}{\ensuremath{D_s^{+} \rightarrow \phi\pi^+}}
\newcommand{\seqphi}{\ensuremath{\phi \rightarrow K^+K^-}}
\newcommand{\seqdstar}{\ensuremath{D^{*+} \rightarrow D^0\pi^+}}
\newcommand{\seqdzero}{\ensuremath{D^0 \rightarrow K^-\pi^+}}
\newcommand{\seqdstard}{\ensuremath{D^{*+} \rightarrow D^+\pi^0}}
\newcommand{\seqd}{\ensuremath{D^+ \rightarrow K^-\pi^+\pi^+}}
\newcommand{\seqlc}{\ensuremath{\Lambda_c^+ \rightarrow pK^-\pi^+}}
\newcommand{\seqlcstar}{\ensuremath{\Lambda_c(2593)^+ \rightarrow \Lambda_c^+\gamma}}
\newcommand{\seqdonezero}{\ensuremath{D_1^{0}\rightarrow D^{*+}\pi^{-}}}
\newcommand{\seqdpronezero}{\ensuremath{D_1^{\prime 0}\rightarrow D^{*+}\pi^{-}}}
\newcommand{\seqdonep}{\ensuremath{D_1^+ \rightarrow D^{*+}\pi^0}}
\newcommand{\seqdpronep}{\ensuremath{D_1^{\prime +} \rightarrow D^{*+}\pi^0}}
\newcommand{\seqtau}{\ensuremath{\tau\rightarrow \mu \overline{\nu}_{\mu} \nu_{\tau}}}

\newcommand{\bddstarpipizero}{\ensuremath{\overline{B}^0\rightarrow D^{*+}\pi^-\pi^0}}
\newcommand{\bddstaraone}{\ensuremath{\overline{B}^0\rightarrow D^{*+}a_1^-}}
\newcommand{\bddpipizero}{\ensuremath{\overline{B}^0\rightarrow D^{+}\pi^-\pi^0}}
\newcommand{\bddaone}{\ensuremath{\overline{B}^0\rightarrow D^{+}a_1^-}}
\newcommand{\bddstarrho}{\ensuremath{\overline{B}^0\rightarrow D^{*+}\rho^-}}
\newcommand{\bddrho}{\ensuremath{\overline{B}^0\rightarrow D^{+}\rho^-}}

\newcommand{\bddpizeromunu}{\ensuremath{\overline{B}^0\rightarrow D^{+}\pi^0\mu^-\overline{\nu}_{\mu}}}
\newcommand{\bpdpimunu}{\ensuremath{B^-\rightarrow D^{+}\pi^-\mu^-\overline{\nu}_{\mu}}}
\newcommand{\bddstarpizeromunu}{\ensuremath{\overline{B}^0\rightarrow D^{*+}\pi^0\mu^-\overline{\nu}_{\mu}}}
\newcommand{\bpdstarpimunu}{\ensuremath{B^-\rightarrow D^{*+}\pi^-\mu^-\overline{\nu}_{\mu}}}
\newcommand{\bpdonezeromunu}{\ensuremath{B^-\rightarrow D_1^{0}\mu^-\overline{\nu}_{\mu}}}
\newcommand{\bpdpronezeromunu}{\ensuremath{B^-\rightarrow D_1^{\prime 0}\mu^-\overline{\nu}_{\mu}}}
\newcommand{\bddonemunu}{\ensuremath{\overline{B}^0\rightarrow D_1^{+}\mu^-\overline{\nu}_{\mu}}}
\newcommand{\bddpronemunu}{\ensuremath{\overline{B}^0\rightarrow D_1^{\prime+}\mu^-\overline{\nu}_{\mu}}}
\newcommand{\bddtau}{\ensuremath{\overline{B}^0\rightarrow D^{+}\tau^-\overline{\nu}_{\tau}}}
\newcommand{\bddstartau}{\ensuremath{\overline{B}^0\rightarrow D^{*+}\tau^-\overline{\nu}_{\tau}}}
\newcommand{\bsdkzero}{\ensuremath{\overline{B}_s\rightarrow D^{+}K^0\mu^-\overline{\nu}_{\mu}}}
\newcommand{\lblcstar}{\ensuremath{\Lambda_b \rightarrow \Lambda_c(2593)^+ \mu^- \overline{\nu}_{\mu}}}

\newcommand{\lblcsstar}{\ensuremath{\Lambda_b \rightarrow \Lambda_c(2625)^+ \mu^- \overline{\nu}_{\mu}}}

\newcommand{\lblcfzero}{\ensuremath{\Lambda_b \rightarrow \Lambda_c^+ f^0\mu^- \overline{\nu}_{\mu}}}
\newcommand{\lblcpizero}{\ensuremath{\Lambda_b \rightarrow \Lambda_c^+ \pi^0\pi^0\mu^- \overline{\nu}_{\mu}}}
\newcommand{\lblcpim}{\ensuremath{\Lambda_b \rightarrow \Lambda_c^+ \pi^+\pi^-\mu^- \overline{\nu}_{\mu}}}
\newcommand{\lblctau}{\ensuremath{\Lambda_b \rightarrow \Lambda_c^+ \tau^- \overline{\nu}_{\tau}}}
\newcommand{\lbsigmaczero}{\ensuremath{\Lambda_b \rightarrow \Sigma_c^0 \pi^+ \mu^- \overline{\nu}_{\mu}}}
\newcommand{\lbsigmacp}{\ensuremath{\Lambda_b \rightarrow \Sigma_c^+ \pi^0 \mu^- \overline{\nu}_{\mu}}}
\newcommand{\lbsigmacpp}{\ensuremath{\Lambda_b \rightarrow \Sigma_c^{++} \pi^- \mu^- \overline{\nu}_{\mu}}}
\newcommand{\bplcpmunu}{\ensuremath{B^- \rightarrow \Lambda_c^+ \overline{p} \mu^- \overline{\nu}_{\mu} }}
\newcommand{\bdlcnmunu}{\ensuremath{\overline{B}^0 \rightarrow \Lambda_c^+ \overline{n} \mu^- \overline{\nu}_{\mu} }}
\newcommand{\mkpi}{\ensuremath{M_{K\pi}}}
\newcommand{\mkpipi}{\ensuremath{M_{K\pi\pi}}}
\newcommand{\mkpipipi}{\ensuremath{M_{K\pi\pi\pi}}}
\newcommand{\mpkpi}{\ensuremath{M_{pK\pi}}}
\newcommand{\mpkpipi}{\ensuremath{M_{pK\pi\pi}}}

\newcommand{\dstarhadk}{\ensuremath{\overline{B}^0 \rightarrow D^{*+}K^-}}
\newcommand{\dhadk}{\ensuremath{\overline{B}^0 \rightarrow D^{+}K^-}}
\newcommand{\lbhadk}{\ensuremath{\Lambda_b \rightarrow \Lambda_c^{+}K^-}}

\newcommand{\bsdspi}{\ensuremath{\overline{B}_s \rightarrow D_s^{+}\pi^-}}

\newcommand{\dstarmu}{\ensuremath{D^{*}\mu}}
\newcommand{\dmu}{\ensuremath{D\mu}}
\newcommand{\lcmu}{\ensuremath{\Lambda_c\mu}}

\newcommand{\bsdsmunu}{\ensuremath{\overline{B}_s\rightarrow D_s^{+}\mu^-\overline{\nu}_{\mu}}}
\newcommand{\incdsmu}{\ensuremath{\overline{B}\rightarrow D_s^{+}\mu^-X}}

\newcommand{\incdstarsemi}{\ensuremath{\overline{B} \rightarrow D^{*+}\mu^- X}}
\newcommand{\incdstarsemie}{\ensuremath{\overline{B} \rightarrow D^{*+}e^- X}}
\newcommand{\dstarsemi}{\ensuremath{\overline{B}^0 \rightarrow D^{*+}\mu^-\overline{\nu}_{\mu}}}
\newcommand{\dstarsemie}{\ensuremath{\overline{B}^0 \rightarrow D^{*+}e^-\overline{\nu}_e}}
\newcommand{\dstarhad}{\ensuremath{\overline{B}^0 \rightarrow D^{*+}\pi^-}}
\newcommand{\incdsemi}{\ensuremath{\overline{B} \rightarrow D^{+}\mu^- X}}
\newcommand{\incdsemie}{\ensuremath{\overline{B} \rightarrow D^{+}e^- X}}
\newcommand{\dsemi}{\ensuremath{\overline{B}^0 \rightarrow D^{+}\mu^-\overline{\nu}_{\mu}}}
\newcommand{\dsemie}{\ensuremath{\overline{B}^0 \rightarrow D^{+}e^-\overline{\nu}_e}}
\newcommand{\dhad}{\ensuremath{\overline{B}^0 \rightarrow D^{+}\pi^-}}
\newcommand{\inclbsemi}{\ensuremath{\overline{B} \rightarrow \Lambda_c^{+}\mu^- 
X}}
\newcommand{\inclbsemie}{\ensuremath{\overline{B} \rightarrow \Lambda_c^{+}e^- 
X}}
\newcommand{\inclcstar}{\ensuremath{\Lambda_b \rightarrow \Lambda_c(2593)^+ \mu^- X}}
\newcommand{\inclcsstar}{\ensuremath{\Lambda_b \rightarrow \Lambda_c(2625)^+ \mu^- X}}
\newcommand{\incsigmaczero}{\ensuremath{\Lambda_b \rightarrow \Sigma_c^0 \pi^+ \mu^- X}}
\newcommand{\incsigmacpp}{\ensuremath{\Lambda_b \rightarrow \Sigma_c^{++} \pi^- \mu^- X}}

\newcommand{\lbsemi}{\ensuremath{\Lambda_b \rightarrow \Lambda_c^{+}\mu^-\overline{\nu}_{\mu}}}
\newcommand{\lbsemie}{\ensuremath{\Lambda_b \rightarrow \Lambda_c^{+}e^-\overline{\nu}_e}}
\newcommand{\lbhad}{\ensuremath{\Lambda_b \rightarrow \Lambda_c^{+}\pi^-}}

\newcommand{\alldstar}{\ensuremath{\overline{B} \rightarrow D^{*+}X}}
\newcommand{\alld}{\ensuremath{\overline{B} \rightarrow D^{+}X}}
\newcommand{\alllc}{\ensuremath{\Lambda_b \rightarrow \Lambda_c^{+}X}}

\newcommand{\bb}{\ensuremath{b\overline{b}}}
\newcommand{\cc}{\ensuremath{c\overline{c}}}
\newcommand{\yile}{\ensuremath{\frac{\sigma_{\Lambda_b}(\pt>6.0){\cal B}(\lbhad)}{\sigma_{\Bd}(\pt>6.0){\cal B}(\dhad)}}}
\newcommand{\tomo}{\ensuremath{\frac{\sigma_{\Lambda_b}(\pt>4.0){\cal B}(\Lb\rightarrow J/\psi \Lambda)}{\sigma_{\Bd}(\pt>4.0){\cal B}(\Bd\rightarrow J/\psi K_s^0)}}}
\newcommand{\rxsec}{\ensuremath{\frac{\sigma_{\Lambda_b}}{\sigma_{\Bd}}}}
\newcommand{\yileab}{\ensuremath{\rho(6)\times\frac{{\cal B}(\lbhad)}{{\cal B}(\dhad)}}}
\newcommand{\bmixdx}{\ensuremath{\overline{B} \rightarrow \Dstar\ (\D, \Lc) X_\mathrm{had}\;\mathrm{anything}}}
\newcommand{\bmixdpimux}{\ensuremath{\overline{B} \rightarrow \Dstar\ (\D, \Lc) X_\mathrm{had}\;l\;\overline{\nu}_l\;\mathrm{anything}}}

\begin{document}

\frontmatter

\School{University of Pennsylvania}
\GSchool{School of Arts and Sciences}
\Title{FIRST MEASUREMENT OF THE RATIO OF BRANCHING FRACTIONS ${\cal B}(\lbsemi)/{\cal B}(\lbhad)$ AT CDF II}  
\Author{Shin-Shan Yu}
\Year{2005}
\Program{Physics and Astronomy}
\Advisor{Nigel S. Lockyer}
\GradChair{Randall D. Kamien}

\signaturepage

\copyrightpage

\clearpage
\vspace*{\fill}
\centerline{\textit{To my grandfather}}
\vspace*{\fill}


\clearpage
\centerline{\textbf{Acknowledgments}}
I would like to thank many people who taught me a tremendous amount 
and who helped me get through my time in graduate school. My thesis adviser, 
Prof.~Nigel~Lockyer, had provided me a full, solid training as a particle 
physicist. The experience of working on the electronics, the detector, and the 
particle identification tool, in addition to the physics analysis, is rare. 
I will always benefit from this in my following career life. 
Prof. Wei-Shu~Hou at 
National Taiwan University motivated me to pursue research at High Energy 
Physics. Rick~Tesarek, being my 
supervisor and mentor at Fermilab, taught me a lot about how to think 
critically about making measurements and at the same time not to dwell on the 
unimportant details. Joel~Heinrich, though far away from Fermilab, was always 
supporting me at Penn. I could phone him any time to ask physics or statistics 
questions. I also gained interesting knowledge outside of physics from him. 
I would like to thank Dmitri~Litvintsev, who helped us performing the 
measurement of the backgrounds and speeded up the progress of this analysis. 

In the first two years of my graduate school, I learned a lot about 
the electronics from Mitch~Newcomer, Rick~Van~Berg, Godwin~Mayers and 
Chuck~Alexander. I remembered the time asking Mitch questions late on 
Friday nights and the time working in Godwin's lab with some homey chatting. 
Walter Kononenko helped me building the ASDQ test station and handled many 
details I tended to forget. After I moved to Fermilab, I spent months 
working on the detector with Dave~Ambrose and Peter~Wittich. We had 
fun going up and down in the lifts, fixing a broken fuse, torn cables and 
many other things. I obtained resources from Dave's library and solved many 
puzzles. Peter always listened to my ranting and provided me some insight as 
a former Penn graduate student. Matthew~Jones and Rolf~Oldeman taught me 
a lot about B physics and asked important questions. 
The former and current Penn CDF graduate students, Chunhui~Chen, Tianjie~Gao, 
Denys~Usynin, Andrew~Kovalev, and Kristian~Hahn, were good companions in the 
CDF trailers and DRL. I also would like to thank Prof.~Joel~Kroll and 
Prof.~Brig~Williams. They had kindly guided me when Nigel was very busy with 
the co-spokesperson responsibility. 
 
I had a nice time taking shifts with the COT group: Aseet~Mukherjee, 
Bob~Wagner, Ken~Schultz, Kevin~Burkett, Robyn~Madrak, Ayana~Holloway, Carter~Hall, Mike~Kirby, JC~Yun, Young-Kee~Kim and Avi~Yagil. Especially, 
Aseet let me understand the physics of drift chamber through many 
interesting debates and discussions. I also enjoyed the time outside of 
IB4 or CDF assembly hall with Robyn, Ayana, Carter and Kirby. 
Barry~Wicklund and Prof.~Marjorie~Shapiro had watched very closely 
my work on $dE/dx$ and shared their rich experience in run I. Bill~Orejudos 
helped writing the software for people to access $dE/dx$ easily. 
 Stefano~Giagu, Mauro~Donega and Diego~Tonelli performed the track-based 
$dE/dx$ calibration and made $dE/dx$ really usable. 
Mat~Martin and Petar~Maksimovi\'{c} patiently answered my questions about the 
fit to the \lbhad\ mode. Guillelmo~G\'omez-Ceballos, Masa~Tanaka, and 
Saverio~D\'{A}uria received my unexpected visits in the office often and 
helped me solving various technical problems. 

My best friend in the US, Keng-hui~Lin, carried a strong, optimistic attitude 
toward the life and that kept me going during the lowest point of my graduate 
student life. I wish I could do the same thing for her. 
I also would like to thank J\'onatan~Piedra and Alberto~Belloni for enlarging my life. 
Most of all, finally, my family in Taiwan, father Shui-Beih~Yu, 
mother Li-Yu~Hsu, and sister Shin-Yun~Yu, always cares about my health and is 
there for me and with me.



\clearpage
\begin{center}
  \textbf{ABSTRACT}

   FIRST MEASUREMENT OF THE RATIO OF BRANCHING FRACTIONS ${\cal B}(\lbsemi)/{\cal B}(\lbhad)$ AT CDF II 
   
   Shin-Shan Yu

   Nigel Lockyer

\end{center}
We present the first measurement of the ratio of 
branching fractions ${\cal B}(\lbsemi)/{\cal B}(\lbhad)$ based on 
171.5~\pbarn\ of $p\overline{p}$ collisions at $\sqrt{s} = 1.96 \rm~TeV$ taken 
with the CDF-II detector. In addition, we present measurements of 
${\cal B}(\dstarsemi)/{\cal B}(\dstarhad)$ and 
${\cal B}(\dsemi)/{\cal B}(\dhad)$, which serve as control samples to
 understand the data and Monte Carlo used for the \Lb\ analysis.
We find the relative branching fractions of the control samples to be:
\[ 
\frac{{\cal B}(\dstarsemi)}{{\cal B}(\dstarhad)} = 
	\rdstarc\ \pm \rdstare\ (stat) \pm \rdstarsyste\ (syst)
	         \pm \rdstarbre\ (BR) \pm \rdstarubre\ (UBR),
\]
and 
\[ 
\frac{{\cal B}(\dsemi)}{{\cal B}(\dhad)} = 
	\rdc\ \pm \rde\ (stat) \pm \rdsyste\ (syst)
	         \pm \rdbre\ (BR) \pm \rdubre\ (UBR),
\]
 which are consistent with the ratios obtained by the Particle Data Group 
at the 0.7 and 1.1 $\sigma$ level, respectively. Finally, we obtain the 
relative \Lb\ branching fraction to be:
\[ 
\frac{{\cal B}(\lbsemi)}{{\cal B}(\lbhad)} = 
	\rlbc\ \pm \rlbe\ (stat) \pm \rlbsyste\ (syst)
	         \rlbbre\ (BR) \pm \rlbubre\ (UBR).
\]
The uncertainties of the three relative branching fractions are from 
statistics, CDF internal systematics, external measured branching ratios and 
unmeasured branching ratios, respectively. We present 
a method to derive \({\cal B}(\lbhad)\) using previous CDF measurements and 
obtain
\[ 
{\cal B}(\lbhad) = \left(\brlbhadc \pm \brlbhade\ (stat \oplus syst) 
 \brlbhadpte\ ({\it \pt}) \right)\%,
\]
where the last uncertainty is due to the measured \Lb\ \pt\ spectrum. 
Combining \({\cal B}(\lbhad)\) with our result, 
we determine the exclusive semileptonic branching 
fraction for the \Lb;
\[
  {\cal B}(\lbsemi)
   =  \left( \brlbsemidatac\ \pm \brlbsemidatae\ (stat)
	\brlbsemidatasyste\ (syst) 
        \pm \brlbsemidatabre\ ({\cal B}(\lbhad)) \right) \%.
\]


\tableofcontents
\listoffigures
\listoftables
  


\mainmatter

\chapter{Introduction}
\label{ch:intro}
  In this dissertation, we measure the properties of the lowest-mass beauty 
baryon, \Lb. Baryons are the bound states of three quarks. Protons and 
neutrons, constituents of atomic nuclei, are the most common baryons. Other 
types of baryons can be produced and studied in the high-energy collider 
environment. 
Three-body dynamics makes baryons composed of low mass quarks difficult to 
study. On the other hand, baryons with one heavy quark simplify the 
theoretical treatment of baryon structure, since the heavy quark can be 
treated the same way as the nucleus in the atom. The \Lb\ is composed of $u$, 
$d$, and $b$ quarks, where the $b$ quark is much heavier than the other 
two. Although, it is accessible, little is known about \Lb. 
In 1991, UA1~\cite{Albajar:1991sq} reconstructed $9\pm 1$ 
\(\Lb\rightarrow J/\Psi \Lambda\) candidates. 
In 1996, ALEPH and DELPHI reconstructed the decay 
\lbhad\ and found only 3-4 candidates~\cite{Abreu:1996mi,Buskulic:1996eq}. 
ALEPH measured a \Lb\ mass of $5614 \pm 21$ \mevcsq, while DELPHI measured 
$5668 \pm 18$ \mevcsq, about 2~$\sigma$ higher. 
Subsequently, CDF-I observed 20 \(\Lb\rightarrow J/\Psi \Lambda\) events~\cite{Abe:1996tr}, confirmed the existence of \Lb\ unambiguously and made a more 
precise measurement of \Lb\ mass, $5621\pm 5$ \mevcsq. 
A recent CDF-II measurement by Korn~\cite{cdfnote:6963} yields $5619.7 \pm 1.7$ \mevcsq, which will significantly improve the current world average, 
 5624$\pm$9 \mevcsq, and resolve the discrepancy of ALEPH and DELPHI.

Several experiments have also measured the product 
of a fragmentation fraction and a branching ratio, such as:  
\(f_{\Lb}{\cal B}(\Lb\rightarrow J/\Psi \Lambda)\)~\cite{Abe:1996tr}, and 
\(f_{\Lb}{\cal B}(\Lb\rightarrow \Lc\mu^-X)\)
~\cite{Abreu:1995me,Barate:1997if}. 
However, branching ratios derived from measurements rely on the knowledge of 
the \Lb\ fragmentation fraction ($f_{\Lb}$), which is defined as the 
probability for a $b$ quark to hadronize into \Lb. Assuming that the \Lb\ 
dominates the production of the beauty baryons, i.e. 
\(f_{\Lb}\cong f_\mathrm{baryon}\), and 
applying the world average \(f_\mathrm{baryon}\) compiled by the Particle Data 
Group (PDG)~\cite{pdg:2004}, we obtain the branching ratios: 
\[
 {\cal B}(\Lb\rightarrow J/\Psi \Lambda)=(4.7\pm 2.8)\times 10^{-4},
\]
and
\[
 {\cal B}(\Lb\rightarrow \Lc\mu^-X)=(9.2\pm 2.1)\times 10^{-2}.
\]
The uncertainties on the above branching ratios are large, about 60$\%$ and 
$20\%$. Several other decays, have been searched for, but either only 1 
candidate was observed (eg: $\Lb\rightarrow \Lc a_1(1260)^-$) or no candidates 
were found (eg: \(\Lb\rightarrow pK^-\)) and an upper limit was set. 
In addition to the mass and branching ratios, the \Lb\ lifetime is an 
important physics quantity. However, the world average lifetime ratio, 
\(\tau(\Lb)/\tau(\Bd)\), is \(0.797\pm 0.052\), in disagreement with the range 
of theoretical predictions: between 0.9 and 1.0~\cite{Anikeev:2001rk}. 
The properties listed above are all that is known about \Lb\ and its decays, 
which motivates us to measure the branching ratios of \Lb. 

\begin{figure}[tbp]
\centering
\includegraphics*[width=150pt,angle=0]
	{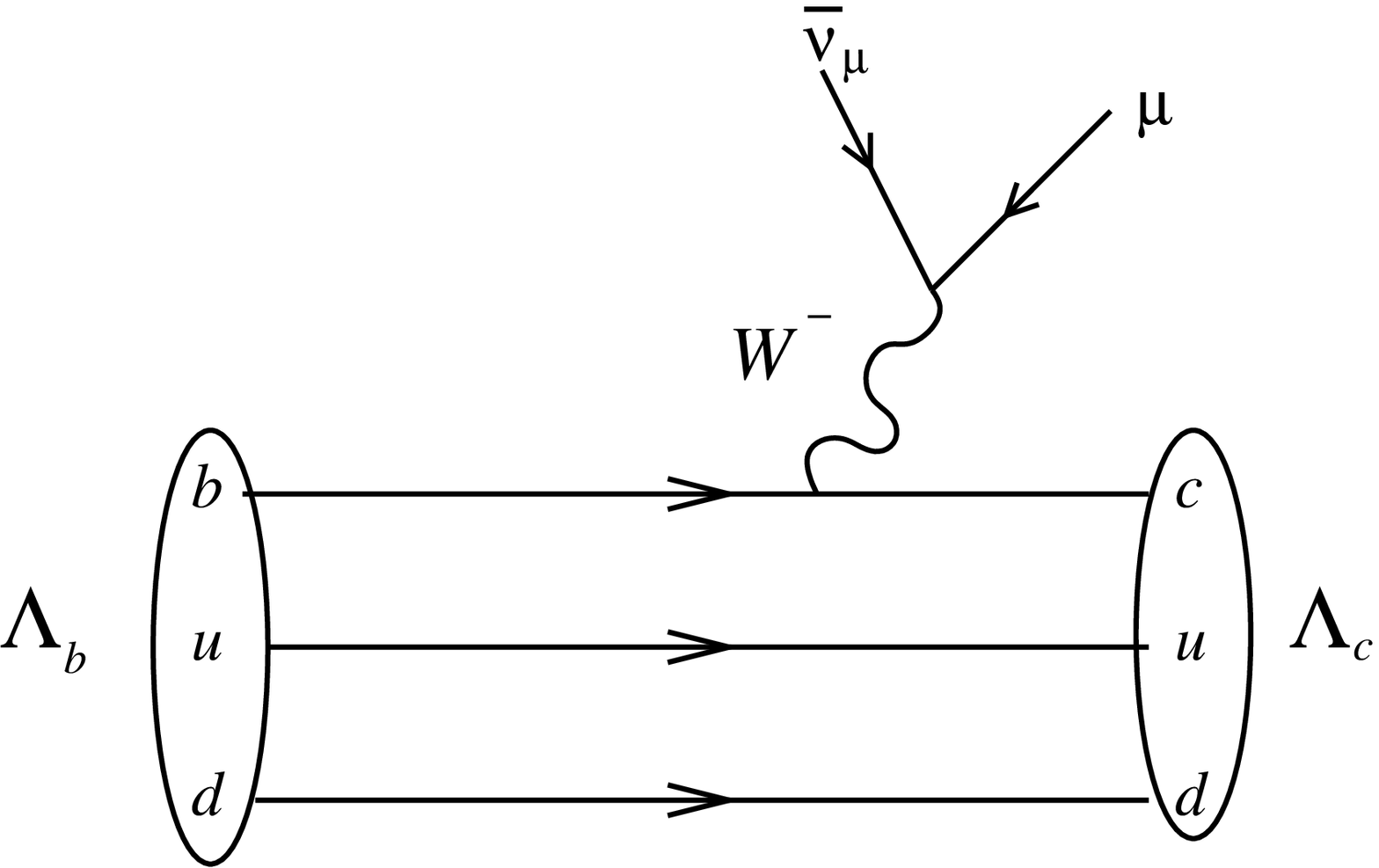}
\includegraphics*[width=150pt,angle=0]{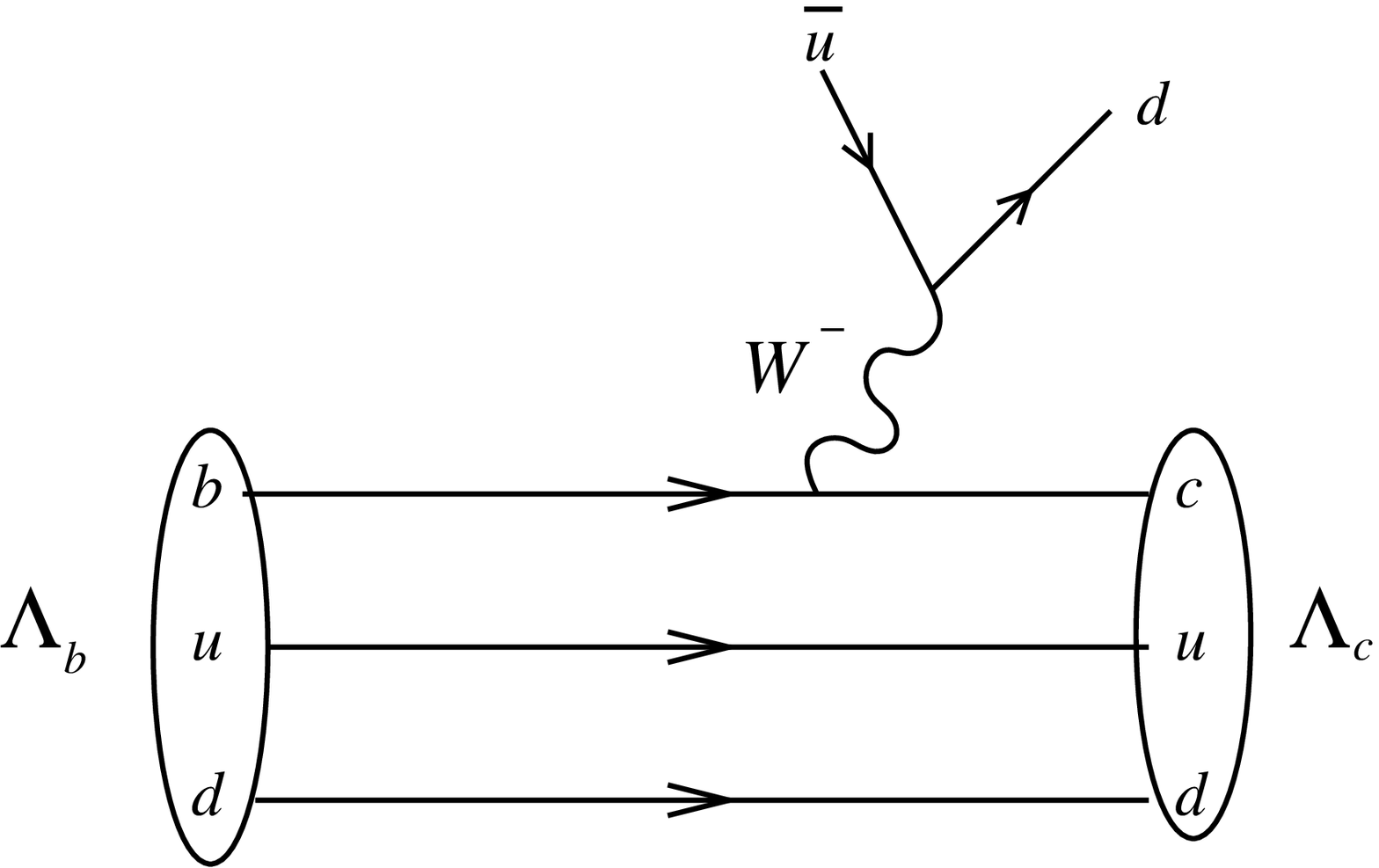}
\caption{Feynman diagram of \lbsemi\ (left) and \lbhad\ (right) decays.}
\label{fig:feynman}
\end{figure}

Currently, the Fermilab Tevatron is the only facility which produces a large 
sample of \Lb. The Heavy Quark Effective Theory (HQET) has been used to predict
 the mass, lifetime, and decay rates of the \Lb. Studying \Lb\ at the 
Tevatron through various measurements allows us to test HQET in different 
aspects. We present a measurement of the relative branching fractions of 
\lbsemi\ to \lbhad. Figure~\ref{fig:feynman} shows that these two decays have 
very similar Feynman diagrams: a $b$ quark decays into a $c$ quark via a 
virtual $W$ boson exchange, and $W$ decay into a muon and an anti-neutrino or 
an $\overline{u}$ and a $d$ quark. The advantage of measuring the ratio of 
branching fractions is that several systematic uncertainties cancel, such as 
those from the trigger and reconstruction efficiencies. To understand the 
$\Lambda_b$ measurement, we perform a similar analysis on the better 
understood $B_d$ decays: we measure ${\cal B}(\dstarsemi)/{\cal B}(\dstarhad)$ 
and ${\cal B}(\dsemi)/{\cal B}(\dhad)$. 

The analysis strategy is as follows: The number of signal events observed in 
the data ($N$) is the product of the number of produced $B$ hadrons ($N_B$), 
the branching ratio (${\cal B}$), the detector acceptance and 
reconstruction efficiency ($\epsilon$) obtained from a Monte Carlo (MC) 
program, i.e. \(N = N_{B}\cdot{\cal B}\cdot\epsilon\). Therefore, the 
\Lb\ (\Bd) relative branching fraction is the yield ratio divided by the 
efficiency ratio since the numbers of $B$ hadrons cancel. For the 
semileptonic mode, 
several backgrounds exhibit a similar signature to the real signal. We 
estimate the amount of these backgrounds ($N_\mathrm{bg}$) and subtract them 
from the observed yield in the data. The formula for extracting the relative 
branching ratio is then expressed as:
\[
 \frac{{\cal B}_\mathrm{semi}}{{\cal B}_\mathrm{had}} = 
(\frac{N_\mathrm{inclusive\;semi}-N_\mathrm{bg}}{N_\mathrm{had}})\times 
 \frac{\epsilon_\mathrm{had}}{\epsilon_\mathrm{semi}}. 
\]
In this dissertation, we measure the following relative branching fractions:
\begin{eqnarray*}
& {\cal B}(\dstarsemi)/{\cal B}(\dstarhad), & \mathrm{where}
	\; \seqdstar, \seqdzero, \\
& {\cal B}(\dsemi)/{\cal B}(\dhad), & \mathrm{where}\; \seqd, \\
& {\cal B}(\lbsemi)/{\cal B}(\lbhad), & \mathrm{where}\; \seqlc.
\end{eqnarray*}
In Chapter~\ref{ch:theory}, a brief overview of the Standard Model and HQET is
 presented. A general description of the Tevatron, the CDF-II detector and 
trigger is found in Chapter~\ref{ch:cdf}. Chapter~\ref{ch:rec} details our 
data sample and event selection. Chapter~\ref{ch:yield} focuses on how the 
signal yields are extracted. The MC simulations for the acceptance and 
efficiency are described and compared with data in Chapter~\ref{ch:mc}. 
Chapter~\ref{ch:bg} involves the estimate of the backgrounds present in the 
semileptonic signal. In Chapter~\ref{ch:result}, the systematic uncertainties 
are first discussed, then the results of the relative branching fractions are 
summarized. Through out the whole dissertation, the analyses of the \Lb\ and 
the \Bd\ control modes are presented in parallel. The charge conjugates of the 
\Lb\ and \Bd\ decays are also included in the reconstruction.  

\chapter{Theoretical Background}
\label{ch:theory}
 This chapter first gives a brief overview of the fundamental particles and 
interactions in Section~\ref{sec-fund}. Then a general idea of the 
Cabibbo-Kobayashi-Maskawa (CKM) matrix and the Heavy Quark Effective Theory 
(HQET) are discussed in Sections~\ref{sec-ckm}--\ref{sec-heavyq}. 

\section{Fundamental Particles and Interactions}
\label{sec-fund}
The ``Standard Model''~\cite{Weinberg:1995mt,Peskin:1995ev} is the 
accepted theory that describes particles with no internal structure and 
their interactions with matter. In the Standard Model, the basic constituents 
of matter are six flavors of spin-1/2 quarks: the down-type quarks 
($d$, $s$, $b$) and the up-type quarks ($u$, $c$, $t$), and six kinds of 
spin-1/2 leptons: the charged leptons, $e$, $\mu$ and $\tau$, and the 
neutrinos $\nu_e$, $\nu_{\mu}$, and $\nu_{\tau}$. 
The magnitude of the electron's electric charge is denoted as ``$e$''. 
The down-type quarks carry electric charge $-\frac{1}{3}e$ and the up-type 
quarks carry electric charge $+\frac{2}{3}e$. The charged leptons have 
electric charge -1$e$ while the neutrinos have zero electric charge. 
The masses of the quarks and leptons exhibit a hierarchy. 
The $u$, $d$ and $s$ quarks are much lighter ($< 200 $ \mevcsq) 
than the $c$, $b$ and $t$ quarks ($\sim$ 1100, 4500 and 175000 \mevcsq). 
This hierarchy is not understood. 
The three charged leptons also have progressively increasing masses: 
$\sim$ 0.51($e$), 106($\mu$), 1777($\tau$) \mevcsq. For the neutrinos, 
currently, only upper limits exist for their masses. 
Separate neutrino types can undergo transitions into one another if at most 
one type of neutrino has zero mass. 
The current best estimates require three-flavor mixing to explain the full 
range of results from the solar neutrino~\cite{Robertson:2005mt,wittich:thesis}, and the atmospheric neutrino experiments
~\cite{Ashie:2005ik,Litchfield:2005np,Ambrosio:2004ig}. 

The Standard Model describes the following three types of interaction 
among quarks and leptons: electromagnetic, weak and strong interactions. 
The gravitational interaction is not described by the Standard Model. The 
gravitational force dominates in the large mass scale, such 
as a galaxy, but has little influence on the scale of quarks and leptons. 
Therefore, it is usually ignored in the fundamental particle interactions.  
The quarks and leptons interact via the exchange of the gauge bosons. The 
Lagrangian of each interaction is invariant under a transformation that 
corresponds to a symmetry group.  The Standard Model is a theory based on the 
symmetry group \(SU(3)\times SU(2) \times U(1)\). Both \(SU(n)\) and \(U(n)\)
 groups are Lie groups, i.e. any element in the group can be represented 
by $m$ fundamental elements or generators~\cite{Kane:1987gb}:
\begin{equation}
\label{eq:lie}
 E = \exp(\sum^m_i \theta_i F_i),
\end{equation}
where $F_i$ is the $i^\mathrm{th}$ generator and $\theta_i$ is the 
``rotation'' angle corresponding to each generator.
Elements of the \(SU(n)\) groups are represented by $n\times n$ unitary 
matrices, $U^{\dag}U=1$, with det $U$ = +1 and have $n^2-1$ generators. 
The theory introduces $n^2-1$ gauge bosons, analogous to the rotation angle in Equation~\ref{eq:lie}. They form a scalar product with the $n^2-1$ generators 
and make the Lagrangian invariant. The \(U(1)\) group is a one dimensional 
unitary group with single generator, where the elements are specified by a 
continuous parameter, $\theta$, and expressed as $e^{i\theta}$. 

The \(U(1)\) group describes the 
electromagnetic interaction among quarks and the charged leptons, via the 
exchange of a massless spin-1 photon. The electromagnetic interaction binds 
the electrons and atomic nuclei together and forms atoms.
The \(SU(2)\) group describes the weak interaction experienced by all the 
fundamental particles, where the gauge bosons are massive spin-1 $W^{\pm}$ and
 $Z^0$. The masses of $W^{\pm}$ and $Z^0$ are about 80 and 91 \gevcsq. A well 
known weak interaction process is the neutron $\beta$-decay: 
$n(udd)\rightarrow p(uud) e\overline{\nu}_e$. The right- and left-handed 
fundamental particles transform differently under $SU(2)$. The right-handed 
quarks and leptons do not couple to $W^{\pm}$ and are singlets under $SU(2)$. 
While the left-handed quarks and leptons are doublets under $SU(2)$ and 
classified into three generations: 
\begin{eqnarray}
L_L & = & 
\left(
\begin{array}[c]{ccc}
 \left(
 \begin{array}[c]{c}
  \nu_e \\
   e 
 \end{array}
  \right)_L,
 &
 \left(
 \begin{array}[c]{c}
  \nu_\mu \\
   \mu
 \end{array}
  \right)_L,
 &
 \left(
 \begin{array}[c]{c}
  \nu_\tau \\
   \tau
 \end{array}
  \right)_L
\end{array}
\right), \nonumber \\\nonumber \\
Q_L & = & 
\left(
\begin{array}[c]{ccc}
 \left(
 \begin{array}[c]{c}
   u \\
   d 
 \end{array}
  \right)_L,
 &
 \left(
 \begin{array}[c]{c}
   c \\
   s
 \end{array}
  \right)_L,
 &
 \left(
 \begin{array}[c]{c}
   t \\
   b
 \end{array}
  \right)_L
\end{array}
\right).
\label{eq:lefthanded}
\end{eqnarray}
The weak interaction allows transitions between quarks of different flavors.
The transitions within the same generation are more favored than those 
across the generations and the coupling strength is given by the CKM matrix 
(see Section~\ref{sec-ckm}). The coupling strength is the same for all 
leptons.

The \(SU(3)\) group describes the strong color interaction among quarks, 
mediated via the exchange of eight massless spin-1 gluons. The quarks carry 
three possible ``chromoelectric charges'': red, green and blue ($R,G,B$), 
which are analogous to the electric charge in the electromagnetic interaction. 
The eight gluons are associated with the color 
combinations:
\begin{center}
  $R\overline{B}, R\overline{G}, B\overline{R}, B\overline{G}, G\overline{R},
  G\overline{B}, (R\overline{R}-G\overline{G})/\sqrt{2}$ and 
  $(R\overline{R}+G\overline{G}-2B\overline{B})/\sqrt{6}$
\end{center}
The strong interaction binds the quarks together to form a colorless state, 
$q\overline{q}$ or $qqq$ ($R\overline{R}$ or $RGB$). The $q\overline{q}$ bound
 state is referred to as ``meson'' and the $qqq$ bound state is referred to as
 ``baryon''. For example, a $\overline{b}d$ bound state is a \Bd\ meson and a 
$udb$ bound state is a beauty baryon, \Lb. Both mesons and baryons are called 
``hadrons''.  
Just as the residual electric field outside of the neutral atoms causes them 
to combine into molecules, the residual color field outside of the protons and
 neutrons forms nuclei. 
 
Each fundamental particle has an associated antiparticle, i.e. 
of which the electric charge, color charge and flavor are reversed, but the 
mass and the spin are the same.  
 In addition, the Standard Model introduces a neutral spin-0 Higgs boson, 
$H^0$, to accommodate the masses of the gauge bosons, quarks and leptons. 
The Higgs boson has not been discovered, yet. The search for the Higgs boson 
remains an important goal of several running and future high energy 
experiments. 

\section{CKM Matrix}
\label{sec-ckm}
The strong interaction conserves the flavor of quarks, and only transitions of 
the same flavor quark will take place (eg: charmness conserved decay, 
$\Dstar\rightarrow \D\pi^0$), while the flavor-changing decays are allowed in 
the electroweak interaction (eg: beauty to charm decay, \dstarhad). The 
Cabibbo - Kobayashi - Maskawa (CKM) matrix
~\cite{Cabibbo:1963yz,Kobayashi:1973fv} in Equation~\ref{eq:ckm} describes the 
coupling in the weak interaction between different flavors of quarks. For 
instance, $V_{cb}$ describes the electroweak coupling strength of the $b$ 
quarks to the $c$ quarks. The CKM matrix represents a unitary transformation 
from the flavor (mass) eigenstates to the weak interaction eigenstates. 
\begin{equation}
V_\mathrm{CKM} \equiv
\left(
\begin{array}[c]{ccc}
V_{ud} & V_{us} & V_{ub} \\
V_{cd} & V_{cs} & V_{cb} \\
V_{td} & V_{ts} & V_{tb} 
\end{array}
\right).
\label{eq:ckm}
\end{equation}
A standard CKM matrix parametrization proposed by Chau,\etal~\cite{Chau:1984fp,
Harari:1986xf,Fritzsch:1986gv,Botella:1985gb}, which is similar to 
Kobayashi - Maskawa's original parametrization~\cite{Kobayashi:1973fv},
has four free parameters: three mixing angles between any two generations, 
$\theta_{12}$, $\theta_{23}$, $\theta_{13}$  and a phase, \(\delta_{13}\);
\begin{eqnarray}
V_\mathrm{CKM} = &
\left(
\begin{array}[c]{ccc}
c_{12}c_{13} & s_{12}c_{13} & s_{13}e^{-i\delta_{13}} \\
-s_{12}c_{23}-c_{12}s_{23}s_{13}e^{i\delta_{13}} 
& c_{12}c_{23}-s_{12}s_{23}s_{13}e^{i\delta_{13}} 
& s_{23}c_{13} \\
s_{12}s_{23}-c_{12}c_{23}s_{13}e^{i\delta_{13}} & 
-c_{12}s_{23}-s_{12}c_{23}s_{13}e^{i\delta_{13}} 
& c_{23}c_{13}
\end{array}
\right) 
\label{eq:ckmpar}\\
\nonumber \\
 \cong &
\left(
\begin{array}[c]{ccc}
c_{12} & s_{12} & s_{13}e^{-i\delta_{13}} \\
-s_{12}c_{23}-c_{12}s_{23}s_{13}e^{i\delta_{13}} 
& c_{12}c_{23}-s_{12}s_{23}s_{13}e^{i\delta_{13}} 
& s_{23} \\
s_{12}s_{23}-c_{12}c_{23}s_{13}e^{i\delta_{13}} & 
-c_{12}s_{23}-s_{12}c_{23}s_{13}e^{i\delta_{13}} 
& c_{23}
\end{array}
\right) 
\label{eq:ckmparapp}
\end{eqnarray}
with \(c_{ij}= \cos\theta_{ij}\) and \(s_{ij} = \sin\theta_{ij}\) where 
$i$, $j$ labels the generations. The matrix elements with simpler forms 
appearing in the first row and third column, have been measured directly in 
decay processes. $c_{13}$ is known to be very close to unity: 
$|c_{13}-1| \sim 10^{-6}$, and this gives an approximation in 
Equation~\ref{eq:ckmparapp}.
 In the Standard Model, the complex phase, $\delta_{13}$, is the origin of the 
Charge-Parity (CP) violation in the weak interaction. We refer 
the reader to the Tevatron Run II B Physics Workshop Report
~\cite{Anikeev:2001rk} for a more detailed description of the weak 
CP violation mechanisms.

Using the world average experimental results of the weak decays as the input, 
and assuming that only three generations exist, with the unitarity, a 
90$\%$ confidence limit can be placed on the amplitude of the matrix elements 
(eg: $|V_{ud}|$);
\begin{equation}
\left(
\begin{array}[c]{ccc}
0.9739\sim 0.9751 & 0.221\sim0.227 & 0.0029\sim0.0045 \\
0.221\sim0.227 & 0.9730\sim0.9744 & 0.039\sim0.044 \\
0.0048\sim0.014 & 0.037\sim0.043 & 0.9990\sim0.9992 
\end{array}
\right)
\label{eqn:ckmvalue}
\end{equation}
As seen in Equation~\ref{eqn:ckmvalue}, quark transitions within the same 
generation are favored over the transitions across generations. 
The latter are called ``Cabibbo suppressed'' decays. 
The goal of several analyses at CDF II, together 
with the experiments {\tt BELLE}~\cite{belle:tdr}, 
{\tt BABAR}~\cite{babar:tdr} and {\tt KTeV}~\cite{Arisaka:1992ch}, etc.
, is to make precise measurements of many matrix 
elements. For example, a measurement of the $\Bd-\overline{B}^0$ oscillation 
frequency can infer the CKM matrix element $V_{td}$. While $V_{ud}$ can 
be obtained by comparing the nuclear $\beta$-decay 
$n\rightarrow pe\overline{\nu}_e$  to muon decay and $V_{us}$ 
can be measured from the decay: $K\rightarrow \pi e \overline{\nu}_e$. By 
making measurements of all the CKM elements, we can determine whether the CKM 
matrix is unitary. If the matrix is not unitary, this would be a signature for 
additional physics not currently described by the Standard Model.

\section{Heavy Quark Effective Theory}
\label{sec-heavyq}
This dissertation presents a measurement of the \Lb\ relative decay rates.
 The transition amplitude (${\cal M}$) that 
describes the decay rate of a \B\ hadron into some final state $f$, can 
be derived by drawing all the possible Feynman diagrams at the quark level 
and summing up all the contributions. The underlying weak interaction is 
simple but the strong interaction that binds the quarks 
into hadrons introduces complications. When the quarks or gluons travel 
over a distance of $1/\lqcd$ or longer, the coupling constant of the strong 
interaction ($\alpha_s$) diverges, so perturbation theory breaks down and 
the nonperturbative effect takes over. For the energy scale of our concern, 
\lqcd\ is around 200 \mevcsq. 

One theoretical tool, the Operator Product Expansion (OPE)
~\cite{Wilson:1969zs}, separates the perturbative from nonperturbative physics
 and ${\cal M}$ becomes: 
\begin{equation}
{\cal M} = -\frac{4 G_F}{\sqrt{2}}V_\mathrm{CKM}\sum_j C_j\bra f|O_j|B\ket
	\left[1+ {\cal O}(\frac{m_b^2}{M_W^2})\right],
\end{equation}
where $j$ indicates the contribution from the $j^\mathrm{th}$ Feynman 
diagram, $G_F$ is the Fermi coupling constant, $V_\mathrm{CKM}$ is the 
CKM matrix element in Equation~\ref{eq:ckm}. The Wilson coefficients
~\cite{Wilson:1969zs}, $C_j$, act as effective coupling 
constants and contain the physics at short distance. The Wilson coefficients 
can be calculated using perturbation theory and are model independent. 
The $\bra f|O_j|B\ket$ is usually referred to as the hadronic matrix 
element, where $O_j$ is a local operator. The hadronic matrix 
elements contain the long distance physics and can only be evaluated using 
nonperturbative methods. Contributions from the higher order operators 
are suppressed by a power of $m_b^2/M_W^2$, where $m_b$ and $M_W$ 
are the masses of the $b$ quark and the $W$ boson.
The Heavy Quark Effective Theory (HQET) significantly
 simplifies the form of the hadronic matrix element. This section gives a 
review of the HQET and shows how the \Lb\ decay rates may be derived using the
 HQET with other theoretical assumptions. 

The HQET stems from the Standard Model and describes the hadrons containing a 
$b$ or $c$ quark. The concept of ``heavy'' is relative. 
In the HQET, the masses of the ``heavy'' $c$, $b$ and $t$ quarks are much 
larger than QCD energy scale, while the masses of the ``light'' $u$, $d$ and 
$s$ quarks are much smaller than \lqcd. 
In the limit of $m_{c,b,t}\gg\Lambda_{QCD}$, a new type of symmetry,
 ``spin-flavor heavy quark symmetry'' arises. The momentum transfer between 
the heavy quark and the light quarks in the hadron system is of the order 
of \lqcd. Or equivalently speaking, the typical size of a hadron 
system is of the order of $\Lambda_{QCD}^{-1}$. 
The change in the heavy quark velocity is then $\sim\Lambda_{QCD}/m_Q$, 
which vanishes when $m_Q$ is infinitely large. 
The velocity of the heavy quark is, therefore, almost unaffected by the strong 
interaction, i.e. the quark-quark interaction terms disappear in the 
Lagrangian. The only strong interaction of a static heavy quark is with gluons 
via choromoelectric charge. This quark-gluon interaction is 
spin-independent. Consequently, the light quark system knows nothing about 
the spin, mass and flavor of the ``nucleus'', i.e. a \B\ hadron at rest 
is identical to a charm hadron at rest regardless of their spin orientations. 

The ``heavy quark symmetry'' implies that we can relate properties of the 
beauty hadrons to those of the charm hadrons. For example, Aglietti~\cite{Aglietti:1991rr} derived a formula to estimate the \Lb\ mass:
$M_{\Lamc} - \frac{1}{4}(M_D + 3M_{D^*})$ = 
$M_{\Lb} - \frac{1}{4}(M_B+ 3M_{B^*})$, which gives $M_{\Lb}\sim$ 5630 \mevcsq,
 in good agreement with the world average, 5624$\pm$9 \mevcsq.  
An analogy can be found in atomic systems, where the isotopes with 
different nuclei have nearly the same chemical properties. 
When performing a calculation of the \B\ or charm hadron mass, decay rate or 
lifetime, we could start from the limit of $m_{c,b,t}\gg\Lambda_{QCD}$. 
Then the correction terms are added in expansion of the power of $1/m_Q$, 
where $m_Q$ is the mass of the heavy quark. The $1/m_Q$ corrections take into 
account finite mass effects and are different for quarks of different masses. 
A more complete description of HQET may be found in 
Manohar\cite{Manohar:2000dt}, Godfrey~\cite{Godfrey:1985xj} and 
Isgur~\cite{Isgur:1989ed}.

The focus of this analysis, examining \Lb\ to \Lamc\ baryon decay, is best 
suited to treatment using HQET since both the initial- and the final- state 
hadrons contain a heavy quark. In addition, the light quark system in a \Lb\ 
baryon is in a spin-0 state; the sub-leading corrections have a simpler form 
than those for the mesons~\cite{Georgi:1990ei}. This analysis concerns 
the relative branching fractions of \lbsemi\ to \lbhad, where the leading 
order Feynman diagrams are shown in Figure~\ref{fig:feynman}.
 Using the Soft Collinear Effective 
Theory (SCET), Leibovich, Ligeti, Stewart, and Wise~\cite{Leibovich:2003tw} 
relate the decay rate of \lbhad\ to \lbsemi. 
The SCET assumes that in the \lbhad\ decay, the pion mass can be neglected 
 and $m_{b}-m_{c}\cong E_{\pi} \gg \lqcd$. Therefore, 
the $\overline{u}d$ quark pair acquires large momentum, remains close 
together and acts as a color dipole (singlet). Within the \Lb, the $b$ quark 
and the diquark ($ud$) form a color dipole since the \Lb\ is colorless;
 the same hold for the $c$ quark and the diquark within the \Lamc. The color 
dipole-dipole interaction is weaker than the color monopole-monopole 
interaction, which 
means the pion interacts weakly with the rest of the system. In the end, the 
\lbhad\ decay factorizes into two subprocesses: the 
hadronization of the \Lamc\ and, completely decoupled, the hadronization of 
the $\pi$. The hadronic matrix element is then expressed as 
\(\bra \Lamc | \overline{c} \gamma^\mu (1 - \gamma^5) b | \Lb \ket \bra \pi |
\overline d \gamma_\mu(1 - \gamma^5)u|0\ket\). 
The first term is common to both hadronic and semileptonic decays and 
can be inferred from the differential decay rate of the semileptonic mode, 
$\frac{d\Gamma(\lbsemi)}{dw}$, at maximal recoil, where $w$ is the scalar 
product of the \Lb\ and \Lamc\ four-velocities, $v$ and $v^{\prime}$:
\begin{eqnarray}
  w & \equiv & v\cdot v^{\prime} \\
     & = & (m_{\Lb}^2+m_{\Lamc}^2 - q^2)/(2m_{\Lb}m_{\Lamc}).
\end{eqnarray}
Here $q^2$ is the four-momentum transfer in the decay. Maximal recoil refers 
to the kinematic configuration when the charged lepton and 
the neutrino momenta are parallel, or equivalently, when $q^2$ is at its 
minimum, $m_l^2$, which is approximately zero. At this configuration, 
semileptonic decay, \lbsemi, has the same kinematics as the hadronic decay, 
\lbhad, since the $q^2$ of the hadronic decay is always $m_{\pi}^2$, which is 
neglected in the SCET, and $q^2\cong 0$. The $\frac{d\Gamma(\lbsemi)}{dw}$ can 
be constructed from six form factors, which are functions of $w$. 
The form factors can be considered as the Fourier 
transformation of the weak charge distribution and describe 
the interaction between the $b$ and the $c$ quarks. 
The second term of the hadronic matrix element is the pion decay constant, 
$f_{\pi}$. The value of $f_{\pi}$ is 131~MeV and was extracted from the decay 
width of $\pi^-\rightarrow \mu^- \overline{\nu}_{\mu}$. 

In a more exact form, the \lbhad\ decay rate is 
\begin{equation}
 \Gamma(\lbhad) = \frac{3\pi^2(C_1+C_2/3)^2|V_{ud}|^2f_{\pi}^2}{m_{\Lb}^2r_{\Lambda}}\left(\frac{d\Gamma(\lbsemi)}{dw}\right)_{w_{max}},
 \label{eq:hadsemi}
\end{equation}
where $C_1$ and $C_2$ are the first two terms of the Wilson coefficients, the 
higher order terms are suppressed, and 
\begin{eqnarray}
r_{\Lambda} & \equiv & m_{\Lamc}/m_{\Lb}. 
\label{eq:rdef}
\end{eqnarray}
The $w_{max}$ corresponds to $q^2=m_{l}^2\simeq m_{\pi}^2\simeq 0$. 
 In the limit of $m_{Q}\gg \lqcd$, the six form factors that describe 
the semileptonic differential decay rate are reduced to one universal 
function, the Isgur-Wise function 
($\zeta(w)$)~\cite{Isgur:1989ed,Isgur:1989vq}, and  
\begin{equation}
\frac{d\Gamma(\lbsemi)}{dw} = \frac{G_F^2 m_{\Lb}^5 |V_{cb}|^2}{24\pi^3}
r_{\Lambda}^3\sqrt{w^2-1}[6w + 6wr_{\Lambda}^2 - 4r_{\Lambda} - 8r_{\Lambda}w^2]\zeta^2(w).
\label{eq:formfactor}
\end{equation}
Note that Equation~\ref{eq:formfactor} proposes an alternative way to measure 
$|V_{cb}|$ using \lbsemi\ decay.

Although HQET reduces the form factors to the Isgur-Wise function, 
it can not predict the functional form of $\zeta(w)$. One functional form 
easy for calculation was suggested by Isgur and Wise~\cite{Isgur:1989ed}:
\begin{equation}
\label{eq:zetaw}
 \zeta(w) = e^{-\rho^2(w-1)}, 
\end{equation} 
where the slope $\rho^2$ has to be calculated using other theoretical 
assumptions. 
Assuming the number of colors ($N_c$) in the baryon is infinitely large, 
Jenkins,~\etal~\cite{Jenkins:1992se} derived \(\rho^2=1.3\). Using the QCD 
sum rules, Huang,~\etal~\cite{Huang:2005me} calculated \(\rho^2=1.35\pm 0.12\).
 A recent DELPHI measurement~\cite{Abdallah:2003gn} gives 
\(\rho^2 = 2.03 \pm 0.46 (stat) {+0.72 \atop -1.00} (syst)\), 
consistent with the numbers from Huang and Jenkins,~\etal. Combining 
Equations~\ref{eq:hadsemi}--\ref{eq:zetaw} and the slope value from Jenkins, 
Leibovich,~\etal\ predict ${\cal B}(\lbhad)$ = 0.45$\%$ and 
${\cal B}(\lbsemi)$ = 6.6$\%$. However, the correction to the large 
$N_c$ limit is of order $1/N_c$, which is 30$\%$ in the case of baryons 
($N_c=3$). The uncertainty from the QCD sum rule is about 10$\%$. 

To summarize, measuring the relative branching fractions of \lbsemi\ to 
\lbhad\ allows us to obtain a ratio free from several experimental systematic 
uncertainties. With the external input of ${\cal B}(\lbhad)$, we could infer 
the ${\cal B}(\lbsemi)$ and vice versa. Finally, the absolute branching ratios 
of \lbhad\ and \lbsemi\ increase our knowledge of the \Lb\ baryon and 
test the validity of Equations~\ref{eq:hadsemi} and \ref{eq:formfactor} which 
are derived from the HQET and the SCET.

\chapter{The CDF-II Detector and Trigger}
\label{ch:cdf}
The Fermilab Tevatron is currently the highest energy accelerator in the 
world. Protons and anti-protons (\ppbar) are accelerated in its~\(\sim\) 6~km
 (4~mile) circumference to be brought into collision with a center of mass 
energy of approximately 1.96~TeV. The collisions take place at the center of 
two detectors: the Collider Detector at Fermilab (CDF-II) and D0-II. These 
two detectors are about 120\degs\ away from each other on the ring as 
indicated in Figure~\ref{fig:acc_complex}. The ``luminosity'' is a measure of 
collision rate normalized by the collision cross section, in unit of 
1/sec$\cdot$cm$^{2}$. The common dimension for the time integrated 
luminosity is ``barn$^{-1}$'', which is 10$^{24}$/cm$^{2}$. During 1992--1995,
 the predecessors of CDF-II and D0-II, CDF-I and D0-I had collected data with a
 time-integrated luminosity of $\sim$110 \pbarn\ (inverse pico-barn) and 
published more than 100 papers. This analysis uses data collected by the 
CDF-II experiment.

Both the accelerator and the collider detectors underwent major upgrades
between 1997 and 2001. The main goals of these upgrades were to increase the 
luminosity of the accelerator, and to collect data samples with an integrated 
luminosity of 2 \fbarn\ (inverse femto-barn) or more. The upgraded Tevatron 
accelerates 36 bunches of $p$ and $\overline{p}$, whereas the previous 
accelerator operated with only 6$\times$6. Consequently, the time between 
collisions (or beam crossings) has decreased from 3.5~$\mu$s to 396~ns for 
the current collider. The new collider configuration required extensive 
detector upgrades at CDF-II to accommodate the shorter bunch spacings.
 In Section~\ref{sec-tevatron}, we give an overview of how the proton and 
anti-proton beams are accelerated to their final center of mass energy of 
1.96 TeV, and collided. We then describe in Sections~\ref{sec-dectector}
--\ref{sec-trigger} the components of the CDF-II detector, and trigger, which 
are used to measure the properties of the particles produced in the \ppbar\ 
collisions.

\section{\ppbar\ acceleration and collisions}
\label{sec-tevatron}
\begin{figure}[tbp]
\centering
\includegraphics[width=300pt,angle=0]{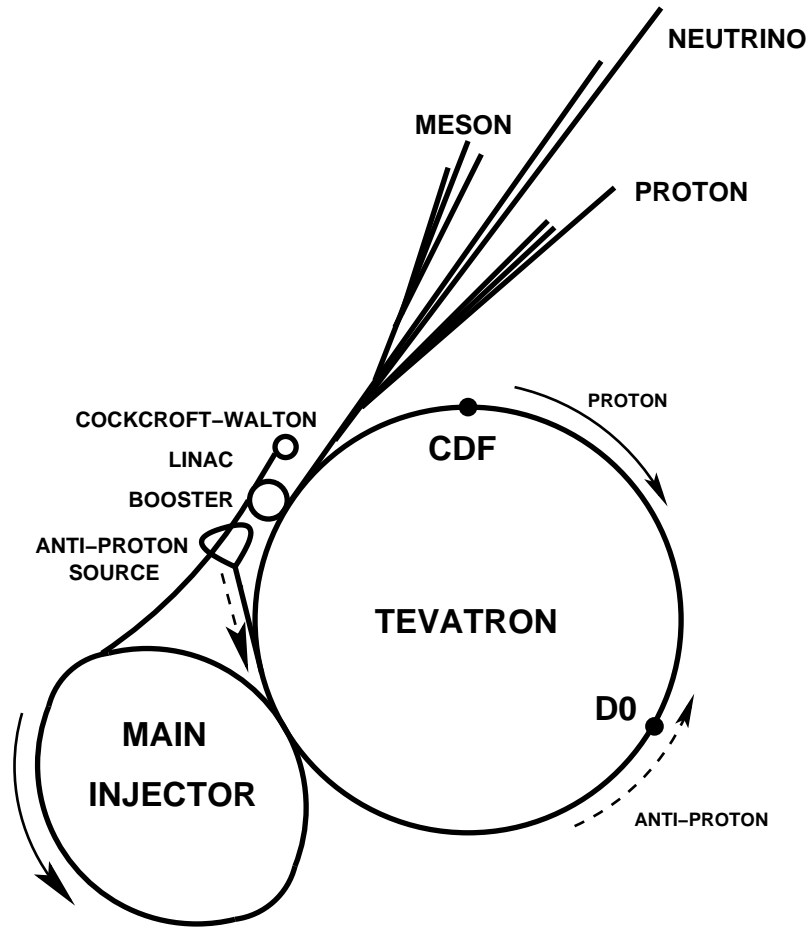}
\caption[Layout of the Fermilab accelerator complex.]
{Layout of the Fermilab accelerator complex. The proton (solid arrow) is 
accelerated at the Cockcroft-Walton, Linac, Booster, Main Injector and 
finally at the Tevatron. The anti-proton (dashed arrow) from the anti-proton 
source is first accelerated at the Main Injector and then at the Tevatron.}
\label{fig:acc_complex}
\end{figure}

In order to create the world's most energetic particle beams, 
Fermilab uses a series of accelerators. The diagram in 
Figure~\ref{fig:acc_complex} shows the paths taken by protons and anti-protons 
from initial acceleration to collision in the Tevatron.
The first stage of acceleration is in the Cockcroft-Walton 
pre-accelerator~\cite{cockcroft:web}
, where H$^{-}$ ions are created from the ionization of 
the hydrogen gas and accelerated to a kinetic energy of 750 keV. 
The H$^-$ ions enter a linear accelerator (Linac)~\cite{Schmidt:1993fz}, 
approximately 500 feet long, where they are accelerated to 400 MeV. The 
acceleration in the Linac is done by a series of ``kicks'' from Radio Frequency
(RF) cavities. The oscillating electric field of the RF cavities groups the 
ions into bunches.
Before entering the next stage, a carbon foil removes the electrons
from the H$^-$ ions at injection, leaving only the protons.
The 400 MeV protons are then injected into the Booster, a 74.5~m-diameter 
circular synchrotron. The protons travel around the Booster about 20,000 
times to a final energy of 8~GeV. 

Protons are then extracted from the Booster into the Main Injector
~\cite{main_injector}, 
where the protons are accelerated from 8~GeV to 150~GeV 
before the injection into the Tevatron. The Main Injector also 
produces 120~GeV protons, where the protons collide with a nickel target, and 
produce a wide spectrum of secondary particles, including anti-protons. 
In the collisions, about 20 anti-protons are produced per one million protons. 
The anti-protons are collected, focused, and then stored in the Accumulator 
ring. Once a sufficient number of anti-protons are produced, they are sent 
to the Main Injector and accelerated to 150~GeV. Finally, both the 
protons and anti-protons are injected into the Tevatron.
The Tevatron, the last stage of Fermilab's accelerator chain, receives 
150~GeV protons and anti-protons from the Main Injector and accelerates
them to 980~GeV. The protons and anti-protons travel around the Tevatron in
opposite directions. The beams are brought to collision at the center of 
the two detectors, CDF-II and D0-II.

\begin{figure}[ht]
\centering
\epsfig{file=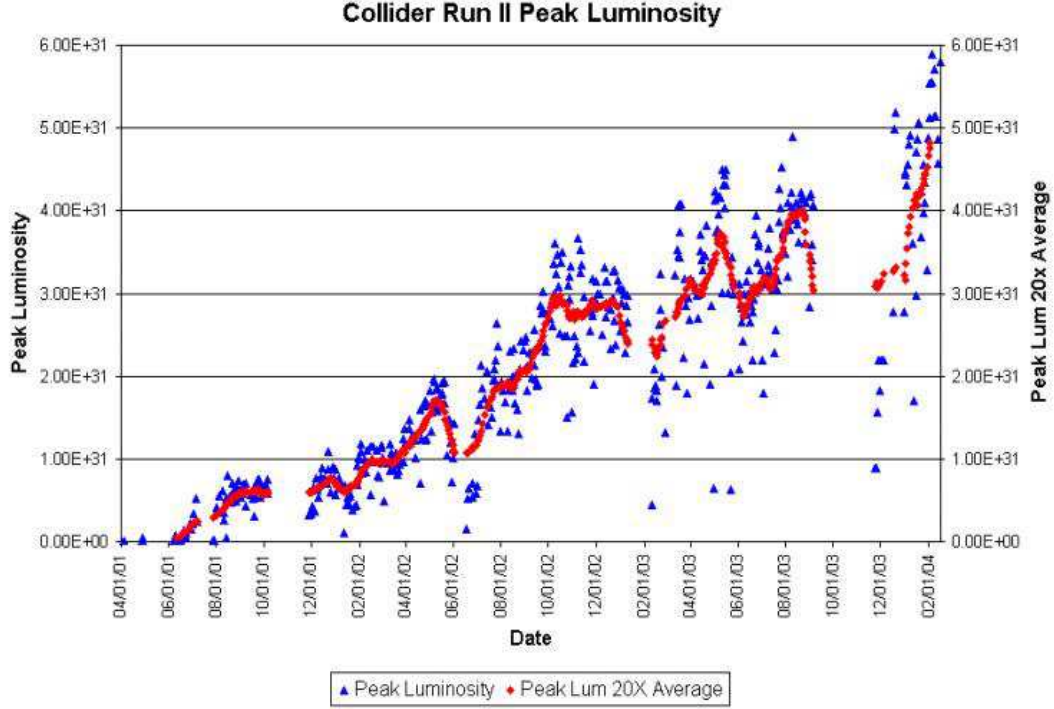, width = 1. \textwidth}
\caption[Peak luminosities for stores collided between April 2001 and February 2004.]{Peak luminosities for stores collided between April 2001 and February 
2004. This analysis uses the data collected from February 2002 to September 2003.}
\label{fig:lumi}
\end{figure}

We use the term ``luminosity'' to quantify the beam particle density and 
the crossing rate. The luminosity in units of cm$^{-2}$s$^{-1}$ 
can be expressed as:
\begin{equation}
{\cal L} = \frac{ f N_B N_{p} N_{\overline{p}} } { 2 \pi ( \sigma_p ^2 +
\sigma_{\overline{p}} ^2 ) } F \left( \frac{ \sigma_l } { \beta ^* } \right)
\end{equation}
where $f$ is the revolution frequency, $N_B$ is the number of bunches,
$N_{p/\overline{p}}$ are the number of protons/anti-protons per bunch, and
$\sigma _{p/\overline{p}}$ are the RMS beam sizes at the interaction
point. $F$ is a form factor which corrects for the bunch shape and depends on
the ratio of $\sigma_l$, the bunch length, to $\beta ^*$, the beta function, at
the interaction point. The beta function is a measure of the beam width, and
is proportional to the beam's $x$ and $y$ extent in phase space.
Figure \ref{fig:lumi} shows the peak luminosities for the stores used 
in this analysis. The collision products are recorded in the CDF-II and D0-II 
detectors. 

\section{The CDF-II Detector}
\label{sec-dectector}
\begin{figure}[tbp]
\centering
 \resizebox{300pt}{!}{\includegraphics*[24pt,24pt][774pt,575pt]
{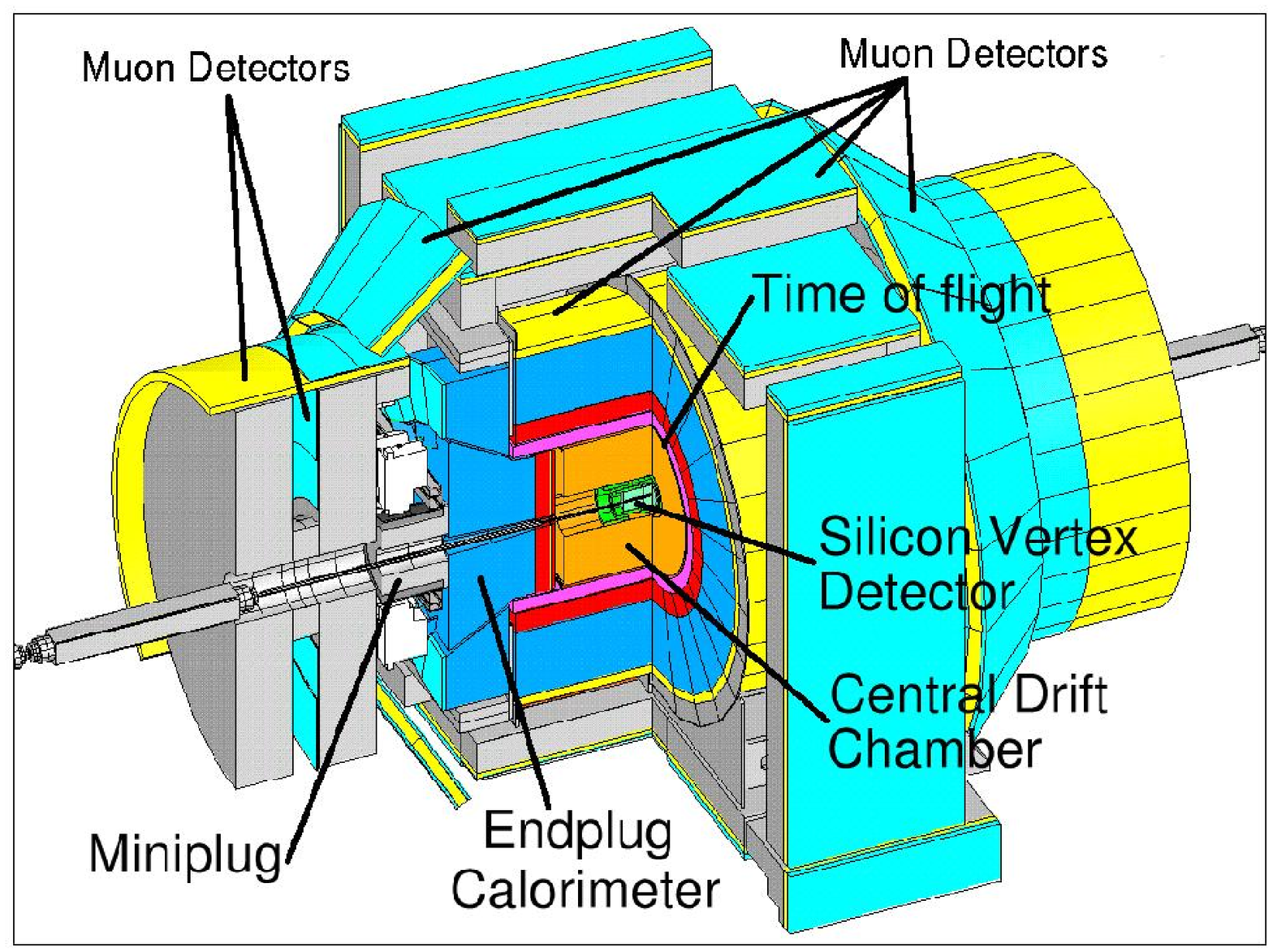}}
\caption[The CDF-II detector with quadrant cut to expose the different 
	sub-detectors.]
	{The CDF-II detector with quadrant cut to expose the different 
	sub-detectors.}
\label{fig:cdf_detector}
\end{figure}
The Collider Detector at Fermilab (CDF) is a general purpose, azimuthally and 
forward-backward symmetric apparatus,  designed to study \ppbar\ collisions at 
the Tevatron. Figure~\ref{fig:cdf_detector} shows the detector and the 
different sub-systems in a solid cutaway view. 

\subsubsection{Standard Coordinates in CDF-II}
\label{sec-cdfcoor}
Because of its barrel-like detector shape, CDF-II uses a cylindrical 
coordinate system ($r, \phi, z$) with the origin at the center of the detector.
 The $z$ axis is along the direction of the proton beam. The $r$ indicates the 
radial distance from the origin and $\phi$ is the azimuthal angle. The \rphi\ 
plane is called the transverse plane, as it is perpendicular to the beam line.
The polar angle, $\theta$, is the angle relative to the $z$ axis. 
 An alternative way of expressing $\theta$, pseudorapidity ($\eta$), is 
defined as:
\begin{equation}
\eta \equiv -\ln \tan (\theta /2).
\end{equation}
The coverage of each CDF-II detector sub-system will be described using 
combinations of $\eta$, $r$, $\phi$ and $z$. 

\subsubsection{Overview}
The CDF-II detector consists of five main detector systems: tracking, particle 
identification (for $e$, $K$, $p$ and $\pi$), calorimetry, muon identification
 and luminosity measurement.  

The innermost system of the detector is the integrated tracking system: a 
silicon microstrip system and an open-cell wire drift chamber, the Central 
Outer Tracker (COT) that surrounds the silicon detector. The tracking system 
is designed to measure the momentum and the trajectory of charged particles. 
Reconstructed particle trajectories are referred to as ``tracks''. 
Multiple-track reconstruction allows us to identify a vertex where either the 
\ppbar\ interaction took place (primary vertex) or the decay of a long-lived 
particle took place (secondary or displaced vertex).

The silicon microstrip detector consists of three sub-detectors in a barrel 
geometry that extends from the radius of $r$= 1.35~cm to $r$= 28~cm and covers 
the track reconstruction in the range of $|\eta| <$ 2. Closest to the beam 
pipe is a single-sided, radiation tolerant silicon strip detector, Layer 00 
(L00), with sensors at $r$=1.35~cm and $r$=1.62~cm. L00 is followed by five 
concentric layers of double-sided silicon sensors (SVX-II) from $r$=2.45~cm to 
10.6~cm.  The outermost silicon detector is the Intermediate Silicon Layers 
(ISL), from $r$=20~cm to $r$=28~cm. L00 only provides $\rphi$ measurements, 
while the SVX-II and ISL provide both \rphi\ and $z$ measurements.

\begin{figure}[tbp]
\begin{center}
\includegraphics[width=350pt,angle=0]{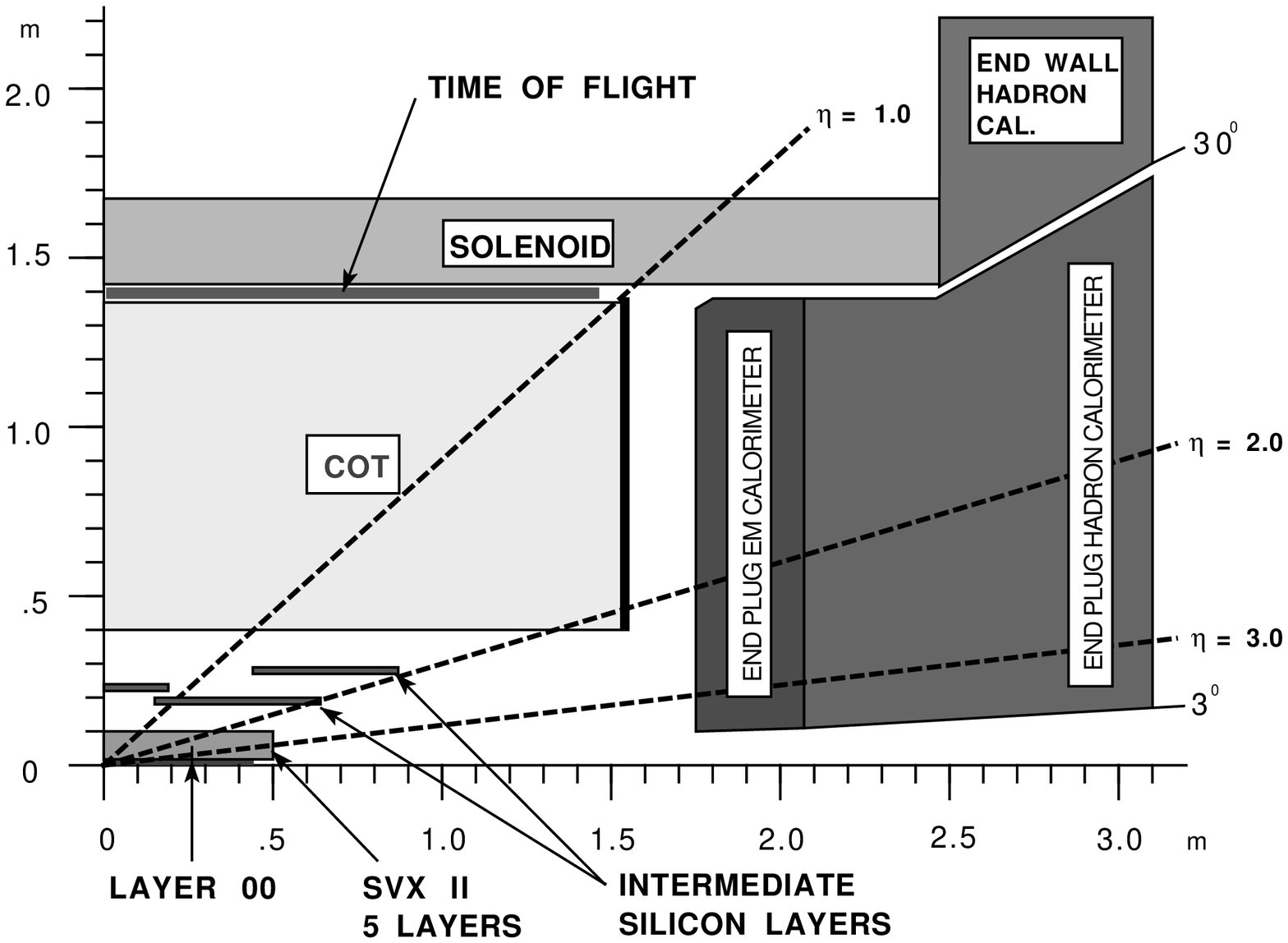}
\end{center}
\caption[A diagram of the CDF-II tracker layout showing the different 
subdetector systems.]{A diagram of the CDF-II tracker layout showing the 
different subdetector systems.}
\label{fig:tracker_layout}
\begin{center}
\includegraphics[width=200pt,angle=0]{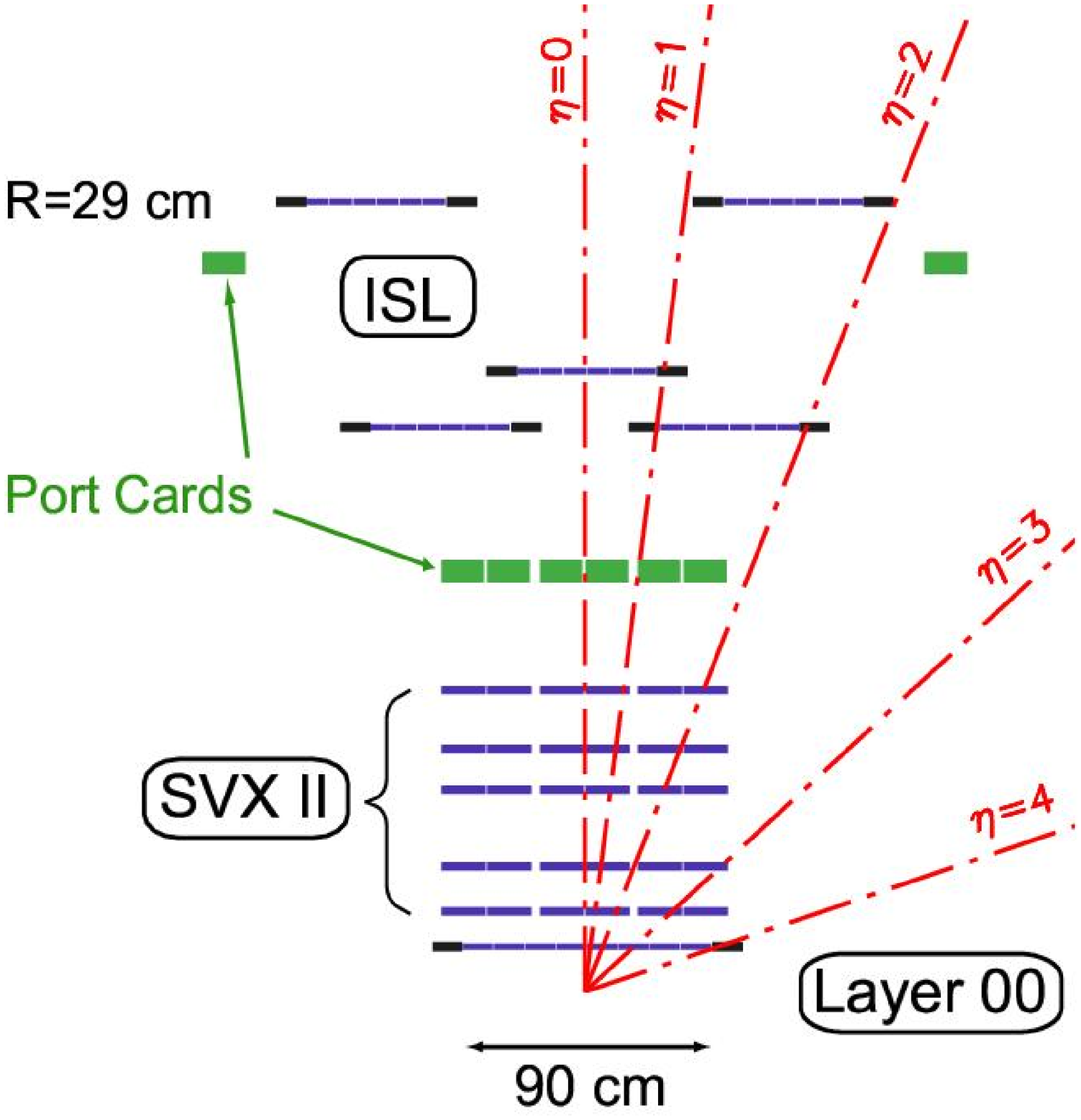}
\end{center}
\caption[Coverage of the different silicon subdetector systems projected into 
         the $r$-$z$ plane]{Coverage of the different silicon subdetector 
	systems projected into the $r$-$z$ plane. The $r$ and $z$ axes have 
	different scales.}
\label{fig:silicover}
\end{figure}

Surrounding the silicon detector is the COT, which covers the radius from 40 
cm to 137~cm and $|\eta| <$ 1. Figure~\ref{fig:tracker_layout} and 
Figure~\ref{fig:silicover} give a $r$-$z$ view of the CDF-II tracker and the 
silicon tracking system, respectively. 

Immediately outside the COT is the Time of Flight system (TOF), which consists
 of 216 scintillator bars, roughly 300~cm in length and with a cross-section of
 $4\times4$~cm$^2$. The bars are arranged into a barrel around the COT outer 
cylinder. The TOF system is designed for the particle identification of charged
 particles with momentum below 2 \gevc. Both the tracking system and the TOF 
are contained in a superconducting solenoid, 1.5~m in radius, 4.8~m in 
length, that generates a 1.4~Tesla magnetic field parallel to the beam axis. 
The solenoid is surrounded by the electromagnetic and hadronic calorimeters, 
which measure the energy of particles that shower when interacting with matter.
 The coverage of the calorimeters is $|\eta|$ $<$ 3. The electromagnetic 
calorimeter is a lead/scintillator sampling device and measures the energy of 
the electrons and photons. The hadronic calorimeter is an iron/scintillator 
device and measures the energy of the hadrons, {\it e.g.}: pions.

\begin{figure}[tbp]
\centering
\includegraphics[width=300pt,angle=0]{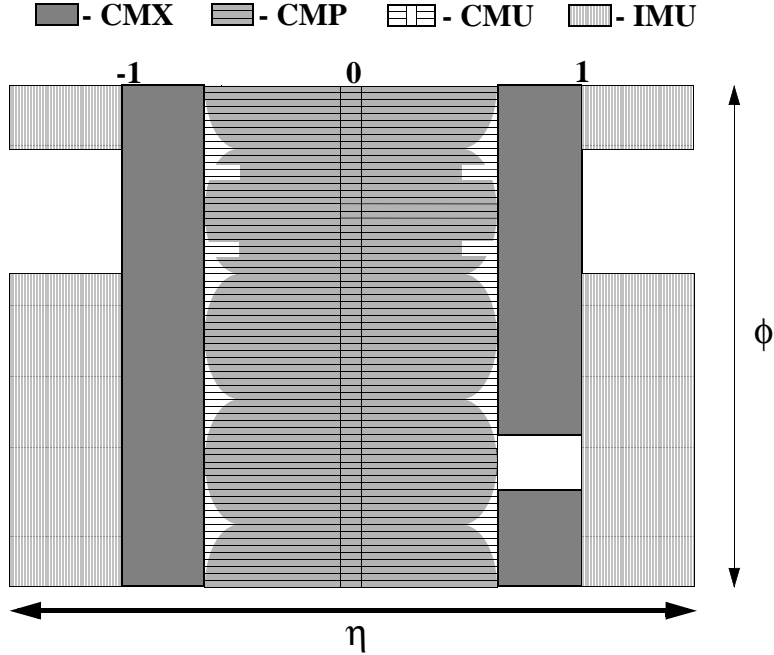}
\caption{Locations of the CDF-II Muon System in the $\eta$-$\phi$ view.}
\label{fig:muoncover}
\end{figure}

The calorimeters are surrounded by the muon detector system. Muons interact 
with matter primarily through ionization. As a result, if a muon is created in 
the collision and has enough momentum, it will pass through the tracking system, TOF, the solenoid and the calorimeters with minimal interaction with the 
detector material. Muon detectors are, therefore, placed radially outside the 
calorimeters. The CDF-II detector has four muon systems covering different 
$\eta$ and $\phi$ regions: the Central Muon Detector (CMU), the Central Muon 
Upgrade Detector (CMP), the Central Muon Extension Detector (CMX), and the 
Intermediate Muon Detector (IMU). Figure~\ref{fig:muoncover} shows the coverage
 of each muon detector in the $\eta$-$\phi$ view. 

At the extreme forward region of the CDF-II detector, 3.75 $<$ $|\eta|$ $<$ 
4.75, two modules of Cherenkov Luminosity Counters (CLC) are placed pointing 
to the center of the interaction region to record the number of \ppbar\ 
interactions. The number of particles recorded in the CLC modules is combined 
with an acceptance of the CLC ($A$) and the inelastic $\ppbar$ cross section 
($\sigma_{in}$) to determine the instantaneous luminosity using the following 
equation:
\begin{equation}
	L = \frac{\mu \cdot f_{BC}}{\sigma_{in}\cdot A}
	\label{eq:cdf_lumi} 
\end{equation}
where $L$ is the instantaneous luminosity, $f_{BC}$ is the rate of bunch 
crossings in the Tevatron and $\mu$ is the average recorded number of $\ppbar$
 interactions per bunch crossing.

This analysis uses SVX-II, COT, CMU and the trigger, which will be described 
in detail in the following sections. More detailed information about each 
sub-detector and the trigger may be found in Bishai\cite{bishai:bxsec}. 
We refer the reader to Affolder and Hill\cite{Affolder:2000tj,Hill:2004qb} 
for a full documentation about the ISL and L00, Acosta\cite{Acosta:2004kc} 
about the TOF, Balka, Bertolucci and Kuhlmann
\cite{Balka:1987ty,Bertolucci:1987zn,Kuhlmann:2003hx} about the calorimeters, 
Artikov\cite{Artikov:2004ew} about the CMP, the CMX and the IMU, and Acosta
\cite{Acosta:2001zu,Acosta:2002hx} about the CLC.

\section{Tracking Systems}
\subsection{Silicon Vertex Detectors II}
\label{sec:silicon}
Silicon tracking detectors are used to obtain precise position measurements
of the path of a charged particle. They present some advantages over the 
gas filled drift chamber. The typical electron-hole creation energy of the 
silicon is about 3~eV, while the ionization energies are about 10--15~eV for 
the drift chamber gas (Argon or Ethane). The increased number of electrons 
gives better energy and position resolutions and signal to noise ratio. The 
fundamental component of the silicon detector is a reverse-biased p-n junction.
The reverse-biased voltage increases the gap between the conduction band and 
the valence band across the p-n junction and reduces the current from the 
thermal excitation. By segmenting the p or n side of the junction into 
``strips'' and reading out the charge deposition separately on every strip, we 
obtain sensitivity to the position of the charged particle. More information 
about the principles of silicon detector may be found in Knoll\cite{knoll:rad}.

\begin{table}[tbp]
\caption[Relevant parameters for the layout of the sensors of different SVX-II 
layers.]{Relevant parameters for the layout of the sensors of different SVX-II 
layers.}
\label{tab:svxiiparms}
\begin{center}
\renewcommand{\tabcolsep}{0.05in}	
\begin{tabular}{lr|c|c|c|c|c}
\hline
\multicolumn{2}{l|}{Property}
	& Layer 0 & Layer 1 & Layer 2 & Layer 3 & Layer 4\\
\hline
\hline
number of $\phi$ strips &         & 256 & 384 & 640 & 768 & 869 \\
number of $Z$ strips    &         & 256 & 576 & 640 & 512 & 869 \\
stereo angle            & (\degs) & 90 & 90  & +1.2 & 90 & -1.2 \\
$\phi$ strip pitch      & (\um)   & 60 & 62  & 60  & 60  & 65  \\
$Z$ strip pitch         & (\um)   & 141 & 125.5 & 60 & 141 & 65 \\
active width            & (mm)    & 15.30 & 23.75 & 38.34 & 46.02 & 58.18 \\
active length           & (mm)    & 72.43 & 72.43 & 72.38 & 72.43 & 72.43 \\
\hline
\end{tabular}
\end{center}
\end{table}

The CDF SVX-II is composed of double-sided silicon sensors. Each side measures 
either the $\phi$ or $z$ coordinates.  The strips on the $\phi$ side run 
axially, while the strips on the $z$ side run either perpendicular to the 
axial strips, or are tilted by 1.2\degs. On the $\phi$ measurement side, 
65~\um\ pitch strips of p-type silicon are implanted near the surface of a 
mild n-type (high purity) silicon bulk. On the $z$ measurement side, strips of 
n-type silicon are implanted on the same high purity bulk. 

\begin{figure}[tbp]
\begin{center}
\resizebox{350pt}{!}{\includegraphics*[0pt,46pt][794pt,474pt]
{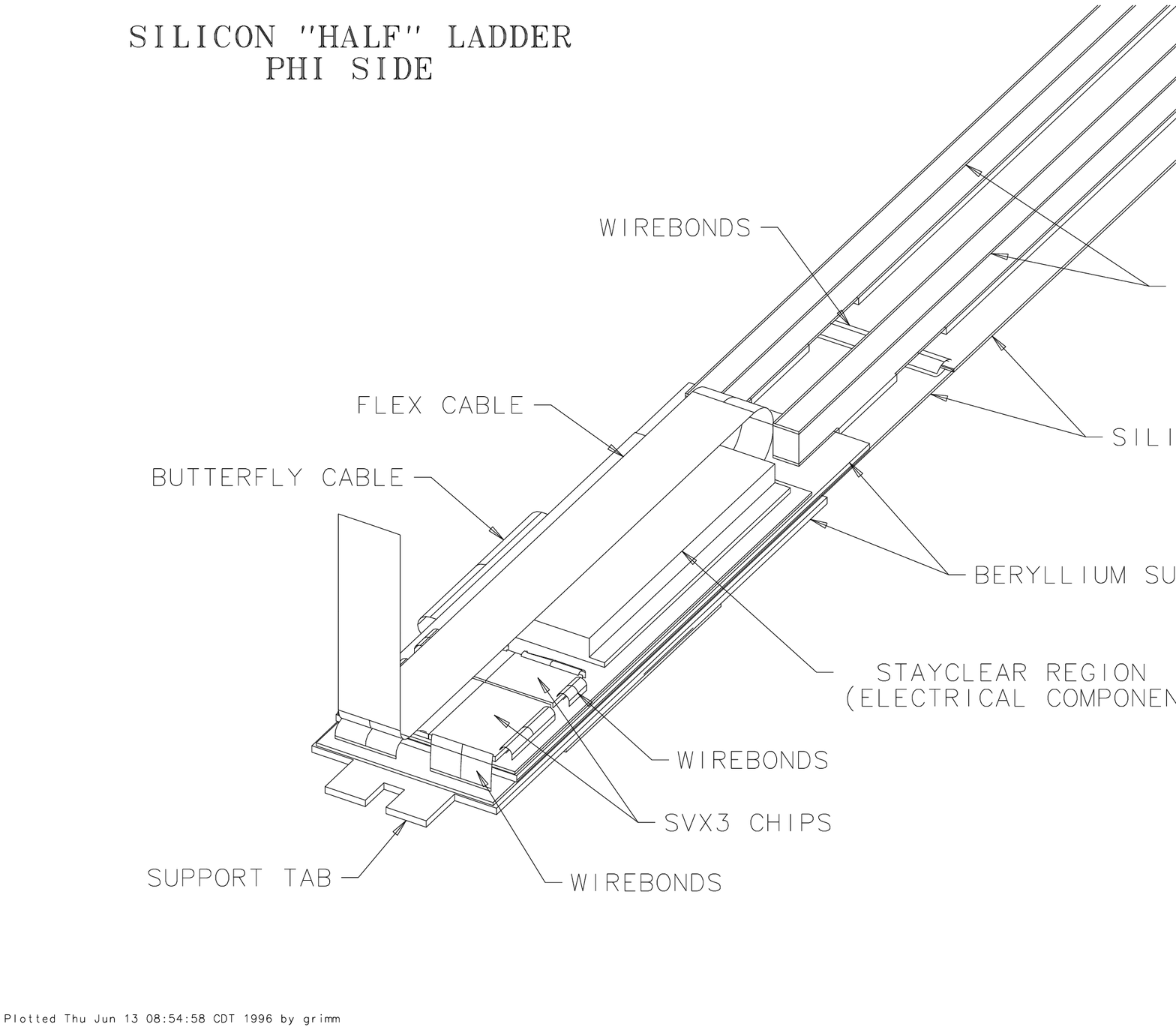}}
\end{center}
\caption{Prospective view of the $\phi$-side of a ladder for the SVX-II.}
\label{fig:layer0}
\begin{center}
\resizebox{200pt}{!}{\includegraphics*[103pt,140pt][568pt,675pt]
{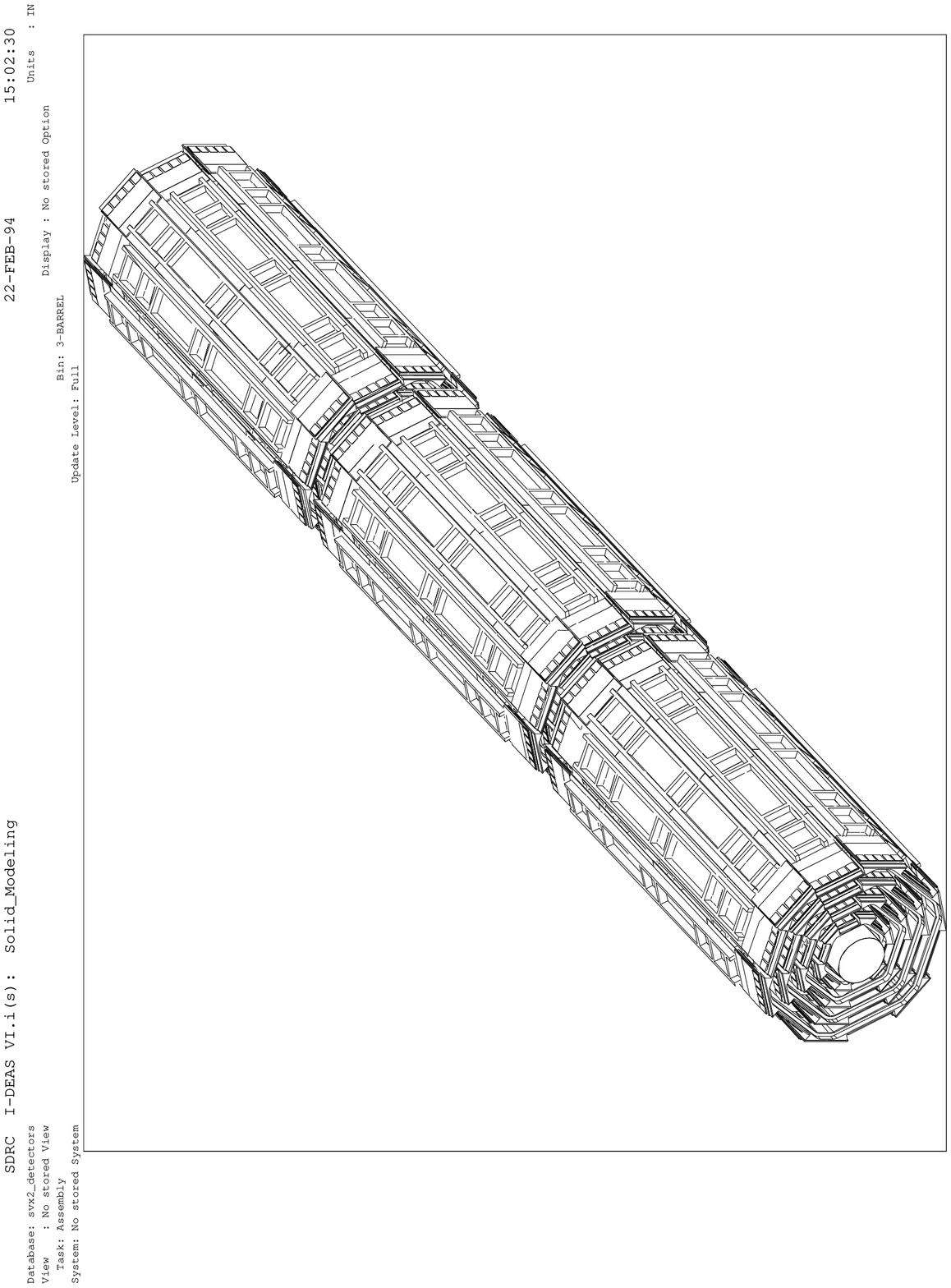}}
\end{center}
\caption{Isometric view of three SVX-II barrels.}
\label{fig:barrel}
\end{figure}

Four silicon sensors are assembled on a ladder. The readout electronics are 
mounted directly to the surface of the silicon sensor at each end of the 
ladder, as shown in Figure~\ref{fig:layer0}.  The ladders are arranged in three, 29~cm long barrels in an approximately cylindrically symmetric configuration.
 The barrels are positioned end-to-end along the beam axis as shown in 
Figure~\ref{fig:barrel}, and segmented azimuthally into 12 wedges (30\degs\ 
each). Each wedge contains 5 layers of silicon sensors. Table
~\ref{tab:svxiiparms} gives the mechanical dimensions of SVX-II. The 
``impact parameter'' is defined as the distance from the closest approach of 
the particle trajectory to the primary vertex. Without ISL and L00, SVX-II 
reaches a impact parameter resolution of $\sim$ 50~\um, which includes the 
30~\um\ contribution from the transverse size of the beam line.

\subsection{Central Outer Tracker}
\label{sec-cot}
\begin{figure}[tbp]
\begin{center}
\includegraphics[width=200pt,angle=0]{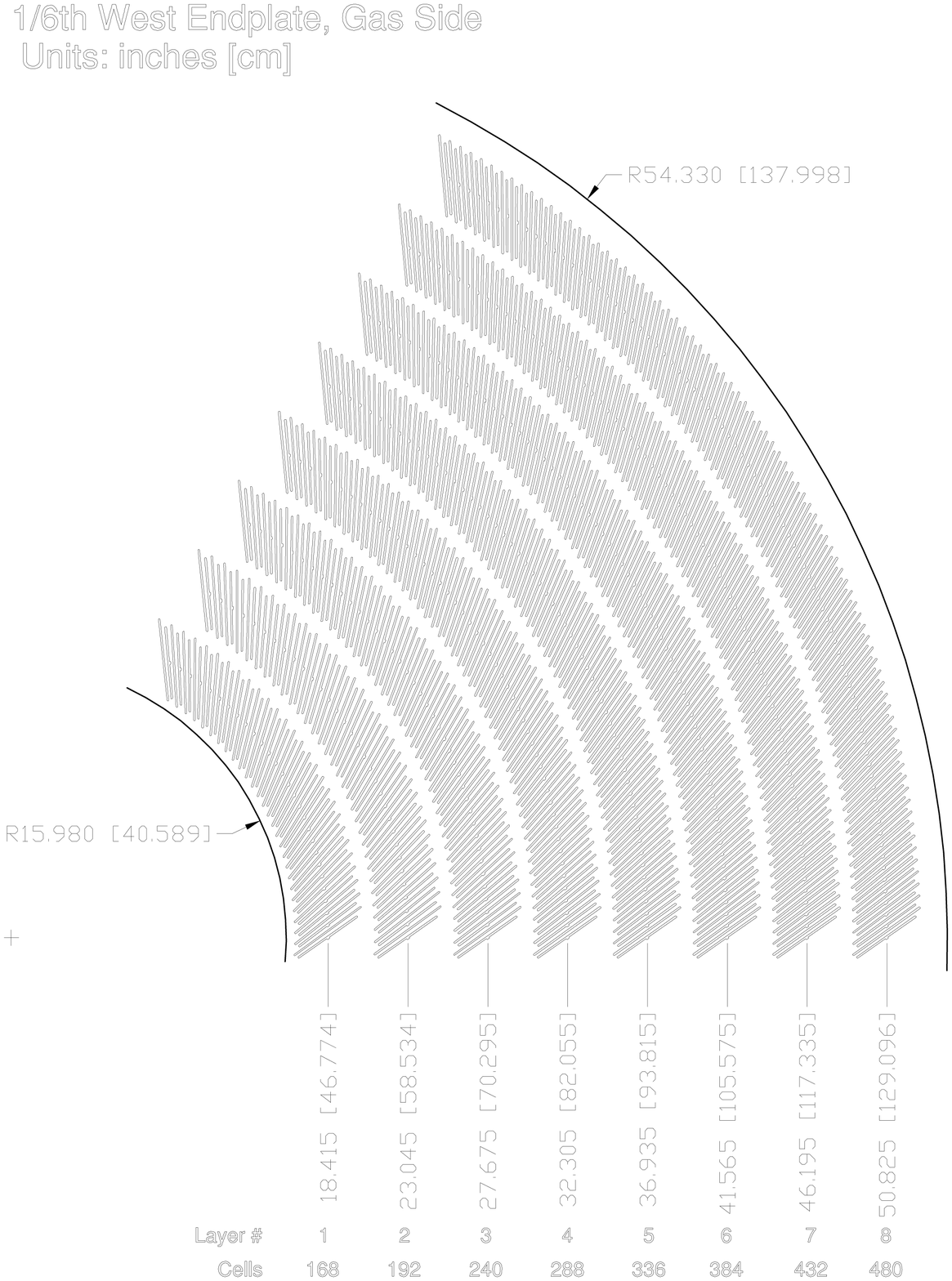}
\caption[Layout of wire planes on a COT endplate.]{Layout of wire planes on a 
COT endplate.
\label{fig:cotendview}
}
\end{center}
\begin{center}
\includegraphics[width=180pt,angle=0]{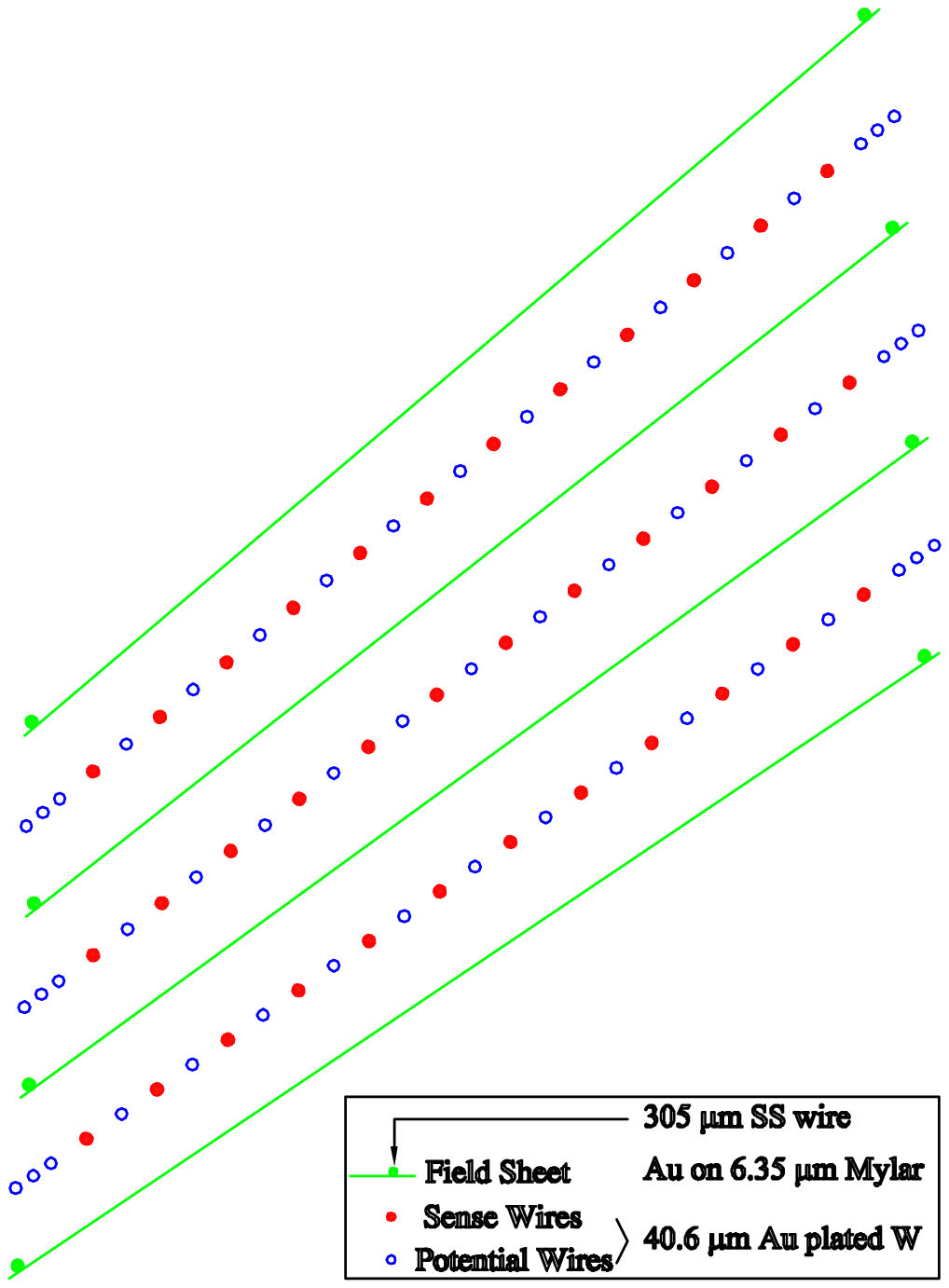}
\caption[Transverse view of three COT cells]{
	Transverse view of three COT cells. Shown are cathode field sheets 
	(solid lines), 12 sense wires (circles with cross) 
	and 17 potential wires (marked by open circles) in each cell.
\label{fig:cotcell}
}
\end{center}
\end{figure}

The Central Outer Tracker (COT) is a cylindrical, open-cell drift chamber. 
The active volume of the COT begins from $r$=43.4~cm to $r$=132.3~cm and spans 
310~cm in the $z$ direction.  The COT contains 96 sense wire layers in the 
radial direction, which are grouped into eight superlayers, as shown in 
Figure~\ref{fig:cotendview}. The maximum drift distance is approximately the 
same for all superlayers. 

Each superlayer is divided in $\phi$ into ``cells''. Figure~\ref{fig:cotcell} 
shows the transverse view of three COT cells. In each cell, 12 sense wires and 
17 potential wires are sandwiched by two grounded field (cathode) sheets. The 
potential wires help shape the electric field near the sense wires. The 
distance between the wires and each field sheet is 0.88~cm. The sense wires 
alternate with the potential wires at a pitch of 0.3556~cm. Each end of the 
cell is closed with two potential wires, the first at the same pitch of 
0.3556~cm, and the second at the pitch of 0.1778~cm. The wires are made of 
40~\um\ of gold plated tungsten and the field sheets are made of 6.35~\um\ 
Mylar with vapor-deposited gold on each side. The entire COT contains 2,520 
cells and 30,240 sense wires. The wires in superlayer 1, 3, 5 and 7 run along 
the $z$ direction (``axial''). The wires in the other superlayers are strung 
at a small angle ($2^\circ$) with respect to the $z$ direction (``stereo'').

The COT chamber is filled with an Argon-Ethane gas mixture and Isopropyl 
alcohol (49.5:49.5:1). The voltages on the sense wires are 2600-3000 volts and 
1000-2000 volts on the potential wires. The field sheets are grounded. These 
voltages have been optimized using Garfield simulation software
~\cite{Veenhof:1998tt} in order to achieve a gas gain of \(2\cdot10^4\) and 
minimize the drift field deviations from wire to wire. When a charged particle 
passes through, the gas is ionized. The applied electric field in the cell 
allows us to collect the electrons from the ionization and to track the 
passage of the charged particle. Electrons drift towards the sense wires with 
a velocity $\sim$ 50~\um/ns. Due to the magnetic field in which the COT is 
immersed, electrons drift at a Lorentz angle of $\sim 35\degs$. Therefore, each
 cell is tilted by $35\degs$ with respect to the radial direction to 
compensate for this effect. After the tilt, the electrons drift approximately 
perpendicular to the radial direction. The maximum electron drift time is 
approximately 177~ns, much shorter than the beam crossings of 396~ns. The 
electric potential in a cylindrical system grows logarithmically with 
decreasing radius. As a result, a limited avalanche via secondary ionization 
is initiated when the electron drifts close to the wire surface. This effect 
provides us a gain large enough to be read out by the electronics with high 
signal to noise. 

Signals on the sense wires are read out by the ASDQ (Amplifier, Shaper, 
Discriminator with charge encoding) chip, which was developed by Bokhari and 
Newcomer~\cite{cdfnote:asdq}. The ASDQ provides input protection, 
amplification, pulse shaping, baseline restoration, discrimination and charge 
measurement. The analog signal arrives at the ASDQ and the output is a digital
pulse. The leading edge gives the arrival time information and the pulse width 
is related to the amount of charge collected by the wire. After calibrating 
the width variations due to the COT geometry, path length of the particle, gas 
gain difference for the 96 wires, the digital width is related to the 
ionization energy loss $dE/dx$, used for particle identification. 
A detailed description of the calibrations performed by the author and 
Donega, Giagu, Tonelli is found in Yu~\cite{yu:dedx} and 
Donega~\cite{cdfnote:6932}.

The ASDQ pulse is then sent through $\sim 105$~cm of micro-coaxial cable, via 
repeater cards to Time to Digital Converter (TDC) boards in the collision hall.
 Hit times are later processed by pattern recognition (tracking) software to 
form helical tracks. The hit resolution of the COT is about 140~\um. 
The transverse momentum (\pt) resolution, \(\sigma_{\pt}/\pt\) is about 
0.15\(\%\cdot\pt\). Table~\ref{t:cotpar} lists the COT parameters. 
A full documentation about the COT may be found in Affolder
\cite{Affolder:2000tj}.

\begin{table}[tbp]
\caption{COT Parameters.}
\label{t:cotpar}
\begin{center}
\begin{tabular}{l|r}
\hline
\hline
Parameter & Value \\
\hline
Gas & Ar/Et/Isopropyl(49.5:49.5:1) \\
Max. Drift distance & 0.88 cm \\
Max. Drift Time & 177 ns \\
Lorentz Angle & $\sim$ 31\degs\ (35 \degs cell tilt) \\
Drift Field & 1.9 kV/cm \\
Radiation Lengths & 1.7 $\%$ \\
Total sense wires & 30,240 \\
Number of cells per SL & 168,192,240,288, \\
                       & 336,384,432,480  \\
Stereo Angle & +2\degs, 0\degs, -2\degs, 0\degs \\
	     & +2\degs, 0\degs, -2\degs, 0\degs \\
\hline
\hline
\end{tabular}
\end{center}
\end{table}

\subsection{Track Reconstruction}
\label{sec-tracking}
\subsubsection{Definition of Track Parameters}
Charged particles moving through a homogeneous solenoidal magnetic field in 
the $-z$ direction follow helical trajectories. To uniquely parameterize a 
helix in three dimensions, five parameters are needed: $C$, $\cot\theta$, 
$d_0$, $\phi_0$ and $z_0$. The projection of the helix is a circle on the 
\rphi\ plane. $C$ is the signed curvature of the circle, defined as  
\(C\equiv\frac{1}{2Q\rho}\), where $\rho$ is the radius of the circle and the 
charge of the particle ($Q$) determines the sign of $C$. The positive charged 
tracks curve counterclockwise in the \rphi\ plane when looking into the $-z$ 
direction and the negative charged tracks bend clockwise. 
The transverse momentum, \pt, is related to $C$, the magnetic field 
(\(B_{magnet}\)), and charge of the particle:
\begin{equation}
 \pt = Q\cdot \frac{1.49898\cdot 10^{-4}\cdot B_{magnet}}{C},
\end{equation}

\begin{figure}[tbp]
 \begin{center}
 \includegraphics[width=300pt,angle=0]{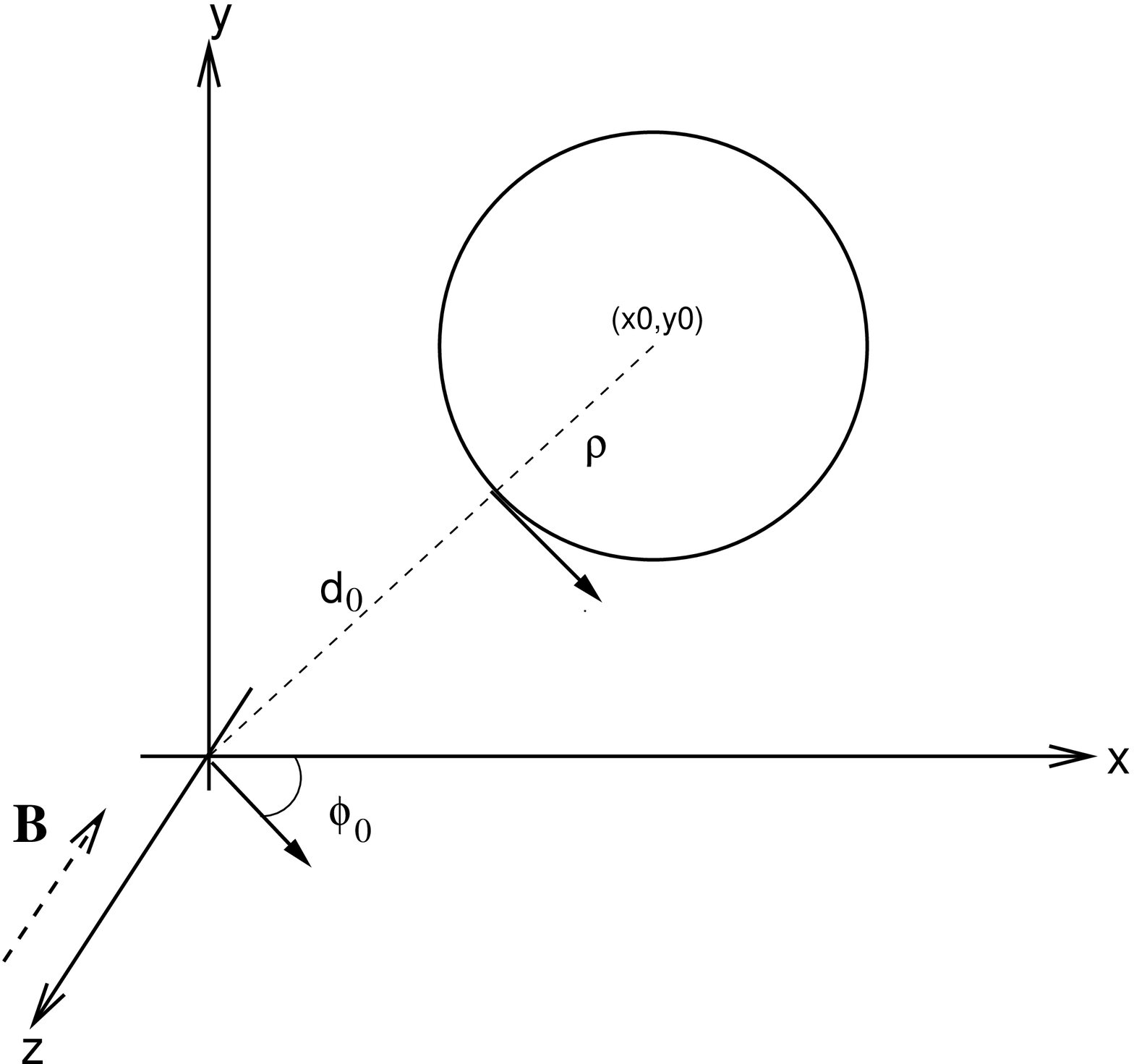}
 \end{center}
\caption{Definition of $d_0$ and $\phi_0$.}
\label{fig:trackpar}
 \begin{center}
 \includegraphics[width=220pt,angle=0]{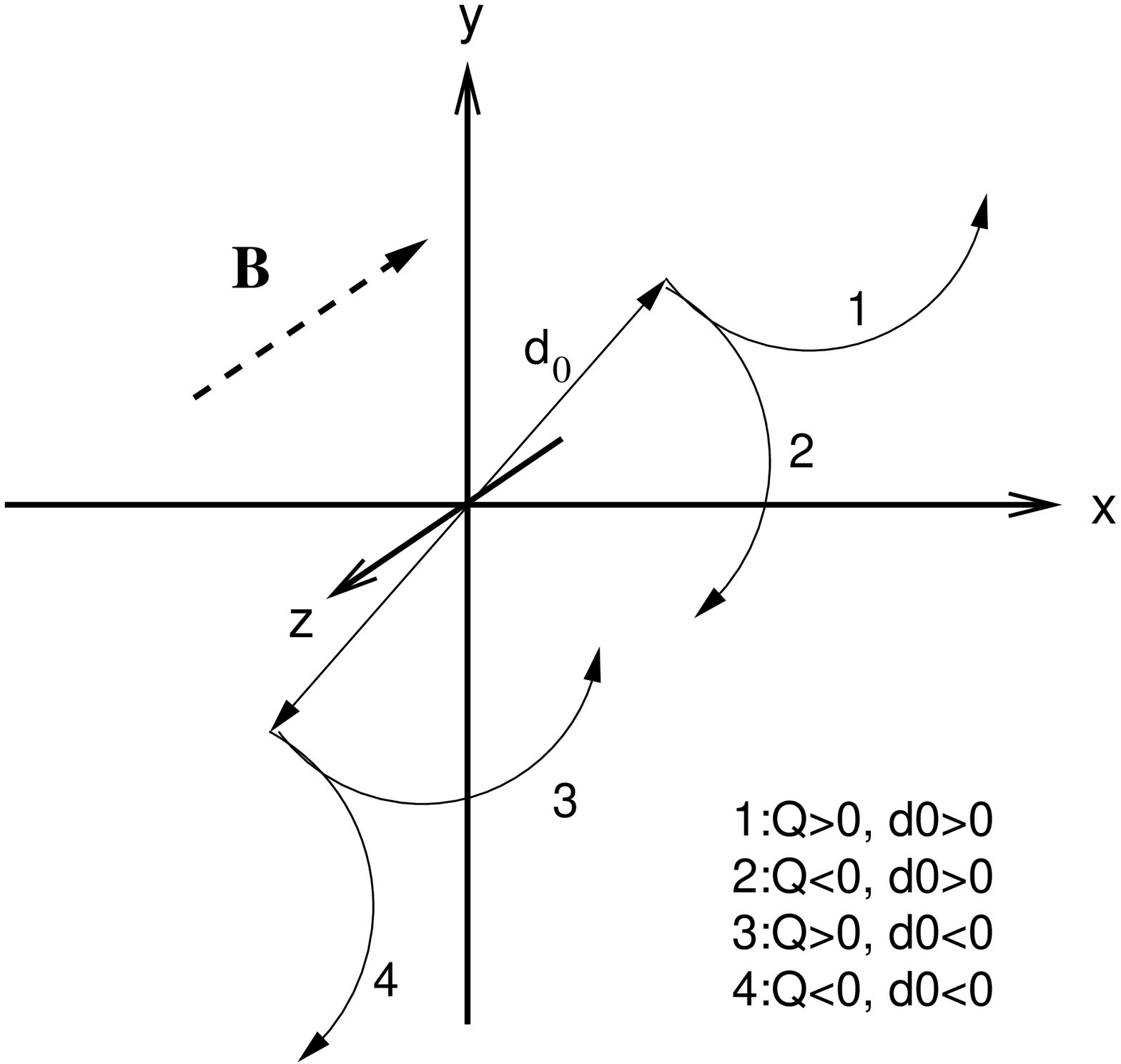}
 \end{center}
\caption[Sign definition of the track impact parameter $d_0$]
	{Tracks of particles with positive/negative charge and 
	positive/negative impact parameters. }
\label{fig:signd0}
\end{figure}

The $\theta$ is the angle between the $z$ axis and the momentum of the 
particle. Therefore, $\cot\theta$ is \(P_z/\pt\), where $P_z$ is the $z$ 
component of the particle momentum. The last three parameters, $d_0$, $\phi_0$,
 and $z_0$, are the $r, \phi$ and $z$ cylindrical coordinates at the point of 
closest approach of the helix to the beam line.  See Figure~\ref{fig:trackpar} 
for the definition of $d_0$ and $\phi_0$. $d_0$ is a signed variable;
\begin{equation}
 d_0 = Q\cdot(\sqrt{x_0^2+y_0^2}-\rho),
\end{equation}
where $(x_0,y_0)$ is the center of the helix circle in the \rphi\ plane, 
Figure~\ref{fig:signd0} illustrates the sign definition of $d_0$.

For decaying particles, we define the displacement \lxy,
\begin{equation}
\lxy = \vec{d} \cdot \hat{\pt}
\end{equation}
where $\vec{d}$ is the displacement of the decay vertex in the transverse
plane, and $\hat{\pt}$ is the unit vector in the direction of $\vec{\pt}$.

\subsubsection{Pattern Recognition Algorithms}
The track reconstruction begins using only the COT information. The first step 
is to look for a circular path in the axial superlayers of the COT. The 
algorithm looks for 4 or more hits in each axial superlayer to form a straight 
line, or ``segments''. The hits on the segment are reconstructed using the 
time difference between when the ionization occurs, $t_0$ (the collision time 
plus the time of flight of the charged particle) and when the signal is picked 
up by the wire (the leading edge time of the digital pulse from the TDC). The 
global time offset, readout time of the wires and cables, electronic channel 
pedestals, charged-based time slewing and non-uniform drift velocities are 
corrected before using the time difference (or drift time) in the tracking.

Once segments are found, there are two approaches to track finding. One 
approach is to link together the segments which are consistent with lying 
tangent to a common circle. The other approach is to constrain its 
circular fit to the beamline, and then add hits which are consistent with
this path. Once a circular path is found in the $\rphi$ plane, segments
and hits in the stereo superlayers are added depending on their proximity to 
the circular fit. This results in a three-dimensional track fit. Typically, if 
one algorithm fails to reconstruct a track, the other algorithm will not. 

Once a track is reconstructed in the COT, it is extrapolated into the
SVX-II. A three-dimensional ``road'' is formed around the extrapolated track, 
based on the estimated errors on the track parameters. 
Starting from the outermost layer, and working inward, silicon clusters 
found inside the road are added to the track. As a cluster is added, a 
new track fit is performed, which modifies the error matrix for the track 
parameters and produces a narrower road. In the first pass of this algorithm,
$\rphi$ clusters are added. In the second pass, stereo clusters are added to 
the track. If there is more than one track with different combinations of 
SVX hits associated with the same COT track, the track with maximum number 
of SVX hits is chosen. 

The track reconstruction efficiency in the COT is $\sim 95\%$ for tracks 
which pass through all 8 superlayers ($\pt \geq 400~\mevc$ ) and 
$\sim 98\%$ for tracks with \pt\ $>$ 10 \gevc.
The SVX track reconstruction efficiency with the COT 
tracks in the denominator is about 93$\%$ for the tracks with at least 
3 SVX \rphi\ hits. A complete description of the COT and the SVX tracking 
is found in Hays\cite{cdfnote:cottracking}. 

\section{Central Muon Detector}
\label{sec-cmu}
\begin{figure}[tbp]
\begin{center}
\includegraphics[width=350pt,angle=0]{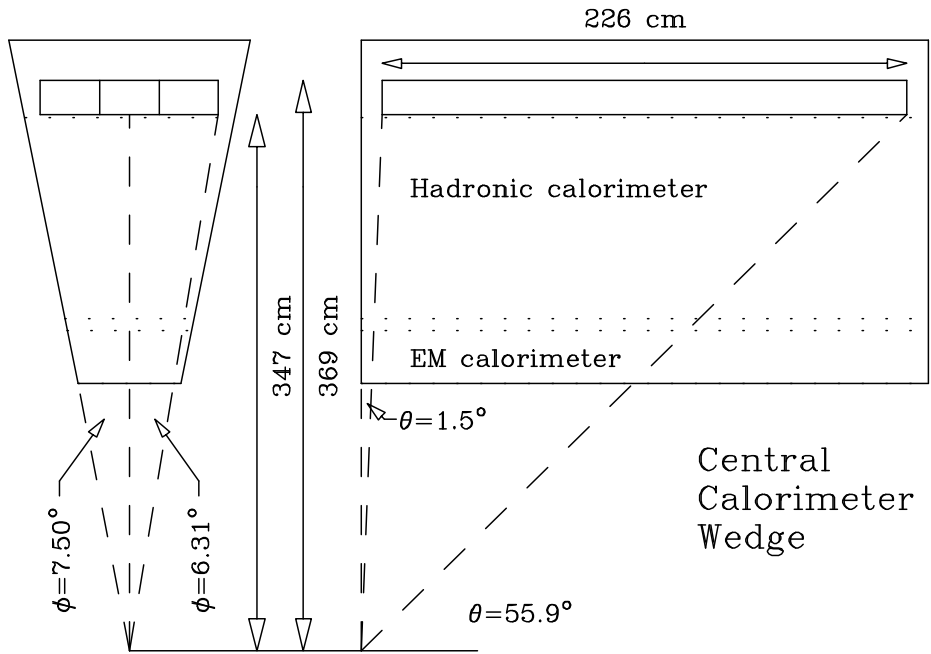}
\end{center}
\caption{Location of the Central Muon Detector (CMU).}
\label{fig:cmupos}
\end{figure}

The Central Muon Detector (CMU) is embedded in the central hadron 
calorimeter wedges at $r$=347 cm as shown in Figure~\ref{fig:cmupos} 
and covers $\eta$ $<$ 0.6. The detector is 
segmented in $\phi$ into 12.6\degs\ wedges, while the calorimeter is segmented 
into 15\degs\ wedges. This leaves a 2.4\degs\ gap between each wedge. In 
addition, there is a gap between the east and west chambers at $\eta$=0. The 
detector is further segmented in $\phi$ into three 4.2\degs\ modules. There 
are 72 modules at the east and west ends of the detector, which gives 
144 modules in total. Each module has 4 layers of 4 rectangular drift cells. 
The dimension of the drift cell is 6.35~cm (x) $\times$ 2.68~cm (y) $\times$ 
226.1 cm (z). Each cell has a 50\um\ stainless steel sense wire in the center. 

The first and the third layers have small offset in the $\phi$ direction from
 the second and the fourth layers, which also means:  
for each $\phi$ module, two sense wires from the alternating layers ( 1 and 3
 or 2 and 4) lie on a radial line. The other two sense wires lie on 
a line with a 2~mm offset from the first two. The ambiguity as to which side 
of the sense wire (in $\phi$) a track passes is resolved by determining which 
two layers of sense wires are hit first. The sense wires of alternating 
$\phi$ cells in the same layer are connected together so to enable readout at 
only one end of the chamber. Signals from the sense wires are discriminated and
 passed on to the same type of TDCs as used by the COT. Short tracks 
reconstructed using the TDC and ADC information are referred to as ``stubs''. 
The muon stubs are matched to the reconstructed tracks to form a muon 
candidate. A $\chi^2$ is computed using the distance between the track and 
the stubs, the difference in the direction of the track and the stub, and 
the covariance matrix of the track. Full documentation about the CMU is found in Ascoli\cite{Ascoli:1987av}. Table~\ref{t:cmupar} lists the parameters of 
the CMU, where the pion interaction lengths and the multiple scattering are 
computed at a reference angle of $\theta$ = 90\degs.

\begin{table}[ht]
\caption{Parameters of the CMU.}
\label{t:cmupar}
\begin{center}
\begin{tabular}{l|r}
\hline
\hline
Parameter & Value \\
\hline 
$\eta$ coverage & $\leq$ 0.6 \\
Drift tube cross-section & 2.68 $\times$ 6.35 cm \\
Drift tube length & 226.1 cm \\
Max drift time & 800 ns \\
Total drift tubes & 2304 \\
Pion interaction lengths & 5.5 \\
Min detectable muon \pt & 1.4 \gevc \\
Multiple scattering resolution & 12cm/p \\
\hline
\hline
\end{tabular}
\end{center}
\end{table}

\section{Triggers}
\label{sec-trigger}
The triggering systems play an important role in the \ppbar\ collider for 
two reasons. First, the collision rate is about 2.5~MHz, which is much higher 
than the rate at which data can be stored on tape, 50~Hz. Second, the total 
hadronic cross-section (including the elastic, inelastic, and diffractive 
processes) is about 75~mb and the \bb\ cross-section is about 1000 times 
smaller, 0.1~mb. Extracting the most interesting physics events from the large 
number of events reduces the cost and time to reconstruct data. 

\begin{figure}[tbp]
\begin{center}
\includegraphics[width=300pt,angle=0]{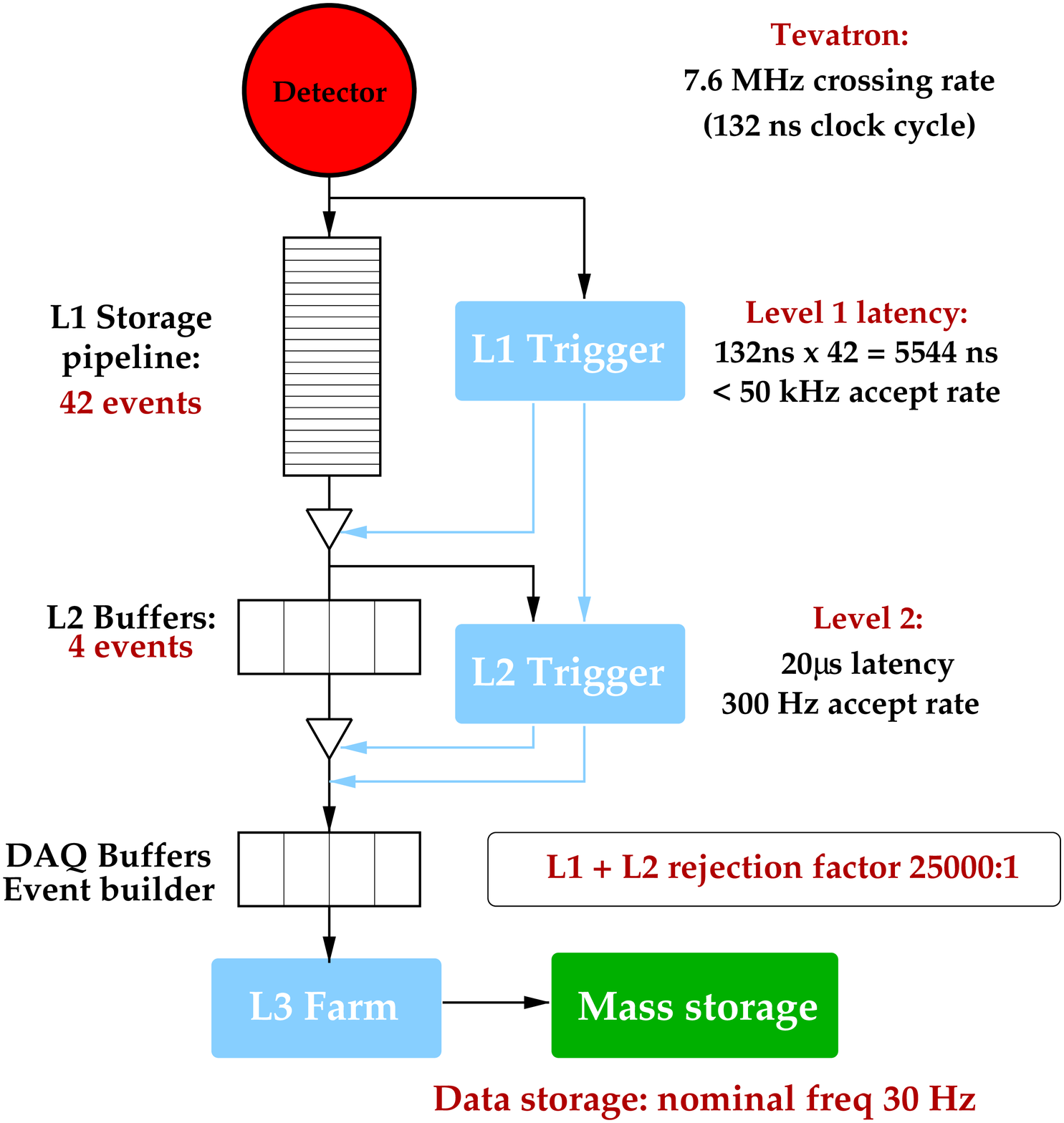}
\caption{Diagram of the CDF-II trigger system.}
\label{fig:pipeline}

\epsfig{file=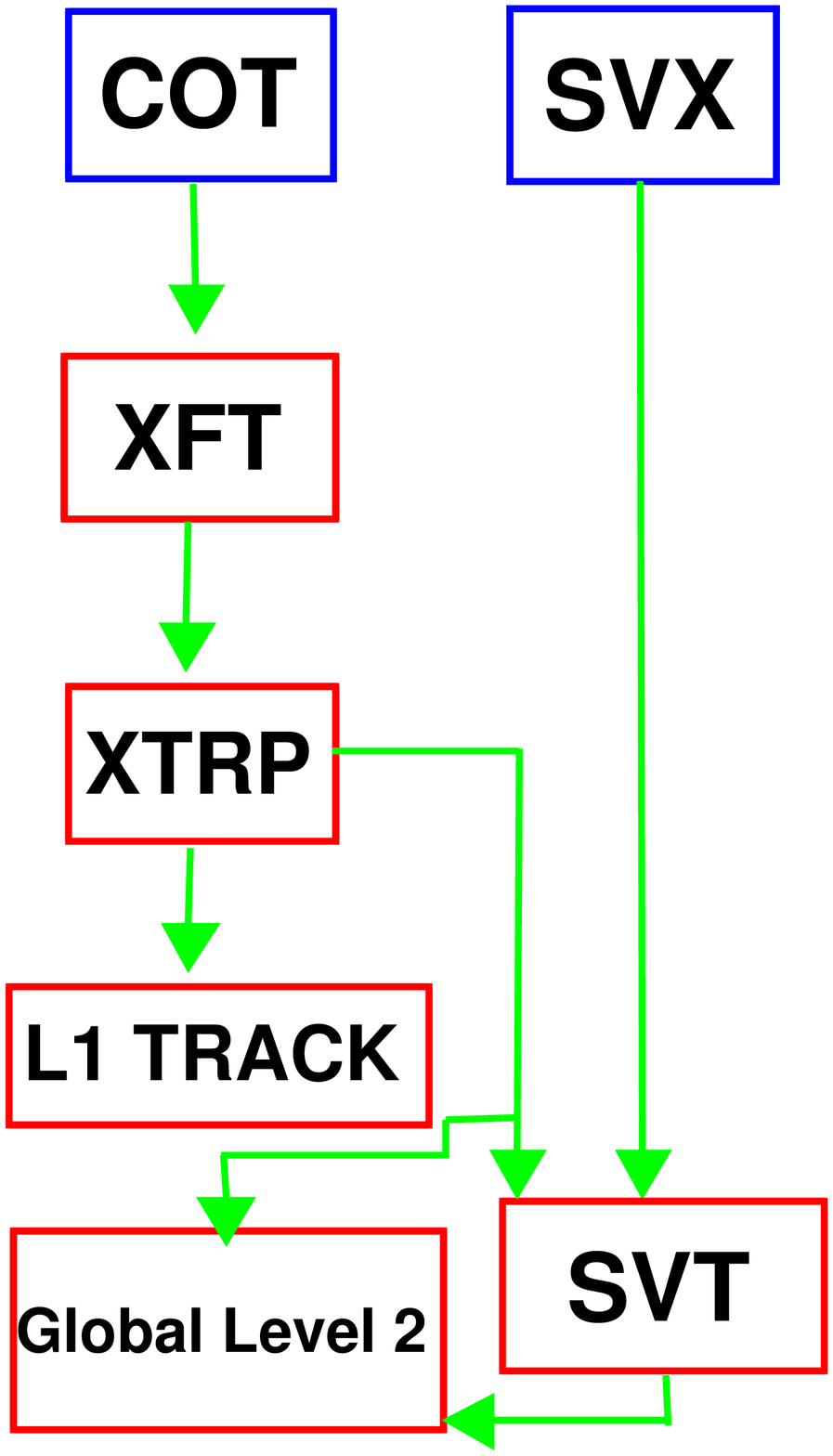, width = 0.4 \textwidth}
\caption{Diagram of the two track trigger path at Level 1 and 2.}
\label{fig:svtpath}
\end{center}
\end{figure}

The goal of the CDF-II triggering system is to be dead-time-less so that 
the system is quick enough to make a decision for every single bunch crossing 
before the next bunch crossing occurs. Each level of the trigger must 
reduce the background to a low enough level so that the rate to the next level 
is not saturated. Each level of the trigger is given an amount of time to make 
a decision about accepting or rejecting an event which depends on the 
complexity of the reconstruction. At the first level (Level 1), a trigger 
decision is made based only on a subset of the detector and quick pattern 
recognition or simple counting algorithms. The second level of the trigger 
(Level 2) does a limited event reconstruction. The third level of the trigger 
(Level 3) uses the full detector information to fully reconstruct events in a 
processor farm. The decision time for Level 1, 2 and 3 is about 5.5~$\mu s$, 
20~$\mu s$ and 1~s, respectively. The event accept rate for Level 1, 2 
and 3 is 40~kHz, 300~Hz and 50~Hz. The delay necessary to make a trigger 
decision is achieved by storing detector readout information in a storage 
pipeline, as shown in Figure~\ref{fig:pipeline}.

 A trigger path is a well defined sequence of Level 1, Level 2 and Level 3 
triggers.  An event coming from one trigger path satisfies the trigger 
requirements at each level. A well defined trigger path eliminates volunteer 
events. A volunteer event is an event which passes a higher level trigger 
requirement but passes a lower level trigger from a different path. 
At CDF II, there are about 100 trigger paths. The trigger path used in this 
analysis is one of the SVT trigger paths, \bcharm. Figure~\ref{fig:svtpath} 
shows a general diagram of the SVT trigger path. 

The strategy of the SVT trigger path is as follows. At Level 1, the eXtremely 
Fast Tracker (XFT) measures the track \pt\ and angle $\phi$.  By cutting on 
\pt\ and $\phi$, most of the inelastic background will be rejected. 
The Extrapolation Unit (XTRP) selects the XFT tracks above a certain \pt\ 
threshold and sends signal to the Level 1 Track Trigger (Level 1 Track). The 
Level 1 Track Trigger counts the number of tracks from the XTRP, if more than 
6 tracks are found, an automatic Level 1 accept is generated. Otherwise, 
depending on the trigger requirements, Level 1 Track Trigger accepts or rejects
 the event. If a Level-1 accept is received, the XFT track information is sent 
to the Level 2 Silicon Vertex Trigger (SVT). At Level 2, SVT uses the SVX-II 
information to obtain impact parameters of the tracks, $d_0$. Requiring 
non-zero impact parameters of tracks will require that they come from decays 
of long-lived particles: charmed and bottom hadrons. The requirements of each 
level of trigger will be described in detail in Section~\ref{sec-path}. 
The trigger components used in this trigger path, XFT, SVT, and Level 3, will 
be discussed in the following text.  We refer the reader to the CDF Run II 
Technical Design Report~\cite{cdf:tdr} for a full descriptions of the trigger 
hardware.

\subsection{The eXtremely Fast Tracker (XFT)} 
The Level 1 trigger decision of the ``two track'' trigger path is based on 
the information from the eXtremely Fast Tracker (XFT). This device is designed 
to measure the momentum of the charged particle using the hit information of 
the 4 COT axial layers. Instead of using the TDC information and a drift 
model to find a track segment as described in Section~\ref{sec-tracking}, 
the XFT uses a fast binary-like algorithm. 

Each hit on the wire is classified as ``prompt'' if the drift time ranges 
from 0 to 66~ns and as ``delayed'' if the drift time ranges from 67 to 
220~ns. Four neighboring COT cells are grouped together when searching for a 
segment in a given superlayer. A track segment in each axial superlayer is 
found by comparing the hit patterns to a list of pre-loaded patterns. The hit 
pattern varies depending on the combination of delayed and prompt hits, and the
 track angle through the cell or the track \pt. The algorithm allows two missed
 hits (2-miss) in each segment for the beginning period of the data used for 
this analysis and tightens the requirement to one missed hit (1-miss) since 
October, 2002. The data used in this analysis in the 2-miss period is about 
26.4~\pbarn, and 124.5~\pbarn\ for the 1-miss period.

\begin{figure}[tbp]
\begin{center}
\includegraphics[width=220pt,angle=0]{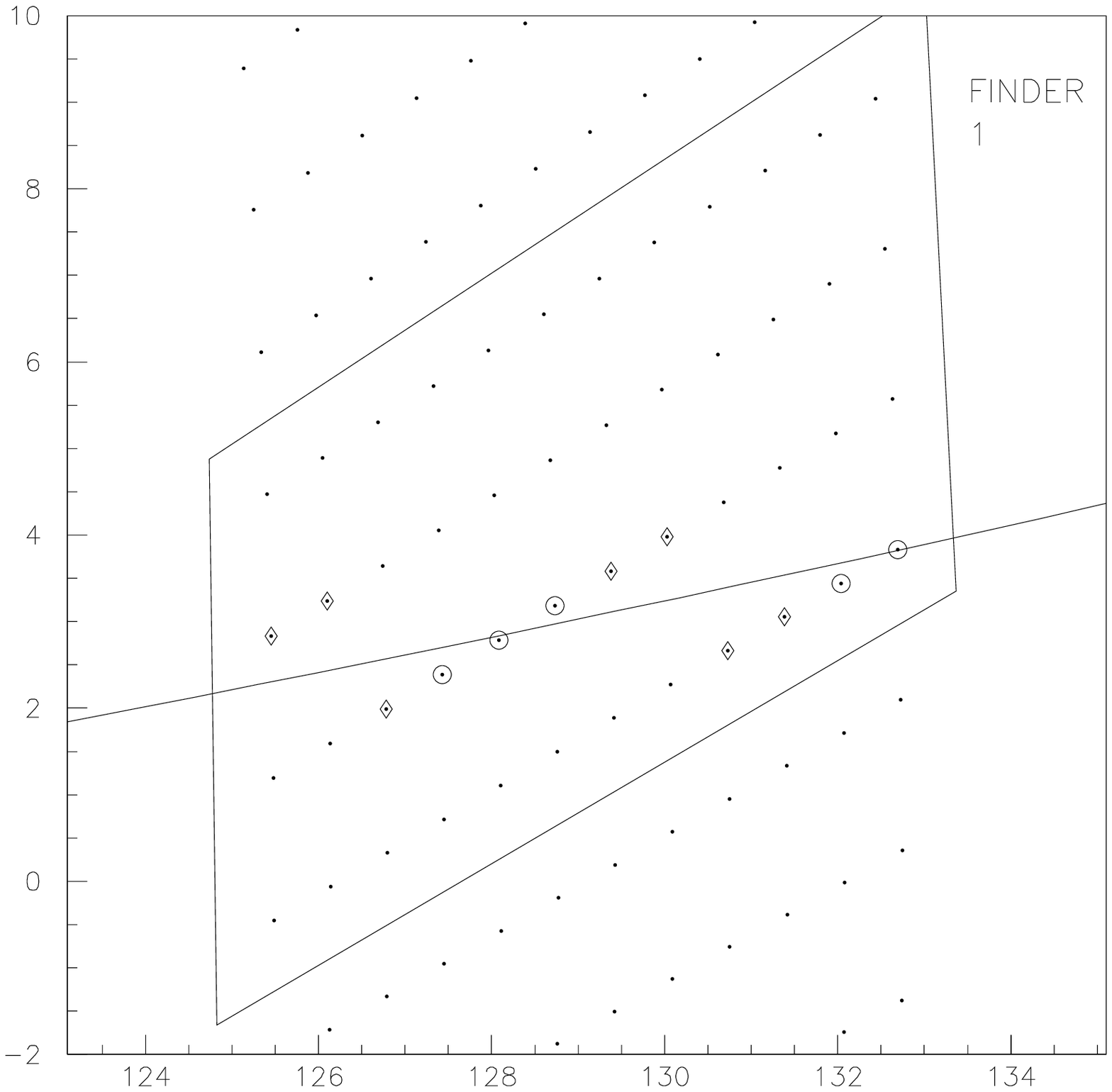}
\caption[Example of a XFT hit pattern]{Close view of a track with \pt\ = 1.5 
\gevc\ in a cell of the 4$^{th}$ COT axial layer from 
Thomson\cite{Thomson:2002xp}. A collection of the prompt hits 
(marked by open circles) and the delayed hits (marked by open diamonds) is 
an example of the XFT hit pattern.\label{fig:xfthit}}
\includegraphics[width=220pt,angle=0]{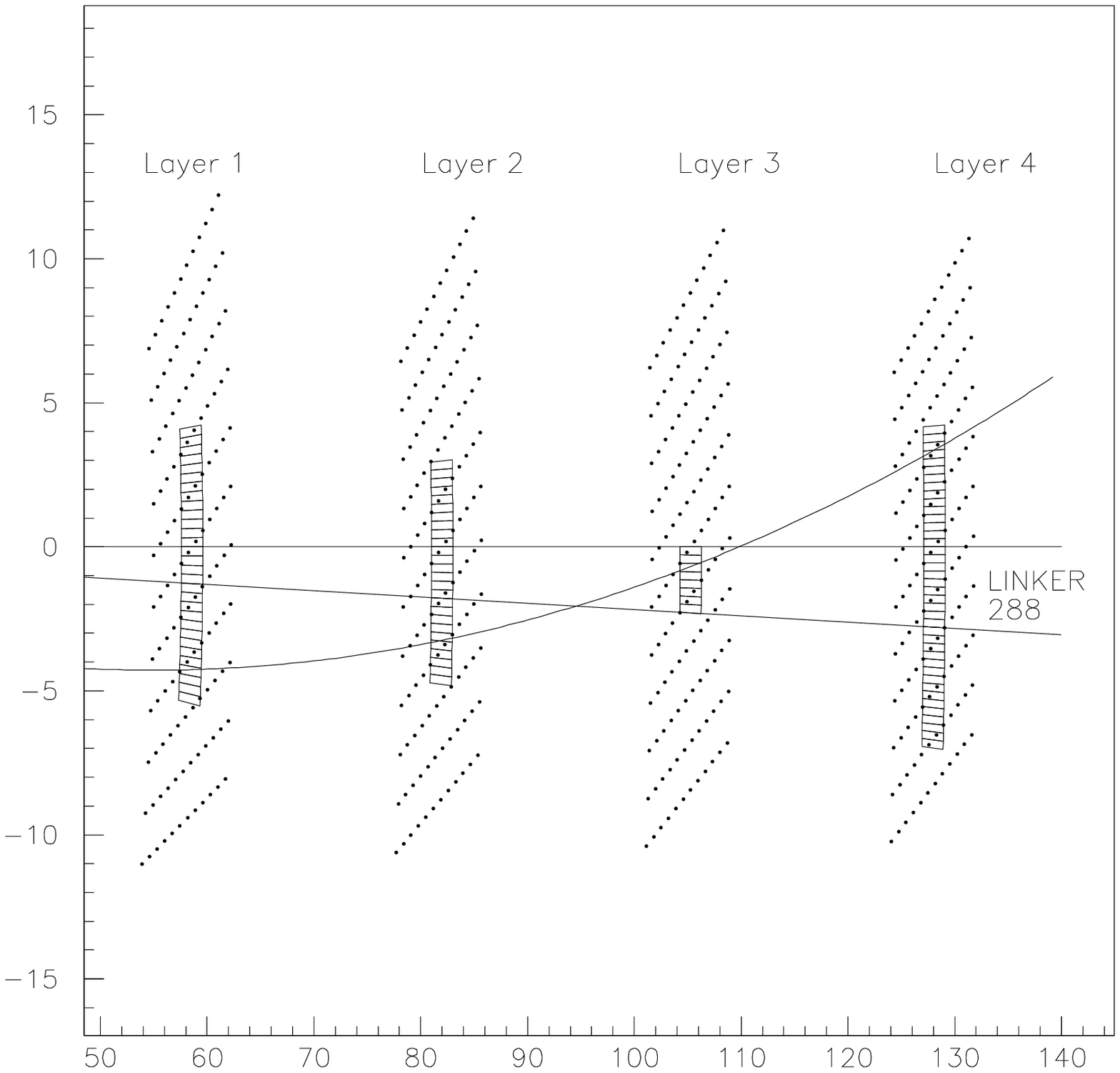}
\caption[Example of a XFT track pattern]{Close view of a track with \pt\ = 1.5 
\gevc\ traversing 4 COT axial layers from 
Thomson\cite{Thomson:2002xp}. Several possible track patterns 
are displayed. \label{fig:xfttrk}}
\end{center}
\end{figure}

Once a segment is found in a superlayer, it it marked as a ``pixel''. XFT 
compares the pixels in all 4 layers to a list of pixel patterns in a 
\(\Delta \phi= 1.25 \degs\) window corresponding to a valid track with 
$\pt \geq$ 1.5 \gevc\ ($\sim$ 2400 roads). The algorithm returns the best 
track.  Figures~\ref{fig:xfthit}--\ref{fig:xfttrk} extracted from 
Thomson\cite{Thomson:2002xp} show an example of the hit and the track pattern 
for a track with \pt\ = 1.5 \gevc. Finally the XFT reports the track $\pt$ and 
$\phi_6$, the angle of the transverse momentum at the sixth superlayer of the 
COT, which is located 106~cm radially from the beamline. 

The XFT efficiency is $\geq$ 96$\%$ for tracks which pass through all 4 axial 
layers. The momentum resolution, $\sigma_{\pt}/\pt$ is about 2$\%$ per \gevc. 
The angular resolution at the sixth superlayer, $\sigma_{\phi_6}$ is about 5 
mR. More detailed information about XFT may be found in Thomson
\cite{Thomson:2002xp}. 

\subsection{The Silicon Vertex Tracker (SVT)} 
At Level 2, the Silicon Vertex Tracker (SVT) combines the Level 1 track 
information, computes the track $d_0$, and improves the measurements of $\pt$ 
and $\phi_0$. The SVT is a new type of Level 2 trigger optimized for B physics
. The heavy flavor particles, such as beauty and charm hadrons, decay at 
positions displaced from the primary vertices. Therefore, their daughter tracks
 tend to have larger impact parameters. The SVT cuts on the minimum track 
impact parameter so we are able to collect large sample rich in heavy flavor. 
The SVT also cuts on the maximum track impact parameter and removes background 
due to the long lived $K_s$, $\Lambda$ or the secondary tracks which come from 
the particles interaction with the beam pipes. 

As mentioned in Section \ref{sec:silicon}, the SVX-II is segmented into 12 
wedges in $\phi$ and three mechanical barrels in $z$. The SVT makes use of 
this symmetry and does tracking separately for each wedge and barrel. 
The XFT track is extrapolated into the SVX-II, forming a ``road''.  Clusters 
of charge on the inner four $\rphi$ layers of the given wedge have to be found
 inside this road. Since June 2003, the requirement is loosened to ask for 
hits from any four \rphi\ layers out the five SVXII layers. This period of 
data corresponds to about 1/3 of the total integrated luminosity used for 
this analysis. The SVT checks if one of the roads in the list of pre-loaded 
patterns is present in the data. The found roads are fed into a linearized 
fitter which returns the measurements of $p_t, \phi_0$ and $d_0$ for the track.
 The Level 1 trigger conditions are confirmed with the improved measurements 
of $\pt$ and $\phi_0$.  An event passes Level 2 selection if there is a track 
pair reconstructed in the SVT and additional requirements on the $d_0$, \lxy, 
and scalar sum \pt\ of the track pair depending on the trigger path.

\begin{figure}[tbp]
\centering
\epsfig{file=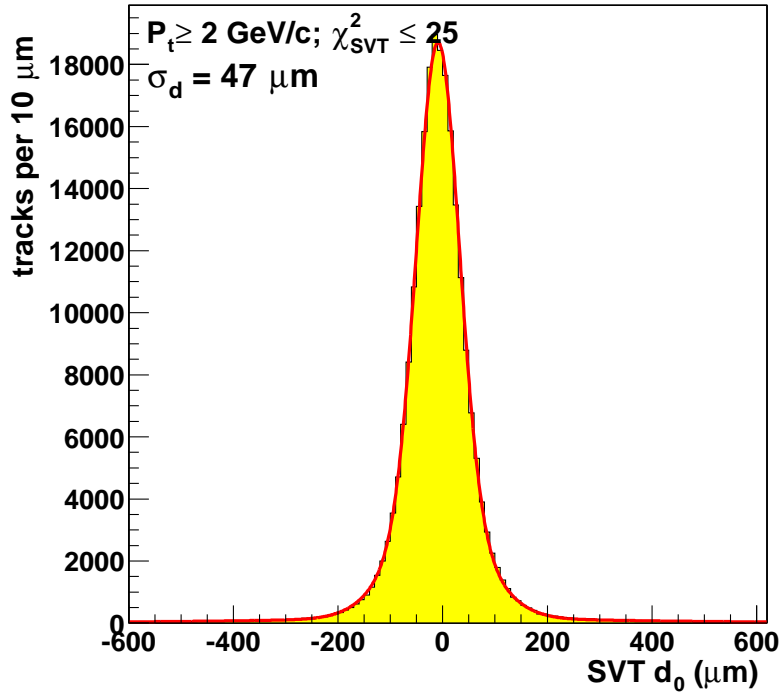, width = 0.7 \textwidth}
\caption{SVT impact parameter resolution.}
\label{fig:svtres}
\end{figure}

Figure \ref{fig:svtres} shows the SVT track impact parameter resolution for 
tracks with $\pt > 2~\gevc$. The width of the Gaussian fit for the distribution
 in Figure \ref{fig:svtres} is 47~\um, which is a combination of the intrinsic 
SVT impact parameter resolution, and the transverse size of the beam line:
\(\sigma_{fit} = \sigma_{SVT} \oplus \sigma_{beam}\), where $\sigma_{beam}$ is 
about 30~\um. Therefore, the intrinsic SVT resolution is about 35~\um. 
Full documentation about SVT is found in Ashmanskas\cite{Ashmanskas:2003gf}.

\subsection{Level 3 Trigger}
The third level of the trigger system is implemented as a Personal Computer 
(PC) farm. The input rate of the Level 3 is roughly 300~Hz. With roughly 
300 CPUs, and one event per CPU, this allocates approximately 1 second to do 
event reconstruction and reach a trigger decision.

\begin{figure}[ht]
\centering
\epsfig{file=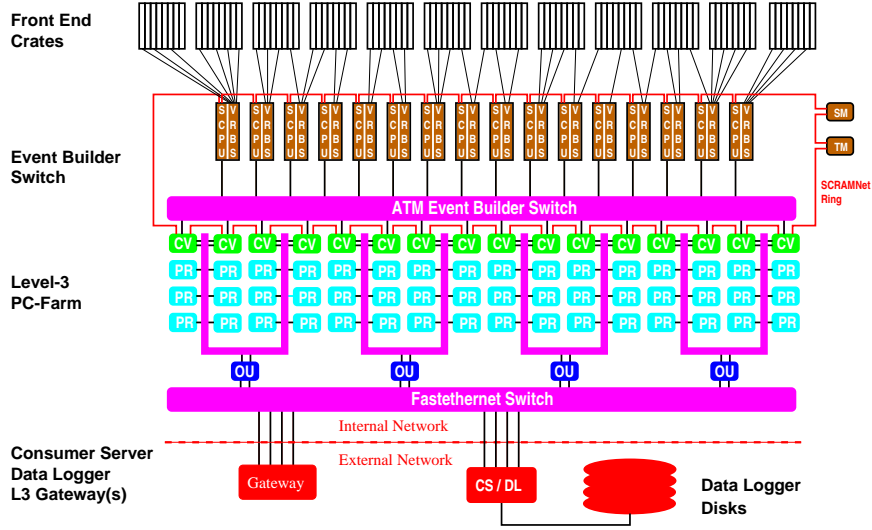, angle = 0, width = 0.8 \textwidth}
\caption[Principle of Event Building and Level 3 Filtering.]
{Event building and Level 3 operating principle: data from the 
 front end crates is prepared by Scanner CPU's (SCPU) and fed into the ATM
 switch. On the other side of the switch, converter nodes (CV)
 assemble events and pass them to processor nodes (PR). Accepted events are 
 passed to output nodes (OU) which send them to the Consumer Server and Data
 Logging systems (CS/DL).}
\label{fig:l3farm}
\end{figure}

Figure \ref{fig:l3farm} shows the principle of the Level-3 farm. The detector 
readout from the Level 2 buffers is received via an Asynchronous Transfer 
Mode (ATM) switch and distributed to 16 ``converter'' node PC's (CV). The main 
task of these nodes is to assemble all the pieces of the same event from the 
different sub-detector systems. The event is then passed via an Ethernet 
connection to a ``processor'' node (PR). Each processor node is a separate 
dual-processor PC. Each of the two CPU's on the node process a single event at 
a time. The processor runs a ``filter'' executable which performs the 
near-final quality reconstruction. If the executable decides to accept an 
event, it is then passed to the ``output'' nodes of the farm (OU). These nodes
 send the event onward to the Consumer Server / Data Logger (CS/DL) system for 
storage first on disk, and later on tape. A full description of the Level 3 
system is found in G\'omez-Ceballos\cite{Gomez-Ceballos:2004jk}.

For the first two thirds of the data used in this analysis, the COT track 
reconstruction algorithms as described in Section~\ref{sec-tracking} are 
performed at Level 3. The COT tracking returns $\pt, z_0, \phi_0$ and $\cot
\theta$ and combines with the $d_0$ measurement from the SVT to 
create a further improved track. The Level 1 and Level 2 trigger conditions 
are confirmed at Level 3 using improved track measurements. For the last 
one third of the data, full SVX-II tracking is available, and the trigger
conditions are repeated using a combined COT/SVX-II fit of the track helices.

\subsection{\bcharm\ Trigger Path}
\label{sec-path}
The \bcharm\ trigger path requires two tracks from the SVT and cuts on the 
$P_T$, the minimum and maximum $d_0$, the differences between the tracks' 
parameters, such as $\Delta \phi_0$, $\Delta z_0$, and \lxy. 
$\Delta \phi_0$ is defined as the opening angle between the track pair in the 
\rphi\ plane. $\Delta z_0$ is the $z_0$ difference of the track pair. \lxy\ is 
the distance of the intersection of two tracks with respect to the SVT beam 
line projected on the direction of the total momentum vector in the \rphi\ 
plane. The cuts at Level 1--3 triggers of {\tt B$\_$CHARM Scenario A} trigger 
path are described below. 

The trigger requirements are:
\subsubsection{Level 1}
  \begin{itemize}
  \item a pair of opposite charged XFT tracks
  \item each XFT track transverse momentum \pt\ $>$ 2.04 \gevc
  \item scalar sum $\pt^1 + \pt^2$ $>$ 5.5 \gevc
  \item $\Delta \phi_0$ $<$ 135 \degs
  \end{itemize}

\subsubsection{Level 2}
\begin{itemize}
  \item a pair of opposite charged SVT tracks
  \item each SVT track satisfies:
  \begin{itemize}
  \item  SVT track fit $\chi^2 <$ 25 
  \item \pt\ $>$ 2 \gevc
  \item  SVT impact parameter: 120 $\mu$m $\leq d_0(SVT) \leq$ 1000 $\mu$m 
  \end{itemize}
  \item scalar sum $\pt^1 + \pt^2$ $>$ 5.5 \gevc
  \item 2\degs\ $<$ $\Delta \phi_0$ $<$ 90\degs\
  \item $\lxy \geq $ 200 $\mu$m, this cut was added starting with run 150010
\cite{lucchesi:bsyield} 
\end{itemize}

\subsubsection{Level 3}
 The following cuts do not change for the whole period of taking data:
\begin{itemize}
 \item each Level 3 track: \pt\ $>$ 2 \gevc\ and $|\eta|$ $<$ 1.2
  \item scalar sum $\pt^1 + \pt^2$ $>$ 5.5 \gevc
  \item 2\degs\ $<$ $\Delta \phi_0$ $<$ 90\degs\
  \item $\lxy \geq $ 200 $\mu$m  
  \item $\Delta z_0$ $<$ 5 cm
\end{itemize}
The following cuts are different for period I and II. All the number of time 
integrated luminosities are after the good run selection.

Period I: 9$^{th}$ February 2002 to 19$^{th}$ May 2003, Runs 138809--163113, 
$\sim$~120~\pbarn.
 \begin{itemize}
 \item Tracking algorithm at Level 3 uses only the COT hits and Level 3 tracks 
should be matched to Level 2 tracks found by the SVT
  \begin{itemize}
  \item Track azimuthal angle difference: $\Delta(\phi_{0L3}-\phi_{0SVT})$ 
	$<$ 0.015 radians
  \item Curvature difference: $\Delta(C_{L3}-C_{SVT})$ $<$ 0.00015 cm$^{-1}$ 
  \end{itemize}
 \item  120 $\mu$m $\leq d_0(L3) \leq$ 1000 $\mu$m, $d_0$ is calculated using 
SVT beamline 
\end{itemize}

Period II: 19$^{th}$ May 2003 to 6$^{th}$ September 2003, Runs 163117--168889, 
$\sim$~50~\pbarn.
\begin{itemize}
  \item Tracking algorithm at Level 3 uses both the SVX and COT hits 
      \begin{itemize}
       \item require at least 3 hits in different SVX layers
       \item no attempt to match Level 3 tracks to SVT tracks
      \end{itemize}
  \item  80 $\mu$m $\leq d_0(L3) \leq$ 1000 $\mu$m, $d_0$ is calculated using 
SVT beamline 
 \end{itemize}

Data derived from the above trigger path were written to the tape for 
subsequent reconstruction and physics analysis.

\chapter{Data Samples}
\label{ch:rec}
Data used in this analysis are collected with the upgraded CDF detector 
from 9$^{th}$ February 2002 to 6$^{th}$ September 2003 and cover runs 138809 
through 168889. This period corresponds to an integrated luminosity of 
$\sim$237 \pbarn. 
In this chapter, we present details of how we arrive at our final data 
sample, optimize the cuts. We describe the 
data sample used for this analysis in Section~\ref{sec-datasample}. 
The cuts that are applied to obtain a clean 
signal with low background are explored in Section~\ref{sec-opt}. 

\section{Data Sample}
\label{sec-datasample}
\subsection{Overview}
\begin{figure}[tbp]
     \begin{center}
        \includegraphics[width=200pt, angle=0]
	{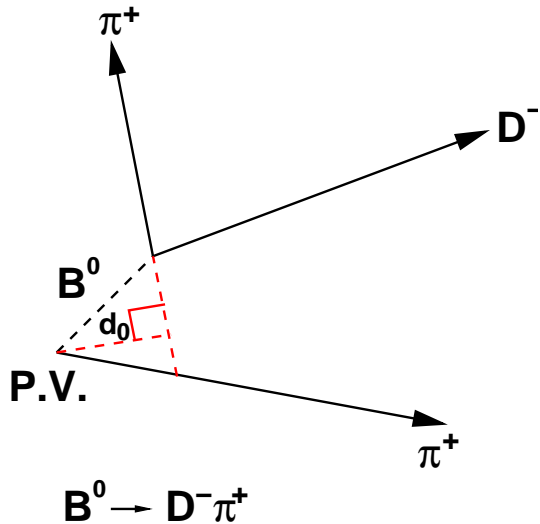}
     \end{center}
        \caption[Example of a \Bd\ decay]
	{Example of a \Bd\ decay: 
	$\pi^+$ from the \Bd\ decay has a larger impact parameter, $d_0$, 
	than that of the track produced at the primary vertex, $\pi^-$.}
     \label{fig:bigd0}
  \end{figure}

 We wish to reconstruct the decays of the \Bd\ and the \Lb. 
The \ctau\ of the \Bd\ and \Lb\ are about 460 and 370 $\mu$m, respectively.
The \ctau\ of \Dzero, \D, \Ds\ and \Lc\ are about 120, 150, 300 and 60 $\mu$m. 
Long lived $K_s$ and $\Lambda$ have much longer lifetimes, 2-8 cm. Therefore, 
a minimum requirement on the distance between the beamline and the secondary 
vertex (decay length) removes contamination of short-lived charm hadrons. 
The daughter particles from the \B\ decays also tend to have a larger impact 
parameter ($d_0$) than the tracks produced at the primary vertex, as 
illustrated in Figure~\ref{fig:bigd0}, but smaller $d_0$ than the daughter 
particles of $K_s$ and $\Lambda$. Consequently, a minimum and a maximum cut on 
the track $d_0$ rejects the background from the primary tracks, $K_s$ and 
$\Lambda$, or secondary tracks from the particle interaction with the detector
 material. The newly developed SVT trigger cuts on the decay length and the 
track $d_0$. It is the best trigger for distinguishing \B\ decays from other 
physics processes. Among all the SVT trigger paths, we find \bcharm\ most 
suitable for this analysis. Section~\ref{sec-path} presents the definition of 
the \bcharm\ trigger path. The goal of any measurement is to have the smallest 
possible uncertainty. By recording data from the same trigger, the systematic 
uncertainties common to both modes cancel. We therefore select all our data 
from the \bcharm\ trigger path.

The data from \bcharm\ are processed with the {\tt Production} executable, 
version 4.8.4, and compressed into the secondary datasets {\tt hbot0h} and 
{\tt hbot1i}. The total size of {\tt hbot0h} and {\tt hbot1i} is about 10 
Terabytes (150M events), which is too big to be analyzed quickly multiple 
times. We apply loose selection cuts and reduce {\tt hbot0h}, {\tt hbot1i} to 
smaller, tertiary datasets. Section~\ref{sec-skim} discusses the data skimming.
 Then we optimize the analysis cuts using the tertiary datasets in 
Sections~\ref{sec-opt}.

\subsection{Data Skimming}
\label{sec-skim}
Before skimming the data, we exclude the runs with an incorrect alignment 
table (152595--154012) in {\tt hbot0h}. The alignment table contains the 
parameters for the positions of the COT and the SVX. The data during runs 
152595--154012 are reprocessed with the correct alignment table and collected 
into {\tt hbot1i}. We further require the following systems declared good by 
the CDF Data Quality Monitoring group: SVX, COT, CMU, Cherenkov Luminosity 
Counters (CLC) and the Level 1--3 triggers. We exclude the runs when SVX is 
off and when there are known high voltage or trigger problems in the COT, CMU 
or SVT. By making these requirements, the Monte Carlo program can better 
reproduce the response of these detectors (see Section~\ref{sec-mccom}). 
After making the good run selection, the integrated 
luminosity reduces from 237~\pbarn\ to 171.5~\pbarn. 

The skimming program starts by storing a set of offline reconstructed tracks 
which satisfy the quality requirements on: \pt, the number of COT hits in the 
axial and stereo layers, the number of SVX \rphi\ hits, and the impact 
parameter. Then, the tracks that are matched to those found by the SVT or to 
muon stubs (CdfMuon) in the CMU, are marked for further use. After saving the 
SVT and muon information, we begin our reconstruction by identifying the charm
 candidates: \Dstar, \D\ and \Lc. 

We first cut on the raw mass of the charm candidates, where the raw mass is 
calculated using the track momentum at the point of closest approach to the 
beam line. We determine the charm (tertiary) vertex by performing a vertex fit 
with the CTVMFT package developed by Marriner~\cite{cdfnote:1996}. CTVMFT 
determines the decay vertex by varying the track parameters of the daughter 
particles within their errors, so that a $\chi^2$ between the track trajectory 
and the points is minimized. We cut on the fitted charm mass and 
$\chi^2_{\rphi}$, where ``$\chi^2_{\rphi}$'' is the $\chi^2$ returned from 
the fit in the \rphi\ plane. 

The charm candidate is then combined with an additional track to form the 
\B\ candidate. The additional track has a minimum $P_T$ requirement of 1.6 
\gevc. Once we have a valid fourth track, we cut on the raw mass of the four 
tracks. The mass window varies depending on whether the fourth track is 
matched to a muon stub. We require that two of the four tracks from the 
reconstructed \B-hadron candidate each matches an SVT track. We confirm the 
trigger by requiring the matched SVT tracks to pass the {\tt Scenario A} cuts 
listed in Section~\ref{sec-path}. We then perform a four-track vertex fit. The 
four-track vertex fit includes a constraint that the tertiary vertex points 
to the secondary vertex in the \rphi\ plane. After the vertex fit, we cut on 
$P_T$ of the charm, the $\chi^2_{\rphi}$, \pt\ and fitted mass 
of the four tracks.  

After applying the requirements discussed above for each signal mode, we 
reduce the secondary datasets, {\tt hbot0h} and {\tt hbot1i}, by a factor of 25
(from 10 to 0.4 Terabytes). The reduced datasets are then written to tape 
for further use. Detailed information about the skimming is found in the 
reference by the author~\etal~\cite{cdfnote:strip}. In Section~\ref{sec-opt}, 
we present our analysis cut optimization with the reduced datasets.

\section{Signal Optimization}
\label{sec-opt}
 From the reduced datasets we reconstruct our signals:
 \begin{itemize}
  \item \dstarhad\ and \incdstarsemi, where \seqdstar, \seqdzero
  \item \dhad\ and \incdsemi, where \seqd
  \item \lbhad\ and \inclbsemi, where \seqlc
 \end{itemize}
 The reconstruction procedure is similar to that described
in Section~\ref{sec-skim} and Yu~\cite{cdfnote:strip}. 
The following cuts are studied more carefully and optimized 
:
\begin{itemize}
\item $\chi^2_{\rphi}$ of \B\ and charm vertex fit 
\item \pt\ of \B\ and charm candidates
\item \ctau\ of \B\ and charm candidates: $\lxy\times\frac{M}{\pt}$. 
\end{itemize}
Our semileptonic signals are larger than the hadronic signals, and 
the statistical uncertainty of the relative branching fraction measurement 
is dominated by the uncertainty of the number of events in the hadronic 
signals. Therefore, we optimize the hadronic mode only and apply the 
optimized cuts to the semileptonic mode. The optimized quantity is the 
significance, $\frac{S}{\sqrt{S+B}}$, where ``$S$'' is the number of signal 
and ``$B$'' is the number of background events. 

\begin{figure}[tbp]
     \begin{center}
        \includegraphics[width=250pt, angle=0]
{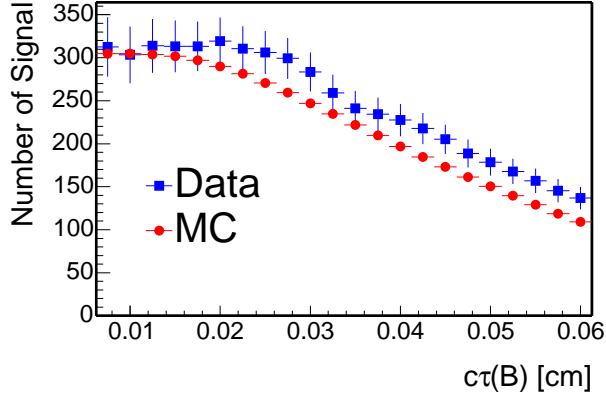}	
     \end{center}
        \caption[Signal optimization: number of signal in the MC and data 
	after applying the normalization factor]
	{Signal optimization: number of signal in the MC and data as a 
	function of the $\ctau(B)$ cut after 
	applying the normalization factor for 
	\lbhad.}
     \label{fig:signalratio}
  \end{figure}

For our optimization, the amount of signal, ``$S$'' comes from a MC as 
described in Section~\ref{sec-mccom}. The reason for using MC signal is 
that the data signal is small and susceptible to statistical biases. We 
generate MC with about 20 times more events than the data and eliminate this
 problem. In order to scale the significance close to the true value measured 
from the data, we apply a normalization factor $f_c$ on the signal MC,
\begin{equation}
f_c = \frac{S_\mathrm{data}}{S_\mathrm{MC}},
\end{equation}
where $S_\mathrm{data}$ and $S_\mathrm{MC}$ are the amount of the signal 
found in the data and MC after applying loose cuts, and 
\begin{equation}
S = f_c \times S_\mathrm{MC}.
\end{equation}
 Note that even though ``$S$'', after scaling, is the same as 
$S_\mathrm{data}$ now, the uncertainty on ``$S$'' is 
$f_c \times \sqrt{S_\mathrm{MC}}$, is much smaller than the uncertainty of 
$S_\mathrm{data}$, $\sqrt{S_\mathrm{data}}$. Figure~\ref{fig:signalratio} 
shows a comparison of the number of signal in the data and in the MC after 
applying the normalization factor.


We evaluate the background beneath the signal peak from the data. 
We first apply loose cuts on each mode to identify a clear \Bd\ or \Lb\ peak;
  \begin{itemize}
   \item $\ctau(B)$ $>$ 50 $\mu$m 
   \item each track \pt\  $>$ 0.5 \gevc 
   \item $\pi$ from the \B\ hadron is CMU fiducial 
   \item for \dstarhad:
       \begin{itemize}
        \item 1.833 $<$ \mkpi\ $<$ 1.893 \gevcsq	
        \item 0.143 $<$ \mkpipi\ - \mkpi\ $<$ 0.148 \gevcsq
	\end{itemize}
   \item for \dhad: 1.8517 $<$ \mkpipi\ $<$ 1.8837 \gevcsq
   \item for \lbhad: 2.269 $<$ \mpkpi\ $<$ 2.302 \gevcsq	
  \end{itemize}
We require that both the muon and pion from the $B$ hadron point within 
CMU fiducial volume because we use the CMU only to identify the muons.
CMU covers the region of pseudo-rapidity ($\eta$) less than 0.6. Making the 
same fiducial requirement for the hadronic mode allows the tracking 
efficiencies from both modes to cancel. 

\begin{figure}[tbp]
     \begin{tabular}{cc}
      \includegraphics[width=180pt, angle=0]
	{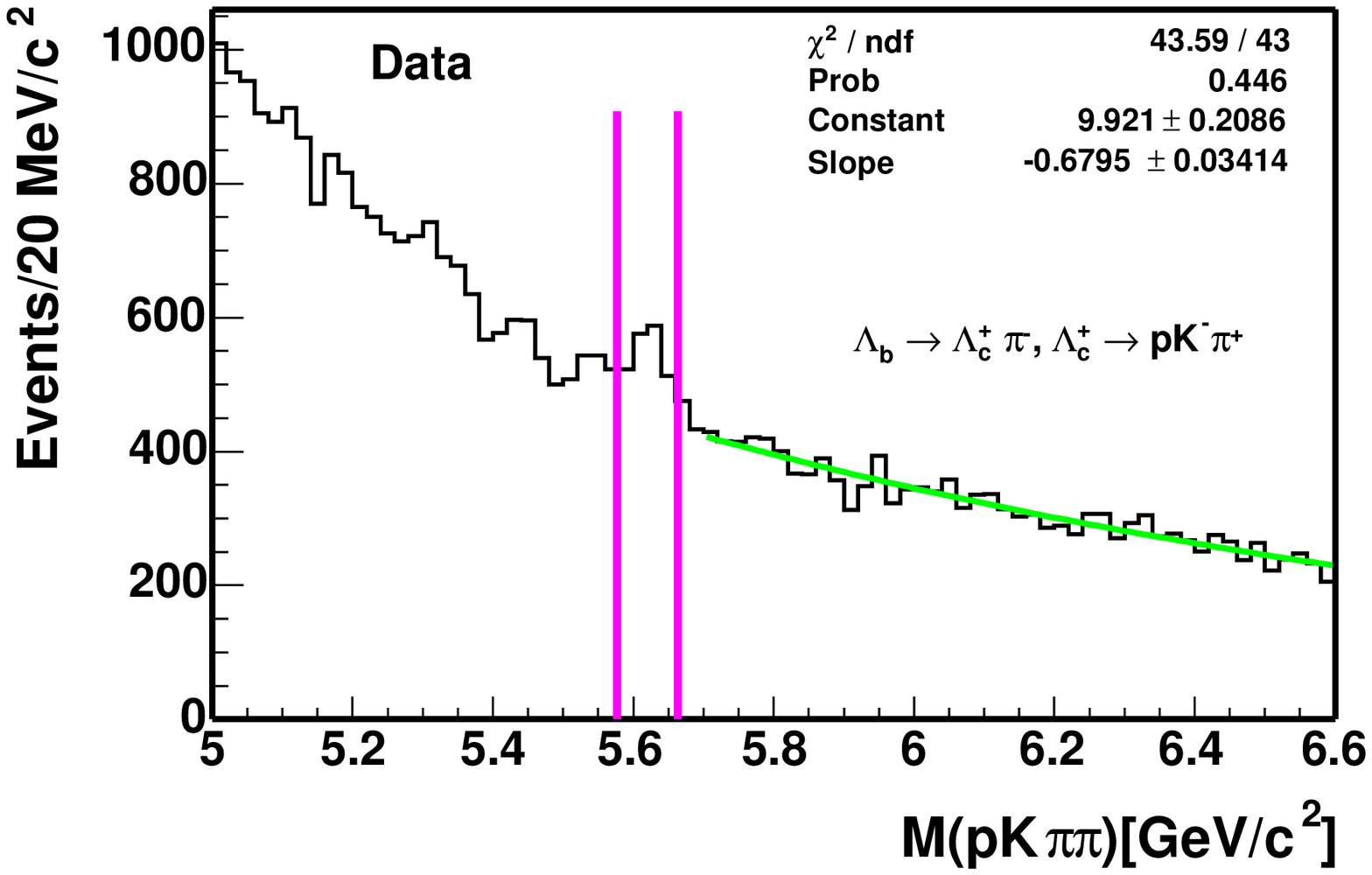} &
        \includegraphics[width=180pt, angle=0]
	{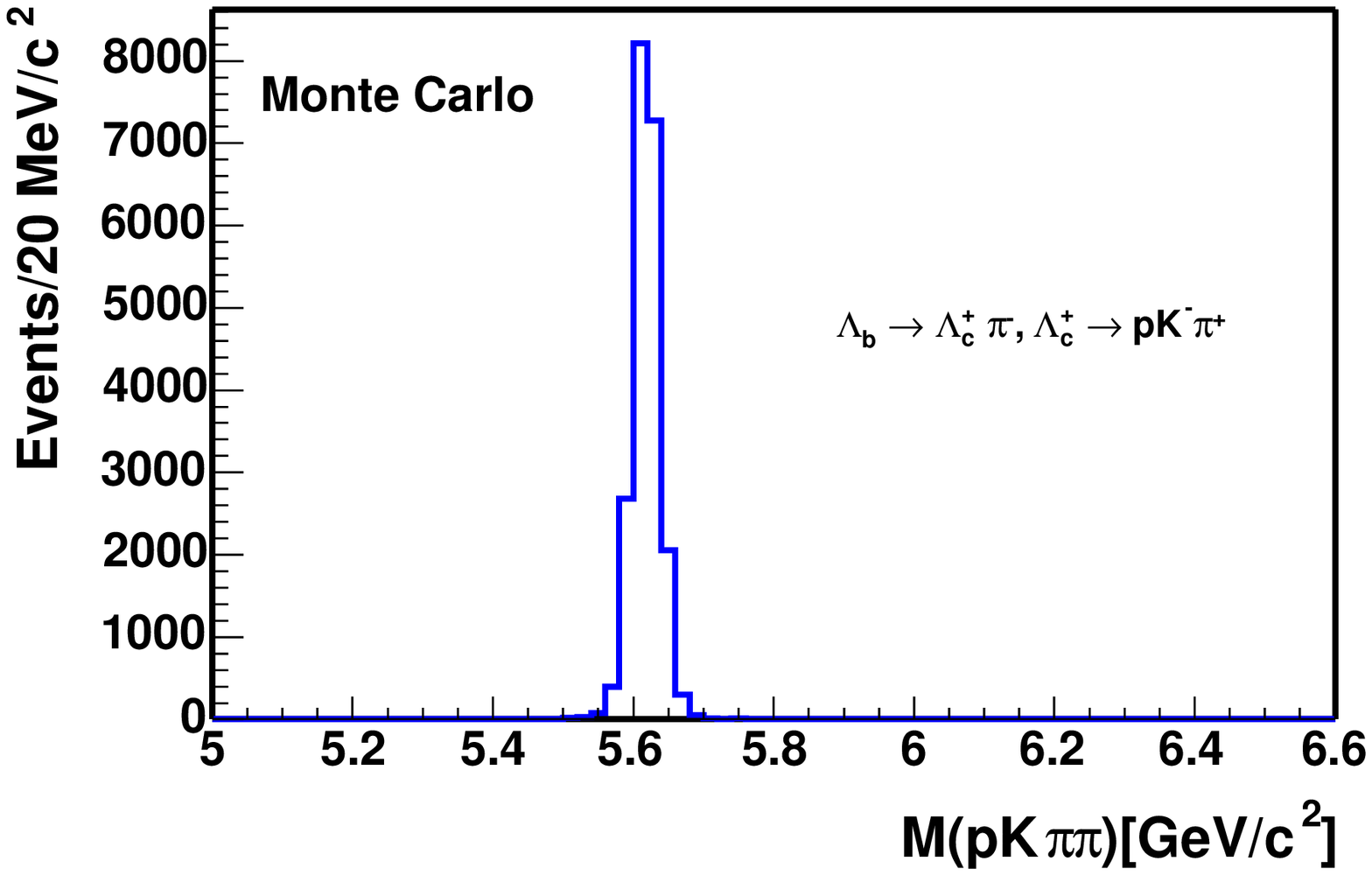}\\		
     \end{tabular}
        \caption[
	$\Lambda_c\pi$ invariant mass for  data and MC]
	{
	$\Lambda_c\pi$ invariant mass for 
         data and MC. 
	Left: data, the pink solid lines 
	indicate the signal region. Right: MC.}
     \label{fig:wrongsigndstar}
  \end{figure}

The backgrounds in the signal and in the upper mass regions are mainly 
combinatorial, and may be described by an exponential function, as we will 
see in Section~\ref{sec-massfit}. Therefore, we fit the upper mass region to 
an exponential function. Finally we extrapolate and integrate the exponential 
over the mass region of $\pm$ 3 $\sigma$ around the signal peak to obtain 
``$B$''. Figure~\ref{fig:wrongsigndstar} shows the \Lb\ mass 
distribution in the data and MC. The figure also shows the signal region we 
define and the upper mass region we fit to an exponential.


\begin{figure}[tbp]
     \begin{center}
      \includegraphics[width=140pt, angle=0]
	{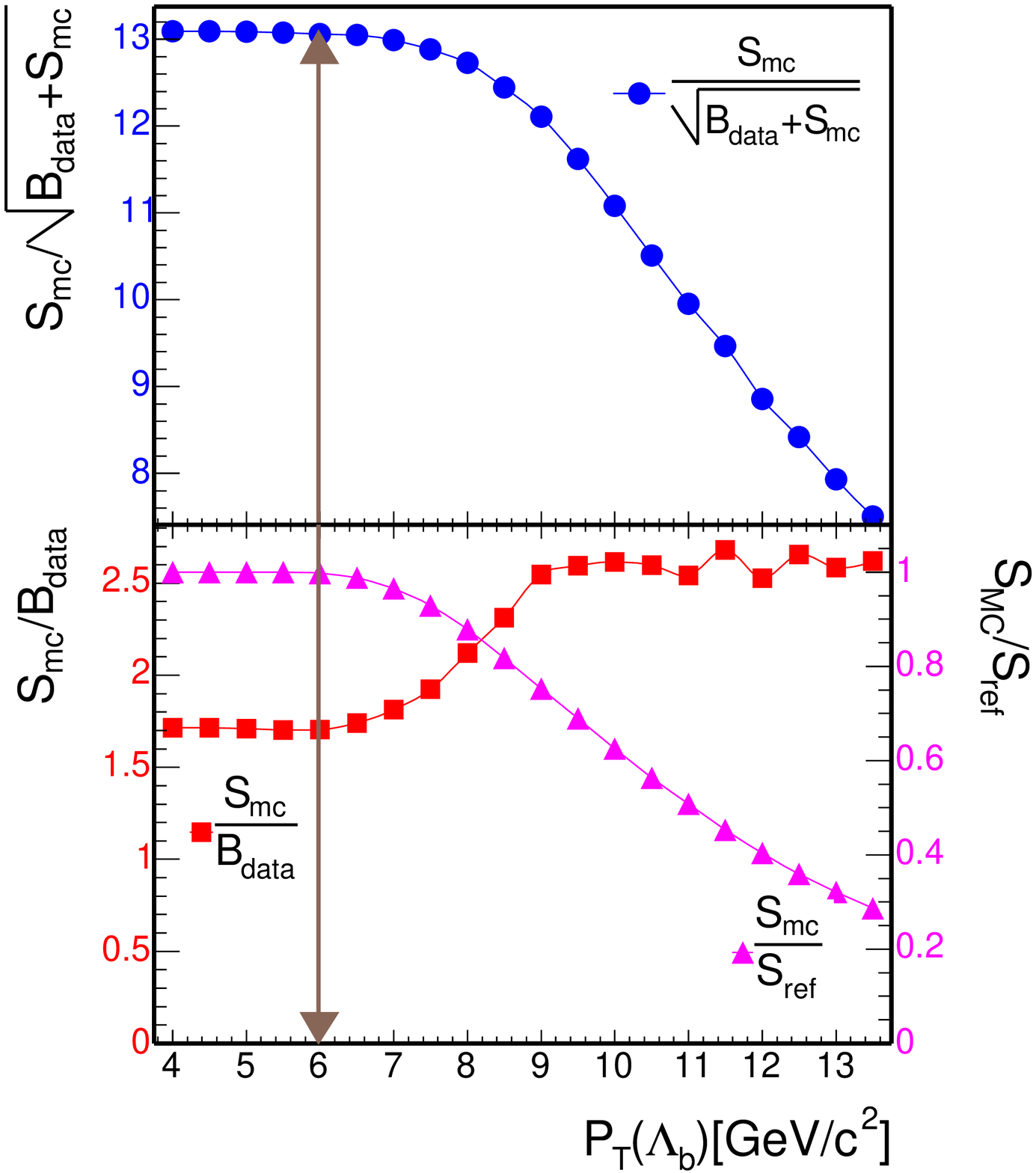}
      \includegraphics[width=140pt, angle=0]
	{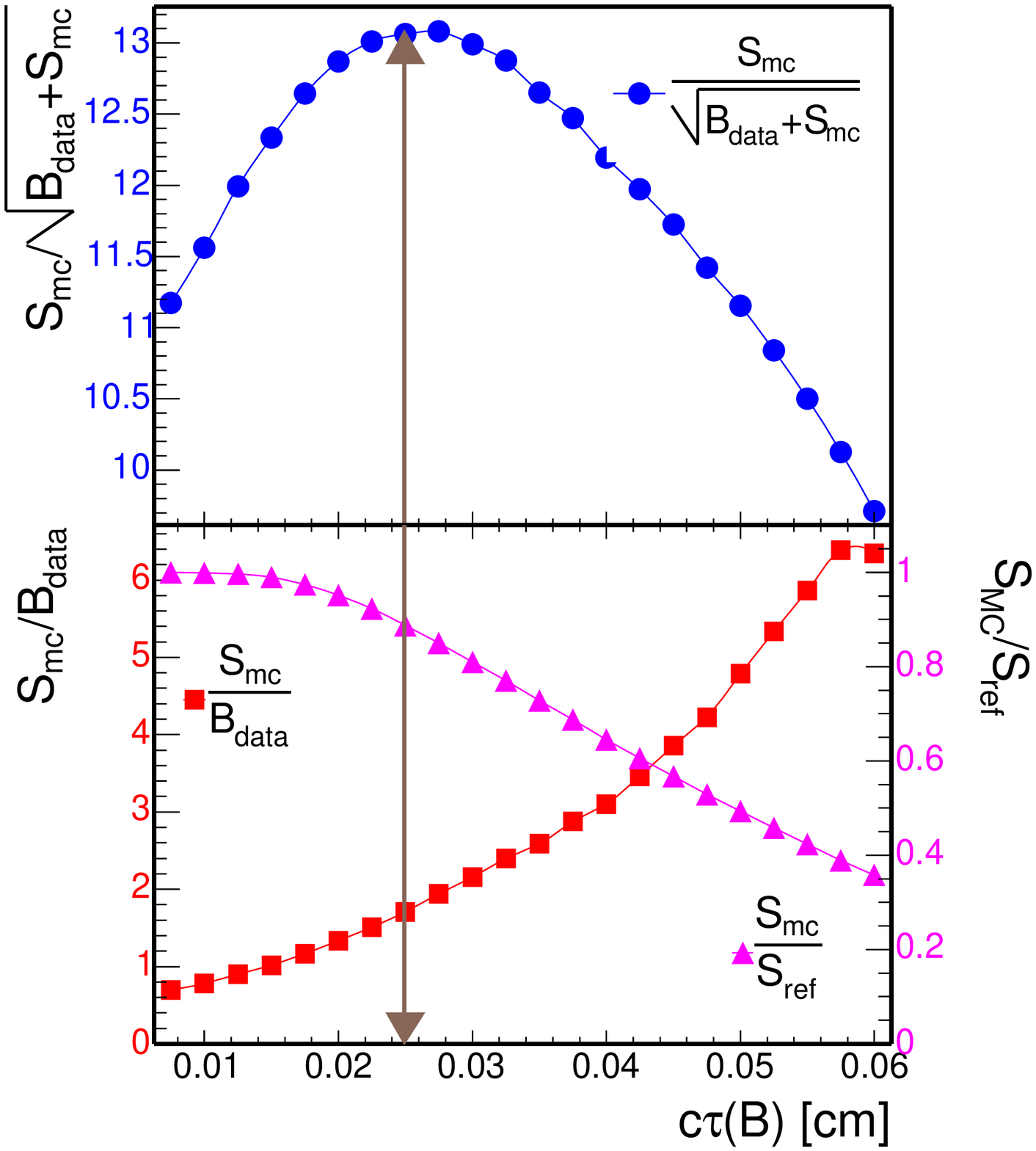}
      \includegraphics[width=140pt, angle=0]
	{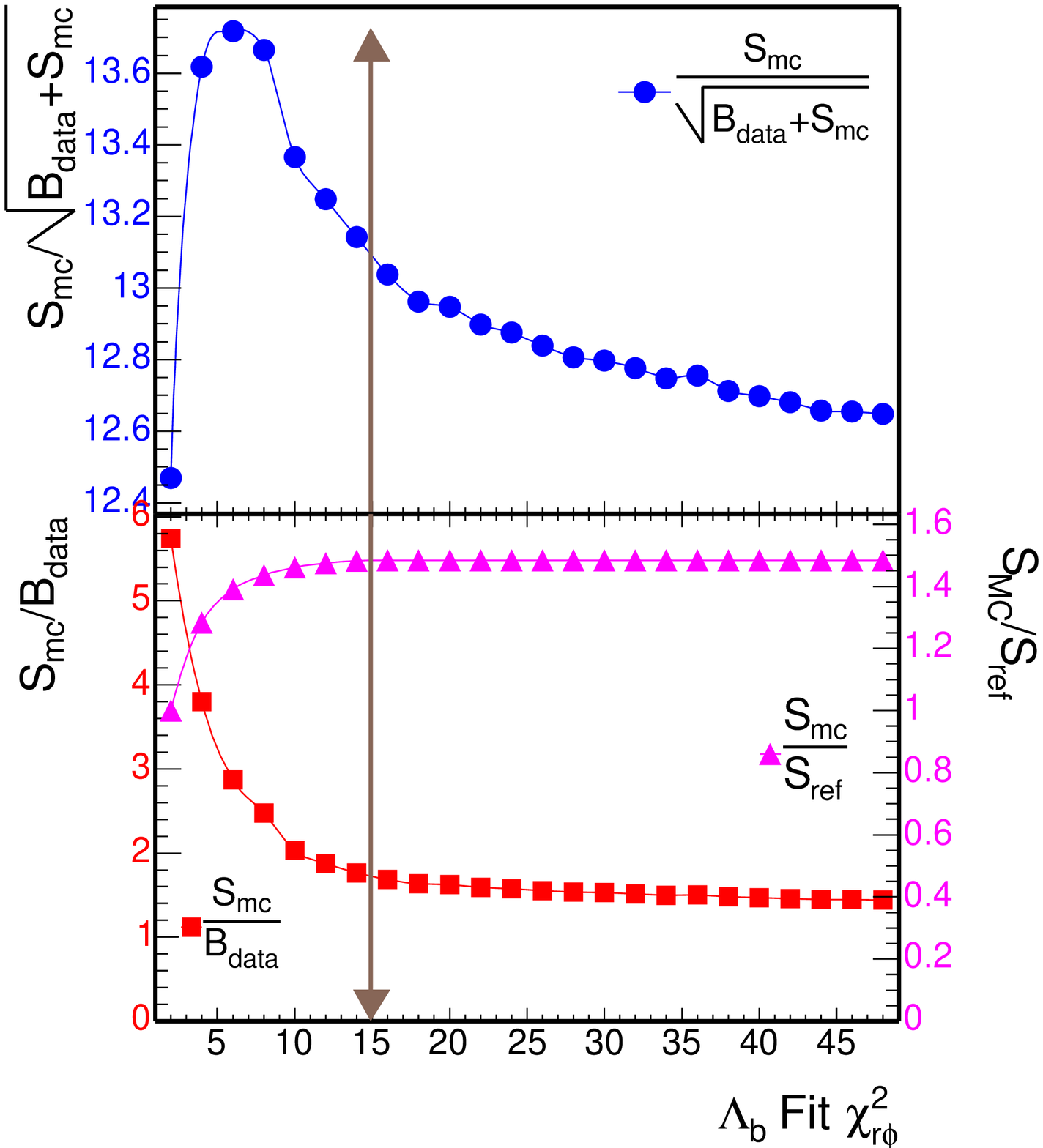}
      \includegraphics[width=140pt, angle=0]
	{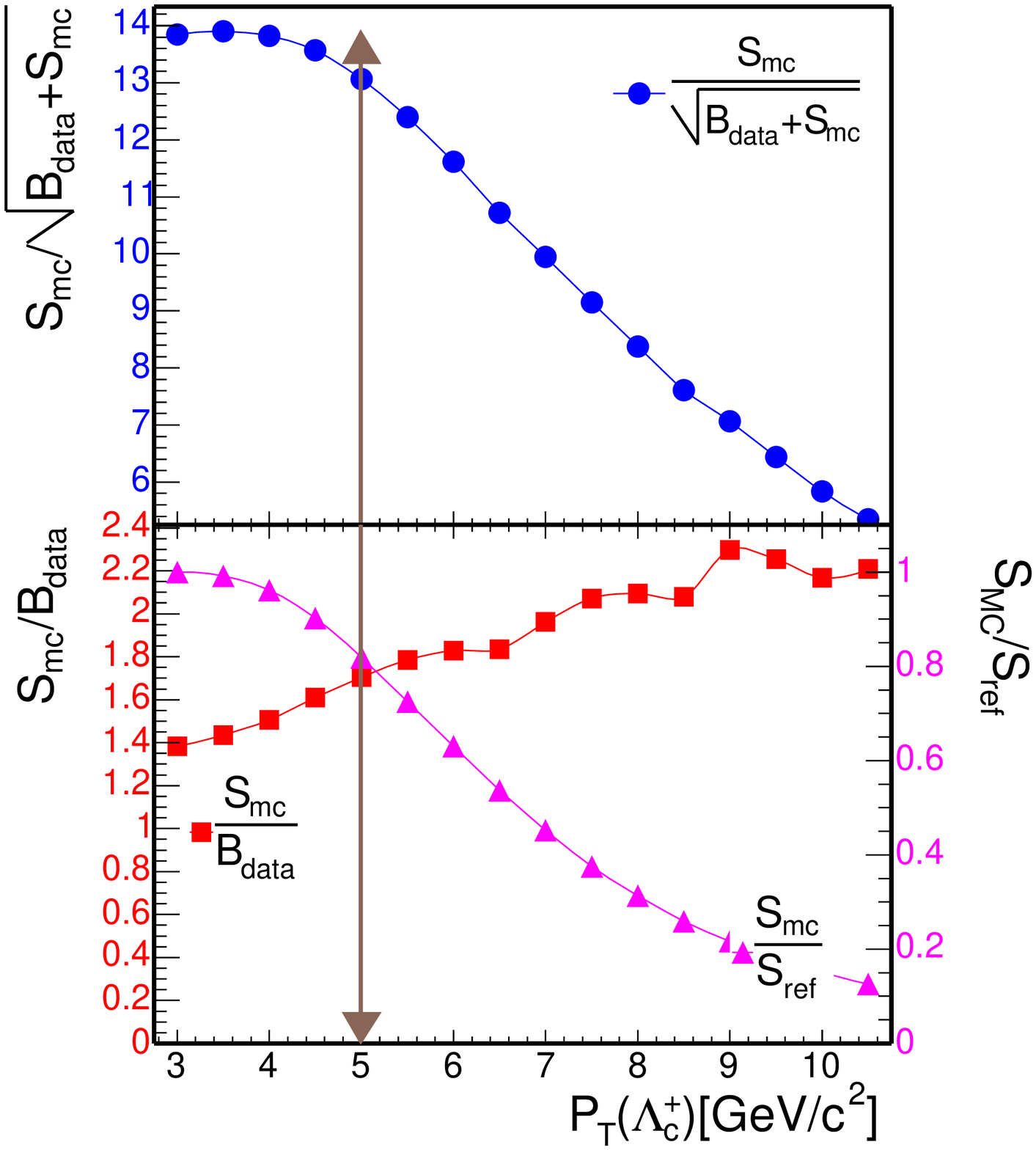}
      \includegraphics[width=140pt, angle=0]
	{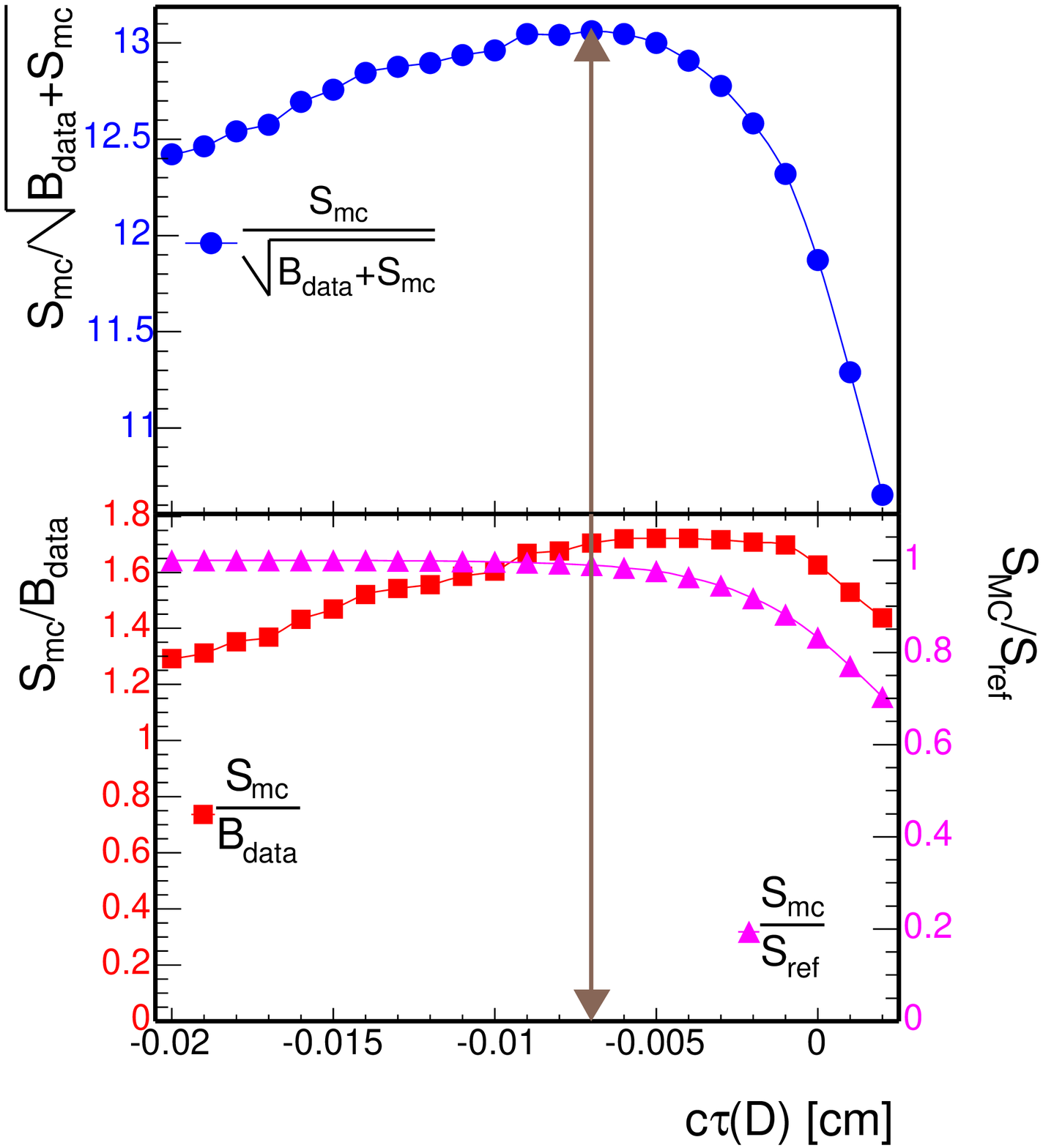}
      \includegraphics[width=140pt, angle=0]
	{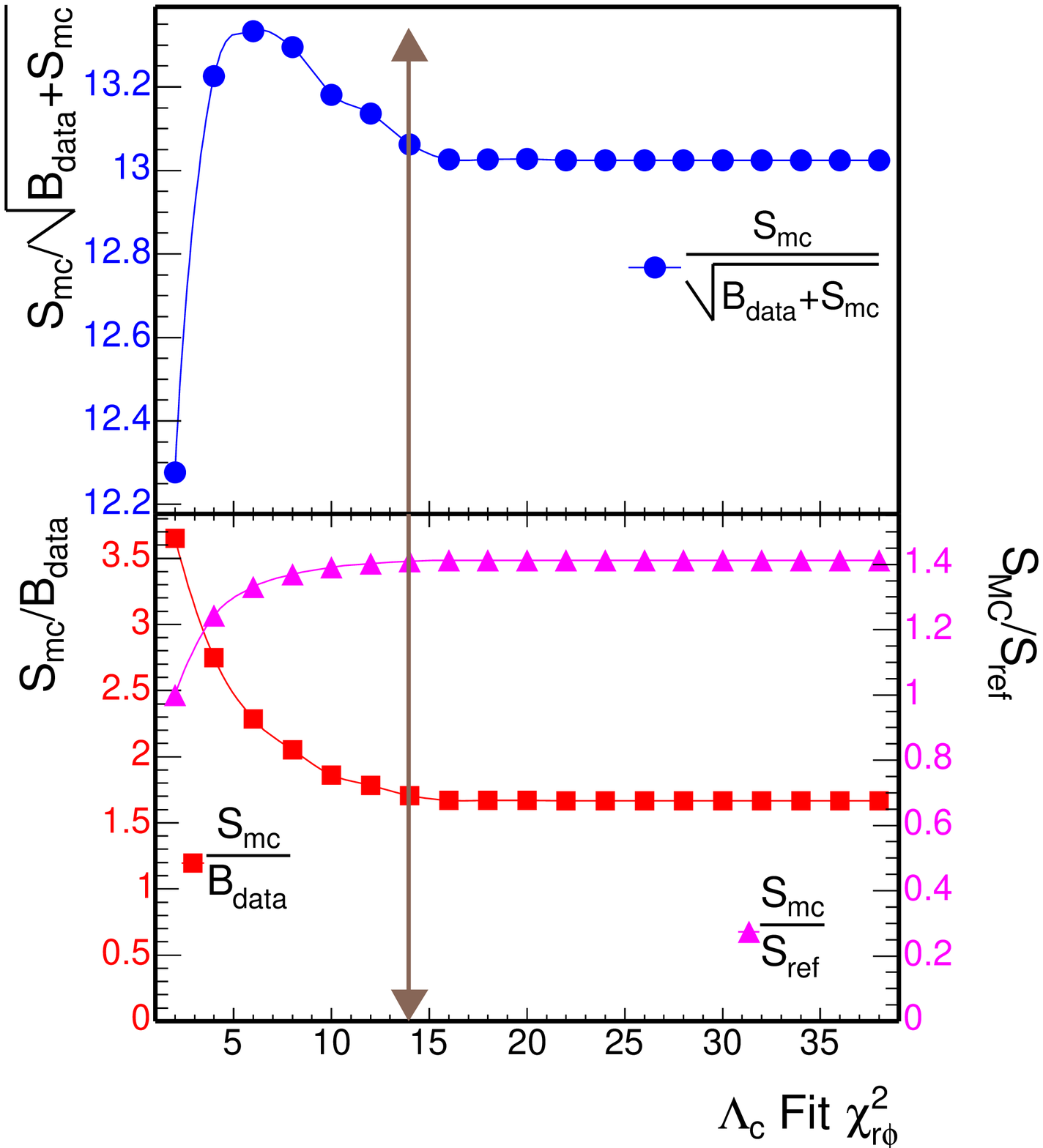}

    \caption[$\frac{S}{\sqrt{S+B}}$, $\frac{S}{B}$ and 
	$\frac{S}{S_\mathrm{ref}}$ for  \alllc\ analysis cuts] 
	{ Significance ($\frac{S}{\sqrt{S+B}}$) marked by circles, 
	signal/background ($\frac{S}{B}$) marked by squares and 
	signal/reference ($\frac{S}{S_\mathrm{ref}}$) marked by triangles 
	for cuts used in \alllc\ analysis.	
	\label{fig:siglc0}}
     \end{center}
  \end{figure}

The optimization follows an iterative procedure which passes through the 
data multiple times. In the first pass, cuts on each variable are scanned 
and optimization points are found. In the second pass, we apply the optimized 
cuts for all but the variable which is being re-optimized. We iterate this 
process several times until the optimization points become stable; usually 
twice is enough. Figures~\ref{fig:siglc0} shows 
$\frac{S}{\sqrt{S+B}}$, $\frac{S}{B}$ and $\frac{S}{S_\mathrm{ref}}$ from 
the optimization of \lbhad\ mode, as a 
function of each cut variable, where $S_\mathrm{ref}$ is the number of signal 
events at the starting point. Tables~\ref{t:fanacut0}--~\ref{t:fanacut1} list 
the final analysis cuts. Note that because the MC and the data $\chi^2_{\rphi}$
 do not agree well, as shown in Section~\ref{sec-datamc}, we choose to make a 
loose cut at the plateau region of the significance. The final analysis cuts 
for the \pt\ of charm hadrons are tighter than the optimization points. The 
tighter cuts arise from the 4 \gevc\ \pt\ threshold applied to the $c$-quark 
in the MC sample for our semileptonic background study 
(see Section~\ref{sec-cbmethod}). This \pt\ threshold makes the reconstruction 
of charm hadrons below 4 \gevc\ inefficient. The MC sample is produced by the 
CDF \B\ group and it would take a prohibitive amount of CPU time to generate a 
new sample more suitable for our analysis. Therefore, we increase the \pt\ cut 
of our charm hadrons to 5 \gevc. As the significance of the charm is a slowly 
varying curve, changing the cuts has little effect on the signal yield.

\begin{figure}[tbp]
 \begin{center}
 \begin{tabular}{cc}
 \includegraphics[width=200pt, angle=0]
	{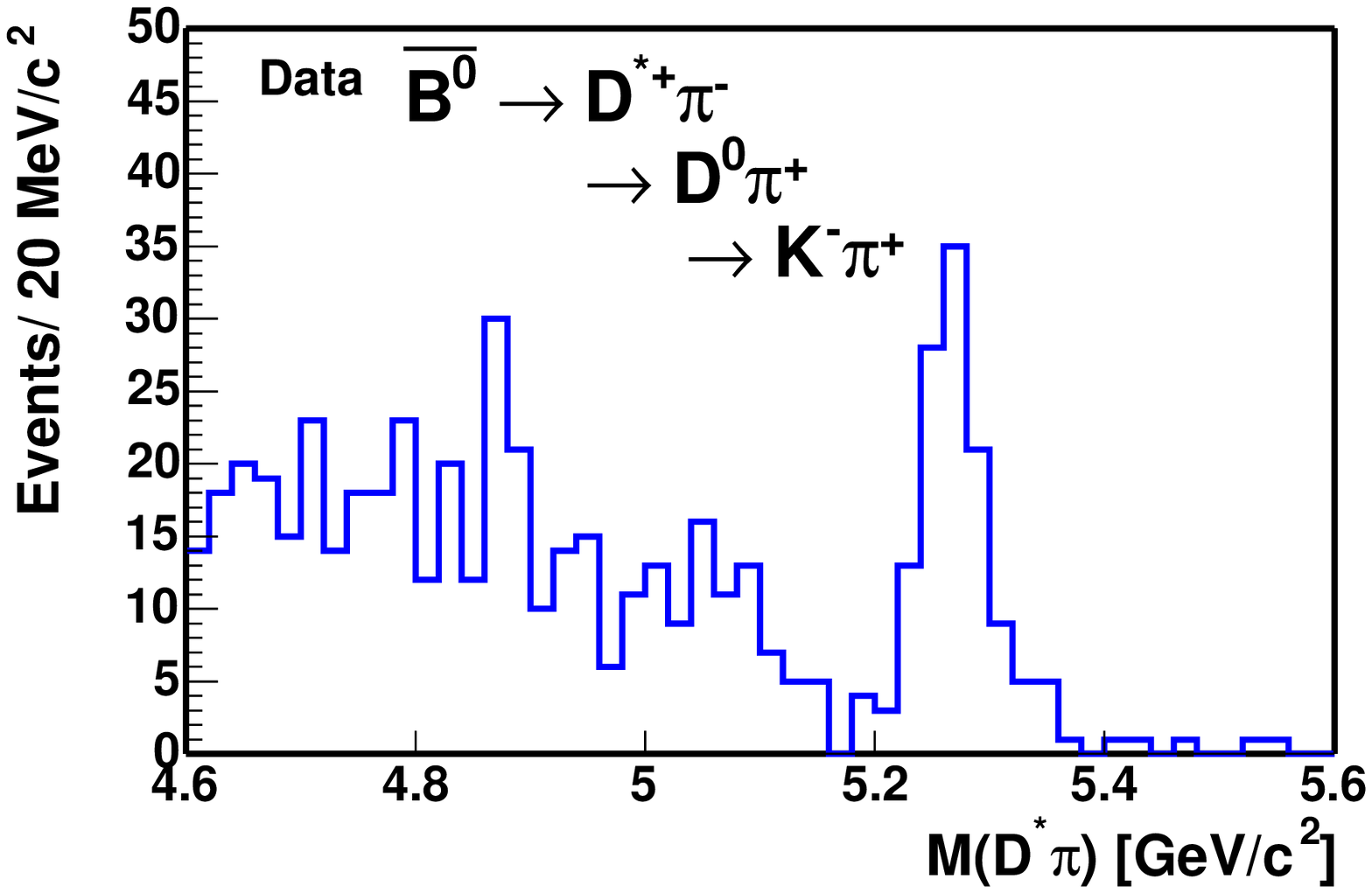} & 
 \includegraphics[width=200pt, angle=0]
	{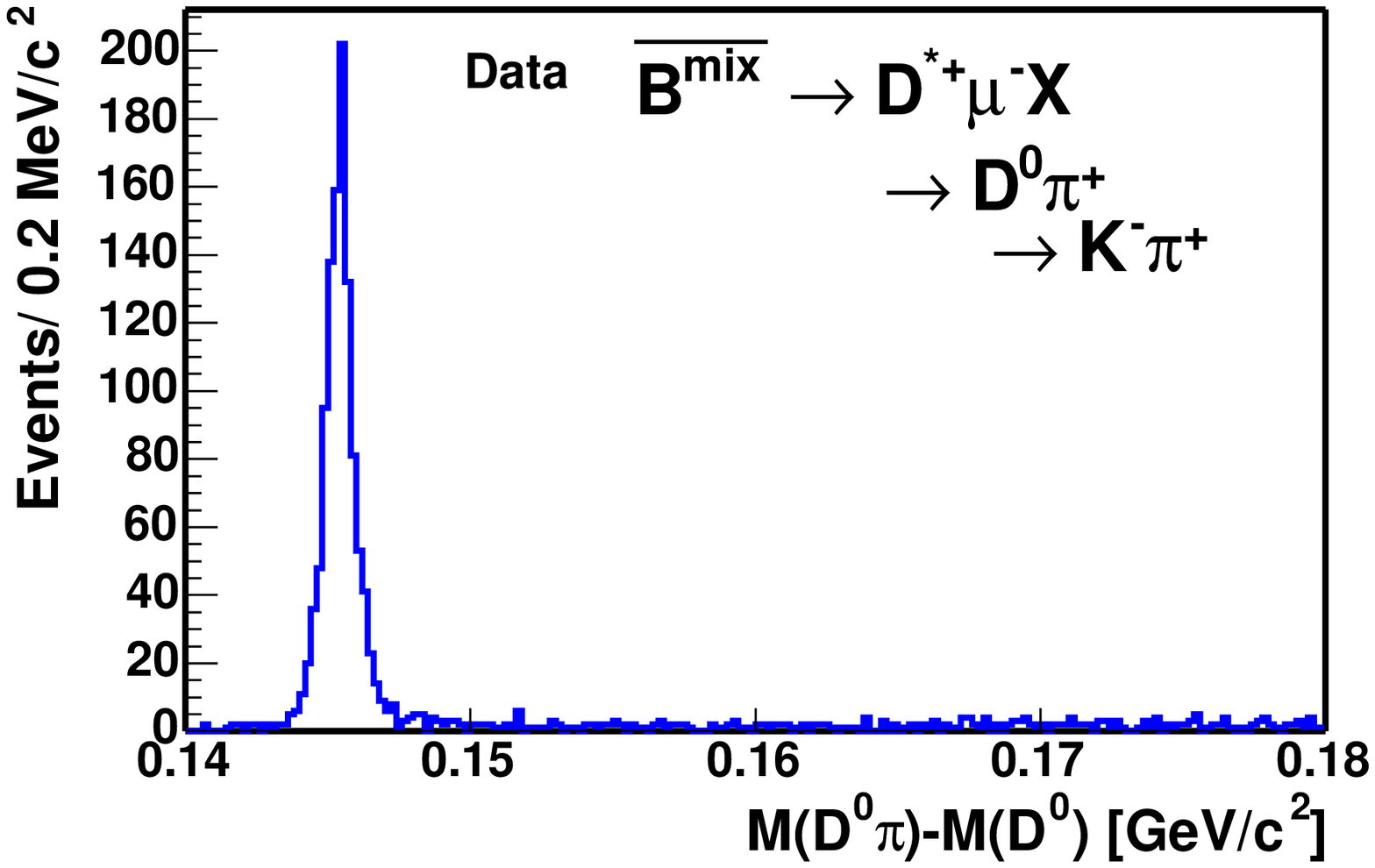}\\
 \includegraphics[width=200pt, angle=0]
	{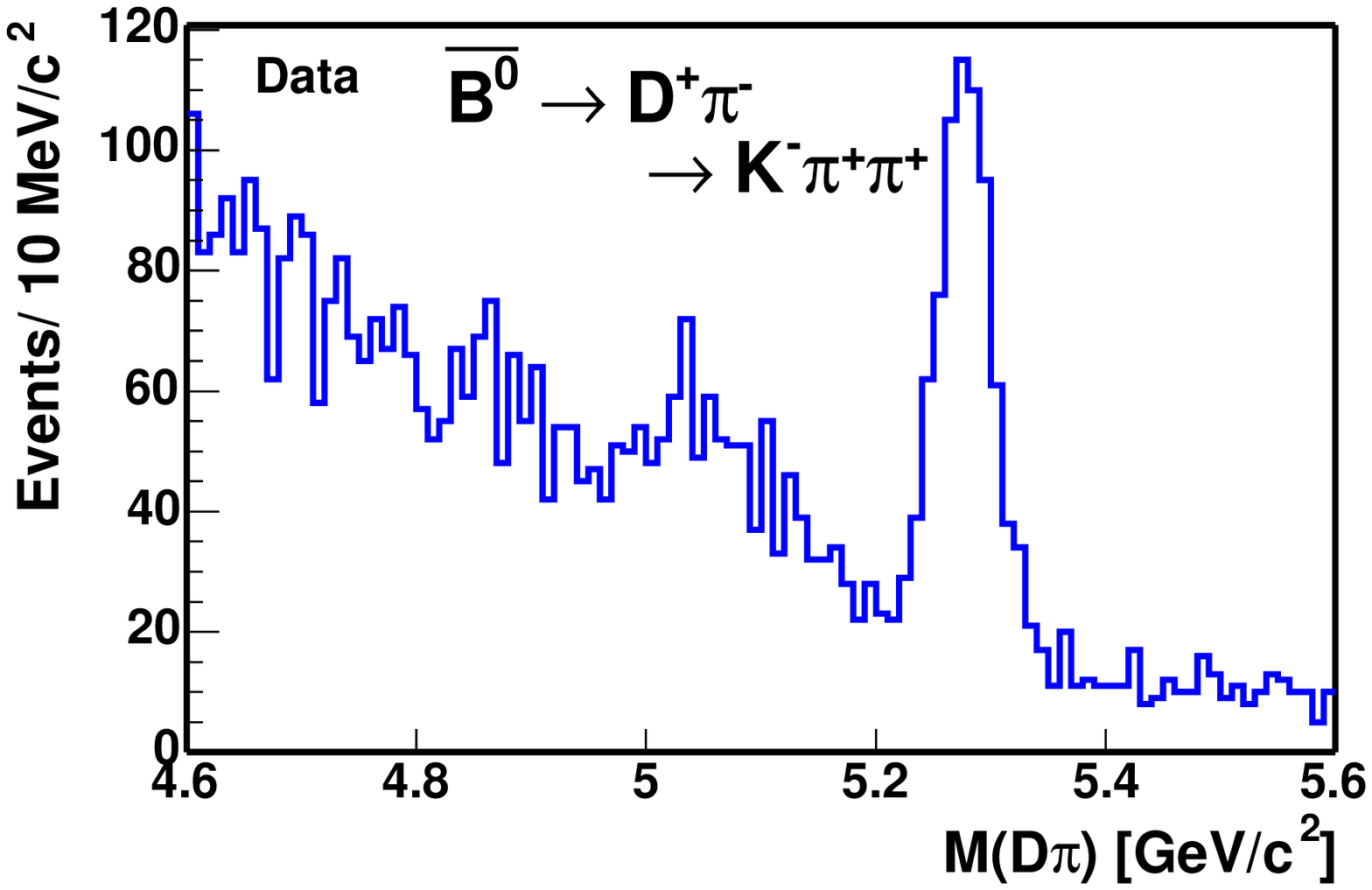}& 
 \includegraphics[width=200pt, angle=0]
	{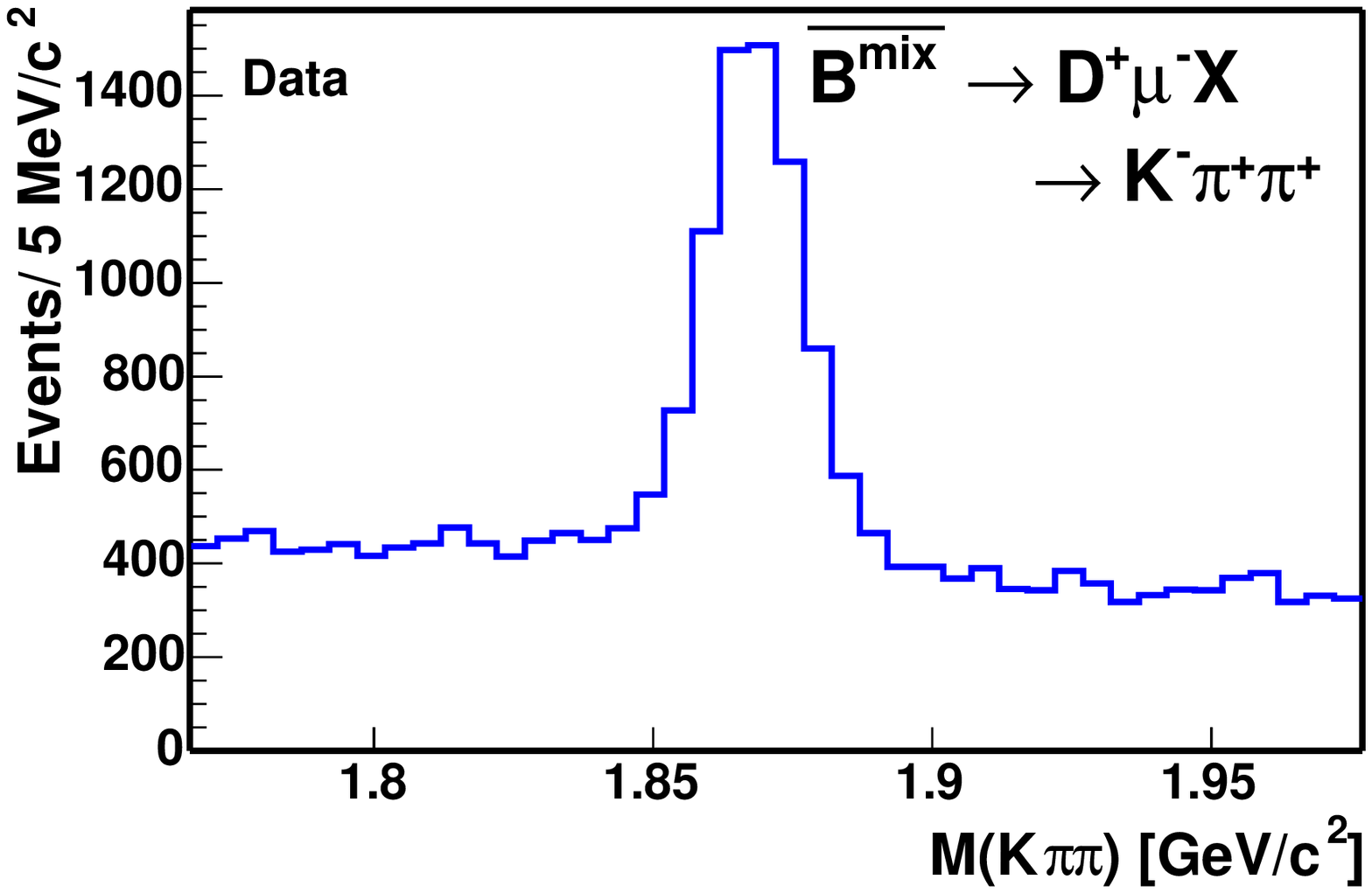}\\
 \includegraphics[width=200pt, angle=0]
	{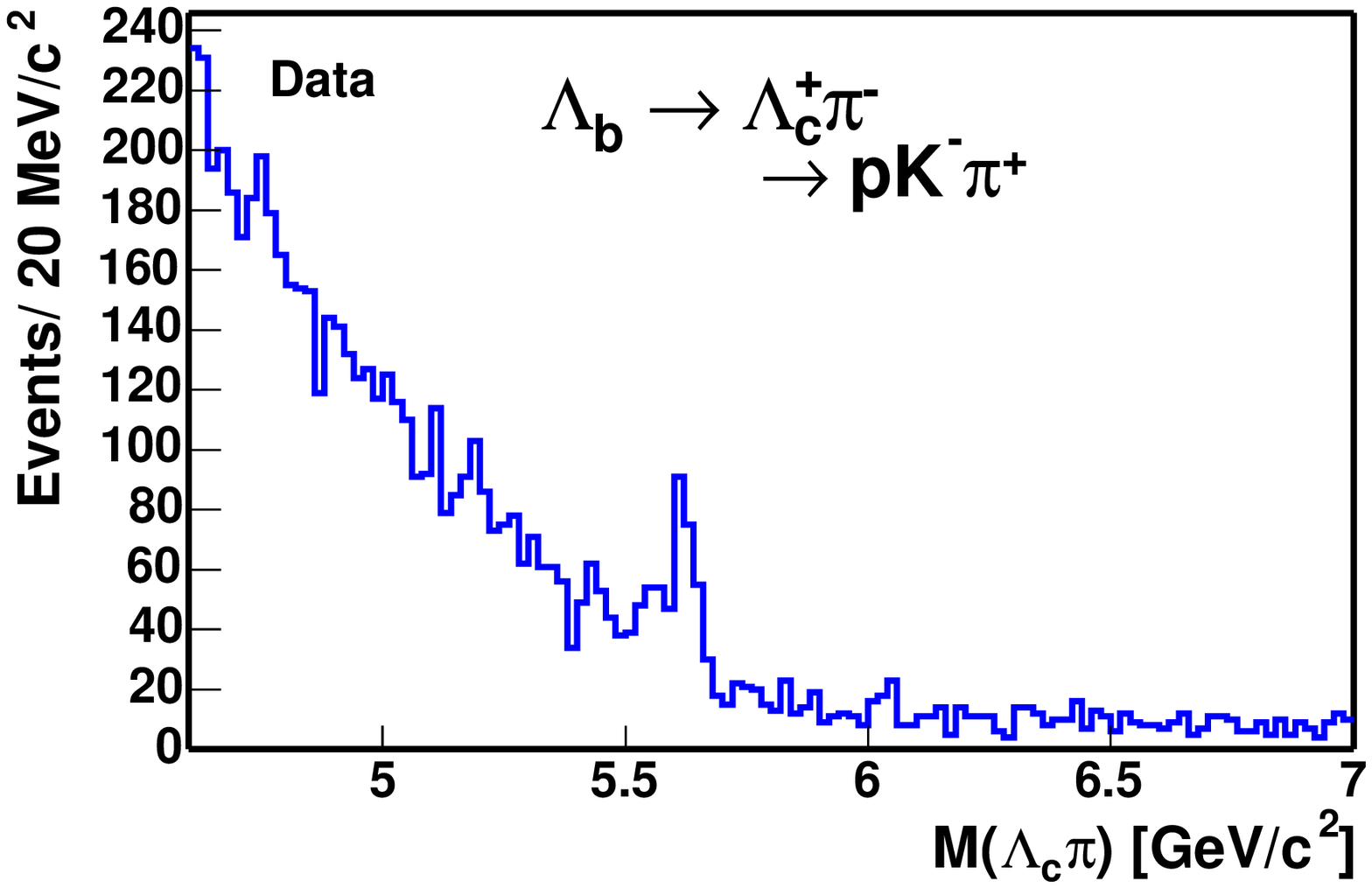} &
 \includegraphics[width=200pt, angle=0]
	{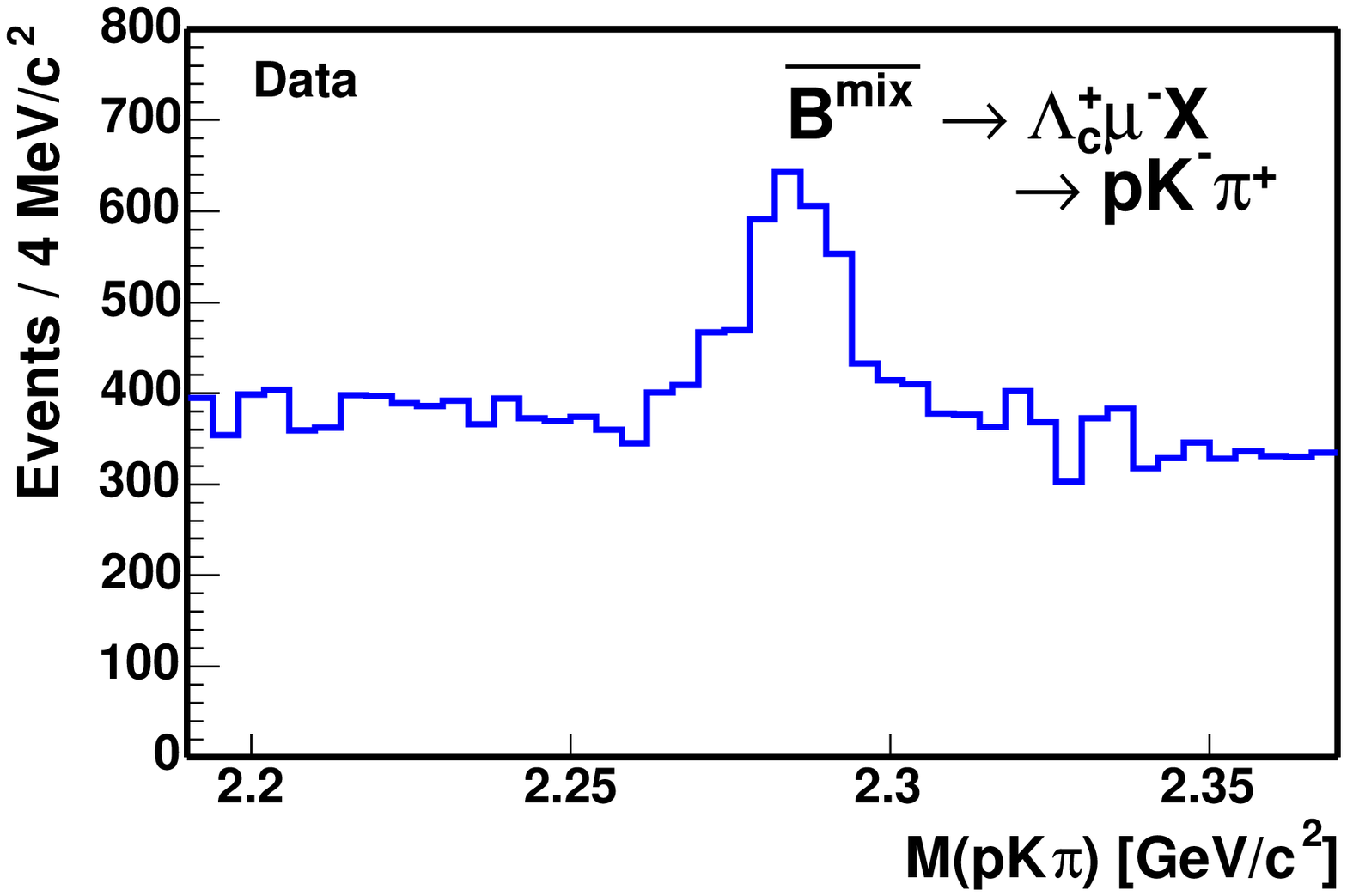} \\
 \end{tabular}
  \caption[$M_{D^*\pi}$, $M_{\Dzero\pi}-M_{\Dzero}$, $M_{D\pi}$, \mkpipi, 
$M_{\Lambda_c}\pi$ and \mpkpi\ after all the analysis cuts]
  {charm+$\pi$ and charm mass spectra from our signals after all cuts.
  From the top left to the bottom right are: $M_{D^*\pi}$, 
	$M_{\Dzero\pi}-M_{D^0}$, 
  $M_{D\pi}$, \mkpipi, $M_{\Lambda_c}\pi$ and \mpkpi.}
 \label{fig:allhisto}
\end{center}
 \end{figure}

In addition to the cuts which are optimized above, we also require that the 
muon and pion from the \B\ hadron each matches an SVT track. Finally, for the 
semileptonic modes, we make cuts on the four track invariant mass 
(eg: $M(\lcmu)$) to reduce the backgrounds from the other $B$ decays, see 
Section~\ref{sec-physicsb} for more details. The signal and sideband 
distribution of each optimized variable after $N-1$ cuts can be found in 
Yu~\cite{cdfnote:7559}.
%
The optimization yields a $S/B$ of 37.6 and 62.8 for the \dstarhad\ and 
\incdstarsemi\ modes, 2.6 and 1.3 for the \dhad\ and \incdsemi\ modes, 
1.6 and 0.3 for the \lbhad\ and \inclbsemi\ modes. 
Figure~\ref{fig:allhisto} shows the charm+$\pi$ (left) and charm (right) mass 
spectra from the hadronic and inclusive semileptonic signals 
in the data after applying the optimized analysis cuts. 

\section{Summary}
We have reconstructed our signals in the data collected from the trigger path
 \bcharm. We have optimized our analysis cuts. In the next chapter, we 
will present the fit to the charm and \B\ hadron mass spectra to obtain the 
number of signal events.


\renewcommand{\arraystretch}{1.6}

 \begin{table}[tbp]
   \caption{Final analysis cuts shared by all the modes.}
   \begin{center}
   \begin{normalsize}
  \label{t:fanacut0}
  \begin{tabular}{|ll|} 
   \hline  
   \hline  
   \multicolumn{2}{|c|}{All} \\
   \hline  
   \pt\ for all tracks & $>$ 0.5 \gevc \\
   $\pi_B$ and $\mu_B$ \pt\ & $>$ 2.0 \gevc \\
   \pt\ of 4 tracks & $>$ 6.0 \gevc \\
   \pt\ of charm hadron & $>$ 5.0 \gevc \\
   $\mu_B$ CMU $\chi^{2}_{x}$ & $<$ 9\\  
   \multicolumn{2}{|l|}{every track exits at COT layer 95} \\
   \multicolumn{2}{|l|}{$\pi_B$ and $\mu_B$ matched to SVT tracks 
    and CMU fiducial} \\
   \hline  
   \hline  
 \end{tabular}
 \end{normalsize}
 \end{center}
 \end{table}

 \begin{table}[tbp]
   \caption{Final analysis cuts for each mode.}
   \begin{center}
   \begin{normalsize}
  \label{t:fanacut1}
  \begin{tabular}{|ll|} 
   \hline  
   \hline  
   \multicolumn{2}{|c|}{\alldstar} \\
   \hline
   $D^0$ VertexFit $\chi^2_{\rphi}$ & $<$ 16 \\
   4 track VertexFit $\chi^2_{\rphi}$ & $<$ 17 \\
   $c\tau$(\Dzero\ $\rightarrow$ $B$) & $>$ -70 $\mu$m \\ 
   $c\tau$($B$ $\rightarrow$ beamspot) & $>$ 200 $\mu$m \\
  \multicolumn{2}{|l|}{1.833 $<$ $M_{K\pi}$ $<$ 1.893 \gevcsq} \\
  \multicolumn{2}{|l|}{3.0$<$ $M_{K\pi\pi\mu}$ $<$5.3 \gevcsq\ for 
	\incdstarsemi}  \\ 
  \multicolumn{2}{|l|}{0.143$<\Delta m<$0.148 \gevcsq\ for \dstarhad} \\

  \hline \hline
   \multicolumn{2}{|c|}{\alld} \\ \hline
   $\D$ VertexFit $\chi^2_{\rphi}$ & $<$ 14 \\
   4 track VertexFit $\chi^2_{\rphi}$ & $<$ 15 \\
   $c\tau$(\D\ $\rightarrow$ $B$) & $>$ -30 $\mu$m \\ 
   $c\tau$($B$ $\rightarrow$ beamspot) & $>$ 200 $\mu$m \\
   \multicolumn{2}{|l|}{3.0$<$ $M_{K\pi\pi\mu}$ $<$5.3 \gevcsq\  for 
\incdsemi} \\ 
  \multicolumn{2}{|l|}{1.8517 $<$ $M_{K\pi\pi}$ $<$ 1.8837 \gevcsq\ for \dhad}
\\ 
   \hline  
   \hline  
   \multicolumn{2}{|c|}{\alllc} \\
   \hline
   $P_T$ of proton & $>$ 2 \gevc \\
   $\Lc$ VertexFit $\chi^2_{\rphi}$ & $<$ 14 \\ 
   4 track VertexFit $\chi^2_{\rphi}$ & $<$ 15 \\
   $c\tau$(\Lc\ $\rightarrow$ \Lb) & $>$ -70 $\mu$m \\
   $c\tau$(\Lb $\rightarrow$ beamspot) & $>$ 250 $\mu$m \\
   \multicolumn{2}{|l|}{3.7$<$ $M_{pK\pi\mu}$ $<$5.64 \gevcsq\ for \inclbsemi}
  \\ 
  \multicolumn{2}{|l|}{2.269$<$ $M_{pK\pi}$ $<$2.302 \gevcsq\  for \lbhad} \\ 
  \hline \hline
 \end{tabular}
 \end{normalsize}
 \end{center}
 \end{table}

\chapter{Signal Yield in the Data}
\label{ch:yield}
In this chapter, we explain how the signal yield in the data is extracted. 
Ideally, if we were capable of fully reconstructing \B\ hadrons in both 
hadronic and semileptonic modes, we would use the \B\ hadron mass distribution 
to obtain the number of signal events (yield). However, a neutrino is missing 
in the semileptonic decay and the invariant mass of ``charm+$\mu$'' is a broad 
spectrum with a shape which is poorly distinguished from the backgrounds. 
Therefore, a proper variable to use for the semileptonic mode is the mass of 
the charm hadron. We extract the yield by fitting the charm+$\pi$ (or charm) 
mass spectra in Figure~\ref{fig:allhisto} to a function which describes both 
the signal and the background. We integrate the signal function to obtain the 
yield. The signal function for all modes is a Gaussian or double-Gaussians. 
The background function varies with the decay mode.     

All our fits use an unbinned, extended likelihood technique. The general 
extended likelihood function ($\cal L$) is expressed as:
\begin{equation}
 \log {\cal L} =  \sum_i\log\{ N_\mathrm{sig}\cdot {\cal S}(m_i)
	+ N_\mathrm{bg}{\cdot B}(m_i) \} - N_\mathrm{sig} - N_\mathrm{bg} 
	+ \log{\cal C},
\label{eq:generallog}
\end{equation}
where $i$ represents $i^{th}$ event, $m$ represents the reconstructed 
charm+$\pi$ (charm) mass. The amounts of signal and background are denoted as 
$N_\mathrm{sig}$ and $N_\mathrm{bg}$, respectively, while ${\cal S}(m)$ 
(${\cal B}(m)$) are the functions which describe the signal (background) mass 
spectrum. The last term in Equation~\ref{eq:generallog}, ${\cal C}$, is a 
Gaussian constraint on one fit parameter, $x$:
\begin{equation} 
 {\cal C} = {\cal G}(x,\mu,\sigma)
	={1\over\sqrt{2\pi}\sigma}e^{-{1\over2}({(x-\mu)\over\sigma})^2},
 \label{eq:gconstraint}
 \end{equation}
where we constrain the variable $x$ around the mean $\mu$. The difference 
of $x$ and $\mu$ follows a Gaussian distribution with an uncertainty $\sigma$. 
The unbinned likelihood fitter calls the ${\tt MINUIT}$ package developed by 
James~\etal~\cite{James:1975dr}. ${\tt MINUIT}$ varies the fit parameters to 
minimizes $-2\cdot\log{\cal L}$ . 

The performance of the fitter was checked on 1000 toy MC samples similar to 
the data distribution. We plot the pull distribution for each parameter, 
i.e. \((x-x_0)/\sigma_x\), where $x$ is the fit value, $x_0$ is the generated 
(input) value, and $\sigma_x$ is the uncertainty from the fit to the toy MC. 
For a large number of toy MC tests, the pull is expected to follow a Gaussian 
distribution. We examine if the fitter returns an output consistent with the 
input, i.e. if the mean of the pull distribution is consistent with zero and 
if the width is consistent with one. Note that the $\mu$ and $\sigma$ of the 
Gaussian constraint in Equation~\ref{eq:gconstraint} are determined from a 
subsidiary measurement using the data and the MC. Therefore, we simulate this 
measurement in the toy MC test, by smearing the mean of the constraint 
with a Gaussian distribution of mean $\mu$ and sigma $\sigma$ in 
Equation~\ref{eq:gconstraint}. In order to evaluate the quality of the fit, 
we also superimpose the fit result on the data histograms and compute a 
$\chi^2$. A complete description about the fitting and the pull 
distributions can be found in Yu~\cite{cdfnote:7559}. 
Remark that as the \B\ hadrons are fully reconstructed in the 
hadronic channels, the yields we extract are the true amount of signal for 
this analysis. The yields we extract for the inclusive semileptonic channels 
include the exclusive signals and indistinguishable backgrounds: such as muon 
fakes, decays from \bb, \cc, or other \B\ hadrons. These backgrounds will be 
estimated in Chapter~\ref{ch:bg} and subtracted in the calculation of the 
relative branching ratios.
  
\section{Mass Fit of the Semileptonic Modes}
\label{sec-semimass}
\subsection{\protect$\mathbf{dstarmu}$ Yield}
 As seen in Figure~\ref{fig:allhisto} (top right), the events with \dstarmu\ 
in the final state have almost no combinatorial background. The combinatorial 
background is reduced largely by requiring \mkpi\ be consistent with the 
world average \Dzero\ mass and cutting on the variable $M_{D^*\mu}$. We fit 
the mass difference $M_{D^0\pi} - M_{D^0}$ instead of $M_{D^0\pi}$, because 
the width of $M_{\Dzero\pi} - M_{\Dzero}$ is significantly narrower than that 
of $M_{D^0\pi}$. The signal to background ratio is thus higher in the signal 
region. The available $Q$ of the \Dstar\ decay is only about 7 \mevcsq, where 
$Q$ is the momentum transferred to the daughters. After the Lorentz boost, 
\Dzero\ carries most of \Dstar's momentum and the bachelor pion from \Dstar\ 
has a lower momentum. Therefore, the width of $M_{\Dzero\pi}$ is similar to 
that of $M_{\Dzero}$. While in the case of $M_{\Dzero\pi} - M_{\Dzero}$, the 
$M_{\Dzero}$ mass resolution is subtracted and only the momentum resolution of 
the soft $\pi$ will contribute to its width.

\begin{figure}[htb]
 \begin{center}
 \includegraphics[width=300pt, angle=0]
	{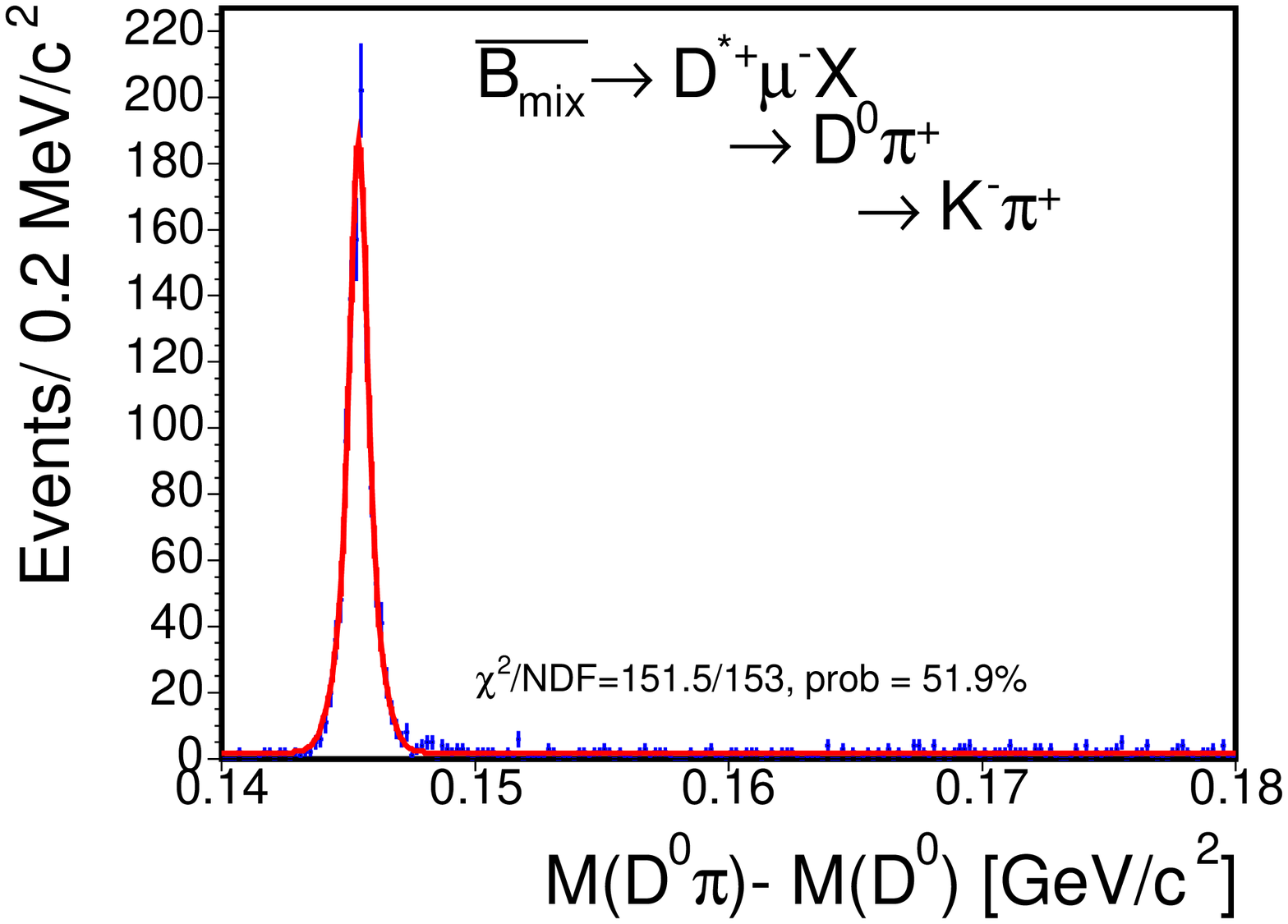}
  \caption[Fit of $M_{\Dzero\pi}- M_{\Dzero}$ from the \dstarmu\ events]
  {$M_{\Dzero\pi}- M_{\Dzero}$ from the \dstarmu\ events fit to a double-
   Gaussian signal and a constant background. The result of the unbinned 
   likelihood fit is projected on the histogram and a $\chi^2$ probability
   is calculated.}
 \label{fig:deltamchi2}
\end{center}
 \end{figure}

The $M_{\Dzero\pi} - M_{D^0}$ distribution is fitted to a double Gaussian 
signal and a constant background. The extended log likelihood function is 
expressed as:
\begin{eqnarray}
 \log{\cal L} & = & \sum_i \log\{N_\mathrm{sig}\cdot[ 
     (1-f_2)\cdot{\cal G}_1(m_i,\mu,\sigma_1) + f_2
	\cdot{\cal G}_2(m_i,\mu,\sigma_2)] \nonumber \\
	& & + N_\mathrm{bg}\cdot\frac{1}{M_\mathrm{max}-M_\mathrm{min}}\}
	-  N_\mathrm{sig} - N_\mathrm{bg}, 
\end{eqnarray}
where $f_2$ is the fraction of the second Gaussian, The mass window 0.14 
$<$ $M_{\Dzero\pi} - M_{\Dzero}$ $<$ 0.18 \gevcsq\ is specified by 
$M_\mathrm{min}$ and $M_\mathrm{max}$. Both Gaussians have the same mean but 
different sigmas. 
Table~\ref{t:dstarmufit} lists the mean, width of 
the pull distribution from 1000 toy MC test and the fit value of each 
parameter from the unbinned likelihood fit to the data. 
%
Figure~\ref{fig:deltamchi2} shows the fit result superimposed on the data 
histogram. We have obtained from the fit:
\begin{displaymath}
 N_{\incdstarsemi} = \ndstarsemi.
\end{displaymath}
 
 \begin{table}[tbp]
   \caption{\dstarmu\ results from the unbinned likelihood fit.}
  \label{t:dstarmufit}
   \renewcommand{\tabcolsep}{0.05in}
   \begin{center}
   \begin{tabular}{|c|lr|r|r|r|} 
     \hline
    \hline
     Index & \multicolumn{2}{|l|}{Parameter} 
	& 1000 toy MC & 1000 toy MC & Data fit value\\	
     & \multicolumn{2}{|l|}{} 
	& pull mean & pull width & \\	
     \hline
     1 & $N_\mathrm{sig}$ &
	& -0.023 $\pm$ 0.031 & 1.006 $\pm$ 0.023 & \ndstarsemi \\
     2 & $f_2$ &
	& 0.002 $\pm$ 0.034 & 1.072 $\pm$ 0.024 & 0.56 $\pm$ 0.10 \\
     3 & $\mu$ & [\gevcsq] 
      & 0.049 $\pm$ 0.033 & 1.044 $\pm$ 0.024 & 0.145410 $\pm$ 0.000016 \\
     4 & $\sigma_1$ & [\gevcsq] 
      & -0.048 $\pm$ 0.033 & 1.052 $\pm$ 0.024 & 0.00031 $\pm$ 0.00004 \\ 
     5 & $\sigma_2$ & [\gevcsq] 
	& 0.011 $\pm$ 0.032 & 1.031 $\pm$ 0.023 & 0.00071 $\pm$ 0.00006 \\
     6 & $N_\mathrm{bg}$ & & 
	0.010 $\pm$ 0.031 & 1.000 $\pm$ 0.022 & 321 $\pm$ 19 \\
     \hline	
      \hline
      \end{tabular}
      \end{center}

  \end{table}

\subsection{\boldmath$\dmu$ Yield}
\begin{figure}[htb]
    \begin{center}
    \includegraphics[width=400pt, angle=0]
	{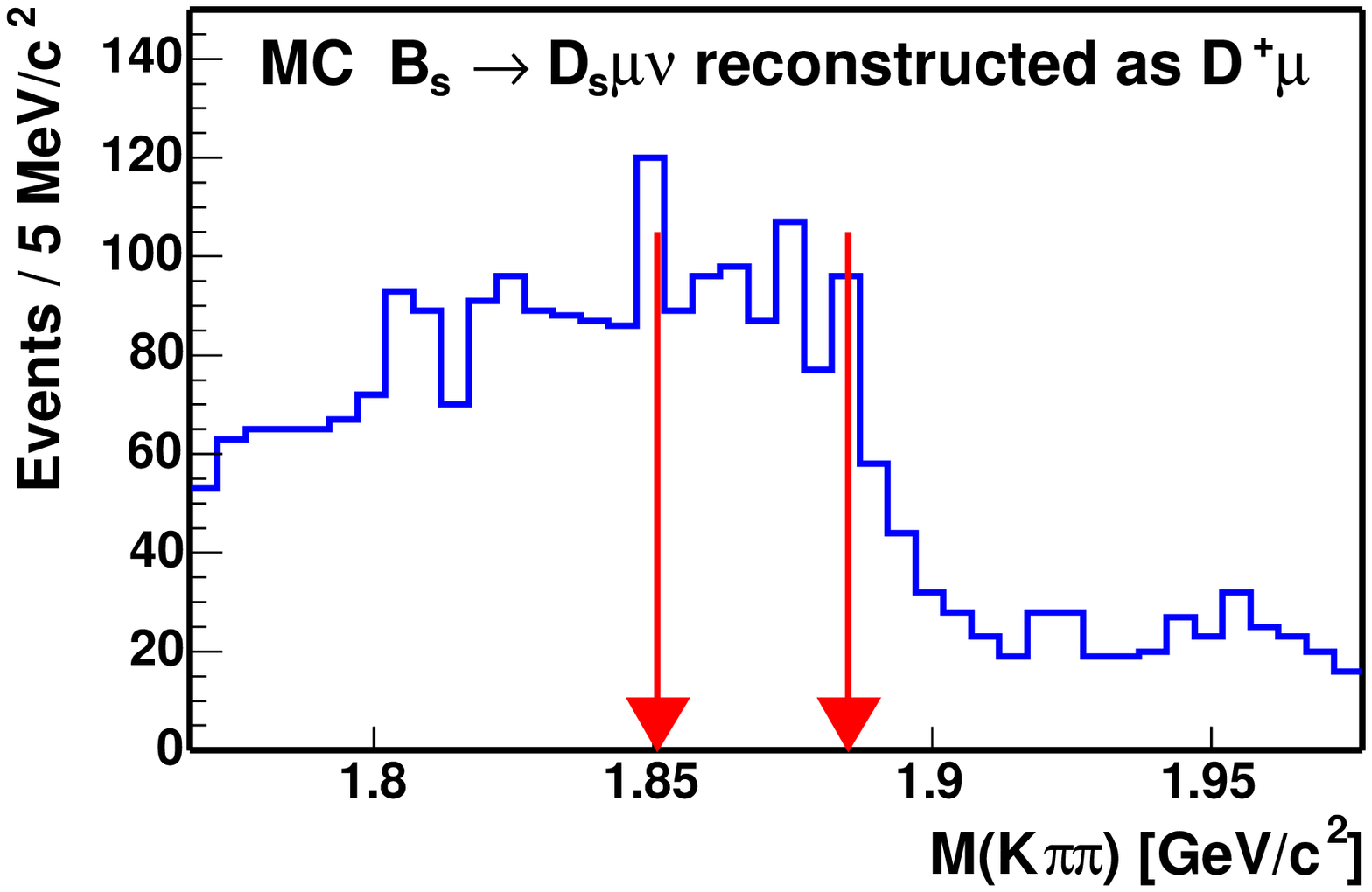}
     \caption[\bsdsmunu\ MC reconstructed as \dmu\ final state]
     {\bsdsmunu\ MC reconstructed as \dmu\ final state.
       Here $D_s$ are forced to decay into the modes listed
      in Table~\ref{t:dsdecays}. The arrows indicate the 3 $\sigma$ 
      \D\ signal region.}
    \label{fig:mismc}
     \end{center}
  \end{figure}

 A first glance of \mkpipi\ in Figure~\ref{fig:allhisto} (middle right) 
might suggest that we could fit \mkpipi\ to a Gaussian signal and a 
first-order polynomial background. But, since we do not apply particle 
identification (PID) in this analysis, the background under the signal 
contains not only the combinatorial background, but also contamination from 
the $D_s$ decays. Not using PID means that a pion mass might be assigned to a 
kaon, and $D_s^+$ may be reconstructed as \D. Figure~\ref{fig:mismc} shows the
 mis-reconstructed \D\ mass spectrum from the \bsdsmunu\ MC, where $D_s$ are 
forced to decay into the final states listed in Table~\ref{t:dsdecays}. 
These final states are selected after a study to identify the dominant $D_s$ 
decays reconstructed in the \D\ mass window. The MC used to assess $D_s$ 
background is produced as described in Section~\ref{sec-mccom}. 

\renewcommand{\arraystretch}{1.2}
 \setdec 00.0000
    	\begin{table}[tbp]
  	\caption[Dominant mis-identified $D_s$ sequential decays in \dmu\
         signal]{Dominant mis-identified $D_s$ sequential decays in \dmu\
         signal. Branching fractions without uncertainties have an upper 
	limit in the PDG.}
	\begin{center}
	\label{t:dsdecays}
 	\begin{tabular}{|l|r@{\,$\pm$\,}lr@{\,$\pm$\,}l|}
 	\hline
        \multicolumn{5}{|c|}{Selected final states of $D_s$ decays}\\	
        \hline
	Mode & \multicolumn{2}{|c}{${\cal B}$ ($\%$)}  
	& \multicolumn{2}{c|}{relative to ${\cal B}(D_s\rightarrow\phi\pi)$}\\
	\hline 
   	\seqds & \dec 3.6 & \dec 0.9 & 
	\multicolumn{2}{c|}{1}\\ 
   	$D_s^+\rightarrow \phi K^+$ & \dec 0.03 & ? & \dec 0.008 & ?\\
   	$D_s^+\rightarrow \eta\pi^+$ 
	& \dec 1.7 & \dec 0.5 & \dec 0.48 & \dec 0.05 \\
        $D_s^+\rightarrow \eta^{\prime}\pi^+$ 	
	& \dec 3.9 & \dec 1.0 & \dec 1.08 & \dec 0.09 \\
	$D_s^+\rightarrow \omega\pi^+$ 
	& \dec 0.28 & \dec 0.11 & \dec 0.077 & \dec 0.025 \\ 
        $D_s^+\rightarrow \rho^0\pi^+$ & \dec 0.04 & ? & \dec 0.011 & ? \\
        $D_s^+\rightarrow \rho^0K^+$ & \dec 0.15 & ?  & \dec 0.042 & ? \\ 
	$D_s^+\rightarrow f_0\pi^+$ 
	& \dec 0.57 & \dec 0.17 & \dec 0.16 & \dec 0.03 \\  
        $D_s^+\rightarrow f_2\pi^+$ 
	& \dec 0.35 & \dec 0.12 & \dec 0.098 & \dec 0.022 \\
        $D_s^+\rightarrow \rho^+\eta$ 
	& \dec 10.8 & \dec 3.1 & \dec 2.98 & \dec 0.44 \\  
        $D_s^+\rightarrow \rho^+\eta^{\prime}$ 
	& \dec 10.1 & \dec 2.8 & \dec 2.78 & \dec 0.41 \\
        $D_s^+\rightarrow \overline{K}^0 \pi^+$ 
        & \dec 0.4 & ? & \dec 0.11 & ?\\
        $D_s^+\rightarrow \overline{K}^{*0} \pi^+$ 
	& \dec 0.65 & \dec 0.28  & \dec 0.18 & \dec 0.06 \\
        $D_s^+\rightarrow K^0 K^+$ 
	& \dec 3.6 & \dec 1.1 & \dec 1.01 & \dec 0.16 \\
        $D_s^+\rightarrow K^{*0} K^+$ 
	& \dec 3.3 & \dec 0.9 & \dec 0.92 & \dec 0.09 \\
        $D_s^+\rightarrow \pi^+\pi^+\pi^-$ 
	& \dec 0.005 & ${+0.022 \atop -0.005}$  & \dec 0.0014 & \dec 0.0007 \\
        $D_s^+\rightarrow K^+K^-\pi^+$ 
	& \dec 0.9 & \dec 0.4 & \dec 0.25 & \dec 0.09 \\
        $D_s^+\rightarrow K^+K^+K^-$ 
	& \dec 0.02 & ? & \dec 0.0056 & ? \\
         \hline
 	\hline
    \end{tabular}
     \end{center}
     \end{table}

We need to include the mis-identified $D_s$ mass shape in our likelihood fit 
so to properly estimate the number of \dmu\ events in the data. Assuming that 
\incdsmu\ has a similar mass spectrum as \bsdsmunu, we could use the MC for 
Figure~\ref{fig:mismc} to obtain the function which describes the line-shape 
of mis-reconstructed $D_s$ mass spectrum. We find the $D_s$ spectrum 
(${\cal F}$) could be described by a constant and a triangular function 
convoluted with a Gaussian (${\cal T}$): 
\begin{equation}
 \label{eq:trigauss}
 {\cal F}(m) =  
	(1-f_\mathrm{trg})\cdot\frac{1}{M_\mathrm{max}-M_\mathrm{min}} +
	f_\mathrm{trg}\cdot{\cal T}(m),
\end{equation}
where $f_\mathrm{trg}$ is the fraction of triangular function, $M_\mathrm{max}$
 and $M_\mathrm{min}$ specify the mass window, 1.767 $<$ \mkpipi\ $<$ 1.977 
\gevcsq, and 
\begin{equation}
{\cal T}(m) = {\frac{2(m-M_0)}{(M_\mathrm{off}-M_0)^2}} \otimes 
{\cal G}(m,M_0,\sigma_\mathrm{trg}).
\end{equation}
Here, $\otimes$ represents convolution, $\cal G$ is the Gaussian and 
$\sigma_\mathrm{trg}$ is the width of ${\cal G}$. 
The triangular function value starts from zero at $M_0$ and increases 
as the mass increases. When the mass reaches $M_\mathrm{off}$, 
the function values is at its maximum and drops precipitously to zero.
A graphical representation of $M_\mathrm{off}$ and $M_0$ may be found in 
Figure~\ref{fig:simpletrifit}. The exact form of ${\cal T}(m)$ is found in 
the appendix of Yu~\cite{cdfnote:7559} derived by Heinrich. 
Figure~\ref{fig:fitmismc} shows the result of the fit to the MC. 

  \begin{figure}[tbp]
    \begin{center}
    \includegraphics[width=350pt, angle=0]
	{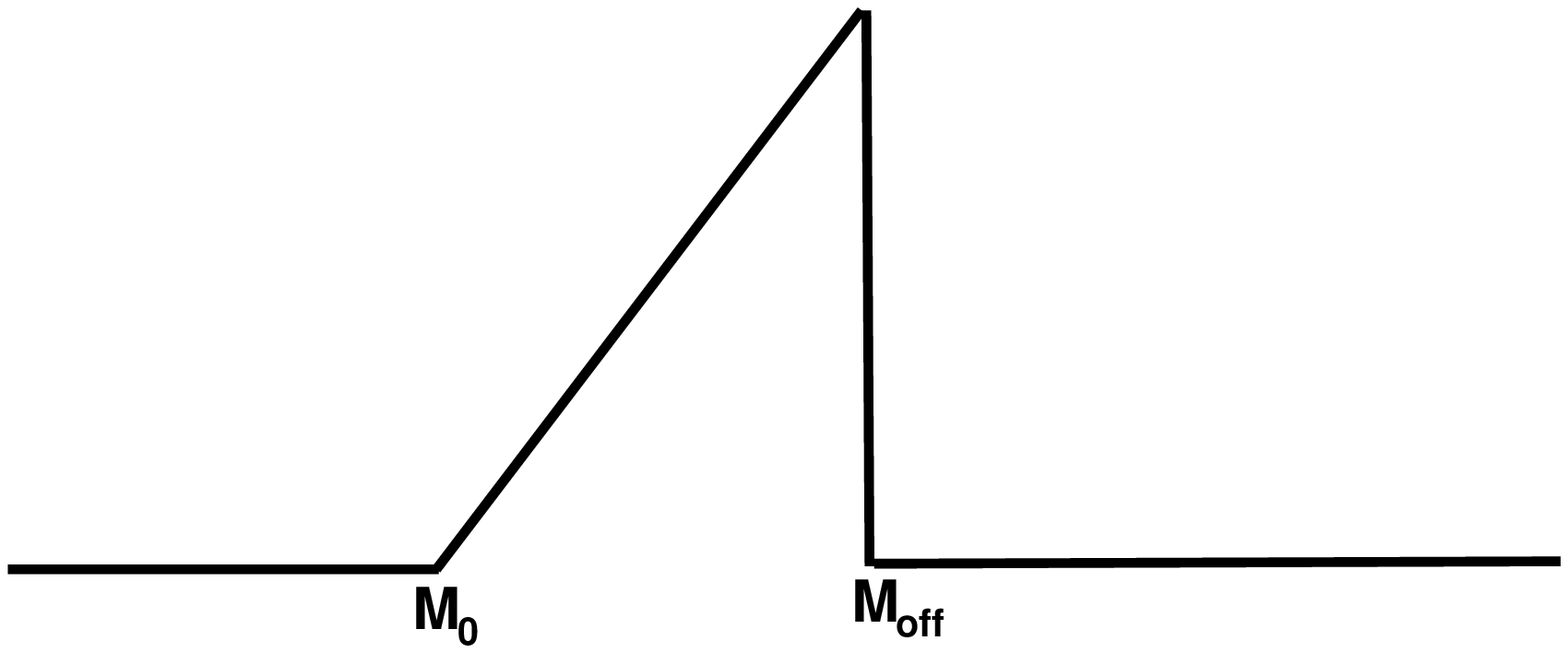}
     \caption[Graphical representation of the triangular function]
     {Graphical representation of the triangular function.}
    \label{fig:simpletrifit}
    \includegraphics[width=350pt, angle=0]
	{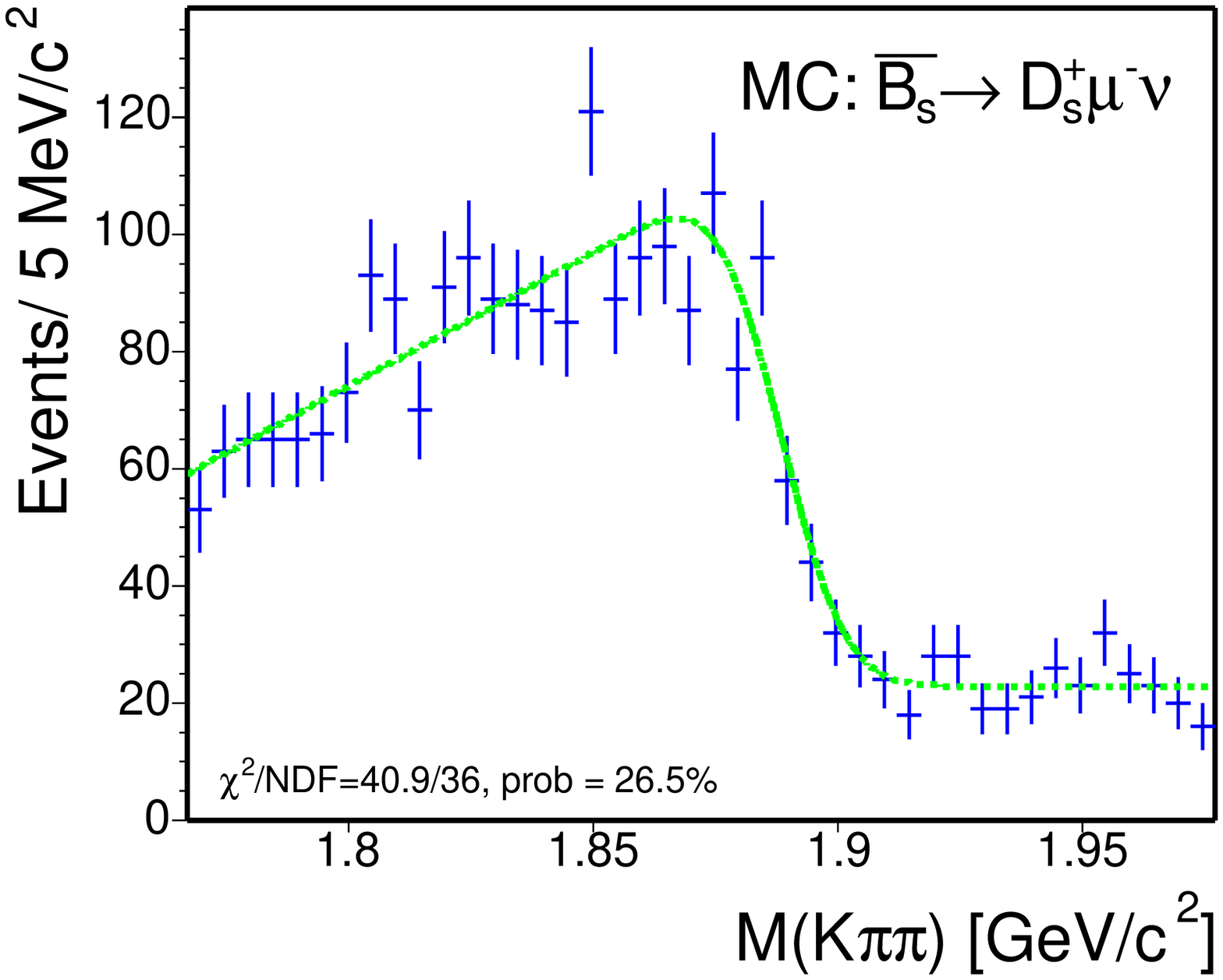}
     \caption[Fit to mis-reconstructed $D_s$ using \bsdsmunu\ MC]
     { Mis-reconstructed $D_s$ from the \bsdsmunu\ MC fit to a constant and 
       a triangular function convoluted with a Gaussian. The dashed curve 
       indicates the result of the unbinned likelihood fit.
      }
    \label{fig:fitmismc}
     \end{center}
  \end{figure}

\begin{figure}[htb]
    \begin{center}
      \includegraphics[width=350pt, angle=0]
	{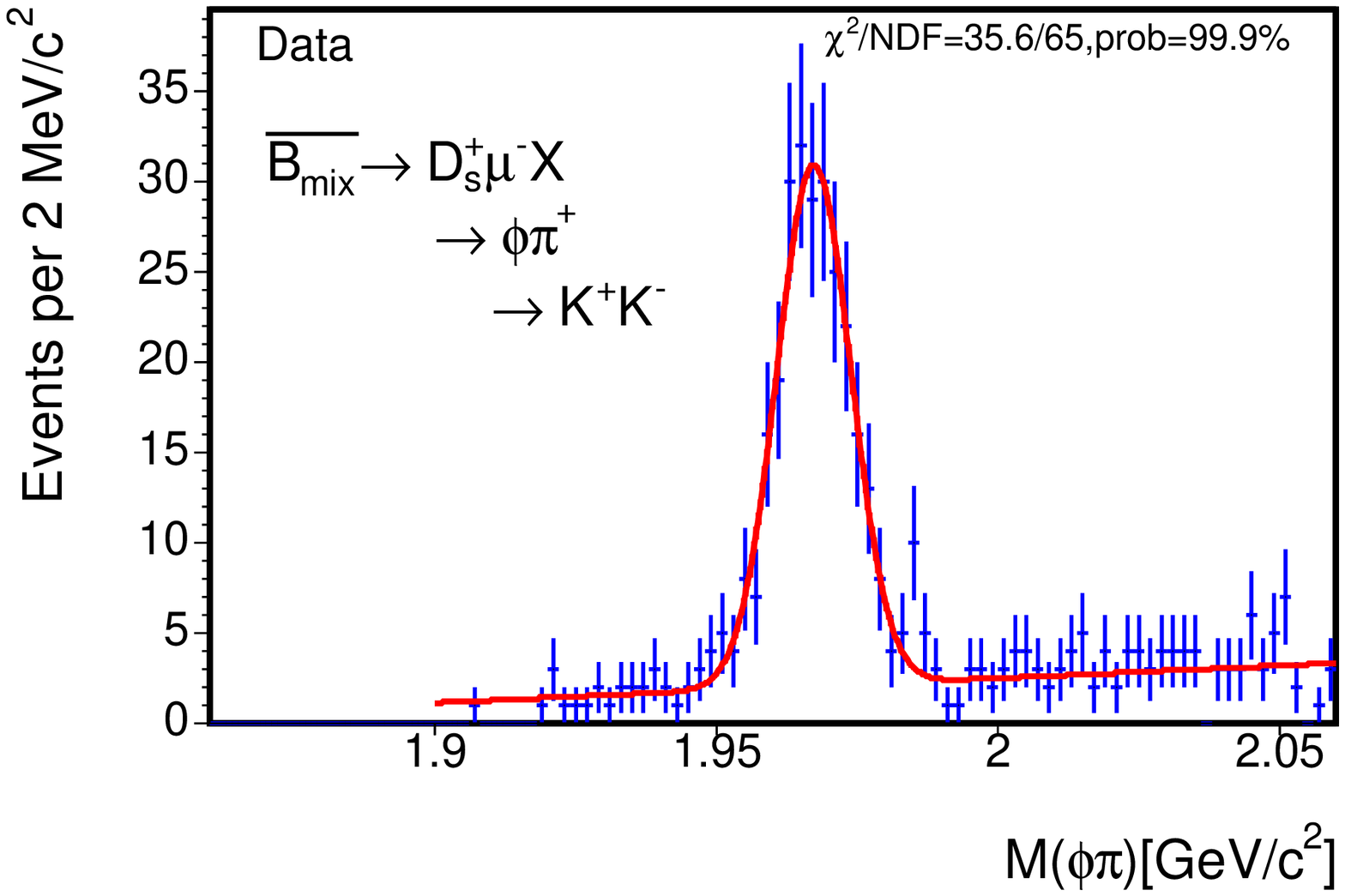}
     \caption[Reconstructed \incdsmu\ in the data]
   {Data: Reconstructed \incdsmu, where \seqds, and \seqphi. 
	\mkpipi\ is required to be between 1.767 and 1.977 GeV/c$^2$. 
        There are 237 $\pm$ 17 events in the peak.}
    \label{fig:bsdataall}
     \end{center}
  \end{figure}

Now with the function form of the \mkpipi\ spectrum from the $D_s$ decays, 
we have to normalize the MC yield to the data. The $D_s$ yield may be obtained 
by reconstructing one of the $D_s$ final states in the data: 
\incdsmu, where \seqds, \seqphi, then using MC to determine the ratio of 
this $D_s$ decay to that of all the $D_s$ decays in Table~\ref{t:dsdecays}, 
$R_{\phi\pi}$: 
\begin{equation}
   R_{\phi \pi} = \frac{N_{\phi\pi}^\mathrm{MC}}{N_\mathrm{all}^\mathrm{MC}}.
\end{equation}
The normalization of $D_s$ is then expressed as:
\begin{equation}
\label{eq:predicted}  
  N_{\incdsmu} = 
  \frac{N_{\incdsmu, D_s\rightarrow \phi\pi, \phi\rightarrow KK} }
	{R_{\phi \pi}}.
 \end{equation}
In order to obtain \(N_{\incdsmu, D_s\rightarrow \phi\pi, \phi\rightarrow KK}\)
 in the data, the same analysis cuts for \dmu\ are applied, except that we 
assign kaon mass to one of the same-sign charged tracks and pion mass to the 
other. We still assign kaon mass to the track which has the opposite charge of 
the other two. In addition, the candidates are required to pass the following 
cuts:
  \begin{itemize} 
  \item 1.767 $<$ \mkpipi\ $<$ 1.977 \gevcsq
  \item $|M_{KK}-1.019| < $ 0.01 \gevcsq
  \end{itemize}
The cut on $M_{KK}$ guarantees that there is no mis-identified \D\ in the 
$D_s^+$ signal we reconstruct. We confirm this by reconstructing $D_s^+$ from 
the \dsemi\ MC and no $D_s$ candidate is found. See Figure~\ref{fig:bsdataall} 
for the \incdsmu\ signal in the data, we find: 
\[
\label{eq:dataphi}
  N_{\incdsmu, D_s\rightarrow \phi\pi, \phi\rightarrow KK}  = 237 \pm 17. 
\]

We then reconstruct the same $D_s$ decay chain in the MC as in the data and 
obtain
\begin{equation}
   R_{\phi \pi} = 0.131 \pm 0.007.
  \label{eq:rphipi}
\end{equation}
Inserting the result of  
$N_{\incdsmu, D_s\rightarrow \phi\pi, \phi\rightarrow KK}$ and 
$R_{\phi\pi}$ into Equation~\ref{eq:predicted}, we have:
\begin{equation}
  N_{\incdsmu} = 1812 \pm 160, 
\label{eq:predictresult}
\end{equation}
The uncertainty in Equation~\ref{eq:predictresult} comes from the fractional 
uncertainties on: \(N_{\incdsmu, D_s\rightarrow \phi\pi, \phi\rightarrow KK}\) 
(7.2$\%$) and $R_{\phi\pi}$ (5$\%$). 

In the unbinned fit, the extended log likelihood function is expressed by the 
sum of two likelihoods: one describing the data and the other describing 
the \bsdsmunu\ MC since we fit the data and MC simultaneously; 
\begin{equation}
 \log{\cal L} = \log{\cal L}^\mathrm{data} + \log{\cal L}^\mathrm{MC}, 
\end{equation}
The likelihood function for the data, \(\log{\cal L}^{data}\), is a sum of a 
signal Gaussian, a first-order polynomial for the combinatorial background 
(${\cal H}$), and the function for the $D_s$ (${\cal F}$, see 
Equation~\ref{eq:trigauss}). A Gaussian constraint on the amount of $D_s$, 
\({\cal C}_{D_s}\), is employed. 
\begin{eqnarray}
 \log{\cal L}^\mathrm{data} & = & 
	\sum_i \log\{N_\mathrm{sig}\cdot{\cal G}(m_i,\mu,\sigma)
	+ N_\mathrm{combg}\cdot {\cal H}(m_i)+  N_{D_s}\cdot{\cal F}(m_i)\}  
       \nonumber \\
      & - & N_\mathrm{sig} - N_\mathrm{combg} - N_{D_s} + \log {\cal C}_{D_s}, 
\label{eq:logdmu}
\end{eqnarray}
where
\begin{eqnarray*}
{\cal H}(m_i) & = & \frac{1}{M_\mathrm{max}-M_\mathrm{min}} 
+ p_1\cdot(m_i- \frac{M_\mathrm{max}+M_\mathrm{min}}{2}), \\
 {\cal C}_{D_s} & = &{\cal G}(N_{D_s}, \mu^{p}, \sigma^{p}).
\end{eqnarray*}
From the prediction of Equation~\ref{eq:predictresult}, we have 
\(\mu^{p}=1812\), and \(\sigma^p=160\).

\begin{figure}[htb]
 \begin{center}
 \includegraphics[width=300pt, angle=0]
  {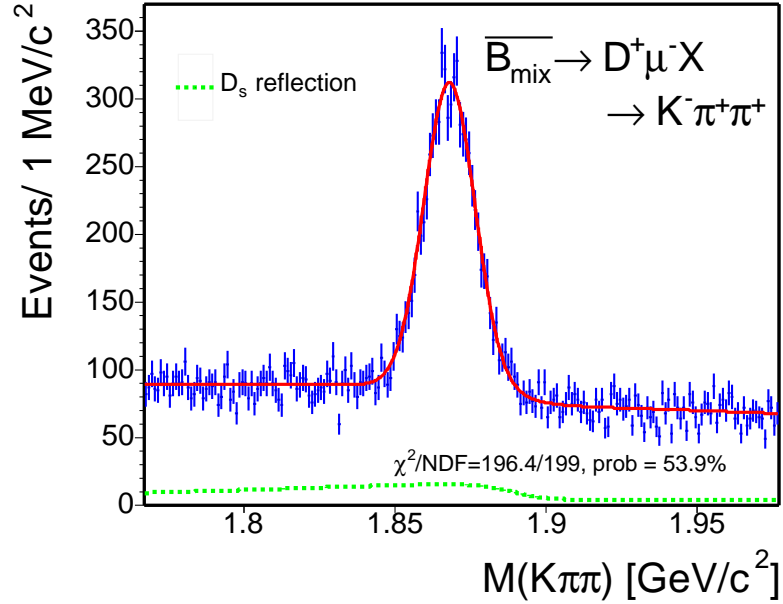}
  \caption[Fit of \mkpipi\ from the \dmu\ events]
  {\mkpipi\ from the \dmu\ events fit to a 
   Gaussian (signal), a first-order polynomial (combinatorial background), 
   and a constant plus a triangular function convoluted with a Gaussian
   (marked by the dashed line). 
    The result of the unbinned likelihood fit is 
    projected on the histogram and a $\chi^2$ probability is calculated.}
 \label{fig:dmchi2}
\end{center}
 \end{figure}

 The likelihood function \(\log{\cal L}^\mathrm{MC}\) is used to fit \bsdsmunu\
 MC and obtain the parameterization of \({\cal F}(m)\). Here the normalization 
does not matter.
 \begin{equation}
 \log{\cal L}^\mathrm{MC}  = \sum_i \log\{{\cal F}(m_i)\}.
 \end{equation}
Table~\ref{t:dmufit} lists the mean, width of the pulls from 1000 toy MC test 
and the result returned from the unbinned likelihood fit to the data. 
Figure~\ref{fig:dmchi2} shows the fit 
result superimposed on the data histogram. We have obtained from the fit:
\begin{displaymath}
 N_{\incdsemi} = \ndsemi.
\end{displaymath}
We also perform a cross-check by removing the constraint on $N_{D_s}$ and 
obtain \(N_{\incdsemi} = 4667 \pm 139\), \(N_{D_s}=2184\pm 620\), 
which are consistent with the result of the constrained fit. The fit 
without constraint has a $\chi^2/\mathrm{NDF}$=197.0/199 and 
probability is 52.7$\%$.


 \begin{table}[tbp]
   \caption{\dmu\ results from the unbinned likelihood fit.}
  \label{t:dmufit}
   \begin{center}
   \renewcommand{\tabcolsep}{0.05in}
   \begin{tabular}{|c|lr|r|r|r|} 
    \hline
     Index & \multicolumn{2}{|l|}{Parameter} 
	& 1000 toy MC & 1000 toy MC & Data fit value\\	
     & \multicolumn{2}{|l|}{} 
	& pull mean & pull width & \\	
     \hline
      1 & $N_\mathrm{sig}$ &
	& -0.012 $\pm$ 0.035 & 1.004 $\pm$ 0.025 & \ndsemi \\
      2 & $\mu$ & [\gevcsq]
	& 0.027 $\pm$ 0.037 & 1.048 $\pm$ 0.027 & 1.8680 $\pm$ 0.0002 \\
      3 & $\sigma$ & [\gevcsq]
	& 0.007 $\pm$ 0.035 & 0.992 $\pm$ 0.025 & 0.0084 $\pm$ 0.0002 \\
      4 & $N_\mathrm{combg}$ &
	&-0.076 $\pm$ 0.038 & 1.073 $\pm$ 0.027 & 15178 $\pm$ 197 \\ 
      5 & $p_1$ & & 0.018 $\pm$ 0.036 & 1.027 $\pm$ 0.026 & -5.2 $\pm$ 0.7 \\
      6 & $N_{D_s}$ & 
	& 0.042 $\pm$ 0.037 & 1.065 $\pm$ 0.027 & 1832 $\pm$ 155 \\
      7 & $f_\mathrm{trg}$ &
	& 0.022 $\pm$ 0.036 & 1.023 $\pm$ 0.026 & 0.617 $\pm$ 0.021 \\
      8 & $M_{0}$ & [\gevcsq]
	& 0.055 $\pm$ 0.035 & 1.007 $\pm$ 0.025 & 1.69 $\pm$ 0.02 \\
      9 & $M_\mathrm{off}$ & [\gevcsq]
	& -0.025 $\pm$ 0.036 & 1.019 $\pm$ 0.026 & 1.888 $\pm$ 0.002 \\ 
      10 &$\sigma_\mathrm{trg}$ & [\gevcsq]
	& -0.035 $\pm$ 0.037 & 1.056 $\pm$ 0.027 & 0.010 $\pm$ 0.002 \\
     \hline	
      \hline
      \end{tabular}
      \end{center}
       \end{table}

\subsection{\boldmath$\lcmu$ Yield}
\begin{figure}[htb]
 \begin{center}
 \includegraphics[width=300pt, angle=0]
	{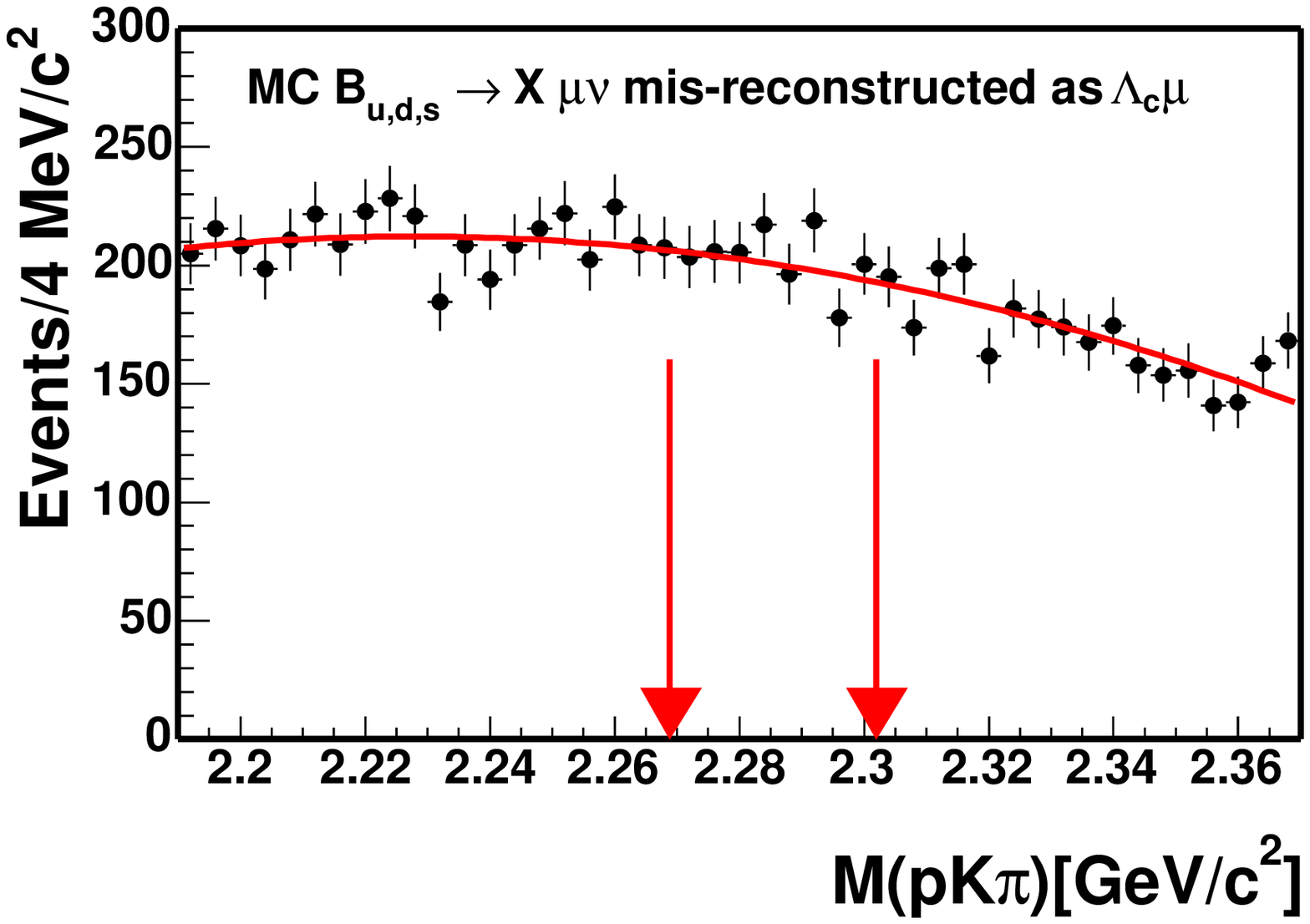}
  \caption[$B\rightarrow X\mu\nu_{\mu}$ MC mis-reconstructed as \lcmu]
  {$B\rightarrow X\mu\nu_{\mu}$ MC mis-reconstructed as \lcmu. The mass
   spectrum fit to a second-order polynomial. $\chi^2/\mathrm{NDF}$ = 36.6/42,
   prob= 70.7$\%$. Two arrows indicate the signal region of \Lc.}
 \label{fig:genlcref}

 \includegraphics[width=300pt, angle=0]
 {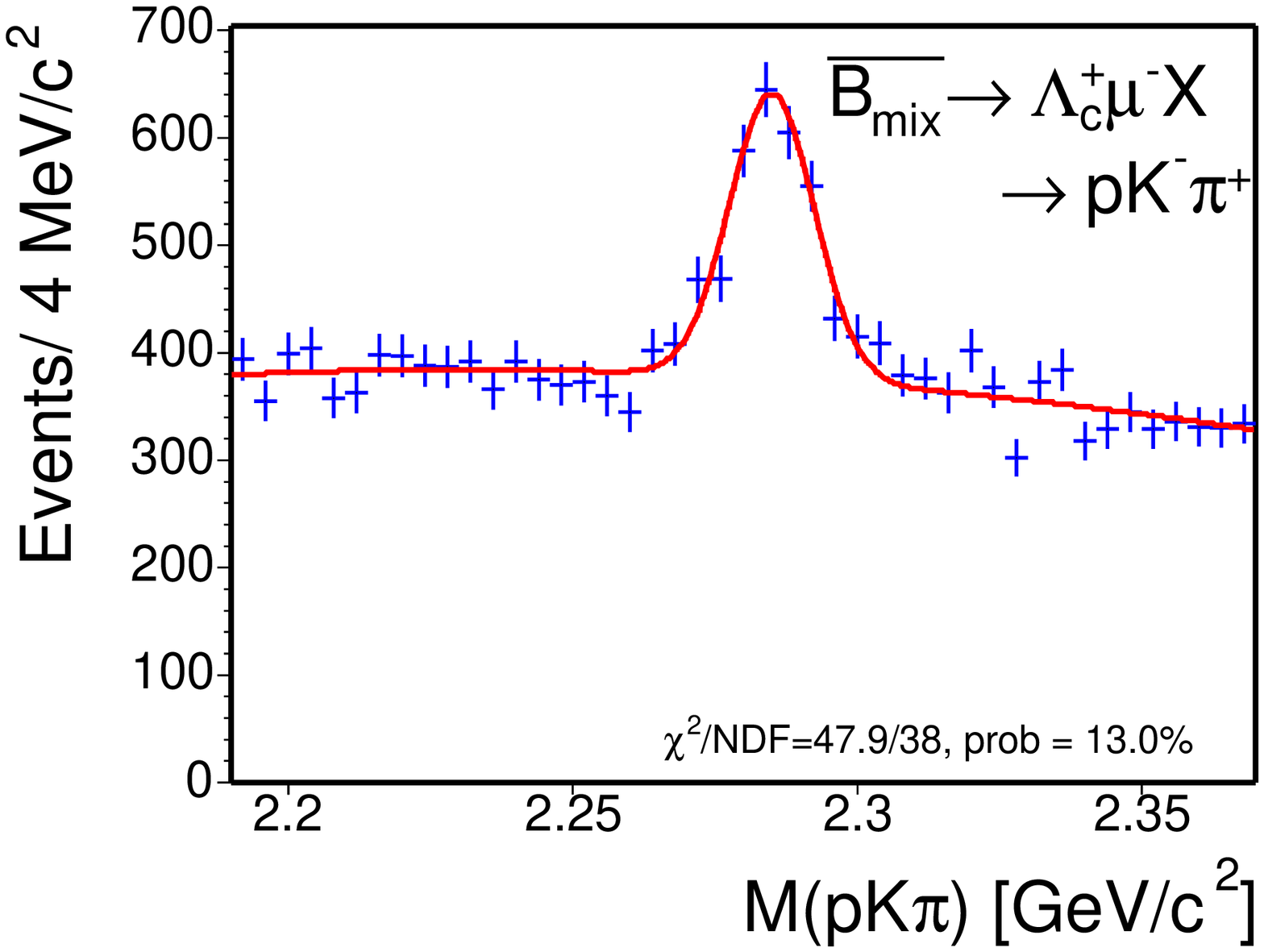}
  \caption[Fit of \mpkpi\ from the \lcmu\ events]
  {\mpkpi\ from the \lcmu\ events fit to a 
   Gaussian (signal), a second-order polynomial (combinatorial + \B\ meson). 
   The result of the unbinned likelihood fit is 
   projected on the histogram and a $\chi^2$ probability is calculated.}
 \label{fig:lcmchi2}
\end{center}
 \end{figure}

When a proton mass is assigned to a kaon or pion, 
numerous \B\ meson to $D$ meson semileptonic decays could be mis-reconstructed 
as a \lcmu\ final state.  
In order to estimate the \B\ meson background shape under our signal, 
we use generator level MC and generate the semileptonic decays ($\mu$ channel) 
of each \B\ meson flavor separately. After applying analysis cuts,  
we add up the mis-reconstructed mass spectrum from each kind of \B\ meson 
according to the production fractions:
\begin{center}
 $b \rightarrow B_d$  =  (39.7 $\pm$ 1.3) $\%$,\\
 $b \rightarrow B_u$  =  (39.7 $\pm$ 1.3) $\%$,\\
 $b \rightarrow B_s$  =  (10.7 $\pm$ 1.1) $\%$. 
\end{center}
 Figure~\ref{fig:genlcref} shows a smooth mass spectrum from the 
generator MC. The shape is best described by a second-order polynomial, with 
$\chi^2/\mathrm{NDF}$ = 36.6/42, prob = 70$\%$. A first-order polynomial fit 
yields $\chi^2/\mathrm{NDF}$ = 56.6/43, prob = 8$\%$. Because the 
combinatorial background may be parameterized by a first-order polynomial, and 
adding a first- to a second-order polynomial gives a second-order polynomial, 
we fit the combinatorial and the \B\ meson background together to a 
second-order polynomial (${\cal H}$). The extended log likelihood function 
could be expressed as:
\begin{equation}
 \log{\cal L}  =  \sum_i \log\{N_\mathrm{sig}\cdot{\cal G}(m_i,\mu,\sigma)
	+ N_\mathrm{bg}\cdot {\cal H}(m_i)\}
         -  N_\mathrm{sig} - N_\mathrm{bg},
\end{equation}
where 
\[
{\cal H}(m_i) = \frac{1}{M_\mathrm{max}-M_\mathrm{min}} + 
	p_1\cdot(m_i-M_\mathrm{mid}) + 
	p_2\cdot(12\cdot(m_i-M_\mathrm{mid})^2 - M_\mathrm{diff}^2). 
\]
Here, $M_\mathrm{max}$ and $M_\mathrm{min}$ specify the \Lc\ mass window:
2.19 $<$ \mpkpi\ $<$ 2.37 \gevcsq. The average of 
$M_\mathrm{max}$ and $M_\mathrm{min}$, or the mid point in the mass window is 
$M_\mathrm{mid}$. The difference of $M_\mathrm{max}$ and $M_\mathrm{min}$ is 
$M_\mathrm{diff}$. 
Table~\ref{t:lcmufit} lists the mean, width of the pulls from the toy MC test 
and the parameter value from the fit to the data. 
Figure~\ref{fig:lcmchi2} shows fit result superimposed 
on the data histogram. We have obtained from the fit:
\begin{displaymath}
 N_{\inclbsemi} = \nlbsemi.
\end{displaymath}


 \begin{table}[htb]
   \caption{\lcmu\ results from the unbinned likelihood fit.}
  \label{t:lcmufit}
   \begin{center}
   \renewcommand{\tabcolsep}{0.05in}
   \begin{tabular}{|c|lr|r|r|r|} 
    \hline
     Index & \multicolumn{2}{|l|}{Parameter} 
	& 1000 toy MC & 1000 toy MC & Data fit value\\	
     & \multicolumn{2}{|l|}{} 
	& pull mean & pull width & \\	
     \hline
      1 & $N_\mathrm{sig}$ &
	& 0.018 $\pm$ 0.030 & 0.997 $\pm$ 0.022 & \nlbsemi \\
      2 & $\mu$ & [\gevcsq] 
	& 0.017 $\pm$ 0.033 & 1.070 $\pm$ 0.024 &  2.2850 $\pm$ 0.0005 \\ 
      3 & $\sigma$ & [\gevcsq] 
	& -0.069 $\pm$ 0.032 & 1.036 $\pm$ 0.023 & 0.0074 $\pm$ 0.0006 \\
      4 & $N_\mathrm{bg}$ &
	& 0.004 $\pm$ 0.031 & 1.021 $\pm$ 0.022 & 16576 $\pm$ 157 \\
      5 & $p_1$ & & 0.010 $\pm$ 0.031 & 1.007 $\pm$ 0.022 & -4.3 $\pm$ 0.8 \\ 
      6 & $p_2$ & & 0.020 $\pm$ 0.031 & 1.012 $\pm$ 0.022 & -3.7 $\pm$ 1.8 \\ 
     \hline	
      \hline
      \end{tabular}
      \end{center}
	\end{table}

\clearpage
\newpage
\section{Mass Fit of the Hadronic Modes}
\label{sec-massfit}
\begin{figure}[tbp]
     \begin{center}
     \begin{tabular}{cl}
        \includegraphics[width=200pt, angle=0]
	{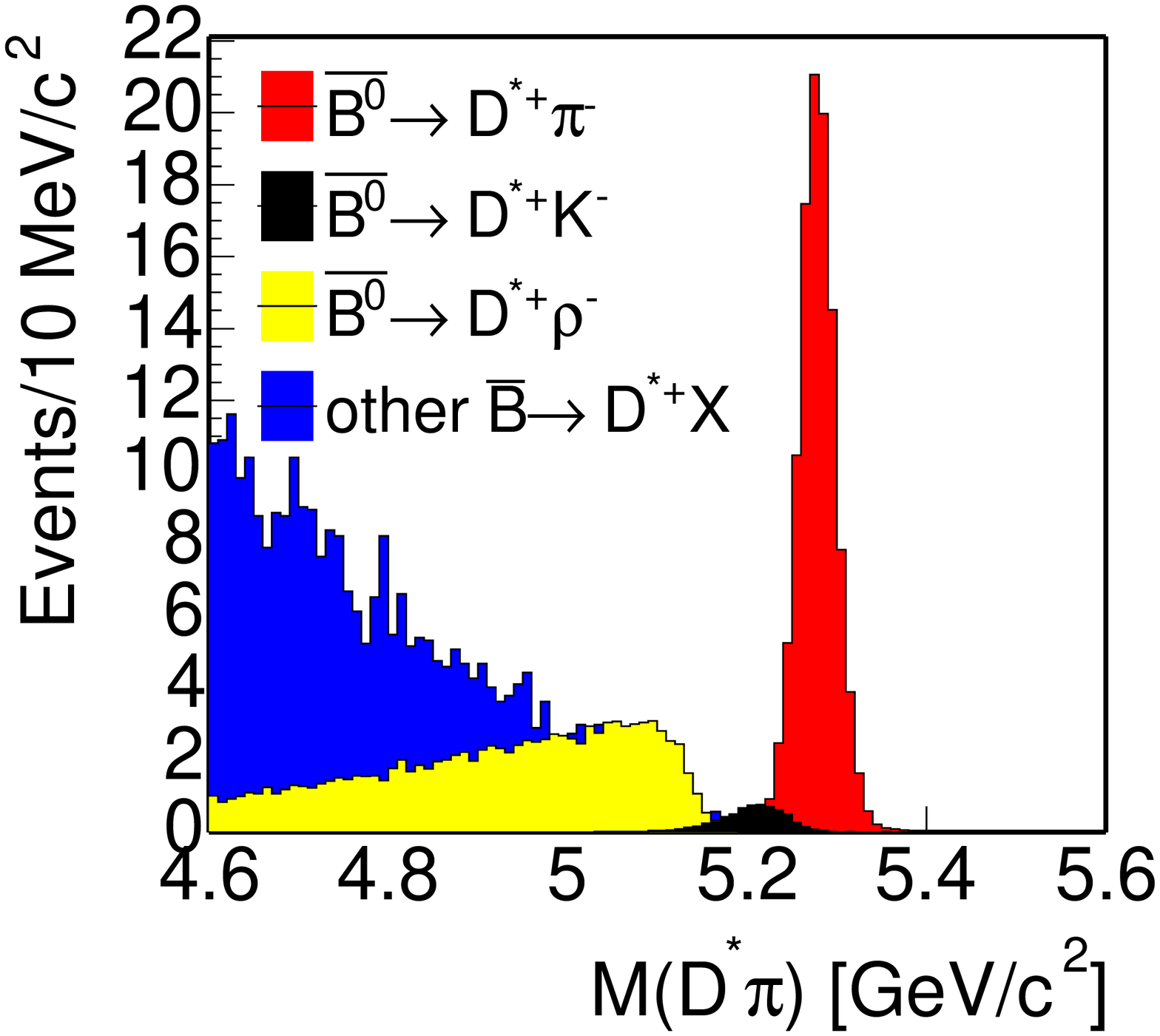} &
        \includegraphics[width=200pt, angle=0]
	{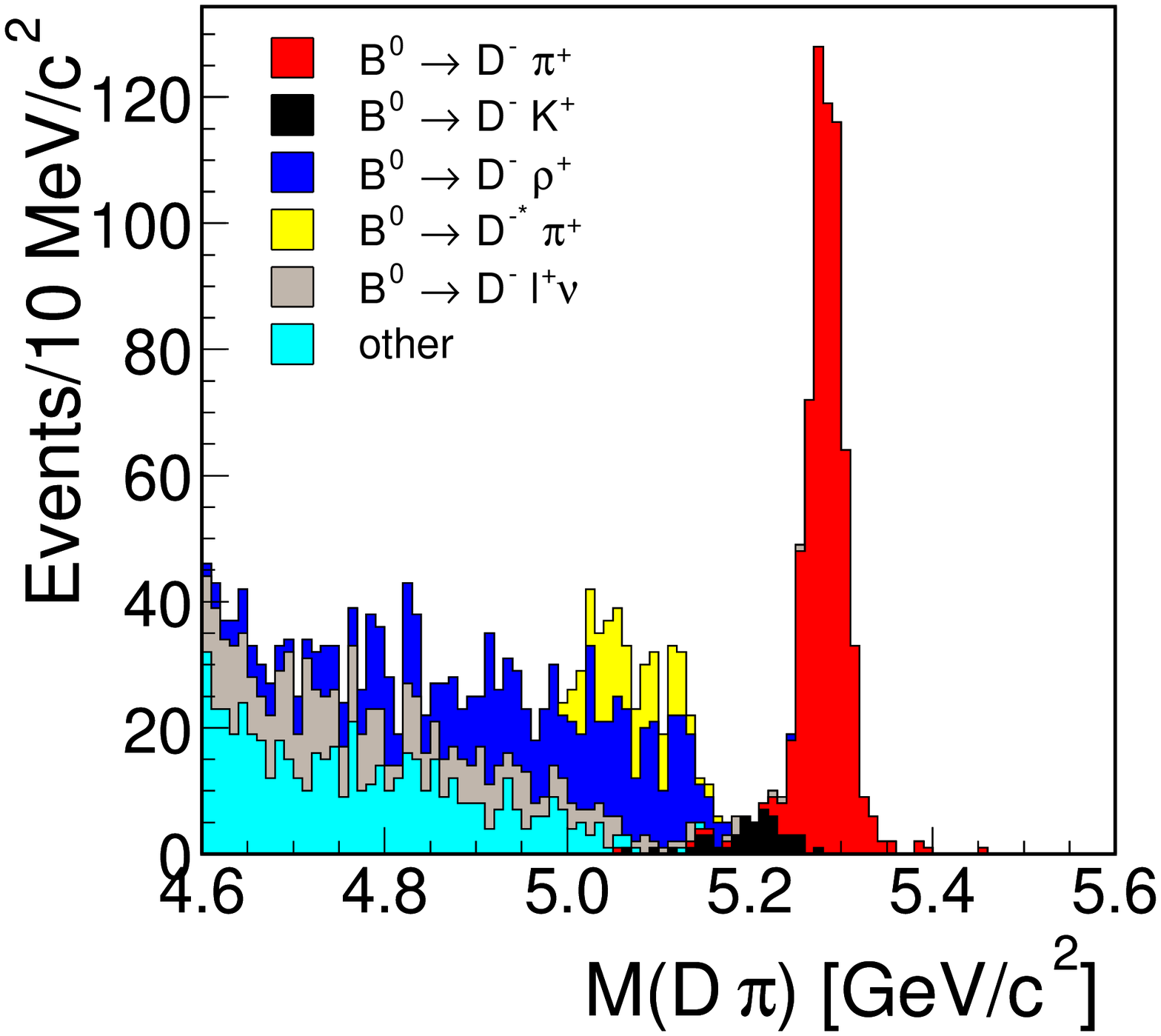}    \\
       \multicolumn{2}{c}{
        \includegraphics[width=200pt, angle=0]
	{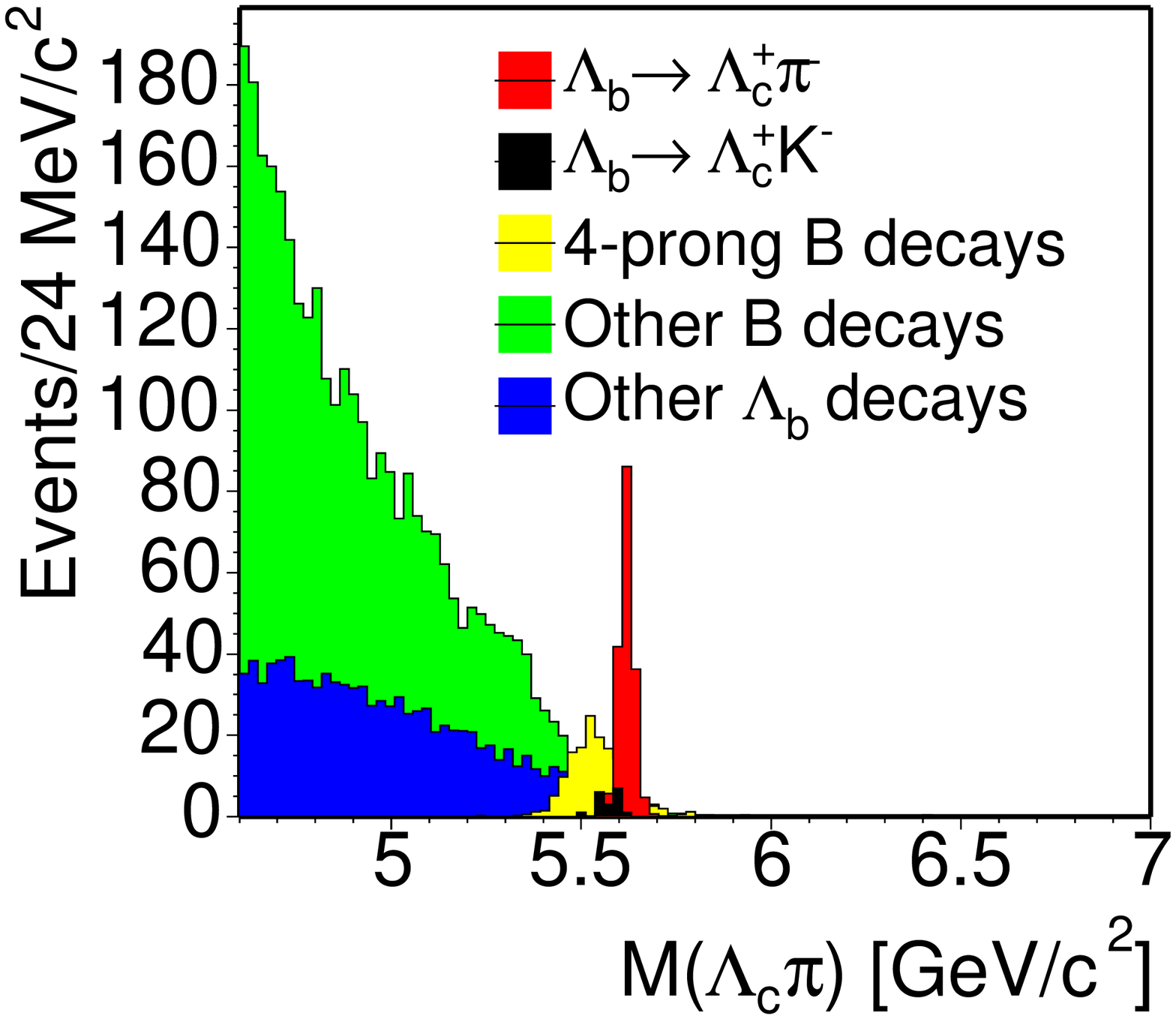}}\\    
  \end{tabular}
 \caption[Background composition in the hadronic modes]
	{Background composition for \dstarhad\ (top), \dhad\ (middle) from
	Furic~\cite{furic:thesis} and \lbhad\ (bottom) from 
        Maksimovi\'{c}~\cite{yile:lblcpi}. Note that the 
	mis-reconstructed \bsdspi\ and \lbhad\ in the $D\pi$ mass is not
        shown.
 \label{fig:bgdecomp}
	}
     \end{center}
 \end{figure}

Figure~\ref{fig:allhisto} shows that to the left of the hadronic signal peak, 
the ``charm+$\pi$'' mass spectrum exhibits an interesting structure. 
In order to extract a correct 
number of events observed in the hadronic channels, we have to take into 
account the background structure when fitting the charm+$\pi$ mass spectrum. 
For the \dhad\ and \lbhad\ modes, we import the \Bd\ and \Lb\ mass functions 
derived in the analyses of Furic~\cite{furic:thesis}, and 
Maksimovi\'{c}~\cite{yile:lblcpi}, respectively. Several parameters that 
describe the background shapes or normalizations are fixed. We find small 
modifications are needed for the numerical values of the fixed parameters in 
the \dhad\ mode, as a few variables we apply cut on are different from those 
in Furic's analysis. We apply our cuts on the MC used in Furic's analysis and 
refit the MC to extract the numbers for our analysis. For the \dstarhad\ mode, 
we produce an inclusive $\B\rightarrow \Dstar X$ MC sample to study the 
background composition. The decay modes with distinguished mass shape are 
separated from the other modes. The decays with similar mass spectra are 
lumped together and fit to the same background function. Figure
~\ref{fig:bgdecomp} shows the \Bd\ and \Lb\ mass spectra from the 
contributions of different decays. 

Our hadronic mass spectra share several common features: It is clear that the 
background from the \B\ hadron decays only contribute to the mass region below 
the signal, while in the data, the upper mass region is composed of 
combinatorial background, which may be described by an exponential or a 
constant. The combinatorial background extends down to the lower \B\ mass 
region as well. In the region 40 to 70 \mevcsq\ below the signal peak, Cabibbo 
suppressed decays, \dstarhadk, \dhadk, \lbhadk, with a branching ratio about 
$8\%$  of our Cabibbo favored signals, produce a small contamination. Going 
further down in the charm+$\pi$ mass, we have partially reconstructed \B\ 
decays from the semileptonic modes, and other mis-identified \B\ hadronic 
decays. 

Note that since both \dstarhad\ and \lbhad\ have low statistics, we 
constrain the widths of their signal Gaussians in the following way: We first 
fit the width of $M_{D\pi}$ ($\sigma^\mathrm{data}_{D\pi}$) from the high 
statistics \dhad\ sample ($\sim$600 events) in the data. Then we multiply 
$\sigma^\mathrm{data}_{D\pi}$ with the MC width ratio:
\(\sigma^\mathrm{MC}_{\Lamc\pi,D^*\pi}/\sigma^\mathrm{MC}_{D\pi}\) and 
predict \(\sigma^\mathrm{data}_{\Lamc\pi,D^*\pi}\).

\subsection{\boldmath{$\dstarhad$} Yield}
 The study from the \alldstar\ MC shows that the background in the lower mass 
region is dominated by the following decays: Cabibbo suppressed decay 
\dstarhadk, \bddstarrho, and the remaining \alldstar. See the texts below for 
the detailed descriptions.
\begin{enumerate}
\item \dstarhadk: fully reconstructed Cabibbo suppressed decays. The mass 
spectrum is a peak about 40 \mevcsq\ below the \dstarhad\ signal, with small 
tails on the lower mass side. The shape is modeled by a lifetime function;
   \begin{equation}
   {\cal E}(m) = \exp(m,\tau_{D^*K})\otimes 
	{\cal G}(m,\mu_{D^*K},\sigma_{D^*K}),
   \label{eq:dstark}
   \end{equation}
where $\tau_{D^*K}$ is the lifetime, $\mu_{D^*K}$ is the zero point of the 
lifetime function also the mean of the Gaussian.  The width of the Gaussian 
also the resolution of the lifetime function is $\sigma_{D^*K}$.
The exact form of ${\cal E}(m)$ is found in the appendix of 
Yu~\cite{cdfnote:7559}.
See Figure~\ref{fig:alldstarmcfit} (top) for the fit to \dstarhadk\ MC. 

\item \bddstarrho, where $\rho^-\rightarrow\pi^0\pi^-$:  modeled by a 
triangular function convoluted with a Gaussian;
\begin{equation}
{\cal T}(m) = {\frac{2(m-M_0^{D^*\rho})}
	{(M_\mathrm{off}^{D^*\rho}-M_0^{D^*\rho})^2}} 
\otimes {\cal G}(m,M_0^{D^*\rho},\sigma_{D^*\rho}).
 \label{eq:dstarrho}
\end{equation}
See Figure~\ref{fig:alldstarmcfit} (middle) for the fit.

\item Continuum: 
remaining \alldstar\ decays partially reconstructed. 
These backgrounds have similar mass spectrum and are group together. The shape 
is modeled by a first-order polynomial with a negative slope and a turn-off 
at \(M_\mathrm{off}^\mathrm{otherB}\); 
when $m < M_\mathrm{off}^\mathrm{otherB}$: 
\begin{equation}
  {\cal H}(m) = \frac{2}{(M_\mathrm{off}^\mathrm{otherB}-M_\mathrm{min})^2}
	\cdot(M_\mathrm{off}^\mathrm{otherB}-m),  
   \label{eq:dstarx}
\end{equation}
and when $m > M_\mathrm{off}^\mathrm{otherB}$: 
\begin{equation}
{\cal H}(m)=0.
\end{equation}
The lowest boundary of the $D^*\pi$ mass window, $M_\mathrm{min}$, is 
4.6 \gevcsq. See Figure~\ref{fig:alldstarmcfit} (bottom) for the fit to these 
MC samples. 
 
\end{enumerate}

 \begin{figure}[tbp]
     \begin{center}
        \includegraphics[width=230pt, angle=0]
	{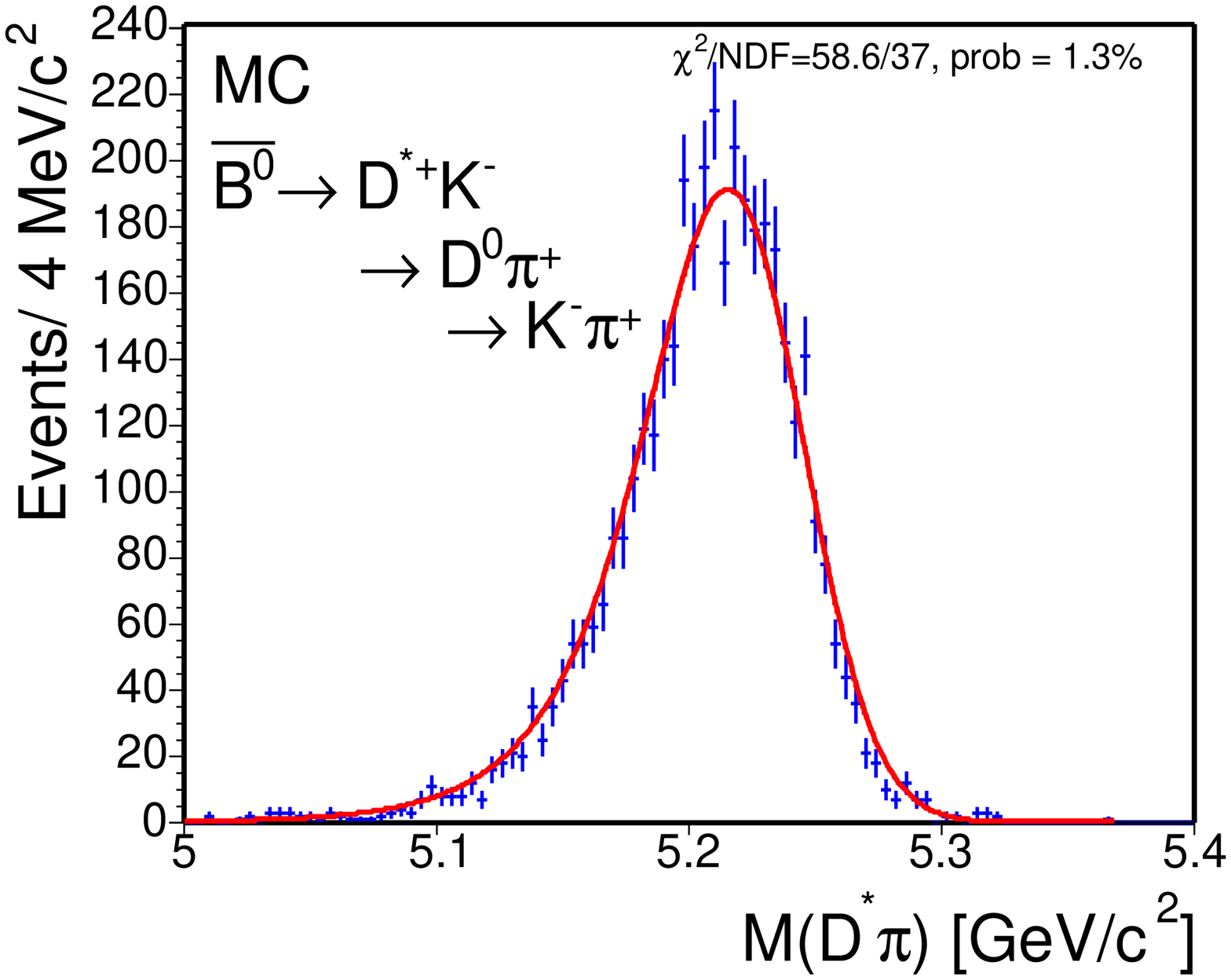}
        \includegraphics[width=230pt, angle=0]
	{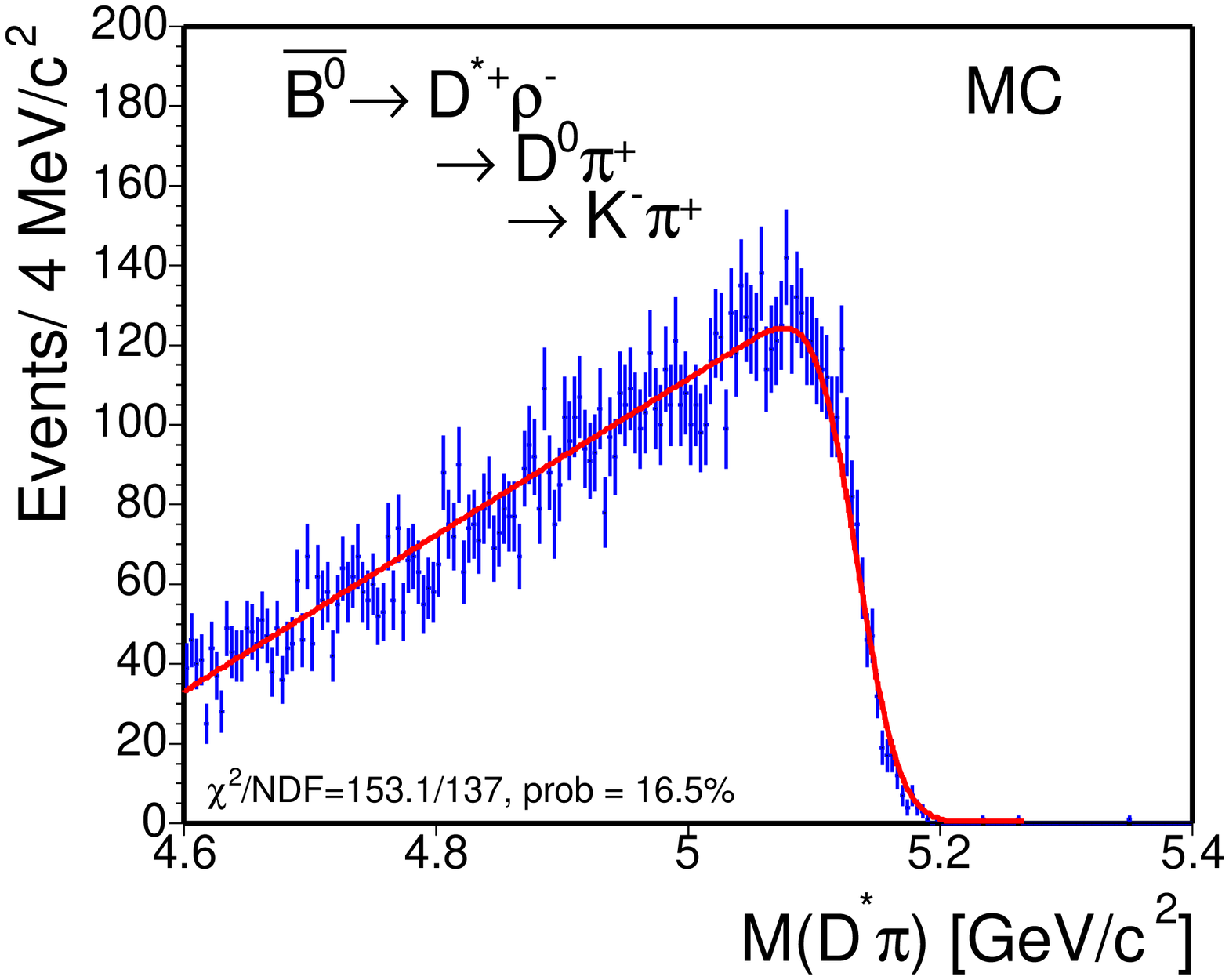}
        \includegraphics[width=230pt, angle=0]
	{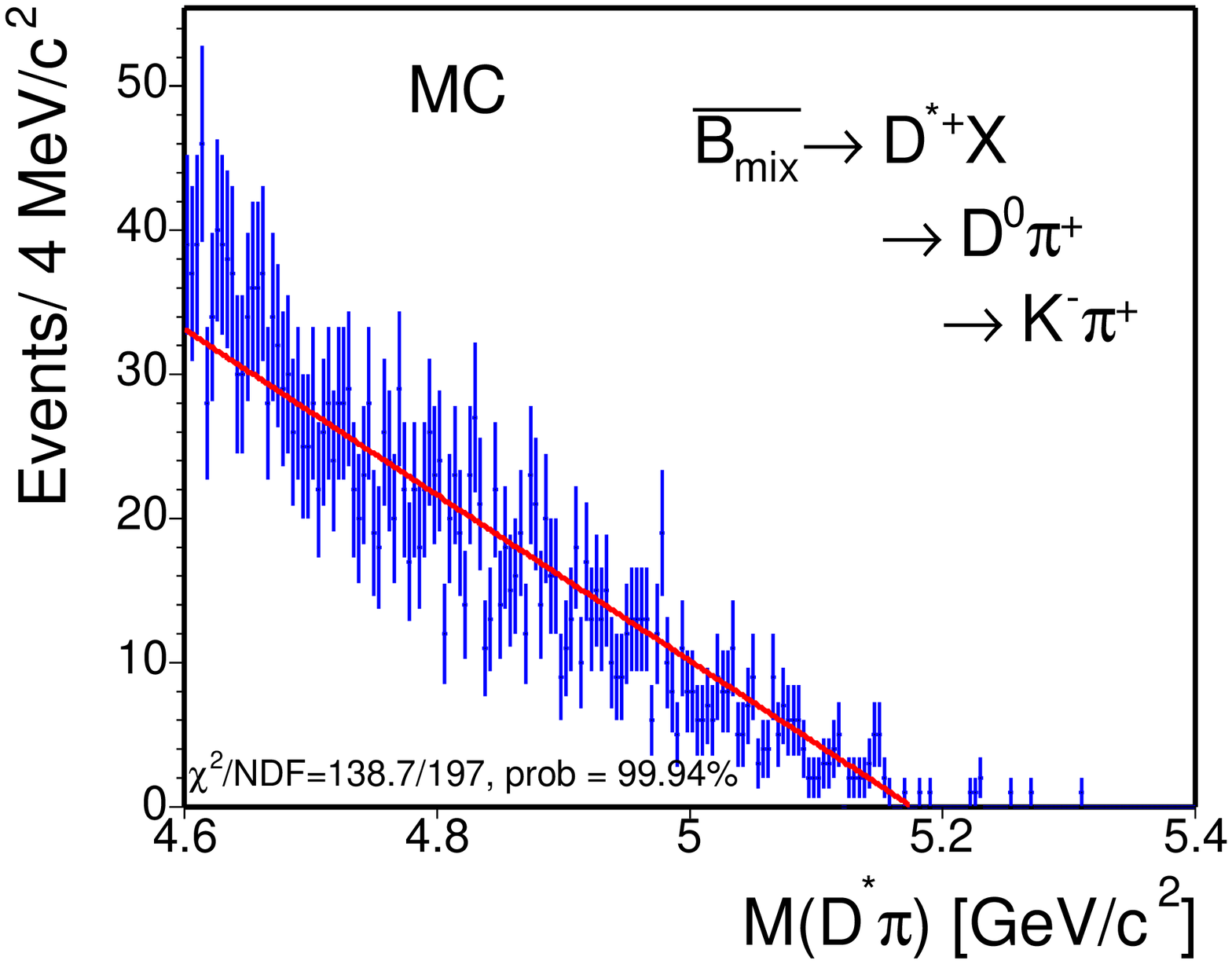}
  	\caption[Fit to $D^*K$, $D^*\rho$ and the remaining \alldstar\ MC 
	(reconstructed as \dstarhad) ]
	{ Various MC samples reconstructed as \dstarhad. 
         From the top to the bottom are \dstarhadk, \bddstarrho, 
	and the remaining \alldstar.  \label{fig:alldstarmcfit}
	}
     \end{center}
 \end{figure}

In the unbinned fit, the extended log likelihood function is expressed by the 
sum of five likelihoods: one describing the data, and the other four 
describing the MC samples from each type of background and the signal: 
\begin{equation}
 \log{\cal L}  = \log{\cal L}^\mathrm{data} + 
	\log{\cal L}^\mathrm{MC}_{D^*\pi}+ \log{\cal L}^\mathrm{MC}_{D^*K} + 
	\log{\cal L}^\mathrm{MC}_{D*\rho}+
	\log{\cal L}^\mathrm{MC}_\mathrm{otherB}, 	 
\end{equation}
The likelihood function \(\log{\cal L}^\mathrm{data}\) is a sum of a signal 
Gaussian, a constant combinatorial background, the functions for $D^*K$ 
(\({\cal E}\)), $D^*\rho$ (\({\cal T}\)), and the continuum (\({\cal H}\)). 
In addition, there is a constraint on each of the following parameters: the 
signal width, relative amount of $D^*K$ to the signal ($f_{D^*K}$), and the 
fraction of $D^*\rho$ in $D^*\rho$ + remaining $\alldstar$ ($f_{D^*\rho}$)
The reason for the last constraint is because \bddstarrho\ and the remaining 
\alldstar\ decays occupy the same mass region. Therefore, the likelihood fit 
converges faster if we constrain $f_{D^*\rho}$. 
\begin{eqnarray}
 \log{\cal L}^\mathrm{data} & = & \sum_i \log\{ N_\mathrm{sig}
	\cdot({\cal G}(m_i,\mu,\sigma)
	+  f_{D^*K}\cdot{\cal E}(m_i)) \nonumber \\
        & & + N_\mathrm{bg}\cdot[f_\mathrm{combg}
	\cdot\frac{1}{M_\mathrm{max}-M_\mathrm{min}} \nonumber \\ 
	& & + (1-f_\mathrm{combg})\cdot[
	f_{D^*\rho}\cdot{\cal T}(m_i) + (1-f_{D^*\rho})\cdot{\cal H}(m_i)] \}
       \nonumber \\
        & - &  N_\mathrm{sig} - N_{D^*K}  - N_\mathrm{bg}
	\nonumber \\
	& + & \log{\cal C}_1 + \log{\cal C}_2 + \log{\cal C}_{\sigma}, 
\label{eq:logdstarpi}
\end{eqnarray}
where ${\cal E}(m_i)$, ${\cal T}(m_i)$ and ${\cal H}(m_i)$ are expressed in 
Equations~\ref{eq:dstark}--\ref{eq:dstarx}. The $M_\mathrm{max}$ and 
$M_\mathrm{min}$ specify the mass window: 4.6 $<$ $M_{D^*\pi}$ $<$ 5.6 \gevcsq.
 The parameters $f_{D^*K}$, $N_\mathrm{bg}$, $f_\mathrm{combg}$ and 
$f_{D^*\rho}$ are defined as follow: 
\begin{eqnarray*}
 f_{D^*K} & \equiv &\frac{N_{D^*K}}{N_\mathrm{sig}}, \\
 N_\mathrm{bg} & \equiv & N_\mathrm{combg} + N_\mathrm{otherB} + N_{D^*\rho},\\
 f_\mathrm{combg} & \equiv & \frac{N_\mathrm{combg}}{N_\mathrm{bg}},\\
 f_{D^*\rho} & \equiv & \frac{N_{D^*\rho}}{N_\mathrm{otherB} + N_{D^*\rho}}.
\end{eqnarray*}

The constraints are expressed as: 
\begin{eqnarray*}
 {\cal C}_1 &= &{\cal G}(f_{D^*K}, \mu_1, \sigma_1),\\
 {\cal C}_2 &= & {\cal G}(f_{D^*\rho}, \mu_2, \sigma_2),\\
 {\cal C}_{\sigma} &= &{\cal G}(\sigma, \mu_p, \sigma_p),
\end{eqnarray*}
where \(\mu_1=0.071\), \(\sigma_1=0.019\), \(\mu_2=0.242\), 
\(\sigma_2=0.008\), \(\mu_p=0.0259\;\gevcsq\), and 
\(\sigma_p=0.0012\;\gevcsq\).  

Here, \(\mu_p\) and \(\sigma_p\) are determined using the \dhad\ signal 
in the data, \dhad\ and \dstarhad\ MC as described earlier. The 
\(\mu_{1}\) and \(\sigma_{1}\) are determined using the world average 
branching ratios, and the efficiencies from the MC listed in 
Table~\ref{t:dstarpibgbr}: 
\begin{equation}
 f_{D^*K} = \frac{{\cal B}(\dstarhadk)}{{\cal B}(\dstarhad)}
	\cdot \frac{\epsilon_{\dstarhadk}}{\epsilon_{\dstarhad}}.
\label{eq:dstarkconstraint}
\end{equation}
The \(\mu_{2}\) and \(\sigma_{2}\) are determined by counting the 
number of reconstructed $D^*\rho$ and the remaining \alldstar\ events in the 
MC after all the analysis cuts.


 \begin{table}[tbp]
   \caption{Branching ratios and relative efficiencies for \dstarhad\ 
 	background.}
  \label{t:dstarpibgbr}
   \begin{center}
   \renewcommand{\tabcolsep}{0.07in}
   \begin{tabular}{l|rr|} 
    \hline
     & \multicolumn{1}{|r}{\dstarhadk} & \multicolumn{1}{r|}{\dstarhad} \\	
     \hline
     {\cal B}($\%$) & 
     \multicolumn{1}{|r}{0.276 $\pm$ 0.021} & 
     \multicolumn{1}{r|}{0.020 $\pm$ 0.005} \\
     $\epsilon$ ratio & 
     \multicolumn{1}{|r}{1} &
     \multicolumn{1}{r|}{1.02 $\pm$ 0.02} \\ 
     \hline	
      $f_{D^*K}$ & \multicolumn{2}{|r|}{0.071 $\pm$ 0.019} \\
      \hline
      \hline

     & \bddstarrho & remaining \alldstar \\
    $ N_{MC}$ & 758 & 2371 \\

     \hline	

      $f_{D^*\rho}$ & \multicolumn{2}{|r|}{0.242 $\pm$ 0.008} \\
      \hline
      \hline
      \end{tabular}
      \end{center}
  \end{table}

 The three likelihoods for the background MC are used to obtain the 
parameterization of \({\cal E}(m)\), \({\cal T}(m)\), and \({\cal H}(m)\). 
The normalizations do not matter here.  
\begin{eqnarray}
 \log{\cal L}^\mathrm{MC}_{D^*K} & = & \sum_i \log {\cal E}(m_i), \\ 
 \log{\cal L}^\mathrm{MC}_{D*\rho} & = & \sum_i \log {\cal T}(m_i),\\
 \log{\cal L}^\mathrm{MC}_\mathrm{otherB} & = & \sum_i \log {\cal H}(m_i).  
\end{eqnarray}
In addition, \(\log{\cal L}^\mathrm{MC}_{D^*\pi}\) is used to obtain the 
reconstructed mass difference between MC and data, $m_\mathrm{diff}$. 
In the \(\log{\cal L}^\mathrm{data}\), all the parameters except the 
normalization and the resolution parameters (\(\sigma\)) for the signal 
Gaussian and the background functions, differ by $m_\mathrm{diff}$ from those 
in the \(\log{\cal L}^\mathrm{MC}\). The resolutions for all the backgrounds 
are kept the same between MC and data, while the resolution of the signal 
Gaussian in the data is a separate free parameter from that in the MC.

\begin{figure}[htb]
 \begin{center}
 \includegraphics[width=300pt, angle=0]
	{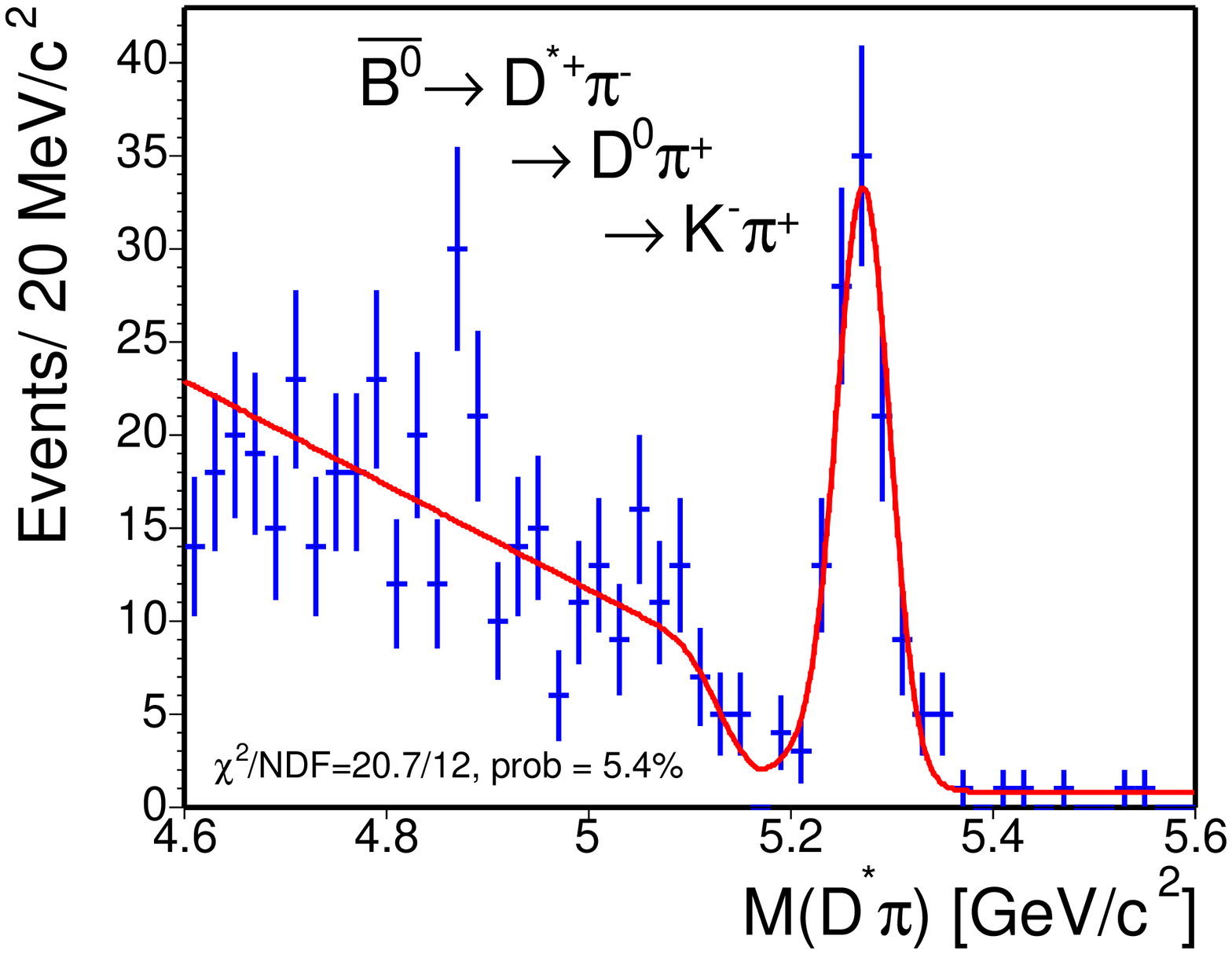}
  \caption[Fit of $M_{D^*\pi}$ from the \dstarhad\ events]
  {$M_{D^*\pi}$ from the \dstarhad\ events is fit to a 
   Gaussian (signal), a constant (combinatorial), and the background 
   functions for the lower mass spectrum as described in the text.
   The result of the unbinned likelihood fit is 
   projected on the histogram and a $\chi^2$ probability is calculated.
   Note that the bins with less than 20 entries are combined.}
 \label{fig:dstarpisignal}
\end{center}
 \end{figure}

We use the total likelihood to fit the data and MC simultaneously. 
Table~\ref{t:dstarpifit} lists the pull means and widths of toy MC test 
and the unbinned likelihood fit result to the data. 
 Figure~\ref{fig:dstarpisignal} shows the fit 
result superimposed on the data histogram. We have obtained from the fit:
\begin{displaymath}
 N_{\dstarhad} = \ndstarhad.
\end{displaymath}
If we remove the constraint on the signal width, we find 
\(N_{\dstarhad} = 110 \pm 11\) and \(\sigma_\mathrm{data} = 0.0295 \pm 0.0033 
\gevcsq\). Removing the constraint on $f_{D^*K}$ gives us \(N_{\dstarhad} = 
107 \pm 11\) and \(f_{D^*K} = 0.053 \pm 0.053\). Removing the constraint on 
$f_{D^*\rho}$ gives us \(N_{\dstarhad} = 107 \pm 11\) and \(f_{D^*\rho} = 
0.38 \pm 0.07\). In conclusion, the un-constrained fits return a value 
consistent with the constrained fit, but with larger uncertainties. 
The fit $\chi^2/\mathrm{NDF}$ are 20.0/12, 20.8/12, 16.5/12 and the fit 
probabilities are 6.7$\%$, 5.4 $\%$, 16.9$\%$ for the three different
 unconstrained fits. 

 \begin{table}[tbp]
   \caption{\dstarhad\ results from the unbinned likelihood fit.}
  \label{t:dstarpifit}
   \begin{center}
   \renewcommand{\tabcolsep}{0.05in}
   \begin{tabular}{|c|lr|r|r|r|} 
    \hline
     Index & \multicolumn{2}{|l|}{Parameter} 
	& 1000 toy MC & 1000 toy MC & Data fit value\\	
     & \multicolumn{2}{|l|}{} 
	& pull mean & pull width & \\	
     \hline
      1 & $N_\mathrm{sig}$ &
	& -0.019 $\pm$ 0.031 & 0.964 $\pm$ 0.022 & \ndstarhad \\
      2 & $\mu$ & [\gevcsq]
	& 0.013 $\pm$ 0.033 & 1.032 $\pm$ 0.024 & 5.2772 $\pm$ 0.0002 \\ 
      3 & $\sigma_\mathrm{MC}$ & [\gevcsq]
	& -0.043 $\pm$ 0.033 & 1.036 $\pm$ 0.024 & 0.0262 $\pm$ 0.0002 \\ 
      4 & $f_{D^{*\rho}}$ & & -0.009 $\pm$ 0.032 
	& 1.006 $\pm$ 0.023 & 0.244 $\pm$ 0.008 \\
      5 & $M_0^{D^*\rho}$ & [\gevcsq]
	& 0.027 $\pm$ 0.033 & 0.943 $\pm$ 0.020 & 4.43 $\pm$ 0.01 \\
      6 & $M_{off}^{D^*\rho}$ & [\gevcsq] 
	& 0.045 $\pm$ 0.032 & 0.993 $\pm$ 0.023 & 5.134 $\pm$ 0.001 \\
      7 & $\sigma_{D^*\rho}$ & [\gevcsq] 
	& -0.077 $\pm$ 0.033 & 1.001 $\pm$ 0.024 & 0.026 $\pm$ 0.001 \\
      8 & $f_\mathrm{combg}$ & &-0.045 $\pm$ 0.032 
	& 0.940 $\pm$ 0.023 & 0.09 $\pm$ 0.03 \\
      9 & $N_\mathrm{bg}$ &
	& -0.010 $\pm$ 0.033 & 1.019 $\pm$ 0.023 & 428 $\pm$ 21 \\
      10 & $M_\mathrm{off}^\mathrm{otherB}$ & [\gevcsq] 
	& -0.220 $\pm$ 0.031 & 0.972 $\pm$ 0.022 & 5.174 $\pm$ 0.004 \\
      11 & $f_{D^*K}$ & & -0.047 $\pm$ 0.033 
	& 1.016 $\pm$ 0.023 & 0.069 $\pm$ 0.018 \\
      12 & $\mu_{D^*K}$ & [\gevcsq]
	& -0.032 $\pm$ 0.033 & 1.024 $\pm$ 0.024 & 5.2345 $\pm$ 0.0009 \\
      13 & $\tau_{D^*K}$ & [\gevcsq]$^{-1}$ 
	&-0.017 $\pm$ 0.032 & 0.986 $\pm$ 0.023 & 0.0287 $\pm$ 0.0009 \\
      14 & $\sigma_{D^*K}$ & [\gevcsq]
	& 0.029 $\pm$ 0.033 & 1.029 $\pm$ 0.024 & 0.0254 $\pm$ 0.0006 \\ 
      15 & $m_\mathrm{diff}$ & [\gevcsq] 
	& -0.034 $\pm$ 0.032 & 0.992 $\pm$ 0.023 & 0.005 $\pm$ 0.003 \\
      16 & $\sigma_\mathrm{data}$ & [\gevcsq]
	& -0.050 $\pm$ 0.031 & 0.971 $\pm$ 0.022 & 0.026 $\pm$ 0.001 \\ 
     \hline	
      \hline
      \end{tabular}
      \end{center}
  \end{table}

\subsection{\boldmath$\dhad$ Yield}
\begin{figure}[htb]
     \begin{center}
     \begin{tabular}{cc}
        \includegraphics[width=200pt, angle=0]
	{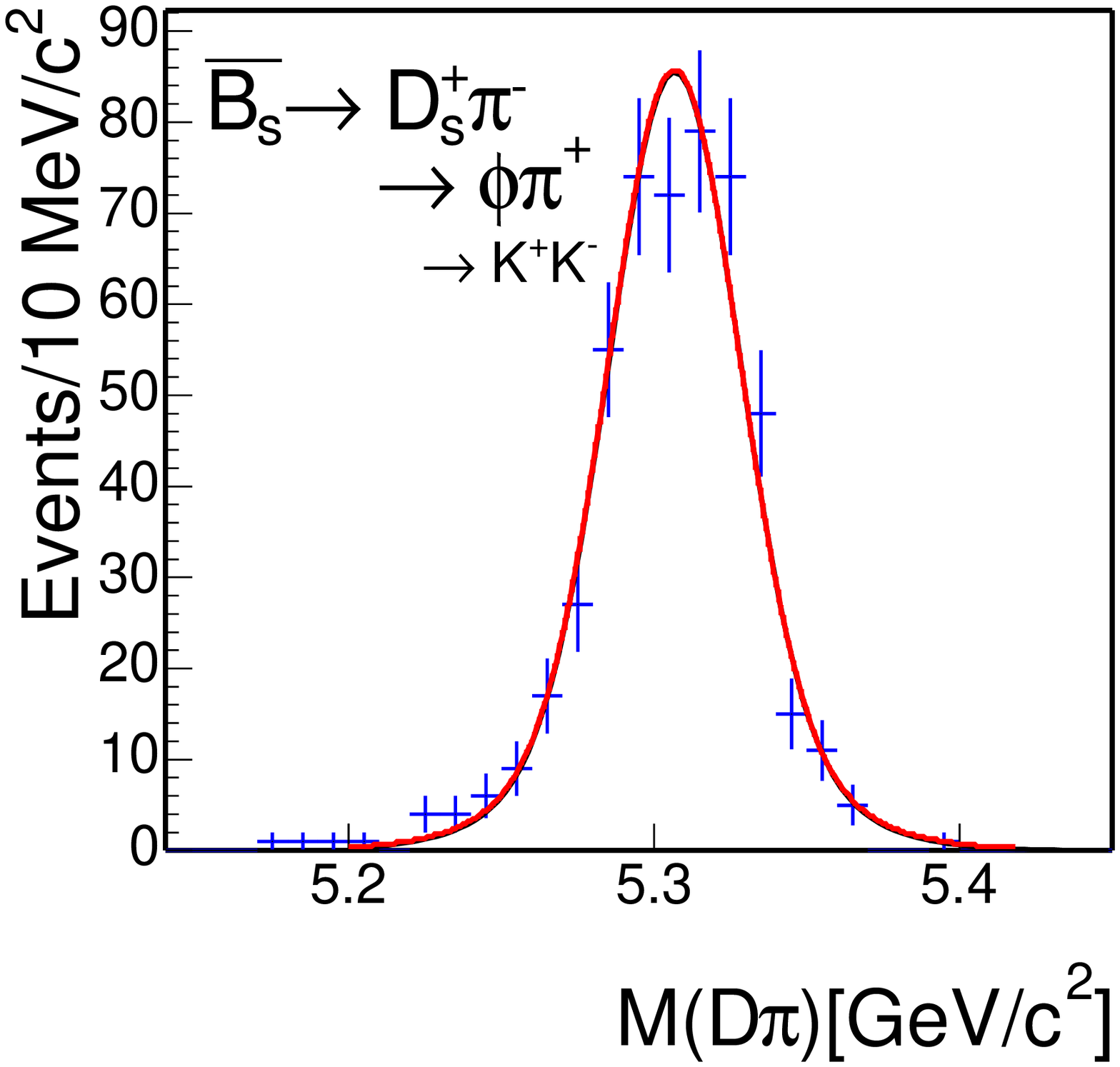} &
        \includegraphics[width=200pt, angle=0]
	{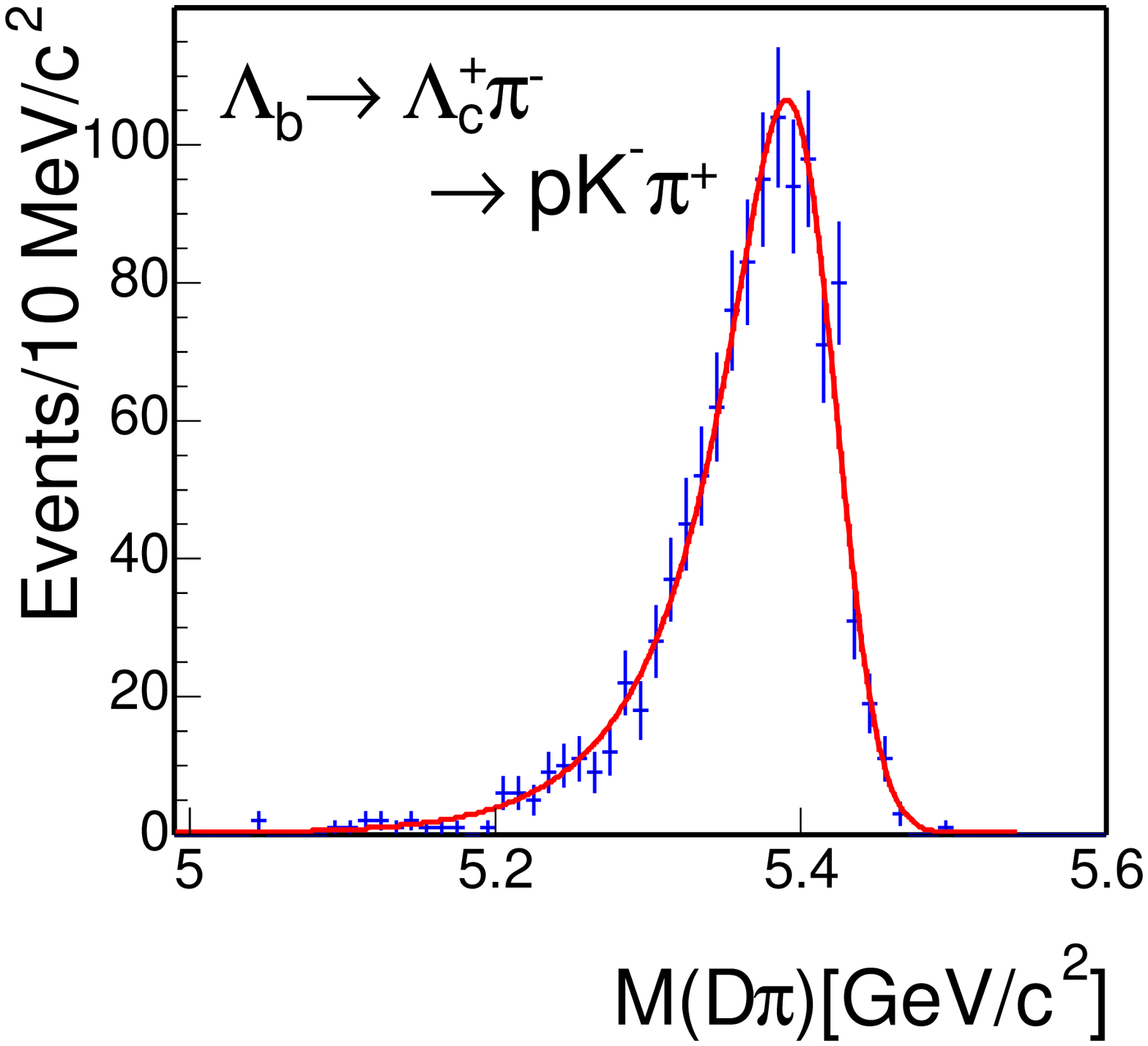}\\
        \includegraphics[width=200pt, angle=0]
	{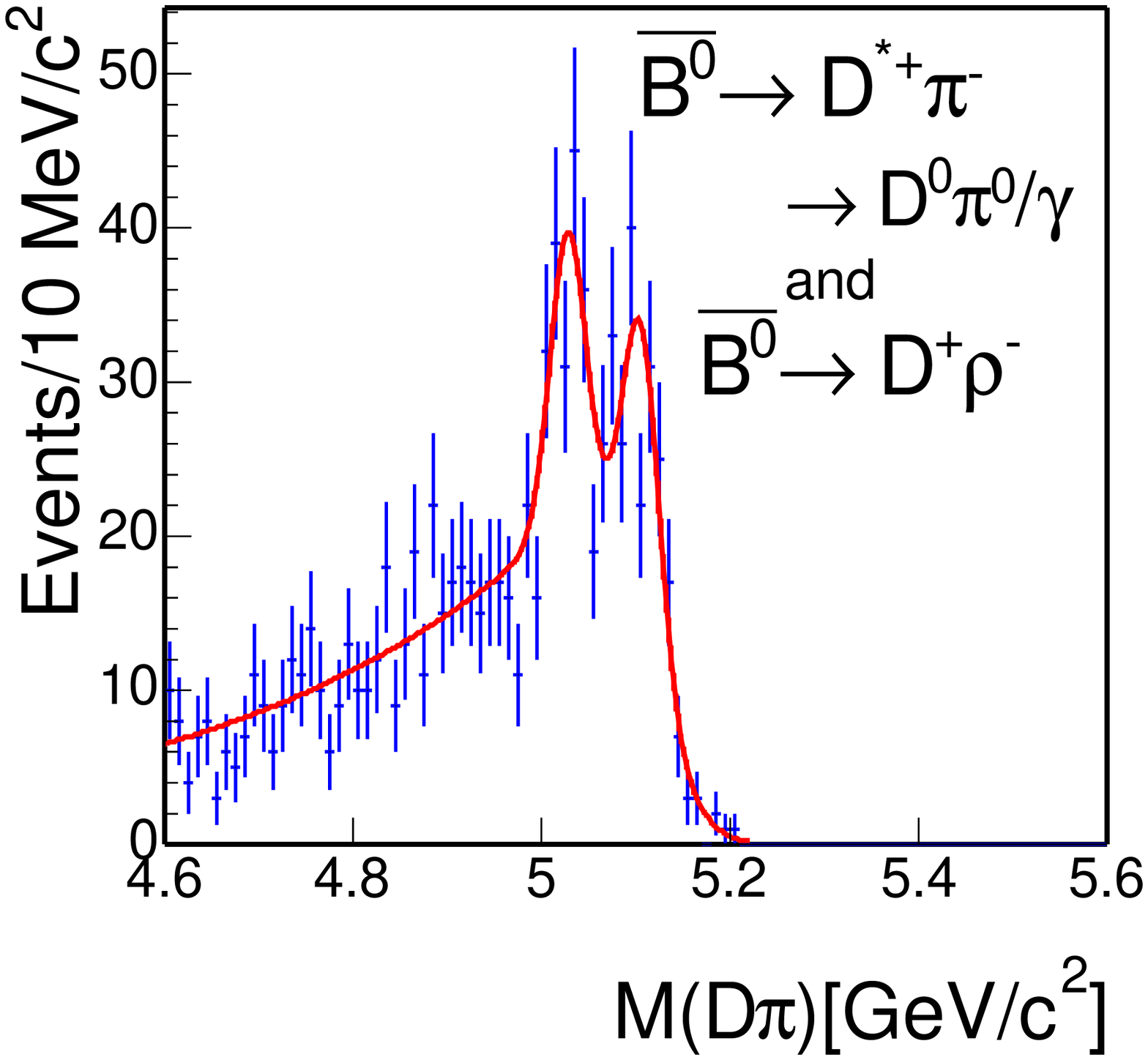} &
        \includegraphics[width=200pt, angle=0]
	{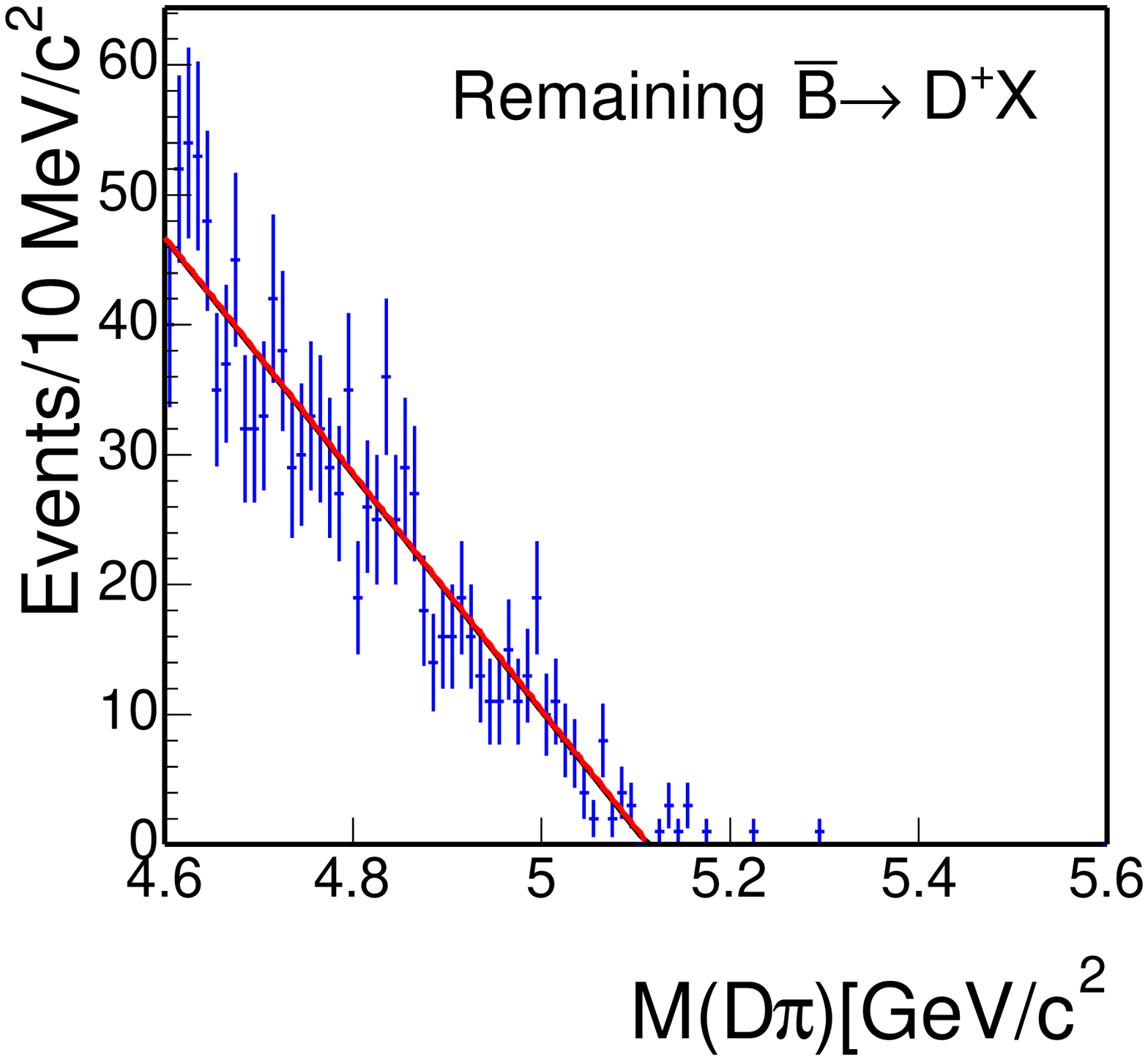} \\
     \end{tabular}
     \end{center}
 \caption[Fit to \Bs, \Lb, $D\rho$, $D^*\pi$ and the remaining \alld\ MC 
	(reconstructed as \dhad) ]
	{ Various MC samples reconstructed as \dhad. 
	From the top left to the bottom
	right are \bsdspi, \lbhad, \dstarhad\ + \bddrho, and the remaining 
	\alld. The fit probabilities are 32.5$\%$, 66.8$\%$, 16.5$\%$ and 32.9
	$\%$.}
     \label{fig:alldmcfit}
 \end{figure}

As noted earlier, we make use of the mass function derived in Furic's analysis 
for the \dhad\ mode.  The parameters which are kept constant in Furic's mass 
function remain constant in our analysis. The following backgrounds contribute 
to the mass spectrum of \Bd\ from Furic's study: Cabibbo suppressed decay 
\dhadk, \dstarhad, \bddrho, remaining \alld\ and the combinatorial background. 
Recent study by Belloni, Martin and Piedra~\etal~\cite{yile:lblcpi}
\cite{cdfnote:7388} shows that the mis-reconstructed \bsdspi\ and \lbhad\ 
produce small contamination in the \dhad\ signal.  See the text below for the 
detailed descriptions. 
\begin{enumerate}
\item \dhadk: 
fully reconstructed Cabibbo suppressed decays. The mass spectrum is a peak 
about 60 \mevcsq\ below the \dhad\ signal. The shape is modeled by a single 
Gaussian; 
 \begin{equation}
 {\cal DK}(m) = {\cal G}(m,\mu-\Delta M_{DK},\sigma_{DK}), 
  \label{eq:dk}
 \end{equation}
where the shift of Gaussian mean from the \dhad\ signal, $\Delta M_{DK}$, and 
the width, $\sigma_{DK}$, are extracted from the MC.

\item \bsdspi, where \seqds\ and \seqphi: 
this decay produces a peak at around 5.31 \gevcsq\ when the pion mass is 
assigned to one of the kaons. The spectrum is modeled by double Gaussians 
with the same mean; 
\begin{equation}
{\cal B_S}(m) = f_1\cdot{\cal G}(m,\mu_{\Bs},\sigma_1) + 
 	(1-f_1)\cdot{\cal G}(m,\mu_{\Bs},\sigma_2), 
 \label{eq:bs}
\end{equation}
where the fraction $f_1$,  $\mu_{\Bs}$, $\sigma_1$ and $\sigma_2$ of each 
Gaussian are obtained from the fit to the MC as shown in 
Figure~\ref{fig:alldmcfit} (top left). 

\item $\lbhad$, where \seqlc: this background produces a broad peak around 
5.4 \gevcsq, the region where the pion mass is mis-assigned to the proton. 
The spectrum is modeled by a lifetime function;
 \begin{equation}
 {\cal L_B}(m) = \exp(m,\tau_{\Lb})\otimes{\cal G}(m,\mu_{\Lb},\sigma_{\Lb}), 
 \label{eq:lb}
 \end{equation}
where $\mu_{\Lb}$ and $\sigma_{\Lb}$ are the zero point and the resolution of 
the lifetime function. See Figure~\ref{fig:alldmcfit} (top right) for the fit 
to the \lbhad\ MC when reconstructed as $\D\pi^-$. 

\item \bddrho, where $\rho^-\rightarrow \pi^0\pi^-$ and \dstarhad\ where 
$\Dstar\rightarrow \D\pi^0$: 
These two backgrounds are combined. The spectrum of \bddrho\ looks like 
\bddstarrho\ in Figure~\ref{fig:alldstarmcfit} (middle) and is modeled by a 
lifetime function. The spectrum of \dstarhad\ is composed of two horns and is 
modeled by two Gaussians with different means. 

The structure of double horns arises for the following reasons: When \dstarhad,
 $\Dstar\rightarrow\D\pi^0$, is reconstructed as $\D\pi^-$, the mass is lower 
than the world average \Bd\ mass due to the missing $\pi^0$. The amount of 
the negative mass shift, $\Delta M$, is determined by the angle between the 
$\pi^0$ and the \Dstar\ flight direction, $d\phi$. Because both \Bd\ and 
$\pi^-$ are scalars (spin=0), to conserve the total angular momentum in the 
decay, the vector particle (spin=1), \Dstar, is transversely polarized. 
The angle $d\phi$ from a transversely polarized \Dstar\ is 
$\cos^2\theta$ distributed and the most probable $d\phi$ is either 
0 or 180 degrees. Therefore, $\Delta M$ is quantized and this forms 
a double-horns spectrum. 

After combing \bddrho\ and \dstarhad, we have:
 \begin{eqnarray}
 {\cal R}(m)& = &(1-f_H)\cdot\exp(m,\tau_\mathrm{ref})	
	\otimes{\cal G}(m,\mu_\mathrm{ref},\sigma_\mathrm{ref}) \nonumber \\
 & + & f_H\cdot(0.5\cdot{\cal G}(m,\mu_\mathrm{ref}-\nu_\mathrm{ref}
	-\delta_\mathrm{ref},\sigma_H) \nonumber \\
  & & + 0.5\cdot{\cal G}(m,\mu_\mathrm{ref}-\nu_\mathrm{ref}+
	\delta_\mathrm{ref},\sigma_H)).
 \label{eq:drho}
 \end{eqnarray}
The exact form of the lifetime function is found in the appendix of Yu~\cite{cdfnote:7559}. The zero point of the lifetime function is 
$\mu_\mathrm{ref}$ and $\nu_\mathrm{ref}$ is the offset of the mid point 
between two horns from the lifetime function. The $\mu_\mathrm{ref}$ and 
$\nu_\mathrm{ref}$ are left free in the likelihood fit to the data. The values 
of the following parameters are extracted from the fit to the MC, as shown in 
Figure~\ref{fig:alldmcfit} (bottom left), and kept constant in the fit 
to the data: the lifetime ($\tau_\mathrm{ref}$), the fraction of horns ($f_H$),
 the half distance between the peak of two horns ($\delta_\mathrm{ref}$),  the 
resolution of the lifetime function ($\sigma_\mathrm{ref}$) and the width of 
both horns ($\sigma_H$).
 
\item Continuum: 
remaining \alld\ decays and partially reconstructed. These backgrounds have 
similar mass spectrum and are group together. The shape is modeled by a 
first-order polynomial with a negative slope and a turn-off at 
\(M_\mathrm{off}\); when $m < M_\mathrm{off}$: 
\begin{equation}
  {\cal H}(m) = \frac{2}{(M_\mathrm{off}-M_\mathrm{min})^2}
	\cdot(M_\mathrm{off}-m),  
  \label{eq:dx}
\end{equation}
and when $m > M_\mathrm{off}$:
\begin{equation}
 {\cal H}(m) =0.
\end{equation}
The lowest boundary of the \(D\pi\) mass window, $M_\mathrm{min}$, is  
4.6 \gevcsq. See Figure~\ref{fig:alldmcfit} (bottom right) for the fit to 
these MC samples from Furic's analysis~\cite{furic:thesis}.

\item combinatorial: modeled by an exponential function. 
 When the slope of the exponential, $p_0$, is not zero, 
 \begin{equation}
  {\cal E_{XP}}(m) = p_0\cdot\frac{e^{-p_0\cdot M_\mathrm{mid}}}
	{e^{-p_0\cdot M_\mathrm{min}}-e^{-p_0\cdot M_\mathrm{max}}}
	\cdot e^{-p_0\cdot(x-M_\mathrm{mid})},
 \label{eq:dcombg}
 \end{equation}
 and when $p_0$ is zero,
 \begin{equation}
  {\cal E_{XP}}(m) = \frac{1}{M_\mathrm{max}-M_\mathrm{min}},
 \label{eq:dcombg2}
 \end{equation}
 where $M_\mathrm{max}$ and $M_\mathrm{min}$ specify the mass window:
 4.6 $<$ \(M_{D\pi}\) $<$ 5.6 \gevcsq\ and $M_\mathrm{mid}$ is the average of 
$M_\mathrm{max}$ and $M_\mathrm{min}$. 

\end{enumerate}


In the unbinned fit, the extended log likelihood function is expressed as 
a sum of a signal Gaussian, the functions for the $DK$ mode (${\cal DK}$), 
\(D_s\pi\) (${\cal B_S}$), \(\Lamc\pi\) (${\cal L_B}$), \(D^*\pi\) plus 
\(D\rho\) (${\cal R}$), the remaining \alld\ decays (${\cal H}$), and 
the combinatorial background (${\cal E_{XP}}$):
\begin{eqnarray}
 \log{\cal L} & = & \sum_i \log\{ 
	N_\mathrm{sig}\cdot[{\cal G}(m_i,\mu,\sigma) +
	f_{DK}\cdot{\cal DK}(m_i) \nonumber \\ 
	& & + f_{\Bs}\cdot{\cal B_S}(m_i) + 
	f_{\Lb}\cdot{\cal L_B}(m_i)] \nonumber \\
        & & +  N_\mathrm{bg}\cdot[(1-f_\mathrm{combg})
	\cdot[(1-f_\mathrm{otherB})\cdot{\cal R}(m_i) 
	+ f_\mathrm{otherB}\cdot{\cal H}(m_i)]\nonumber \\	
        & & + f_\mathrm{combg}\cdot{\cal E_{XP}}(m_i)]\}
       \nonumber \\
        & - &  N_\mathrm{sig}\cdot(1+f_{DK}+f_{\Bs}+f_{\Lb}) 
	-  N_\mathrm{bg}, 
\end{eqnarray}
where ${\cal DK}(m_i)$, ${\cal B_S}(m_i)$, ${\cal L_B}(m_i)$, ${\cal R}(m_i)$,
  ${\cal H}(m_i)$ and ${\cal E_{XP}}(m_i)$ are expressed in 
Equations~\ref{eq:dk}--\ref{eq:dcombg2}. The fractions $f_{DK}$, $f_{\Bs}$ 
and $f_{\Lb}$ are the ratios of $N_{DK}$, $N_{\Bs}$ and $N_{\Lb}$ to the 
signal, $N_\mathrm{sig}$. The total amount of combinatorial background, the 
backgrounds from the $D\rho$, $D^*\pi$, and the remaining \B\ decays is 
denoted as $N_\mathrm{bg}$. The parameters $f_\mathrm{combg}$ and 
$f_\mathrm{otherB}$ are defined as follows:
\begin{eqnarray*}
f_\mathrm{combg} & \equiv & \frac{N_\mathrm{combg}}{N_\mathrm{bg}}, \\
f_\mathrm{otherB} & \equiv & \frac{N_\mathrm{otherB}}
	{N_\mathrm{otherB}+ N_{D^*\pi}+ N_{D\rho}}.
\end{eqnarray*}
All the fractions and ratios here except $f_\mathrm{combg}$ are kept constant 
in the likelihood fit. The \dhadk\ fraction, $f_{DK}$, is determined from the 
world average branching ratios;
 \begin{equation}
 f_{DK} = \frac{{\cal B}(\dhadk)}{{\cal B}(\dhad)}.
 \label{eq:fdk}
 \end{equation}
Table~\ref{t:dkpar} lists the values of the branching ratios in 
Equation~\ref{eq:fdk}. We have \(f_{DK}= 0.073 \pm 0.023\). 

The \Bs\ fraction, $f_{\Bs}$, is obtained using the formula:
\begin{eqnarray}
 f_{\Bs} & = &\frac{f_s}{f_d}\cdot \frac{{\cal B}(\bsdspi)}{{\cal B}(\dhad)}
\cdot\frac{{\cal B}(\seqds){\cal B}(\seqphi)}{{\cal B}(\seqd)}\nonumber\\ 
 & & \cdot\frac{\Gamma(D_s^+\rightarrow K^+K^-\pi^-)}
	{\Gamma(D_s^+\rightarrow \phi(K^+K^-)\pi^-)}\cdot
 	\frac{\epsilon_{\bsdspi}^{MC}}{\epsilon_{\dhad}^{MC}},
\label{eq:bsdspipredict}
\end{eqnarray}
where the branching ratios are from the 2004 PDG and the CDF II measurement 
\(\frac{f_s}{f_d}\cdot \frac{{\cal B}(\bsdspi)}{{\cal B}(\dhad)}\) 
by Furic~\cite{furic:thesis}. 
The efficiency ratio is obtained by applying our $D\pi$ analysis cuts on
 the \Bs\ MC.  Inserting the numbers listed 
in Table~\ref{t:dkpar} into Equation~\ref{eq:bsdspipredict}, we 
obtain \(f_{Bs}=0.006\pm0.001\). Note that the uncertainties from the 
branching ratios of $\phi$, $D_s$, and $D$ decays vanish after multiplying 
Furic's result with the ratio: 
\(\frac{{\cal B}(\seqds){\cal B}(\seqphi)}{{\cal B}(\seqd)}\).

The \Lb\ fraction, $f_{\Lb}$, is obtained using a similar formula;
\begin{equation}
 f_{\Lb} = \yile \times \frac{{\cal B}(\seqlc)}{{\cal B}(\seqd)}\times\frac{\epsilon_{\lbhad}^{MC}}{\epsilon_{\dhad}^{MC}}, 
\label{eq:flb}
\end{equation}
where the product of the first and the second terms come from 2004 PDG and 
CDF II measurements by Le,~\etal~\cite{yile:lblcpi}. The uncertainties from 
the branching ratios of \Lamc\ and $D$ decays vanish in Equation~\ref{eq:flb}. 
The efficiency ratio 
is obtained using the \lbhad\ MC. The value of \(f_{\Lb}\) is 
then \(0.031\pm0.005\). Table~\ref{t:dkpar} lists the numerical values of 
Le's result and the MC efficiency.
Finally, $f_\mathrm{otherB}$ is obtained using Furic's 
$\overline{B}\rightarrow \D X$ MC. We apply our analysis cuts and 
count the number of $D^*\pi+D\rho$ and the remaining \alld\ events. 
We find \(f_\mathrm{otherB} = 0.569 \pm 0.011.\)

\begin{figure}[htb]
 \begin{center}
 \includegraphics[width=300pt, angle=0]{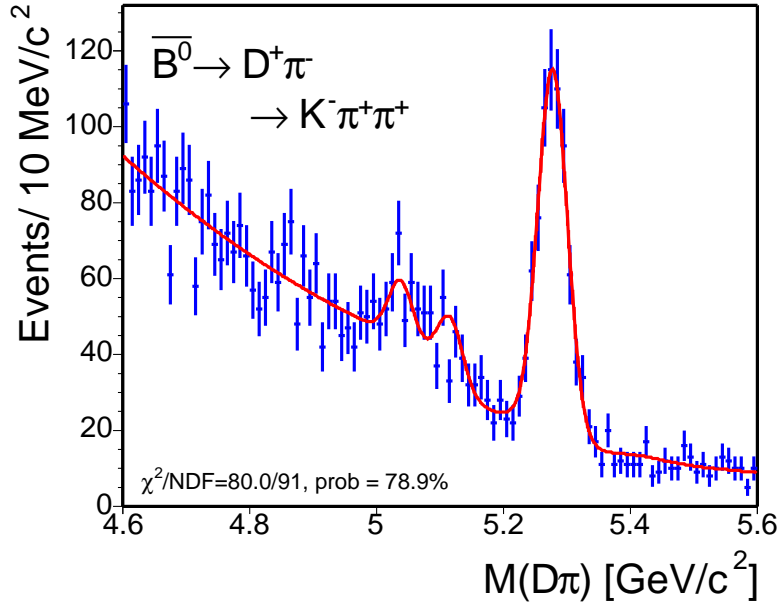}
  \caption[Fit of $M_{D\pi}$ from the \dhad\ events]
  {$M_{D\pi}$ from the \dhad\ events is fit to a 
   Gaussian (signal), an exponential (combinatorial), and the background 
   functions for the lower mass spectrum as described in the text.
   The result of the unbinned likelihood fit is 
   projected on the histogram and a $\chi^2$ probability is calculated.}
 \label{fig:dpisignal}
\end{center}
 \end{figure}

Table~\ref{t:dpifix} lists the constant parameters with their values and 
uncertainties obtained from the fit to the MC. 
Table~\ref{t:dpifit} lists the mean, width of the pulls from the toy MC test 
and the value of each fit parameter from the fit to the data. 
Figure~\ref{fig:dpisignal} shows the fit result superimposed on the data 
histogram. We have obtained from the fit:
\begin{displaymath}
 N_{\dhad} = \ndhad.
\end{displaymath}

 \begin{table}[htb]
   \caption{Fixed parameters in the \dhad\ unbinned likelihood fit.}
    \setdec 0.0000
  \label{t:dpifix}
   \begin{center}
   \renewcommand{\tabcolsep}{0.03in}
   \begin{tabular}{l|l|r@{\,$\pm$\,}l}
    \hline
     Parameter & Meaning & \multicolumn{2}{|r}{Value} \\	
    \hline
     \hline
      $f_{DK}$ & $N_{\dhadk}/N_{\dhad}$ & \dec 0.073 & \dec 0.023 \\
      $\Delta M_{DK}$ & mass shift of \dhadk\ [\gevcsq]  
	& \dec 0.067 & \dec 0.006 \\
      $\sigma_{DK}$ & width of \dhadk [\gevcsq] 
	& \dec 0.032 & \dec 0.009 \\	
      $f_{\Bs}$ & $N_{\bsdspi}/N_{\dhad}$
 	        & \dec 0.006 & \dec 0.001 \\
      $\mu_{\Bs}$ & mean of \Bs\ background [\gevcsq] 
                & \dec 5.307 & \dec 0.001 \\ 
       $f_1$ & fraction of the narrow \Bs\ Gaussian 
	& \dec 0.773 & \dec 0.002 \\
      $\sigma_1$ & width of the narrow \Bs\ Gaussian [\gevcsq] 
	& \dec 0.021 & \dec 0.002 \\
       $\sigma_2/\sigma_1$ & width ratio of the \Bs\ Gaussians 
	& \dec 1.8 & \dec 0.3 \\
       $f_{\Lb}$ & $N_{\lbhad}/N_{\dhad}$ & \dec 0.031 & \dec 0.005 \\
       $\mu_{\Lb}$ & mean of \Lb\ [\gevcsq] 
	& \dec 5.416 & \dec 0.002  \\ 
       $\sigma_{\Lb}$ & width of \Lb\ background [\gevcsq] 
	& \dec 0.024 & \dec 0.002 \\
       $\tau_{\Lb}$ & lifetime of \Lb\ background [\gevcsq$^{-1}$]  
	& \dec 0.052 & \dec 0.002 \\ 
       $\tau_\mathrm{ref}$ & lifetime of $D\rho$ background 
	[\gevcsq$^{-1}$] 
	& \dec 0.36 & \dec 0.06 \\
       $\sigma_\mathrm{ref}$  & width of $D\rho$ background [\gevcsq]&
	\dec 0.039 & \dec 0.008 \\
       $f_H$ & fraction of $D^*\pi$ horns & \dec 0.20 & \dec 0.06 \\ 
       $\delta_\mathrm{ref}$  & distance between two horns [\gevcsq]
	& \dec 0.039 & \dec 0.003 \\
       $\sigma_{H}$ &  width of the horns [\gevcsq] &
	\dec 0.019 & \dec 0.003 \\
       $f_\mathrm{otherB}$ & fraction of the remaining \alld 
	& \dec 0.569 & \dec 0.011 \\
       $M_\mathrm{off}$ & cut off for \alld\ mass [\gevcsq] 
	& \dec 5.112 & \dec 0.007 \\
     \hline	
      \hline
      \end{tabular}
      \end{center}
  \end{table}

\begin{table}[htb]
\renewcommand{\arraystretch}{1.3}
\caption{Parameter values used to determine $f_{DK}$, $f_{\Bs}$ and $f_{\Lb}$.}
\label{t:dkpar}
\begin{center}
\begin{tabular}{l|r|}
\hline\hline
 ${\cal B}(\dhadk)$ & (2.0 $\pm$ 0.6)$\times 10^{-4}$ \\
 ${\cal B}(\dhad)$  & (2.76 $\pm$ 0.25)$\times 10^{-3}$ \\
 \hline
  $f_{DK}$ & 0.073 $\pm$ 0.023 \\
\hline\hline
$\frac{f_s}{f_d}\cdot \frac{{\cal B}(\bsdspi)}{{\cal B}(\dhad)}$ 
& 0.35 $\pm$ 0.05({\em stat}) $\pm$ 0.02 ({\em syst}) $\pm$ 0.09 ({\em BR})\\
${\cal B}(\seqds)$ &  (3.6 $\pm$ 0.9)$\%$ \\
${\cal B}(\seqphi)$ & (49.1 $\pm$ 0.6)$\%$ \\
${\cal B}(\seqd)$ & (9.2 $\pm$ 0.6)$\%$ \\
$\frac{\Gamma(D_s^+\rightarrow K^+K^-\pi^-)}{\Gamma(D_s^+\rightarrow \phi(K^+K^-)\pi^-)}$ & 0.81 $\pm$ 0.08 \\
$\epsilon_{\bsdspi}^{MC}/\epsilon_{\dhad}^{MC}$ & 0.071 $\pm$ 0.004 \\
\hline
$f_{\Bs}$ & 0.006 $\pm$ 0.001\\
\hline\hline
\yile & 0.82 $\pm$ 0.08 ({\em stat}) $\pm$ 0.11 ({\em syst}) $\pm$ 0.22 ({\em BR}) \\
${\cal B}(\seqlc)$ & (5.0 $\pm$ 1.3)$\%$ \\
$\epsilon_{\lbhad}^{MC}/\epsilon_{\dhad}^{MC}$ & 0.069 $\pm$ 0.002 \\
\hline
$f_{\Lb}$ & 0.031 $\pm$ 0.005 \\
\hline\hline
\end{tabular}
\end{center}

   \caption{\dhad\ results from the unbinned likelihood fit.}
  \label{t:dpifit}
   \begin{center}
   \renewcommand{\tabcolsep}{0.05in}
   \begin{tabular}{|c|lr|r|r|r|} 
    \hline
     Index & \multicolumn{2}{|l|}{Parameter} 
	& 1000 toy MC & 1000 toy MC & Data fit value\\	
     & \multicolumn{2}{|l|}{} 
	& pull mean & pull width & \\	
     \hline
        1 & $N_\mathrm{sig}$ & 
	& 0.012 $\pm$ 0.035 & 1.021 $\pm$ 0.026 & \ndhad \\
        2 & $\mu$ & [\gevcsq]
	& -0.026 $\pm$ 0.034 & 0.989 $\pm$ 0.025 & 5.278 $\pm$ 0.001 \\ 
        3 & $\sigma$ & [\gevcsq]
	& -0.040 $\pm$ 0.035 & 1.015 $\pm$ 0.026 & 0.0235 $\pm$ 0.0012 \\ 
        4 & $N_\mathrm{bg}$ &
	& 0.017 $\pm$ 0.034 & 0.990 $\pm$ 0.025 & 4049 $\pm$ 67 \\
        5 & $\mu_\mathrm{ref}$ & [\gevcsq] 
	& -0.036 $\pm$ 0.036 & 1.037 $\pm$ 0.026 & 5.145 $\pm$ 0.015 \\
        6 & $\nu_\mathrm{ref}$ & [\gevcsq]
	& -0.037 $\pm$ 0.038 & 1.085 $\pm$ 0.028 & 0.068 $\pm$ 0.020 \\ 
        7 & $f_\mathrm{combg}$ &
	& -0.145 $\pm$ 0.034 & 0.988 $\pm$ 0.025 & 0.583 $\pm$ 0.044 \\ 
        8 & $p_0$ & 
	& -0.051 $\pm$ 0.034 & 0.976 $\pm$ 0.024 & 1.75 $\pm$ 0.15 \\
     \hline	
      \hline
      \end{tabular}
      \end{center}
	\end{table}

\clearpage

\subsection{\boldmath$\lbhad$ Yield}
\begin{figure}[tbp]
     \begin{center}
      \begin{tabular}{cc}
        \includegraphics[width=200pt, angle=0]
	{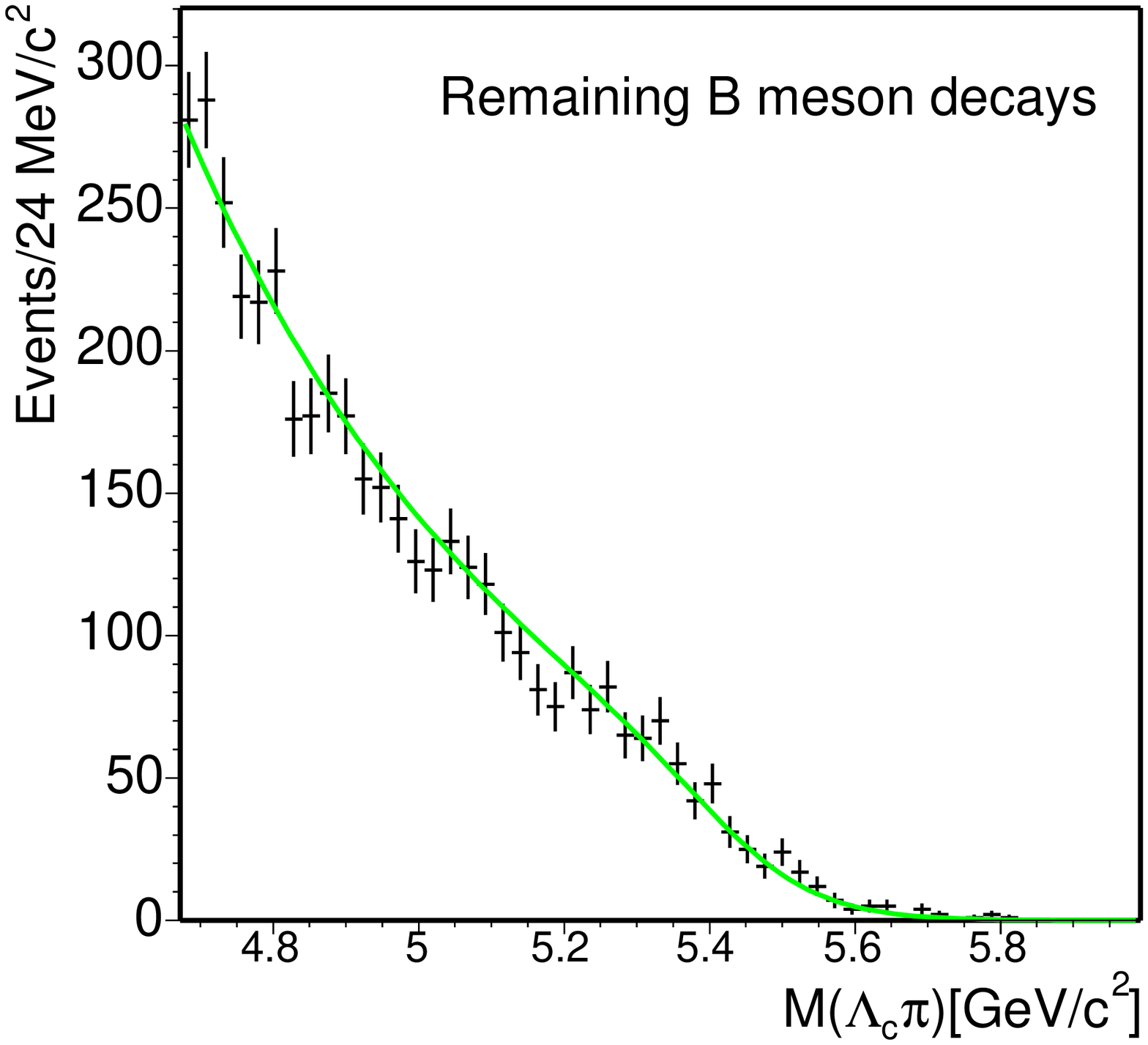} & 
        \includegraphics[width=200pt, angle=0]
	{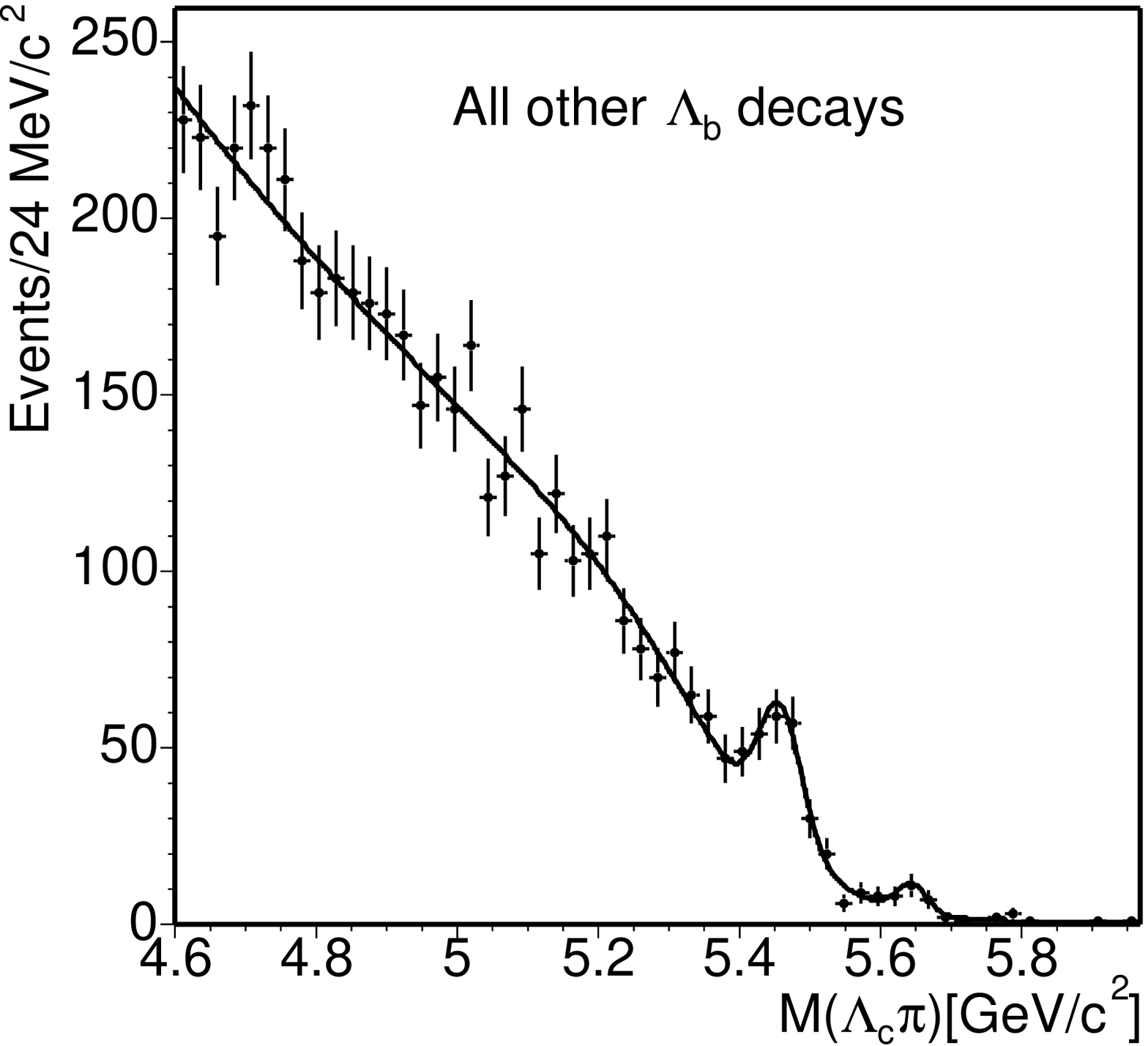} \\ 
      \multicolumn{2}{c}{
        \includegraphics[width=300pt, angle=0]
	{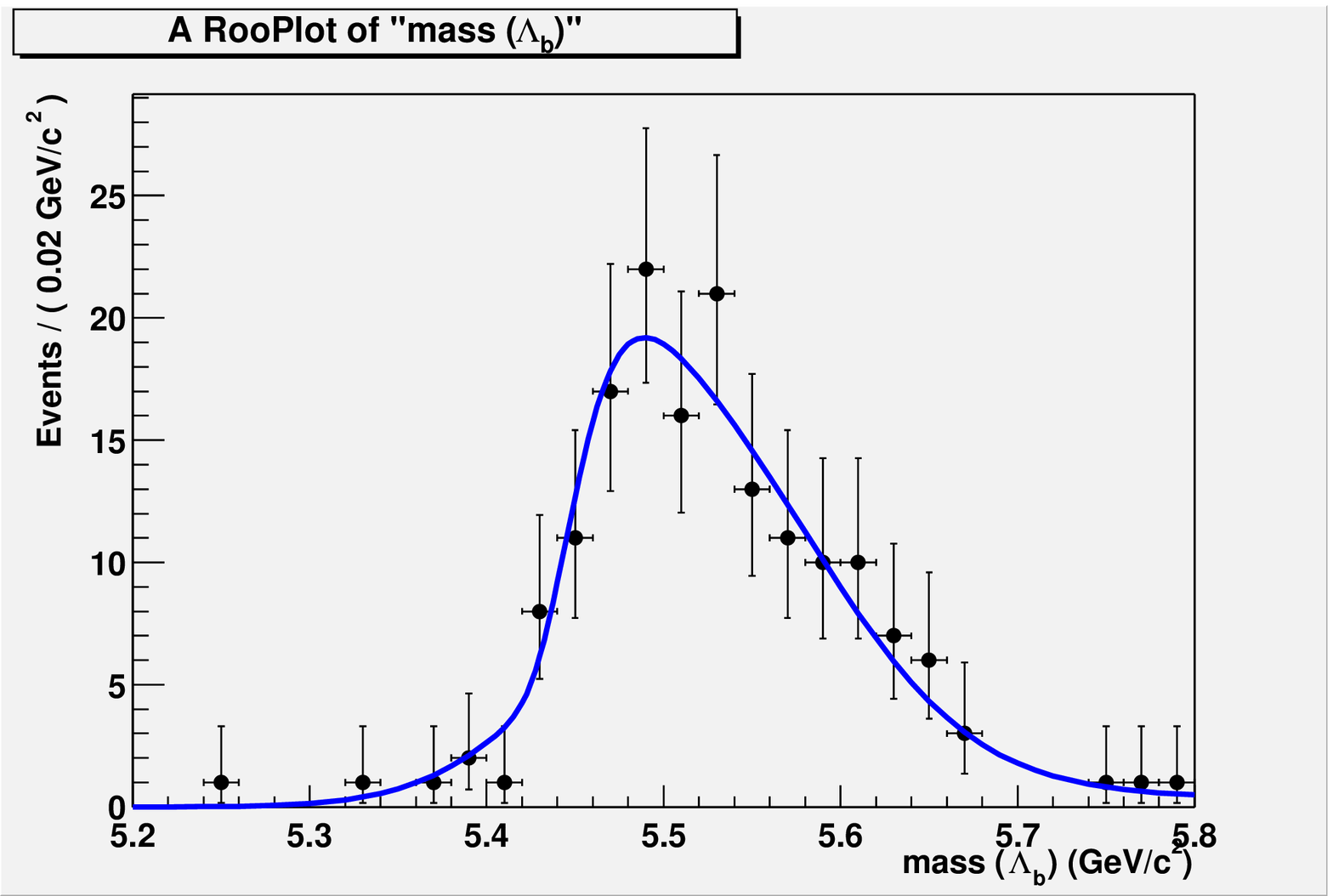} }\\
      \end{tabular}
     \end{center}
 \caption[Fit to MC of four-prong \B\ meson decays, the remaining \B\ meson 
	decays and the remaining \Lb\ decays (reconstructed as \lbhad) ]
	{ Various MC samples reconstructed as \lbhad. From the top left 
	to the bottom are the remaining \B\ 
	meson decays, the remaining \Lb\ decays and four-prong \B\ meson 
	decays.} 
     \label{fig:alllbmcfit}
 \end{figure}

We use the mass function derived in the analysis of Maksimovi\'{c}
for the fit to \lbhad\ data~\cite{yile:lblcpi}. We cross-check the values of 
the background shape parameters by applying our analysis cuts on the MC used 
in the Maksimovi\'{c} analysis. We find the same numbers can be 
used for this analysis. The following backgrounds contribute to the mass 
spectrum of \Lb\ from their study: Cabibbo decay \lbhadk, four-prong 
mis-identified \B\ meson, the remaining \B\ meson decays, the remaining \Lb\ 
decays and the combinatorial background.
\begin{enumerate}
\item \lbhadk: fully reconstructed Cabibbo suppressed decays. 
The shape is modeled by two Gaussians of different mean and width; 
\begin{equation}
{\cal L_CK}(m) = f_1\cdot{\cal G}(m,\mu_{\Lambda_cK}^1,\sigma_1) + 
 	(1-f_1)\cdot{\cal G}(m,\mu_{\Lambda_cK}^2,\sigma_2), 
 \label{eq:lck}
\end{equation}
where $f_1$, $\mu_{\Lambda_cK}^1$, $\sigma_1$, $\mu_{\Lambda_cK}^2$ and
 $\sigma_2$ are from the fit to the MC.

\item mis-identified four-prong \B\ mesons: all the \B\ mesons with four 
tracks in the final states and fully reconstructed. \dhad\ contributes 
about 50$\%$ of this type of background. Since these decays have similar 
final state as our \lbhad\ signal, they produce a distinguished peak to the 
left of the signal Gaussian. This background (\({\cal B}_{4{\cal PRONG}}\)) 
is modeled by the sum of a Landau (\({\cal L_{AND}}\)) and a Gaussian function:
\begin{equation}
{\cal B}_{4{\cal PRONG}}(m) = f_\mathrm{L}\cdot{\cal L_{AND}}
	(m,\mu_\mathrm{BPL},\sigma_L) + 
	(1-f_\mathrm{L})\cdot{\cal G}(m,\mu_\mathrm{BPG},\sigma_G), 
 \label{eq:b4prong}
\end{equation}
where $f_\mathrm{L}$, $\mu_\mathrm{BPL}$ and $\sigma_L$ are the fraction, 
mean and the width of Landau distribution. The mean and the width of the 
Gaussian are denoted as $\mu_\mathrm{BPG}$ and $\sigma_G$. These parameters 
are extracted from fit to the MC as shown in Figure~\ref{fig:alllbmcfit} 
(bottom).

\item remaining \B\ meson decays: this background (\({\cal O_B}\)) 
spectrum is modeled by the sum of an exponential function and a product of 
a bifurcated Gaussian (\({\cal B_F}\)) with a step-down function:
\begin{equation}
 {\cal O_B}(m) = {\cal E_{XP}}(m) + 
	f_\mathrm{bifg}
	\cdot{\cal B_F}(m,\mu_\mathrm{ob},\sigma_\mathrm{ob}^{L},
	\sigma_\mathrm{ob}^{R})
	\cdot(1-\frac{1}{1+e^{(\mu_{obst}-m)/a_0^\mathrm{ob}}})
\label{eq:otherB}
\end{equation}
where ${\cal E_{XP}}(m)$ is expressed in Equations
~\ref{eq:dcombg}--\ref{eq:dcombg2}. The parameters $f_\mathrm{bifg}$, 
$\mu_\mathrm{ob}$, $\sigma_\mathrm{ob}^L$, and $\sigma_\mathrm{ob}^R$ are the 
fraction, mean, left sigma, right sigma of bifurcated Gaussian. 
The step-down function parameters, \(\mu_{obst}\) and \(a_0^\mathrm{ob}\), 
together with the parameters for the bifurcated Gaussian, are 
extracted from the MC as shown in Figure~\ref{fig:alllbmcfit} (top left).
The exact form of the bifurcated Gaussian is found in the appendix 
of Yu~\cite{cdfnote:7559}.

\item remaining \Lb\ decays: this background (\({\cal O_L}\)) spectrum is 
modeled by the sum of two Gaussians and the product of a bifurcated 
Gaussian and a step-down function
\begin{eqnarray}
 {\cal O_L}(m) & = & f_1^\mathrm{ol}\cdot{\cal G}(m,\mu_{1}^{olg},
	\sigma_1^\mathrm{ol}) + 
 	f_2^\mathrm{ol}\cdot{\cal G}(m,\mu_2^{olg},
	\sigma_2^\mathrm{ol}) \nonumber \\
 & & + \cdot{\cal B_F}(m,\mu_\mathrm{ol},\sigma_\mathrm{ol}^{L},
	\sigma_\mathrm{ol}^{R})
	\cdot(1-\frac{1}{1+e^{(\mu_\mathrm{olst}-m)/a_0^\mathrm{ol}}})
\label{eq:otherLb}
\end{eqnarray}
where the parameters in the function are from the fit to the MC as shown in 
Figure~\ref{fig:alllbmcfit} (top right).

\item combinatorial background: described by an exponential function 

\end{enumerate}

In the unbinned fit, the extended log likelihood function is expressed as 
a sum of a signal Gaussian and the functions for \(\Lamc K\) (\({\cal L_CK}\)),
 four-prong \B\ meson (\({\cal B}_{4{\cal PRONG}}\)), remaining 
\B\ meson decays (\({\cal O_B}\)), remaining \Lb\ decays (\({\cal O_L}\)) 
and the combinatorial background (\({\cal E_{XP}}\)). In addition, there 
is a constraint on the width of the signal Gaussian determined using the 
\dhad\ data, \dhad\ and \lbhad\ MC as described earlier. 
\begin{eqnarray}
 \log{\cal L} & = & \sum_i \log\{ 
	N_\mathrm{sig}\cdot[{\cal G}(m_i,\mu,\sigma) +
	f_{\Lamc K}\cdot{\cal L_CK}(m_i)] 
	+ N_\mathrm{B4prong}\cdot{\cal B}_{4{\cal PRONG}}(m_i) \nonumber \\ 
	& & + N_\mathrm{OB}\cdot{\cal O_B}(m_i) 
	+ N_\mathrm{OL}\cdot{\cal O_L}(m_i) 
	+ N_\mathrm{combg}\cdot{\cal E_{XP}}(m_i)\}
       \nonumber \\
        & - &  N_\mathrm{sig}\cdot(1+f_{\Lamc K})
	- N_\mathrm{B4prong}-N_\mathrm{OB}-N_\mathrm{OL} 
	- N_\mathrm{combg}  
	 +  \log {\cal C}_{\sigma}, 
\end{eqnarray}
where ${\cal L_CK}(m_i)$, ${\cal B}_{4{\cal PRONG}}(m_i)$, ${\cal O_B}(m_i)$, 
${\cal O_L}(m_i)$ and ${\cal E_{XP}}(m_i)$ are expressed in 
Equations~\ref{eq:lck}--\ref{eq:otherLb} and Equations~\ref{eq:dcombg}
--\ref{eq:dcombg2}. The fraction $f_{\Lamc K}$ is defined as:
\begin{equation}
 f_{\Lamc K} = \frac{N_{\lbhadk}}{N_{\lbhad}},
\end{equation}
and is fixed to 0.08; the number is suggested by the branching ratio of the 
Cabibbo suppressed relative to the Cabibbo favored decay in the \B\ meson 
system. The mean and the sigma of the Gaussian 
constraint for the signal width are \(0.0231\;\gevcsq\), and 
\(0.0012\;\gevcsq\), respectively. 

\begin{figure}[htb]
     \begin{center}
      \includegraphics[width=300pt, angle=0]{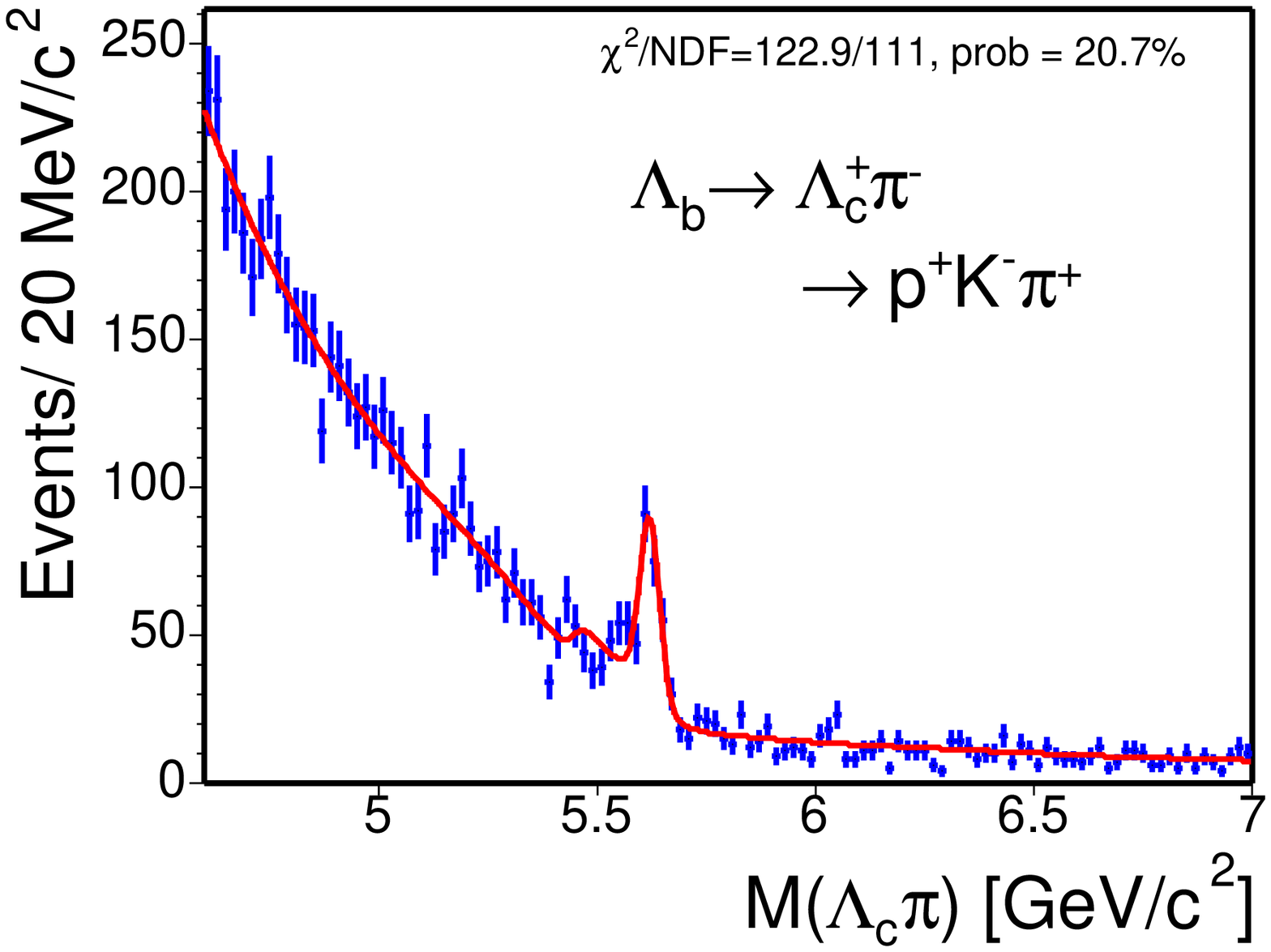}
     \end{center}
     \caption[Fit of $M_{\Lamc\pi}$ from the \lbhad\ events]
  {$M_{\Lamc\pi}$ from the \lbhad\ events is fit to a 
   Gaussian (signal), an exponential (combinatorial), and the background 
   functions for the lower mass spectrum as described in the text.
   The result of the unbinned likelihood fit is 
   projected on the histogram and a $\chi^2$ probability is calculated.}

 \label{fig:lcpisignal}
 \end{figure}

Table~\ref{t:lcpifit} lists the mean, width of the pulls from the toy MC test 
and the fit result to the data. 
Table~\ref{t:lcpifix} lists the values of the constant parameters. 
Figure~\ref{fig:lcpisignal} shows the fit result superimposed on the data 
histogram. We have obtained from the fit:
\begin{displaymath}
 N_{\lbhad} = \nlbhad.
\end{displaymath}
We also cross-check by removing the constraint on the signal width and obtain 
\(N_{\lbhad} = 177 \pm 22\), and \(\sigma = 0.022 \pm 0.004\), 
which are consistent with the result in Table~\ref{t:lcpifit}. 
The fit without constraint has a $\chi^2/\mathrm{NDF}$ of 123.2/111 and fit 
probability of 20.2 $\%$.


 \begin{table}[tbp]
   \caption{\lbhad\ results from the unbinned likelihood fit.}
  \label{t:lcpifit}
   \begin{center}
  \renewcommand{\tabcolsep}{0.05in}
   \begin{tabular}{|c|lr|r|r|r|} 
    \hline
     Index & \multicolumn{2}{|l|}{Parameter} 
	& 1000 toy MC & 1000 toy MC & Data fit value\\	
     & \multicolumn{2}{|l|}{} 
	& pull mean & pull width & \\	
     \hline
     1 & $N_\mathrm{sig}$ & 
	& 0.007 $\pm$ 0.032 & 0.995 $\pm$ 0.023 & \nlbhad \\
     2 & $\mu$ & [\gevcsq] 
	& 0.021 $\pm$ 0.033 & 1.031 $\pm$ 0.024 & 5.621 $\pm$ 0.003 \\
     3 & $\sigma$ & [\gevcsq]  
	& 0.026 $\pm$ 0.031 & 0.976 $\pm$ 0.022 & 0.023 $\pm$ 0.001 \\
     4 & $N_\mathrm{B4prong}$ &
	& 0.002 $\pm$ 0.032 & 1.018 $\pm$ 0.023 & 150 $\pm$ 32 \\
     5 & $N_\mathrm{OB}$ &
	& 0.038 $\pm$ 0.033 & 1.046 $\pm$ 0.024 & 3170 $\pm$ 291 \\
     6 & $N_\mathrm{OL}$ & 
	& -0.048 $\pm$ 0.033 & 1.030 $\pm$ 0.023 & 962 $\pm$ 324 \\
     7 & $N_\mathrm{combg}$ & 
	& -0.023 $\pm$ 0.032 & 1.013 $\pm$ 0.023 & 1971 $\pm$ 171 \\ 
     8 & $p_0$ & 
	& -0.027 $\pm$ 0.032 & 1.010 $\pm$ 0.023 & 0.63 $\pm$ 0.10 \\
     \hline	
      \hline
      \end{tabular}
      \end{center}
	\end{table}	

 \begin{table}[tbp]
   \caption{Fixed parameters in the \lbhad\ unbinned likelihood fit.}
  \label{t:lcpifix}
   \begin{center}
    \renewcommand{\tabcolsep}{0.05in}
   \begin{tabular}{l|l|r} 
    \hline
     Parameter & Meaning & Value\\	
    \hline
    \hline
  $f_{\Lamc K}$ & $N_{\lbhadk}/N_{\lbhad}$ &  0.080 \\
  $f_1$ & fraction of the narrow $\Lambda_CK$ Gaussian &  0.902 \\
  $\mu_{\Lambda_cK}^1$ 
	& mean of the narrow $\Lambda_CK$ Gaussian [\gevcsq] & 5.573\\
  $\sigma_1$ 
	& width of the narrow $\Lambda_CK$ Gaussian [\gevcsq] & 0.029 \\ 
  $\mu_{\Lambda_cK}^2$ 
	& mean of the wide $\Lambda_CK$ Gaussian [\gevcsq] & 5.529 \\
  $\sigma_2$ 
	& width of the wide $\Lambda_CK$ Gaussian [\gevcsq] & 0.075 \\
  $f_\mathrm{L}$ & fraction of the Landau, 4-prong & 0.413 \\
  $\mu_\mathrm{BPL}$ & mean of the Landau, 4-prong [\gevcsq] & 5.486 \\
  $\sigma_L$ & width of the Landau, 4-prong [\gevcsq] & 0.025 \\
  $\mu_\mathrm{BPG}$ & mean of the Gaussian, 4-prong [\gevcsq] & 5.526 \\
  $\sigma_G$ & width of the Gaussian, 4-prong [\gevcsq] & 0.078 \\
  $s_0$ & slope of the exponential, other \B & 2.180 \\
  $f_\mathrm{bifg}$ & fraction of the bifurcated Gaus, other \B & 0.106\\
  $\mu_\mathrm{ob}$ & mean of the bifurcated Gaus, other \B\ [\gevcsq] 
	& 5.598 \\
  $\sigma_\mathrm{ob}^L$ 
	& left $\sigma$ of the bifurcated Gaus, other \B\ [\gevcsq]& 10.0 \\
  $\sigma_\mathrm{ob}^R$ 
	& right $\sigma$ of the bifurcated Gaus, other \B\ [\gevcsq]& 4.800 \\
  $\mu_{obst}$ & mean of ``step-down'', other \B\ [\gevcsq] & 5.436 \\
  $a_0^\mathrm{ob}$ & slope of the ``step-down'', other \B & 0.079 \\

  $\mu_\mathrm{ol}$ & mean of the bifurcated Gaus, other \Lb\ [\gevcsq]
	& 3.469 \\
  $\sigma_\mathrm{ol}^L$ & left $\sigma$ of the bifurcated Gaus, other \Lb\ 
[\gevcsq] & 10.0 \\
  $\sigma_\mathrm{ol}^R$ & right $\sigma$ of the bifurcated Gaus, other \Lb\ 
[\gevcsq] & 1.236 \\
  $\mu_\mathrm{olst}$ & mean of ``step-down'', other \Lb\ [\gevcsq] & 5.451 \\
  $a_0^\mathrm{ol}$ & slope of ``step-down'', other \Lb\ [\gevcsq] & 0.091 \\
  $f_1^\mathrm{ol}$ & fraction of first Gaus, other \Lb\ & 0.0005 \\
  $\mu_1^\mathrm{ol}$ & mean of first Gaus, other \Lb\ & 5.644 \\
  $\sigma_1^\mathrm{ol}$ & width of first Gaus, other \Lb\ & 0.019 \\
  $f_2^\mathrm{ol}$ & fraction of second Gaus, other \Lb\ & 0.0034 \\
  $\mu_2^\mathrm{ol}$ & mean of second Gaus, other \Lb\ & 5.459 \\ 
  $\sigma_2^\mathrm{ol}$ & width of second Gaus, other \Lb\ & 0.030 \\
    \hline
    \hline
      \end{tabular}
      \end{center}
  \end{table}

\clearpage
\newpage
\section{Summary}
\label{sec-yieldsummary}
Using the unbinned, extended log likelihood technique, we fit the charm 
and \B\ hadron mass spectra to obtain the number of events. 
The yield for each mode is listed below. The performance of the fitter is 
validated using 1000 toy MC test for each mode. In general, the mean 
of each pull distribution from the toy MC test is consistent with zero and 
the width is consistent with one. For the fit parameter with a pull mean
deviated from zero and a width deviated from unity, the fitter only 
indicates a less than 1$\%$ bias on the central value. Besides, 
these fit parameters are not correlated with the number of signal events 
and do not affect the yield we obtain. The fit result to the data is also 
superimposed on the data histograms and a $\chi^2$ is computed. We have 
obtained good $\chi^2$ for each mode. 

   \begin{center}
   \setdec 0000.0
   \begin{tabular}{|l|r@{\,$\pm$\,}l|} 
    \hline
     Mode & \multicolumn{2}{|c|}{Yield}\\	
     \hline
      \dstarhad & \dec \ndstarhadc. & \dec \ndstarhade. \\
      \incdstarsemi & \dec \ndstarsemic. & \dec \ndstarsemie. \\
      \dhad & \dec \ndhadc. & \dec \ndhade. \\
      \incdsemi & \dec \ndsemic. & \dec \ndsemie. \\ 
      \lbhad & \dec \nlbhadc. & \dec \nlbhade. \\
      \inclbsemi & \dec \nlbsemic. & \dec \nlbsemie. \\
     \hline	
      \hline
      \end{tabular}
      \end{center}

\chapter{Monte Carlo Samples, Acceptance and Efficiencies}
\label{ch:mc}
With the raw yield in hand, we now turn to the correction which
must be applied to obtain the value of the ratio of branching fractions,
that is the acceptance, trigger and reconstruction efficiency which may 
only be calculated using a Monte Carlo program.
The Monte Carlo (MC) simulation plays a crucial role in this analysis. 
In addition to the acceptance and efficiencies for our signals and 
backgrounds, as described in Section~\ref{sec-eff} and Chapter~\ref{ch:bg},
the MC is used for the optimization of signals in Section~\ref{sec-opt}. 
MC is also used to find out the function form that describes the 
mass spectrum of the background due to partial- or mis-reconstruction in 
Section~\ref{sec-massfit}.
In this chapter, we first explain the components of Monte Carlo samples and 
show that, in general, the MC reproduces the data. Then we present the
acceptance, trigger and reconstruction efficiencies obtained from the MC.

\section{Monte Carlo Simulation Components}
\label{sec-mccom}
There are several components in the MC simulation:
\begin{itemize}
\item production and decay of the \B\ hadrons
\item detector simulation
\item trigger simulation
\end{itemize}

\subsubsection{Production and Decay of \B\ Hadrons}
\begin{figure}[tbp]
     \begin{center}
        \includegraphics[width=180pt, angle=0]
	{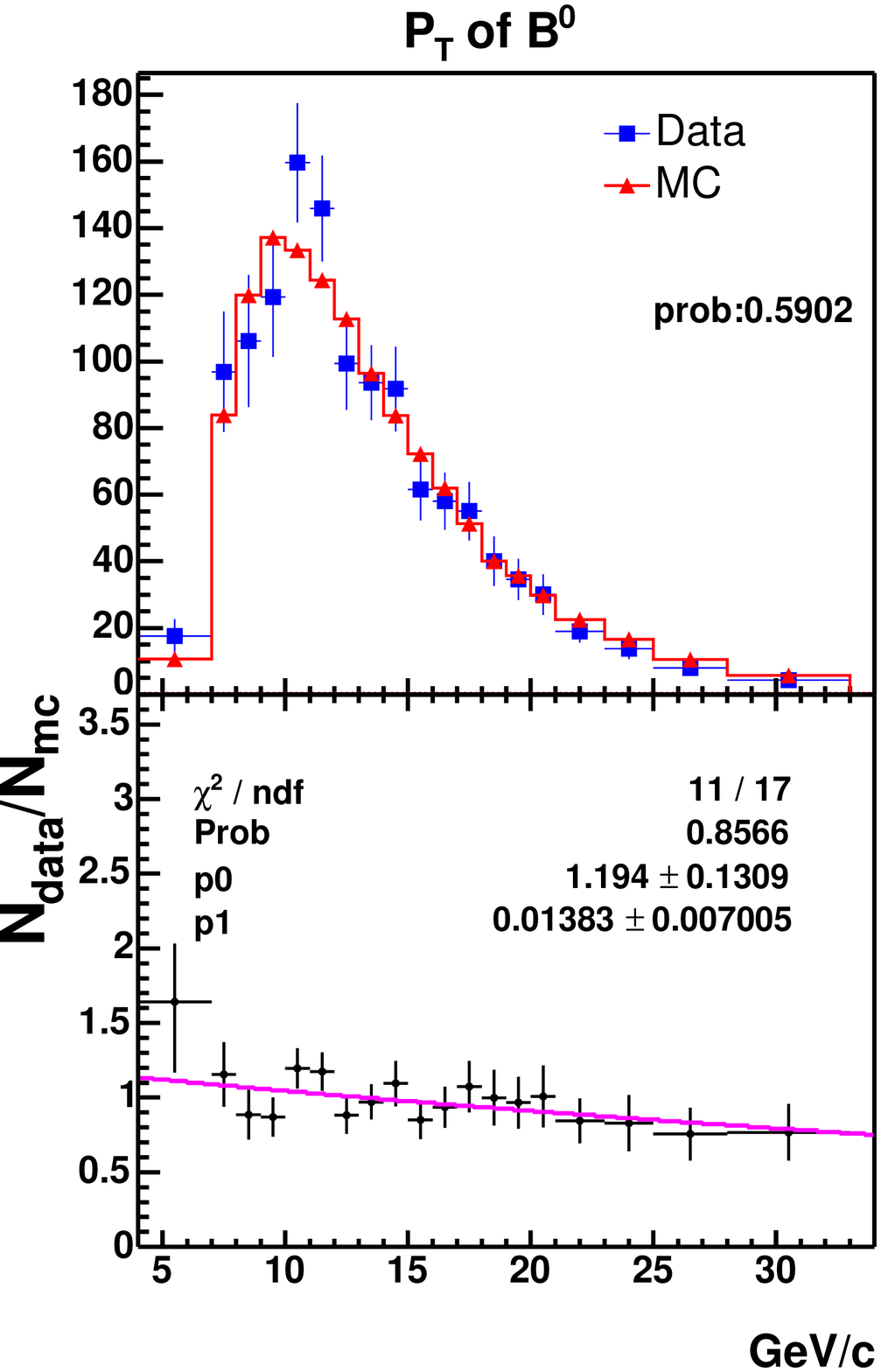}
        \includegraphics[width=180pt, angle=0]
	{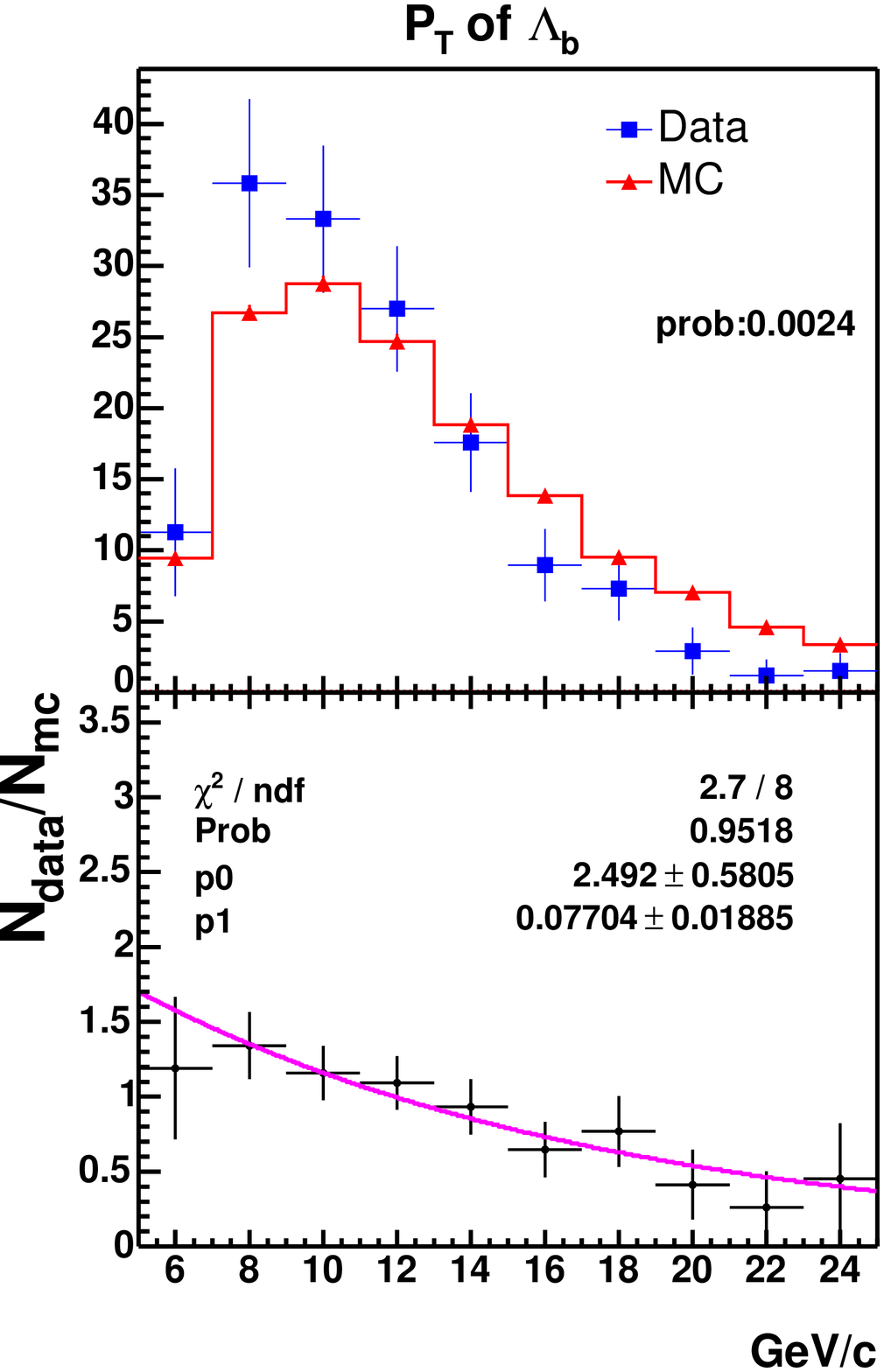}
        \caption[Comparison of the reconstructed \Bd\ and \Lb\ \pt\ spectrum 
	between 
	\bgen\ and data]
	{Comparison of reconstructed \Bd\ (left) and \Lb\ (right) \pt\ 
         spectrum between data and \bgen\ (MC).
         The top figures show the \pt\ distribution while the bottom 
         figures show the ratio data/MC. The curves in the bottom figures
         are the result of an exponential fit to the ratio.  
         It is evident that the MC \pt\ spectrum is harder than that 
	of the data.}
     \label{fig:bptbefore}
        \includegraphics[width=280pt, angle=0]{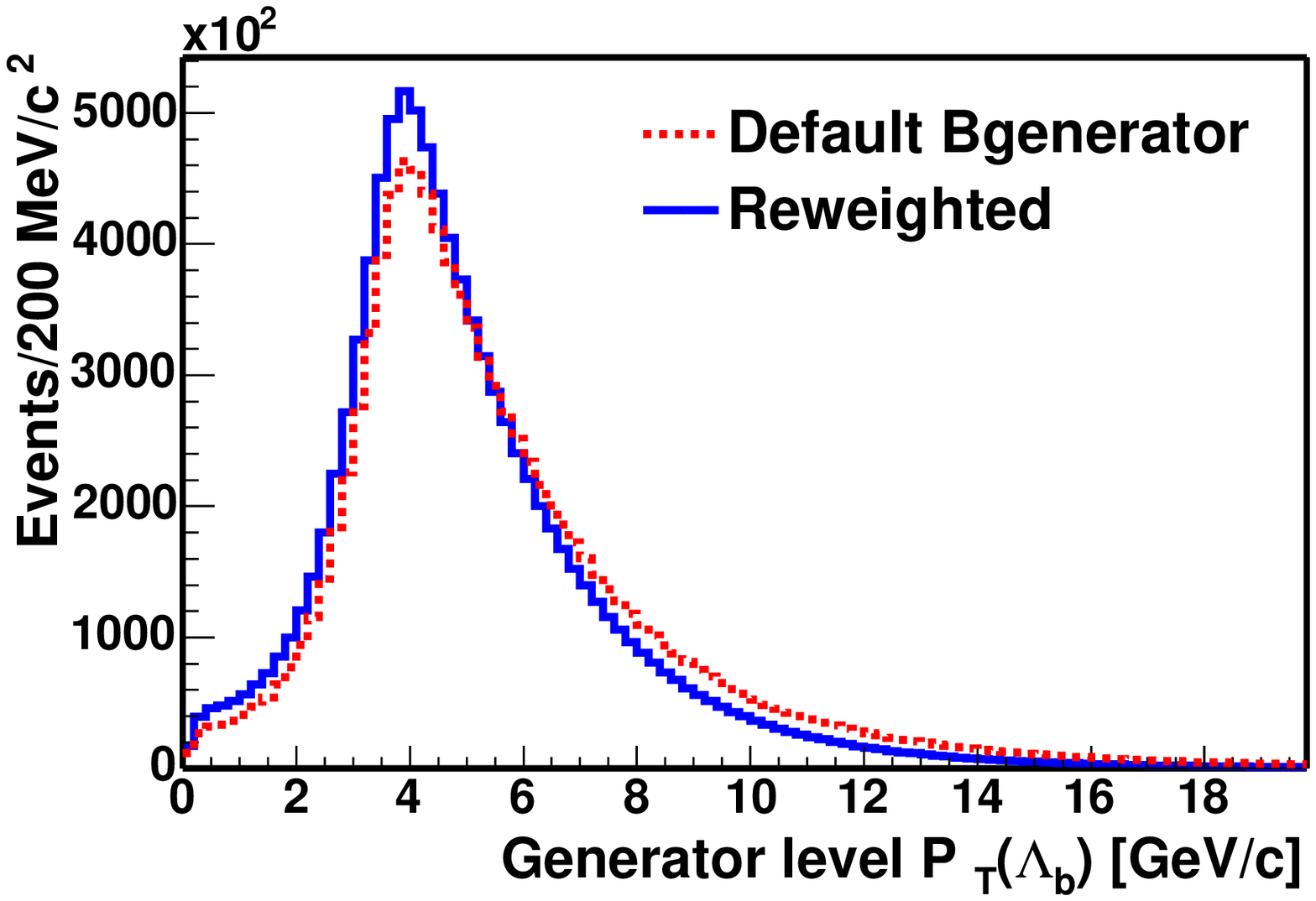} 
        \caption[Generator level \Lb\ \pt\ spectra before and after the 
	reweighting]
	{Generator level \Lb\ \pt\ spectra before and after the 
	reweighting.      
	\label{fig:lbptgen}}
     \end{center}
	  \end{figure}

We use two types of event generators: \bgen~\cite{mit:bgen} and 
\pythia~\cite{pythia:manual}. The \bgen\ is the primary generator used in this
analysis for calculating the acceptance and efficiencies of our signals
and most backgrounds. \bgen\ generates a single $b$-quark according to the 
$\pt(b)$ spectrum which follows the NLO calculation by Nason, Dawson, and 
Ellis (NDE)~\cite{Nason:1989zy}. The rapidity of the $b$ quark, $y(b)$, is 
generated according to $1+ q(\pt)$, where 
\begin{equation}
q(\pt) = -0.0456 - 0.00289 \pt.
\end{equation}
The mass of the $b$ quark is set to 4.75 \gevcsq. For the \B\ meson MC sample,
the b-quarks are generated with a $P_T$ threshold of 4.0 GeV/c over the range 
in rapidity $|y| < 2.5$, and then fragmented into \B\ mesons with the CDF 
default Peterson fragmentation parameter~\cite{Peterson:1982ak}, $\epsilon_B$,
 set to 0.006. Figure~\ref{fig:bptbefore} shows a small discrepancy in the 
reconstructed $\pt(\Bd)$ between data and MC. The slope of the data to MC 
ratio is about 2 $\sigma$ away from zero. The MC events which survive the 
trigger simulation, reconstruction and the analysis cuts, will be re-weighted 
according to the ratio numerically, i.e. we multiply each event with the ratio,
 $w$. We then calculate the efficiencies using the re-weighted MC events;
\begin{equation}
 R_\mathrm{pass} = \sum_i^{N_\mathrm{pass}}w_i 
\label{eq:weightr}
\end{equation}
\begin{equation}
\label{eq:numweff}
\epsilon = \frac{R_\mathrm{pass}}{N_\mathrm{gen}}
\end{equation}

Figure~\ref{fig:bptbefore} also shows a 4$\sigma$ discrepancy in the 
reconstructed \pt(\Lb) between data and MC from the \bgen. As the discrepancy 
is significant, in order to correctly assess the acceptance and efficiency of 
the \Lb, the fragmentation process inside \bgen\ has to be turned off. 
The \Lb\ needs to be generated 
directly with a \pt\ spectrum which reproduces the data. This spectrum is 
obtained in the following way: We first obtain the default generated \Lb\ 
\pt\ spectrum from the \bgen. 
Then, the default generated $\pt(\Lb)$ is re-weighted with the exponential 
slope of the ratio data/MC shown in Figure~\ref{fig:bptbefore}, using the 
``acceptance-rejection (Von Neumann)'' method~\cite{pdg:mctech}.
See Figure~\ref{fig:lbptgen} for the \Lb\ \pt\ spectra before and after our 
reweighting. We also confirm that the reconstructed $\pt(\Lb)$ from the MC 
using the re-weighted spectrum reproduces the data, see 
Figure~\ref{fig:mcdatalcpi0}.

The other event generator, \pythia, is a program for the generation of 
collisions at high energies. \pythia\ simulates physics processes using 
leading-order matrix elements, supplemented by the initial and final state 
radiation. The program also includes fragmentation and hadronization of the 
quarks and gluons in the final state. Unlike \bgen, \pythia\ includes the beam 
remnants that are left when a parton from the beam particle is removed to 
participate in the hard QCD interaction. \pythia\ provides more realistic 
simulation of an event than \bgen, and produces multi-particle final states 
similar to the hadron collider data. However the generation using the \pythia\
 is also more time consuming than the \bgen. This makes \pythia\ inefficient 
to understand the acceptance and efficiency of a single decay mode. Therefore, 
\pythia\ has been used in this analysis only to study the background from \bb\
 and \cc\ decays.

After the event generation, the hadrons are allowed to decay using the \evtgen\
 software package developed by Lange and Ryd~\cite{Lange:2001uf}. This package 
is maintained by {\tt BABAR} and mainly tuned by the results from the 
experiments at the $\Upsilon(4S)$ resonance. The decay model and branching 
ratios for \Bd\ and $B^+$ are well described but not necessarily those of 
the $B_s$ and the B baryons. As a proper decay model for the \Lb\ 
semileptonic decays is not implemented in the \evtgen\ yet, we use a flat 
phase space to decay the \Lb\ first. Then, we will apply a scaling factor on 
the acceptance after taking into account the effect of the semileptonic form 
factors (see Section~\ref{sec-eff}).

\subsubsection{Detector, Trigger Simulation, Production and Reconstruction}
The particles from the output of \bgen\ and \evtgen\ are then run 
through a full (``realistic") simulation of the CDF detector and trigger.
The software version for the simulation is 4.11.2 with patches which
implement the most up-to-date configuration of SVT. The geometry and
response of the detectors active and passive components are simulated using
the {\tt GEANT} software developed by Brun, Hagelberg, Hansroul, and 
Lassalle~\cite{Brun:1978fy}. Most of the detector subsystems, like COT and 
CMU, are assumed to be in a time-independent and perfect condition, which 
means there are no dead channels 
and the high voltages are constantly at full value. Selecting the data when 
these systems are in good condition helps to ensure that MC reproduces the 
data, see Section~\ref{sec-datasample}. Because the SVX active coverage
and the configuration for the XFT and SVT systems change on various occasions 
(see Section~\ref{sec-trigger}), we divide the data taking period into eight 
sub-periods, where the detector and trigger performance is constant. 
We generate our MC samples for these eight sub-periods by choosing the 
runs with maximum number of L3 triggered events as the representative runs. 
Each run has its own parameters for the performance of the detector and 
triggers. For the sub-periods with large integrated luminosity, we choose 
more representative runs so that each run corresponds to a period 
with integrated luminosity around 3--6~\pbarn. See Table~\ref{t:mcrun} for 
the representative runs in the MC. The number of generated events is 
proportional to the integrated luminosity of the sub-period each run 
represents. The positions of the beamline for each run is taken directly from 
the database and simulated in the MC. 

After the detector and trigger simulation, the MC events are run
through a trigger decision program, {\tt svtfilter}. {\tt svtfilter} takes
the information from the simulated SVT data and makes the 
{\tt B$\_$CHARM Scenario A} requirements described in Section~\ref{sec-path}. 
The events which pass {\tt svtfilter} are processed with the same 
{\tt Production} executable (version 4.9.1hpt3) as that which is run on the 
data. The {\tt Production} executable reconstructs higher level objects, such 
as electrons, muons, tracks and missing energy, from the simulated detector and
 trigger data. The resulting MC events have the same structure and format as 
the data and are then run through the same analysis program described in 
Section~\ref{sec-datasample}.

        \renewcommand{\arraystretch}{0.9}
        \begin{table}[tbp]
	\caption{Simulated Runs in the MC sample.}
	\label{t:mcrun}
	\begin{footnotesize}
       \begin{center}
 	\begin{tabular}{|c|cr|l|} 
	\hline  
         Range & Run & $\int {\cal L}\; dt$ (\pbarn) & Comment \\
        \hline
        138809--143000 & 140129 & 3.4 & Scenario A implemented \\
        \hline
        143001--146000 & 145005 & 4.0 & Tevatron incident \\
        \hline
        146001--149659 & 148824 & 4.2 & SVX coverage improved \\
	               & 149387 & 2.9 &                       \\
        \hline
        149660--150009 & 149663 & 0.6 & SVT optimization (coverage+patterns)\\
        \hline
        150010--152668 & 150820 & 4.1 & \lxy\ $>$ 200 $\mu$m cut added \\
                       & 151844 & 3.7 & \\
                       & 152520 & 3.5 & \\
        \hline
        152669--156487 & 152967 & 3.6 & XFT from 2-miss to 1-miss \\
                       & 153327 & 3.7 & \\
                       & 153447 & 3.7 & \\
                       & 153694 & 2.4 & \\
                       & 154452 & 4.2 & \\
                       & 154654 & 4.9 & \\
                       & 155364 & 4.3 & \\
                       & 155795 & 2.5 & \\
                       & 155895 & 3.6 & \\
                       & 156116 & 3.7 & \\
                       & 156484 & 2.6 & \\
        \hline
       	159603--164302 & 160230 & 3.7 & data taken after the shutdown  \\
                       & 160441 & 3.4 & \\
                       & 160823 & 3.7 & \\
                       & 161029 & 3.8 & \\
                       & 161379 & 3.3 & \\
                       & 161678 & 3.9 & \\
                       & 162130 & 3.6 & \\
                       & 162393 & 3.6 & \\
                       & 162498 & 5.6 & \\
                       & 162631 & 5.7 & \\
                       & 162857 & 4.4 & \\
                       & 163064 & 3.7 & \\
                       & 163431 & 4.3 & \\
        \hline
       164303--167715  & 164451 & 4.6 & SVT change from 4/4 to 4/5\\
                       & 164844 & 3.5 & \\
                       & 165121 & 2.9 & \\
                       & 165271 & 3.9 & \\
                       & 165412 & 3.6 & \\
                       & 166008 & 6.0 & \\
                       & 166063 & 2.9 & \\
                       & 166567 & 5.2 & \\
                       & 166662 & 5.3 & \\
                       & 167053 & 5.9 & \\
                       & 167186 & 2.2 & \\
                       & 167506 & 4.0 & \\
                       & 167551 & 2.7 & \\
	\hline 
        Total & & 170.9  & \\
	\hline \hline
    	\end{tabular}
        \end{center}
        \end{footnotesize}
        \end{table}


\section{Monte Carlo and Data Comparison}
\label{sec-datamc}
\begin{figure}
   \begin{center}
      \includegraphics[width=280pt, angle=0]{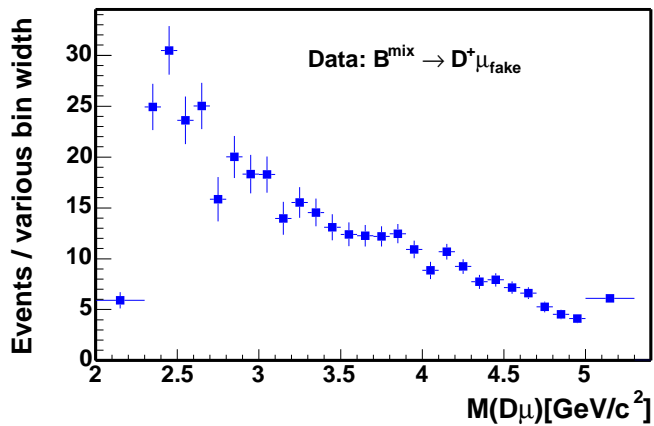} 
        \caption[$M_{D\mu}$ from the muon fake]
	{$M_{D\mu}$ from the $B\rightarrow \D \mu_{fake}$ data. 
         The distribution is sideband subtracted using \mkpipi. 
     \label{fig:fakedmum}}
	
    \end{center}
\end{figure}

To confirm that the simulation accurately 
reproduces the data, we compare various reconstructed distributions from the MC
 with the same distribution from the data. To ensure a fair comparison, the 
combinatorial background present in the signal region of data has to be removed
. We perform a sideband subtraction for the \dstarhad, \incdstarsemi, \dhad\ 
and \incdsemi\ decays. For the \lbhad\ and \inclbsemi\ decays, a sideband 
subtraction can not remove all the backgrounds in the signal region as 
explained later in the text and in Section~\ref{sec-massfit}. Instead, a signal
 distribution of variable ``X'' is obtained by fitting $M_{\Lambda_c\pi}$ and 
\mpkpi\ to get the number of signal events in bins of variable ``X''. For all 
the semileptonic modes, we include the MC samples of the physics backgrounds 
described in Section~\ref{sec-physicsb}. The distribution from each physics 
background is scaled according to the assumed or measured branching ratio for 
that background. In addition, the distribution of each compared variable from 
the fake muons is subtracted from the data. The distribution from the fake 
muons is obtained by reconstructing the ``fake muon-charm'' final state as 
described in Section~\ref{sec-fakemu}. The combinatorial background in the 
``fake muon-charm'' is removed using the same method as described above for 
the real muon. See Figure~\ref{fig:fakedmum} for the $M_{\dmu}$ from the 
muon fakes. 

\begin{figure}[tbp]
     \begin{center}
        \includegraphics[width=200pt, angle=0]
	{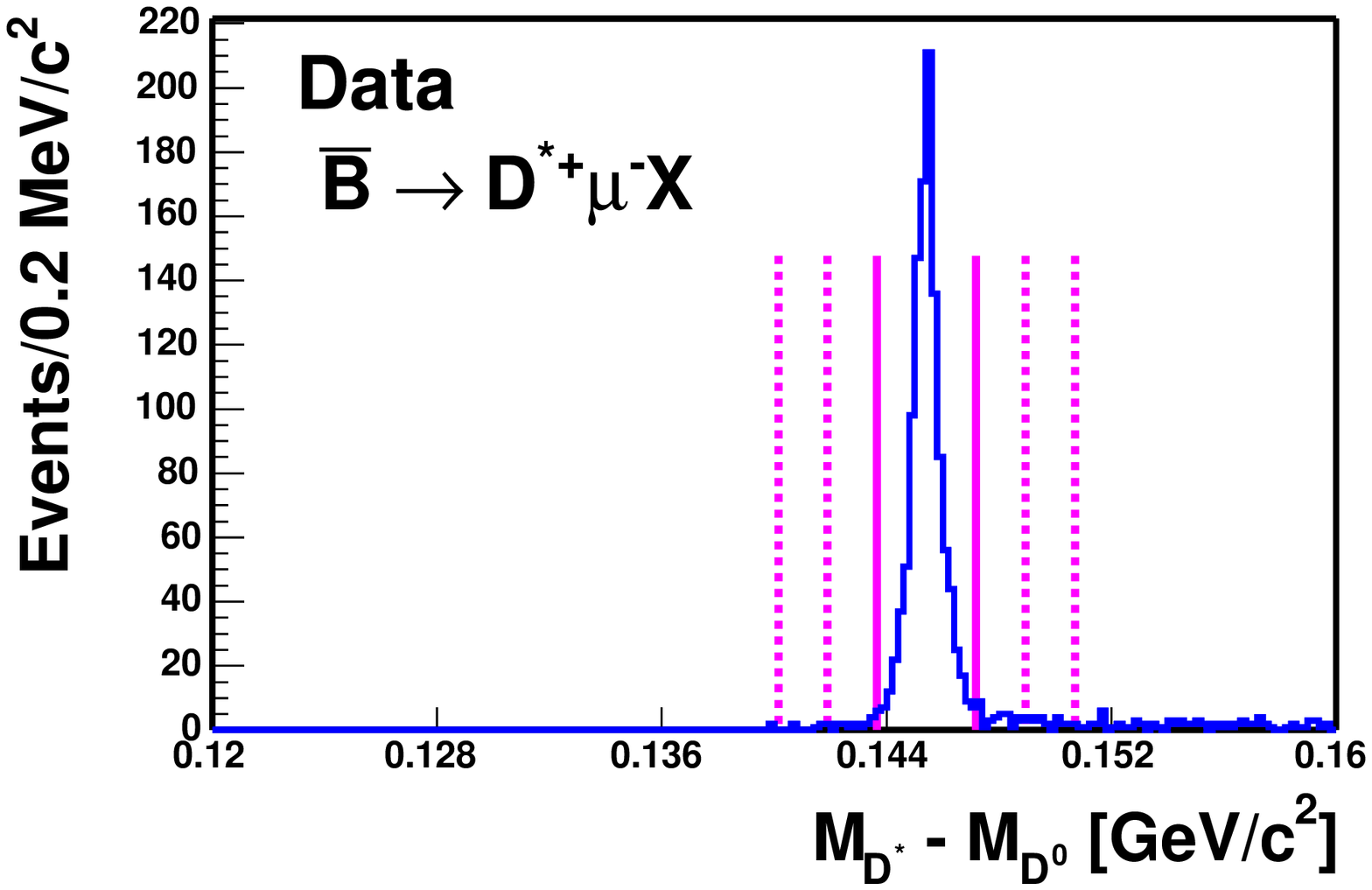}
        \includegraphics[width=200pt, angle=0]
	{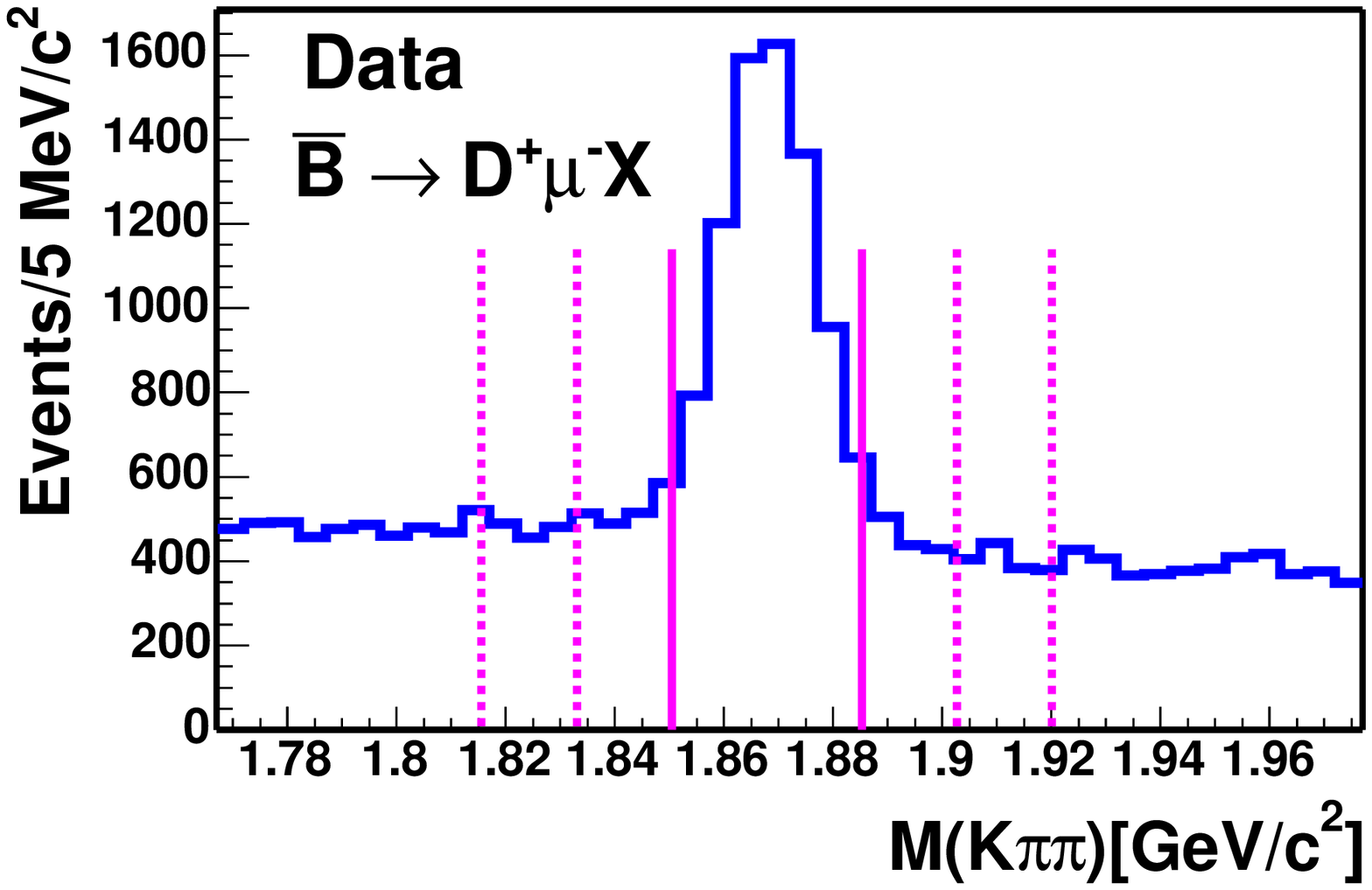}
        \caption[Definition of the signal and sideband regions in the 
\incdstarsemi\ and \incdsemi\ data]
{Invariant mass difference $\mkpipi-\mkpi$ (left) and invariant mass \mkpipi\ 
(right) showing our reconstructed \incdstarsemi\ and \incdsemi\ signals.
The vertical solid (dashed) lines indicate the signal (sideband) regions.} 
     \label{fig:semiregion}
     \end{center}
  \end{figure}

For the \B\ meson semileptonic channels, the mass difference between \Dstar\ 
and \Dzero\ in the \incdstarsemi\ mode (\mkpipi\ - \mkpi),
and mass of \D\ in the \incdsemi\ mode (\mkpipi), are used as 
the variables to perform the sideband subtraction. The signal region for
both these modes is defined as: 
\begin{equation}
|M - M_\mathrm{PDG}| < 2 \sigma,
\end{equation}
 and the sideband region is defined as: 
\begin{equation}
4 \sigma < |M - M_\mathrm{PDG}| < 6 \sigma.
\end{equation}
The background function is assumed to be a straight line. Therefore, the 
amount of background in our signal region is the same as that in 
our sideband regions. We obtain a clean signal distribution by subtracting
the histogram in the sideband region from the histogram in the signal region. 
Figure~\ref{fig:semiregion} displays 
the signal and sideband regions of \mkpipi\ - \mkpi\ and \mkpipi.

\begin{figure}[tbp]
  \begin{center}
      \includegraphics[width=200pt, angle=0]
	{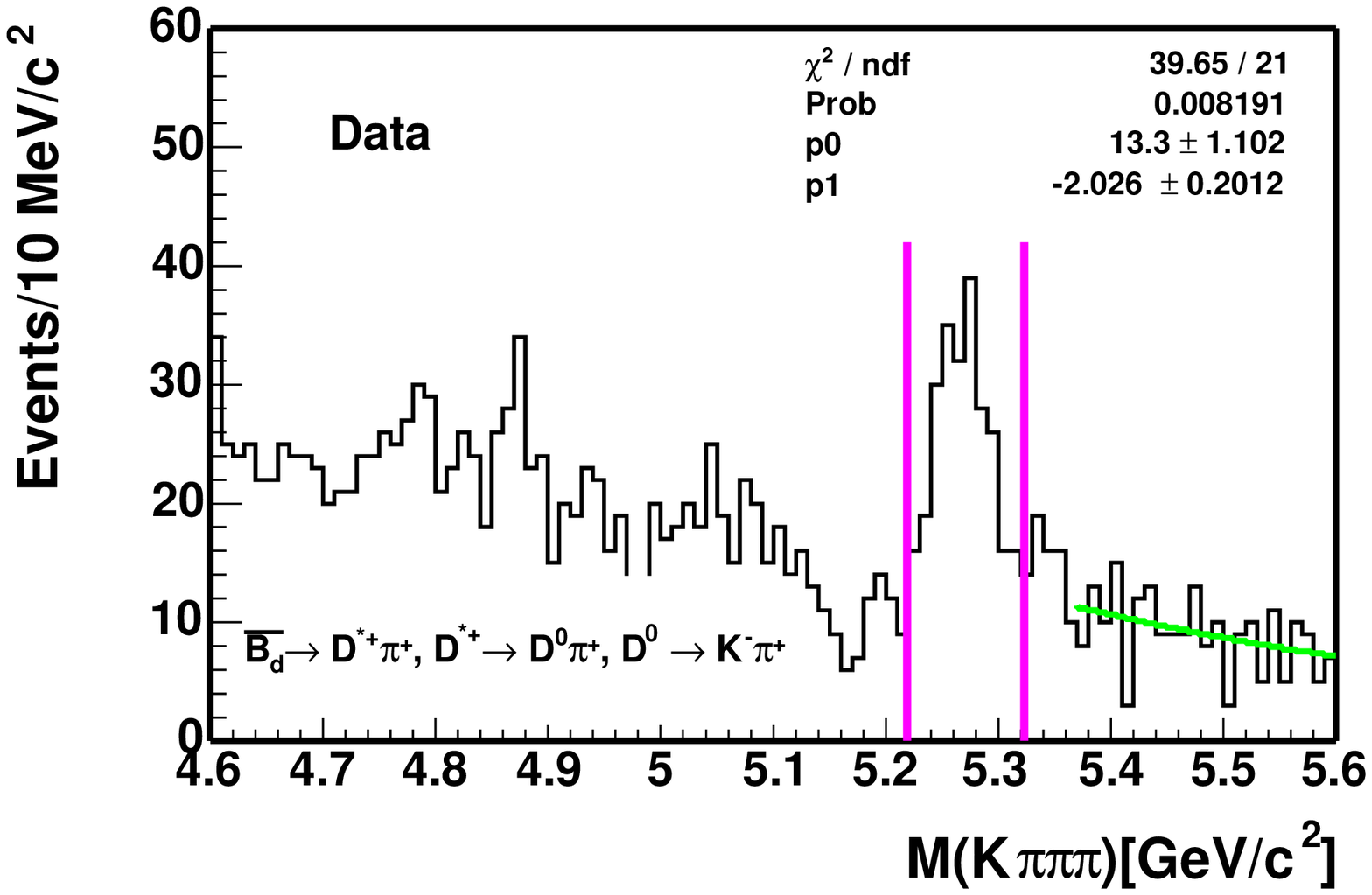} 
      \includegraphics[width=200pt, angle=0]
	{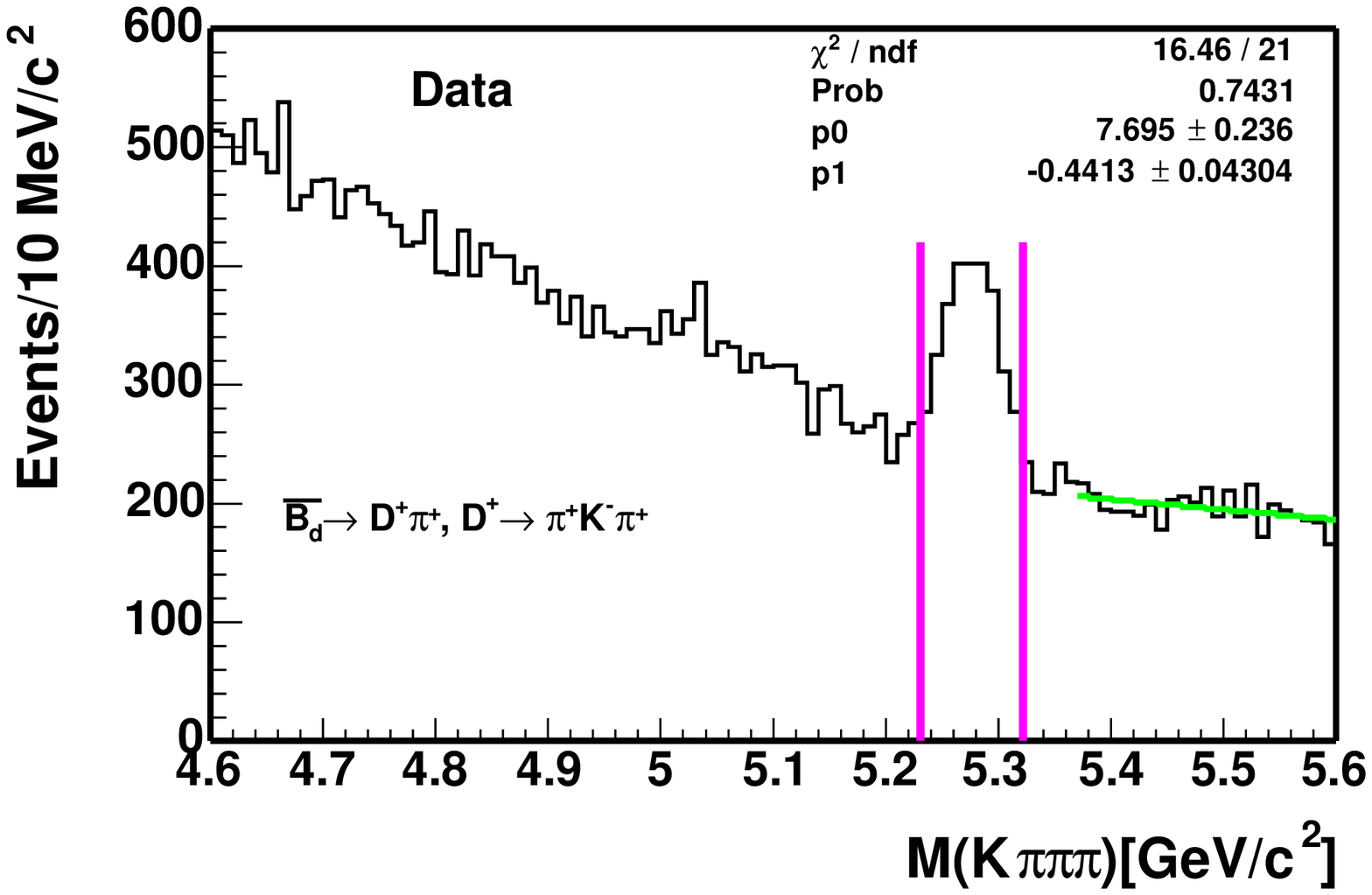}
        \caption[Definition of the signal and upper mass regions in the 
         \dstarhad\ and \dhad\ data]
	{Invariant mass \mkpipipi\ for \dstarhad\ (top) and \dhad\ (bottom)
         signals. The vertical solid lines indicate the signal region. 
	 The upper mass regions on the right are fitted to exponential
         functions.}
     \label{fig:hadregion}
  \end{center}
  \end{figure}

For the \B\ meson hadronic modes, we use the upper mass sideband above 
the signal peak to perform the sideband subtraction. The lower mass region 
below the signal peak consists of both combinatorial background and 
partially reconstructed \B\ decays. However, the background in the
signal region and in the upper mass region above the peak is 
mainly combinatorial as shown in Figures~\ref{fig:dstarpisignal}
--~\ref{fig:dpisignal}. 
We have learned in Section~\ref{sec-massfit} that the combinatorial 
background is adequately described by an exponential function. Therefore, we 
fit the upper mass region to an exponential function. We further extrapolate 
the exponential to the signal region and obtain the ratio of the background in
 our signal region to that in our upper mass sideband, $R_\mathrm{bg}$. The 
histogram of the compared variable extracted from the upper mass sideband is 
scaled by $R_\mathrm{bg}$ and subtracted from the histogram in the signal 
region. See Figure~\ref{fig:hadregion} for the \Bd\ mass signal region we 
define and the upper mass region we fit to an exponential.

\begin{figure}[tbp]
     \begin{center}
\renewcommand{\tabcolsep}{0.03in}
      \begin{tabular}{ccc}
\includegraphics[width=133pt, angle=0]{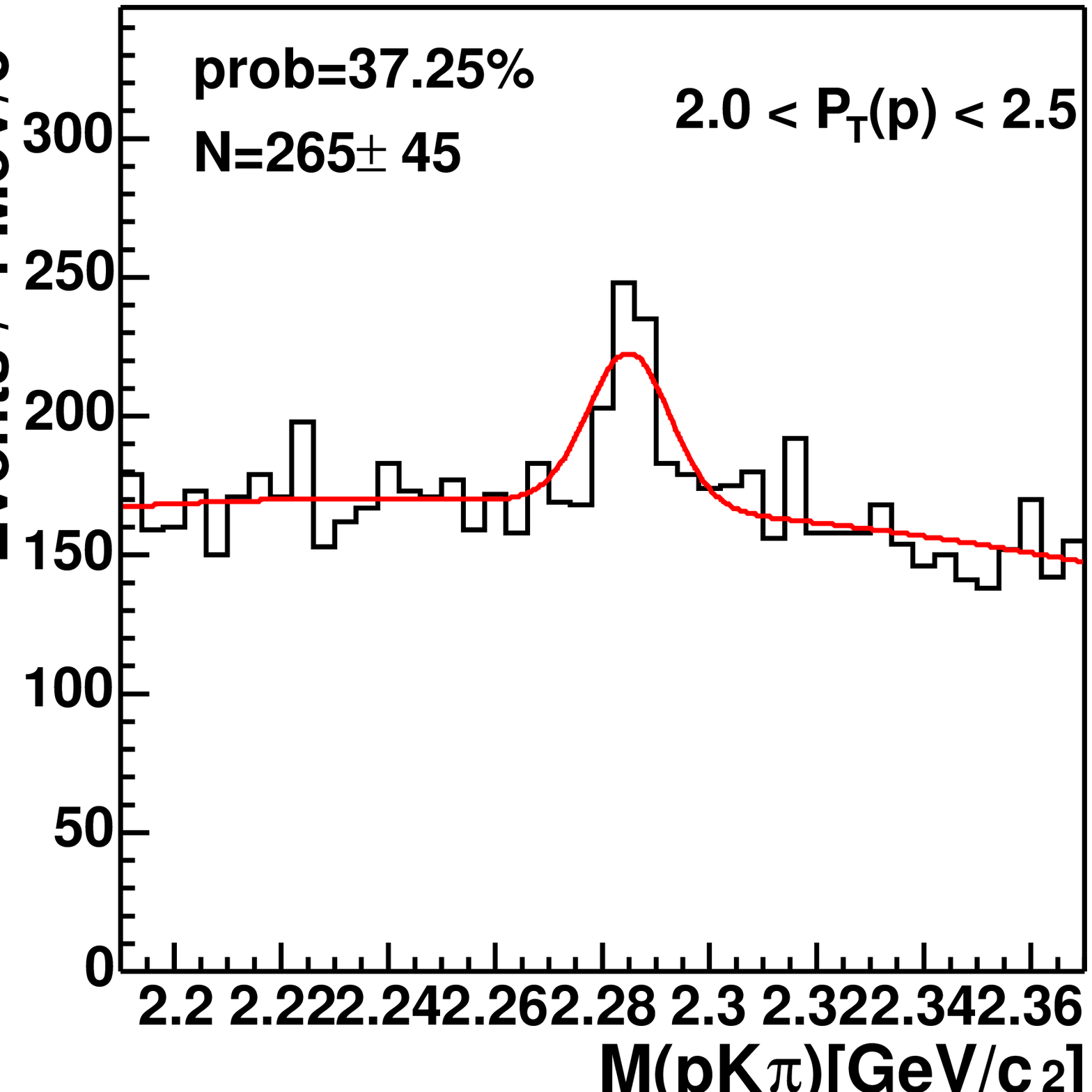}&
\includegraphics[width=133pt, angle=0]{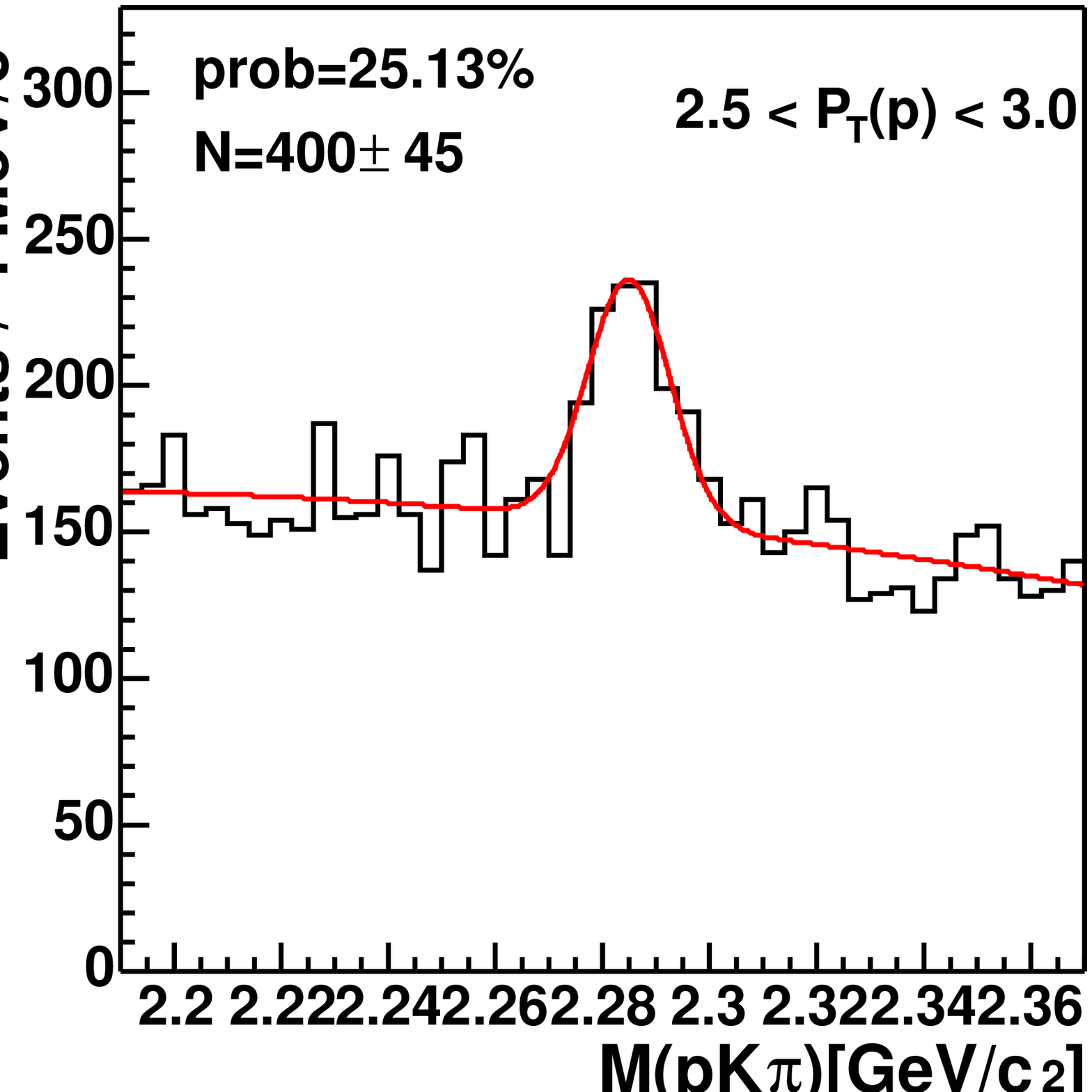}&
\includegraphics[width=133pt, angle=0]{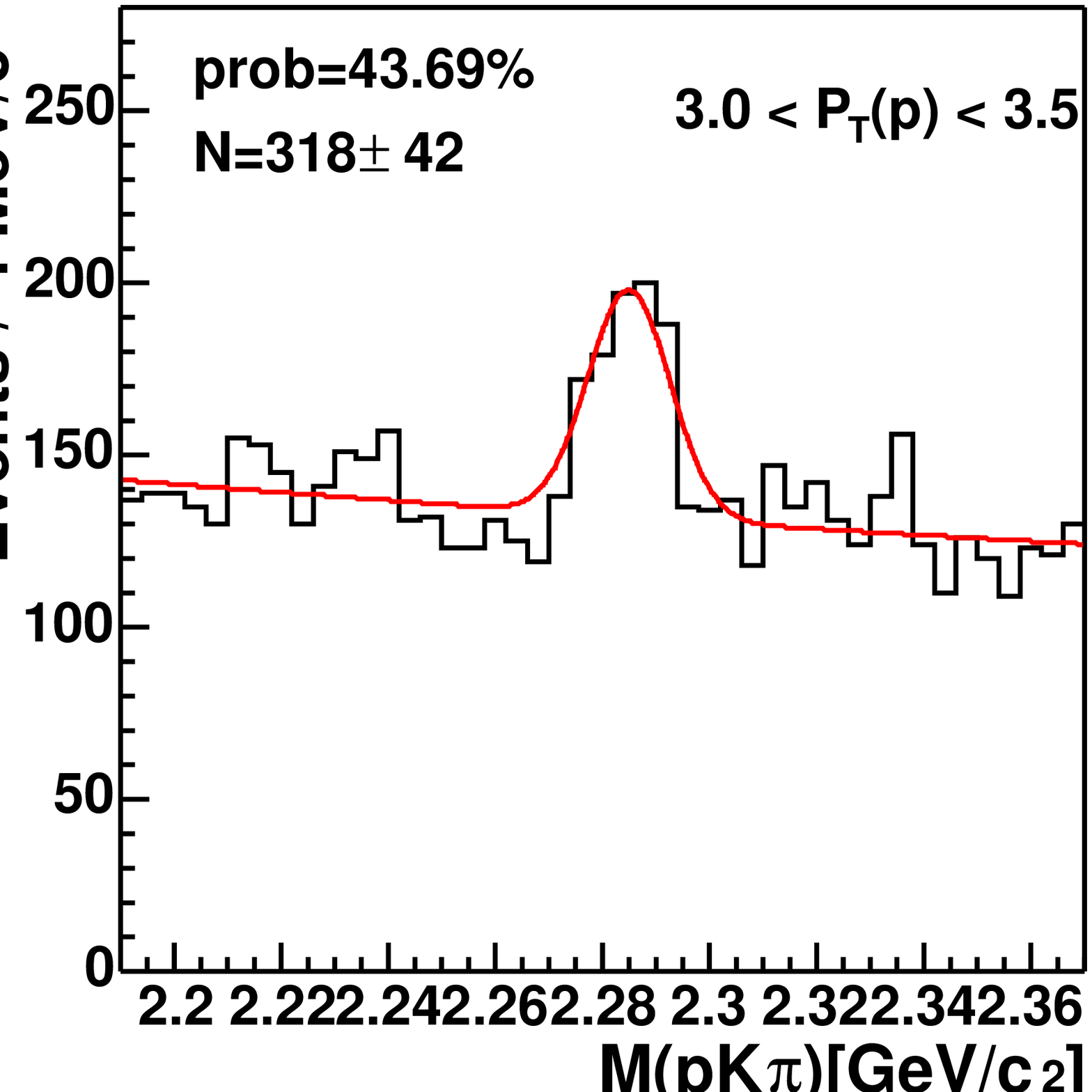}\\
\includegraphics[width=133pt, angle=0]{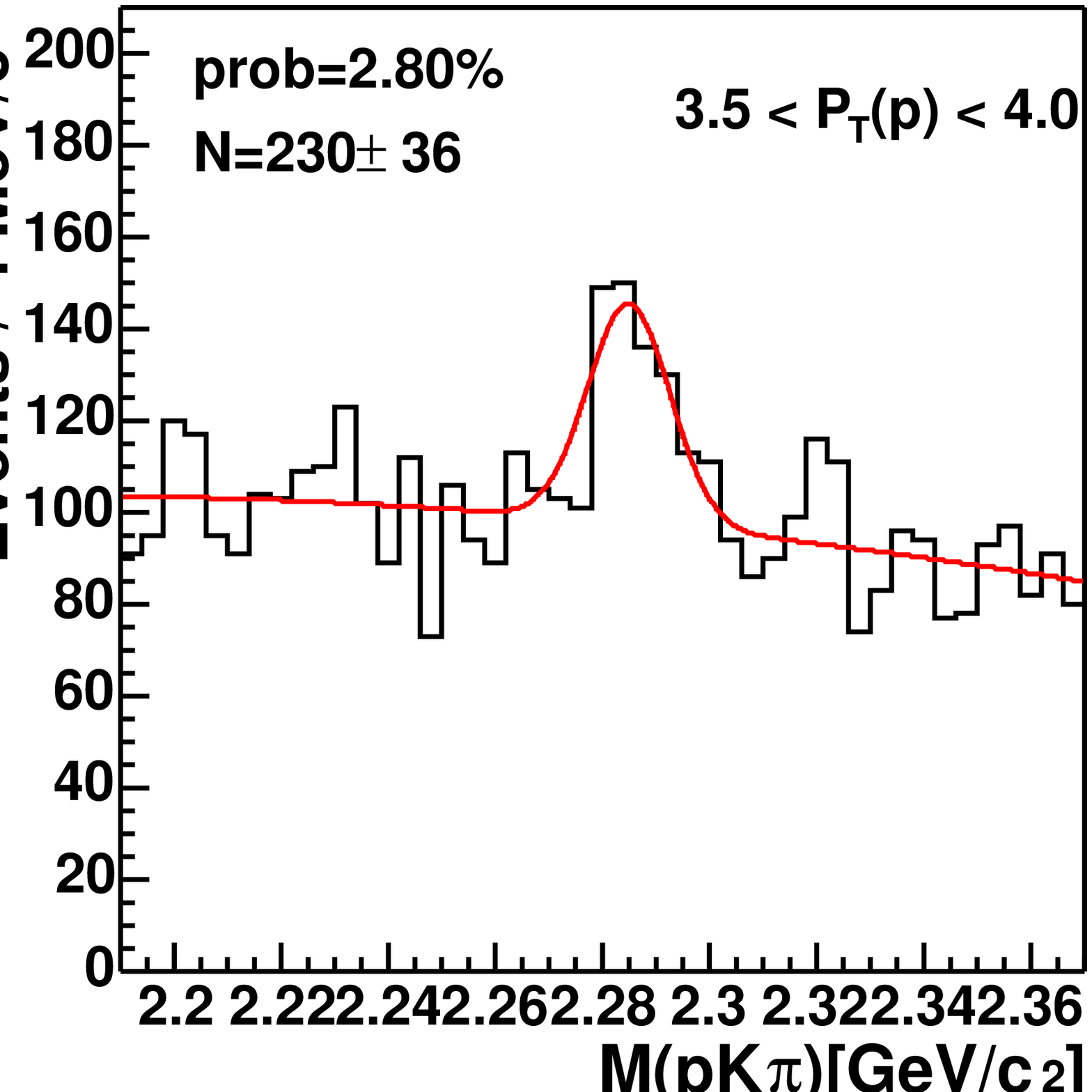}&
\includegraphics[width=133pt, angle=0]{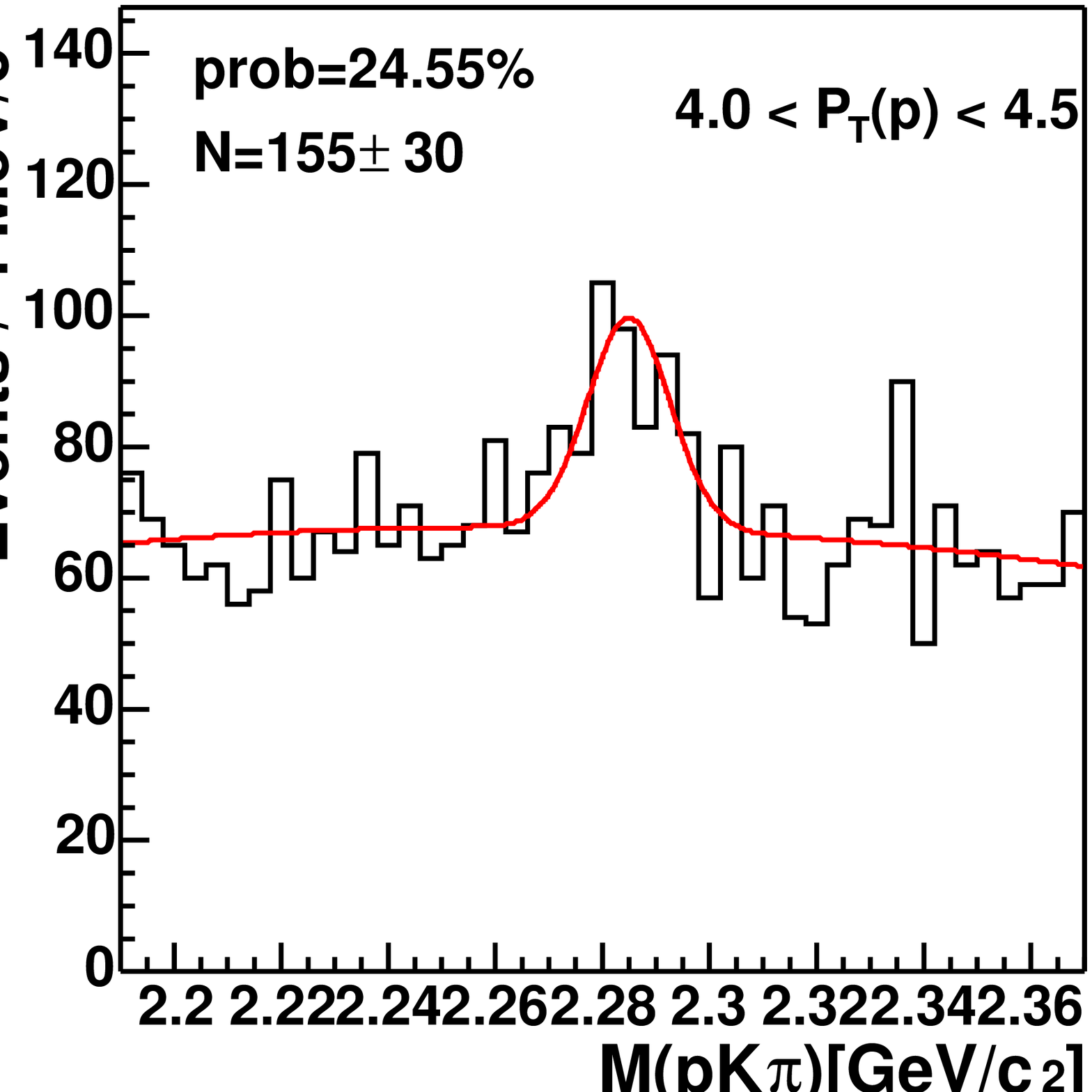}&
\includegraphics[width=133pt, angle=0]{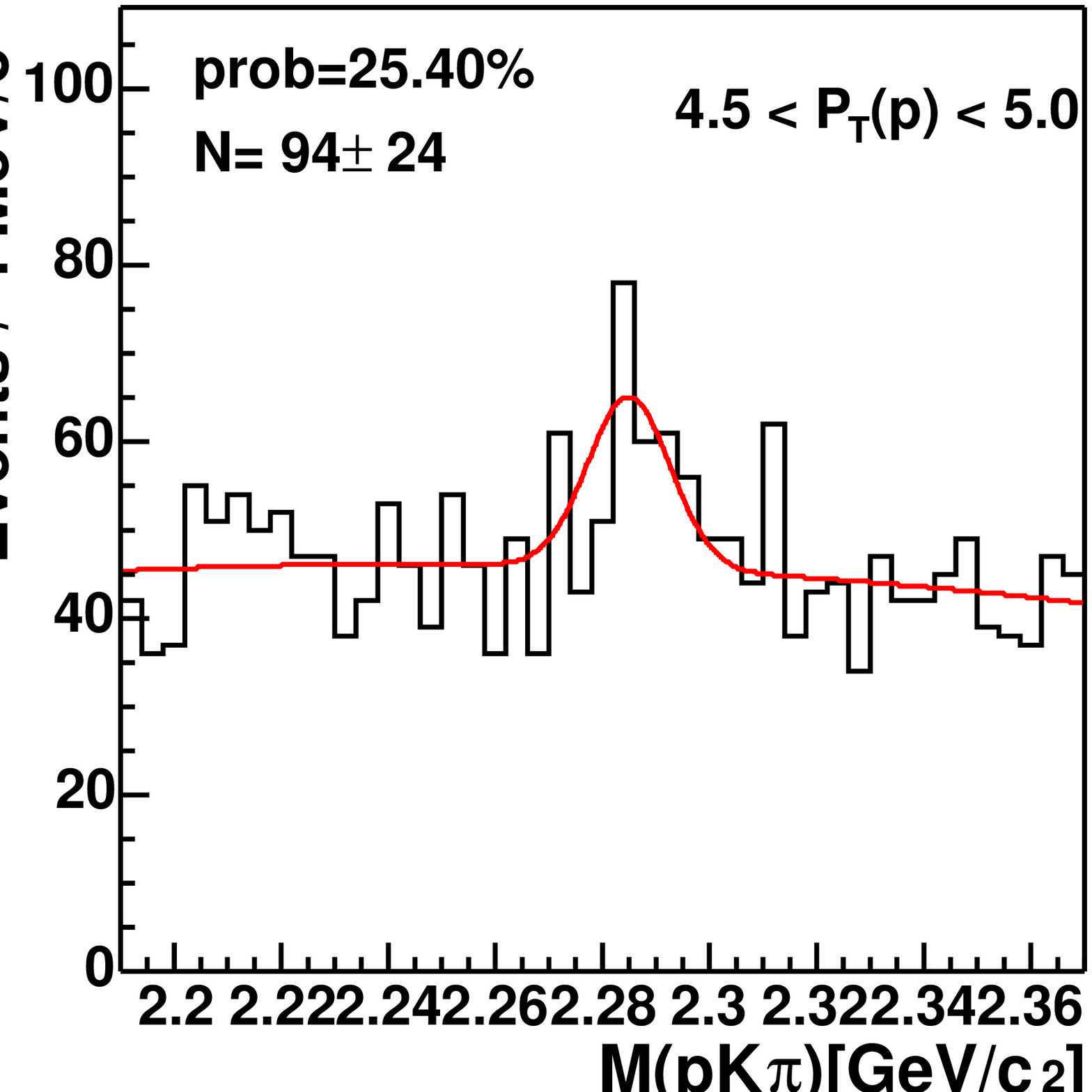}\\
\includegraphics[width=133pt, angle=0]{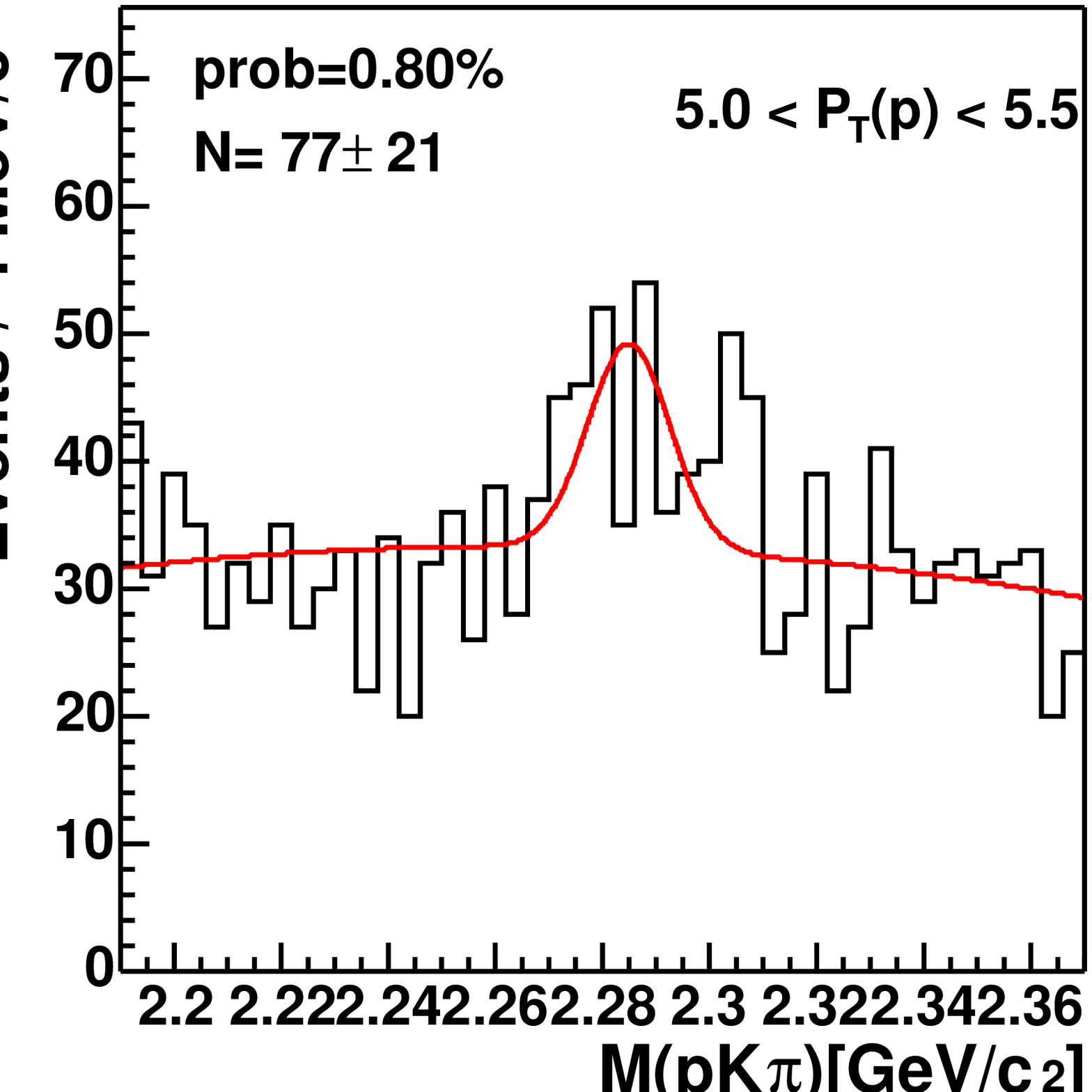}&
\includegraphics[width=133pt, angle=0]{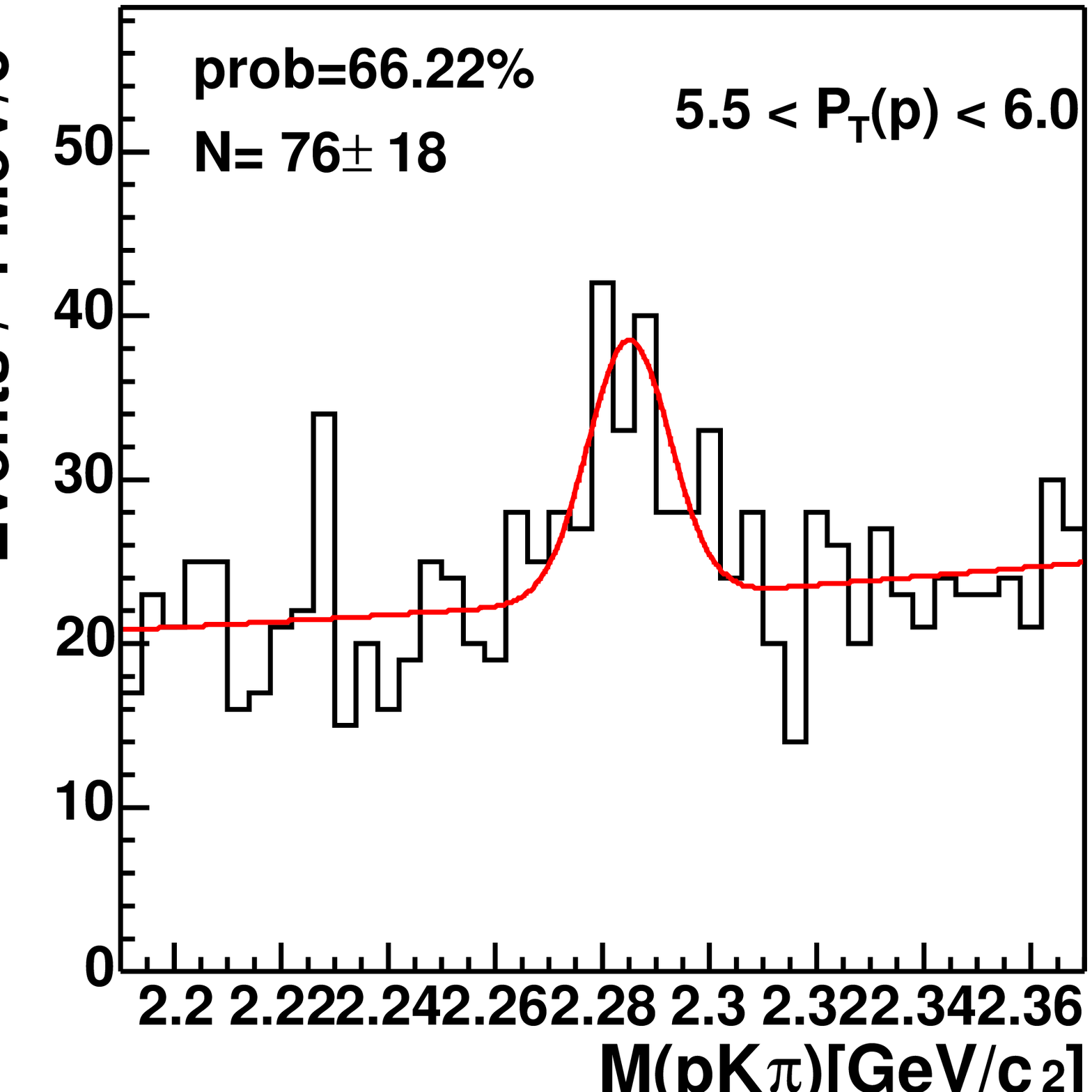}&
\includegraphics[width=133pt, angle=0]{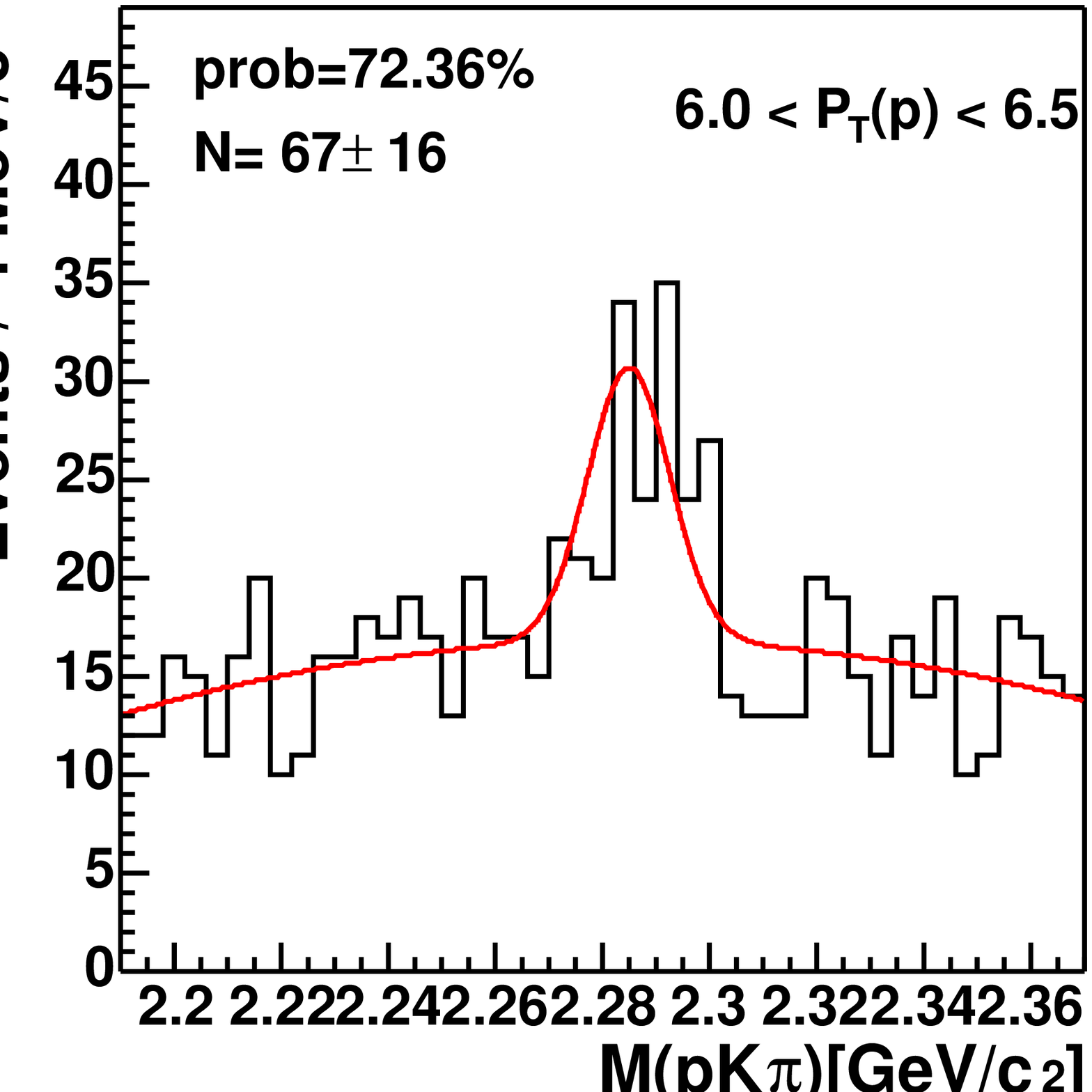}\\
\includegraphics[width=133pt, angle=0]{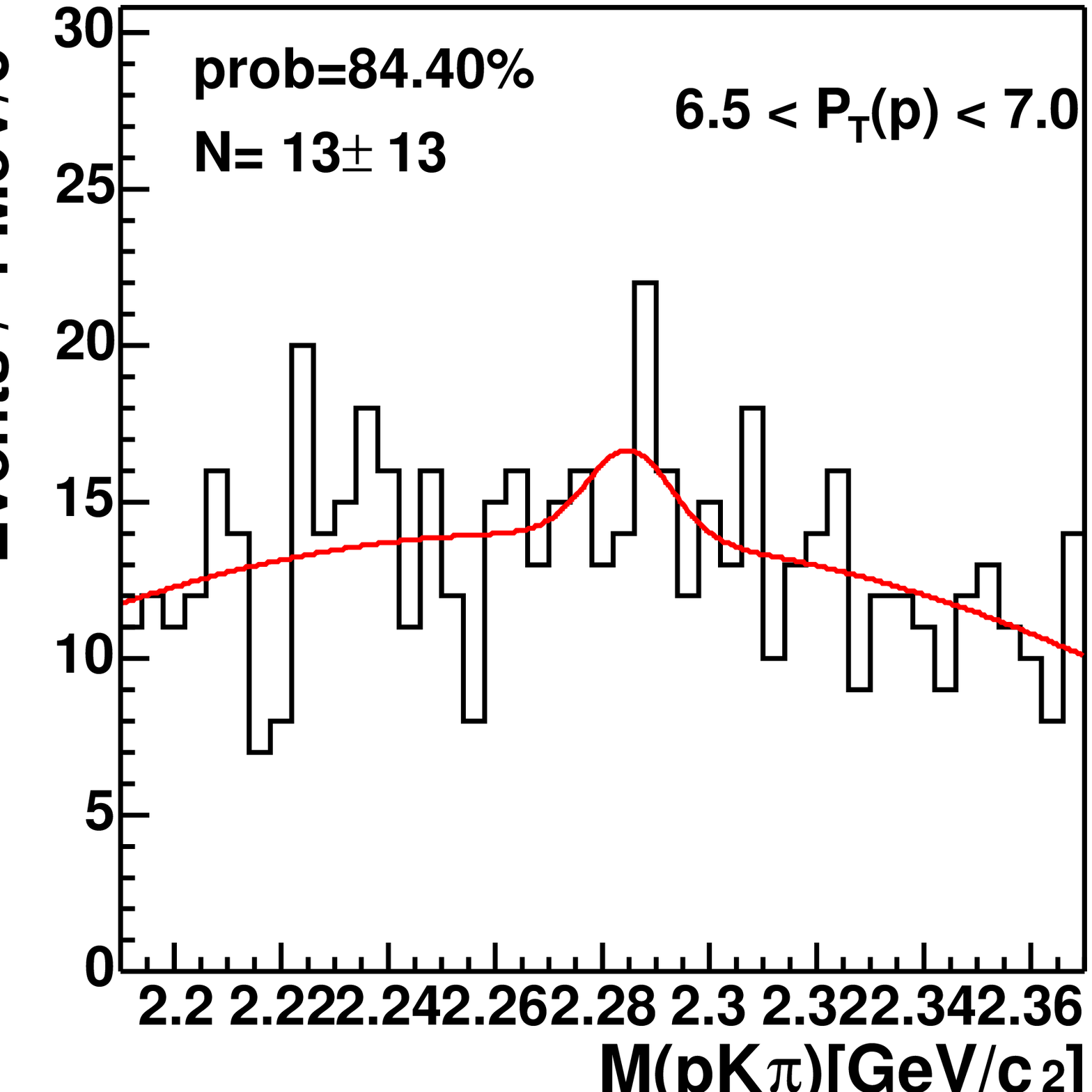}&
\includegraphics[width=133pt, angle=0]{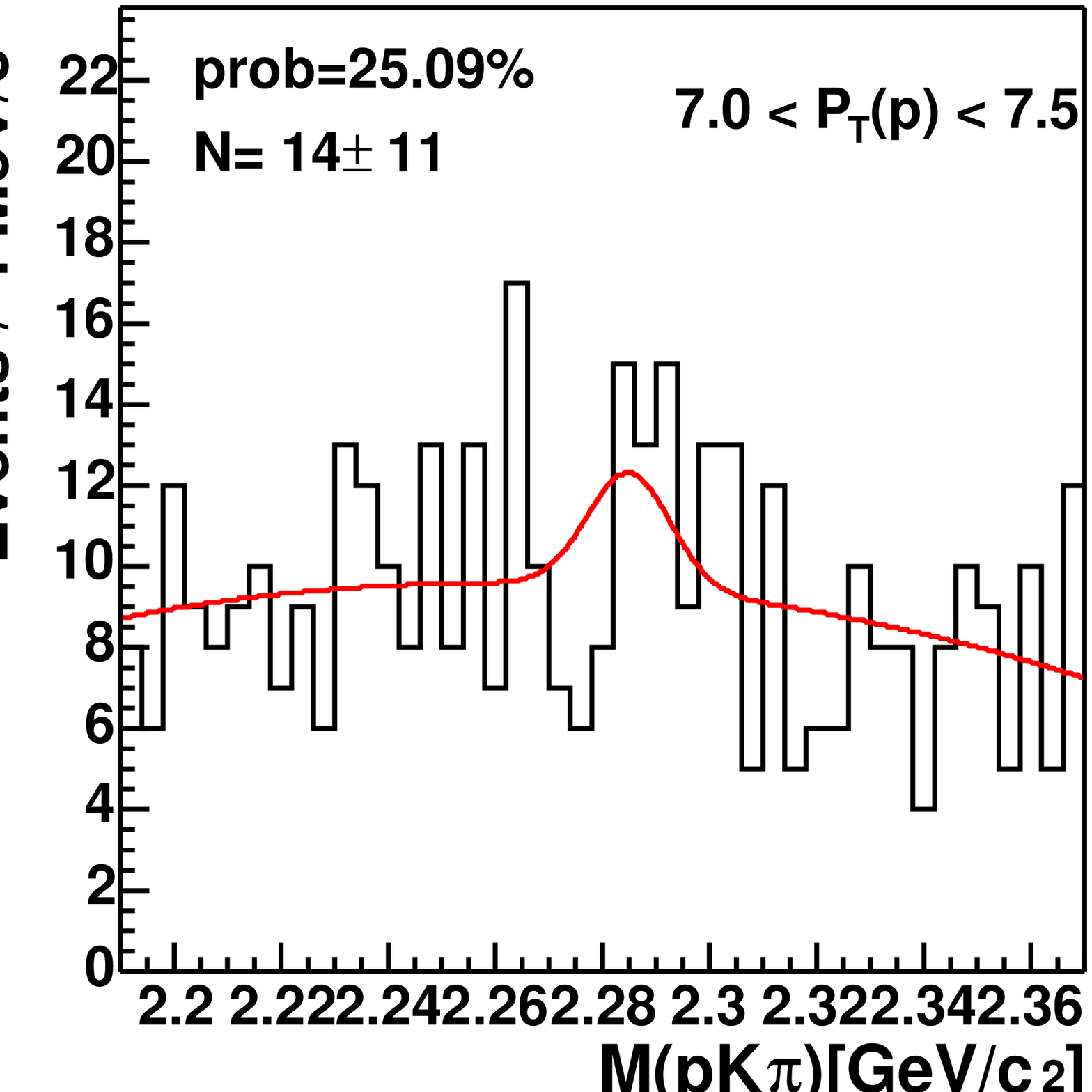}&
\includegraphics[width=133pt, angle=0]{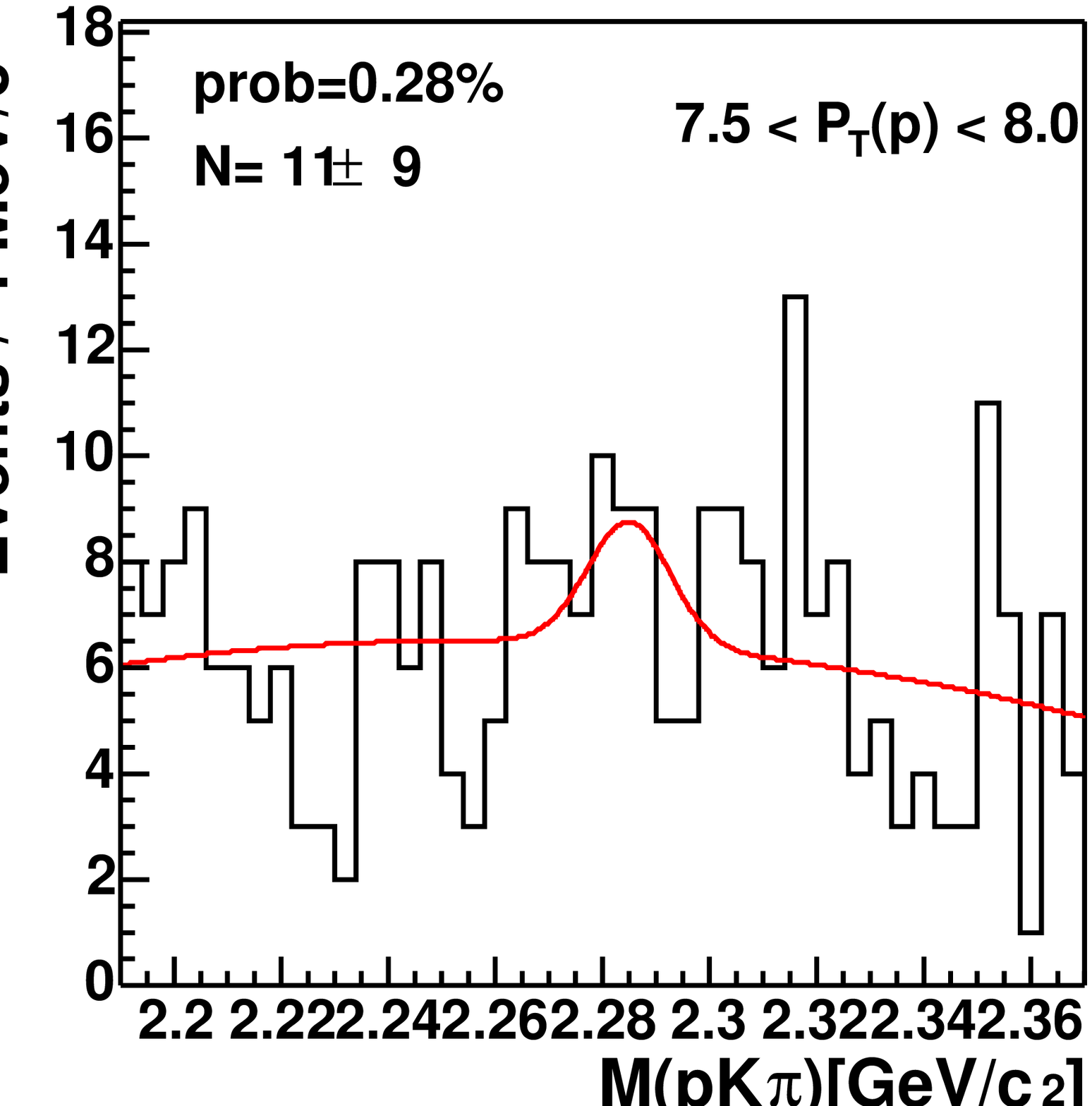}\\
      \end{tabular}
        \caption[Example of \Lc\ mass fit for the MC and data comparison]
	{Example of \Lc\ mass fit for the MC and data comparison. The variable
         to compare is the \pt\ of proton, from 2 to 8 \gevc, in bins of 0.5
         \gevc. \mpkpi\ is fitted to a signal Gaussian and a second-order 
	polynomial.}
     \label{fig:lccountfit}
     \end{center}
  \end{figure}

\begin{figure}[tbp]
     \begin{center}
\renewcommand{\tabcolsep}{0.02in}
      \begin{tabular}{cc}
\includegraphics[width=180pt, angle=0]{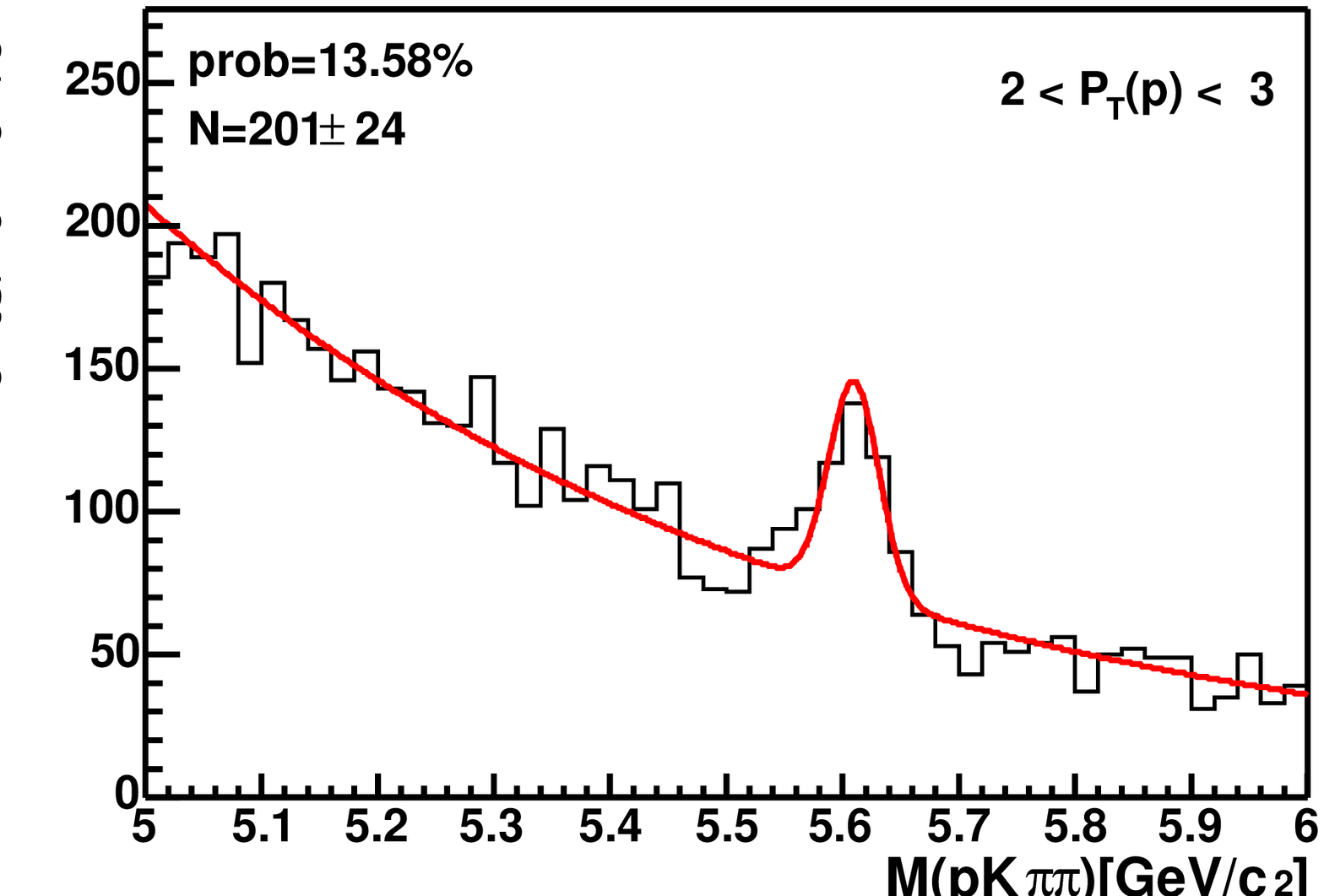}&
\includegraphics[width=180pt, angle=0]{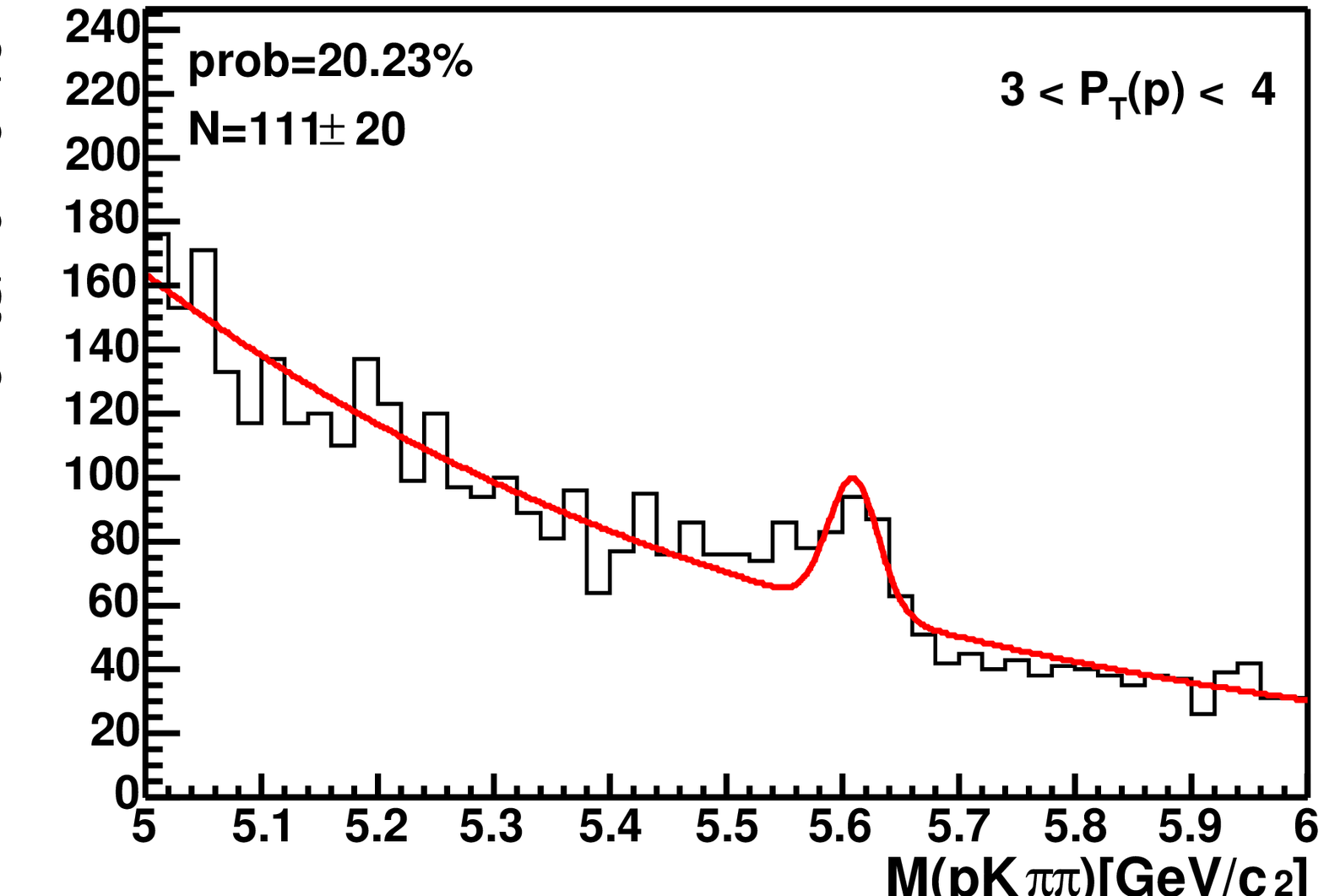}\\
\includegraphics[width=180pt, angle=0]{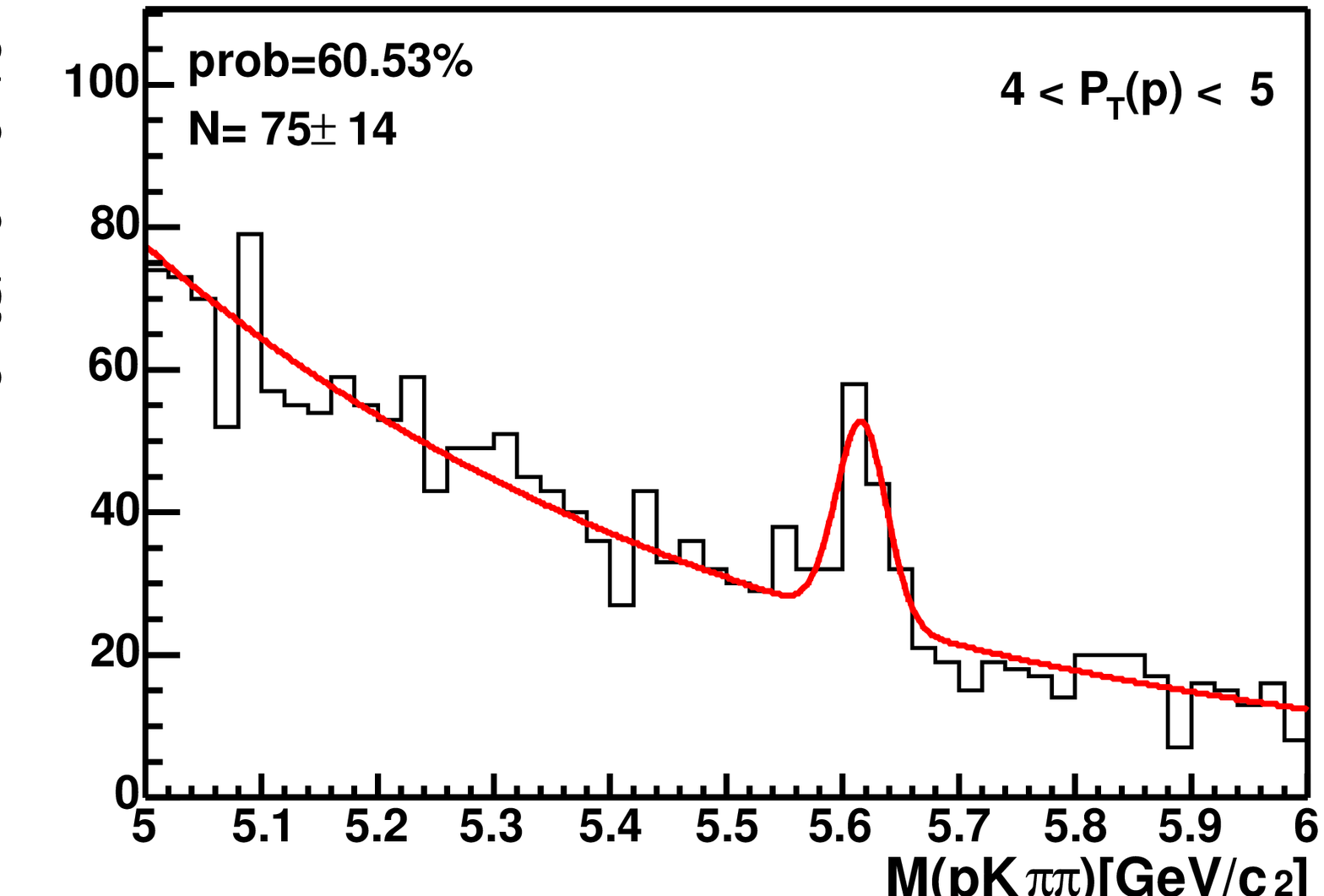}&
\includegraphics[width=180pt, angle=0]{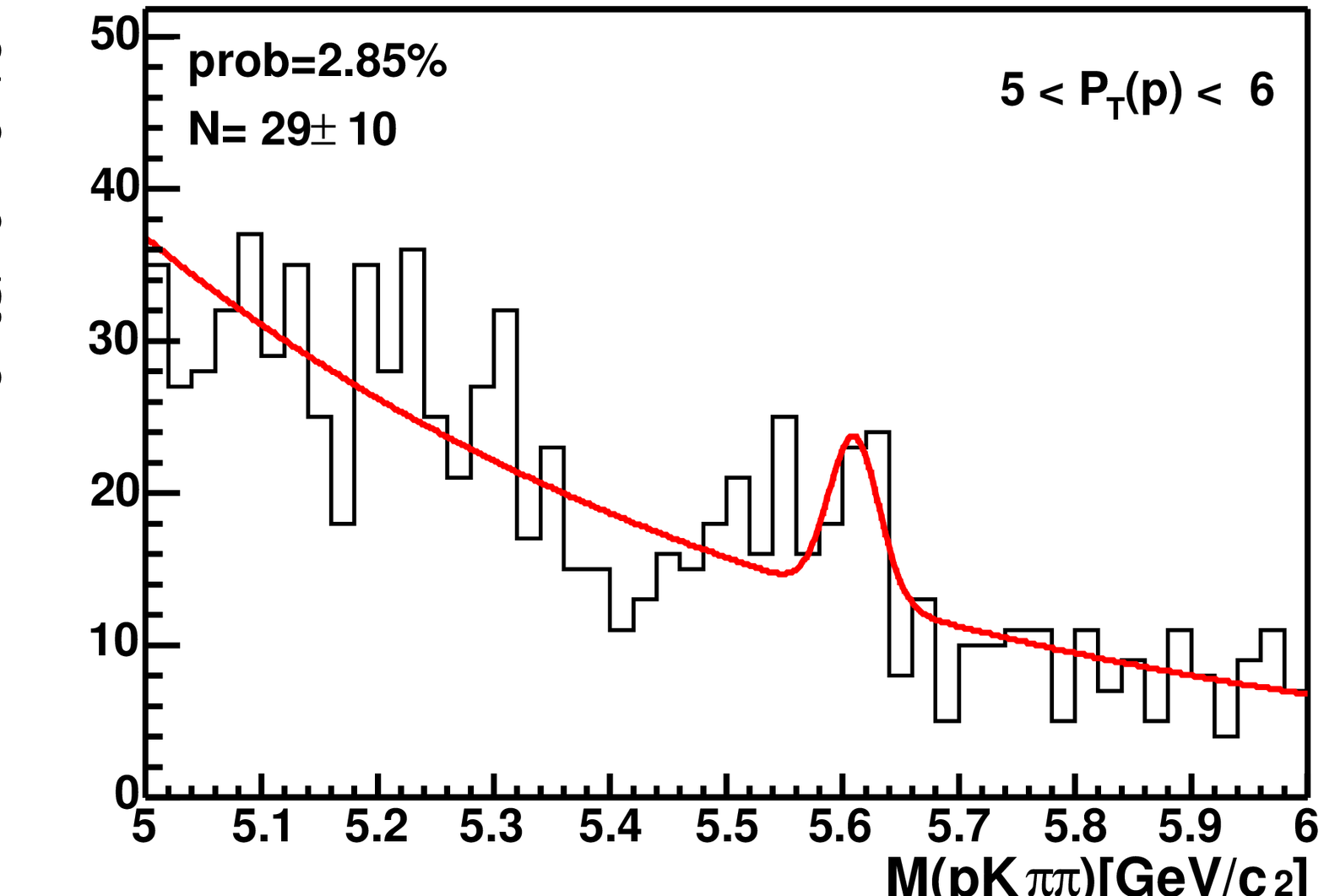}\\
\includegraphics[width=180pt, angle=0]{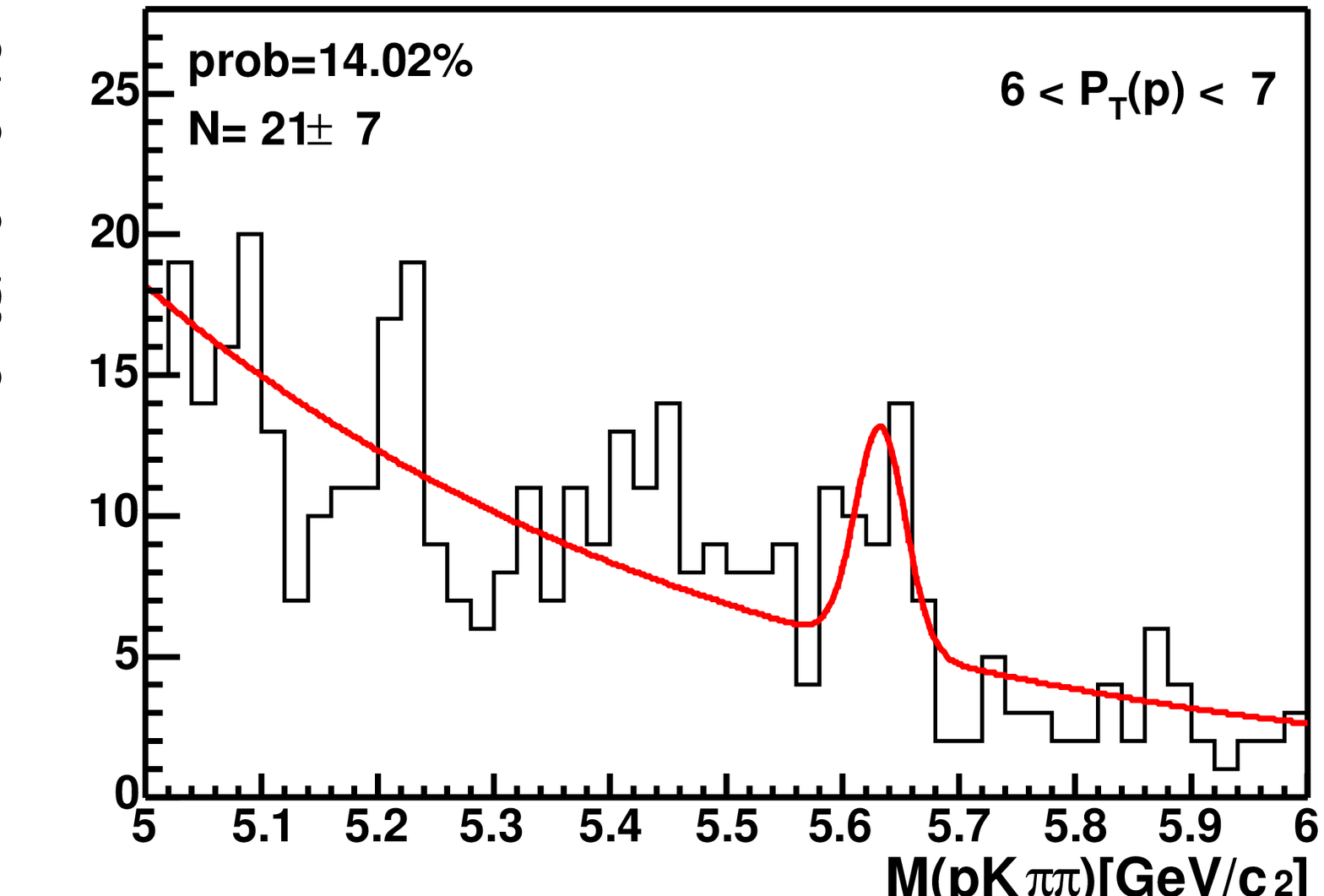}&
\includegraphics[width=180pt, angle=0]{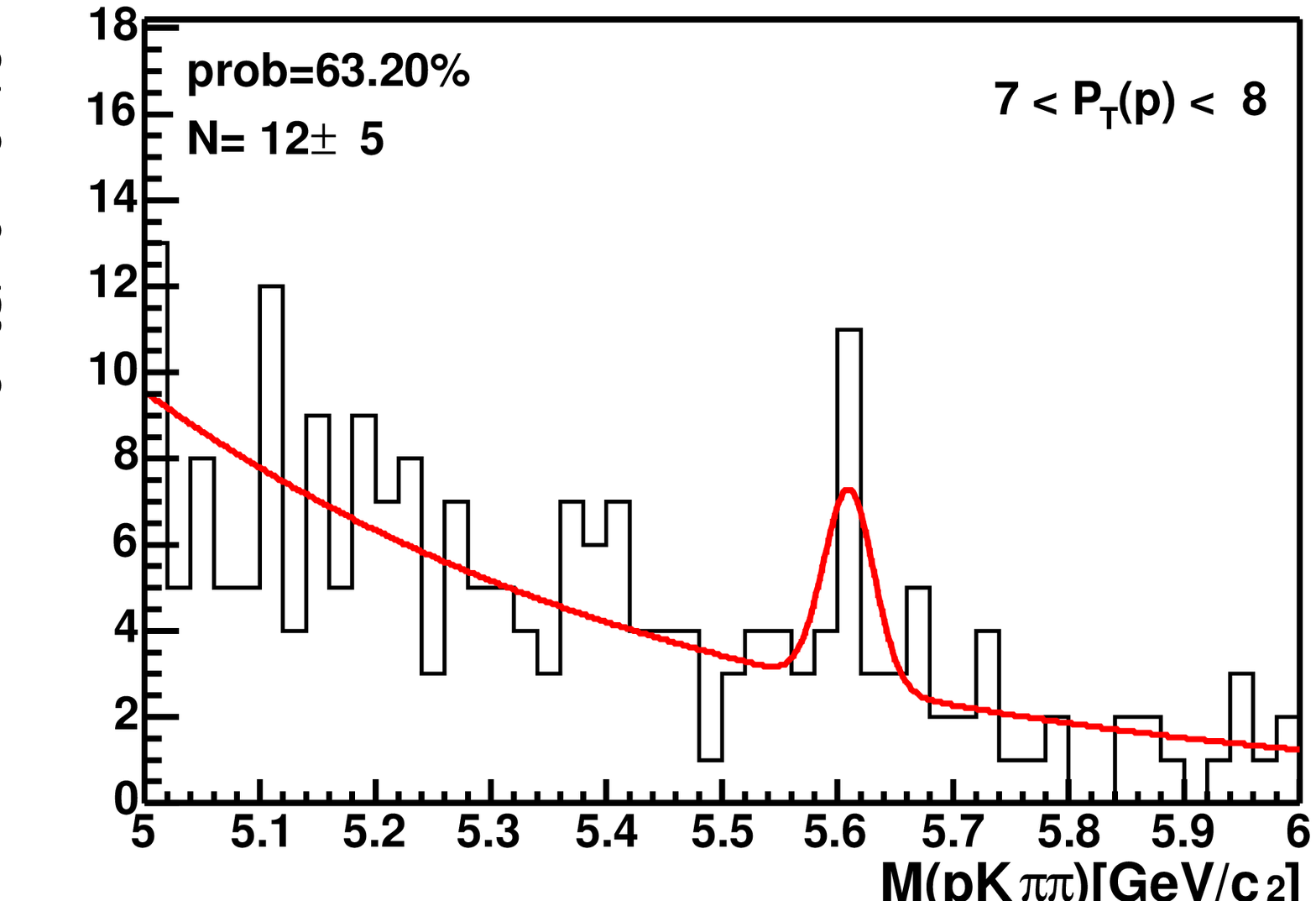}\\
\includegraphics[width=180pt, angle=0]{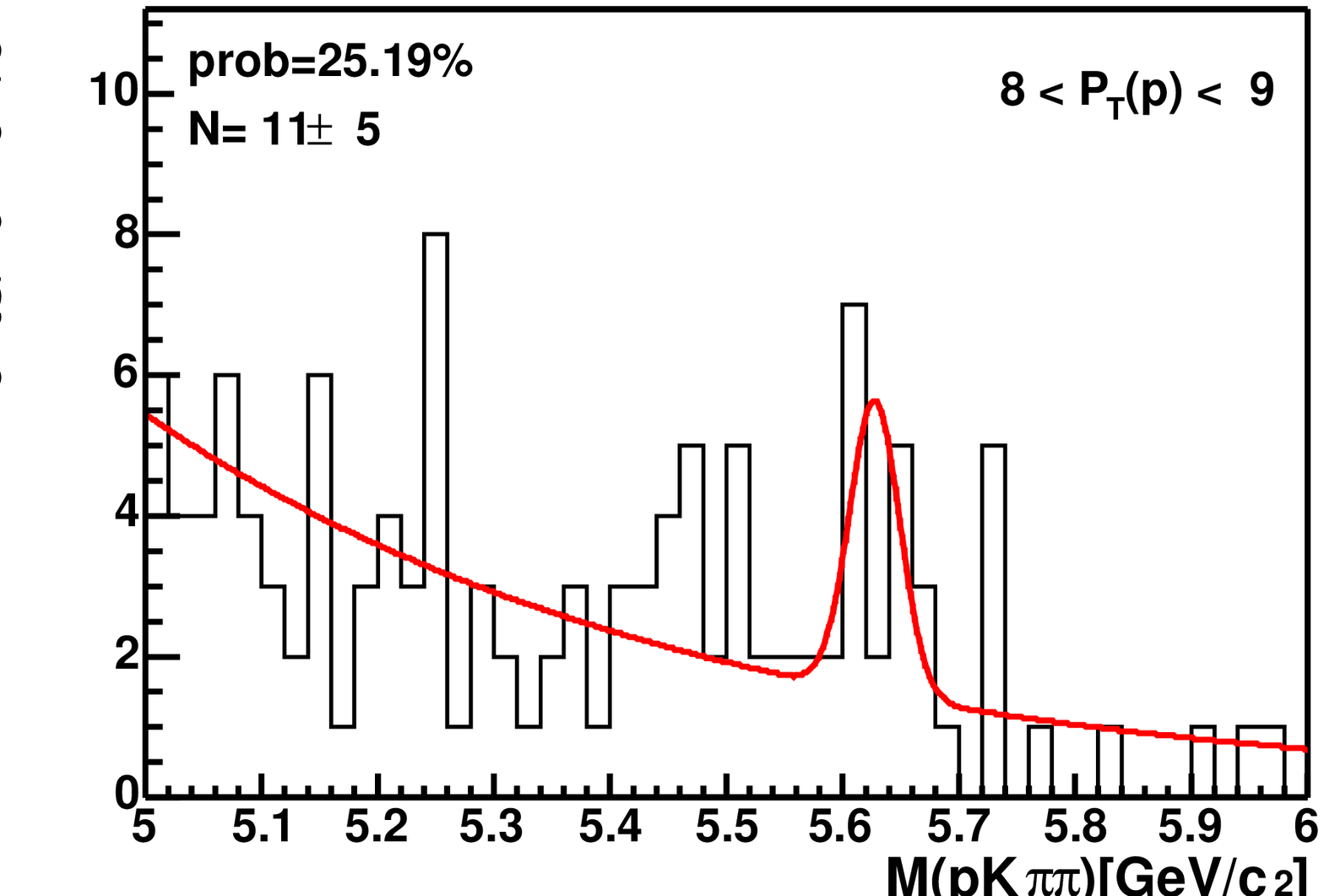}&
\includegraphics[width=180pt, angle=0]{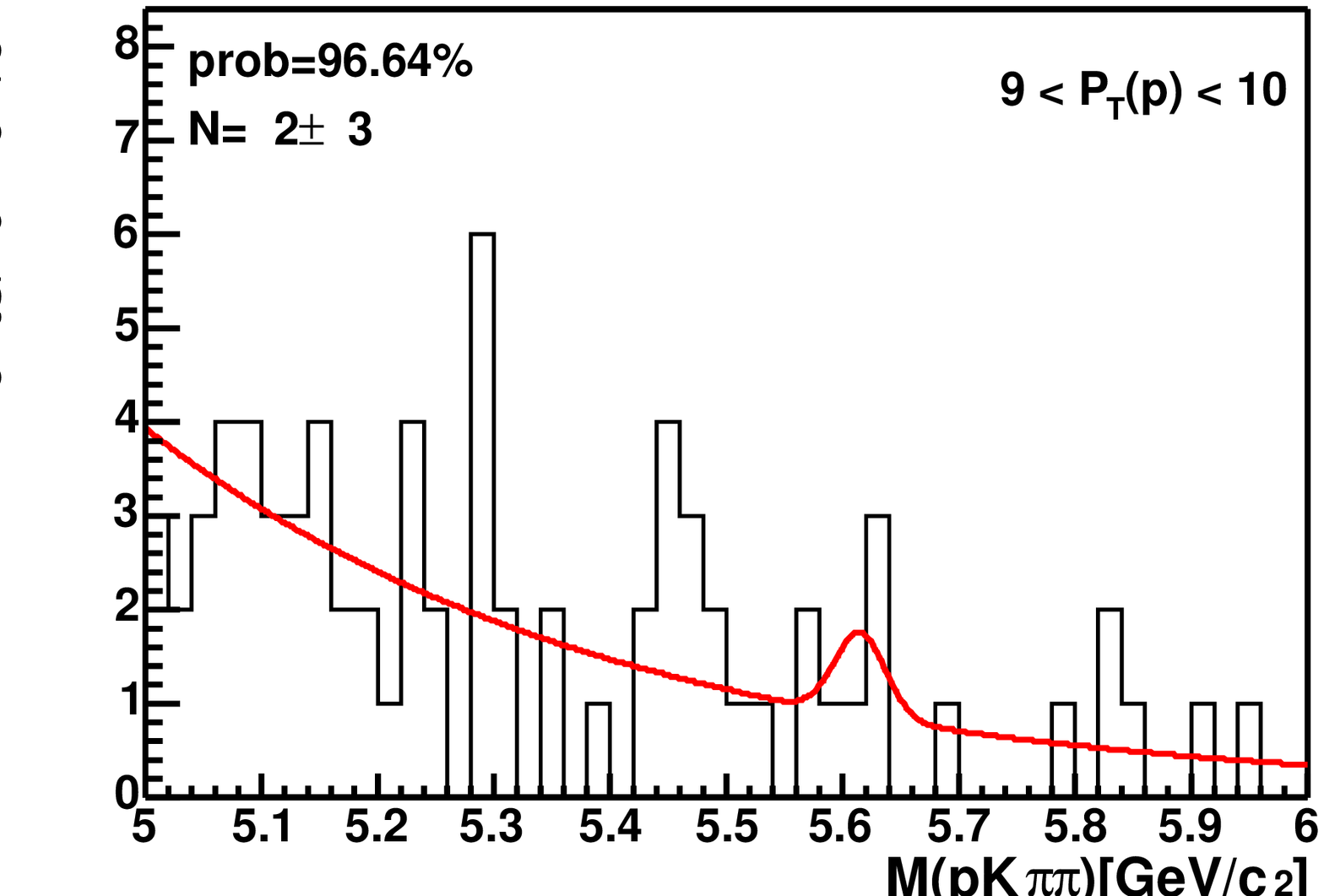}\\
      \end{tabular}
        \caption[Example of \Lb\ mass fit for the MC and data comparison]
	{Example of \Lb\ mass fit for the MC and data comparison. The variable
         to compare is the \pt\ of proton, from 2 to 10 \gevc, in bins of 1
         \gevc. \mpkpipi\ is fitted to a signal Gaussian and an
	exponential background.}
     \label{fig:lbcountfit}
     \end{center}
  \end{figure}

\begin{figure}[tbp]
     \begin{center}
        \includegraphics[width=190pt, angle=0]{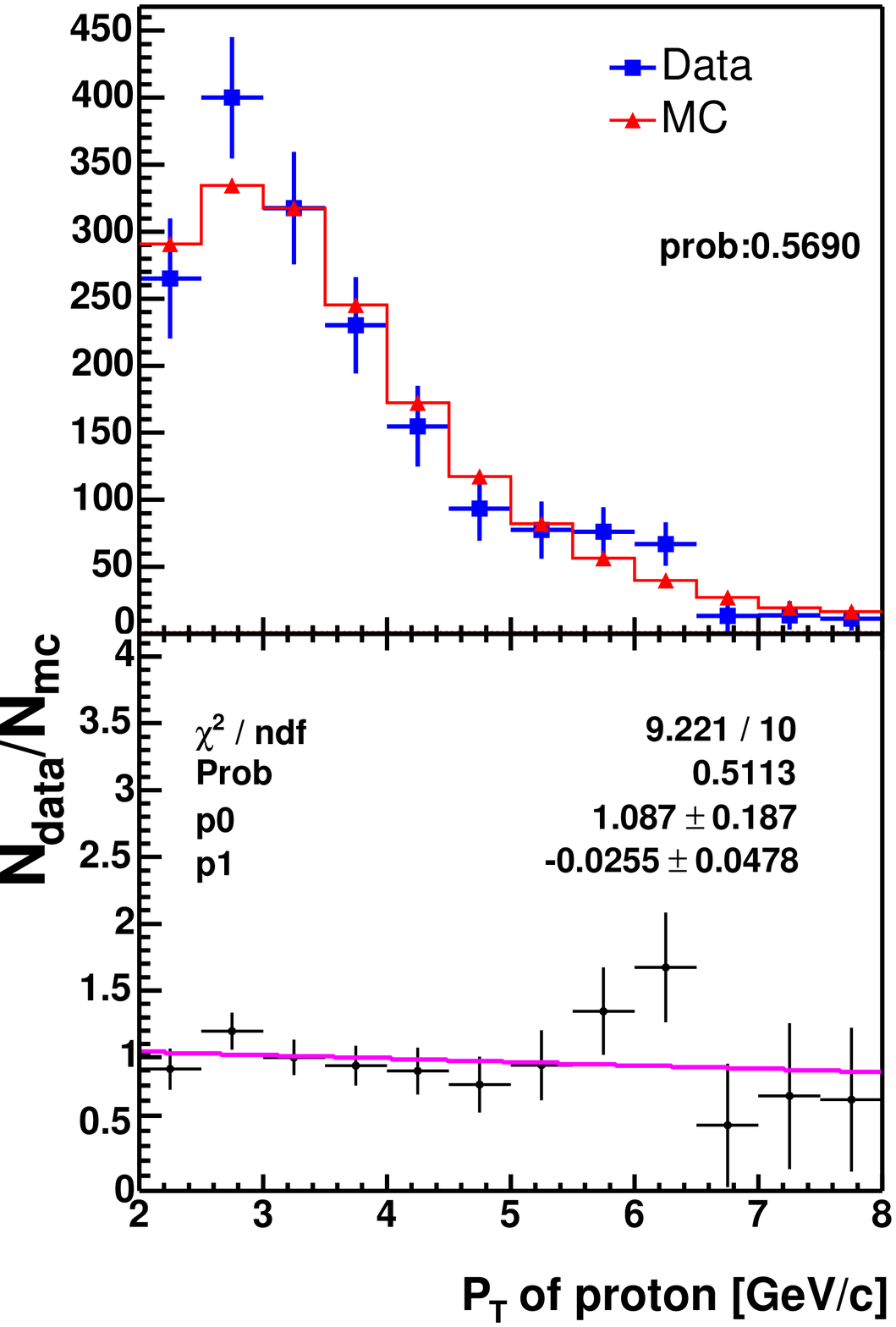}
        \includegraphics[width=190pt, angle=0]{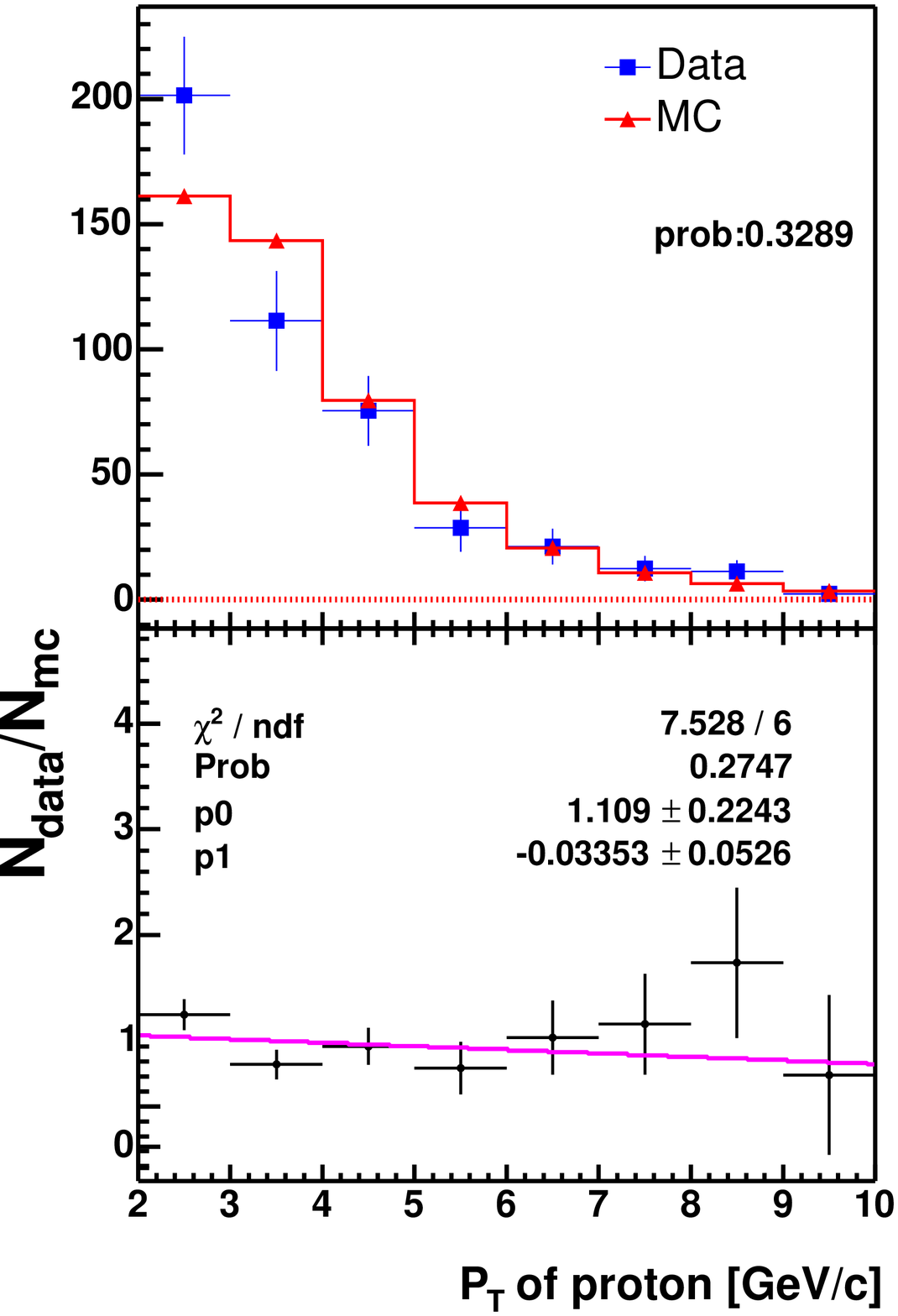}
        \caption[MC and data comparison of \pt(proton)]
        {MC and data comparison of \pt(proton). The data points come 
	from the fit to \mpkpi\ in Figure~\ref{fig:lccountfit} (left) 
	and the fit to \mpkpipi\ in Figure~\ref{fig:lbcountfit} (right).}
     \label{fig:pptcomp}
        \includegraphics[width=300pt, angle=0]
	{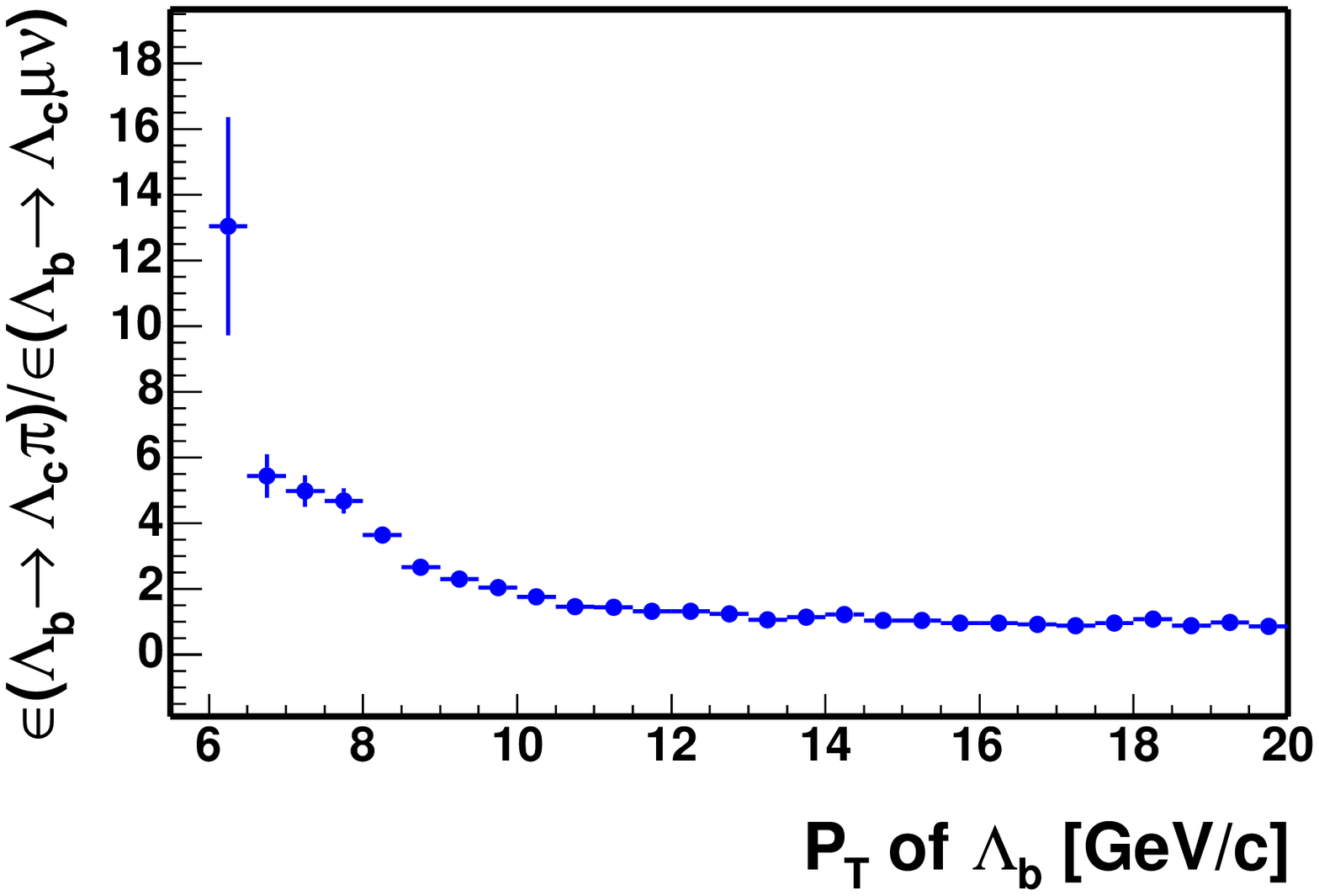}
     \end{center}
        \caption[MC Efficiency ratio of \lbhad\ to \lbsemi\ vs \pt(\Lb)]
        {MC Efficiency ratio of \lbhad\ to \lbsemi\ as a function
	of the transverse momentum of \Lb.}
     \label{fig:ratiopt}
  \end{figure}

For the \inclbsemi\ and \lbhad\ modes, there are non-negligible backgrounds 
under the signal peak from the reflections due to a mis-assignment of the mass 
for one of the particles, see Chapter~\ref{ch:yield} for more details. 
This type of background has a different behavior from the combinatorial 
background in the sideband region. Since a background-free sideband subtraction
 is difficult to perform, we choose to fit the number of signal events in each 
bin of the variables which we want to compare. For the number of \inclbsemi\ 
candidates, the \mpkpi\ distribution is fitted to a signal Gaussian and a 
second-order polynomial background as shown in Figure~\ref{fig:lccountfit}. 
For the number of \lbhad candidates, the \mpkpipi\ distribution is fitted 
to a simplified model: a Gaussian signal and an exponential background,
as shown in Figure~\ref{fig:lbcountfit}. Note that although the \Lb\ fit model 
is simplified, the systematic uncertainty due to the naive model is no more 
than $3\%$ of the number of signal events in each bin compared with the 
$15\%$  statistical uncertainty. The widths of \mpkpi\ and \mkpi\ are fixed to
 the values obtained from the full statistics when doing the fit. 
Figure~\ref{fig:pptcomp} shows the data and MC comparison using the fit values 
obtained from Figures~\ref{fig:lccountfit}--~\ref{fig:lbcountfit}.

 When comparing the MC and data distributions, if the number of data signal 
events in one bin is less than 20, that bin is combined with the next bin until
 the sum of the events is over 20. Then a $\chi^2$ is computed, 
\begin{equation}
 \label{eq:chi2}
 \chi^2 = \sum_{i}^{n} \frac{(N_{{\mathrm MC}}(i)-N_{{\mathrm data}}(i))^2}
	{\sigma_{{\mathrm MC}}(i)^2 + \sigma_{{\mathrm data}}(i)^2}
\end{equation}
where $i$ stands for $i^{th}$ bin and total number of bins in a histogram
is $n$. The number of degree freedom is $n-1$. For the \inclbsemi\ and
\lbhad\ modes, a $\chi^2$ is also calculated except that the bin width of each
 variable is fixed in this case. Besides the $\chi^2$ test, we also plot the 
ratio data/MC. We fit the ratio to a first-order polynomial and check if the 
slope, ${\cal M}$, is consistent with zero. 

\begin{figure}[tbp]
     \begin{center}
      \includegraphics[width=130pt, angle=0]{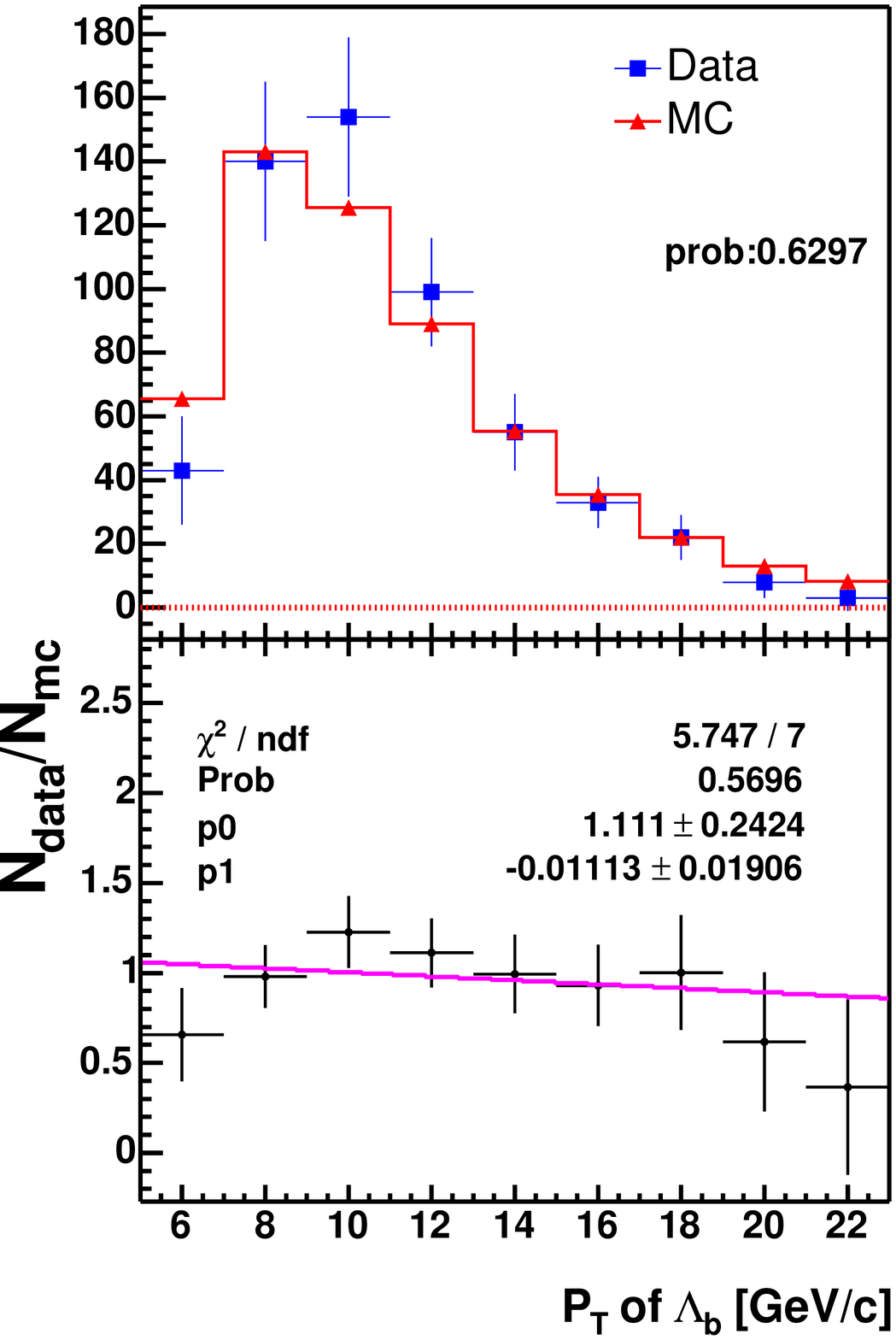}
        \includegraphics[width=130pt, angle=0]{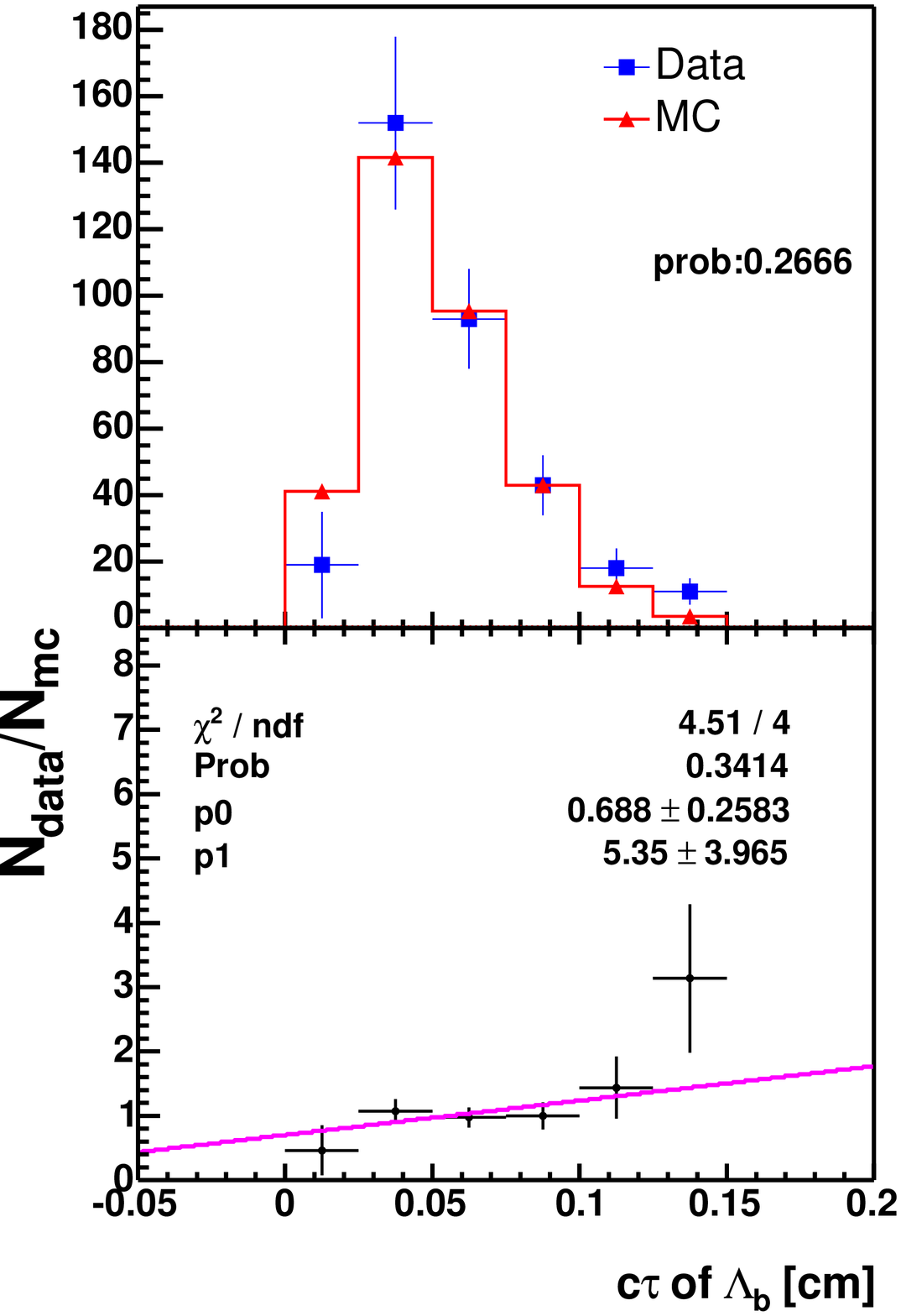}
        \includegraphics[width=130pt, angle=0]{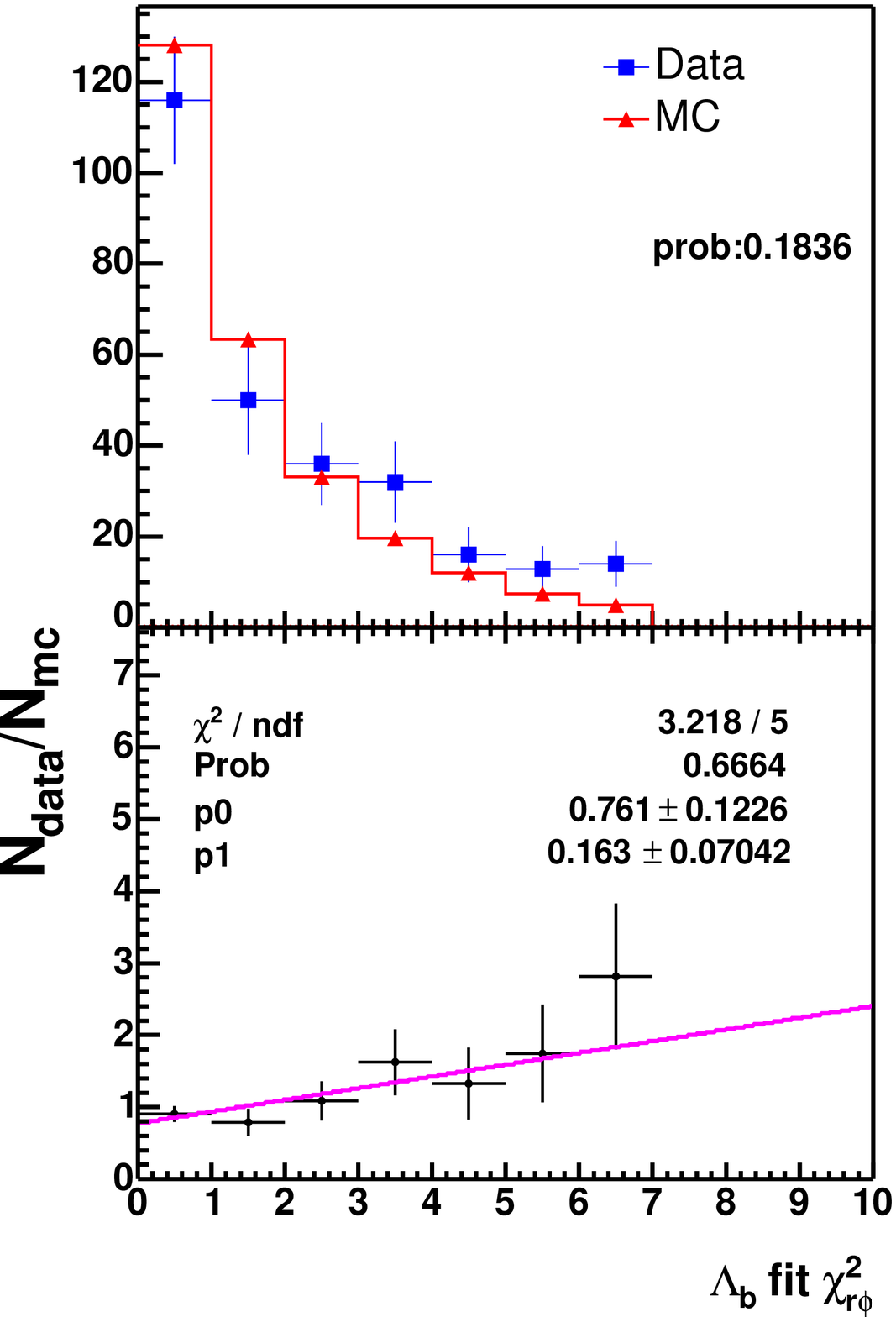}
      \includegraphics[width=130pt, angle=0]{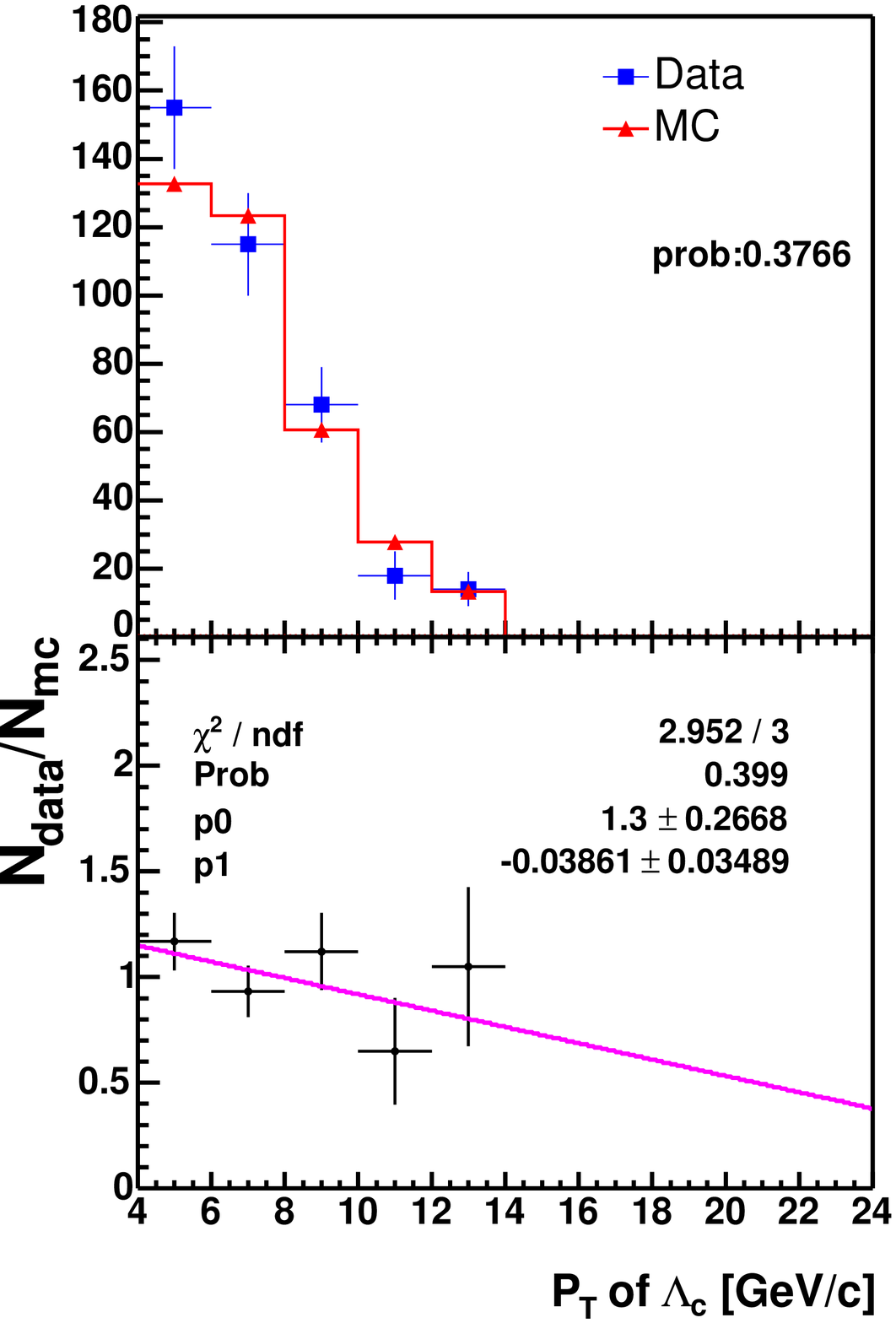}
        \includegraphics[width=130pt, angle=0]{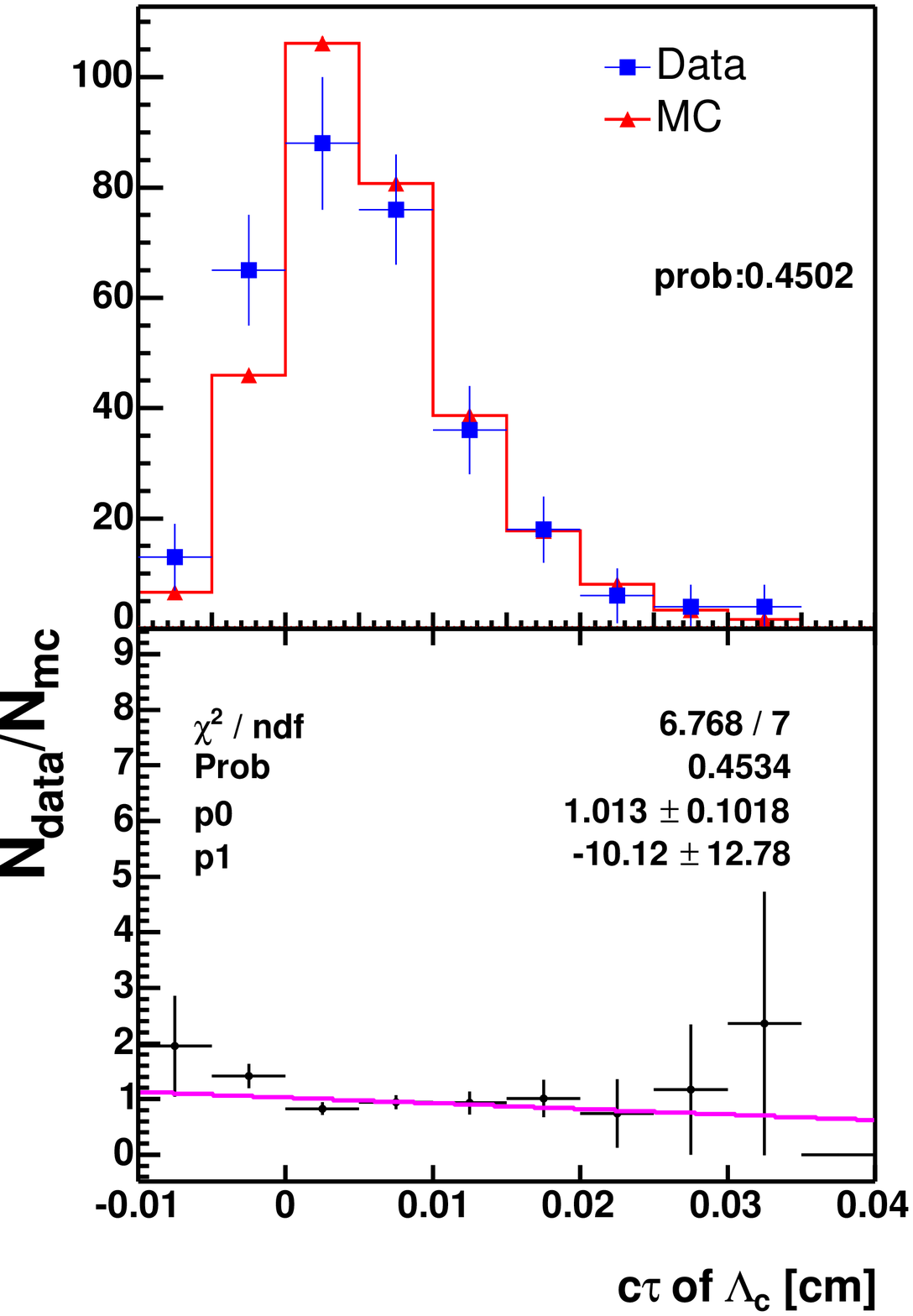}
        \includegraphics[width=130pt, angle=0]{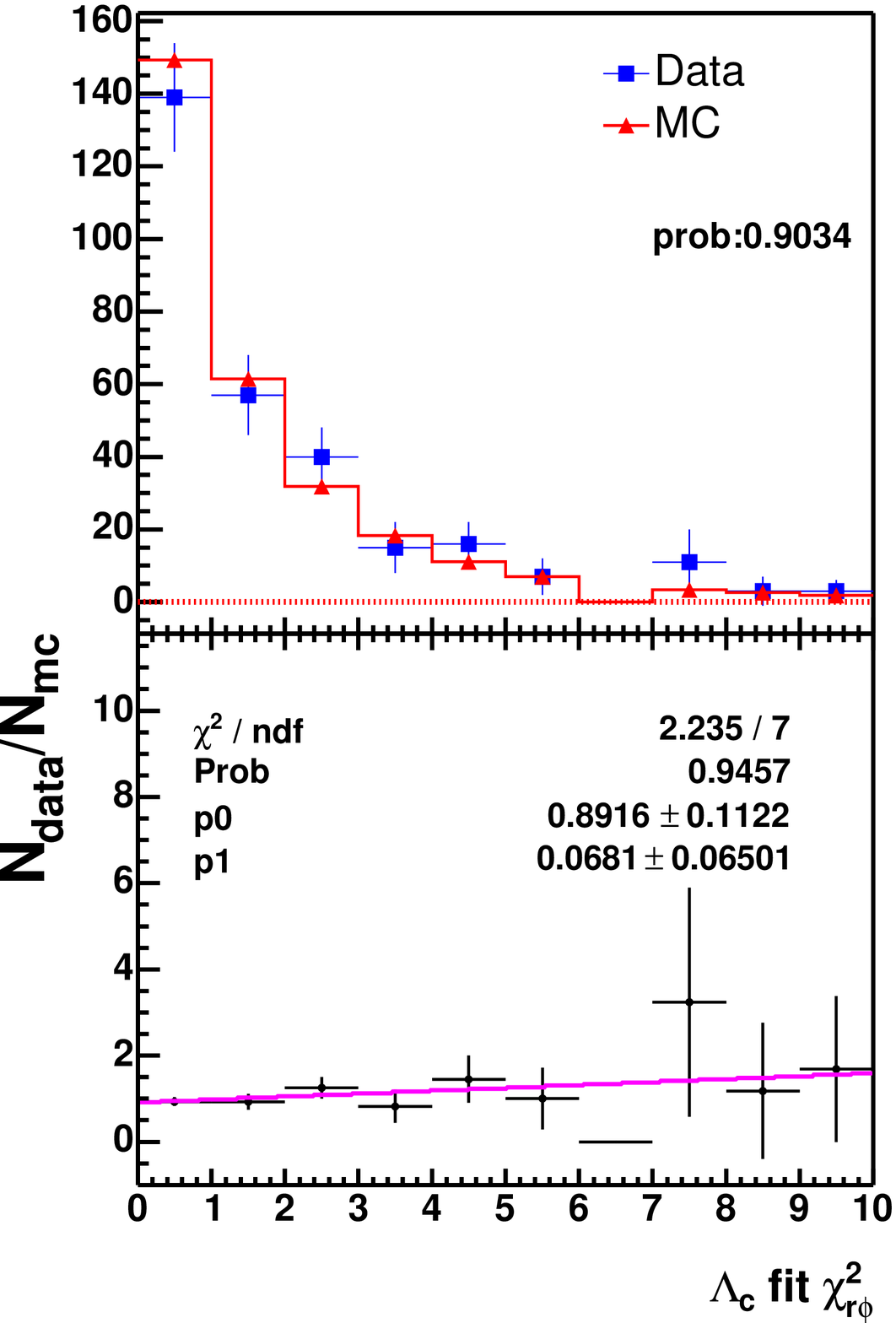}
	\caption[\pt, \ctau, $\chi^2_{\rphi}$
             of \B\ and charm in MC and data (\lbhad)]
	{\lbhad\ MC and data comparison: from the top left to the bottom
	right are: \pt(\Lb), \ctau(\Lb), vertex fit 
	$\chi^2_{\rphi}$ for the \Lb\ vertex, \pt(\Lc), \ctau(\Lc), 
        and vertex fit $\chi^2_{\rphi}$ for the \Lc\ vertex.}
 	\label{fig:mcdatalcpi0}
     \end{center}
  \end{figure}

\begin{figure}[tbp]
     \begin{center}
      \includegraphics[width=130pt, angle=0]{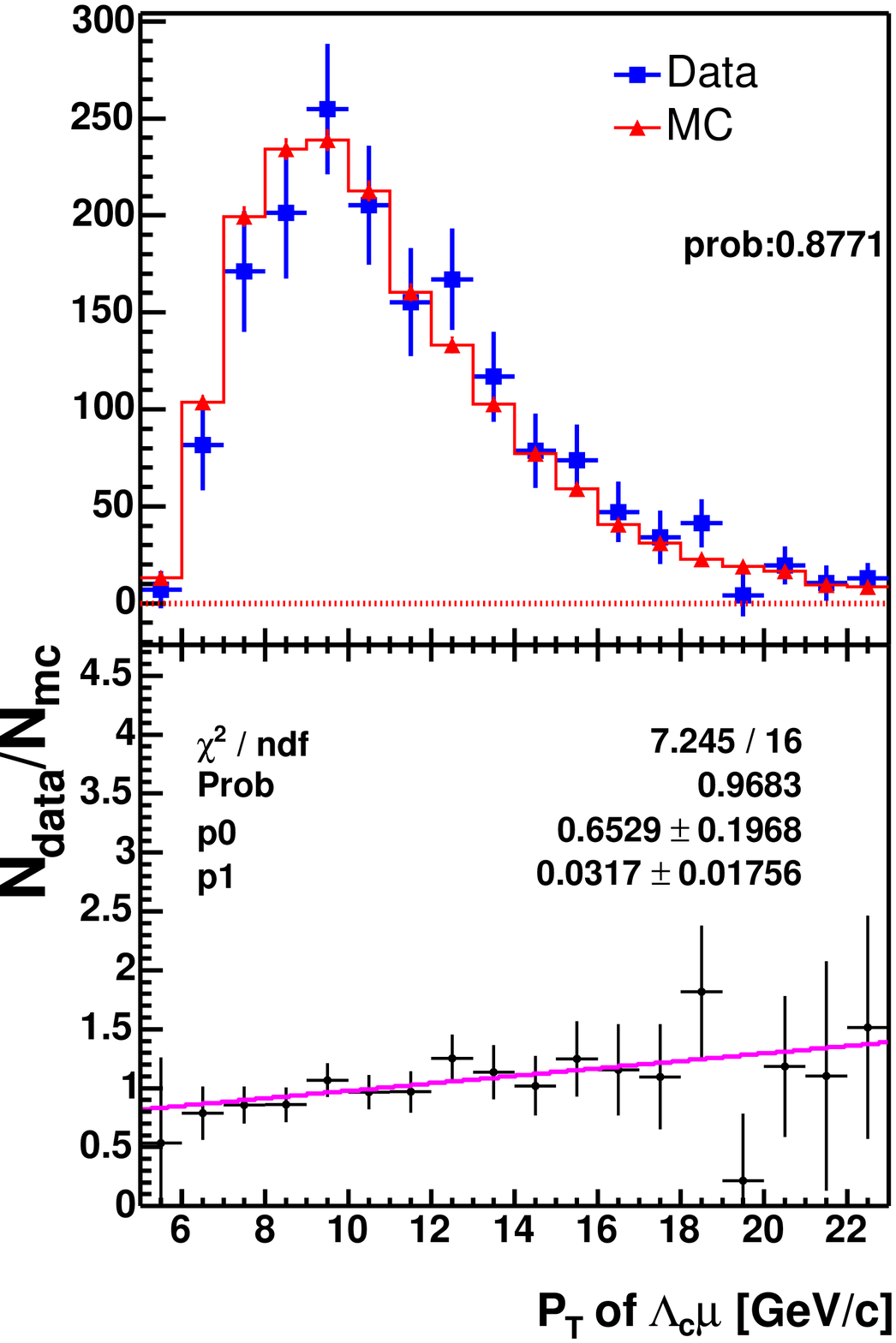}
        \includegraphics[width=130pt, angle=0]{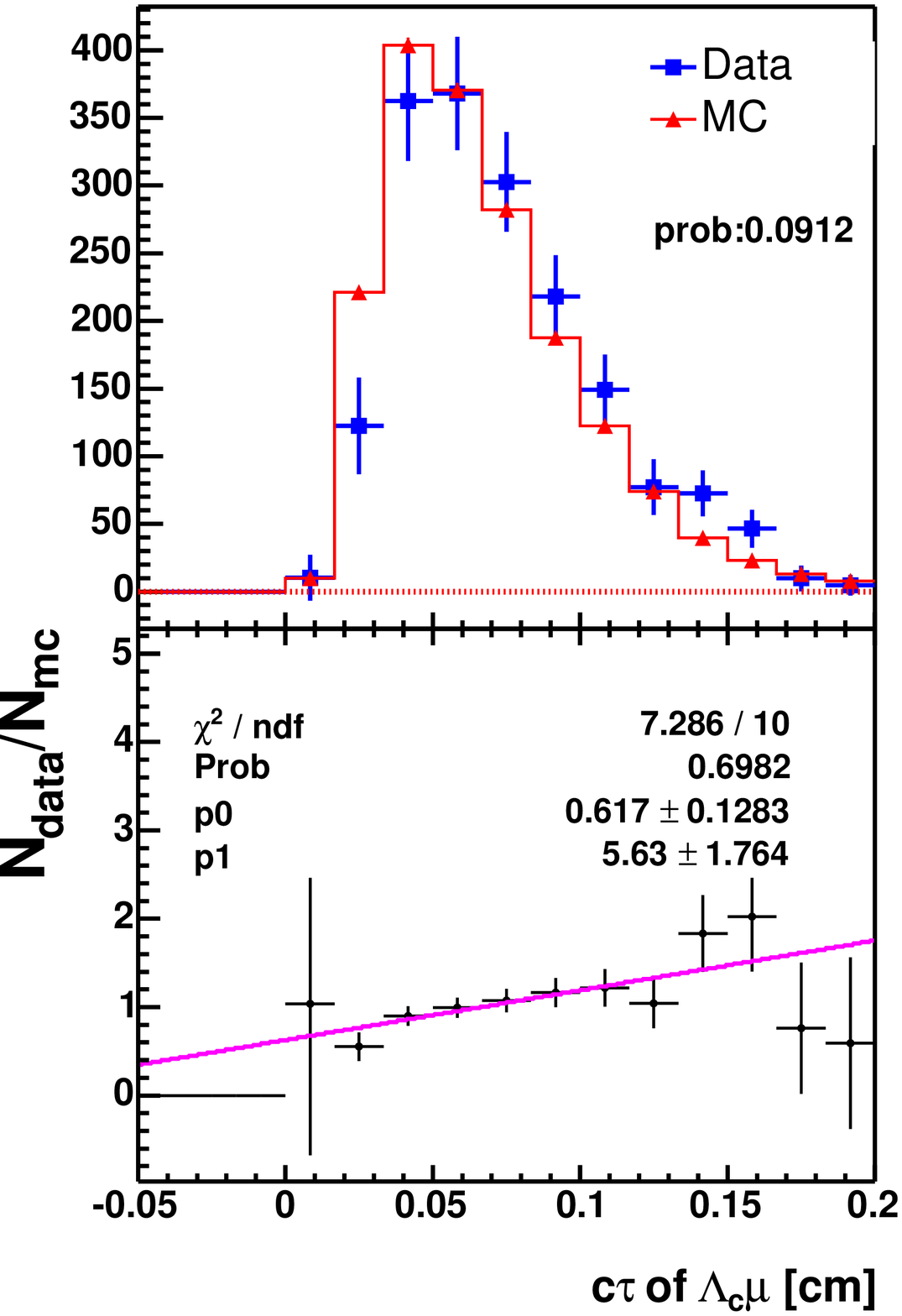}
      \includegraphics[width=130pt, angle=0]{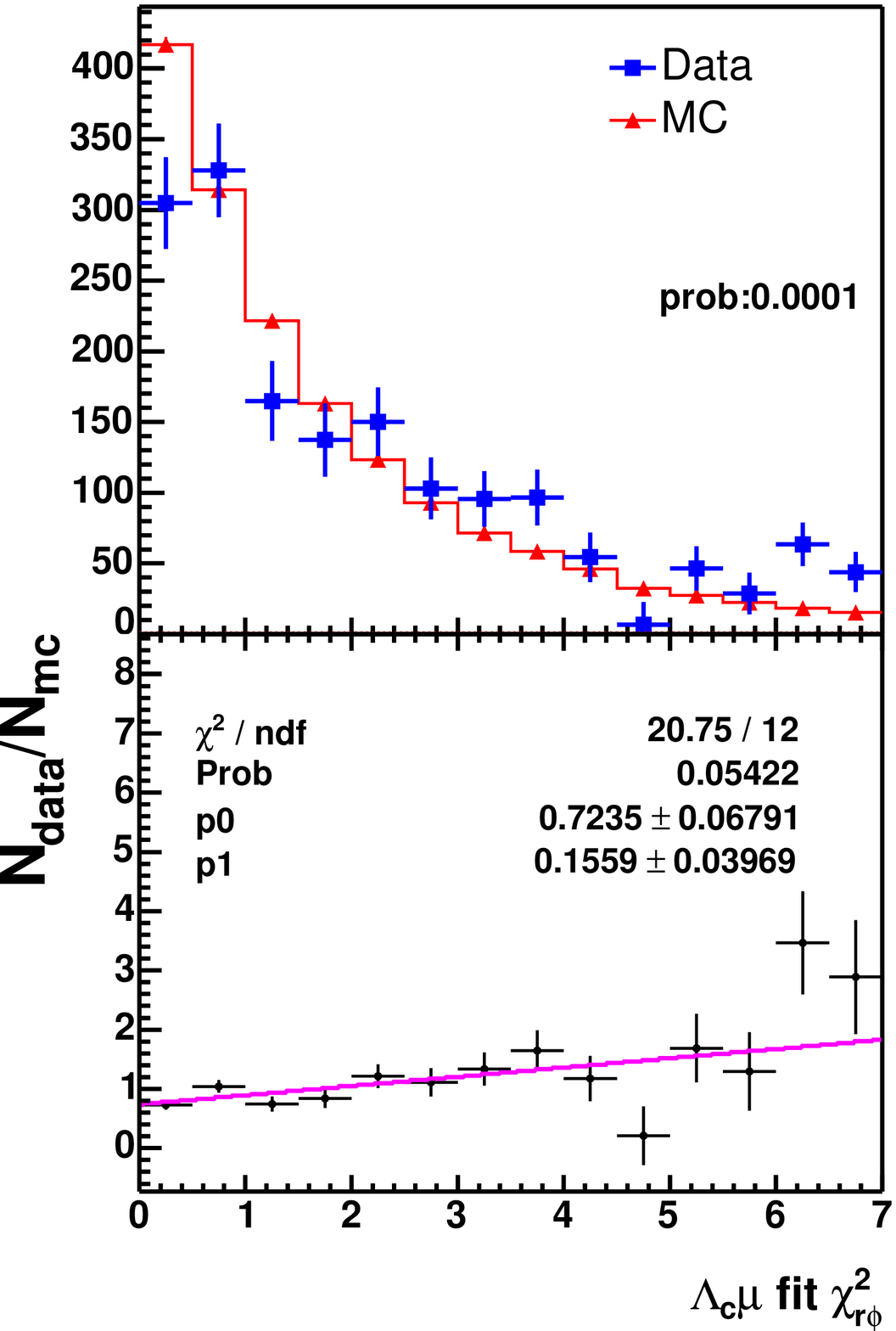}
      \includegraphics[width=130pt, angle=0]{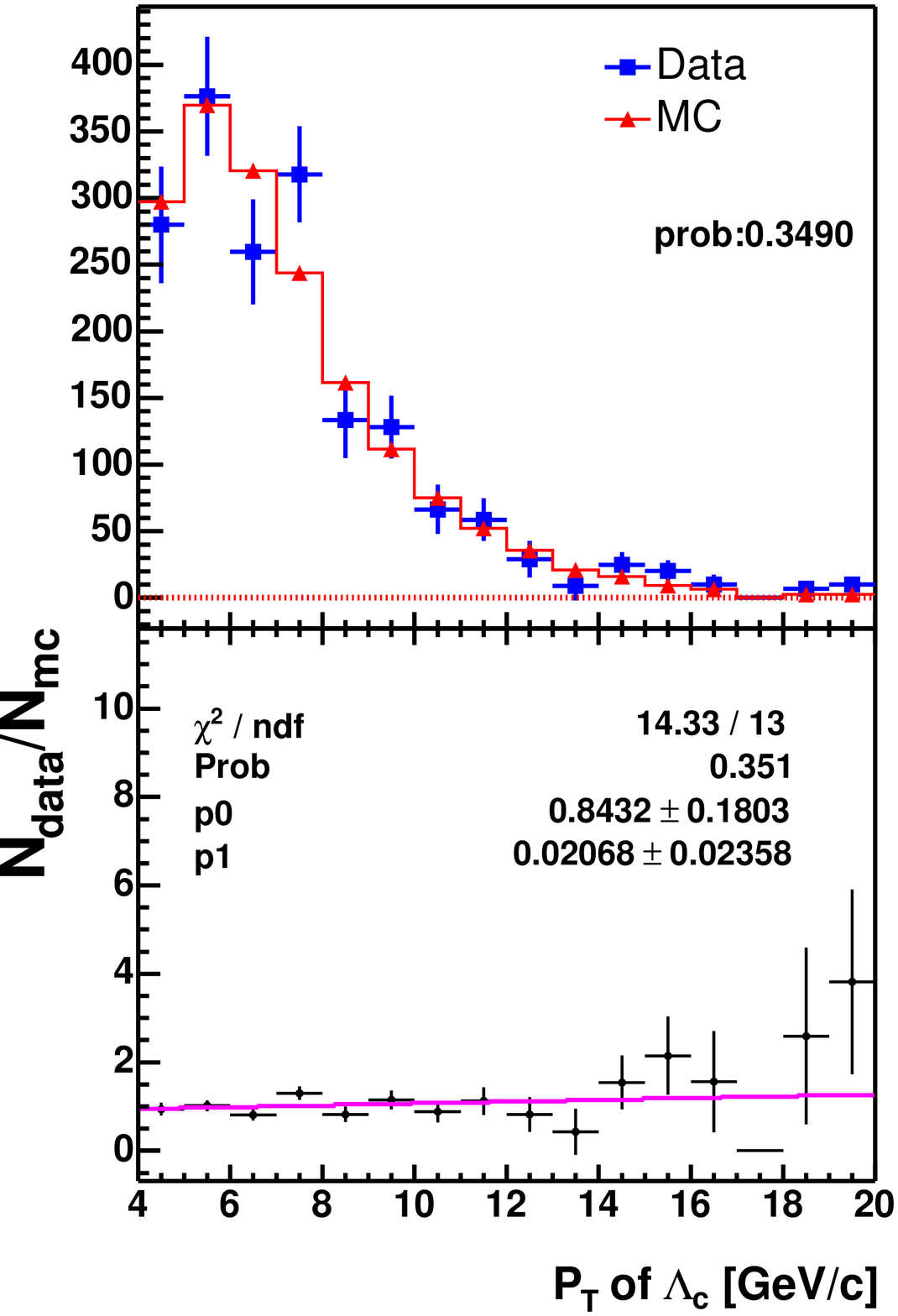}
        \includegraphics[width=130pt, angle=0]{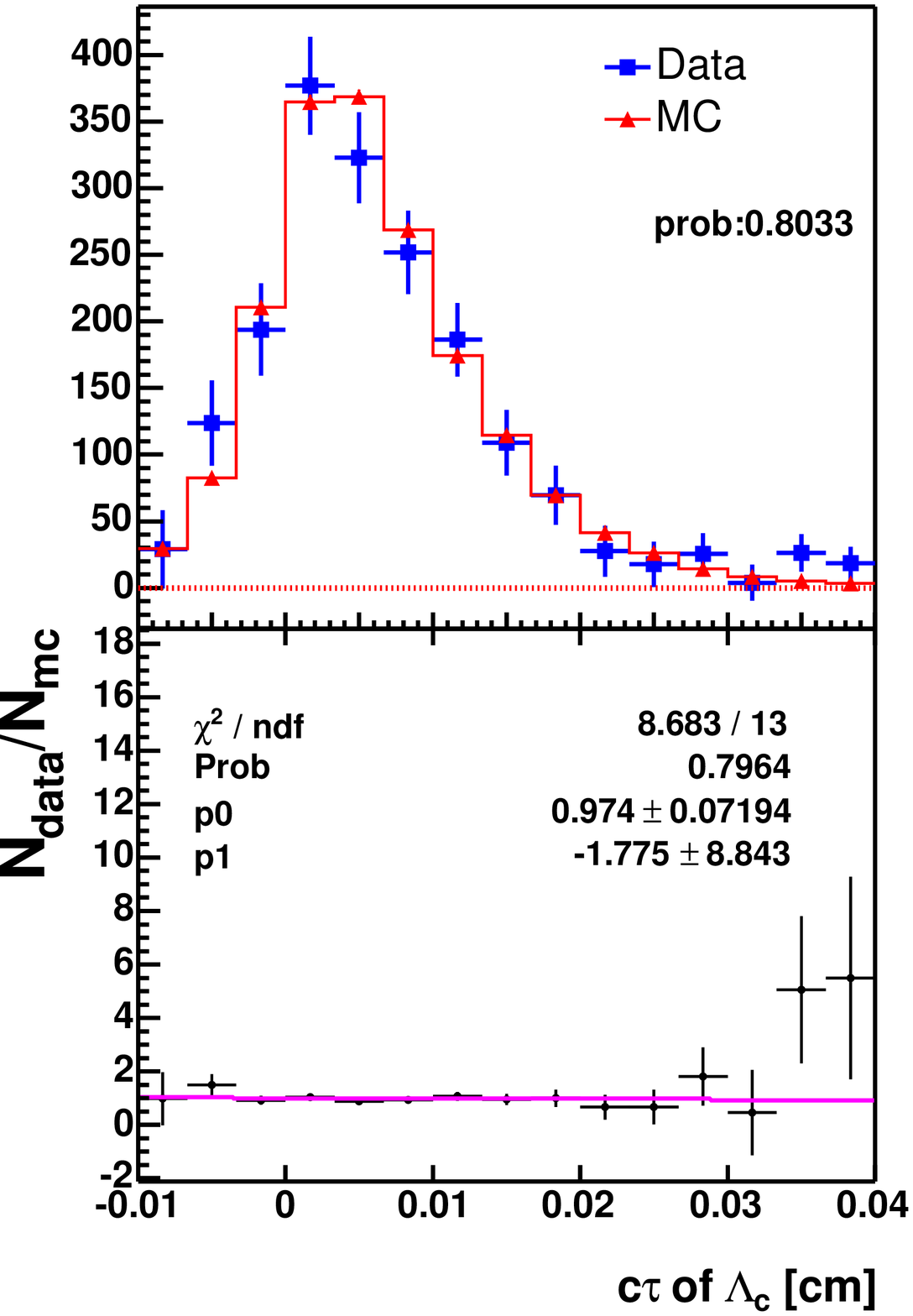}
      \includegraphics[width=130pt, angle=0]{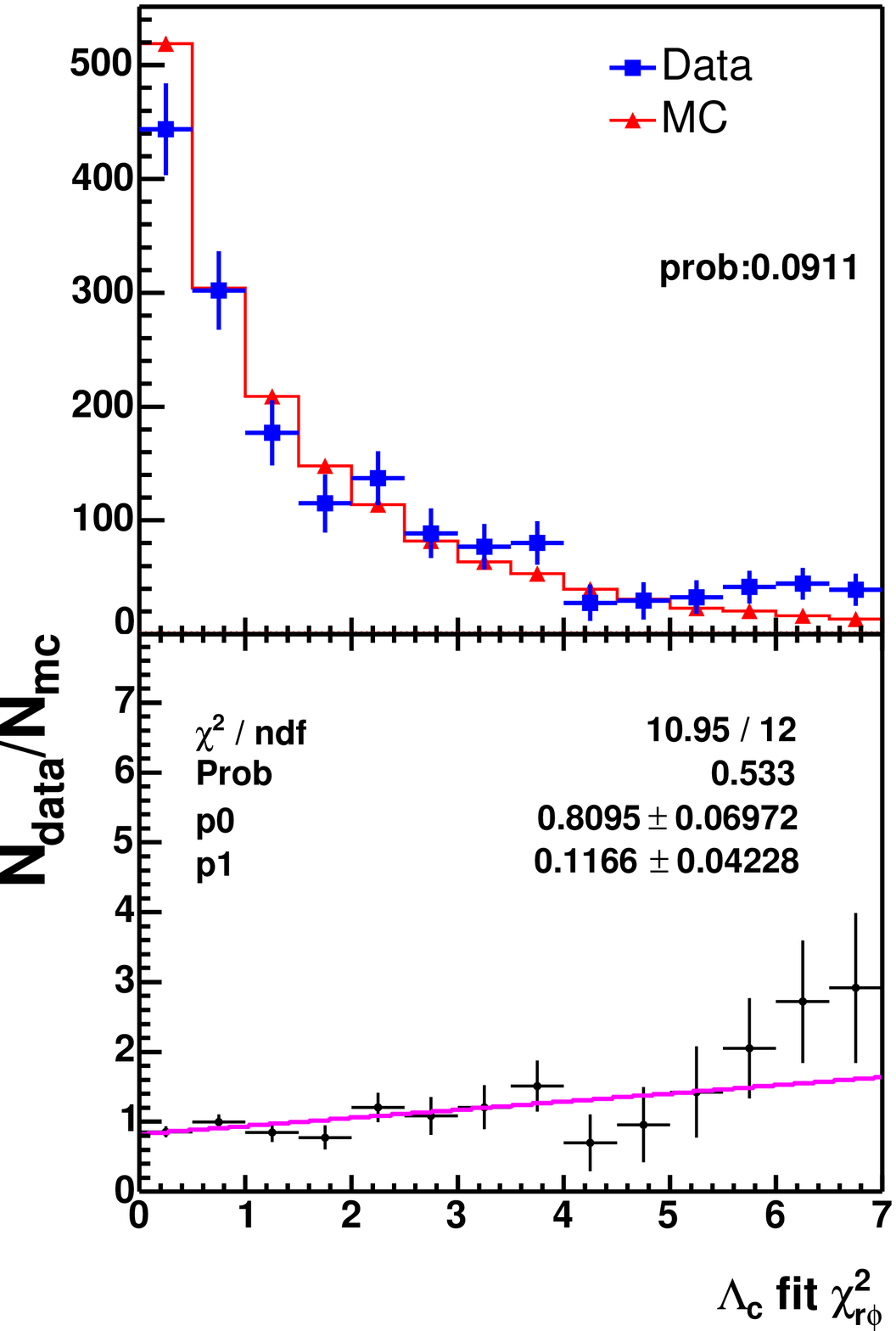}
	\caption[\pt, \ctau, $\chi^2_{\rphi}$
             of \lcmu\ and charm in MC and data (\inclbsemi)]
	{\inclbsemi\ MC and data comparison: from the top left to the bottom
	right are: \pt(\lcmu), \ctau(\lcmu), vertex fit 
	$\chi^2_{\rphi}$ for the \lcmu\ vertex, \pt(\Lc), \ctau(\Lc), 
        and vertex fit $\chi^2_{\rphi}$ for the \Lc\ vertex.}
 	\label{fig:mcdatalcsemi0}
     \end{center}
  \end{figure}

\begin{figure}[tbp]
     \begin{center}
      \includegraphics[width=160pt, angle=0]{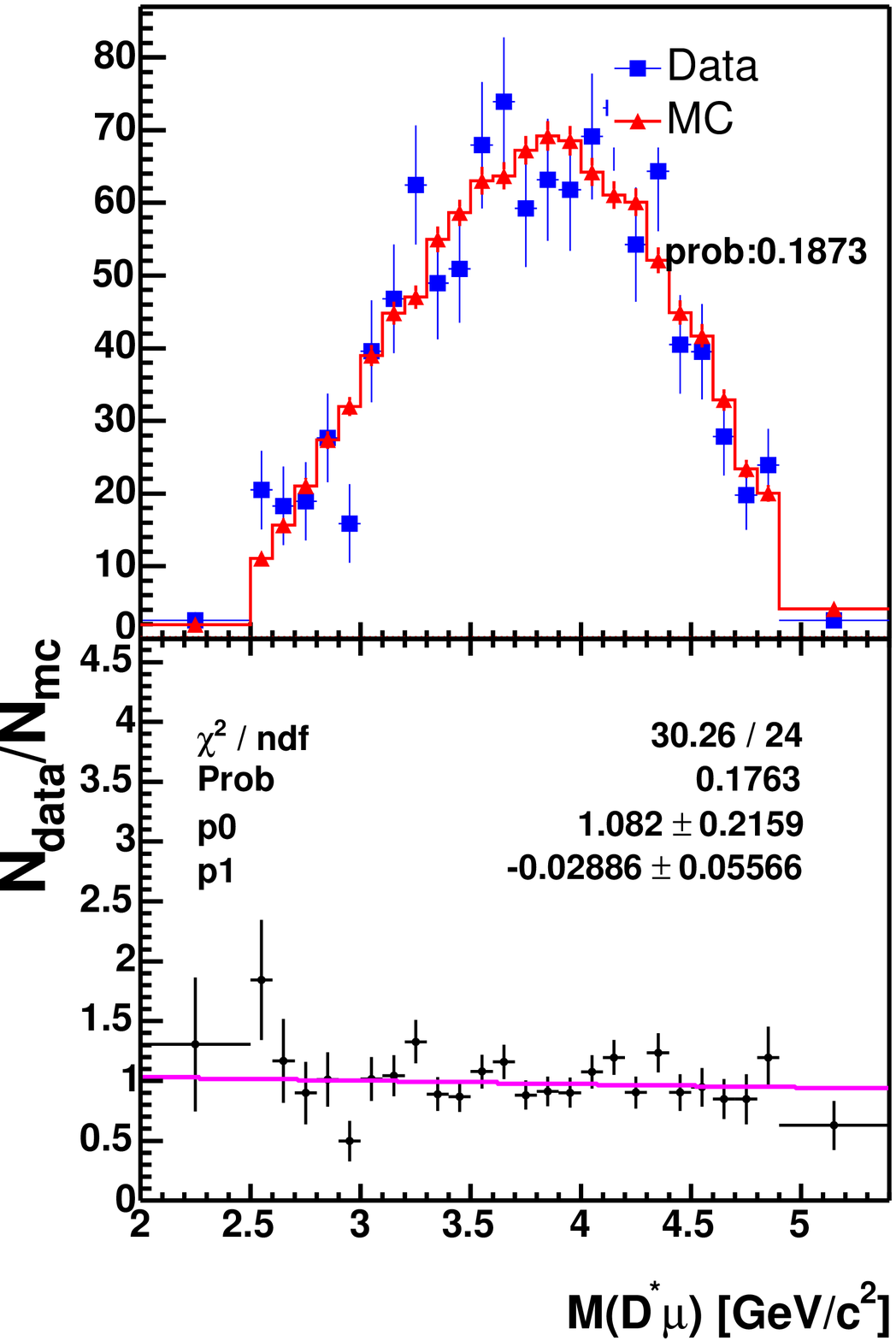}
      \includegraphics[width=160pt, angle=0]{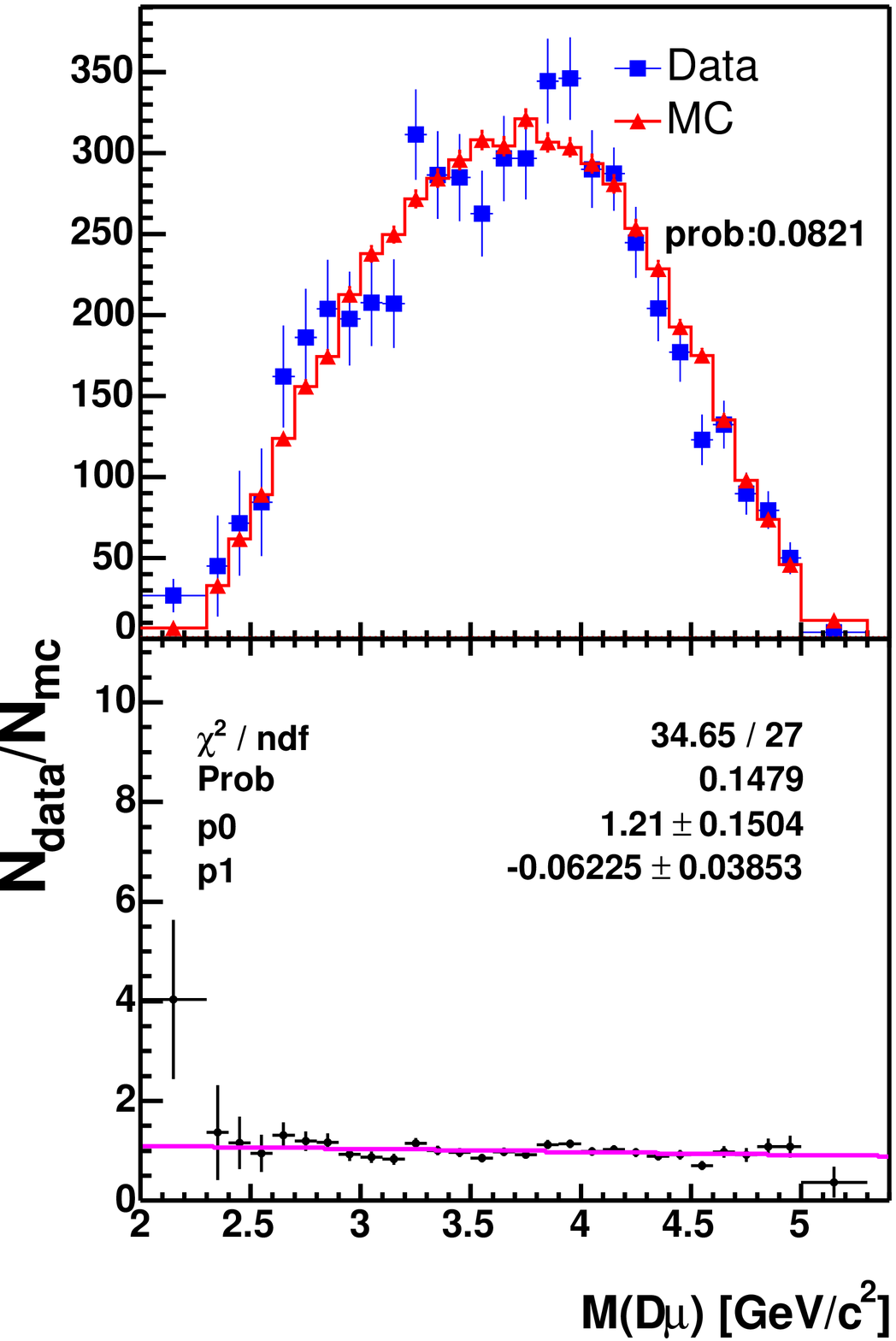}
      \includegraphics[width=160pt, angle=0]
	{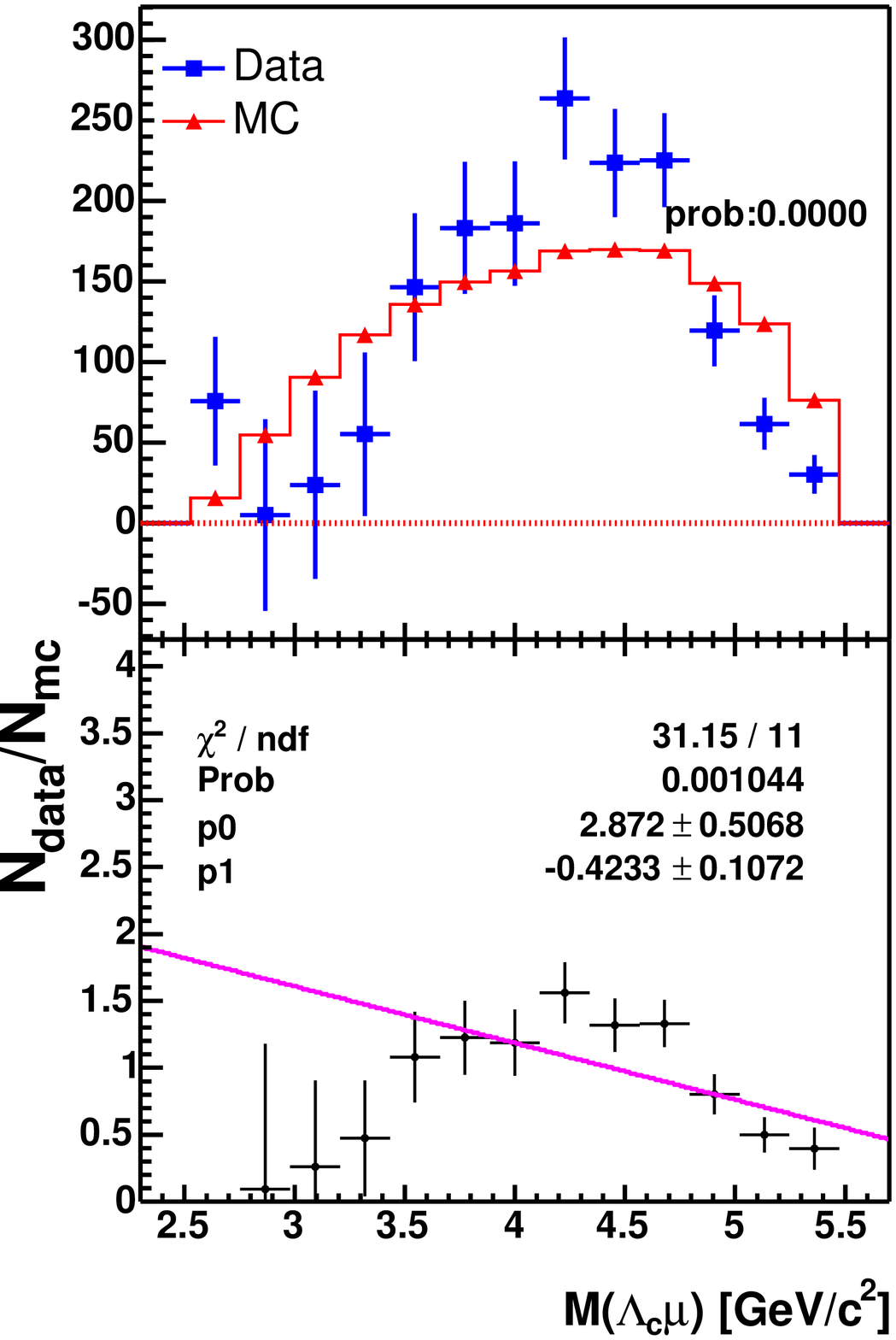}
      \includegraphics[width=160pt, angle=0]
	{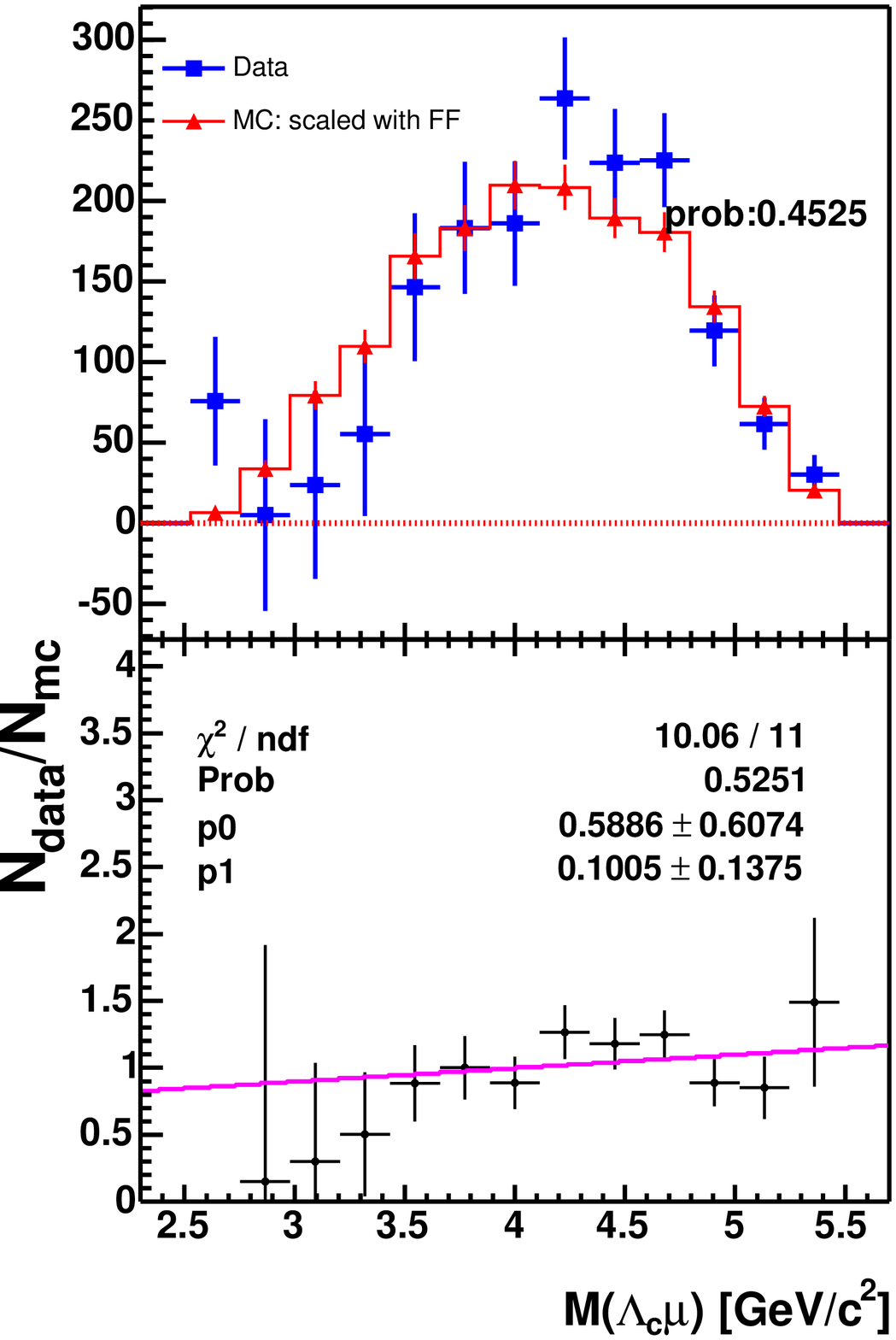}
	\caption[$M(\dstarmu)$, $M(\dmu)$ and $M(\lcmu)$ in MC and data]
	{four track invariant mass MC and data comparison: 
	from the top left to the bottom	right are: $M_{\dstarmu}$, 
	$M_{\dmu}$, $M_{\lcmu}$ (phase space MC without scaling), $M_{\lcmu}$ 
	(phase space MC after scaling).}
 	\label{fig:mcdatalcsemi1}
     \end{center}
  \end{figure}

 In the first pass, we find discrepancies in the \pt\ spectra of \Bd\ and \Lb\ 
between MC and data (see Figure~\ref{fig:bptbefore}). As the semileptonic modes
 are three-body decays and the hadronic modes are two-body decays, the 
efficiency of the trigger and analysis \pt\ cut depends strongly on the \pt\ 
of \B\ hadron (see Figure~\ref{fig:ratiopt}).
We decide to reweight the \pt\ spectra of \Bd\ and \Lb\ as described in 
Section~\ref{sec-mccom}. 
Figures~\ref{fig:mcdatalcpi0}--~\ref{fig:mcdatalcsemi0} show the 
comparison between MC and data for the analysis cut variables of the 
$\Lb\rightarrow \Lamc$ modes. Figure~\ref{fig:mcdatalcsemi1} shows the 
comparison for the $M_{\dstarmu}$, $M_{\dmu}$, and the 
$M_{\lcmu}$ from the phase space MC before and after 
multiplying each bin entry with a scaling factor.  The scaling factor is 
obtained by dividing the $M_{\lcmu}$ distribution from the form factor 
weighted (see Section~\ref{sec-eff}) by that from the 
phase space generator-level MC. The agreement of the MC $M_{\lcmu}$ 
distribution with that from the data has significantly improved 
after applying the scaling factor. 
In addition, we compare 
the efficiency of the $M_{\lcmu}$ cut in the MC and data, given that the 
other (N-1) analysis cuts are applied. The contribution of the fake muons 
and physics backgrounds are included. Since there are uncertainties from 
the fit to both data and MC, the efficiency is defined as:
\begin{equation}
 \epsilon = \frac{n}{N},
\end{equation}
where $n$ is the number of events after making all the analysis cuts and 
$N$ is the number of events after making the $N-1$ cuts. The uncertainty 
on the efficiency is derived by Heinrich~\cite{heinrich:comm}:
\begin{equation}
 \sigma_{\epsilon} = \sqrt{ (\epsilon\cdot\frac{\sigma_N}{N})^2 + 
	(1-2\epsilon)\cdot(\frac{\sigma_n}{N})^2},
\end{equation}
where $\sigma_N$ and $\sigma_n$ are the uncertainties from the fit.
We find the data give an efficiency of 0.77$\pm$0.04, while
 the form factor weighted MC gives an efficiency of 0.72$\pm$0.05, which 
is in good agreement with the data efficiency. 
%
Other distributions important for this analysis may be found in 
Yu\cite{cdfnote:7559}. 

In general, the MC describes the data well except for 
the pseudo \ctau\ of \lcmu, and the $\chi^2_{\rphi}$ of 
the \B\ and charm vertex fits. For the disagreement in the vertex fit 
$\chi^2_{\rphi}$, as it is beyond the scope of this analysis to scale the 
measurement errors in the MC, we choose to make a loose cuts on the data 
$\chi^2_{\rphi}$. In Section~\ref{sec-check}, 
we perform a cross-check of the relative branching ratio variation by dividing 
the data into two subsets, according to the cuts on the pseudo \ctau\ of 
\lcmu, $\chi^2_{\rphi}$ and other variables. We do not see significant 
inconsistency. 
Therefore, we do not assign any systematic uncertainties for these variables. 

\clearpage

\clearpage


\clearpage
\newpage
\section{Acceptance, Trigger and Reconstruction Efficiencies of Signal}
\label{sec-eff}
\begin{figure}[htb]
 \begin{center}
 \includegraphics[width=200pt, height=200pt,angle=0]{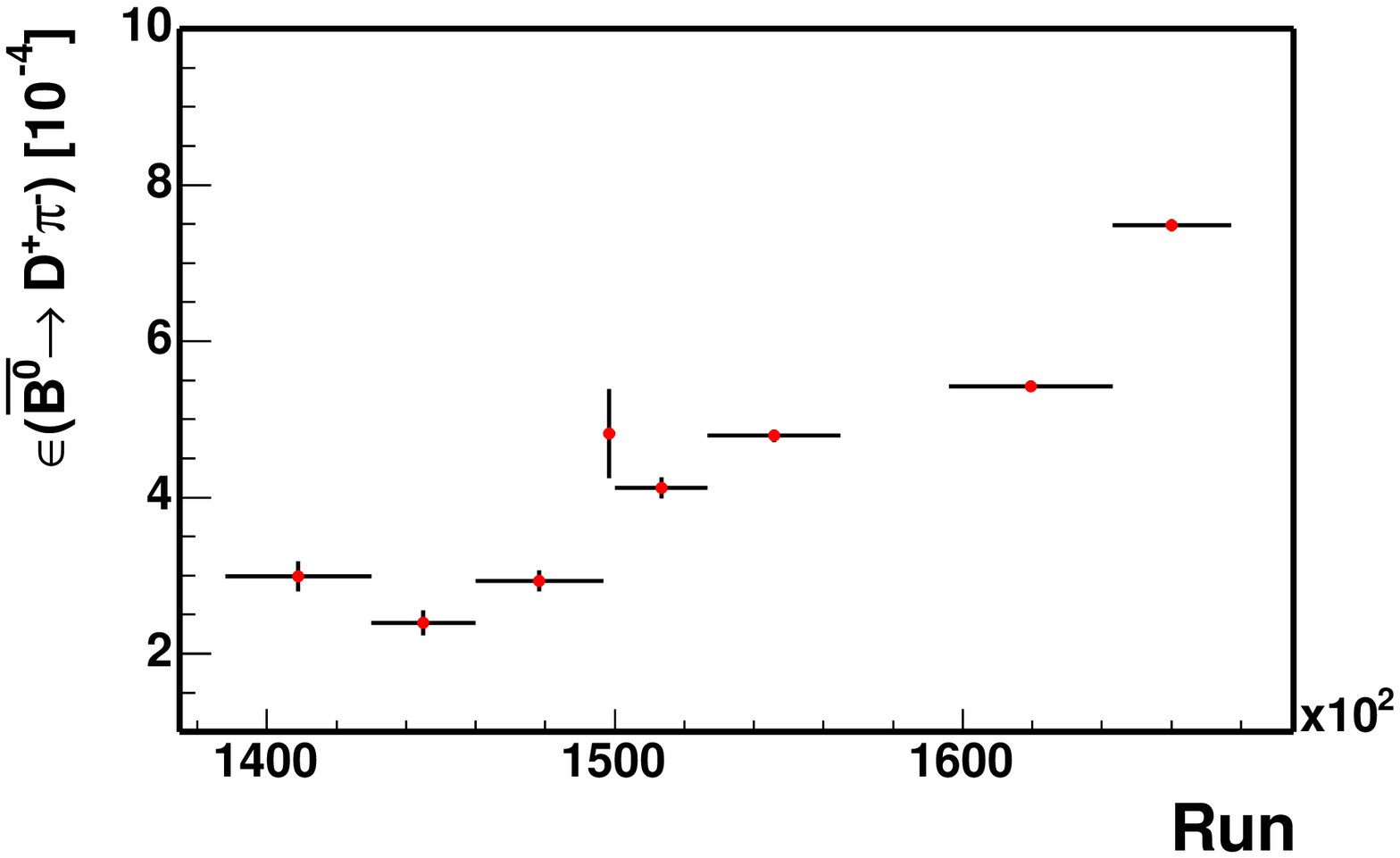}  
 \includegraphics[width=200pt, height=200pt,angle=0]{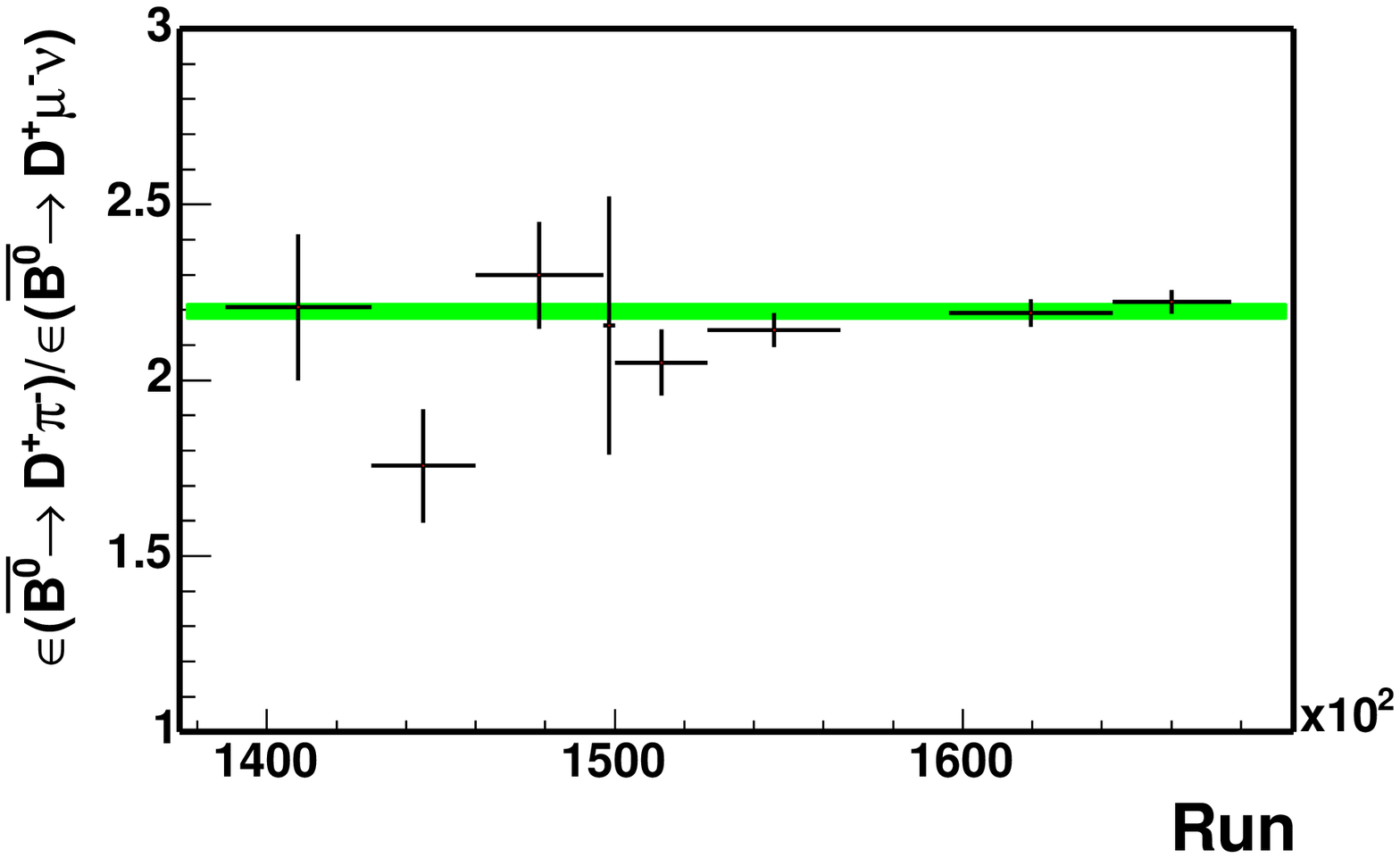}
  \caption[Total efficiency from the \dhad\ MC and the ratio to that from the 
	\dsemi\ MC]
 {\dhad\ MC total efficiency (left) and the ratio of that to the \dsemi\ MC 
	efficiency (right) in eight different hardware configurations. 
	The shaded area represents the average efficiency ratio 
	including the uncertainty.}
 \label{fig:effrun2}
 \includegraphics[width=200pt, height=200pt, angle=0]
	{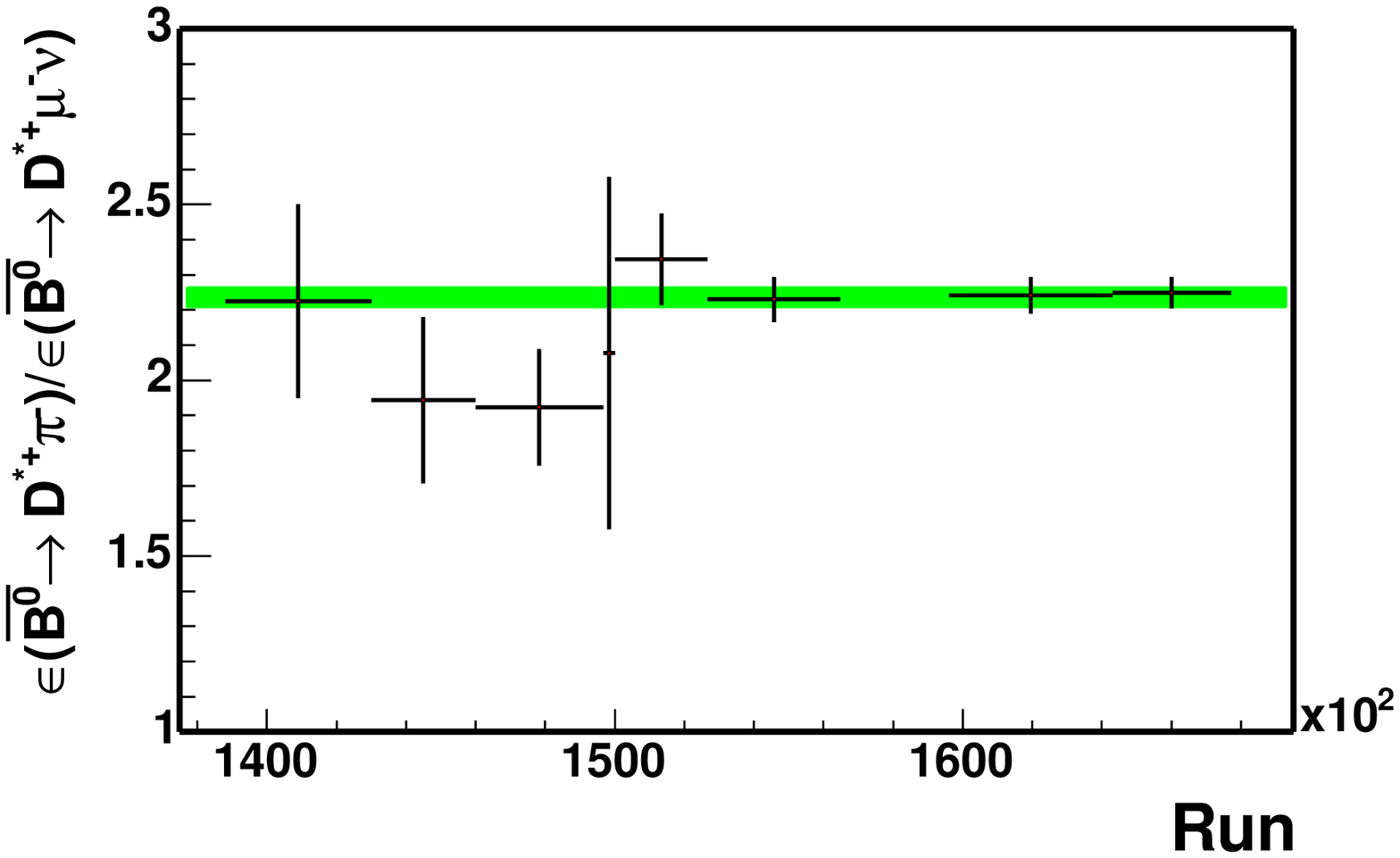}  
 \includegraphics[width=200pt, height=200pt, angle=0]{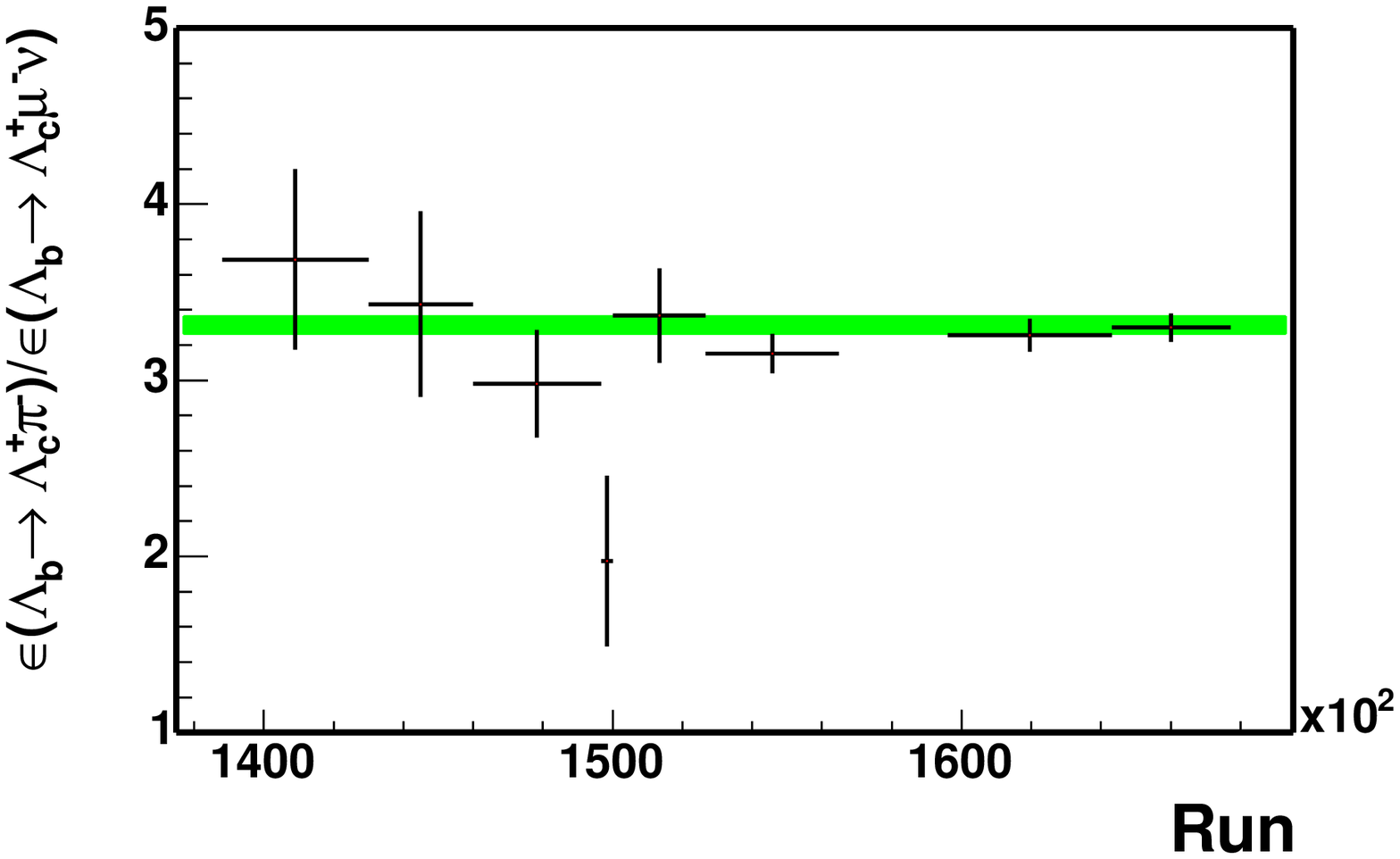}
  \caption[Total efficiency ratio of \dstarhad\ to \dstarsemi\ and 
	\lbhad\ to \lbsemi\ MC as a function of 
	run number]
 {Total efficiency ratio of \dstarhad\ to \dstarsemi\ (left) and 
	\lbhad\ to \lbsemi\ (right) MC in eight different
	hardware configurations. The shaded area represents the total average 
	efficiency ratio including the uncertainty.}
 \label{fig:effrun3}
\end{center}
 \end{figure}

We obtain the product of the acceptance, trigger and reconstruction 
efficiencies using the MC described in Section~\ref{sec-mccom}.  The total
efficiency is defined as: the number of reconstructed events surviving the 
trigger simulation and analysis cuts, divided by the total number of events 
generated. Efficiencies for the backgrounds are found in Chapter~\ref{ch:bg}. 
Our data could be divided into eight sub-periods under different trigger and 
hardware configurations, see Table~\ref{t:mcrun}. In this analysis, 
because the final states are nearly alike, we expect that the ratio of 
the efficiencies to be independent of the detector, trigger and calibration 
effects. To confirm this, we divide our signal MC samples into eight 
sub-periods and calculate their efficiencies and the ratio of hadronic to
semileptonic modes. Tables~\ref{t:effdstar}--~\ref{t:efflc} list the 
efficiencies for our signals. Figure~\ref{fig:effrun2}--\ref{fig:effrun3} 
show that the absolute efficiency may vary dramatically in 
each period, however the efficiency ratio of the hadronic to the 
semileptonic mode is quite stable. 

\begin{figure}[tbp]
 \begin{center}
  \renewcommand{\tabcolsep}{0.02in}
 \begin{tabular}{cc}
 \includegraphics[width=220pt, angle=0]{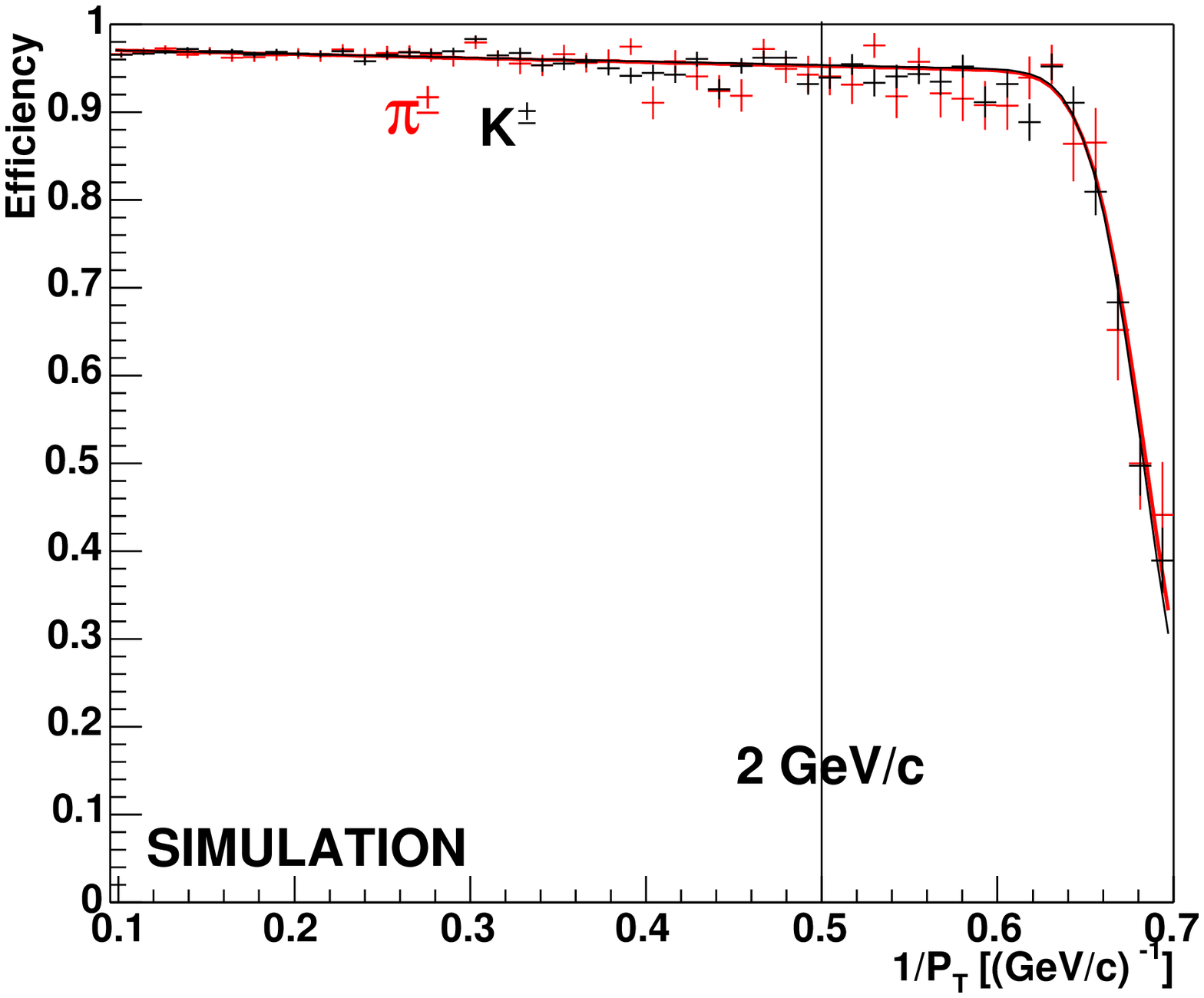} & 
 \includegraphics[width=180pt, angle=0]{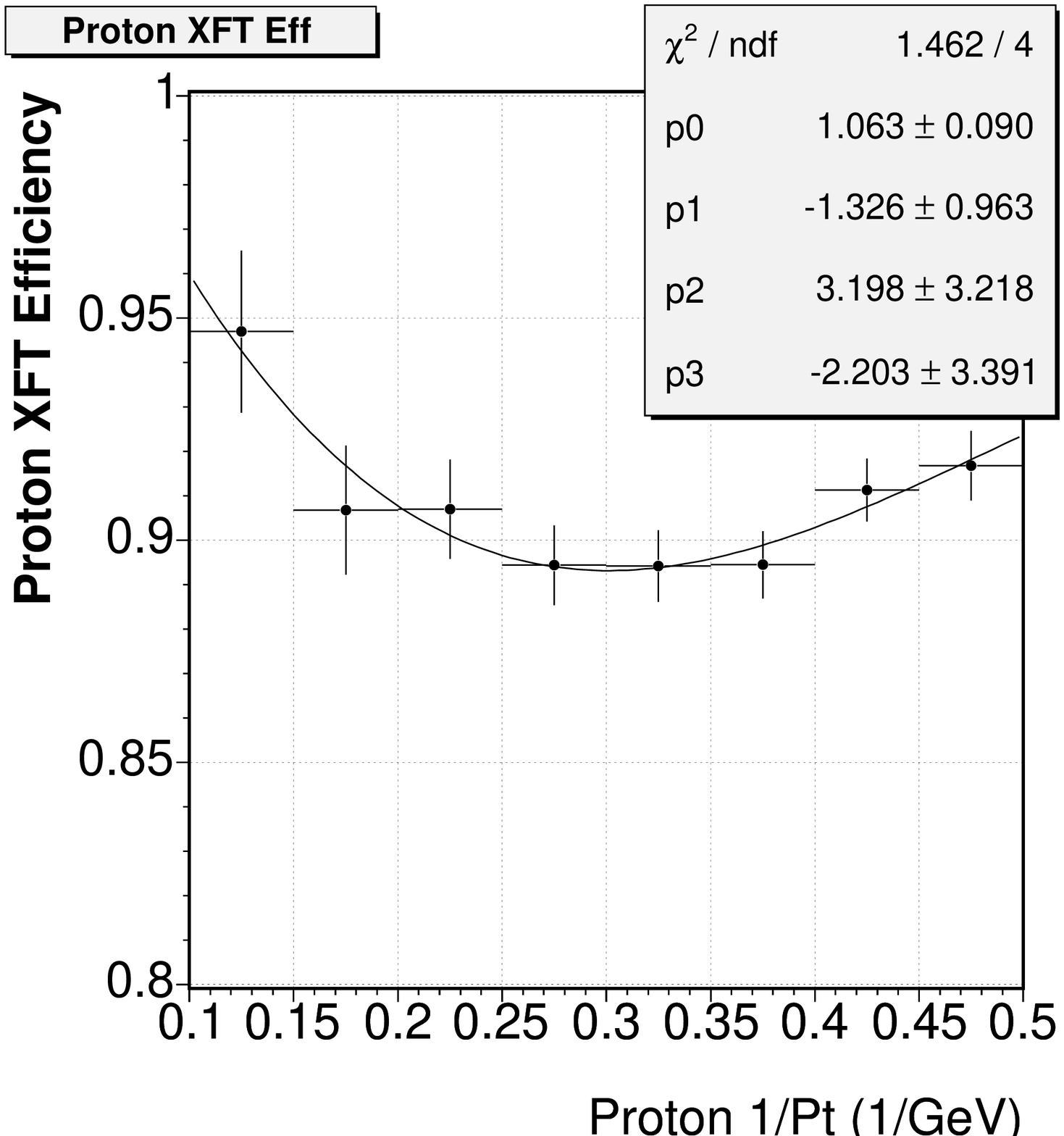} \\
 \multicolumn{2}{c}{\includegraphics[width=280pt, angle=0]{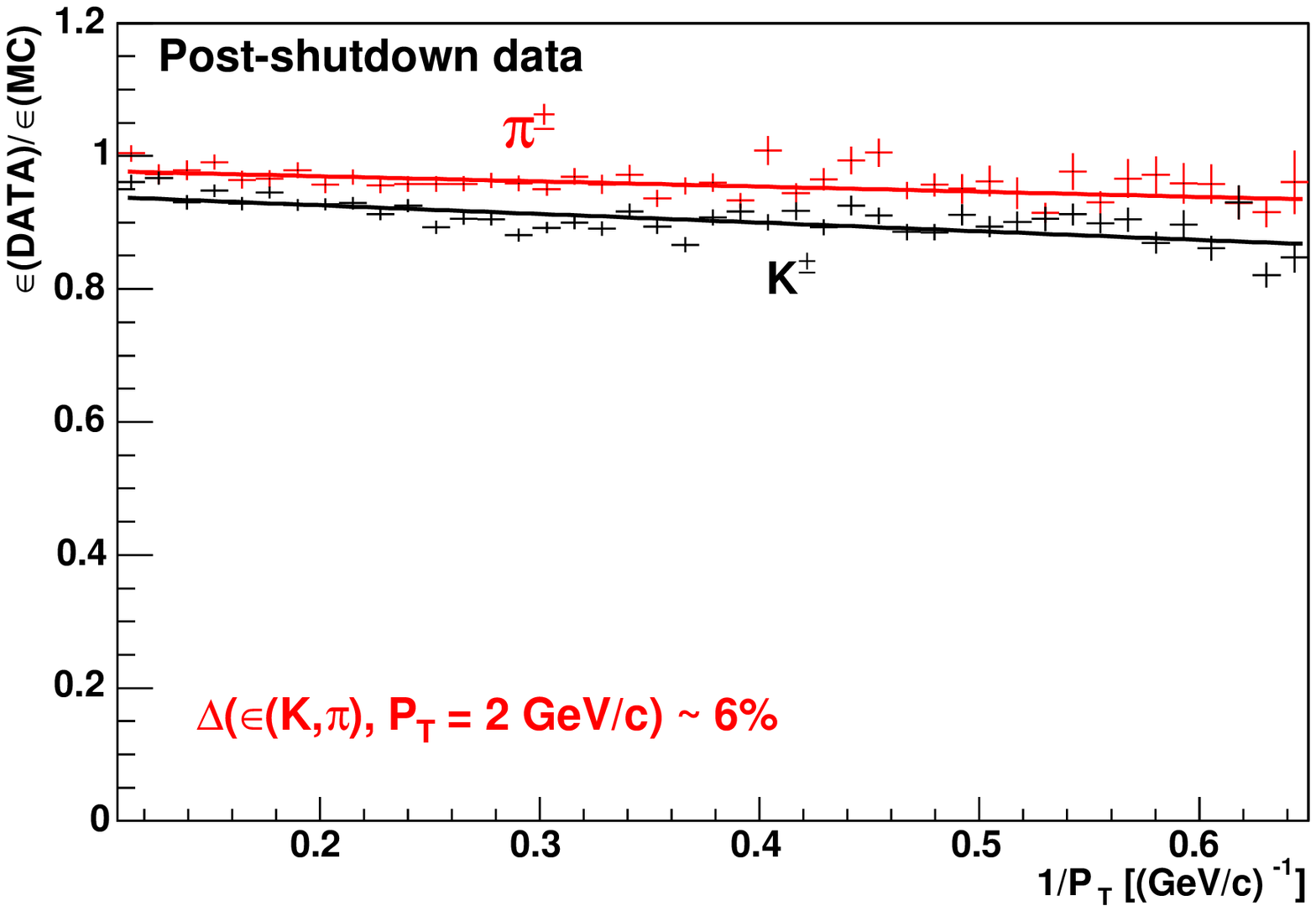}} \\  
  \end{tabular} 
  \caption[$p$, $K$ and $\pi$ XFT efficiency scaling factor and 
	  the XFT efficiency in the MC]
  { Giagu and Rescigno~\cite{cdfnote:6391} find that $K$ and $\pi$ XFT 
    efficiencies in the MC (top left) are identical. 
   The relative proton (top right) and $K$, $\pi$ (bottom) 
   XFT efficiencies between MC and data, in bins of 1/\pt, are fit 
    to a third order polynomial by Herndon~\cite{cdfnote:7301}, and to a first 
   order polynomial by Giagu and Rescigno~\cite{cdfnote:6391}, respectively. 
 \label{fig:xftpkpi}
    }
\end{center}
 \end{figure}

Note that because Bloom and Dagenhart~\cite{cdfnote:6347} find a difference in 
the CMU muon reconstruction efficiency between MC and data, we apply a scaling 
factor on the efficiencies of the semileptonic signals and backgrounds:
\(
 R = 0.986 \pm 0.003.
\)
In addition, Giagu, Herndon and 
Rescigno~\cite{cdfnote:6391}\cite{cdfnote:7301} notice that there are 
differences in the XFT efficiencies for the charged kaons, pions, and protons, 
when the XFT configuration is switched to the ``1-miss'' mode, i.e. when the 
tracking algorithm in the XFT requires at least 11 hits of 12 wires from each 
COT axial superlayer. 
The COT frond-end electronics requires a minimum input charge from the 
ionization of the incident particle. At a fixed momentum, protons and kaons 
deposit less charge than the pions, have more hits below 
the electronics threshold, and fail the stringent 
XFT ``1-miss'' requirement. Therefore, in general, 
the proton and kaon XFT efficiencies are lower than that of the pion.
Figure~\ref{fig:xftpkpi}(top) shows that kaon and pion XFT efficiencies are 
identical in the MC and need to be corrected. Giagu, Herndon and Rescigno 
measure the ratio, data/MC for the XFT efficiencies of pions, kaons and 
protons, as shown in Figure~\ref{fig:xftpkpi}. 
We reweight the MC events numerically according to the ratio:
\begin{eqnarray*}
 C_{\pi} &= &1.002 - \frac{0.067}{\pt}, \\
 C_K &= & 0.969- \frac{0.094}{\pt}, \\      
 C_p &= & 1.06 -\frac{1.3}{\pt} + \frac{3.2}{\pt^2} 
	-\frac{2.2}{\pt^3},
\end{eqnarray*}
where $\pt$ is the transverse momentum of the track that 
passes the trigger cuts in our reconstruction program. 

\begin{figure}[tbp]
\begin{center}
 \includegraphics[width=200pt,height=200pt,angle=0]{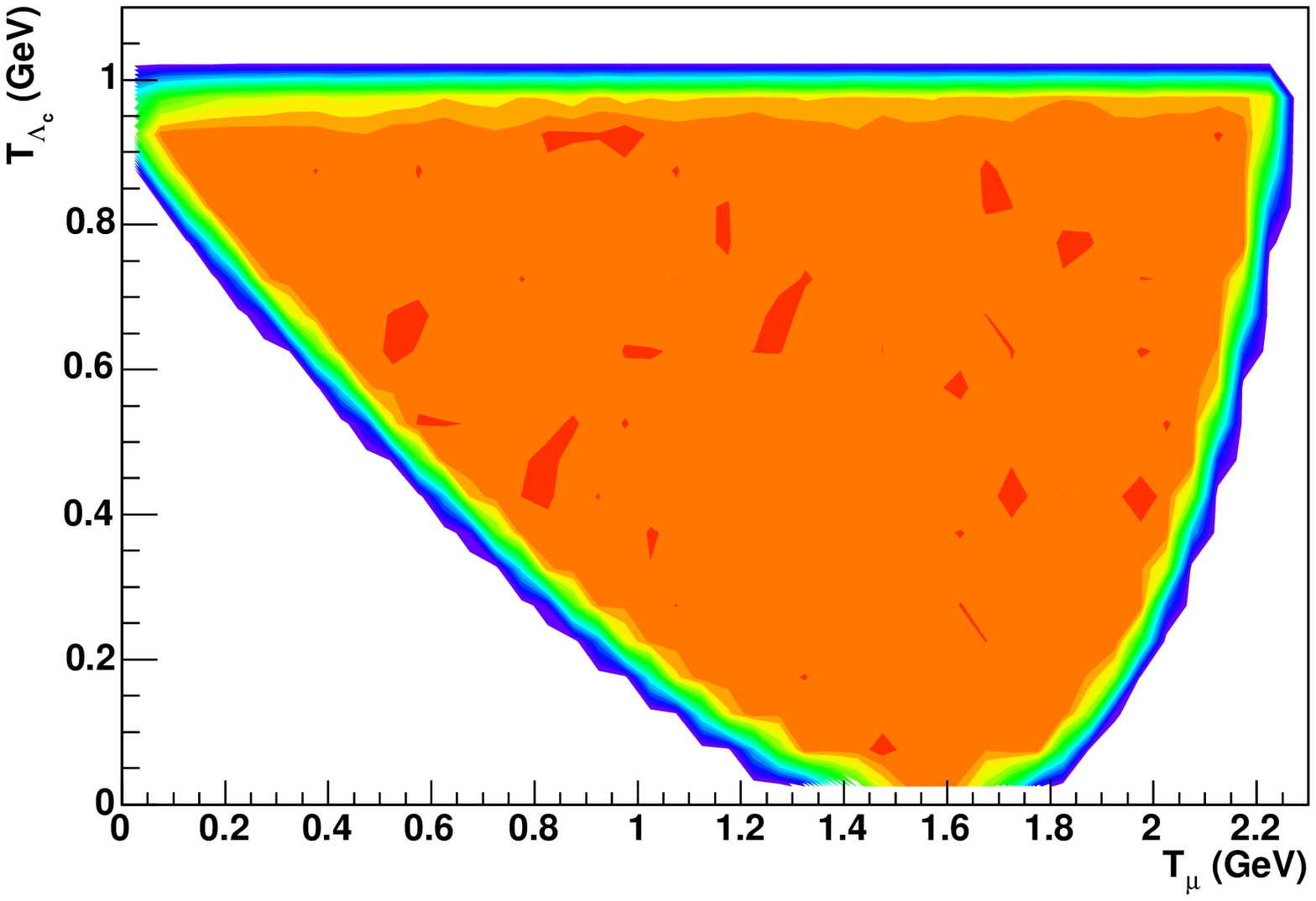}
 \includegraphics[width=200pt,height=200pt,angle=0]{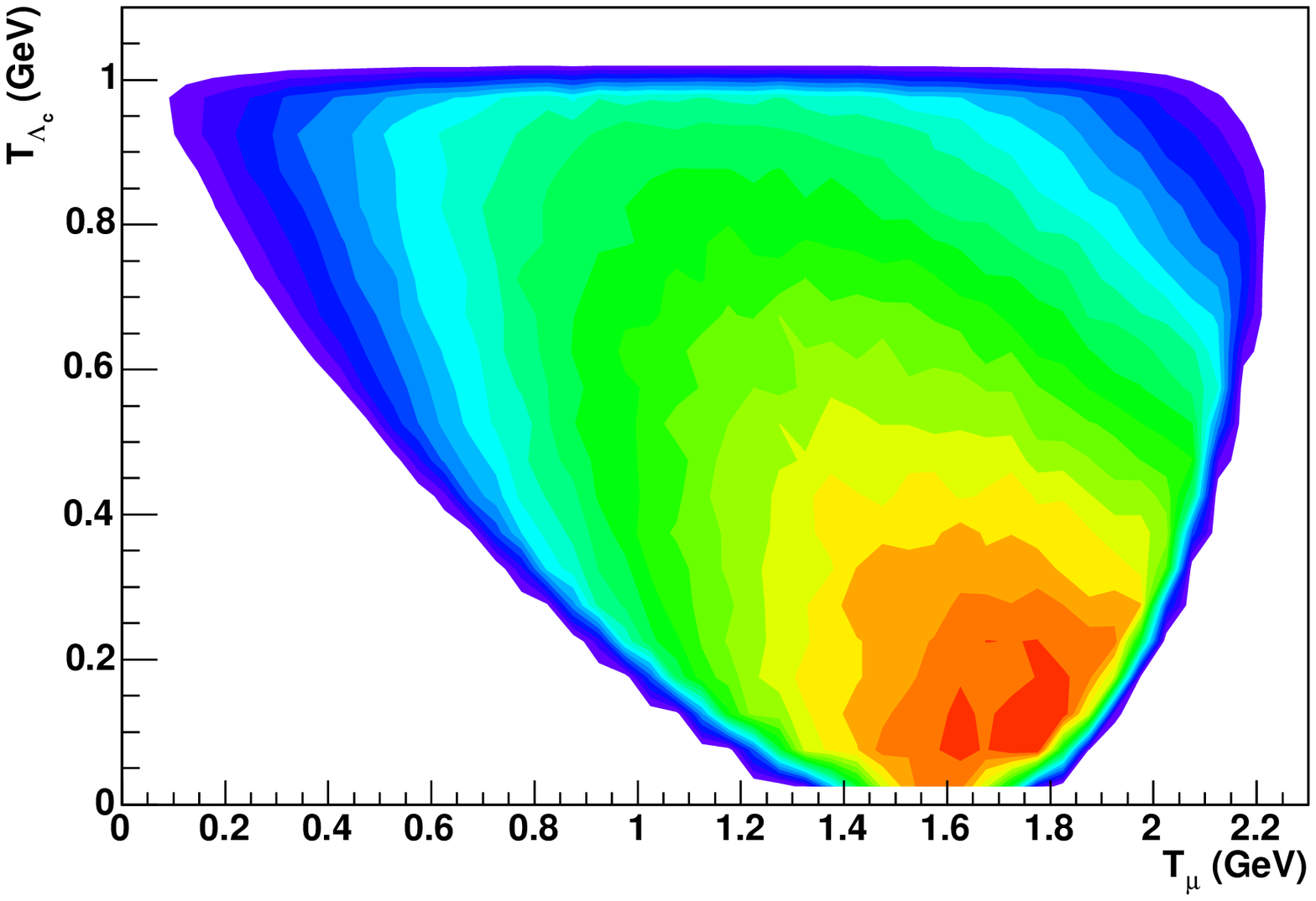}
 \includegraphics[width=200pt,height=200pt,angle=0]{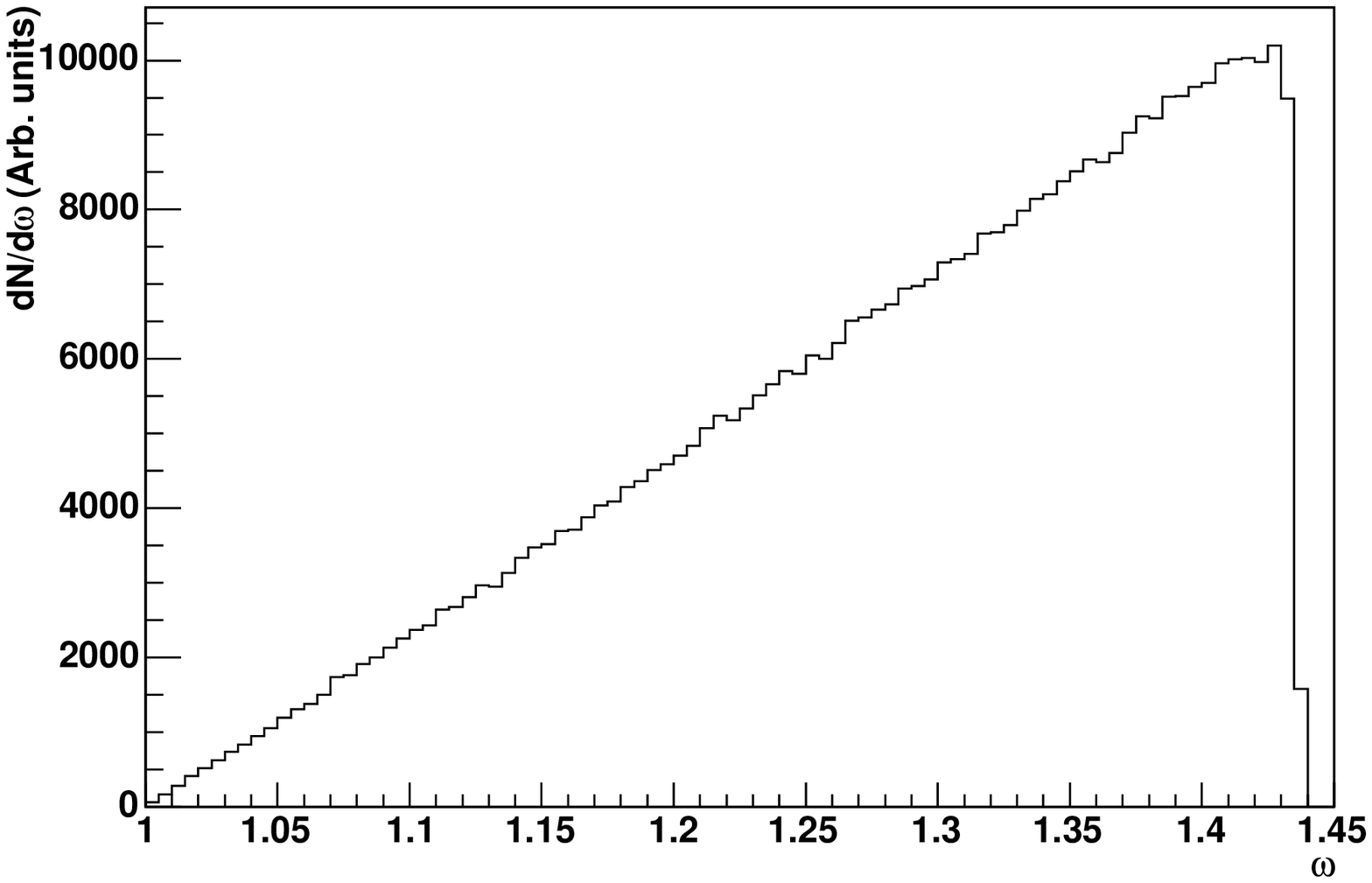}
 \includegraphics[width=200pt,height=200pt,angle=0]{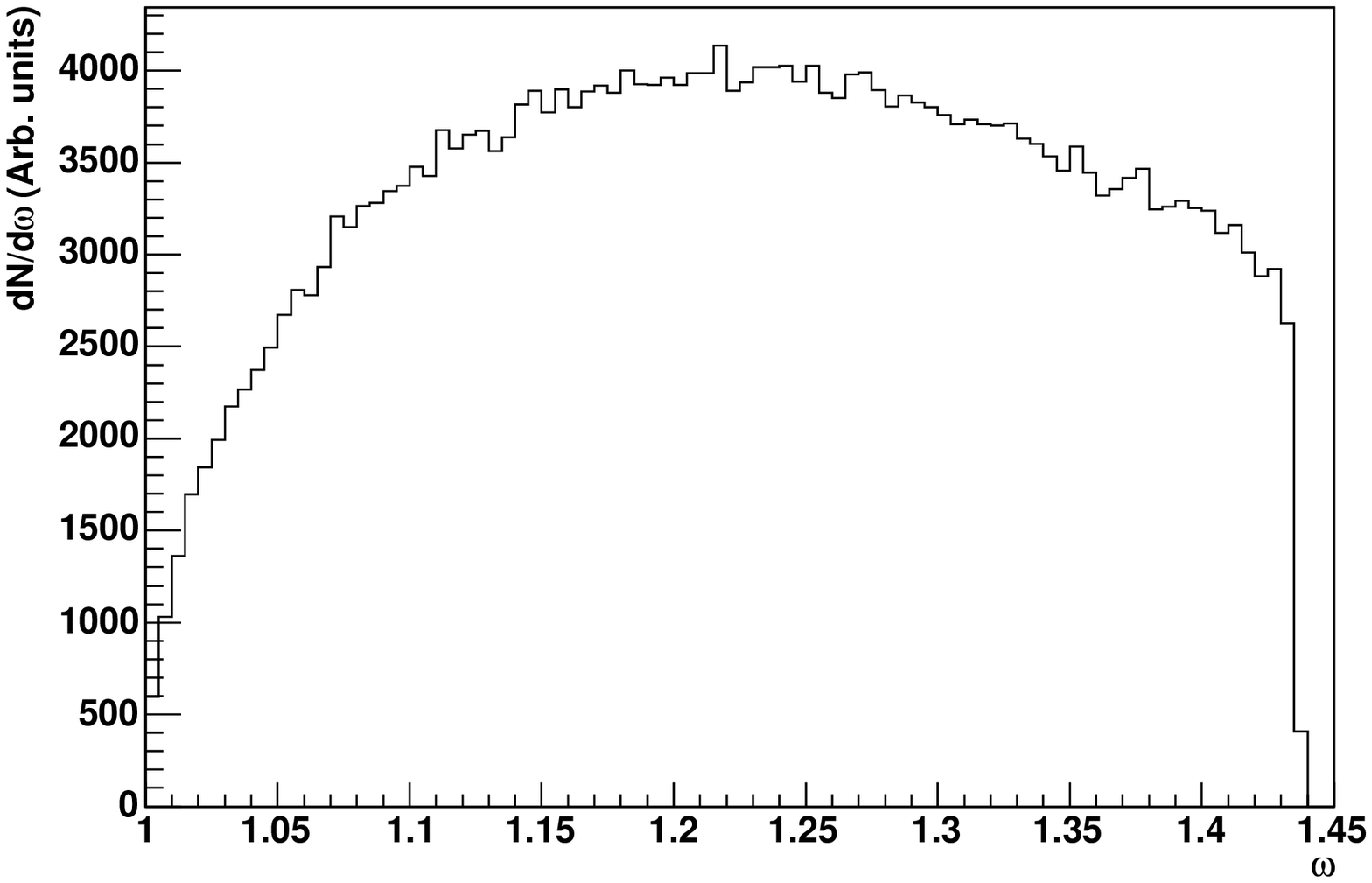}
  \caption[The phase space and the $w$ distribution from 
	the phase space MC before 
	and after reweighting according to the semileptonic form factors]
 {The phase space (top) and the $w$ (bottom) distribution obtained from the 
 \lbsemi\ phase space MC,
  before (left) and after (right) reweighting the events according to the 
  semileptonic form factors from Huang\cite{Huang:2005me}.
 \label{fig:dndw}}
\end{center}
 \end{figure}

Finally, one additional scaling factor has to be applied on all 
the \Lb\ decays with $\Lamc\mu$ in the final state. 
We have mentioned in Section~\ref{sec-mccom} that a phase space 
decay model was used for these decays. In a phase space, the 
event density in the $w$-$\cos\theta$ plane is a constant within the 
kinematic boundary. The $w$ is the scalar product of the \Lb\ and \Lamc\ 
four-velocities, and $\theta$ is the angle between the muon and the neutrino 
momentum vectors in the \Lb\ rest frame. 
The form factors that describe the strong interaction in the \Lb\ 
semileptonic decay modify the event distribution in the phase space and 
change the fraction of events accepted. 
We obtain the scaling factor in the following way: Using the 
``acceptance-rejection'' method, we reweight the generator-level 
\lbsemi\ MC according to:
 \begin{equation}
   f_c = \frac{d\Gamma(\lbsemi)}{dw}\cdot\frac{T(\cos\theta,w)}{P(w)}, 
  \label{eq:formscale}
 \end{equation}
where the differential semileptonic decay rate, \(\frac{d\Gamma(\lbsemi)}{dw}\), is obtained from Huang\cite{Huang:2005me}. The $T(\cos\theta,w)$ includes 
the $W$ spin effect and describes the correlation between the $\mu$ and $\overline{\nu}_{\mu}$, and 
\begin{equation}
   P(w) = \int_{{\cos\theta}_\mathrm{min}(w)}^{{\cos\theta}_\mathrm{max}(w)}T(\cos\theta,w)d\cos\theta.   
\end{equation}
Here, ${\cos\theta}_\mathrm{max}$ and ${\cos\theta}_\mathrm{min}$ specify 
the kinematic range and are functions of $w$. 
Figure~\ref{fig:dndw} shows the phase space and the $w$ 
distribution from the phase space and the form factor reweighted MC. 
Then, we apply generator-level analysis cuts to obtain the acceptance. 
We further divide this acceptance by that from the phase space MC and 
obtain a scaling factor of 0.994$\pm$0.025, where the uncertainty is 
dominated by the size of the MC sample.

 \begin{normalsize}
    	\begin{table}[ht]
  	\caption{Total efficiency and ratios for \dstarhad\ and \dstarsemi.}
        \label{t:effdstar}
	\begin{center}
 	\begin{tabular}{|c|r|r|r|r|} 
	\hline
        Run Range
	& $\int {\cal L}\; dt$ 
	& $\epsilon_{\dstarhad}$
	& $\epsilon_{\dstarsemi}$
        & $\epsilon$ Ratio\\ 
	& (\pbarn) & ($10^{-4}$) & ($10^{-4}$) & \\	
	\hline
	138809--143000 
	& 3.4 & 1.72 $\pm$ 0.15 & 0.77 $\pm$ 0.07 & 2.22 $\pm$ 0.28\\
	143001--146000 
	& 4.0 & 1.42 $\pm$ 0.12 & 0.73 $\pm$ 0.06 & 1.94 $\pm$ 0.24\\	
	146001--149659 
	& 7.1 & 1.57 $\pm$ 0.10 & 0.82 $\pm$ 0.05 & 1.92 $\pm$ 0.17\\
        149660--150009 
	& 0.6 & 2.35 $\pm$ 0.40 & 1.13 $\pm$ 0.19 & 2.08 $\pm$ 0.50\\
	150010--152668 
	& 11.3 & 2.65 $\pm$ 0.10 & 1.13 $\pm$ 0.05 & 2.34 $\pm$ 0.13\\
	152669--156487 
	& 39.2 & 2.74 $\pm$ 0.06 & 1.23 $\pm$ 0.03 & 2.23 $\pm$ 0.07\\
	159603--164302 
	& 52.7 & 3.10 $\pm$ 0.05 & 1.38 $\pm$ 0.02 & 2.24 $\pm$ 0.05\\
	164303--167715 
	& 52.7 & 4.27 $\pm$ 0.06 & 1.90 $\pm$ 0.03 & 2.25 $\pm$ 0.05\\
	\hline
	Total average & 171.0 & 3.22 $\pm$ 0.03 & 1.44 $\pm$ 0.01 & 2.24 $\pm$ 0.03
 \\ 
	\hline
        \hline
  	\end{tabular}
       \end{center}

  	\caption{Total efficiency and ratios for \dhad\ and \dsemi.}
        \label{t:effd}
	\begin{center}
 	\begin{tabular}{|c|r|r|r|r|} 
	\hline
        Run Range
	& $\int {\cal L}\; dt$ 
	& $\epsilon_{\dhad}$
	& $\epsilon_{\dsemi}$
        & $\epsilon$ Ratio\\
	& (\pbarn) & ($10^{-4}$) & ($10^{-4}$) & \\	
	\hline
	138809--143000 
	& 3.4 & 2.99 $\pm$ 0.19 & 1.35 $\pm$ 0.09 & 2.21 $\pm$ 0.21 \\
	143001--146000 
	& 4.0 & 2.39 $\pm$ 0.16 & 1.36 $\pm$ 0.09 & 1.76 $\pm$ 0.16\\	
	146001--149659 
	& 7.1 & 2.93 $\pm$ 0.13 & 1.28 $\pm$ 0.06 & 2.30 $\pm$ 0.15 \\
        149660--150009 
	& 0.6 & 4.82 $\pm$ 0.57 & 2.23 $\pm$ 0.27 & 2.16 $\pm$ 0.37\\
	150010--152668 
	& 11.3 & 4.12 $\pm$ 0.14 & 2.01 $\pm$ 0.06 & 2.05 $\pm$ 0.09\\
	152669--156487 
	& 39.2 & 4.79 $\pm$ 0.08 & 2.24 $\pm$ 0.04 & 2.14 $\pm$ 0.05\\
	159603--164302 
	& 52.7 & 5.43 $\pm$ 0.07 & 2.48 $\pm$ 0.03 & 2.19 $\pm$ 0.04\\
	164303--167715 
	& 52.7 & 7.49 $\pm$ 0.08 & 3.37 $\pm$ 0.04 & 2.22 $\pm$ 0.03\\
	\hline
	Total average 
	& 171.0 & 5.67 $\pm$ 0.04 & 2.58 $\pm$ 0.02 & 2.20 $\pm$ 0.02 \\
	\hline \hline
  	\end{tabular}
       \end{center}
  	\caption{Total efficiency and ratios for \lbhad\ and \lbsemi.}
        \label{t:efflc}
	\begin{center}
 	\begin{tabular}{|c|r|r|r|r|} 
	\hline
        Run Range
	& $\int {\cal L}\; dt$
	& $\epsilon_{\lbhad}$
	& $\epsilon_{\lbsemi}$
        & $\epsilon$ Ratio\\ 
	& (\pbarn) & ($10^{-4}$) & ($10^{-4}$) & \\	
	\hline
	138809--143000 
	& 3.4 & 1.84 $\pm$ 0.15 & 0.50 $\pm$ 0.06 & 3.69 $\pm$ 0.51\\
	143001--146000 
	& 4.0 & 1.23 $\pm$ 0.12 & 0.36 $\pm$ 0.04 & 3.43 $\pm$ 0.53\\	
	146001--149659 
	& 7.1 & 1.42 $\pm$ 0.09 & 0.48 $\pm$ 0.04 & 2.98 $\pm$ 0.30\\
        149660--150009 
	& 0.6 & 2.23 $\pm$ 0.39 & 1.13 $\pm$ 0.19 & 1.98 $\pm$ 0.48\\
	150010--152668 
	& 11.3 & 2.13 $\pm$ 0.12 & 0.63 $\pm$ 0.04 & 3.37 $\pm$ 0.27\\
	152669--156487 
	& 39.2 & 2.37 $\pm$ 0.05 & 0.75 $\pm$ 0.02 & 3.15 $\pm$ 0.11 \\
	159603--164302 
	& 52.7 & 2.67 $\pm$ 0.05 & 0.82 $\pm$ 0.02 & 3.26 $\pm$ 0.09 \\
	164303--167715 
	& 52.7 & 3.76 $\pm$ 0.06 & 1.14 $\pm$ 0.02 & 3.30 $\pm$ 0.08\\
	\hline
	Total average 
	& 171.0 & 2.86 $\pm$ 0.03 & 0.86 $\pm$ 0.01 & 3.31 $\pm$ 0.05\\
	\hline \hline
  	\end{tabular}
       \end{center}
	\end{table}
	\end{normalsize}

\section{Summary}
We have described the procedure of generating MC samples and compared 
the MC distributions with those in the data. In general, the MC and data 
are in good agreement. We also obtain the signal efficiencies from the MC. 
It is confirmed that the 
efficiency ratios from both the \Lb\ and \Bd\ modes are insensitive to 
the time variation of beam lines and SVT trigger configurations.

\chapter{Backgrounds of the Semileptonic Modes}
\label{ch:bg}
 Unlike the $e^+e^-$ collider experiments, {\tt BELLE}~\cite{belle:tdr} 
and {\tt BABAR}~\cite{babar:tdr}, the initial energies of the \B\ hadrons are 
unknown in CDF. For the hadronic channels, such as \dhad, the \B\ meson is 
fully reconstructed using the momenta of the daughter particles (\D\ and 
$\pi^-$) at the point of the parent decay. However, for the semileptonic 
channels, such as \dsemi, without the information of initial energies, it is 
difficult to constrain the momenta of the missing neutrinos and fully 
reconstruct the \B\ candidates. Therefore, the $D^*\mu$, $D\mu$ and 
$\Lambda_c\mu$ combinations we observe in the data consist of the exclusive 
semileptonic signals, \dstarsemi, \dsemi, \lbsemi, in the presence of other 
backgrounds. These backgrounds arise from three sources:   
\begin{itemize}
\item physics backgrounds: \B\ hadron decays into similar final states, 
a charm hadron, a real muon and other tracks.
\item muon fakes: a charm hadron and a track which fakes a muon.
\item \bb\ and \cc: two \B\ or charm hadrons from the \bb\ and \cc\ pairs 
decay into a \Dstar\ (\D, \Lc) and a muon, respectively.
\end{itemize}
The goal is to measure the relative partial decay widths of the exclusive 
semileptonic decays to hadronic decays. The backgrounds listed above should be 
subtracted from the observed inclusive semileptonic signal in the data. 
The ratio of branching fractions is then calculated as follows:
\begin{equation}
\label{eq:bigf}
\frac{{\cal B}_\mathrm{semi}}{{\cal B}_\mathrm{had}} = 
(\frac{N_\mathrm{inclusive\;semi}-N_\mathrm{physics}- N_\mathrm{fake\mu}
-N_{c\overline{c},\;b\overline{b}}}{N_\mathrm{had}})\cdot 
	\frac{\epsilon_\mathrm{had}}{\epsilon_\mathrm{semi}}, 
\end{equation}
where ${\cal B}$ stands for the branching ratio, $\epsilon$ is the efficiency 
from the MC. We estimate the contribution from the physics and \bb, \cc\ 
backgrounds, using the efficiencies from the MC and the branching ratios from 
the Particle Data Group (PDG)~\cite{pdg:2004}. We normalize the backgrounds to 
the observed number of events in the fully reconstructed hadronic signal 
in the data,
\begin{equation}
\label{eq:bgexample}
N_{\mathrm{physics}\;(\bb,\cc)} =  N_\mathrm{had}\cdot \sum 
\frac{ {\cal B}_{i}\cdot \epsilon_i}{{\cal B}_\mathrm{had} \cdot 
       \epsilon_\mathrm{had}}.
\end{equation}
Substituting Equation~\ref{eq:bgexample} into Equation~\ref{eq:bigf}, 
$N_\mathrm{had}$ cancels. The estimate of \bb, \cc\ and physics background 
contributions is then free from the uncertainties in the hadronic yields. 
In the case of fake muons, we subtract the fake muon candidates measured 
in the data directly. 

The branching ratios of \Bd\ hadronic modes in Equation~\ref{eq:bgexample} 
come from the world average in the PDG. For the \Lb\ mode, we extract 
${\cal B}(\lbhad)$ from the recent CDF result, \yile\ by Le, 
\etal~\cite{yile:lblcpi}.  Because Le's measurement requires the ratio of 
the \Lb\ to \Bd\ production cross sections as an input, we correct for the 
ratio in Section~\ref{sec-lbxsec} and obtain ${\cal B}(\lbhad)$.
Sections~\ref{sec-physicsb}--~\ref{sec-ccbb} estimate the 
amount of backgrounds in the semileptonic signal. We will show that the 
dominant signal contamination is from the physics background.  The second 
largest background arises from muon fakes. The smallest background source is 
from \bb\ and \cc. 

\section{Correction of \rxsec\ and ${\cal B}(\lbhad)$}
\label{sec-lbxsec}
\begin{figure}[tbp]
  \begin{center}
\renewcommand{\tabcolsep}{0.05in}
  \begin{tabular}{cc}
  \includegraphics[width=200pt, angle=0]{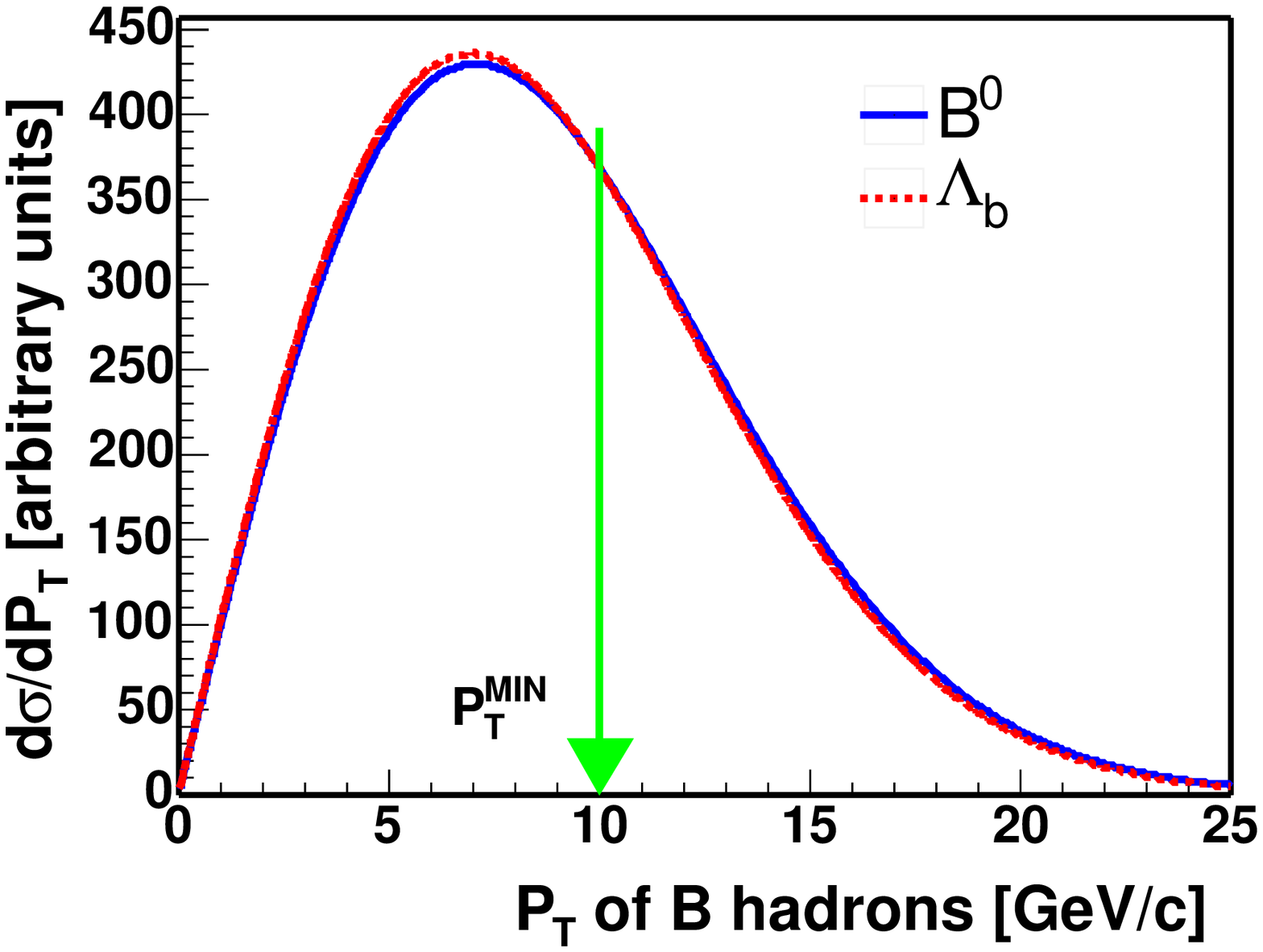} &
  \includegraphics[width=200pt, angle=0]{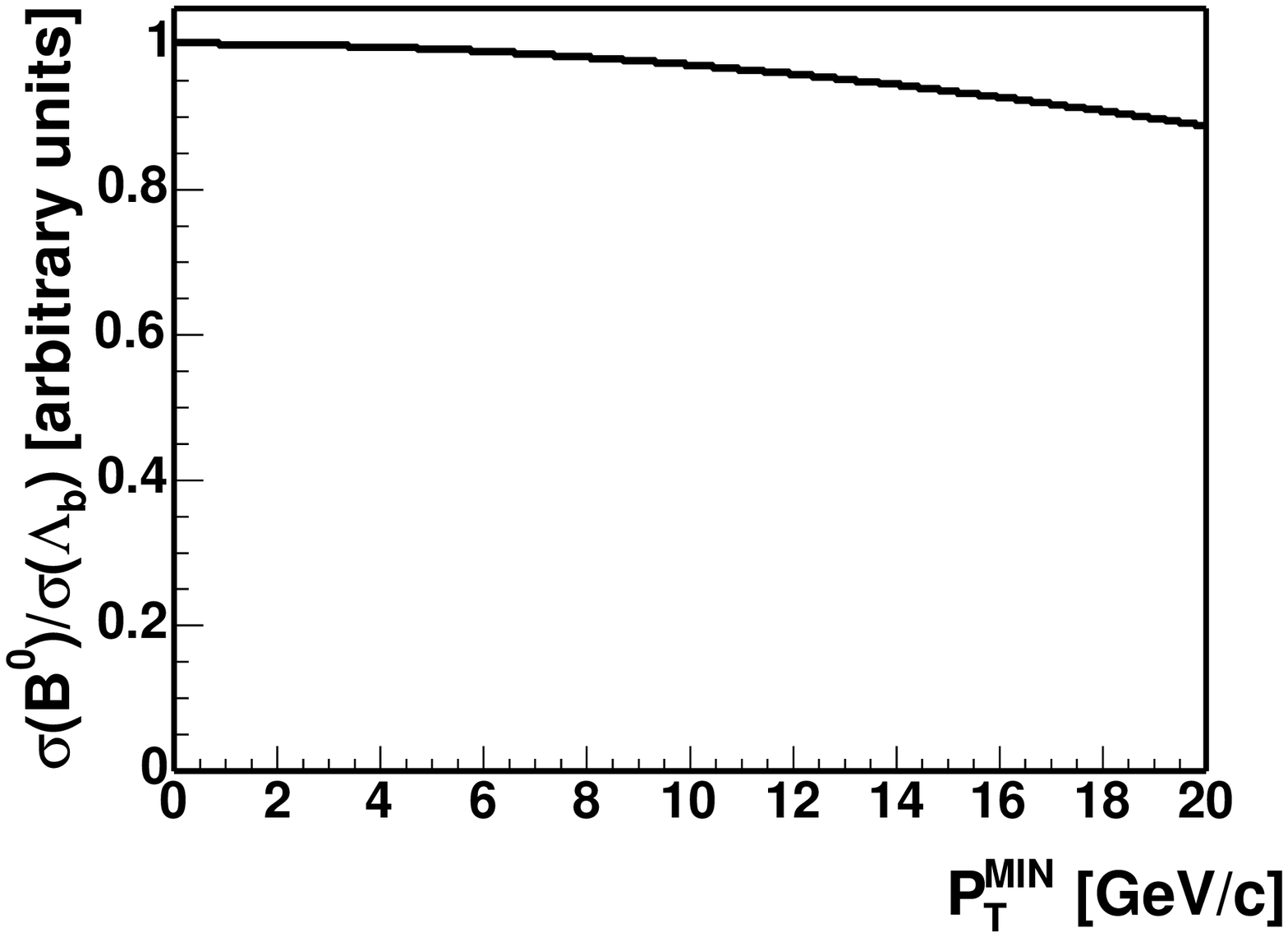} \\
  \includegraphics[width=200pt, angle=0]{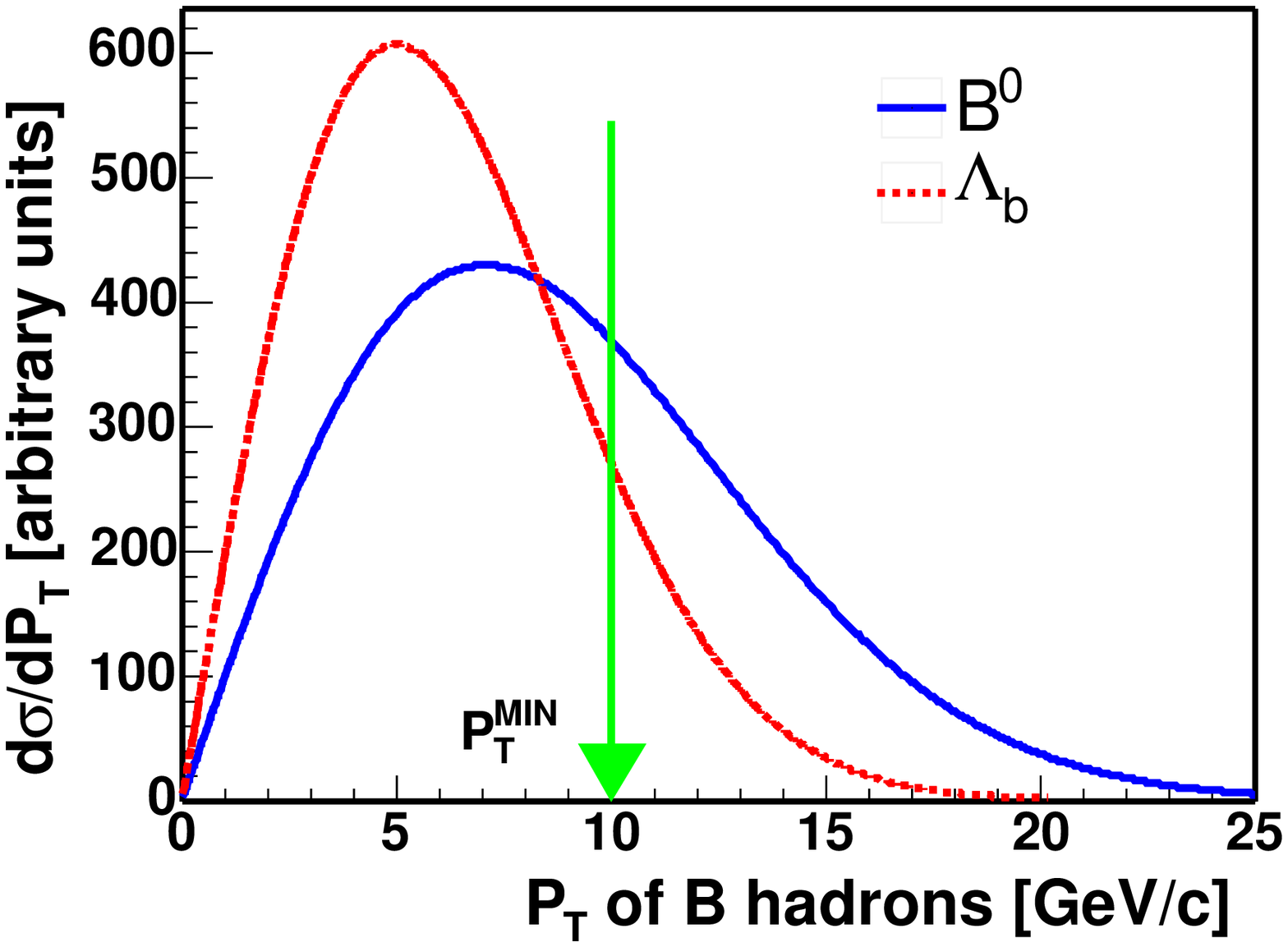} &
  \includegraphics[width=200pt, angle=0]{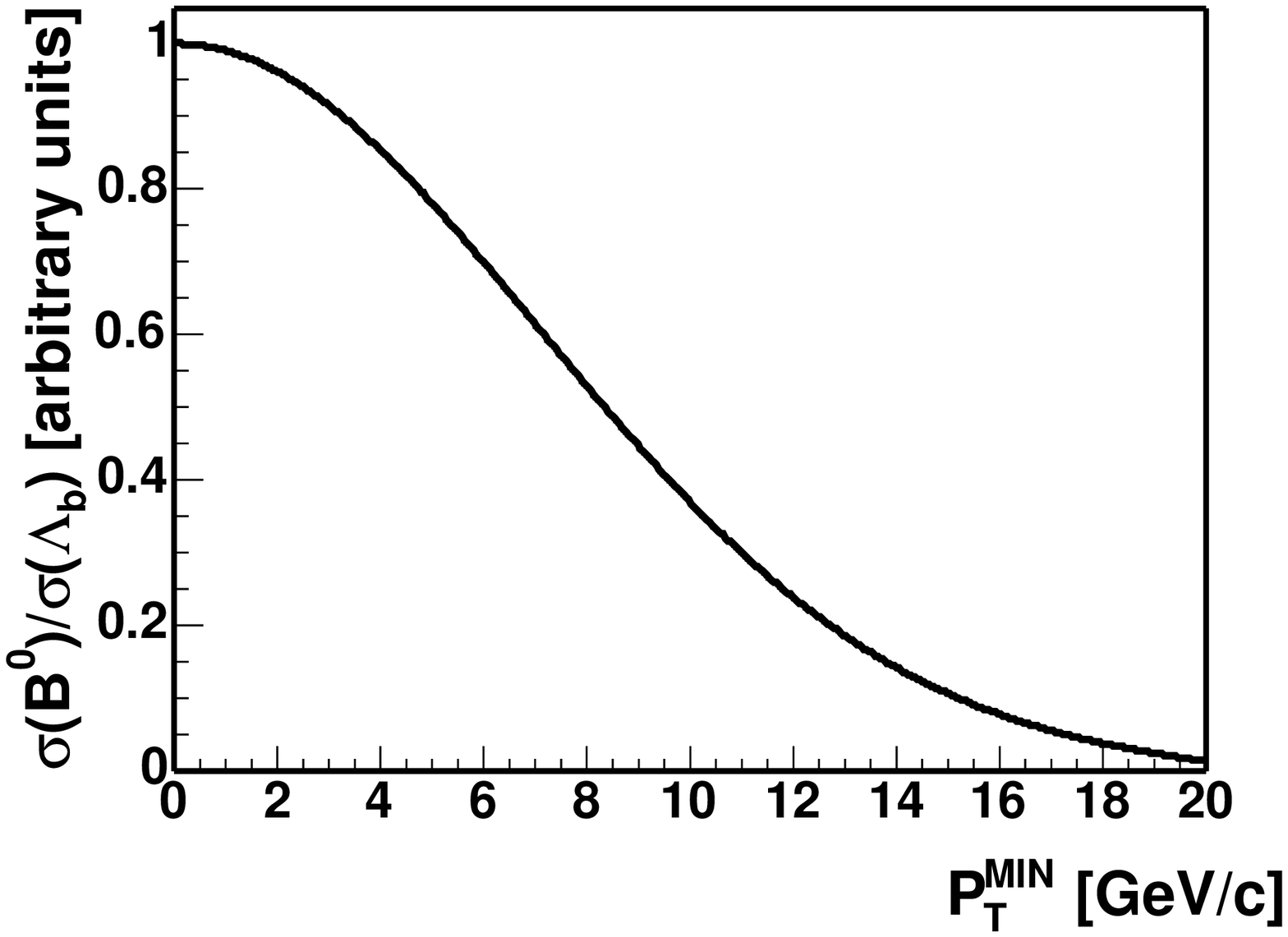} \\
  \end{tabular}
 \caption[Dependence of the Production ratio on the \B\ hadron \pt\ threshold]
 {\pt\ spectrum of \Bd\ and \Lb\ (left) and the dependence of the production 
  ratio on the \pt\ threshold, $P_T^{\mathrm{MIN}}$ (right). 
  The top figures show the case
  where both hadrons have an almost identical spectrum. The ratio of the 
  integrated areas underneath the spectrum, from $\pt^{\mathrm{MIN}}$
  and above, does not depend much on the value of $\pt^{\mathrm{MIN}}$. 
  The bottom figure shows a softer \Lb\ \pt\ spectrum. The ratio of 
  the integrated areas will depend strongly on the value of 
  $\pt^{\mathrm{MIN}}$.}
 \label{fig:teachxsec} 
 \end{center}
 \end{figure}

CDF has made a number of measurements of the relating information from the 
decay of \B-hadron to that of another with similar topology, such as
\begin{center}
\yile~\cite{yile:lblcpi} or \tomo~\cite{cdfnote:6643}. 
\end{center}
In order to extract the individual branching fraction, one must have the 
knowledge of the production cross section. Since no production cross section 
measurements exist for \Lb, one must infer the cross-section using other means.
 Using the total $b$-quark cross-section~\cite{bishai:bxsec}, the 
fragmentation fraction ($f_u$, $f_d$, $f_s$ and $f_\mathrm{baryon}$), 
and assuming the composition of the $b$-baryons is dominated by \Lb, i.e.  
$f_{\Lb}\cong f_\mathrm{baryon}$, we may infer the production cross-section of
 any \B-hadron species. 
However, the fragmentation fractions assume the entire \pt\ spectrum is 
measured for both the particles (\Bd\ and \Lb) in the above ratios. If the 
\pt\ spectrum is incorrect, the kinematic acceptance used in the above ratios 
will be over- or under-estimated. In addition, since most analyses at CDF 
require a \pt\ threshold to improve the signal to background ratio, any 
difference in the \pt\ spectrum between the particles participating in the 
ratio will modify the effective fragmentation fraction. 
Figure~\ref{fig:teachxsec} illustrates this effect for a very small difference 
in the spectrum (top) and a large difference in the spectrum (bottom). 
In this section, we derive corrections for the production cross sections 
\rxsec\ to account for their different \pt\ spectra and overestimated 
kinematic acceptance from the previous measurements.

To simplify the notation, 
we define the relative production cross sections of \Lb\ to \Bd\ as $\rho$, 
 \begin{equation}
\rho(x) \equiv 
\frac{\sigma_{\Lb}(\pt\ > x\; \gevc)}{\sigma_{\Bd}(\pt\ > x\; \gevc)}, 
 \end{equation}
where $x$ is the \pt\ threshold of the \B\ hadrons in \gevc. 
 When estimating the backgrounds for our semileptonic signals, we choose
to use the hadronic signals as the normalization because they are fully 
reconstructed, and free from indistinguishable physics backgrounds.
Efficiencies and branching ratios of the hadronic modes are required when
performing the normalization, see Equation~\ref{eq:bgexample}. We use Monte
Carlo to calculate the efficiencies and external measurements of hadronic 
decays for the branching ratios. While precise measurements of 
${\cal B}(\dstarhad)$ and ${\cal B}(\dhad)$ exist~\cite{pdg:2004}, there is 
not yet a direct measurement of ${\cal B}(\lbhad)$. The CDF Run II measurement 
of \yileab\ by Le,~\etal~\cite{yile:lblcpi} provides an input but requires a 
good understanding of $\rho(6)$. In addition, we need to know $\rho(4)$ to 
normalize the \B\ meson background, $B^{0,+}\rightarrow \Lc \mu X$ in the 
inclusive $\Lamc\mu$ signal, to \lbhad,
  \begin{equation}
 \frac{N_{B^{0,+}\rightarrow \Lc \mu X}}{N_{\lbhad}} = \frac{1}{\rho(4)}\cdot 
 \frac{{\cal B}_{B^{0,+}\rightarrow \Lc \mu X}\cdot 
  \epsilon_{B^{0,+}\rightarrow \Lc \mu X}}
  {{\cal B}_{\lbhad} \cdot \epsilon_{\lbhad}}.
 \end{equation}
A lower minimum \pt\ requirement of 4 \gevc\ is applied here to cover the 
entire kinematic range of the B hadron semileptonic decays, see 
Section~\ref{sec-lcphysicsb} for more details.

The fragmentation fractions $f_d$, $f_u$, $f_s$ and $f_\mathrm{baryon}$ in the 
PDG are defined as the probability for a $b$ quark to hadronize into 
\Bd, $B^+$, $B_s^0$ or baryons,
\begin{eqnarray}
 f_d & \equiv & {\cal B}(\overline{b}\rightarrow \Bd), \\
 f_u & \equiv & {\cal B}(\overline{b}\rightarrow B^+), \\
 f_s & \equiv & {\cal B}(\overline{b}\rightarrow B_s^0), \\
 f_\mathrm{baryon} & \equiv & {\cal B}(\overline{b}\rightarrow 
	\mathrm{ b-baryon}), 
\end{eqnarray}
with the assumption that
\begin{equation}
 f_d + f_u + f_s + f_\mathrm{baryon} = 1.
\end{equation}
The PDG~\cite{Abbaneo:2001bv} determines $f_\mathrm{baryon}$ by combining the 
measurements of mean lifetimes, \Bd\ mixing parameters and branching ratios 
by LEP, SLD and Taylor in CDF-I~\cite{Affolder:1999iq}. 
Nevertheless, it is still questionable whether the $f_\mathrm{baryon}$ should 
be the same for the LEP and the Tevatron experiments. 
As the collider and the detector environments for 
Taylor's analysis are most similar to those for this analysis and 
we have access to the details of the measurement, 
we discuss the corrections for Taylor's $f_\mathrm{baryon}/f_{d}$ result, 
0.236 $\pm$ 0.084.

\subsubsection{Correcting the Kinematic Acceptance}
Taylor's analysis uses electron-charm final states, such as \incdstarsemie, 
\incdsemie, and \inclbsemie\ to measure $f_\mathrm{baryon}$.  
Accurate \Lb\ and \Bd\ \pt\ spectra from the fully reconstructed decays were 
not available at the time. The MC samples~\cite{taylor:thesis} in Taylor's 
analysis for calculating the acceptance and efficiencies were generated 
using the default settings in the package \bgen\ as described in 
Section~\ref{sec-mccom}. We have shown in Section~\ref{sec-datamc} 
that the \Lb\ and \Bd\ \pt\ spectra from the \bgen\ are stiffer than those 
measured in the data. This leads to an over-estimate of the acceptance and 
efficiencies, particularly for the \Lb\ decays. One must correct for this 
effect first.

For technical reasons, it is impossible to repeat a full CDF Run I 
detector simulation. Consequently, we obtain the correction using 
generator-level MC. We generate \Bd\ and \Lb\ with the default \bgen\ 
\pt\ spectra as described in Section~\ref{sec-mccom}. Then we decay \Bd\ and 
\Lb\ using the {\tt QQ} software package~\cite{cleo:qq} as in Taylor's 
analysis. We apply the cuts listed in Table~\ref{t:taylorcut} on the generator 
level quantities. These cuts mimic those applied in Taylor's analysis as much 
as possible. We divide the number of events which pass the cuts by the number 
of events with \Bd\ (\Lb) \pt\ $>$ 10 \gevc\ and rapidity $|y|$ $<$ 2.0, to 
obtain the acceptance and efficiencies for the exclusive semileptonic decays: 
\dsemie, and \lbsemie. 
 We repeat the same process using the corrected \Bd\ and \Lb\ \pt\ spectra as 
described in Section~\ref{sec-mccom}, except that we use {\tt QQ} to decay
\Lb\ and \Bd\ to be consistent with Taylor's analysis. The production 
cross-section ratio derived from the Taylor's analysis 
(\(\rxsec^\mathrm{Taylor}\)) could be expressed as a ratio of the number of 
signal events divided by the ratio of the product of the branching ratio and 
efficiency for the \lbsemie\ mode, to the same expression for the \dsemie\ 
mode:
\begin{equation}
\rxsec^\mathrm{Taylor}= N_R \cdot {\cal B}_R \cdot \epsilon_R^\mathrm{Taylor},
\end{equation}
where we use the following shorthand notation:
\begin{eqnarray}
\label{eq:correffnotation}
N_R & = & \frac{N_{\lbsemie}}{N_{\dsemie}},\\
{\cal B}_R & = & \frac{{\cal B}(\dsemie)}{{\cal B}(\lbsemie)},\\
\epsilon_R & = & \frac{\epsilon_{\dsemie}}{\epsilon_{\lbsemie}}. 
\end{eqnarray}
Using the same notation, we could express the corrected ratio as:
\begin{equation}
\rxsec^\mathrm{corrected}= N_R \cdot {\cal B}_R \cdot 
\epsilon_R^\mathrm{corrected} = \rxsec^\mathrm{Taylor} \cdot C_{\epsilon} 
= \frac{f_d}{f_\mathrm{baryon}} \cdot C_{\epsilon} 
\end{equation}
where
\begin{equation}
 C_{\epsilon} =  \epsilon_R^\mathrm{corrected}/\epsilon_R^\mathrm{Taylor}. 
\label{eq:cepsilon}
\end{equation}
This gives us the first correction factor, 
$C_{\epsilon}$=1.81 $\pm$ 0.04 ({\em stat}) ${+0.42 \atop -0.22}$ ({\em \pt}).
 The last uncertainty comes from the uncertainty of the exponential 
slope used to obtain the corrected \Lb\ \pt\ spectrum as described in 
Section~\ref{sec-mccom}. Table~\ref{t:taylorcut} also gives the 
value of $C_{\epsilon}$ after each cut is applied. Note that using the 
generator level MC, we do not obtain the reconstruction efficiencies for 
Taylor's analysis. However, the reconstruction efficiency is 
the same for both \(\rxsec^\mathrm{Taylor}\) and \(\rxsec^\mathrm{corrected}\) 
and cancels in Equation~\ref{eq:cepsilon}.

\renewcommand{\arraystretch}{1.2}
 \begin{table}[tbp]
 \caption{Generator-level cuts as Taylor's analysis.}
  \label{t:taylorcut}
   \begin{normalsize}
   \begin{center}
   \setdec 0.000
   \begin{tabular}{|l|l|r@{\,$\pm$\,}r|} 
   \hline 
   Parameter & Cut Value & \multicolumn{2}{|l|}{$C_{\epsilon}$} \\ 
   \hline
   \hline 
   $\pt(e)$ & $>$ 7 \gevc & \dec 1.206 & \dec 0.006 \\
   Transverse energy $E_T(e)$ & $>$ 8 \gevcsq & \dec 1.332 & \dec 0.009 \\
   $|\eta(e)|$ & $<$ 1.1  & \dec 1.343 & \dec 0.012 \\
   \hline 
   all tracks \pt\ & $>$ 0.4 \gevc & \dec 1.420 & \dec 0.015 \\
   daughters of charm $|d_0|/\sigma_{d_0}$ & $>$ 1.5 
	& \dec 1.428 & \dec 0.015 \\
   $\pt(\pi)$ & $>$ 0.5 \gevc & \dec 1.428 & \dec 0.015 \\
   $\pt(K)$ & $>$ 1.2 \gevc & \dec 1.588 & \dec 0.020 \\
   $\pt(p)$ & $>$ 2.0 \gevc & \dec 1.808 & \dec 0.026 \\
   \hline 
   $\lxy(D,\Lamc)/\sigma_{\lxy}$ & $>$ 1 & \dec 1.777 & \dec 0.037 \\
   $M(De)$ & $<$ 5.0 \gevcsq & \dec 1.777 & \dec 0.037 \\ 
   $M(\Lamc e)$ & $<$ 5.3 \gevcsq & \dec 1.805 & \dec 0.039 \\
   \hline \hline 
   \end{tabular}
   \end{center}
   \end{normalsize}   
 \end{table}

\subsubsection{Correction due to the Difference in the \pt\ Threshold}
\begin{figure}[tbp]
  \begin{center}
\includegraphics[width=180pt, angle=0]{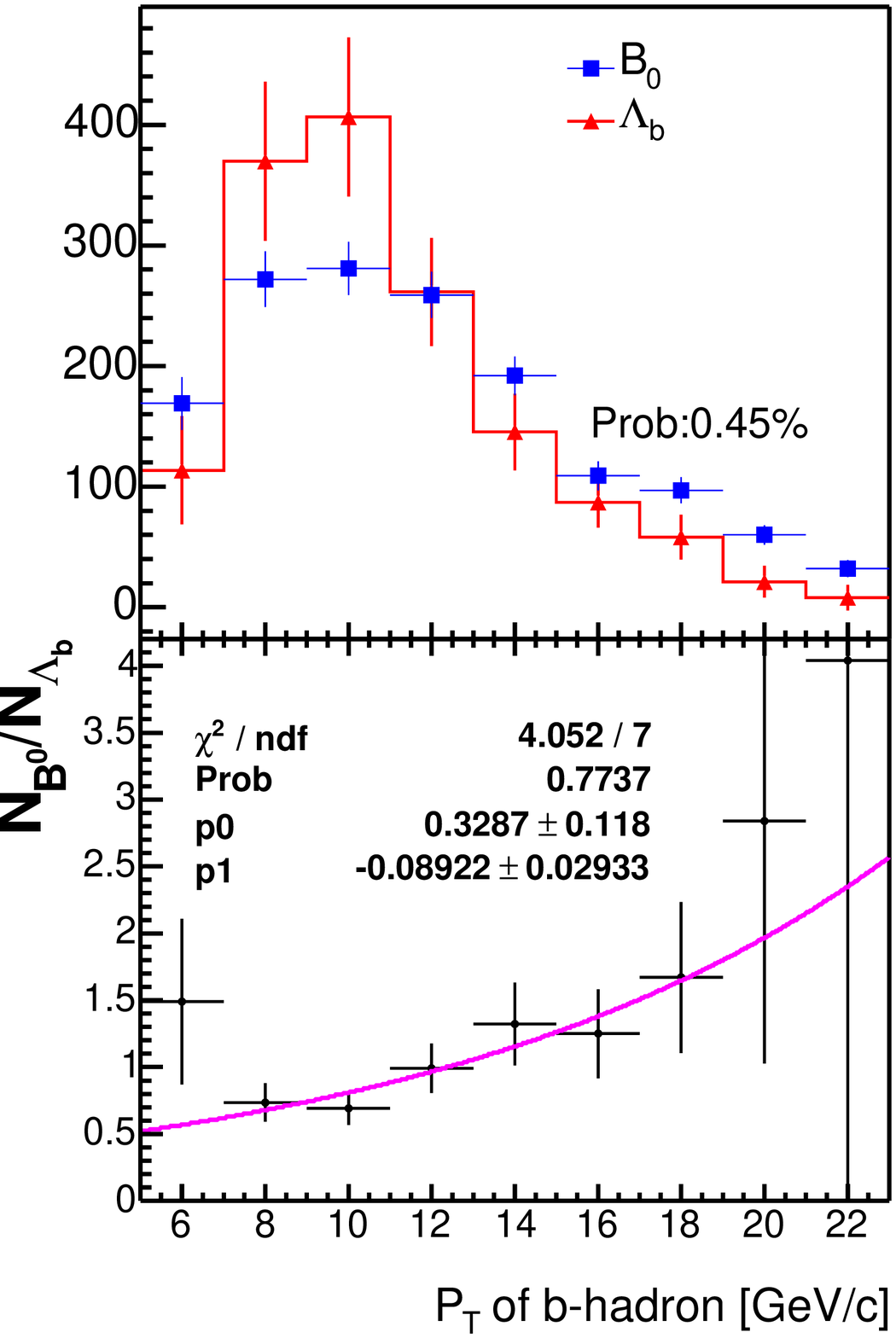}
 \caption[Comparison of \Bd\ and \Lb\ \pt\ spectra measured in the data]
 {Comparison of \Bd\ and \Lb\ \pt\ spectra measured in the data. The positive
 slope (2 $\sigma$ away from zero) of the ratio of \Bd\ to \Lb\ histograms
 indicates that $\pt(\Bd)$ is harder than $\pt(\Lb)$.}
 \label{fig:bptlbpt} 
\includegraphics[width=280pt, angle=0]{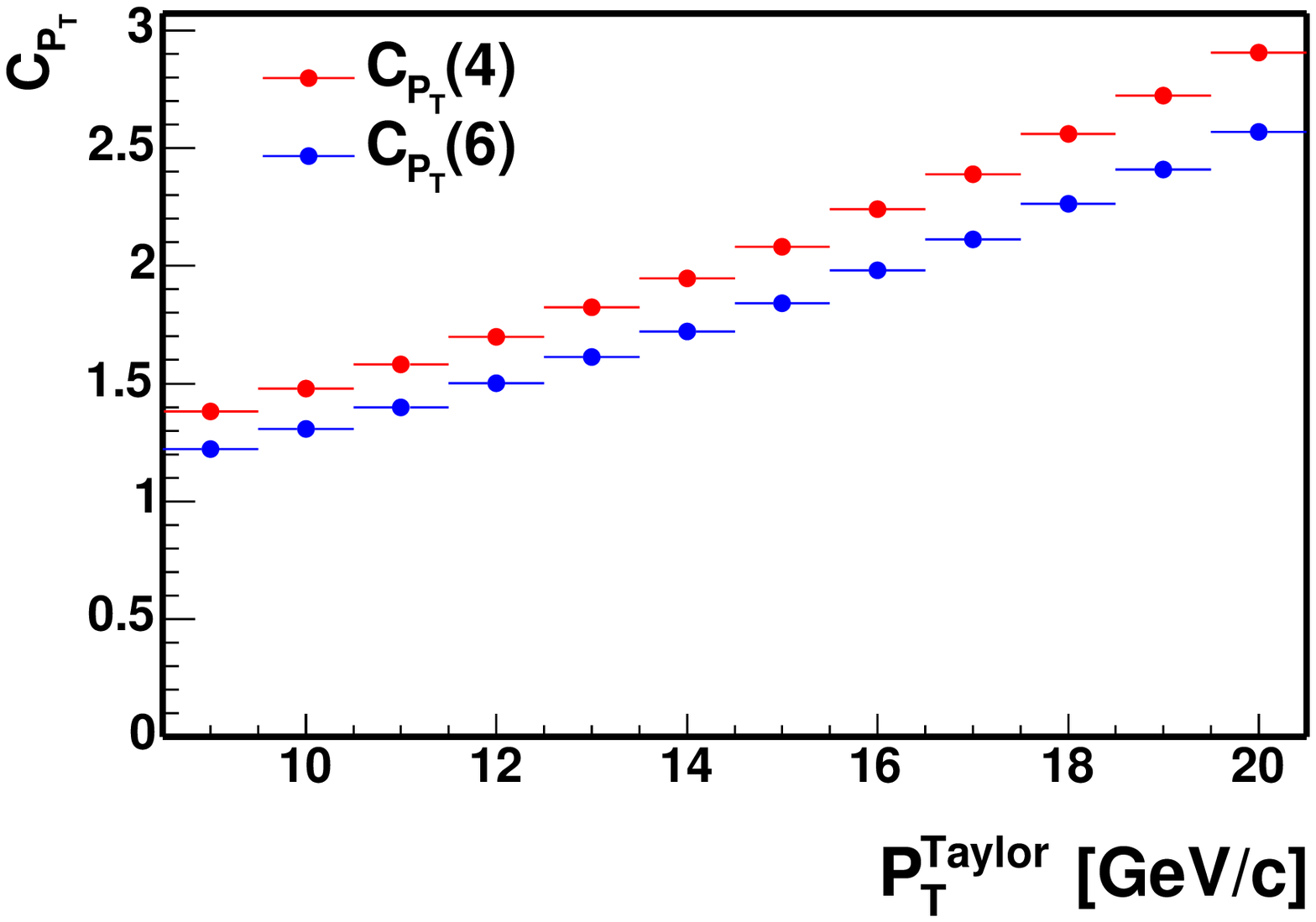}
 \caption[$C_{\pt}$ vs. $\pt^{\mathrm{Taylor}}$]
 {\rxsec\ correction factor, $C_{\pt}$, as a function of 
  $\pt^{\mathrm{Taylor}}$. 
  The two curves represent the correction factor for two \pt\ thresholds of 
  \B\ hadrons: 4~\gevc\ and 6~\gevc, respectively. 
  $\pt^{\mathrm{Taylor}}$ is the \pt\ threshold of \B\ hadrons in Taylor's 
analysis. }
 \label{fig:corrxsec} 
 \end{center}
 \end{figure}

The second correction is due to a difference in the \pt\ threshold 
of Taylor's and our analysis. Data used in Taylor's analysis 
come from an electron trigger which cuts on the transverse energy, $E_T$, of 
electron below 8 \gevcsq\ and probe the \B\ hadrons with \pt\ greater 
than 10 \gevc. Our data come from a two track trigger with a looser \pt\ 
requirement and extend the minimum \pt\ of \B\ hadrons down to 4 \gevc.
If the \pt\ spectra of \Bd\ and \Lb\ are almost identical, the value of 
$\rho$ will be independent of the \pt\ threshold of the \B\ hadrons, as 
shown in Figure~\ref{fig:teachxsec} (top). If the \pt\ spectra of \Bd\ and 
\Lb\ are quite different, the value of $\rho$ will strongly depend on the 
\pt\ threshold, as shown in Figure~\ref{fig:teachxsec} (bottom).
Figure~\ref{fig:bptlbpt} shows that \Lb\ \pt\ spectrum is softer than 
that of the \Bd. Therefore, we need to apply another correction factor,
 $C_{\pt}$, on the previous efficiency corrected Taylor's result, 
$\rho(10)=\frac{f_\mathrm{baryon}}{f_d}\cdot C_{\epsilon}$, to obtain 
$\rho(4)$ and $\rho(6)$ for this analysis. 

The $\rho(x)$ is expressed as follows:
\begin{equation}
\label{eq:xseccorr}
 \rho(x)  = \frac{f_\mathrm{baryon}}{f_d}\cdot C_{\epsilon}\cdot C_{\pt}(x),
\end{equation}
where 
\begin{equation}
  C_{\pt}(x) = \frac{N_{\Lb}(\pt> x)}{N_{\Bd}(\pt>x)}/
\frac{N_{\Lb}(\pt>\pt^{\mathrm{Taylor}})}{N_{\Bd}(\pt>\pt^{\mathrm{Taylor}})}, 
\end{equation} 
Here, $x$ and $\pt^{\mathrm{Taylor}}$ stand for the \pt\ thresholds of the \B\ 
hadrons in our and Taylor's analysis. We obtain $C_{\pt}(x)$ using the 
generator level MC produced with the corrected \pt\ spectra of \Lb\ and \Bd. 
About 20M events of \Lb\ and \Bd\ are generated. No additional cuts are 
applied except that all the \B\ hadrons have rapidity less than 2.0. We count 
the number of \B\ hadrons above 4 and 6 \gevc, and divide that 
by the number of B hadrons above certain $\pt^{\mathrm{Taylor}}$. 
A scan of $\pt^{\mathrm{Taylor}}$ from 
9 \gevc\ to 20 \gevc\ is performed but the value at 10 \gevc\ is used in 
the analysis. We have $C_{\pt}(4)$ = 1.480 $\pm$ 0.002 ({\em stat}) 
${+0.190 \atop -0.172}$ ({\em \pt}), 
and $C_{\pt}(6)$ = 1.309 $\pm$ 0.002 ({\em stat}) ${+0.111 \atop -0.105}$ 
({\em \pt}), where the last uncertainty comes from the uncertainty of the \Lb\ 
\pt\ spectrum. 
Figure~\ref{fig:corrxsec} presents $C_{\pt}(4)$ and $C_{\pt}(6)$ as a 
function of $\pt^{\mathrm{Taylor}}$.

After applying corrections $C_{\epsilon}$ and $C_{\pt}$, we calculate 
$\rho(4)$ and $\rho(6)$ to be:
\begin{eqnarray}
\label{eq:rho4}
\rho(4) & = & 0.63 \pm 0.23 (stat\oplus syst) {+0.24 \atop -0.14} (\pt),\\
\label{eq:rho6}
\rho(6) & = & 0.56 \pm 0.20 (stat\oplus syst) {+0.19 \atop -0.11} (\pt). 
\end{eqnarray}
Then, we could go back to CDF Run II measurements of \yileab\ and extract
${\cal B}(\lbhad)$:
\begin{equation}
\label{eq:lbhadbr}
{\cal B}(\lbhad) =  G \cdot \frac{1}{\rho(6)} \cdot {\cal B}(\dhad), 
\nonumber 
\end{equation}
where
 \begin{equation}
 G  =  \yileab.
\end{equation}
The values for each of the parameters are listed in Table~\ref{t:xsecvalue}. 
We find 
\[
{\cal B}(\lbhad) = \left(\brlbhadc\ \pm \brlbhade\ (stat \oplus syst)  \brlbhadpte\ (\pt)\right) \%,
\] 
which is in good agreement with the prediction by Leibovich,~\etal~\cite{Leibovich:2003tw}, 0.45$\%$. 

As noted earlier, ${\cal B}(\lbhad)$ will be used later for the normalization 
of the amount of physics, \bb\ and \cc\ backgrounds to the observed number of 
events in our hadronic signals.  Several variables, such as hadronic to 
semileptonic efficiency ratios, $C_{\epsilon}$, $C_{\pt}$ and $G$, depend on 
the \Lb\ \pt\ spectrum and are correlated. We would like to study the 
systematics on the relative branching fractions from these variables 
simultaneously. Consequently, the uncertainty due to the \Lb\ \pt\ spectrum is 
separated from the other systematic and statistical uncertainties. 

 \begin{table}[tbp]
 \caption{Parameters for calculating ${\cal B}(\lbhad)$.}
 \label{t:xsecvalue}
 \begin{normalsize}
  \begin{center}
  \begin{tabular}{cl}  
 \hline
  Parameter & Value \\
  \hline \hline 
  G 
& 0.82 $\pm$ 0.25 ({\em stat $\oplus$ syst}) $\pm$ 0.06 ({\em \pt})\\
 $\frac{f_d}{f_\mathrm{baryon}}$ & 4.2 $\pm$ 1.5  \\
 $C_{\pt}(6)$ & 1.309 $\pm$ 0.002  ({\em stat}) ${+0.111 \atop -0.105}$ ({\em \pt}) \\
 $C_{\epsilon}$ & 1.81 $\pm$ 0.04 ({\em stat}) ${+0.42 \atop -0.22}$ ({\em \pt}) \\
 ${\cal B}(\dhad)$ &  (2.76 $\pm$ 0.25)$\cdot$ 10$^{-3}$ \\
 \hline
 ${\cal B}(\lbhad)$ ($\%$) &  \brlbhadc\ $\pm$ \brlbhade\ ({\em stat} $\oplus$
 {\em syst}) \brlbhadpte\ ({\em \pt}) \\
 \hline\hline
\end{tabular}
 \end{center}
 \end{normalsize}
 \end{table}

\section{Physics Backgrounds}
\label{sec-physicsb}
  Physics backgrounds come from the decays of \B\ hadrons into similar final 
state as our semileptonic signal: a \Dstar\ (\D, \Lc), a $\mu^-$ and missing 
particles. Branching ratios and efficiencies of these physics decays are 
needed to normalize the background contribution to the observed number of 
hadronic signal events in the data;
\begin{equation}
 \frac{N_\mathrm{physics}}{N_\mathrm{had}} = 
	\frac{ \sum \; {\cal B}_{i}\cdot \epsilon_i}
	 {{\cal B}_\mathrm{had} \cdot \epsilon_{had}}.
 \label{eq:physicsb}
 \end{equation}
For the backgrounds to the \dstarsemi\ and \dsemi\ decays, 
 we find the modes which give similar final states as our semileptonic 
signals in the decays listed in the Particle Data Group (PDG) summary
~\cite{pdg:2004} and the default decay table inside \evtgen\ 
package~\cite{Lange:2001uf}. Many decays of \B\ and $D$ mesons have been 
measured by other experiments, such as {\tt CLEO}, {\tt BELLE} and {\tt BABAR}.
 These measurements serve as inputs to the \evtgen\ decay package. Since 
{\tt BELLE} and {\tt BABAR} also use the \evtgen\ package, they have 
included decay modes into \evtgen\ which have not yet been measured and 
estimate the branching ratios. For the backgrounds to the \lbsemi\ decay, 
none of the \B\ hadrons decays to \lcmu\ final states have been measured 
in the CDF-I and other experiments, or estimated
inside \evtgen. We use the results from the preliminary measurements by 
Litvintsev,~\etal~\cite{cdfnote:7546} and the prediction of \pythia\ to 
obtain the background branching ratios. 

After we have a list of decays which share similar final states 
as our signals, we use the generator level simulation to estimate the 
composition of the inclusive semileptonic signal from each physics 
background. Details of this procedure maybe be found in 
the study by Tesarek,~\etal\cite{cdfnote:6599}\cite{cdfnote:7545}.
The decays which contribute $\geq$ 1$\%$ to the semileptonic signal 
after trigger-like and the four track invariant mass 
$M_{D^*(D,\Lambda_c)\mu}$ cuts are selected for further consideration. 
We generate each selected decay separately and run through
the full CDF detector simulation as described in 
Section~\ref{sec-mccom}. Then we run the same signal reconstruction
program used for the data on the MC and divide the number of reconstructed 
events by the number of generated events to obtain the efficiency.


\subsection{Physics backgrounds of \dstarsemi\ and \dsemi}
\label{sec-dphysicsb}
 Physics backgrounds of \dstarsemi\ and \dsemi\ fall into two categories:
\begin{enumerate}
 \item Semileptonic decays of \Bd, \Bu, \Bs, which include either 
additional particles (eg: \bddpizeromunu) or a higher mass charm meson 
with subsequent decay into our charm signal (eg: \dstarsemi, \seqdstard)
  \item Hadronic \Bd\ decays into two charm mesons, one charm meson
decays as our charm final state, the other charm meson decays semileptonically
(eg: $\overline{B}^0\rightarrow \D D_s^-$, 
$D_s^- \rightarrow \phi \mu^- \overline{\nu}_{\mu}$)
 \end{enumerate}
 
\begin{figure}[tbp]
    \begin{center}
    \resizebox{350pt}{!}{\includegraphics*[80pt,130pt][476pt,713pt]
{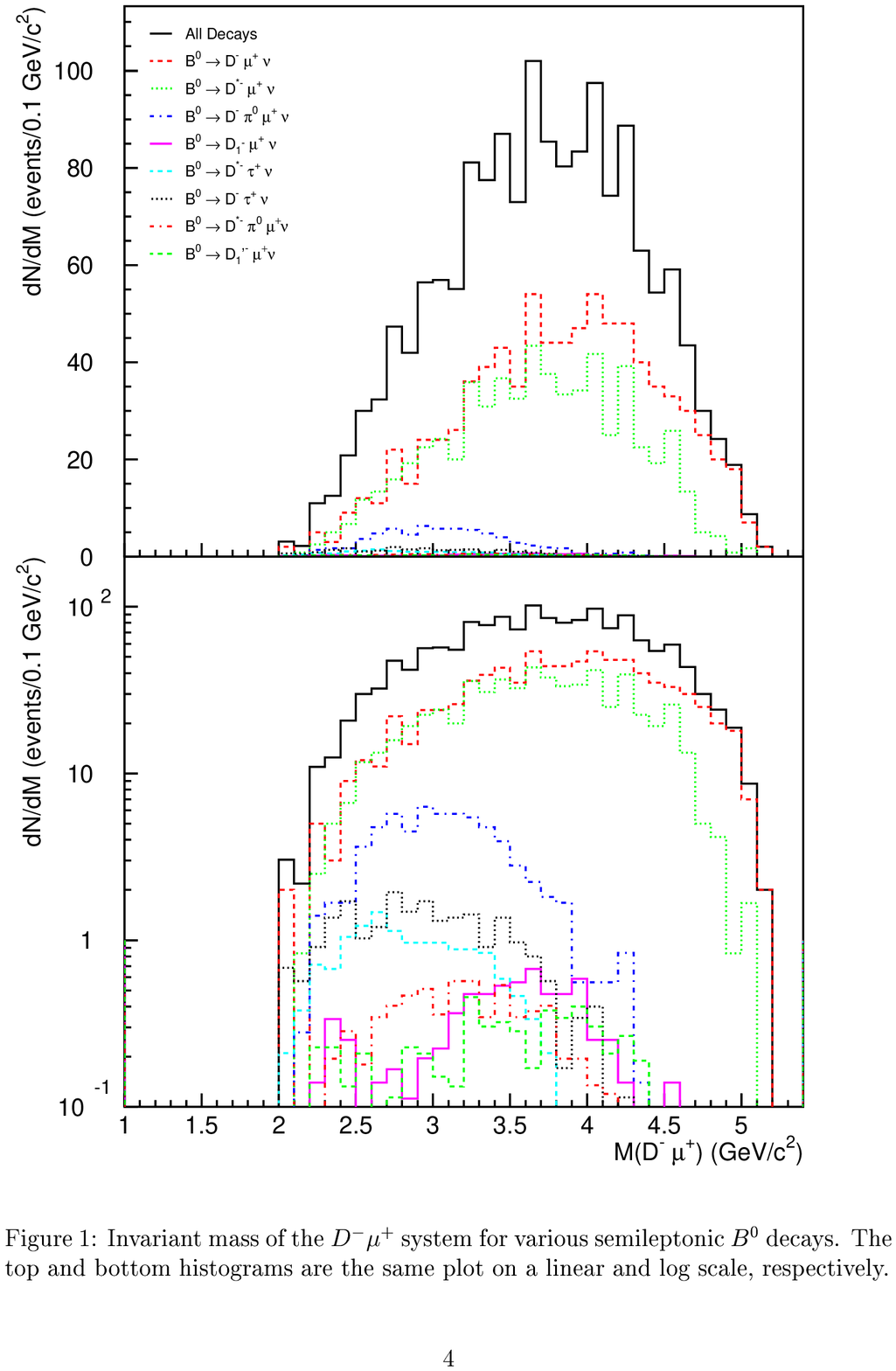}}
     \caption[Invariant mass of $\D\mu^-$ for the signal and physics 
backgrounds]{Invariant mass of $\D\mu^-$ for the signal and physics 
backgrounds from semileptonic \Bd\ decays~\cite{cdfnote:6599}. 
The top and bottom histograms are the same plot on a linear and log scale, 
respectively. 
Note that the backgrounds are concentrated in the low mass region. The 
signal to background ratio is larger at the higher mass region.}
     \label{fig:m4track}
     \end{center}
  \end{figure}

Tables~\ref{t:physicsdstar}--~\ref{t:physicsd} summarize the 
physics background in \dstarsemi\ and \dsemi\ which contribute
$\geq$ 1$\%$. The second column in the table lists the measured or estimated
branching ratios. All the numbers in parentheses are estimated uncertainties 
(100$\%$ for $B$ and 5$\%$ for charm) for the unmeasured branching fractions. 
The third column lists their efficiencies relative to the 
hadronic signals with statistical errors. The fourth column lists
the normalization of each background relative to the exclusive semileptonic
signal. The last column lists the number of events from each background
after multiplying the relative branching ratio and efficiencies with
the number of hadronic signal in the data, as expressed in 
Equation~\ref{eq:physicsb}. The uncertainty in the last column only includes
the statistical uncertainty of the hadronic yield. 
When normalizing backgrounds from \Bu\ and \Bs\ to \Bd\ signals, the following 
fragmentation fractions quoted in the 2004 PDG are used:
\begin{eqnarray}
 b \rightarrow B_d & = & (39.7 \pm 1.3) \%, \nonumber \\
 b \rightarrow B_u & = & (39.7 \pm 1.3) \%, \nonumber \\
 b \rightarrow B_s & = & (10.7 \pm 1.1) \%. \nonumber 
\end{eqnarray}
Note that most of the backgrounds which contribute $\geq$ 1$\%$ share
the same Feynman diagram as our semileptonic signal: 
\begin{itemize}
\item resonant mode: a spectator quark and a $b$ to $c$ quark transition via 
a virtual $W$ boson exchange (eg: background of \dsemi, \dstarsemi)
\item non-resonant mode: with additional $u\overline{u}$ or $d\overline{d}$ 
quark pair created from the vacuum (eg: \bddpizeromunu)
\end{itemize}
 Note that the non-resonant modes tend to have smaller branching ratios and
smaller efficiencies than the resonant mode.
The dominant background of \dstarsemi\ is \bpdonezeromunu\ where \seqdonezero. 
The total physics background in Table~\ref{t:physicsdstar} is 
about 15$\%$ of \incdstarsemi\ events in the data after all the cuts.
The dominant background of \dsemi\ is \dstarsemi\ where \seqdstard.
The total physics background in Table~\ref{t:physicsd} contributes 
about 40$\%$ of \incdsemi\ events in the data after all the cuts.
As shown in Figure~\ref{fig:m4track}, a cut on the invariant mass of
\Dstar(\D)$\mu^-$ can reduce or eliminate the background from \Bd, \Bu\ 
decaying semileptonically to more particles or a higher mass charm state.

\renewcommand{\arraystretch}{1.0}
\renewcommand{\tabcolsep}{0.05in}
 \begin{normalsize}
 \begin{table}[tbp]
 \setdec 00.000
   \begin{center}
 \caption{Physics backgrounds in \dstarsemi.}
 \label{t:physicsdstar}
   \begin{tabular}{|l|r@{\,$\pm$\,}r|r|r|r@{\,$\pm$\,}r|}
    \hline
     \multicolumn{1}{|c|}{Mode}  
     & \multicolumn{2}{|c|}{BR ($\%$)}	
     & \multicolumn{1}{|c|}{$\epsilon$ ratio}
     & \multicolumn{1}{|c|}{Norm}
     & \multicolumn{2}{|c|}{$N_{event}$} \\ 
     \hline	
     \hline	
	\dstarhad
	& \dec 0.276 & \dec 0.021
	&  1
	& -- &  \ndstarhadc &  \ndstarhade \\

     \hline	
	\incdstarsemi
	&  \multicolumn{2}{|c|}{--}
	&  --
	&  -- &  \ndstarsemic & \ndstarsemie \\
     \hline
      \dstarsemi	
	& \dec 5.44 & \dec 0.23
	&  0.447 $\pm$ 0.006
	& 1.000 & \multicolumn{2}{|c|}{--} \\
     \hline

         \bpdonezeromunu
	&  \dec 0.56 & \dec 0.16
	&  0.356 $\pm$ 0.008
	&  0.055 & 51 & 5 \\
         $   \hspace{36pt}   \hookrightarrow D^{*+}\pi^{-}$
	&  \dec 66.67 & \dec (3.33)
	& 
	&  & \multicolumn{2}{|c|}{}\\
     \hline	
 	
        \bddonemunu
	&  \dec 0.56 & \dec (0.56)
        &  0.349 $\pm$ 0.008
	&  0.027 & 25 &  3 \\
         $\hspace{36pt} \hookrightarrow D^{*+}\pi^{0}$
	&  \dec 33.33 & \dec (1.67)
	& 
	& & \multicolumn{2}{|c|}{} \\
     \hline	

	\bpdpronezeromunu		
	&  \dec 0.37 & \dec (0.37)
	&  0.351 $\pm$ 0.008
        &  0.036 & 33 & 4 \\
         $\hspace{36pt} \hookrightarrow D^{*+}\pi^{-}$
	& \dec 66.67 & \dec (3.33)
	& 
	& &\multicolumn{2}{|c|}{} \\
     \hline	

        \bddpronemunu
	&  \dec 0.37 & \dec (0.37)
	&  0.336 $\pm$ 0.008
        &  0.017 & 16 &  2 \\	
        $\hspace{36pt} \hookrightarrow D^{*+}\pi^{0}$
	& \dec 33.33 & \dec (1.67)
	& 
	& & \multicolumn{2}{|c|}{}\\

     \hline	
  	\bpdstarpimunu
	&  \dec 0.20 & \dec (0.20)
	& 0.242 $\pm$ 0.007
	& 0.020 & 19 & 2 \\
     \hline

	\bddstarpizeromunu	
	&  \dec 0.10 & \dec (0.10)
	&  0.239 $\pm$ 0.006
	&  0.010 & 9 & 1 \\

     \hline	
  	
	\bddstartau	
	&  \dec 1.60 & \dec (1.60)
	&  0.136 $\pm$ 0.005
	&  0.016 & 15 &  2 \\

        $ \hspace{54pt}    \hookrightarrow \mu^{-}\overline{\nu}_{\mu}$
	& \dec 17.36 & \dec 0.06
	& 
	&  & \multicolumn{2}{|c|}{} \\
	\hline
        \hline	
  	\end{tabular}

 
\renewcommand{\arraystretch}{1.1}
\renewcommand{\tabcolsep}{0.05in}
 \caption{Physics backgrounds in \dsemi.}
 \label{t:physicsd}
   \begin{tabular}{|l|r@{\,$\pm$\,}r|r|r|r@{\,$\pm$\,}r|}
    \hline
     \multicolumn{1}{|c|}{Mode}  
     & \multicolumn{2}{|c|}{BR ($\%$)}	
     & \multicolumn{1}{|c|}{$\epsilon$ ratio}
     & \multicolumn{1}{|c|}{Norm}
     & \multicolumn{2}{|c|}{$N_{event}$} \\ 
     \hline	
     \hline	
	\dhad
	&   \dec 0.276 & \dec 0.025
	&  1.000
	& --  & \ndhadc & \ndhade \\   	
     \hline		
	\incdsemi
	&  \multicolumn{2}{|c|}{--}
	&  --
	& -- & \ndsemic & \ndsemie \\
     \hline	
	\dsemi
	&  \dec 2.14 & \dec 0.20
	& 0.455 $\pm$ 0.004 
	& 1.000 & \multicolumn{2}{|c|}{--} \\
     \hline	
	\dstarsemi	
	&   \dec 5.44 & \dec 0.23
	&   0.372 $\pm$ 0.005
        &   0.671 &  1373 & 71 \\
        $\hspace{36pt} \hookrightarrow D^{+}\pi^{0}/\gamma$
	& \dec 32.30 & \dec 0.64
	&  & & \multicolumn{2}{|c|}{}\\

     \hline	
	\bddpizeromunu	
	&  \dec 0.30 & \dec (0.30)
	& 0.165 $\pm$ 0.004
	& 0.051 & 104 &  5 \\ 
     \hline	

  	\bpdpimunu	
	&  \dec 0.60 & \dec (0.60)
	& 0.165 $\pm$ 0.004
	& 0.102 & 208 & 11 \\
     \hline	
         \bpdonezeromunu
	&  \dec 0.56 & \dec 0.16
	& 0.278 $\pm$ 0.005
	& 0.034 & 70 &  4 \\
         $   \hspace{36pt}   \hookrightarrow D^{*+}\pi^{-}$
	&  \dec 66.67 & \dec (3.33)
	& 
	&  & \multicolumn{2}{|c|}{} \\
        $\hspace{60pt} \hookrightarrow D^{+}\pi^{0}/\gamma$
	& \dec 32.30 & \dec 0.64
	& 
	& & \multicolumn{2}{|c|}{}\\
     \hline	
 	\bpdonezeromunu	
        &  \dec 0.37 & \dec (0.37)
	& 0.273 $\pm$ 0.005
        & 0.022 & 46 &  3 \\
         $\hspace{36pt} \hookrightarrow D^{*+}\pi^{-}$
	& \dec 66.67 & \dec (3.33)
	& 
	& & \multicolumn{2}{|c|}{} \\
        $\hspace{60pt} \hookrightarrow D^{+}\pi^{0}/\gamma$
	& \dec 32.30 & \dec 0.64
	& 
	& & \multicolumn{2}{|c|}{} \\
     \hline

	\bddtau
	&  \dec 0.70 & \dec (0.70)
	&  0.100 $\pm$ 0.004
	& 0.013 & 26 & 1 \\

        $ \hspace{54pt}    \hookrightarrow \mu^{-}\overline{\nu}_{\mu}$
	& \dec 17.36 & \dec 0.06
	& 
	& & \multicolumn{2}{|c|}{} \\
	\hline

	\bsdkzero	
	&  \dec 0.30 & \dec (0.30)
	&  0.137 $\pm$ 0.005
	& 0.011 & 23 & 1 \\
     \hline	
     \hline		
	
  	\end{tabular}
	\end{center}
 	\end{table}
 \end{normalsize}

\subsection{Physics backgrounds of \lbsemi}
\label{sec-lcphysicsb}
 Physics backgrounds of \lbsemi\ fall into two categories:
\begin{enumerate}
 \item Other semileptonic decays of \Lb, which either include additional
particles (eg: \lblcpim) or include
a higher mass charm baryon with subsequent decay into our charm 
signal (eg: \lblcstar, \seqlcstar)
  \item Baryonic semileptonic decays of $B_{u,d,s}$, which decay into
\Lc\ or higher mass charm baryon, a proton or a neutron and leptons 
(eg: \bplcpmunu)
 \end{enumerate}
\subsubsection{Other \Lb\ Semileptonic Decays}
 We have learned from Section~\ref{sec-dphysicsb} that these physics 
backgrounds should have the same or similar Feynman diagrams as \lbsemi. 
The observation of $\Lambda_c(2593)^+$ with spin $\frac{1}{2}$ and 
$\Lambda_c(2625)^+$ with spin $\frac{3}{2}$~\cite{Edwards:1994ar}
\cite{Albrecht:1997qa} indicates the existence of decays \lblcstar\ and 
\lblcsstar. Leibovich and Stewart~\cite{Leibovich:1997az} 
predict a range of the relative decay widths of \Lb\ to these two excited 
\Lamc\ decays to the inclusive semileptonic decay, 
$\Lb \rightarrow \Lc \mu \; \mathrm{anything}$:
\begin{eqnarray}
0.083 < \frac{\Gamma(\lblcstar)}{\Gamma(\Lb \rightarrow \Lc \mu \; 
	\mathrm{anything})} < 0.248, \nonumber \\
0.079 < \frac{\Gamma(\lblcsstar)}{\Gamma(\Lb \rightarrow \Lc \mu \; 
	\mathrm{anything})} < 0.166.\nonumber 
\end{eqnarray}
However, the wide range gives a systematic uncertainty as large as 100$\%$. 
In addition, the following decays have similar Feynman diagrams
 as \lblcstar\ and \lblcsstar:
\begin{center}
\begin{tabular}{ll}
 \lblcfzero, & \lbsigmaczero, \\
 \lblcpizero, & \lbsigmacp, \\
 \lblcpim, & \lbsigmacpp. \\
\end{tabular}
\end{center}
Corresponding decays in the $\tau$ channel can also produce contamination 
in our semileptonic signal, as seen in Section~\ref{sec-dphysicsb}. 

In order to reduce the systematic uncertainties due to the branching 
ratios of these backgrounds, Litvintsev reconstructed the following 
decays~\cite{cdfnote:7546}:
\begin{center}
\begin{tabular}{ll}
\lblcstar, & \lblcsstar,\\ 
\lbsigmaczero, & \lbsigmacpp, \\
\end{tabular}
\end{center}
using the data processed with the {\tt Production} executable, 
version 5.3.1, and compressed into the secondary datasets {\tt xbhd0d} and 
{\tt xbhd0e}. The cuts applied in Litvintsev's analysis are similar to the 
ones for this analysis. After taking into account the efficiency difference 
between the reconstructed backgrounds and the \lbsemi, we extract the relative 
branching ratios of these backgrounds to the \lbsemi:
\begin{eqnarray}
\frac{{\cal B}(\lblcstar)}{{\cal B}(\lbsemi)} &
	= &(4.7\pm 1.6)\times 10^{-2},\\
\frac{{\cal B}(\lblcsstar)}{{\cal B}(\lbsemi)} &
	= & (7.9\pm 1.5)\times 10^{-2},\\
\frac{{\cal B}(\lbsigmaczero)}{{\cal B}(\lbsemi)} & 
	= & (4.2\pm 1.6)\times 10^{-2},\\
\frac{{\cal B}(\lbsigmacpp)}{{\cal B}(\lbsemi)} & 
	= & (4.2\pm 1.6)\times 10^{-2}.
\end{eqnarray}
Assuming the isospin symmetry, we can infer 
\[\frac{{\cal B}(\lbsigmacp)}{{\cal B}(\lbsemi)} = (4.2\pm 1.6)\times 10^{-2}.
\]
However, we need the input of ${\cal B}(\lbsemi)$ to obtain the absolute 
branching fractions of the decays listed above. Assuming heavy quark symmetry, 
we expect the semileptonic decay width for all \B\ hadrons are the same, 
i.e. \(\Gamma^\mathrm{semi}_{\Lb} = \Gamma^\mathrm{semi}_{\Bd}= 
\Gamma^\mathrm{semi}_{\Bu} = \Gamma^\mathrm{semi}_{\Bs}\). Then, the 
semileptonic branching ratios of the \B\ hadrons, 
$\Gamma^\mathrm{semi}/\Gamma^\mathrm{total}$, only vary due to the lifetime 
difference which result in a difference in $\Gamma^\mathrm{total}$. 
Since the \Lb\ decays to a spin-$\frac{1}{2}$ \Lamc, we expect contributions 
from both S and P wave amplitudes. Taking a weighted average of 
${\cal B}(\overline{B}^0\rightarrow \D \mu \nu + \Dstar \mu \nu)$ and 
${\cal B}(B^-\rightarrow \Dzero \mu \nu + \rightarrow D^{*0} \mu \nu)$, where the branching ratios of the \D\ and \Dzero\ (\Dstar\ and $D^{*0}$) 
 final states correspond to the S (P) wave amplitudes, we obtain 
7.83 $\pm$ 0.26 $\%$. We further scale the number by the ratio of lifetimes, 
$\frac{\tau_{\Lb}}{<\tau_{B}>}=0.80$, and estimate ${\cal B}(\lbsemi)$=
(\brlbsemic\ $\pm$ \brlbsemie)$\%$. For the ${\cal B}(\lblctau)$, we 
multiply ${\cal B}(\lbsemi)$ by the ratio of phase space 
$\frac{Ph. Sp. (\lblctau)}{Ph. Sp. (\lbsemi)} =0.277$.

Adding the branching fractions of \Lb\ to $\Lamc$, $\Lamc(2596)$, 
$\Lamc(2625)$, $\Sigma_c^{0,+,++}$ semileptonic decays in the $\mu$ 
and $\tau$ channels after correcting for the 
${\cal B}(\seqtau)=17.36\pm 0.06 \%$, we get 8.2$\%$. 
The branching ratio of the inclusive \Lb\ semileptonic decays 
in the 2004 PDG is:
\[
{\cal B}(\Lb \rightarrow \Lc \mu \; \mathrm{anything}) = 9.2 \pm 2.1 \%. 
\]
We fill the difference, 1.0$\%$, with the following decays: 
\begin{center}
\begin{tabular}{lll}
 \lblcfzero, & 
 \lblcpizero, & 
 \lblcpim, \\
\end{tabular}
\end{center}
where the branching fraction of \lblcpim\ is estimated to be twice of 
\lblcpizero\ based on the isospin invariance. 
A more detailed description about the estimate of these \Lb\ semileptonic 
decays can be found in Tesarek~\cite{cdfnote:7545}.
Table~\ref{t:physicslc} summarizes the 
physics background from the \Lb\ semileptonic decays discussed above and 
their relative efficiencies to the hadronic signal.
The dominant backgrounds are \lblcstar\ and \lblcsstar. Total physics
background in Table~\ref{t:physicslc} is about 9.2$\%$ of the \inclbsemi\ 
events in the data.

\renewcommand{\arraystretch}{1.1}
\begin{normalsize}
 \begin{table}[tbp]
 \setdec 000.000
 \caption{Physics backgrounds in \lbsemi\ from other \Lb\
semileptonic decays.}
 \label{t:physicslc}
   \begin{center}
   \begin{tabular}{|l|r@{\,$\pm$\,}r|r|r|r@{\,$\pm$\,}r|}
    \hline
     \multicolumn{1}{|c|}{Mode}  
     & \multicolumn{2}{|c|}{BR ($\%$)}	
     & \multicolumn{1}{|c|}{$\epsilon$ ratio}
     & \multicolumn{1}{|c|}{Norm}
     & \multicolumn{2}{|c|}{$N_{event}$} \\ 
     \hline	
     \hline	
 	 \lbhad	
	& \brlbhadc & 0.21  
	&  1
	&  -- &  \nlbhadc & \nlbhade\\

     \hline	
	
         \inclbsemi	
	&  \multicolumn{2}{|c|}{--}	 
	&  --
	&  -- & \nlbsemic & \nlbsemie \\

     \hline	
	 \lbsemi
	&   \brlbsemic & \brlbsemie 
	&   0.300 $\pm$ 0.004
	&  1 & \multicolumn{2}{|c|}{--} \\

     \hline	
	\lblcstar
	&  \brlcstarc & \brlcstare
	&  0.196 $\pm$ 0.003
	& 0.031 & 26 & 3 \\
         $\hspace{36pt} \hookrightarrow \Sigma_c^{++}\pi^{-}$
	& \dec 24. & \dec 7. 
	& 
	& & \multicolumn{2}{|c|}{} \\
        $\hspace{60pt} \hookrightarrow \Lc \pi^+$ 
	& \multicolumn{2}{|l|}{ \dec 100. } 
	& 
	& &\multicolumn{2}{|c|}{} \\
         $\hspace{36pt} \hookrightarrow \Sigma_c^{0}\pi^{+}$ 
	& \dec 24. & \dec 7. 
	& 
	& &\multicolumn{2}{|c|}{} \\
        $\hspace{60pt} \hookrightarrow \Lc \pi^-$
	& \multicolumn{2}{|l|}{ \dec 100. } 
	& 
	& &\multicolumn{2}{|c|}{} \\
         $\hspace{36pt} \hookrightarrow \Sigma_c^+\pi^{0}$
	& \dec 24. & \dec (1.2) 
	& 
	& &\multicolumn{2}{|c|}{} \\
        $\hspace{60pt} \hookrightarrow \Lc \pi^0$
	& \multicolumn{2}{|l|}{ \dec 100. } 
	& 
	& & \multicolumn{2}{|c|}{} \\
        $\hspace{36pt} \hookrightarrow \Lc \pi^+ \pi^-$
	& \dec 18. & \dec 10. 
	& & & \multicolumn{2}{|c|}{}\\
        $\hspace{36pt} \hookrightarrow \Lc \pi^0 \pi^0$
	& \dec 9. & \dec (0.45)
	& & & \multicolumn{2}{|c|}{}\\
        $\hspace{36pt} \hookrightarrow \Lc \gamma$
	& \dec 1. & \dec (0.05)
	& & & \multicolumn{2}{|c|}{}\\
     \hline	
  	\lblcsstar
	&  \brlcsstarc & \brlcsstare
        &  0.191 $\pm$ 0.003
	& 0.050 & 42 &  4  \\
       $\hspace{36pt} \hookrightarrow \Lc \pi^+ \pi^-$
	& \dec 66. & \dec (3.3) 
	& & & \multicolumn{2}{|c|}{}\\
        $\hspace{36pt} \hookrightarrow \Lc \pi^0 \pi^0$
	& \dec 33. & \dec (1.7)
	& & & \multicolumn{2}{|c|}{} \\
        $\hspace{36pt} \hookrightarrow \Lc \gamma$
	& \dec 1. & \dec (0.05)
	& & & \multicolumn{2}{|c|}{}\\
     \hline
	\lblcfzero
	&  \brfzero & (\brfzero)
	& 0.023 $\pm$ 0.002
        & 0.003 & 2.6 & 0.3\\
     \hline	
	\lblcpim
	&  \brlcpipi & (\brlcpipi)
	& 0.032 $\pm$ 0.002 
        & 0.009 & 7 &  1 \\
     \hline	
	\lblcpizero
	&  \brfzero & (\brfzero)
	& 0.033 $\pm$ 0.002 
        & 0.004 & 3.6 & 0.4 \\
    \hline
	\lbsigmaczero
	&  \brsigcc & \brsigce
	& 0.081 $\pm$ 0.004
	& 0.011 & 10 & 1 \\
        $\hspace{36pt} \hookrightarrow \Lc \pi^-$
	& \multicolumn{2}{|l|}{ \dec 100. } 
	& 
	& &\multicolumn{2}{|c|}{}\\
     \hline	
  	\lbsigmacp
	&  \brsigcc & \brsigce
	&  0.072 $\pm$ 0.004
	& 0.010 & 8 & 1 \\
        $\hspace{36pt} \hookrightarrow \Lc \pi^0$
	& \multicolumn{2}{|l|}{ \dec 100. } 
	& 
	& &\multicolumn{2}{|c|}{}\\
     \hline	
	\lbsigmacpp
	&  \brsigcc & \brsigce
	&  0.077 $\pm$ 0.004
	& 0.011 & 9 &  1 \\
        $\hspace{36pt} \hookrightarrow \Lc \pi^+$
	& \multicolumn{2}{|l|}{ \dec 100. } 
	& 
	& &\multicolumn{2}{|c|}{} \\
     \hline
	\lblctau
	& \brlcsemitau & (\brlcsemitau)
	& 0.040 $\pm$ 0.003
	& 0.007 & 5 &  1 \\
       $ \hspace{48pt}    \hookrightarrow \mu^{-}\overline{\nu}_{\mu}$
	& \dec 17.36 & \dec 0.06
	& 
	& &\multicolumn{2}{|c|}{}\\
	\hline
	\hline
  	\end{tabular}
       \end{center}
	\end{table}
\end{normalsize}

\subsubsection{\B\ Meson Baryonic Semileptonic Decays}
While there are branching ratio measurements of the \B\ baryonic hadronic 
decay, eg: $\overline{B}^0 \rightarrow \Lc \overline{p} \pi^+ \pi^-$,
there is only an upper limit for the semileptonic decay of $B_u/B_d$ mixture:
\begin{eqnarray}
 {\cal B}(\Bd/B^+ \rightarrow \Lambda_c^- p e \nu_{e}) < 0.15 \%. \nonumber
\end{eqnarray}
In order to obtain a list of \B\ meson baryonic semileptonic decays 
that could contribute to our signal, we make use of the predictions from the 
\pythia. We generate \Bd, \Bu, and \Bs\ mesons using \pythia. We force the 
mesons to decay semileptonically and let \pythia\ handle the fragmentation. 
We count the number of events for each specific \B\ meson to \Lc\ decay, 
%
and find the maximum contributions are from the decays \bplcpmunu\ and 
\bdlcnmunu. 
We assume the upper limit for the muon-neutron or muon-proton final state 
should be the same as those decays with a proton and electron in the final 
state (see above). The value of this upper limit is then used for each 
branching fraction of \bplcpmunu\ and \bdlcnmunu.


We obtain the efficiencies for these two decays from the MC. 
Since we find the \pt\ spectra of \B\ mesons and \Lb\
are quite different, it is least ambiguous to calculate
the quantity:
 \begin{equation}
 \frac{N_{\bplcpmunu}}{N_{\lbhad}} = 
 \frac{\sigma_{\Lb}(\pt>4.0)}{\sigma_{\Bd}(\pt>4.0)} \cdot
  \frac{{\cal B}(\bplcpmunu) \cdot 
  \epsilon_{\bplcpmunu}}
 {{\cal B}(\lbhad) \cdot \epsilon_{\lbhad}}, 
 \end{equation}
where the production cross section ratio is the $\rho(4)$ in 
Equation~\ref{eq:rho4}. We use a low \pt\ threshold (\pt\ $>$ 4 \gevc) because 
we wish to accurately assess the acceptance of the \B\ hadron after applying
the reconstruction requirements. Specifically we are concerned
about the case where the neutrino is emitted in the direction opposite
to the direction that \B\ hadron is traveling. This case increases
the \pt\ of the remaining daughters and may make their total \pt\ greater 
than the \pt\ of the \B\ hadron. 
Therefore, the denominator of the efficiency is the number of events
with \B\ mesons or \Lb\ \pt\ $>$ 4 \gevc\ and rapidity $<$ 2.0.
The numerator is the number of events which pass all the trigger and
analysis cuts. 
Table~\ref{t:physicslc2} summarizes the \B\ meson to \Lc$\mu^-$ 
backgrounds. The contribution of \Bd\ and $B^+$ in the \inclbsemi\ events 
is about 0.4$\%$ each. 

\renewcommand{\arraystretch}{1.1}
\renewcommand{\tabcolsep}{0.04in}
\begin{normalsize}
 \begin{table}[tbp]
 \caption{Physics backgrounds in \lbsemi\ from \B\ mesons.}
 \label{t:physicslc2}
  \setdec 0000.000
   \begin{center}
   \begin{tabular}{|l|r@{\,$\pm$\,}r|r|r|r@{\,$\pm$\,}l|}
    \hline
     \multicolumn{1}{|c|}{Mode} 
     & \multicolumn{2}{|c|}{BR ($\%$)}
     & \multicolumn{1}{|c|}{Relative $\epsilon$}
     & \multicolumn{1}{|c|}{Norm}
     & \multicolumn{2}{|c|}{$N_{event}$}\\
     \hline	
     \hline	
	\lbhad &  \brlbhadc & 0.21 &  1 & --
        &  \dec \nlbhadc.  & \dec \nlbhade. 
        \\
     \hline	
         \inclbsemi\	
	&  \multicolumn{2}{|r|}{--}
	&  -- & --
	&  \dec \nlbsemic. & \dec \nlbsemie. \\

     \hline	
	 \lbsemi\
	&   \brlbsemic & \brlbsemie 
	&   0.265 $\pm$ 0.004 & 1
	& \multicolumn{2}{|c|}{--} \\

     \hline	
	\bplcpmunu
	&  \dec 0.15 & \dec (0.15)
	&  0.045 $\pm$ 0.002 & 0.006
	&  \dec 4.7 & \dec 0.5 \\
        
     \hline	
	\bdlcnmunu
	&  \dec 0.15 & \dec (0.15)
	&  0.044 $\pm$ 0.002 & 0.006
	&  \dec 4.6 & \dec 0.5 \\
        
  	\hline
  	\hline
  	\end{tabular}
       \end{center}
	\end{table}
\end{normalsize}

\section{Fake Muons}
\label{sec-fakemu}

 
  Another source of background originates from a charm 
hadron together with a hadron track ($\pi$, $K$, proton) misidentified 
as a muon.  A hadron is misidentified as a muon when it has higher energy and
punches through the hadron calorimeter, or when it decays into a muon 
before being stopped in the hadron calorimeter via decays like 
$\pi^+ \rightarrow \mu^+ \nu_{\mu}$, $K^+ \rightarrow \mu^+ \nu_{\mu}$
or $K^+ \rightarrow \pi^0 \mu^+ \nu_{\mu}$. Physics processes that 
generate these hadrons are direct production, inelastic collisions
with the detector material, fragmentation or the decays of charm 
and \B\ hadrons. 
Fragmentation is the process by which a $b$ or $c$ quark combines with 
additional quarks and gluons to form a $q\overline{q}$ or $qqq$ bound state. 
Fake muons from the first three categories tend to have a softer
\pt\ spectrum than the real muons from \B\ decays. A tighter \pt\ cut on 
the muon candidate largely removes these backgrounds. 
Fake muons from the charm hadrons which are produced at the primary vertex 
are also suppressed. For the reason that we require the muon candidate
should be matched to an SVT track with a $d_0$ greater than 120 $\mu$m, 
while fake muons from the promptly produced charm tend to have smaller impact 
parameter. Also for the reason that we require the charm hadron and the muon 
candidate to form a vertex displaced from the beam line and make a strict 
requirement on the pseudo \ctau.
\begin{equation}
 \mathrm{pseudo}\; \ctau\ =  \frac{M_{B}}{\pt(\mathrm{charm} + \mu)}
\cdot \lxy.
\end{equation}
 Here $\pt(\mathrm{charm} +\mu)$ is the total transverse momentum of 
charm hadron and the muon. 

Therefore, our principle source of fake muons comes from two types
of \B\ hadron decays:
\begin{itemize}
\item \bmixdx: hadronic decays of any \B\ hadrons, where $X_\mathrm{had}$ is
$\pi$, $K$ or proton which fakes the muon.
\item \bmixdpimux: semileptonic \B\ decay into a charm, a hadron 
track $X_\mathrm{had}$, and any leptons ($e$, $\mu$, $\tau$). The muon
is not reconstructed but $X_\mathrm{had}$ fakes the muon.
\end{itemize}
In this section, we estimate the fake muon contamination for our three 
signals: \dstarsemi, \dsemi\ and \lbsemi. Note the charge conjugates of 
the modes listed are also included.

\subsection{Background Estimate}
\label{sec-fakemumethod}
\begin{figure}[tbp]
    \begin{center}
 \renewcommand{\tabcolsep}{0.01in}
  \begin{tabular}{cc}
  \includegraphics[width=200pt, angle=0]{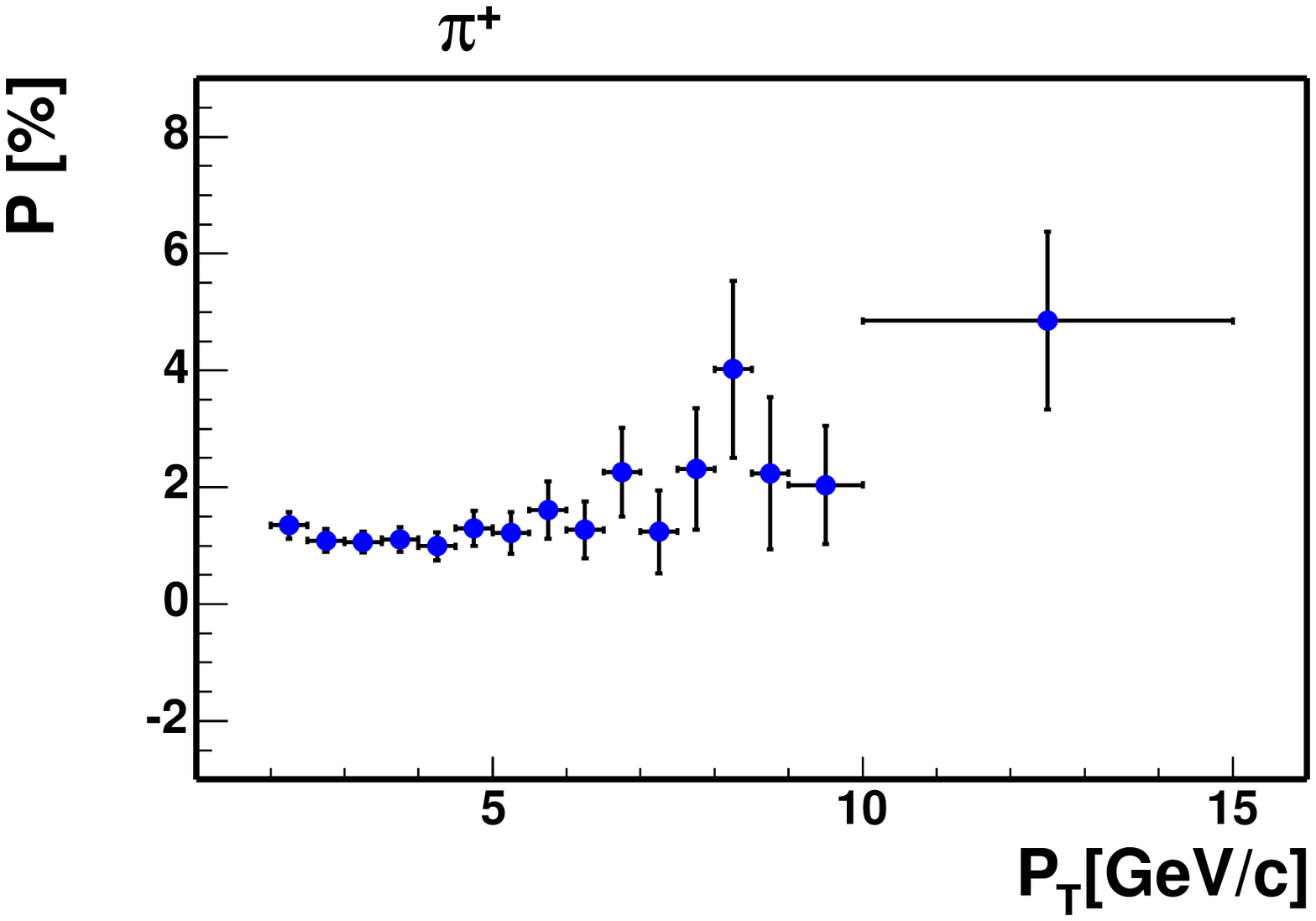} &
  \includegraphics[width=200pt, angle=0]{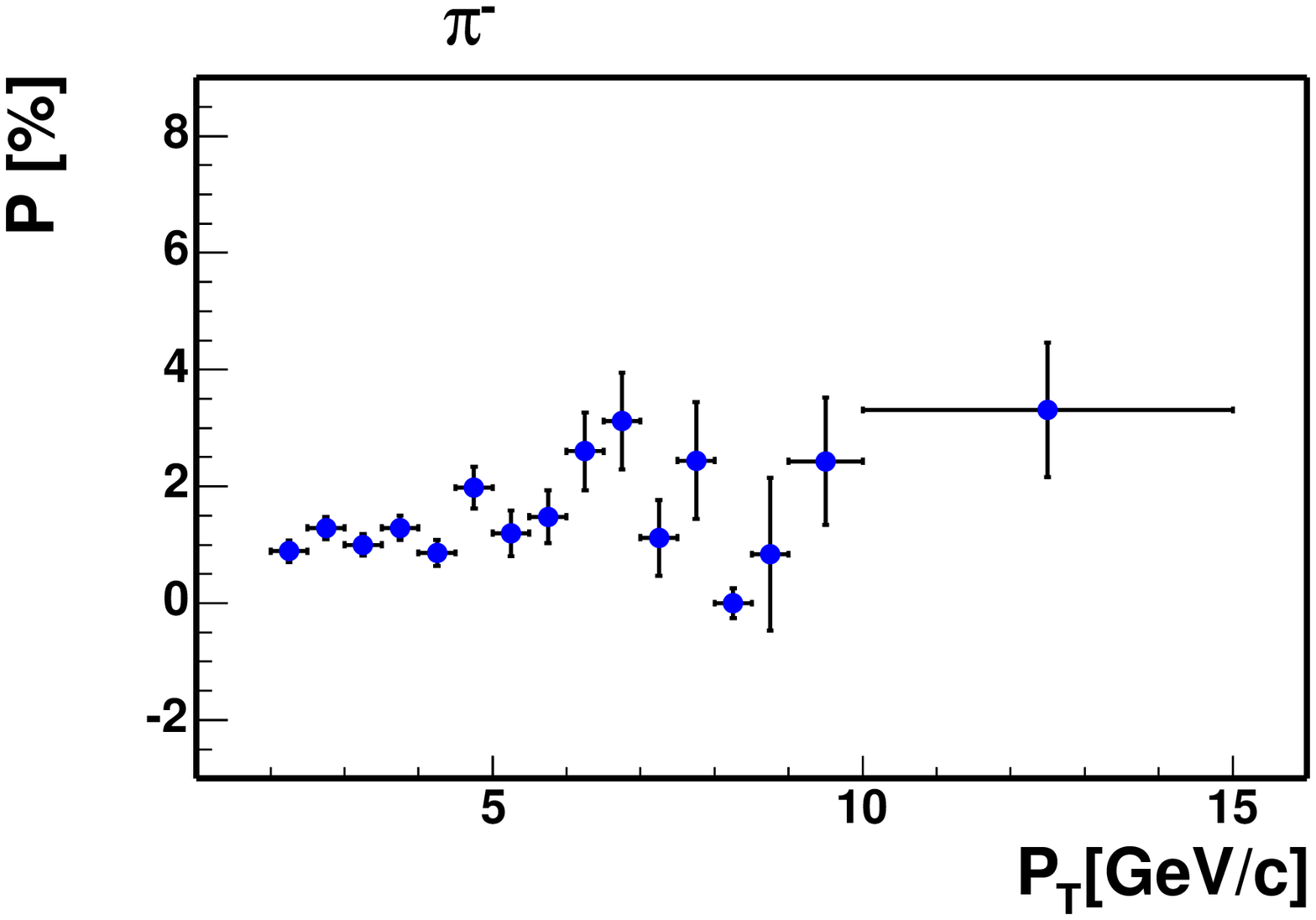}\\
  \includegraphics[width=200pt, angle=0]{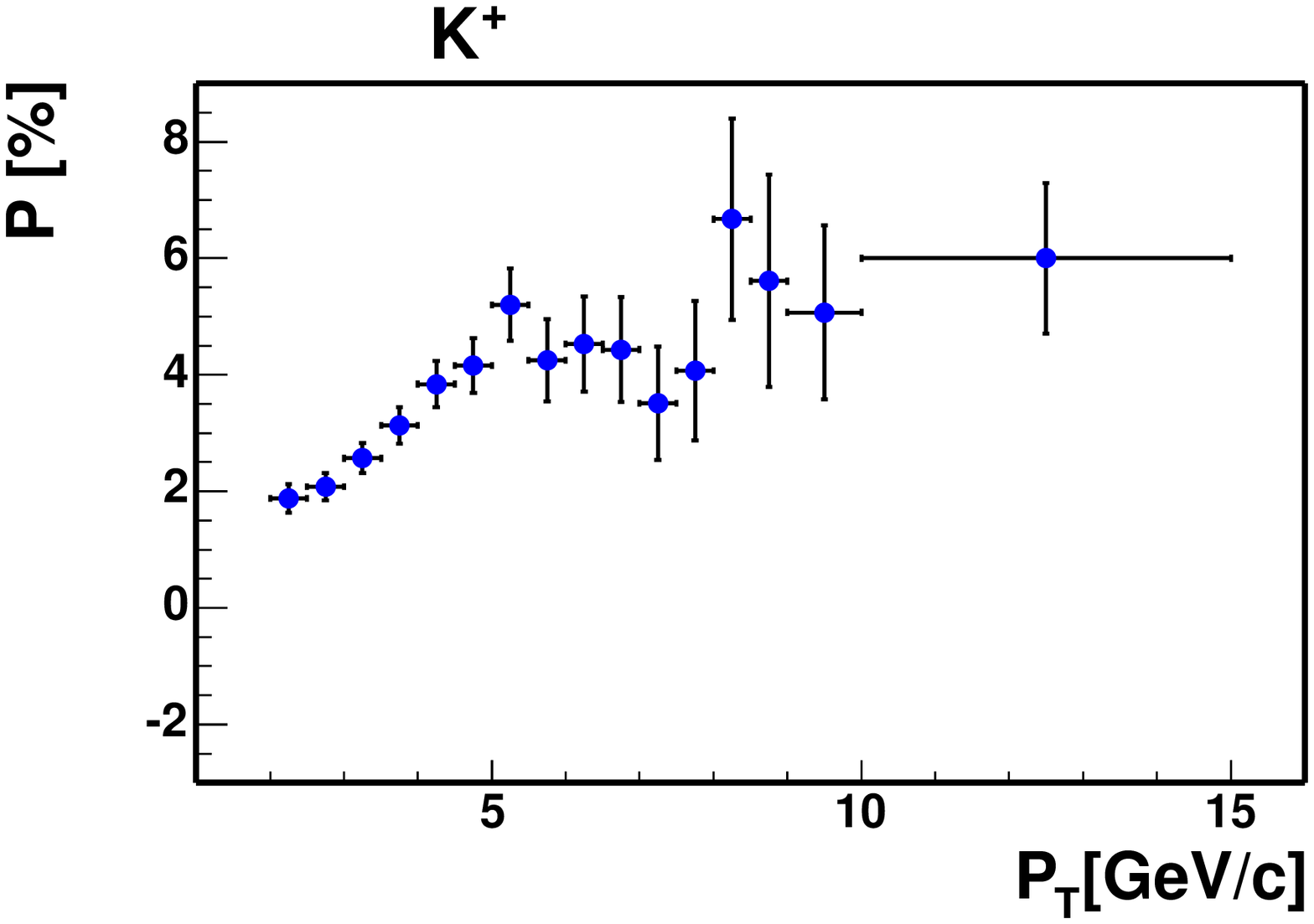} & 
  \includegraphics[width=200pt, angle=0]{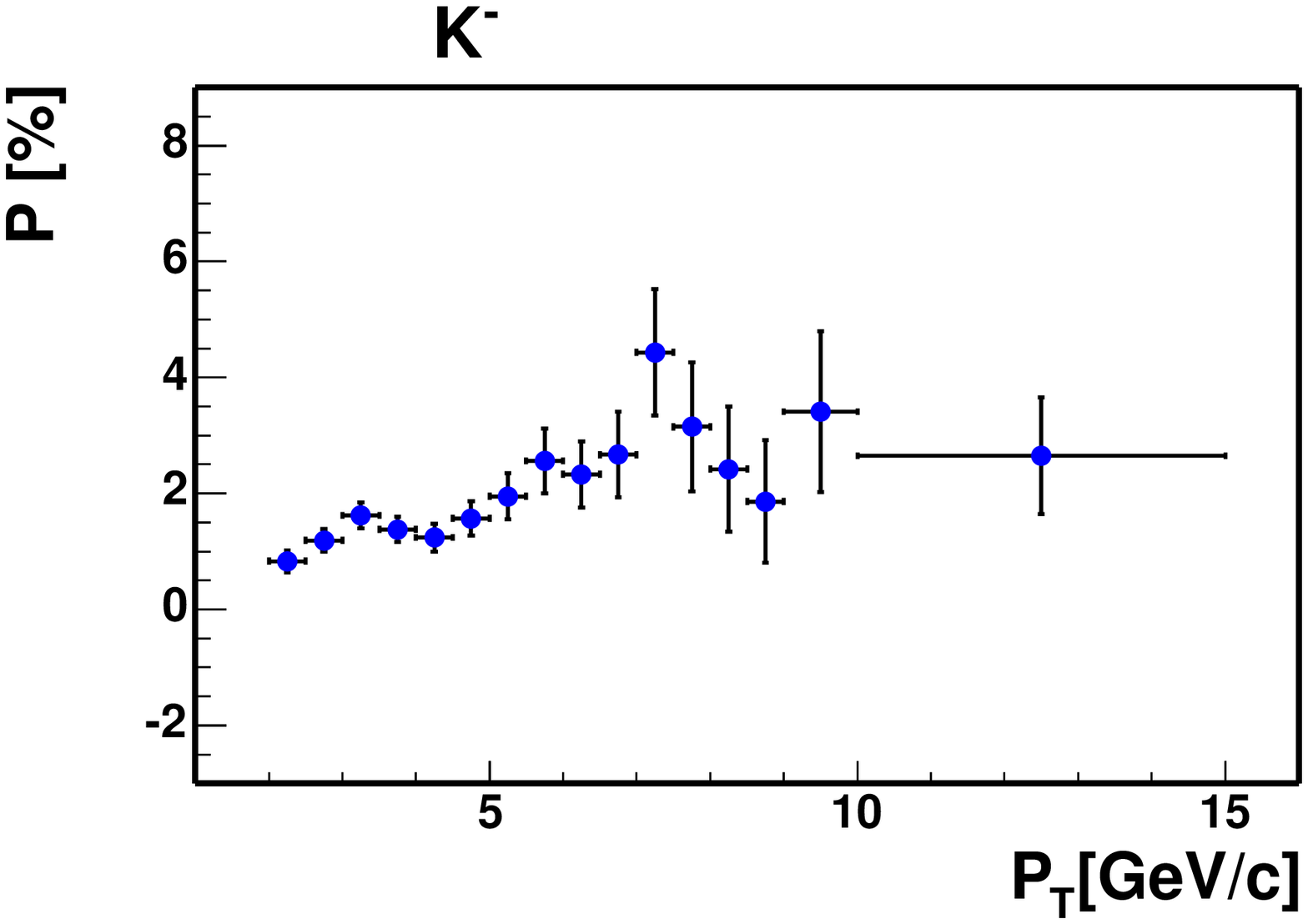}\\
  \includegraphics[width=200pt, angle=0]{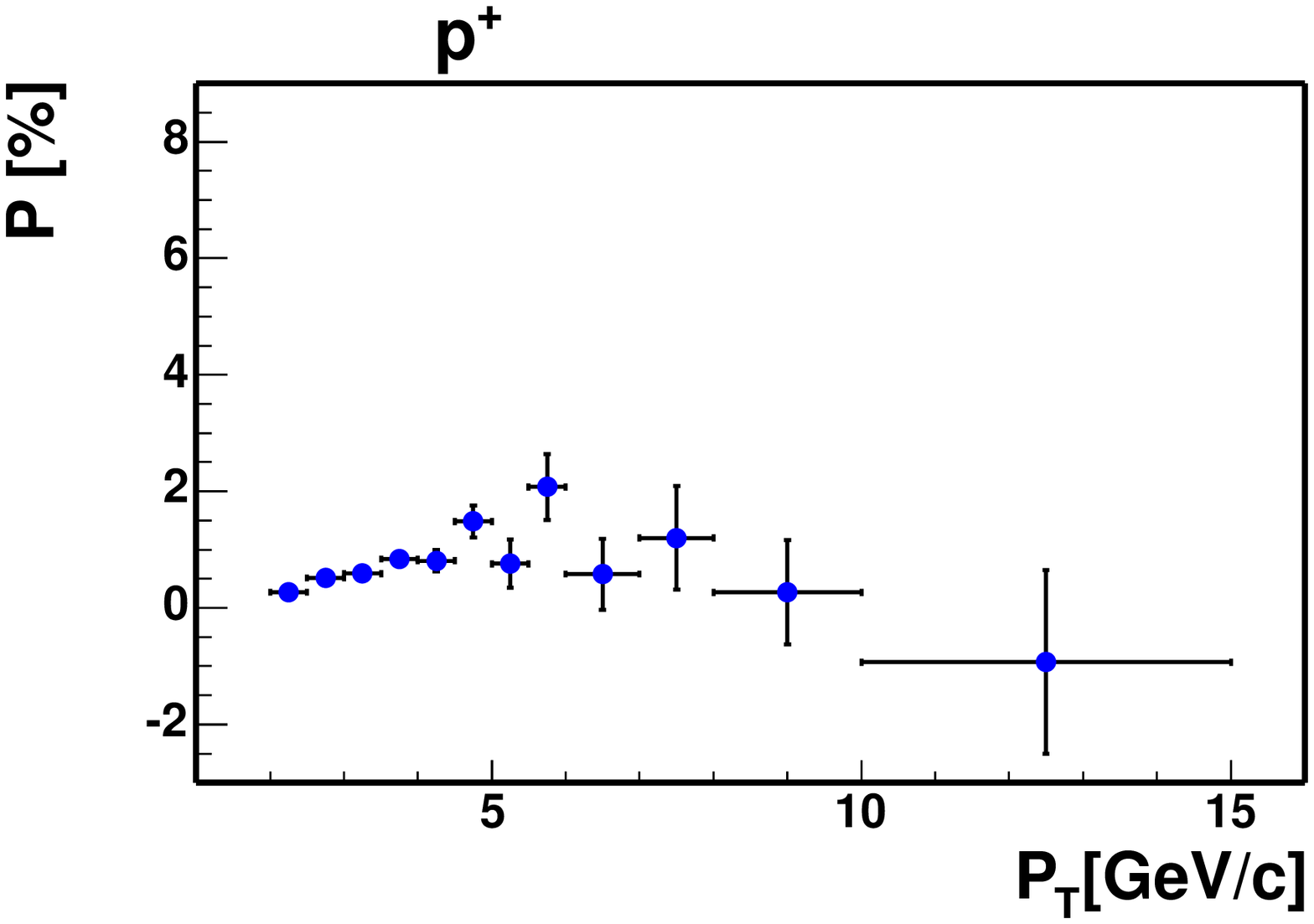} &
  \includegraphics[width=200pt, angle=0]{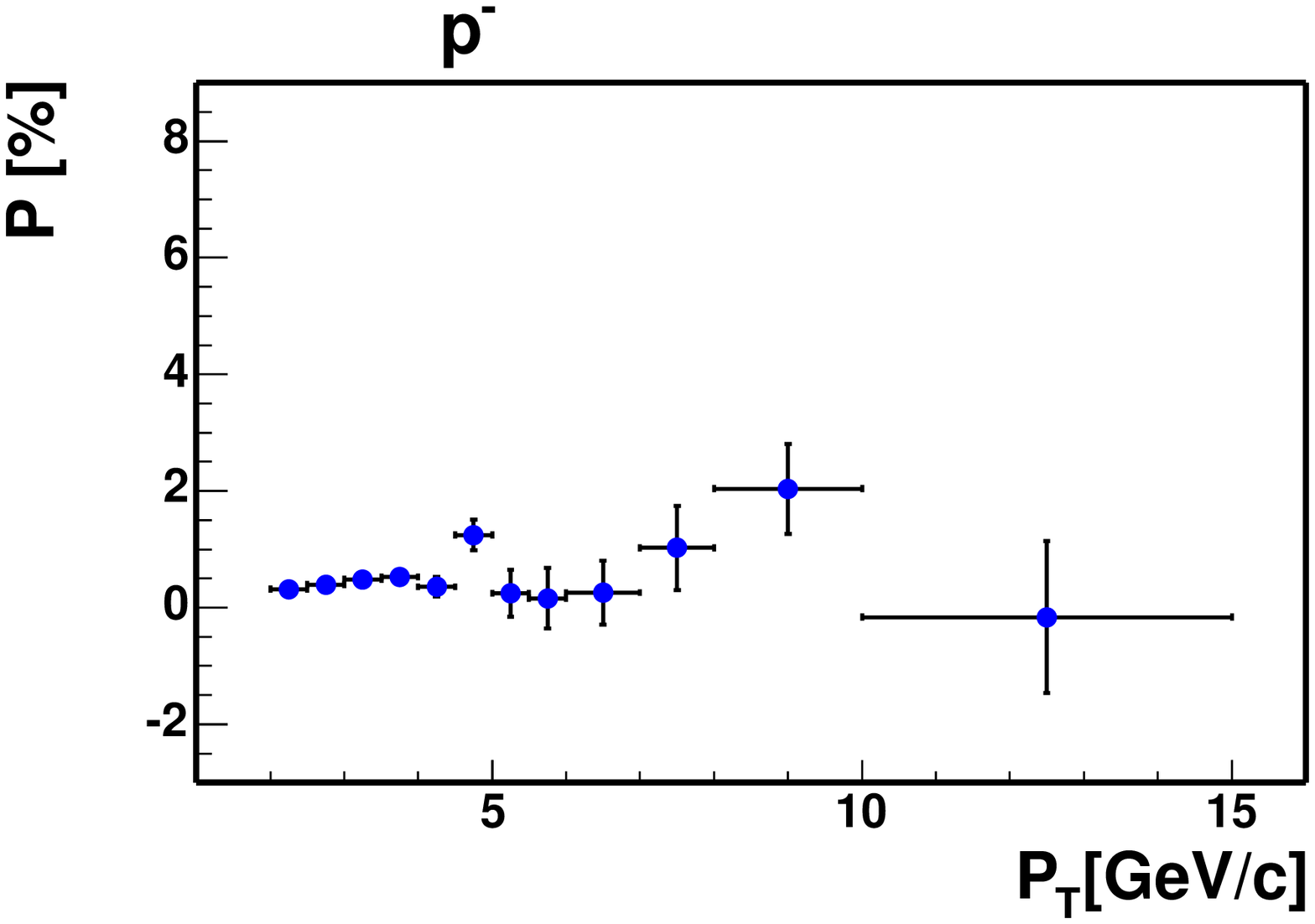}\\
  \end{tabular}
   \caption[Pion, kaon and proton fake probabilities]
{The probability for a pion, kaon or proton being misidentified as a muon 
in bins of transverse momentum (\pt) from the measurements by Ashmanskas, Harr
~\cite{ash:dmm} and Litvintsev\cite{dmitri:prepare}.
From the top left to the bottom right are $\pi^+$, $\pi^-$, $K^+$, $K^-$, $p$ 
and $\overline{p}$ fake probabilities.}
     \label{fig:fakerate}
     \end{center}
  \end{figure}

We use two methods to estimate the amount of contamination from fake muons in
our semileptonic signal. Each method uses a different way to obtain the 
number of (hadron track, charm hadron) candidates in our data. Both methods 
apply the previous CDF measurements of the probabilities for a real pion, kaon
 and proton to be misidentified as a muon. These measurements are performed 
by Ashmanskas and Harr~\cite{ash:dmm} using a pion and kaon sample from the 
\seqdstar decays, where \seqdzero, and by Litvintsev~\cite{dmitri:prepare} 
using a proton sample from the $\Lambda\rightarrow p^+ \pi^-$ decays.
The fake probability ${\cal P}_{\pi}$ (${\cal P}_K$, ${\cal P}_p$) is 
defined as the number of pions (kaons or protons) that pass the following 
muon identification cuts divided by the total number of pions 
(kaons or protons) inside the fiducial volume of CDF Central Muon 
Detector (CMU) and matched to an SVT track.
 \begin{itemize}
 \item The track is fiducial to the CMU and matched to an SVT track
 \item The track is associated with hits in the CMU
 \item The matching $\chi^2$ between the track and the hits in the CMU
	is less than 9.
 \end{itemize}
Figure~\ref{fig:fakerate} shows the ${\cal P}_{\pi}$, ${\cal P}_K$  
 measured in sixteen and ${\cal P}_{p}$ measured in twelve 
transverse momentum bins for positive and negative charged tracks, separately.

\subsubsection{Method I}
\begin{figure}[htb]
    \begin{center}
   \includegraphics[width=350pt, angle=0]{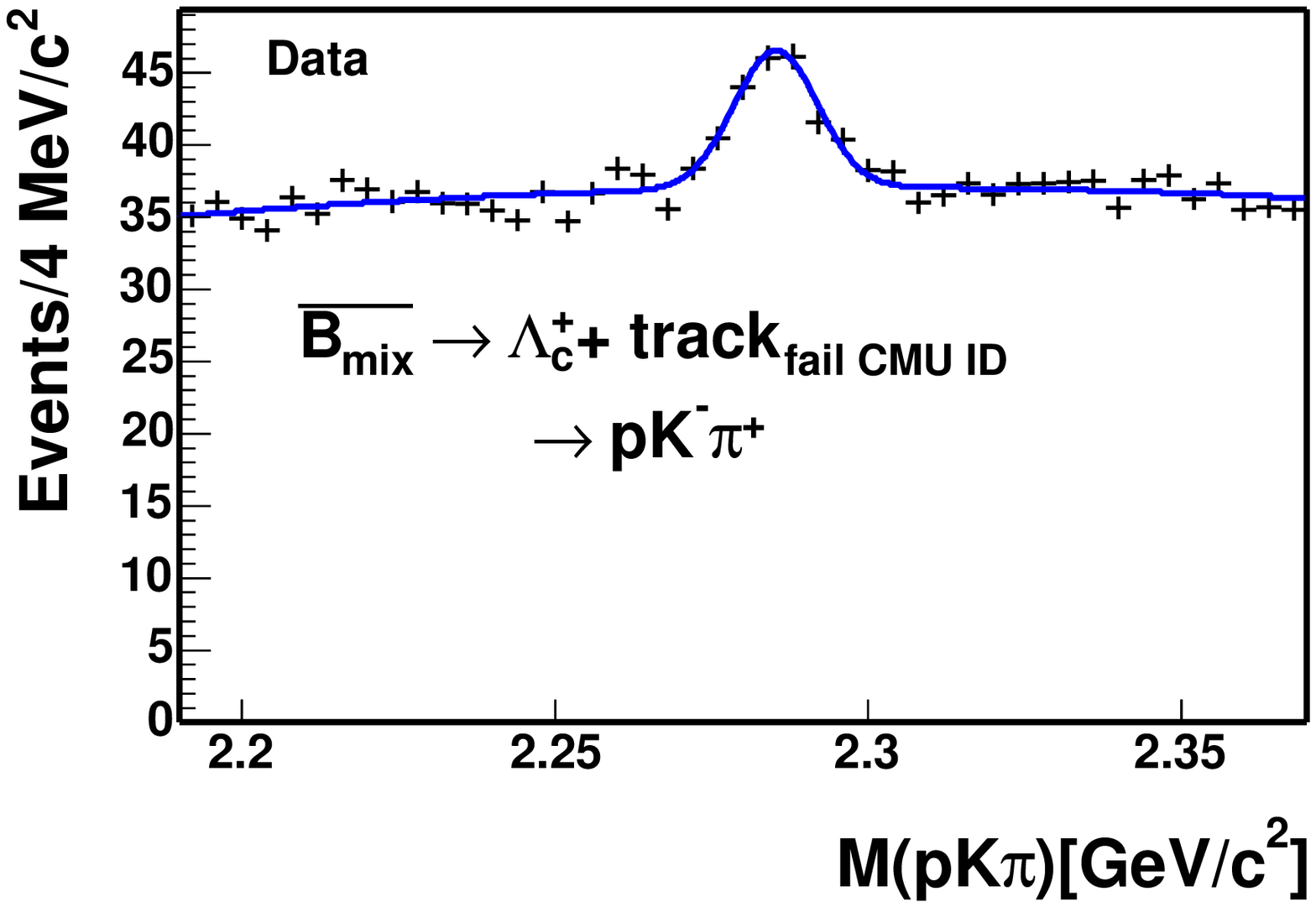}
     \caption[\mpkpi\ from $\overline{B}\rightarrow \Lc \mu_\mathrm{fake}$]
   { Fit of $\overline{B}\rightarrow \Lc \mu_\mathrm{fake}$ yield 
     after weighting the charged track which fails the muon ID cut with an 
     average muon fake probability. There 40~$\pm$~6 events in the peak. Fit 
     $\chi^2/\mathrm{NDF}$ = 55.6/39, probability = 4.1$\%$. A sideband 
     subtraction yields 44~$\pm$~25 events in the signal peak. 
     \label{fig:lcmfake}}
     \end{center}
  \end{figure}

The first method uses data to obtain the number of (hadron track,charm hadron)
 candidates, then Monte Carlo to determine the ratio of pions,
kaons and protons in the hadron tracks. We run the same signal reconstruction
program on the secondary datasets {\tt hbot0h} and {\tt hbot1i}.
We do not use the skimmed tertiary datasets (see Yu~\cite{cdfnote:strip} ) as 
the samples are biased by requiring at least one track in
the event matched to a muon stub in the muon detector. 
We look for a charged track which fails the muon identification 
cuts (TRK$^{fail}$). TRK$^{fail}$ and a charm hadron should form 
a displaced vertex and pass the same analysis cuts we apply to the signal. 
Each event is weighted with the fake probability (${\cal P}_\mathrm{avg}$) 
according to the momentum and the charge of TRK$^{fail}$. We then
 fit the weighted charm hadron 
mass distribution, i.e. $\mkpipi - \mkpi$, 
\mkpipi\ and \mpkpi, using the same functions as described in 
Section~\ref{sec-semimass}, to obtain the signal contamination from the 
fake muons. Since an event-weighted likelihood fit will not give a
proper error for the yield, a binned $\chi^2$ fit is performed.
${\cal P}_\mathrm{avg}$ is a weighted average of pion, kaon and proton 
fake probability (${\cal P}_{\pi}$, ${\cal P}_K$, ${\cal P}_p$). 
The weight $R_i$ is determined by the fraction of pions, kaons and  
protons in the \bmixdpimux\ and \bmixdx\ MC after 
analysis cuts: 
\begin{equation}
 \label{eq:fr}
  {\cal P}_\mathrm{avg} =  
	R_{\pi}{\cal P}_{\pi} + R_K{\cal P}_K + R_p{\cal P}_p,
\end{equation}
where 
 \[ R_{i} =  \frac{N_{i}}{N_{\pi}+ N_K+ N_p},\]
and  $i$ is $\pi$, $K$ or proton. 
The Monte Carlo is generated as described in Section~\ref{sec-mccom}. 
Decays of \Bd, $B^+$, $B_s$ and \Lb\ are generated
separately and decay tables include all the possible decays.
Each kind of \B\ hadron gives different $R_i$ and is weighted with 
the product of the production fractions, total branching ratios and 
the number of generated events. 
Table~\ref{t:bigfake1} summarizes the pion to kaon ratio and the number of 
fake muon candidates before and after weighting the events in our three 
different signals. See Figure~\ref{fig:lcmfake} for 
the weighted mass distribution of the \Lb\ mode.   

The uncertainties of the number of fake muon candidates 
come from three sources: 1. the uncertainty from the binned $\chi^2$ fit, 2. 
the uncertainties on the pion, kaon and proton fractions due to the finite 
Monte Carlo sample size, the uncertainties of the branching ratios and 
production fractions, and 3. 
the uncertainty on the measured fake probability. 
For the last source, we vary 
the fake rate in each momentum bin $\pm$ 1 sigma,  
independently. We then add the systematic shifts in quadrature to get the 
accumulative uncertainty. 
The number of fake muons using this method is about 3--5$\%$ of the 
inclusive semileptonic signals in the data.

\subsubsection{Method II}
The second method relies on the MC and the external input of the
branching ratios from the Particle Data Group (PDG)~\cite{pdg:2004}.
MC is run through the same reconstruction program for the data. 
Then we apply the same cuts as signal reconstruction and obtain the 
trigger and reconstruction efficiencies of \bmixdx\ and \bmixdpimux\ decays. 
We weight the MC events which pass the analysis cuts with the fake 
probability according to the momentum, the charge and the particle type 
of the track, $X_\mathrm{had}$. 
The particle identification of the track, $X_\mathrm{had}$, is
obtained by matching the hits on the reconstructed track with those on
the input simulated track. Together with the efficiency of hadronic mode, 
branching ratios of our hadronic signals, \bmixdx\ and \bmixdpimux\ 
from the PDG, we normalize the background to the observed number of 
hadronic signals in the data,
 \begin{equation}
  \frac{N_{\mathrm{fake}\;\mu}}{N_\mathrm{had}} = 
	\frac{{\cal B}(\overline{B} \rightarrow \Dstar\ (\D, \Lc) X) 
	\cdot \epsilon_{\mathrm{fake}\;\mu}}{f_{d,\mathrm{baryon}}
	\cdot {\cal B}_\mathrm{had} \epsilon_\mathrm{had}}.
 \label{eq:fake}
 \end{equation}
Equations~\ref{eq:brd}--~\ref{eq:brd2} use \D\ as an example to
show how we derive the \bmixdx\ and \bmixdpimux\ branching ratios 
from the existing information in the PDG.
\begin{eqnarray}
 \label{eq:brd}
 {\cal B}(\overline{B}\rightarrow \D X_\mathrm{had}\; l \nu_l\;
	\mathrm{anything})
 & = &\frac{7}{3}\cdot {\cal B}(\overline{B} \rightarrow \D \pi^+ \mu \; 
	\mathrm{ anything}) \nonumber \\
& + & \frac{7}{3}\cdot {\cal B}(\overline{B} \rightarrow \D \pi^- \mu \; 
	\mathrm{ anything}),  \\
 \label{eq:brd2}
 {\cal B}(\overline{B} \rightarrow \D X_\mathrm{had}\; \mathrm{anything}) 
 &= & {\cal B}(\overline{B} \rightarrow \D \mathrm{ anything}) \nonumber \\
 & -&  \frac{7}{3}\cdot{\cal B}(\overline{B} \rightarrow \D \mu \; 
	\mathrm{ anything}),
\end{eqnarray}
where the factor, $\frac{7}{3}$, comes from the fact that the 
branching ratios of muon and electron channels are equal and the branching 
ratio of the tau channel is scaled down by the ratio of the phase space,
$\sim \frac{1}{3}$. Therefore, we have to scale up the branching ratio
of the muon channel by $1+1+\frac{1}{3}=\frac{7}{3}$ to get the total
branching ratio of all the lepton channels.

Table~\ref{t:bigfake2} summarizes the parameters used to calculate the number 
of fake muon events, where the decay \bmixdx\ is denoted as mode ``1'' and 
\bmixdpimux\ is denoted as mode ``2'' in the table. 
The uncertainties on the number of fake muons originate from: 
the uncertainty on the hadronic yield, the relative efficiency, the uncertainty on 
the fake rate and the relative branching ratios. 
The dominant uncertainty is from the relative branching ratios. 
The number of fake muon backgrounds from method I is consistent
with the result using method II. We use the results of method I in the 
calculation of our final result of the relative branching ratios. In general, 
the fraction of fake muons is about 5$\%$ of the total semileptonic yield in 
the data.

\renewcommand{\arraystretch}{1.5}
  \begin{table}[tbp]
   \caption{Parameters for the number of fake muons: Method I.}
  \label{t:bigfake1}
   \begin{center}
   \begin{small}
  \begin{tabular}{|r|r|r|r|} 
   \hline
   & \dstarsemi\ & \dsemi\  & \lbsemi\  \\ \cline{2-4}
    N before weighting & 2953 $\pm$ 57  &  15343 $\pm$ 303 & 3560 $\pm$ 198\\
    $R_{\pi}$ & 0.937 $\pm$ 0.009 
              & 0.909 $\pm$ 0.005 
              & 0.71 $\pm$ 0.16 \\
    $R_{K}$ & 0.063 $\pm$ 0.009
            & 0.091 $\pm$ 0.005
            & 0.05 $\pm$ 0.08 \\
    $R_{p}$ & -- & -- & 0.24 $\pm$ 0.16\\ \cline{2-4} 
    $N_{\mathrm{fake}\;\mu}$ & 
	44 $\pm$ 3 & 
        230 $\pm$ 19 & 
   	40 $\pm$ 9 \\ 
    \hline 
 \end{tabular}
 \end{small}
 \end{center}

   \caption{Parameters for the number of fake muons: Method II.}
  \label{t:bigfake2}
   \begin{center}
   \begin{small}
  \begin{tabular}{|r|r|r|r|} 
   \hline
   & \dstarsemi\ & \dsemi\  & \lbsemi\  \\ \cline{2-4}
    ${\cal B}_\mathrm{had}$ $\%$ & 0.276 $\pm$ 0.021  & 0.276 $\pm$ 0.025 & 
	\brlbhad \\
    ${\cal B}_1$ $\%$
         & 10.9 $\pm$ 2.1 & 17.7 $\pm$ 2.4 & 4.8 $\pm$ 3.0 \\
    $\frac{\epsilon_1}{\epsilon_\mathrm{had}}$ 
      & 0.0038 $\pm$ 0.0004 &  0.0022 $\pm$ 0.0002& 0.0029 $\pm$ 0.0003\\ 
    ${\cal B}_2$ $\%$  
	& 1.3 $\pm$ 0.3 & 1.8 $\pm$ 0.6 & $<$ 1.23  \\
    $\frac{\epsilon_2}{\epsilon_\mathrm{had}}$ 
      &0.0005 $\pm$ 0.0002 & 0.0010 $\pm$ 0.0002  & 0.0002 $\pm$ 0.0001\\ 
    $N_\mathrm{had}$ &  \ndstarhad & \ndhad & \nlbhad \\ \cline{2-4} 	
    $N_{\mathrm{fake}\;\mu}$ &	
	45 $\pm$ 11  & 
	220 $\pm$ 41 & 
	28 $\pm$ 34 \\ 
   \hline
 \end{tabular}
 \end{small}
 \end{center}
 \end{table}

\subsubsection{Like-sign Combination}
Note that we do not use the like sign combination  
(i.e. the charm hadron and the muon have the same sign of charges) to 
estimate the fake muon background for two reasons: First, two different 
\B\ hadrons from the \bb\ in the event can produce a real muon and a real 
charm of the same 
charge sign when the \B\ hadrons in 
the event have opposite flavors and one \B\ hadron decays semileptonically.
Second, the two track trigger, used for this analysis, requires a pair of 
tracks with opposite charges. The trigger requirement greatly reduces
the number of like-sign (wrong-sign) candidates and introduces large 
statistical errors for the number of fake muons.

\subsubsection{Fake muons from \bb\ and \cc}
One type of fake muons is not included in the previous subsections. These
fake muons stem from \bb, \cc\ to two \B\ or charm hadrons then decay into 
a charm signal, a hadron track misidentified as a muon and other missing 
particles. 
A study at the generator level for the \dsemi\ mode is done using the \bb\ and 
\cc\ \pythia~\cite{pythia:manual} Monte Carlo datasets as described in 
Section~\ref{sec-cbmethod}. We apply analysis-like cuts on the Monte Carlo.
We weight the events that pass the cuts with the muon fake probability 
according to the ``muon'' candidate momentum, 
charge and the true particle identification: a kaon, a pion or a proton.
Then we compare the number of weighted events with the number of charm 
hadron and real muon combinations, i.e, a \bb\ and \cc\ background as
described in Section~\ref{sec-ccbb}. We find that fake muons from \bb\
and \cc\ is about 10$\%$ of the \bb\ and \cc\ background with real muons.
See Table~\ref{t:fakeccbb}.
From Section~\ref{sec-ccbb}, we show that the \bb\ and \cc\ background with
real muons is at the 1$\%$ level. Therefore, we conclude that \bb\ and \cc\ 
background with fake muons is about or less than 0.1$\%$ and can be ignored.

\renewcommand{\arraystretch}{1}
  \begin{table}[tbp]
   \caption{Fake muons from \bb\ and \cc.}
  \label{t:fakeccbb}
   \begin{center}
   \begin{normalsize}
  \begin{tabular}{|l|r|r|} 
   \hline 
   & \bb & \cc \\ 
   \hline
    N$_{gen}$ &  43454949 &   89718181 \\
    Real muon N$_{pass}$ &  15 &  35 \\
    Fake muon N$_{pass}$ &  1.8 &  0.4 \\
   \hline
 \end{tabular}
 \end{normalsize}
 \end{center}
 \end{table}


      
\section{\bb\ and \cc\ Backgrounds}
\label{sec-ccbb}
\begin{figure}[tbp]
  \begin{center}
\includegraphics[width=180pt, angle=0]
	{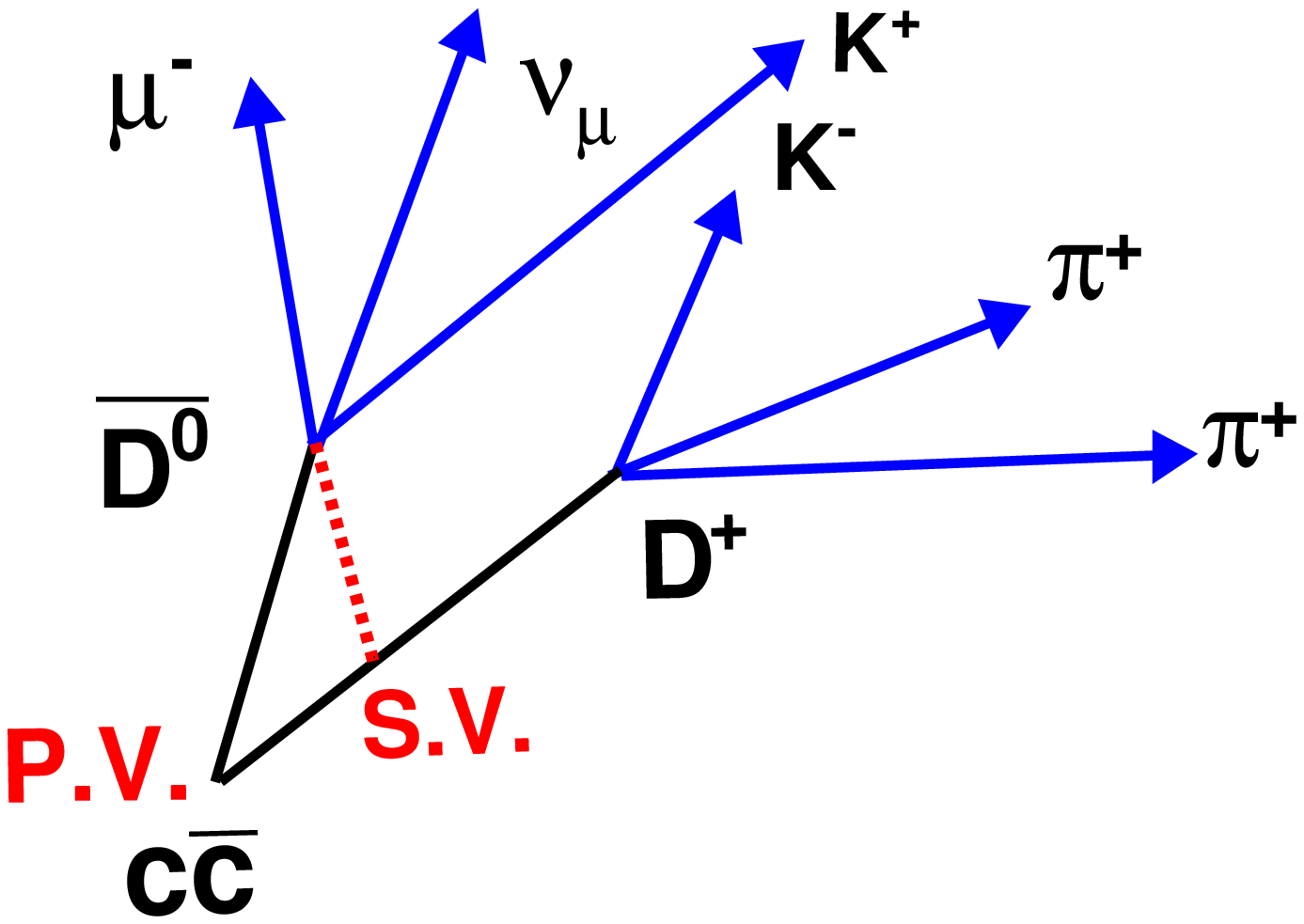}
\includegraphics[width=210pt, angle=0]
	{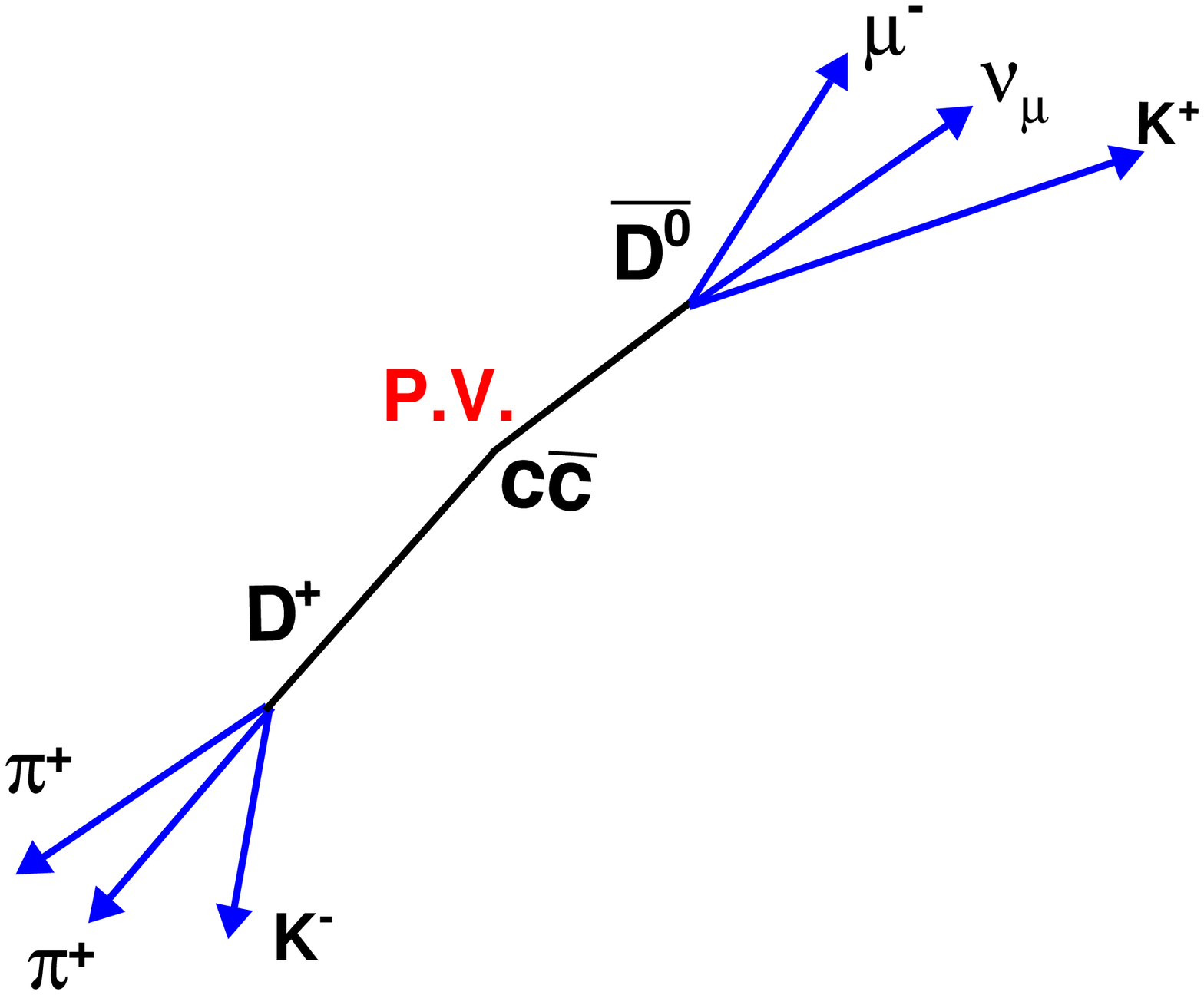}
 \caption[Charm hadrons from \cc\ with small and big $\Delta \phi$]
	{Charm hadrons from \cc\ with small (left) and big (right) 
	$\Delta \phi$. In the left figure, the muon from the semileptonic 
	decay of $\overline{D}^0$ and the \D\ forms a secondary vertex 
	and fake our \dsemi\ signal. In the right figure, $\Delta \phi$ 
	between two charm hadrons is too big and the daughters can not form 
	a secondary vertex.}
 \label{fig:cbexample} 
 \end{center}
 \end{figure}

 When the azimuthal angle ($\Delta \phi$) between \bb\ or \cc\ quark pair is 
small, daughters of two heavy flavor hadrons from the fragmentation of \bb\ or
 \cc\ appear to come from the same decay vertex, see Figure~\ref{fig:cbexample}
. Here $\Delta \phi$ is defined as the opening angle in the plane perpendicular
 to the proton and antiproton beam axis. If one hadron decays semileptonically,
 and the other hadron decays into a charm final state, such as \seqdstar, 
\seqd, and \seqlc, the muon from the semileptonic decay, together with the 
charm, may fake our semileptonic signal. Production mechanisms and an 
estimate of the amount of \bb\ and \cc\ backgrounds are discussed below.

\subsection{\bb\ and \cc\ Production Mechanism}
\label{sec-cbprod}
 In $p\overline{p}$ collisions, the b or c quarks may be single or pair 
produced by the electroweak and the strong (QCD) processes.
 The b or c quark production cross-section for the electroweak process 
$\sigma \cdot {\cal B} (p\overline{p} \rightarrow W \rightarrow bc)$ is 
around 0.01~$\mu b$ and is derived from the CDF measurement of the inclusive 
W cross-section by Halkiadakis, \etal~\cite{Acosta:2004uq}. The \bb\ and \cc\ 
production cross-sections for the QCD process are around 50 and 200$\mu b$ 
respectively from the \pythia~\cite{pythia:manual} Monte Carlo, when the total
 transverse momenta of the hard scattering, i.e. the part of the interaction 
with the largest momentum scale, is greater than 5 \gevc\ and at least one b 
or c quark has \pt\ $>$ 4.0 \gevc, pseudo-rapidity $\eta$ $<$ 1.5.
The \bb\ and \cc\ production rates from the electroweak process 
are about five thousand times smaller than the QCD processes. Therefore, only
pair production by the QCD processes are discussed here. 

\begin{figure}[tbp]
  \begin{center}
\resizebox{350pt}{!}{\includegraphics*[120pt,150pt][520pt,360pt]
	{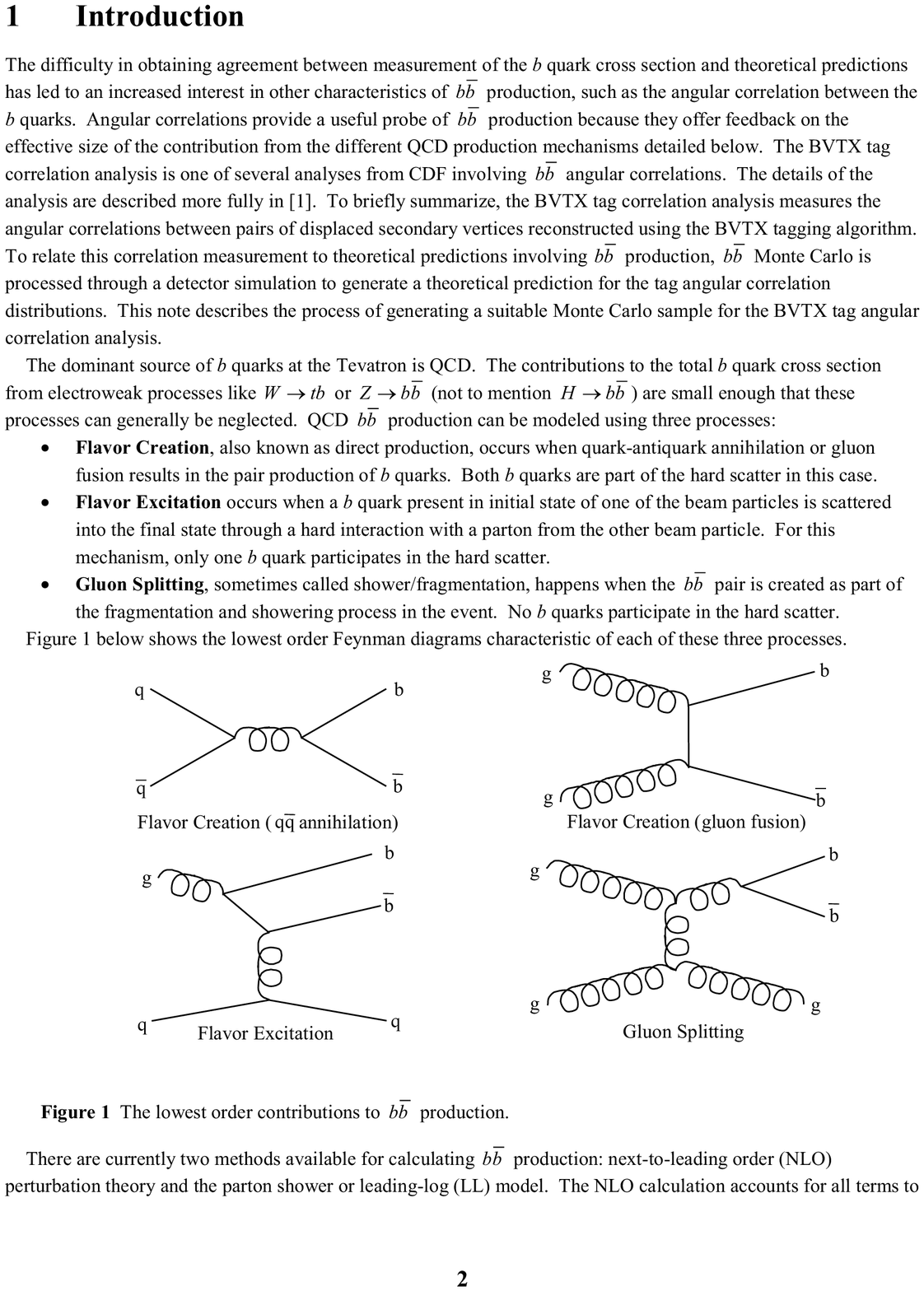}}
    \caption[representative lowest order Feynman diagrams of flavor 
creation, flavor excitation and gluon splitting]{
Representative lowest order Feynman diagrams (without loops or radiative 
corrections) of quark annihilation, gluon fusion, flavor excitation and 
gluon splitting. Details of these processes may be found in 
Lannon\cite{lannon:bpythia}.}
  \label{fig:cbproc}
  \end{center}
  \end{figure}

Figure~\ref{fig:cbproc} shows the leading and next-to-leading order Feynman 
diagrams for \bb(\cc) production by the QCD processes from 
Lannon\cite{lannon:bpythia}. The QCD process that contributes the production 
at leading order is flavor creation; which includes quark anti-quark 
annihilation ($q\overline{q} \rightarrow$ \bb\ or \cc) and gluon fusion 
($gg \rightarrow$ \bb\ or \cc). 
The distribution of the azimuthal angle ($\Delta \phi$) between two produced 
b(c) quarks peaks at 180 degrees. The reason is that the $q$ and 
$\overline{q}$ (or $g$ and $g$) come from the proton and the anti-proton 
separately. The initial total momenta of the gluon $q\overline{q}$ pair is 
zero. The \bb\ (\cc) pair are produced back-to-back to balance the momentum in 
the final state.

 The next-to-leading order (NLO) processes, flavor excitation and gluon 
splitting, contribute at the same level as the flavor creation~\cite{lannon:bpythia}. Flavor excitation refers to the following process: The gluons within 
one of the beam particles in the initial state split into a \bb\ (\cc) pair. 
One of b(c) quarks is scattered out of the initial state into the final state 
by a gluon or a light quark from the other beam particle. The other b(c) quark
 is not involved in the hard scattering process. The $\Delta \phi$ of \bb\ 
(\cc) from flavor excitation is more evenly distributed than flavor creation.

\begin{figure}[tbp]
  \begin{center}
\resizebox{350pt}{!}{\includegraphics*[164pt,270pt][454pt,450pt]
	{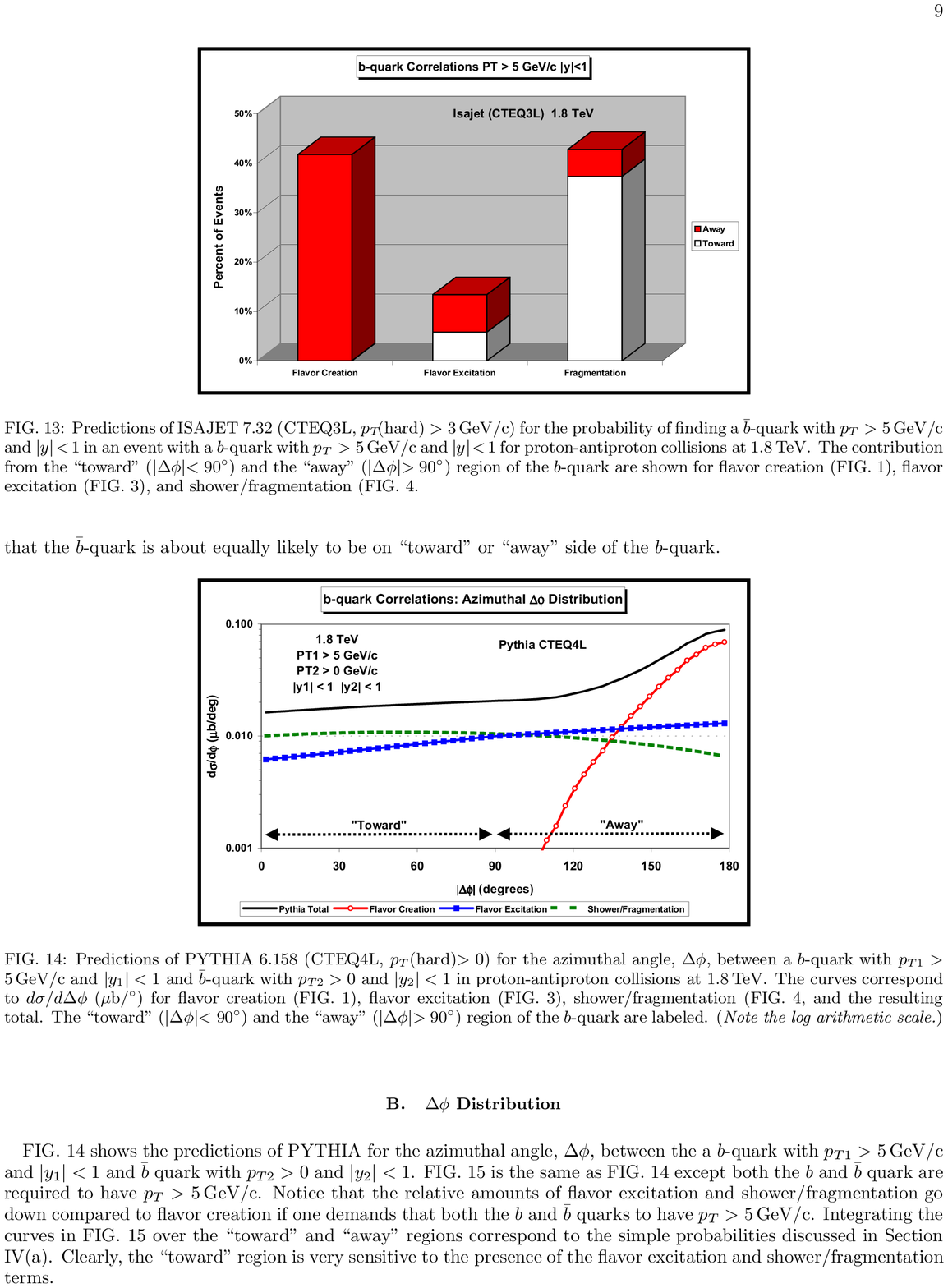}}
  \end{center}
  \begin{center}
\resizebox{350pt}{!}{\includegraphics*[164pt,558pt][452pt,738pt]
	{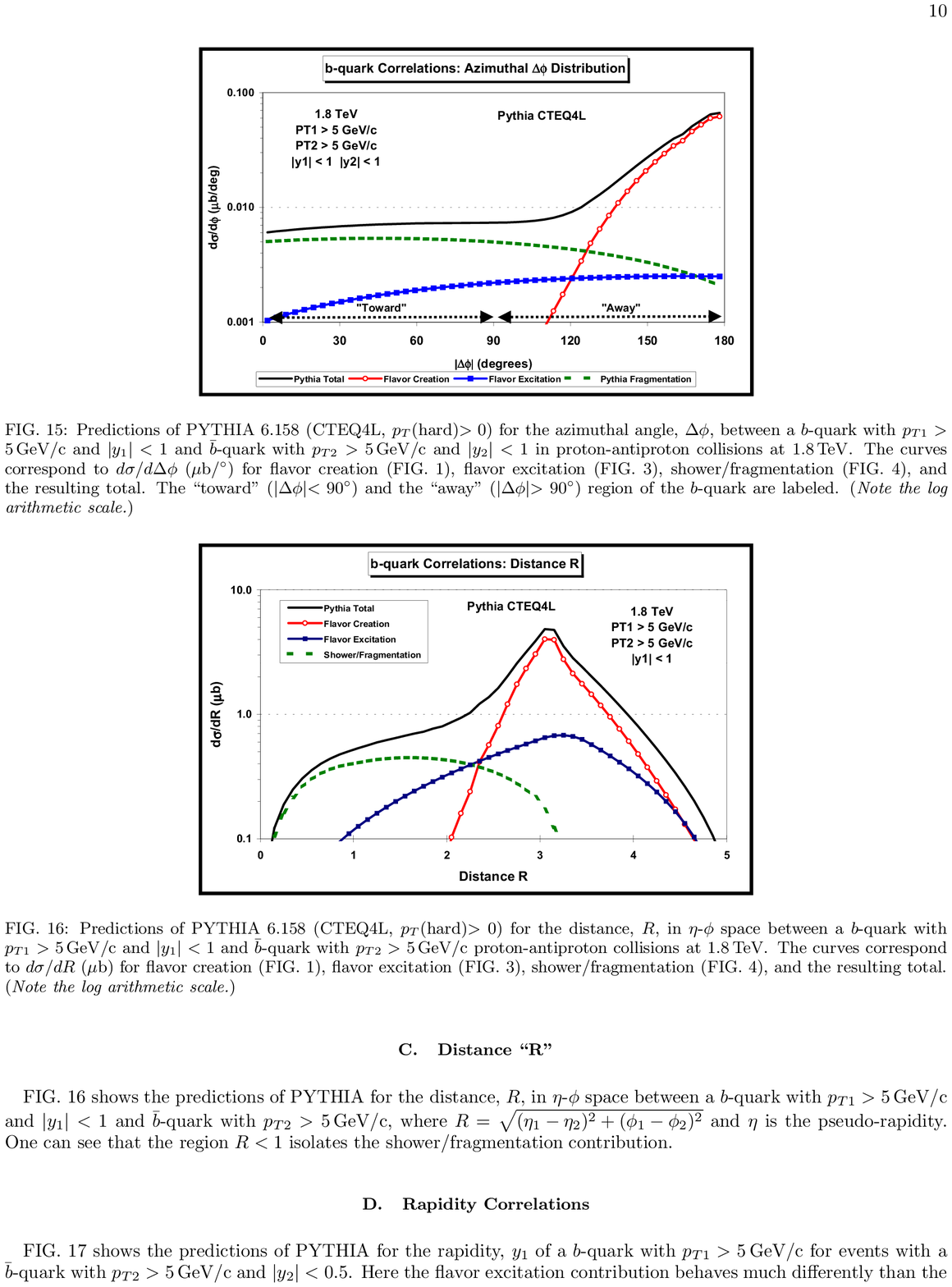}}
  \end{center}
    \caption[{\tt PYTHIA CTEQ4L} predictions of the azimuthal angle between b 
quarks]
{$\Delta \phi$ between $b$ and $\overline{b}$ from the study of Field
~\cite{rfield:pythia} using {\tt PYTHIA CTEQ4L}. 
Both b quarks have $|y|$~$<$~1. In the top plot, one b quark has \pt\ $>$ 5 
\gevc\ and the other b quark does not have any \pt\ cut. In the bottom plot, 
both b quarks have \pt\ $>$ 5 \gevc.}
  \label{fig:cbdeltaphi1}
  \end{figure}

Gluon splitting refers to the process when no b(c) quarks are involved in the 
hard scattering. One gluon in the final state splits into \bb\ (\cc) pair. If 
the gluon is soft, $\Delta \phi$ will be a flat distribution. If the gluon is 
hard, the daughters of the gluon, \bb\ (\cc) tend to move co-linearly and have 
small $\Delta \phi$.  Figure~\ref{fig:cbdeltaphi1} shows the azimuthal angle 
distribution between two b quarks from the study of Field~\cite{rfield:pythia} 
using {\tt PYTHIA CTEQ4L} prediction.

\subsection{Background Estimate}
\label{sec-cbmethod}
 The amount of \bb\ and \cc\ background is normalized to the number of
events observed in the hadronic modes in the 
the data,
 \begin{equation}
 \frac{N_{\bb, \cc}}{N_\mathrm{had}} = \frac{\sigma_{\bb, \cc} \cdot \sum_{i} 
	\sum_{j} f^i {\cal B}_j \epsilon_j}{\sigma_{\Bd, \Lb} \cdot 
	{\cal B}_\mathrm{had} \epsilon_\mathrm{had}}.
\label{eq:ccbb}
\end{equation}
Here, $i$ represents the species of b(c) hadrons and $j$ represents the decay 
modes which could contribute to \bb\ and \cc\ backgrounds. $f^i$ 
stands for the production fraction ratio for species $i$. 
${\cal B}_j$ and $\epsilon_j$ are the branching ratio and the efficiency of 
$j^{th}$ decay mode. The following subsections detail the
methods to estimate 
$\sigma_{\bb, \cc} \cdot \sum_{i} \sum_{j} f^i {\cal B}_j \epsilon_j$ and 
$\sigma_{\Bd, \Lb} \cdot {\cal B}_\mathrm{had} \epsilon_\mathrm{had}$ in 
Equation~\ref{eq:ccbb}. We do not use 
detector and trigger simulations to obtain the efficiencies for the following 
reasons: First, detector and trigger simulations are time and CPU intensive. 
Second, we will find the contribution of this background is quite small 
compared with the other backgrounds. Third, we care about 
the efficiency ratio of the background to the signal, not the absolute 
efficiency. Our studies show that generator level Monte Carlo 
gives a good approximation. For instance, the relative efficiency
$\frac{\epsilon(\lbhad)}{\epsilon(\lbsemi)}$ is  3.31 $\pm$ 0.05 from the full 
detector simulation and 3.23 $\pm$ 0.01 from the generator level  
simulation. The difference is only about 2.5$\%$. Similar results are
obtained from the relative efficiencies of our other signals.

\subsubsection{Background: $\sigma_{\bb, \cc} \cdot \sum_{i} \sum_{j} f^i {\cal B}_j \epsilon_j$}
Our estimate of 
$\sigma_{\bb, \cc} \cdot \sum_{i} \sum_{j} f^i {\cal B}_j \epsilon_j$ relies 
heavily on the Monte Carlo. We use \pythia\ version 6.2~\cite{pythia:manual} 
and to generate \bb\ and \cc\ events, we include the QCD processes mentioned 
in Section~\ref{sec-cbprod}: flavor creation, flavor excitation and gluon 
splitting (MSEL=1). We further require the \pt\ of the hard scattering be 
greater than 5 \gevc. Events with $b$ quarks \pt\ greater than 4.0 \gevc\ and 
pseudo-rapidity less than 1.5 are collected into the {\tt nbot90} sample. 
Events with $c$ quarks which satisfy the same kinematic cuts are 
collected into the {\tt nbota0} sample. Note that {\tt nbot90} and 
{\tt nbota0} have small overlap when both $b$ and $c$ quarks are produced 
and are above the \pt\ and pseudo-rapidity thresholds. Details of 
{\tt nbot90} and {\tt nbota0} datasets could be found in~\cite{bmc:nbot90}.  
\pythia\ \bb\ and \cc\ cross-sections are used for $\sigma_{\bb, \cc}$.
 
The product of the efficiency, branching ratio and production fraction from 
all the modes, $\sum_{i} \sum_{j} f^i {\cal B}_j \epsilon_j$, is obtained
using the following steps: First, we identify the \bb\ (\cc) production 
mechanism to which we are most sensitive. In this pass, we only study the 
background that forms a $D^+\mu^-$ signature since \D\ has longer lifetime
than \Dstar, \Lc\ and \dsemi\ suffers larger \cc\ background 
contamination compared to the other two modes. We need to achieve higher
accuracy for the estimate of $\sum_{i} \sum_{j} f^i {\cal B}_j \epsilon_j$. 
Consequently, for the \bb\ background, we re-decay {\tt nbot90} sample ten 
times, i.e. we re-use the same kinematic distribution of the parent 
hadrons from {\tt nbot90} ten times but decay the hadrons with independent 
random numbers and force the decay \seqd. For the \cc\ background, we force
the decay of \seqd\ and require that all the negative charged charm hadron 
decay semileptonically. 

Then, a generator level two track trigger filter 
({\tt SvtFilter}) is applied. We further identify any combination of \D\ and 
a muon which passes the generator-level, analysis-like cuts found in 
Table~\ref{t:gencut}. In order to avoid double counting (count \cc\
as \bb\ background in {\tt nbot90} and \bb\ as \cc\ background in 
{\tt nbota0}) due to the overlap of {\tt nbot90} and {\tt nbota0} samples, 
the ancestors of the muon and charm hadrons are retrieved by tracing the true 
information from the generator. If both muon and charm hadron come from the 
same \B\ hadron, the combination is rejected. If the muon and charm hadron 
come from different \B\ hadrons, the combination is categorized into \bb\ 
background, otherwise, the combination is categorized as a \cc\ background. 
We find that for both \bb\ and \cc, more than 90$\%$ of the events that 
pass the cuts are from gluon splitting. Therefore, we are most sensitive to 
the ``gluon splitting'' mechanism. Table~\ref{t:sensitive} summarizes the 
background contributions from different production processes.

\renewcommand{\arraystretch}{1.1}
  \begin{table}[tbp]
   \caption{Summary of \bb\ and \cc\ production mechanisms and our relative
sensitivity for reconstructing the event in our semileptonic sample.}
  \label{t:sensitive}
   \begin{center}
   \begin{normalsize}
  \begin{tabular}{|l|r|r|} 
   \hline 
   & \bb\ background & \cc\ background\\ 
   \hline
    N$_{gen}$ & 219093011 & 21996889 \\
    N$_{pass}$ & 75 & 62 \\
    N$_{gluon}$ & 70 & 57 \\
    N$_{excitation}$ & 5 & 5 \\
    N$_{creation}$ & 0 & 0 \\
   \hline
    f$_{gluon} (\%)$ & 93 $\pm$ 3 & 92 $\pm$ 3\\
    f$_{excitation} (\%)$ & 7 $\pm$ 3 & 8 $\pm$ 4 \\
    f$_{creation} (\%)$ & 0 & 0 \\
   \hline
 \end{tabular}
 \end{normalsize}
 \end{center}
 \end{table}

Second, we filter the gluon splitting events and re-decay the b(c) hadrons in
{\tt nbot90} and {\tt nbota0}  
ten times with the procedure described above. For the \bb\ background 
estimate, we let
all the b hadrons and negative charged charm hadrons decay freely,
but force the decays of the positive charged charm hadrons in two ways:
 \begin{itemize}
 \item \seqdstar, \seqdzero\ for the background of \dstarsemi
 \item \seqd\ and \seqlc\ for the background of \dsemi\ and \lbsemi
 \end{itemize}
Then the {\tt SvtFilter} and the cuts listed in Table~\ref{t:gencut} are 
applied. We divide the number of reconstructed events by the number of 
generated events and get $\sum_{i} \sum_{j} f^i {\cal B}_j \epsilon_j$.
Table~\ref{t:cbbb} lists the parameters for the \bb\ background.

For the \cc\ background estimate, we force the decays of both 
positive and negative charged charm hadrons. The positive charged charm 
hadrons are forced to decay into the modes listed above. The negative charged 
charm hadrons are forced to decay into semileptonic modes individually for
$D^-$, $\overline{D}^0$, $D_s^-$ and $\Lambda_c^-$. 
As the semileptonic decay modes of these four charm hadrons are all
different, we separate the events into four classes denoted by the parent 
charm particles. 
After applying {\tt SvtFilter} and the cuts listed in Table~\ref{t:gencut}, 
we obtain
$\sum_{j} f^i {\cal B}_j \epsilon_j$ for each class. Then we
multiply the semileptonic branching ratios for each kind of charm hadron
with its $\sum_{j} f^i {\cal B}_j \epsilon_j$ and sum them up to get
the total amount of $\sum_{j} f^i {\cal B}_j \epsilon_j$  for the \cc\ 
background, 
\begin{eqnarray}
\sum_{i} \sum_{j} f^i {\cal B}_j \epsilon_j(total) & = & \sum_{i} \sum_{j} f^i {\cal B}_j \epsilon_j(D^-){\cal B}(D^-\rightarrow \mu X) \nonumber \\ & & + 
\sum_{i} \sum_{j} f^i {\cal B}_j \epsilon_j(\overline{D}^0){\cal B}(\overline{D}^0\rightarrow \mu X) \nonumber \\ & & + 
\sum_{i} \sum_{j} f^i {\cal B}_j \epsilon_j(D_s^-){\cal B}(D_s^-\rightarrow \mu X) \nonumber \\ & & +
\sum_{i} \sum_{j} f^i {\cal B}_j \epsilon_j(\Lambda_c^-){\cal B}(\Lambda_c^-\rightarrow \mu X).
\label{eq:cc}
\end{eqnarray}
Table~\ref{t:cbcc} lists the parameters for the \cc\ background.

In both \bb\ and \cc\ background estimates, since we force the positive 
charged charm hadron to decay into the same final state as our charm signals, 
we have to multiply the 
final result by two to include the contribution from both charge states. 
Table~\ref{t:cbfinal} lists $N_{\bb}$ and $N_{\cc}$ in our three different 
signals after multiplying the ratio in Equation~\ref{eq:ccbb} with the
observed number of events in the hadronic signals.

\renewcommand{\arraystretch}{1.2}
  \begin{table}[h]
   \caption{Generator-level analysis-like cuts for \bb\ and \cc\ background
	study.}
  \label{t:gencut}
   \begin{center}
   \begin{normalsize}
  \begin{tabular}{|l|l|} 
   \hline 
   Parameter & Cut Value \\
   \hline
   \hline
  \pt\ of all tracks & $>$ 0.5 \gevc         \\
  \pt\ of $\mu$ ($\pi_B$) & $>$ 2.0 \gevc               \\
   \hline
  $\eta$ of all tracks & $<$  1.2              \\
  $\eta$ of $\mu$ ($\pi_B$) & $<$ 0.6                     \\
   \hline
  \pt\ of four tracks & $>$ 6.0 \gevc              \\
  \pt\ of charm hadron & $>$ 5.0 \gevc             \\
   \hline
  \ctau\ of four tracks 
	& $>$ 200 $\mu$m (B), $>$ 250 $\mu$m ($\Lambda_b$)\\
  \ctau\ of charm hadron 
	& $>$ -70 $\mu$m ($D^{*}$, $\Lambda_c$), $>$ -30 $\mu$m ($D^+$) \\
   \hline
  \multicolumn{2}{|l|}{3.0 $<$ $M_{D^(*)\mu}$ $<$ 5.5 \gevcsq} \\  
  \multicolumn{2}{|l|}{3.7 $<$ $M_{\Lambda_c\mu}$ $<$ 5.7 \gevcsq} \\ 
  \multicolumn{2}{|l|}{$\mu$ ($\pi_B$) match to a SVT track} \\ 
  \multicolumn{2}{|l|}{charm hadron and $\mu$ ($\pi_B$) 
have opposite charge signs}\\
 \multicolumn{2}{|l|}{2 out of 4 tracks of \B\ candidate pass two track trigger cuts} \\ 
  \hline 
 \end{tabular}
 \end{normalsize}
 \end{center}
 \end{table}

\renewcommand{\arraystretch}{1.3}
  \begin{table}[tbp]
   \caption{Parameters used for \bb\ background estimate.}
  \label{t:cbbb}
   \begin{center}
   \begin{normalsize}
  \begin{tabular}{l|r|r|r|} 
   \hline
     \pythia\ $\sigma_{\bb}$ ($\mu$b) & \multicolumn{3}{c|}{49.6} \\
   \hline
   & $\bb\ \rightarrow \dstarmu$  & $\bb\ \rightarrow \dmu$ &
    $\bb\ \rightarrow \lcmu$ \\ \cline{2-4}
    $N_\mathrm{gen}$ & 221606748 & 221619610 & 221619610\\
     $N_\mathrm{pass}$ & 43  &  80 &  9 \\
    $\sum_{i} \sum_{j} f^i {\cal B}_j \epsilon_j$ (10$^{-7}$) 
	& 1.9 $\pm$ 0.3 
	& 3.6 $\pm$ 0.4 
	& 0.41 $\pm$0.14\\ 
    \hline
    2 $\cdot$ $\sigma_{\bb}$ $\cdot$ $\sum_{i} \sum_{j} f^i {\cal B}_j \epsilon_j$ (10$^{-5}$ $\mu$ b) & 1.9 $\pm$ 0.3 &  3.6 $\pm$ 0.4 & 0.41 $\pm$ 0.14 \\
   \hline
 \end{tabular}
 \end{normalsize}
 \end{center}

   \caption{Parameters used for \cc\ background estimate.}
  \label{t:cbcc}
   \begin{center}
   \begin{normalsize}
  \begin{tabular}{l|r|r|r|} 
   \hline
    \multicolumn{1}{l|}{\pythia\ $\sigma_{\cc}$ ($\mu$b)} & \multicolumn{3}{c|}{198.4}  \\
   \hline
    & $\cc\ \rightarrow \dstarmu$  & $\cc\ \rightarrow \dmu$ &
     $\cc\ \rightarrow \lcmu$ \\ \cline{2-4}
     $N_\mathrm{gen}$ & 720741510 & 698988700 & 698988700\\  
    \hline
    \multicolumn{1}{l|}{$N_\mathrm{pass}$} & 214 & 396 & 7  \\ \cline{2-4} 
    \multicolumn{1}{l|}
	{$D^{-}$ : ${\cal B}(D^- \rightarrow \mu X)$ = 14.22 ($\%$)}
	  & 117  & 205  &  4 \\
    \multicolumn{1}{l|}	
	{$\overline{D}^0$ : ${\cal B}(\overline{D}^0 \rightarrow \mu X)$ = 
	6.15 ($\%$)}  
    & 76  & 157  & 2  \\
   \multicolumn{1}{l|}{$D_s^-$ : ${\cal B}(D_s^- \rightarrow \mu X)$ = 
	13.32 ($\%$)} &  19   & 34  & 1  \\ 
   \multicolumn{1}{l|}
	{$\Lambda_c^-$ : ${\cal B}(\Lambda_c^- \rightarrow \mu X)$ = 
	4.5 ($\%$)}  & 2 & 0 & 0\\
    \cline{2-4}
    $\sum_{i} \sum_{j} f^i {\cal B}_j \epsilon_j$ (10$^{-8}$) & 
    3.3 $\pm$ 0.6 & 6.2 $\pm$ 1.0 & 0.12 $\pm$ 0.05	
 \\ 
   \hline 
    2 $\cdot$ $\sigma_{\cc}$ $\cdot$ 
	$\sum_{i} \sum_{j} f^i {\cal B}_j \epsilon_j$ (10$^{-5}$ $\mu$ b) 
	& 1.3 $\pm$ 0.2  
	& 2.5 $\pm$ 0.4 
	& 0.047 $\pm$ 0.020 \\
   \hline
 \end{tabular}
 \end{normalsize}
 \end{center}
 \end{table}

\subsubsection{Hadronic signal: $\sigma_{\Bd, \Lb} \cdot {\cal B}_\mathrm{had}
 \epsilon_\mathrm{had}$}
\label{sec-cbsignal}
 In order to normalize the background to the observed number of events in 
the hadronic mode, the \Bd\ or $\Lambda_b$ production cross-section, 
the efficiency and the branching ratio of the hadronic signal, 
have to come from external input or must be calculated using MC 
(see Equation~\ref{eq:ccbb}). We obtain $\sigma_{\Bd}$ by 
multiplying the previous CDF $\sigma_{\Bu}$ measurement by Keaffaber,~\etal
~\cite{run1:bpxsec} with the production fraction ratios, \(f_d/f_u\), 
from the 2004 PDG\cite{pdg:2004}. \Bd\ decay branching ratios are also 
obtained from the PDG. The product of $\sigma_{\Lambda_b}$ and 
${\cal B}(\Lambda_b\rightarrow\Lambda_c^+\pi^-)$ is obtained by multiplying 
the CDF measurement of \yile\ by Le, \etal~\cite{yile:lblcpi} with the 
$\sigma_{\Bd}$ we derive and the PDG ${\cal B}(\dhad)$.

Since we reconstruct both b and anti-b hadrons in the data,
we should multiply the measured cross-section by two.
For the efficiencies, we use the MC to generate and decay \B\ hadrons 
into our signals as described in Section~\ref{sec-mccom}. 
The CDF $\sigma_{\Bu}$ measurement is restricted to the 
$B^+$ with \pt\ greater than 6 \gevc\ and rapidity 
($y$) less than 1.0. Therefore, the denominator of the efficiency is 
the number of events in which the \B\ hadrons have \pt\ $>$ 6 \gevc\ and $
|y|$ $<$ 1.0. 
The numerator of the efficiency is the number of events which pass the cuts in 
Table~\ref{t:gencut} except the cut on four track invariant mass. 
Table~\ref{t:cbhad} lists the 
parameters that are used to calculate 
$\sigma_{\Bd, \Lb} \cdot {\cal B}_\mathrm{had} \epsilon_\mathrm{had}$.

\subsubsection{Semileptonic signal}
  Table~\ref{t:cbhad} also lists the efficiency and branching ratio of the
semileptonic signal mode for a comparison with the background $N_{\bb, \cc}$.
 In the case of the \Lb, we lack the external input for the branching ratio of
\lbsemi. Therefore, instead of the exclusive mode, we list the branching 
ratio of the inclusive mode: \inclbsemi\ from the 2004 PDG as 
an upper bound. We multiply the ratio $\rho(6)$ 
from Section~\ref{sec-lbxsec} with the Keaffaber $\sigma_{\Bu}$ result
to get $\sigma_{\Lb}$ for \pt\ greater than 6.0 \gevc.
The efficiency of \lbsemi\ is listed for a comparison with 
$\epsilon_{\bb, \cc}$. Note that the amount of \bb\ and \cc\
background relative to the signal is around 1$\%$.  The numbers from
three different modes should not be compared directly without multiplying
the branching ratios of the charm decays.

 
\renewcommand{\arraystretch}{1.5}
  \begin{table}[tbp]
   \caption{Parameters used to calculate $\sigma_{\Bd, \Lb}$, ${\cal B}_{had} \epsilon_{had}$ and~${\cal B}_{semi} \epsilon_{semi}$.}
  \label{t:cbhad}
   \begin{center}
   \begin{normalsize}
 \renewcommand{\doublerulesep}{0.03in}
  \begin{tabular}{l|r|r|r|} 
   \hline
    $\sigma_{B^+}$ ($\mu$b) & 
     \multicolumn{3}{c|}{ 3.6 $\pm$ 0.6} \\

   \hline
   & \alldstar\ & \alld\ & \alllc\ \\ \cline{2-4}
    $f_x/f_u$  & 1.00 $\pm$ 0.04 & 1.00 $\pm$ 0.04 & 0.25 $\pm$ 0.04\\
    $\sigma_{B^0,\Lb}$ ($\mu$b) & 3.6 $\pm$ 0.6 & 3.6 $\pm$ 0.6
    &  2.2 $\pm$ 0.5 \\ 
   \hline
   & \dstarhad\ & \dhad\ & \lbhad\ \\ \cline{2-4}
    \yile 
	& -- & -- & 0.82 $\pm$ 0.26 \\

    ${\cal B}_\mathrm{had}$ ($\%$) & 0.276 $\pm$ 0.021  &  0.276 $\pm$ 0.025
	& -- \\
    $\sigma(\Lb){\cal B}(\lbhad)$ ($\mu$b) & -- & -- & 0.008 $\pm$ 0.003 \\

    $N_\mathrm{gen}$ & 4242100 &  4242100 & 39999996\\

    $N_\mathrm{pass}$ & 70147 & 130433 & 843693\\

    $\epsilon_\mathrm{had}$ (10$^{-2}$) & 1.654 $\pm$ 0.006 & 
    3.075 $\pm$ 0.008 & 2.109 $\pm$ 0.002 \\
    \hline
    2$\sigma_{\Bd, \Lb}{\cal B}_{had} \epsilon_{had}$ (10$^{-5}$ $\mu$b) &  33 $\pm$ 6  &   
    61 $\pm$ 12  & 34 $\pm$ 13 \\
   \hline
   & \dstarsemi\ & \dsemi\ & \lbsemi\ \\ \cline{2-4}
    ${\cal B}_\mathrm{semi}$ ($\%$) & 5.44 $\pm$ 0.23 
    &  2.14 $\pm$ 0.20 & 9.2 $\pm$ 2.1 ($\%$)\\
    $N_\mathrm{gen}$ & 4242100 & 4242100& 39999996\\
    $N_\mathrm{pass}$ & 32620 & 66854 & 264484\\
    $\epsilon_\mathrm{semi}$  (10$^{-2}$) 
	& 0.769 $\pm$ 0.004 & 1.576 $\pm$ 0.006
    & 0.661 $\pm$ 0.001 \\
   \hline
     2$\sigma_{\Bd, \Lb}{\cal B}_\mathrm{semi} \epsilon_\mathrm{semi}$  
	(10$^{-5}$ $\mu$b) &  300 $\pm$ 50  & 240 $\pm$ 50 & 270 $\pm$ 90 \\
   \hline
 \end{tabular}
 \end{normalsize}
 \end{center}
 \end{table}

\renewcommand{\arraystretch}{1.2}
  \begin{table}[tbp]
   \setdec 0000.00
   \caption{The amount of \bb\ and \cc\ background.}
  \label{t:cbfinal}
   \begin{center}
   \begin{normalsize}
   \begin{tabular}{l|r@{\,$\pm$\,}l|r@{\,$\pm$\,}l|r@{\,$\pm$\,}l|}
   \hline
   & \multicolumn{2}{|c|}{\alldstar} 
   & \multicolumn{2}{|c|}{\alld} 
   & \multicolumn{2}{|c|}{\alllc} \\ \cline{2-7}
   $N_\mathrm{had}$ & \dec \ndstarhadc. & \dec \ndstarhade. 
   & \dec \ndhadc. & \dec \ndhade. & \dec \nlbhadc. & \dec \nlbhade. \\
   $N_\mathrm{semi}$ & \dec \ndstarsemic. & \dec \ndstarsemie. 
   & \dec \ndsemic. & \dec \ndsemie.  & \dec \nlbsemic. & \dec \nlbsemie. \\
   $N_{\bb}$ & \dec 6. & \dec 0.6 & \dec 34. & \dec 2. 
   & \dec 2.1 & \dec 0.2 \\
   $N_{\cc}$ & \dec 4. & \dec 0.4  & \dec 23. & \dec 1. 
   & \dec 0.2 & \dec 0.03 
  \\
  \hline
 \end{tabular}
 \end{normalsize}
 \end{center}
 \end{table}


\subsection{Comparison of Data and MC Cross Section}
\label{sec-xsec}
\begin{figure}[tbp]
  \begin{center}
\includegraphics[width=180pt, angle=0]{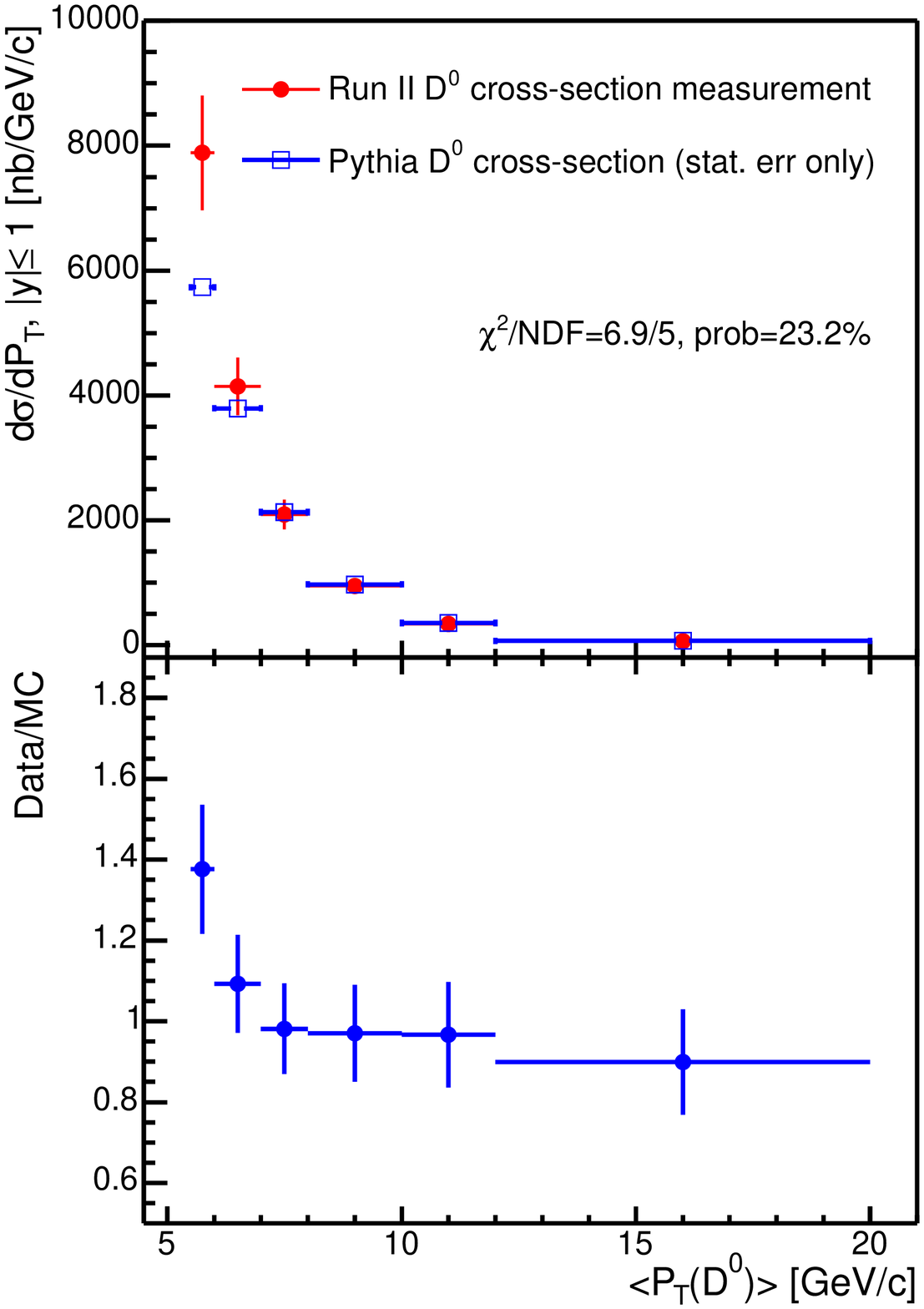}
\includegraphics[width=180pt, angle=0]{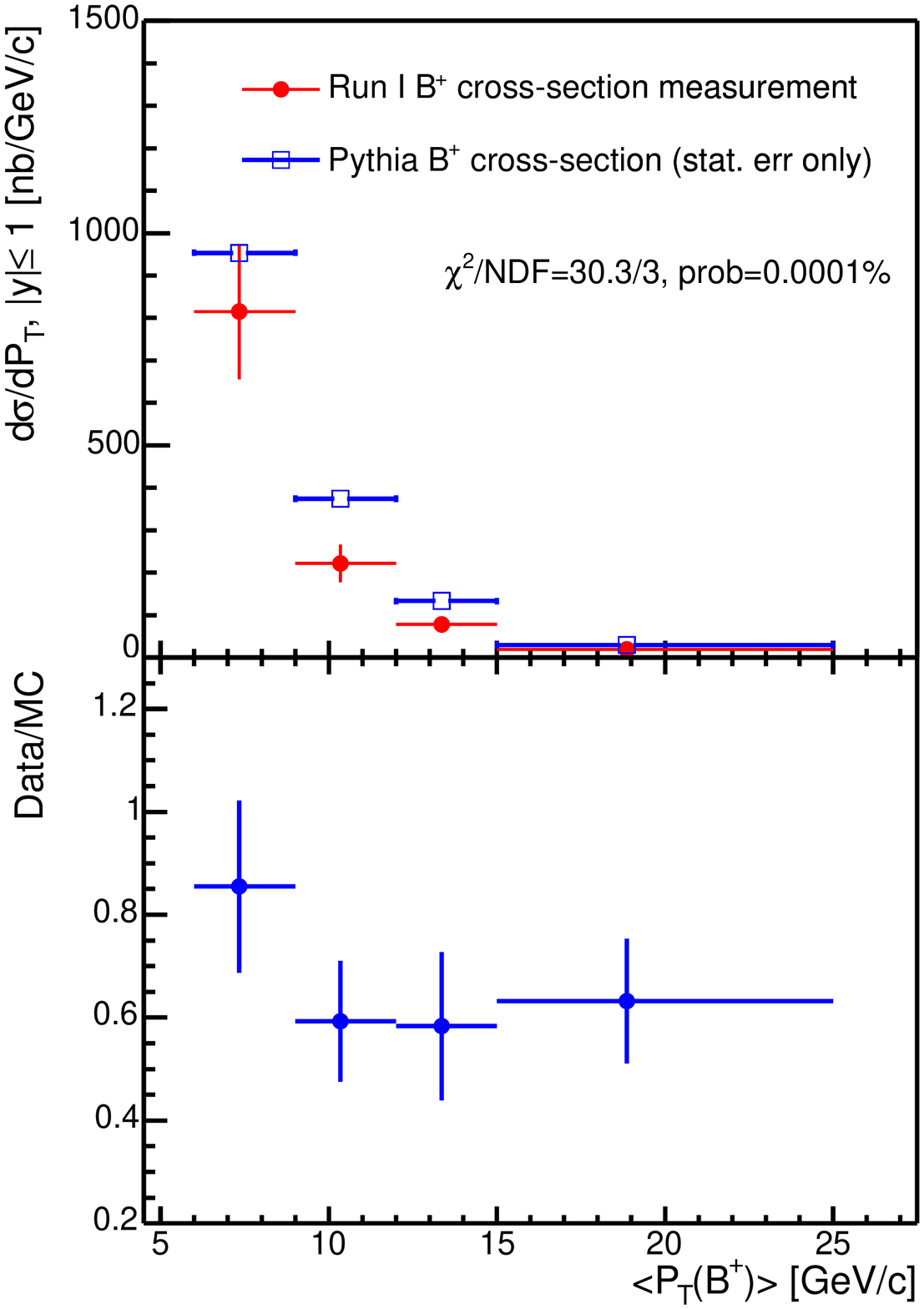}
\includegraphics[width=180pt, angle=0]{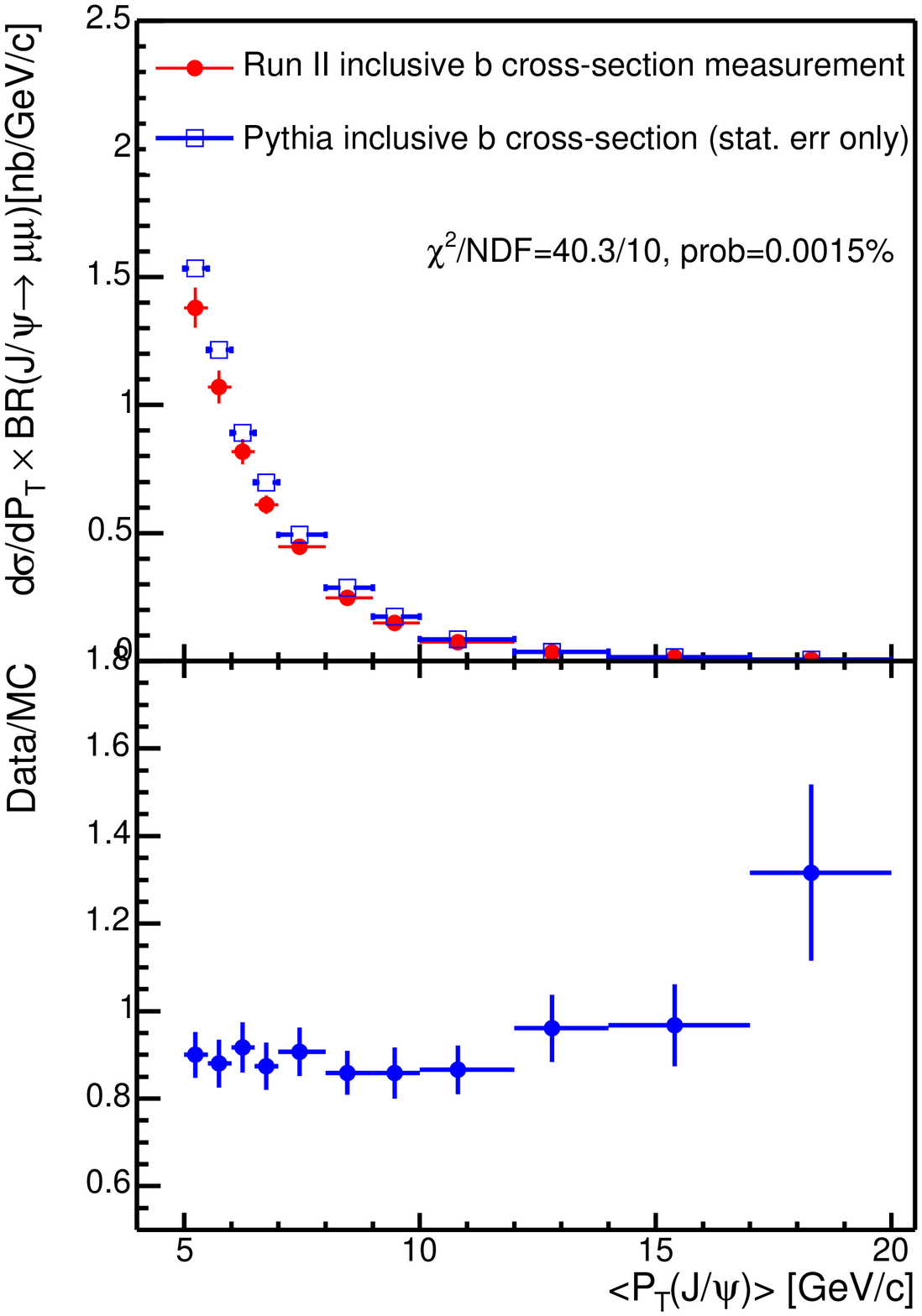}
 \caption[\Dzero, \Bu, and inclusive $b$ differential cross-section for MC 
	and data]
  {\Dzero\ (top left), \Bu\ (top right), and inclusive $b$ (bottom) 
   differential cross-sections. The upper plot in each 
   figure shows the differential cross-section for data (closed circles) by  
   Chen~\cite{cchen:dxec}, Keaffaber~\cite{run1:bpxsec}, 
   and Bishai~\cite{bishai:bxsec} and MC (open squares).
   The lower plot in each figure shows the data to MC ratio.}
 \label{fig:xseccomp} 
 \end{center}
 \end{figure}

 While there are precise measurements of the single charm hadron, $B^+$ and
inclusive b hadron cross-sections, there are no accurate 
measurements of the total \bb\ and \cc\ cross-section ($\sigma_{\bb}$,
$\sigma_{\cc}$) at the Tevatron, yet. 
To understand how well \pythia\ predicts $\sigma_{\bb}$ and $\sigma_{\cc}$,
we cross-check indirectly by comparing the ``differential cross-section'' 
of $D^0$, $B^+$ and inclusive b hadrons in \pythia\ with CDF Run I and II 
measurements by Chen~\cite{cchen:dxec}, Keaffaber\cite{run1:bpxsec}, and 
Bishai~\cite{bishai:bxsec}~\etal. 
We count the number of $D^0$, $B^+$ or b hadrons from {\tt nbot90} and 
{\tt nbota0} in bins of $\pt(D^0)$, $\pt(B^+)$ and $\pt(J/\psi)$. The bin width
and the \pt\ ranges are the same as Chen, Keaffaber and Bishai analyses. 
We divide the number of hadrons in each \pt\ bin by the total number 
of generated events. Then we multiply Pythia assumed $\sigma_{\bb}$ and 
$\sigma_{\cc}$ (see Tables~\ref{t:cbbb}--~\ref{t:cbcc}) to get the cross 
section of hadrons in each \pt\ bin. We further divide the number by the 
bin width to obtain the ``differential cross-section''. The
agreements between Monte Carlo and data cross-sections are generally within 
10$\%$ for charm hadrons and 40$\%$ for \B\ hadrons 
(see Figure~\ref{fig:xseccomp}).

Besides the total cross-section of \bb\ and \cc, the ratio of gluon splitting 
relative to the other two processes, flavor creation and flavor excitation,
also affects the amount of \bb\ and \cc\ backgrounds. Previous
CDF Run I measurement of \bb\ azimuthal production correlations by
Lannon~\cite{lannon:bcorel} concludes that {\tt Pythia} gives 
reasonable prediction
of the relative \bb\ production rates from the three processes.
However, due to the lack of measurements of the \cc\ relative production rates,
we do not yet have a comparison of the fraction of \cc\ gluon splitting 
between Monte Carlo and data. Therefore,
we assign 100$\%$ uncertainty when calculating the systematic errors for
the estimate of \bb\ and \cc\ backgrounds. As the contribution of \bb\ and
\cc\ background is at the 1$\%$ level, the systematic errors from 100$\%$
uncertainty is also about 1$\%$. 

\subsection{Comparison of Data and MC Impact Parameter} 
\begin{figure}[tbp]
    \begin{center}
      \includegraphics[width=170pt, angle=0]{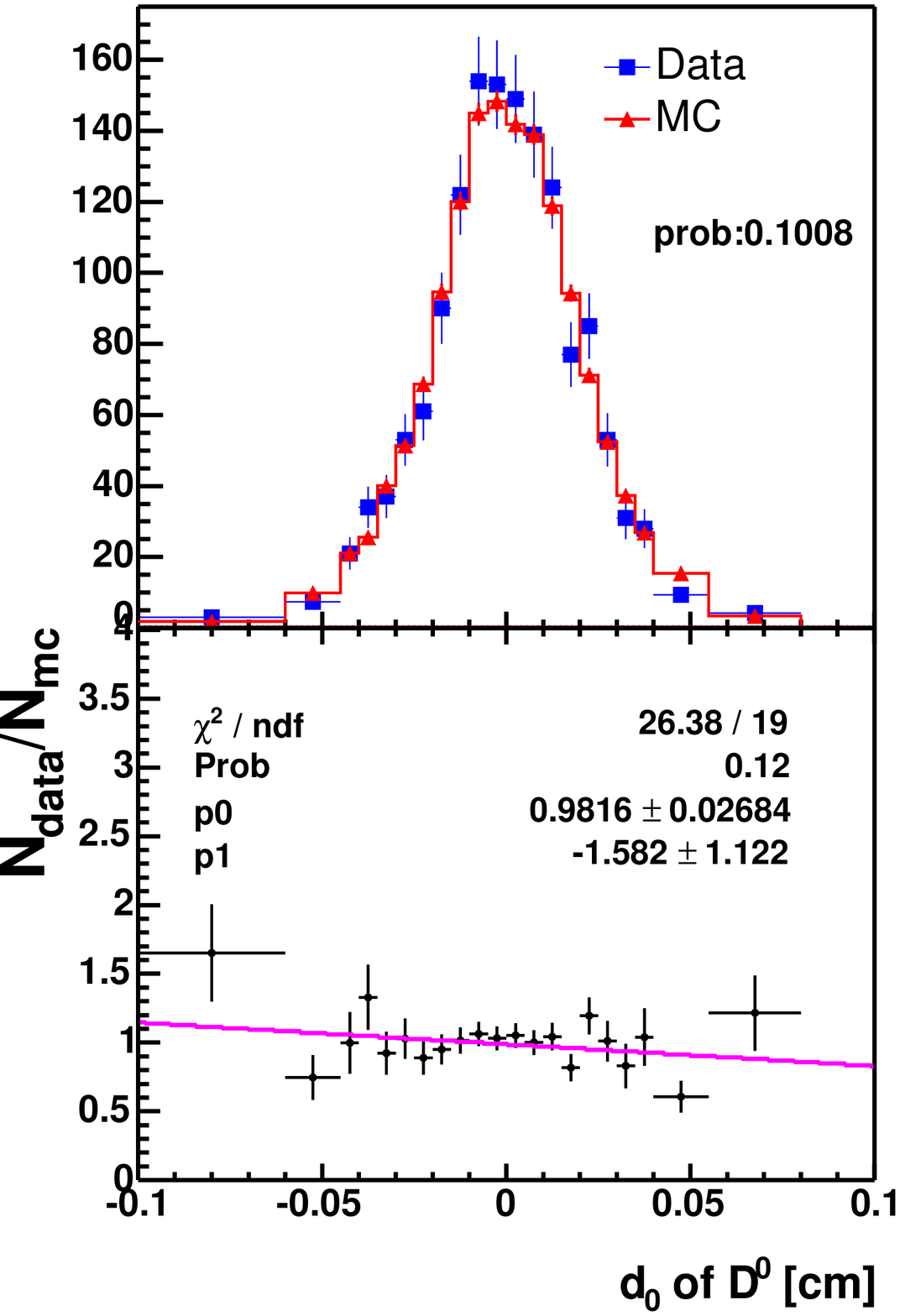}
      \includegraphics[width=170pt, angle=0]{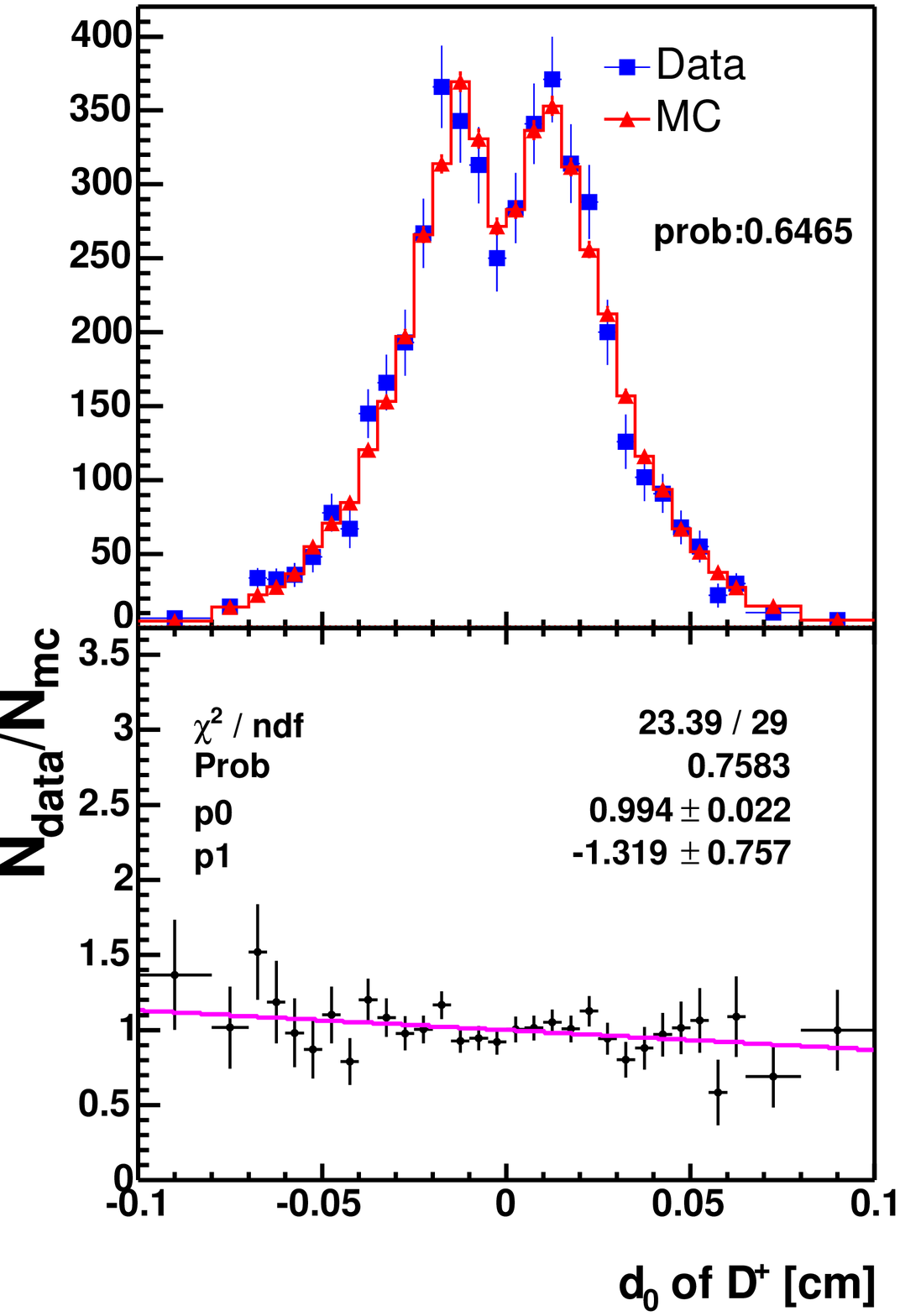}
      \includegraphics[width=170pt, angle=0]{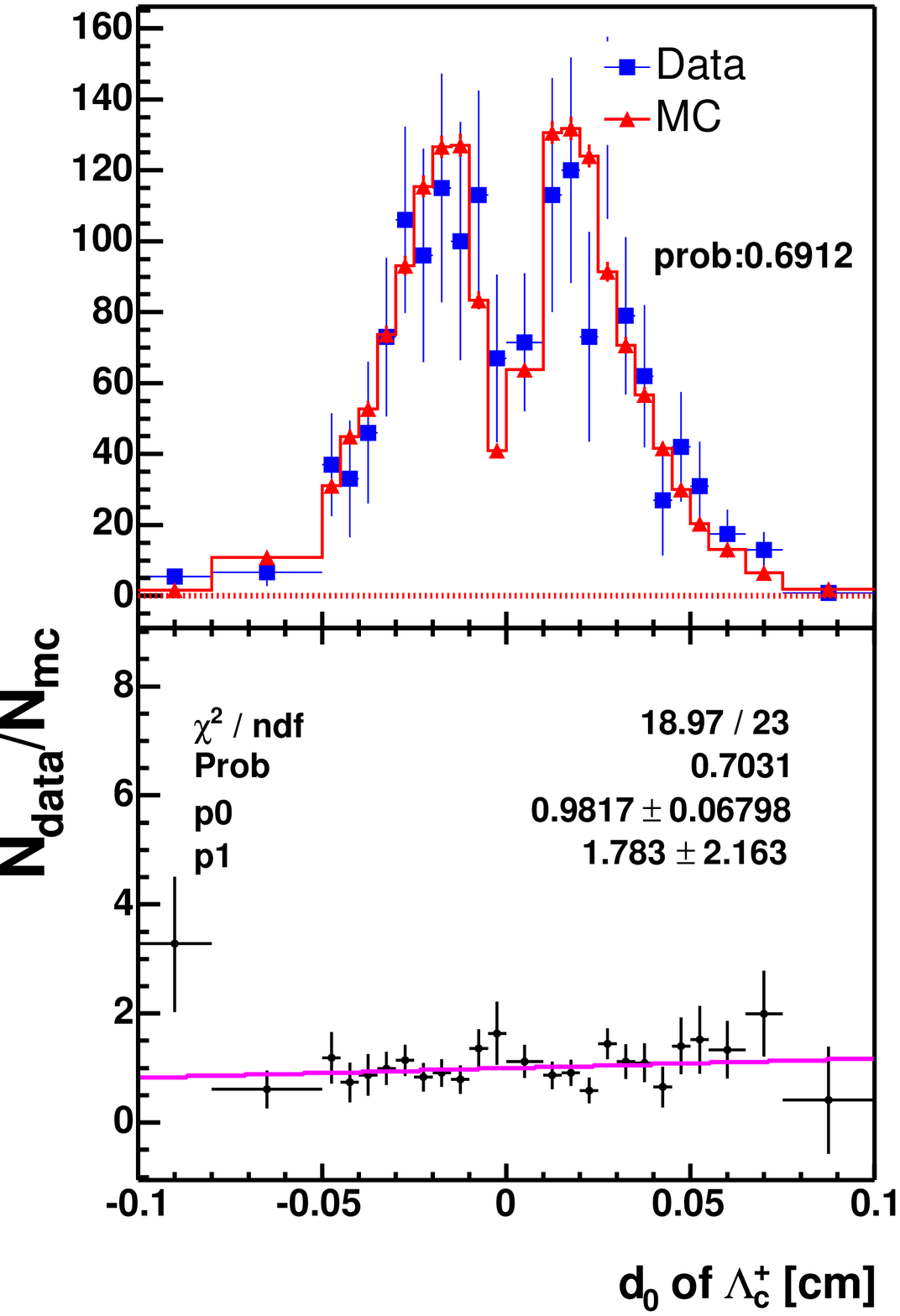}
     \caption[Impact parameters of \Dzero, \D\ and \Lc]
{Impact parameters of the charm hadrons, from the top left to the bottom are
\Dzero, \D\ and \Lc: MC and data comparison. The
good agreement of the MC with the data indicates that background from 
the promptly produced charm (\cc) is negligible.}
     \label{fig:dimpact}
     \end{center}
  \end{figure}

We compare the distribution of the impact parameter
of charm hadrons with respect to the beam spot in MC and data. 
Figure~\ref{fig:dimpact} shows a good agreement of the MC with the data. 
No excess of charm hadrons with small $d_0$ is found in the data.
This indicates that the promptly produced charm from \cc\ has negligible
contribution to the background in the semileptonic \B\ decays, which
is consistent with our estimate using \pythia.

        
\section{Background Summary}
The fraction of each type of background in our semileptonic signal 
is summarized below. The dominant signal contamination is from the 
physics background. The second largest background arises from muon fakes. 
The smallest background source is from \bb\ and \cc. 
\vspace{-12pt}
\begin{table}[htb]
\setdec 00.00
\caption{Summary of the backgrounds to the semileptonic modes.}
\label{t:bgsumall}
\begin{center}
\begin{tabular}{c|rrr|}
  & \multicolumn{3}{c|}{$N_\mathrm{bg}/N_\mathrm{inc\;semi}$ ($\%$)} \\
\hline
Background Type & \multicolumn{1}{c}{\dstarmu} & \multicolumn{1}{c}{\dmu} &
	\multicolumn{1}{c|}{\lcmu} \\ 
\hline
  Physics  &  \dec 15. & \dec 40. & \dec 9.8 \\
  Muon fakes & \dec 4.3 & \dec 4.9 & \dec 3.2 \\
  \bb\ and \cc\ & \dec 0.9 & \dec 1.2 & \dec 0.2 \\
\hline
  Total & \dec 20.2 & \dec 46.1 & \dec 13.8 \\
\hline\hline
\end{tabular}
\end{center}
\end{table}

\chapter{Relative Branching Fraction Results and Systematics}
\label{ch:result}
Recall that the equation for the ratio of branching fractions is:
\begin{equation}
\label{eq:bigf2}
\frac{{\cal B}_\mathrm{semi}}{{\cal B}_\mathrm{had}} = 
(\frac{N_\mathrm{inclusive\;semi}-N_\mathrm{physics}- N_\mathrm{fake\mu}
-N_{c\overline{c},\;b\overline{b}}}{N_\mathrm{had}})
\times \frac{\epsilon_\mathrm{had}}{\epsilon_\mathrm{semi}}, 
\end{equation}
where the yields, $N_\mathrm{had}$ and $N_\mathrm{inclusive\;semi}$, are 
obtained in Chapter~\ref{ch:yield}. In Section~\ref{sec-eff}, the 
efficiencies, $\epsilon_\mathrm{had}$ and $\epsilon_\mathrm{semi}$ are 
determined and in Chapter~\ref{ch:bg}, the backgrounds to our semileptonic 
signals are estimated. We are now in a position to calculate the 
relative branching fractions. The relative branching fractions with the 
statistical uncertainties only are:
\begin{eqnarray*}
\frac{{\cal B}(\dstarsemi)}{{\cal B}(\dstarhad)} & = &\rdstar, \\
\frac{{\cal B}(\dsemi)}{{\cal B}(\dhad)} & = & \rd, \\
\frac{{\cal B}(\lbsemi)}{{\cal B}(\lbhad)} & = &\rlb.
\end{eqnarray*}
In the remainder of this chapter, we first discuss and estimate the systematic 
uncertainties. Then, we show the result of the relative branching fractions for
 each mode. Finally, implications of our measurements are discussed and a 
conclusion is given at the end of the chapter.
\section{Systematic Uncertainties}
\label{sec-sys}
\subsection{Sources of Systematics}
Systematic uncertainties in our measurements may arise from the 
difference in the semileptonic and hadronic decays, from the lack of knowledge
 of certain backgrounds, and from the uncertainties on the external 
information. Most of the sources are common to all the decay modes. Systematic 
uncertainties which affect only one mode are discussed separately. To simplify 
the notation, we define our measurement of the relative branching fractions as 
$R$:
\[
 R \equiv \frac{{\cal B}_\mathrm{semi}}{{\cal B}_\mathrm{had}},
\]
and $\sigma_R$ is the systematic uncertainty. We also denote the term, 
``branching ratio'' in the text, as \br, while the branching ratio of one 
specific mode is denoted as ${\cal B}(\mathrm{mode})$, eg: ${\cal B}(\dhad)$. 

\subsubsection{Mass Fitting}
\begin{itemize}
\item \incdstarsemi\ and \inclbsemi: The mass functions are general 
and cover all the possible backgrounds. Because the functions do not 
involve any external \br\ or MC efficiencies, we do not assign systematic 
uncertainty for the mass fitting of these two modes.
\item \incdsemi: The uncertainties on the $D_s$ decay \br\ can modify the 
misreconstructed $D_s$ mass spectrum in \incdsemi. In addition, the mean of 
the Gaussian constraint for the amount of $D_s$ background in 
Equation~\ref{eq:logdmu}, $\mu_p$, also changes accordingly. We study this 
effect by varying the following numbers in Table~\ref{t:dsdecays} $\pm$ 1 
$\sigma$ independently: the \br\ of \seqds, and the \br\ of each selected 
$D_s$ decay relative to ${\cal B}(\seqds)$, because these $D_s$ \br\ were 
measured relative to the $\phi\pi$ mode~\cite{pdg:2004}. The corresponding 
$D_s$ background shape and the $\mu_p$ are re-evaluated for each change of 
$D_s$ \br. The changes in the yield are added in quadrature to get the 
accumulative variation. Table~\ref{t:dmusys} summarizes the yield variations. 
Modes which are not listed give identical results to the central value.
We find total $\Delta(N_{\incdsemi})$ = 33 events and $\sigma_R$ = 0.13.
 Note that a few $D_s$ decays in Table~\ref{t:dsdecays} only have an upper 
limit in the PDG and the estimated values in the \evtgen\ decay table are used.
 We assign 100$\%$ uncertainty for these modes. 

\begin{table}[h]
\caption{\incdsemi\  yield change due to the variation of $D_s$ \br.}
\label{t:dmusys}
\begin{center}
\renewcommand{\tabcolsep}{0.2in}
\begin{tabular}{|l@{\,$\pm$\,}r|}
\hline\hline
 \multicolumn{2}{|r|}{$\Delta(N_{\incdsemi})$} \\
\hline
$D_s^+\rightarrow K^+K^-\pi^+$  & 23  \\
$D_s^+\rightarrow \phi K^+$ & 2\\
$D_s^+\rightarrow \eta\pi^+$ &  2 \\
$D_s^+\rightarrow \eta^{\prime}\pi^+$  &  3 \\
$D_s^+\rightarrow \omega\pi^+$ &  1 \\
$D_s^+\rightarrow \rho^0K^+$ &  1 \\
$D_s^+\rightarrow f_2\pi^+$ &  1 \\
$D_s^+\rightarrow \rho^+\eta$  &  3 \\
$D_s^+\rightarrow \rho^+\eta^{\prime}$ &  2 \\
$D_s^+\rightarrow K^0 K^+$ & 2 \\
$D_s^+\rightarrow K^{*0} K^+$ &  22 \\
$D_s^+\rightarrow \overline{K}^{*0} \pi^+$ &  2\\
\hline
Total & 33 \\
\hline\hline
\end{tabular}
\end{center}
\end{table}

\item \dstarhad: the composition of the remaining \alldstar\ background 
can affect the shape of its mass spectrum, and its ratio to the $D^*\rho$ 
background. The latter changes the mean of the Gaussian constraint, $\mu_2$, 
in Equation~\ref{eq:logdstarpi}. We study the systematics by varying the \br\ 
of \bddstarrho\ and the dominant modes in the remaining \alldstar\ background. 
The change of signal yield from each variation of \br\ is listed in 
Table~\ref{t:dstarpibrsys}. The accumulative yield change is only 
${+0.1 \atop -0.2}$ events, which is insignificant. Therefore, we do not 
assign systematic uncertainty for the mass fitting of this decay mode.

 \begin{table}[tbp]
  \caption{\dstarhad\ yield change due to the variation of the background \br.}
  \label{t:dstarpibrsys}
   \begin{center}
   \renewcommand{\arraystretch}{1.3}
   \setdec 00.000
   \begin{tabular}{|l|l@{\,$\pm$\,}r |r|}
    \hline
     Mode &
     \multicolumn{2}{|r|}{${\cal BR}$ ($\%$)} & $\Delta N$\\	
     \hline
      \dstarsemie & \dec 5.44 & \dec 0.23 & $<$ 0.1 
	\\
      \dstarsemi & \dec 5.44 & \dec 0.23 & $<$ 0.1 \\ 
      \bddstarpipizero & \dec 0.7 & ? & $<$ 0.1 \\ 
      \bddstaraone & \dec 1.30 & \dec 0.27 & $<$ 0.1 \\ 
      \bddstarrho & \dec 0.68 & \dec 0.09 & ${+0.1 \atop -0.2}$  \\
      \hline
      \multicolumn{3}{|l|}{Total}  &  ${+0.1 \atop -0.2}$ \\
      \hline\hline
      \end{tabular}
      \end{center}
  \end{table}

\item \dhad: 
the systematic uncertainties come from three sources: the normalizations of 
the Cabibbo suppressed decay, \Bs\ and \Lb\ backgrounds, the uncertainties of 
the background function fit to the MC, and the \br\ of the backgrounds. 
We study the effect of the first two sources in the following way:
We vary each constant parameter including the normalizations and the 
shape parameters in Table~\ref{t:dpifix} $\pm$ 1 $\sigma$, independently. 
The changes of yield ($\Delta N$) are listed in Table~\ref{t:dpifixsys}. 
The normalizations of the backgrounds from the Cabibbo suppressed decay, 
\Bs\ and \Lb\ decays are independent from the shape parameters. But,   
several shape parameters for the same background are correlated, as shown in 
Tables~\ref{t:bscorr}--\ref{t:drhocorr}. 
In order to take into account the correlation properly, 
the correlated shape parameters are grouped together. We calculate the 
product of the correlation coefficient matrix ($\mathbf{M}$), with the row 
and column vectors of $\mathbf{\Delta N}$, to obtain a total systematic 
uncertainty. For instance, the systematic uncertainty from the \Bs\ background 
shape parameters is:
\begin{equation}
 \sigma_{N}^2 =
\left(
\begin{array}[c]{cccc}
\Delta N_{\mu} & \Delta N_{\sigma_1} & \Delta N_{f_1} 
& \Delta N_{\frac{\sigma_2}{\sigma_1}} \\
\end{array}
\right)
\mathbf{M}
\left(
\begin{array}[c]{c}
\Delta N_{\mu} \\ 
\Delta N_{\sigma_1} \\ 
\Delta N_{f_1} \\
\Delta N_{\frac{\sigma_2}{\sigma_1}} \\
\end{array}
\right),
\end{equation}
where $\mathbf{M}$ is a $4\times 4$ correlation coefficient matrix returned 
from the fit to the \Bs\ MC (see Table~\ref{t:bscorr}). The value 
of $\Delta N$ for each parameter is listed in Table~\ref{t:dpifixsys}.

For the systematics associated with the \br, we vary the 
\br\ of \dstarhad, \bddrho\ and the dominant modes in the remaining \alld\ 
backgrounds, $\pm$~1~$\sigma$ independently.
We re-fit the background shapes using the MC, fix the shape parameters and 
re-fit the data. Table~\ref{t:dpibrsys} lists the signal yield change due to 
the variation of the \br. Table~\ref{t:dpifixsystotal} summarizes the signal 
yield change from the variation of the shape parameters and the \br. 
These changes are added in quadrature to get the accumulative difference. 
The total change in the yield is 13 events, which modifies $R$ by 0.38. 


 \begin{table}[tbp]
   \caption{\dhad\ yield change due to an independent variation of the 
	fixed parameter value.}
  \label{t:dpifixsys}
   \begin{center}
   \renewcommand{\tabcolsep}{0.07in}
   \begin{tabular}{l|l|@{\,$\pm$\,}l}
    \hline
    \hline
      Parameter & \multicolumn{2}{|r}{$\Delta N$} \\	
     \hline
      $f_{DK}$ & $N_{\dhadk}/N_{\dhad}$ & 4.97 \\
      $\Delta M_{DK}$ & mass shift of \dhadk & 3.82 \\
      $\sigma_{DK}$ & width of \dhadk & 0.82 \\	
      $f_{\Bs}$ & $N_{\bsdspi}/N_{\dhad}$ & 0.51 \\
      $\mu_{\Bs}$ & mean of \Bs\ background & 0.02 \\ 
       $f_1$ & fraction of the narrow \Bs\ Gaussian & 0.00 \\
      $\sigma_1$ & width of the narrow \Bs\ Gaussian & 0.10 \\
       $\sigma_2/\sigma_1$ & width ratio of the \Bs\ Gaussians & 0.04 \\
       $f_{\Lb}$ & $N_{\lbhad}/N_{\dhad}$ & 0.22 \\
       $\mu_{\Lb}$ & mean of \Lb & 0.22\\ 
       $\sigma_{\Lb}$ & width of \Lb\ background & 0.02 \\
       $\tau_{\Lb}$ & lifetime of \Lb\ background & 0.21 \\ 
       $\tau_\mathrm{ref}$ & lifetime of $D\rho$ background & 2.62 \\
       $\sigma_\mathrm{ref}$  & width of $D\rho$ background & 0.27 \\
       $f_H$ & fraction of $D^*\pi$ horns & 6.38 \\ 
       $\delta_\mathrm{ref}$  & distance between two horns & 1.04 \\
       $\sigma_{H}$ &  width of the horns & 2.70 \\
       $f_\mathrm{otherB}$ & fraction of the remaining \alld & 1.90\\
       $M_\mathrm{off}$ & cut off for \alld\ mass & 1.02 \\
      \hline\hline
      \end{tabular}
      \end{center}
	\end{table}

 \begin{table}[tbp]
 \caption{\dhad\ yield change due to the variation of the background \br.}
  \label{t:dpibrsys}
   \begin{center}
   \renewcommand{\arraystretch}{1.3}
   \setdec 00.000
   \begin{tabular}{|l|l@{\,$\pm$\,}r |r|}
    \hline
     Mode &
     \multicolumn{2}{|r|}{${\cal BR}$ ($\%$)} & $\Delta N$\\	
     \hline
      \bddrho & \dec 0.77 & \dec 0.13 & ${+0.5 \atop -3.4}$ \\
      \dstarhad & \dec 0.276 & \dec 0.021 & ${+2.2 \atop -1.3}$ \\
      \dsemie & \dec 2.14 & \dec 0.20 & ${+0.6 \atop -0.2}$ \\
      \dsemi & \dec 2.14 & \dec 0.20 & ${+1.0 \atop -0.4}$ \\
      \dstarsemie & \dec 5.44 & \dec 0.23 & ${+0.1 \atop -0.2}$ \\
      \dstarsemi & \dec 5.44 & \dec 0.23 & $\pm$ 0.2 \\
      \bddpipizero & \dec 0.1 & ? & $\pm$ 0.5 \\ 
      \bddstarpipizero & \dec 0.7 & ? & $\pm$ 1.0 \\
      \bddaone & \dec 0.60 & \dec 0.33 & $\pm$ 3.0 \\
      \bddstaraone & \dec 1.30 & \dec 0.27 & $\pm$ 0.1 \\
      \bddstarrho & \dec 0.68 & \dec 0.09 & $\pm$ 0.6 \\
      \hline
      \multicolumn{3}{|l|}{Total}  & $\pm$ 4.5 \\
      \hline\hline
      \end{tabular}
      \end{center}

   \caption{Systematic uncertainty on the \dhad\ yield from each 
	independent parameter group.}
  \label{t:dpifixsystotal}
   \begin{center}
   \renewcommand{\tabcolsep}{0.07in}
   \begin{tabular}{|l@{\,$\pm$\,}r |}
    \hline
     \multicolumn{2}{|r|}{$\Delta N$}\\	
     \hline
      $f_{DK}$  & 5.0 \\ $DK$ shape & 3.9 \\
      $f_{\Bs}$ & 0.5 \\ \Bs\ shape & 0.1 \\
       $f_{\Lb}$ & 0.2 \\ \Lb\ shape & 0.4  \\
       $D\rho+D^*\pi$ shape & 9.9 \\ $f_\mathrm{otherB}$ & 1.9\\
       $M_\mathrm{off}$ & 1.0 \\  ${\cal BR}$ & 4.5\\
      \hline
      Total  & 12.8 \\
      \hline\hline
      \end{tabular}
      \end{center}

  \end{table}

 \begin{table}[tbp]
  \caption{Correlation coefficients returned from the fit to \Bs\ MC.}
  \label{t:bscorr}
   \begin{center}
   \begin{tabular}{|l|rrrrr|} 
     \hline
    \hline
	& N & $\mu_{\Bs}$ & $\sigma_1$ & $f_1$ & $\sigma_2/\sigma_1$ \\
     \hline
	N & 1.000 & & & & \\
        $\mu_{\Bs}$ & 0.024 & 1.000 & & & \\ 
        $\sigma_1$ & -0.003 & -0.084 & 1.000 & & \\
        $f_1$ & 0.007 & -0.004 & 0.847 & 1.000 & \\
        $\sigma_2/\sigma_1$ & 0.023 & 0.133 & 0.268 & 0.647 & 1.000 \\
     \hline	
      \hline
      \end{tabular}
      \end{center}

  \caption{Correlation coefficients returned from the fit to \Lb\ MC.}
  \label{t:lbcorr}
   \begin{center}
   \begin{tabular}{|l|rrrr|} 
     \hline
    \hline
	& N & $\mu_{\Lb}$ & $\sigma_{\Lb}$ & $\tau_{\Lb}$ \\
     \hline
	N & 1.000 & & &   \\
      $\mu_{\Lb}$ & 0.000 & 1.000 & & \\
      $\sigma_{\Lb}$ & 0.000 & -0.624 & 1.000 &  \\
      $\tau_{\Lb}$ & 0.000 & 0.699 & -0.508 & 1.000 \\
     \hline	
      \hline
      \end{tabular}
      \end{center}

  \caption{Correlation coefficients returned from the fit to $D^*\pi$ and 
	$D\rho$ MC.}
  \label{t:drhocorr}
   \begin{center}
   \begin{tabular}{|l|rrrrrrrr|} 
     \hline
    \hline
     & $N$ & $\tau_{ref}$ & $\mu_{ref}$ & $\sigma_{ref}$ & $f_H$ 
	& $\delta_{ref}$ 
	& $\sigma_H$ & $\nu_{ref}$ \\
     \hline
     $N$ & 1.000 & & & & & & & \\	
     $\tau_\mathrm{ref}$ & 0.000 & 1.000 & & & & & & \\
     $\mu_\mathrm{ref}$ & 0.000 & 0.013 & 1.000 & & & & & \\
     $\sigma_\mathrm{ref}$ & 0.000 & 0.169 & -0.841 & 1.000 & & & & \\
     $f_H$ & 0.000 & 0.549 & -0.688 & 0.720 & 1.000 & & & \\
     $\delta_\mathrm{ref}$ 
	& 0.000 & 0.294 & 0.284 & -0.429 & -0.435 & 1.000 & & \\
     $\sigma_H$ & 0.000 & 0.407 & -0.507 & 0.507 & 0.699 & -0.381 & 1.000 & \\
     $\nu_\mathrm{ref}$ 
	& 0.000 & 0.029 & 0.971 & -0.786 & -0.624 & 0.261 & -0.455 & 1.000 \\
     \hline	
      \hline
      \end{tabular}
      \end{center}
  \end{table}


\item \lbhad: we follow the same scheme applied by 
Martin~\cite{cdfnote:6953}. Using a generic \B-decay MC, we first extract the 
top twenty largest contributing modes in the mass 
region 5.3 $<$ $M_{\Lamc\pi}$ $<$ 6.0 \gevcsq, from each type of background: 
four-prong \B\ meson, the 
remaining \B\ meson decays, and the remaining \Lb\ decays. Each dominant 
decay contributes $N_\mathrm{base}^i$ events. Then, we generate a new 
distribution for each dominant mode, according to the shape determined from a 
large single-decay MC. The normalization of the new distribution 
is first Gaussian fluctuated with a mean $N_\mathrm{base}^i$, a sigma of 
$\Delta(\br)/(\br)$ and then Poisson fluctuated.  For the 
measured decays, $\Delta(\br)$ is the uncertainty reported in the PDG. 
For the unmeasured \B\ meson decays, $\Delta(\br)$ is assumed to be 
three times the uncertainty of the closest equivalent mode in the measured 
\B\ meson decays. For the unmeasured \Lb\ decays, $\Delta(\br)$ is
 hypothesized to be ${+100 \atop -50}\%$ of the \br. 
These Gaussian and Poisson fluctuated distributions are then re-combined with 
the other non-dominant modes. The combined background mass spectrum is 
refitted and the newly derived shape parameters are fixed in the fit to 
the data. The whole procedure is repeated 1000 times with different 
random seeds for the Gaussian and Poisson fluctuations. We plot the
 distribution of the \lbhad\ yield and record the RMS as the change in 
the yield due to the variation of the \br. Figure~\ref{fig:b4prongsig} shows 
an example of the \lbhad\ yield distribution from the \br\ variation of the 
four-prong \B\ meson background. 
In addition, we vary the fraction of the Cabibbo suppressed mode, 
$f_{\Lamc K}$, ${+100 \atop -50}\%$ and record the yield change. 
Table~\ref{t:lcpifitsys} summarizes the change of \lbhad\ yield. 
The accumulative $\sigma_R$ is 0.63.

\begin{table}[tbp]
   \caption{\lbhad\ yield change due to the variation of the background \br.}
  \label{t:lcpifitsys}
   \begin{center}
   \renewcommand{\tabcolsep}{0.07in}
   \begin{tabular}{|lr|}
    \hline
     \multicolumn{2}{|r|}{$\Delta N$}\\	
     \hline
      \lbhadk  & ${+1.1 \atop -2.8}$  \\ 
      four-prong \B\ meson decays & $\pm$ 2.9 \\
      remaining \B\ meson decays  & $\pm$ 0.9 \\
      all the other \Lb\ decays & $\pm$ 2.8  \\
      \hline
      Total  & ${+4.3 \atop -5.0}$ \\
      \hline\hline
      \end{tabular}
      \end{center}
  \end{table}

 \begin{figure}[htb]
     \begin{center}
        \includegraphics[width=250pt, angle=0]
	{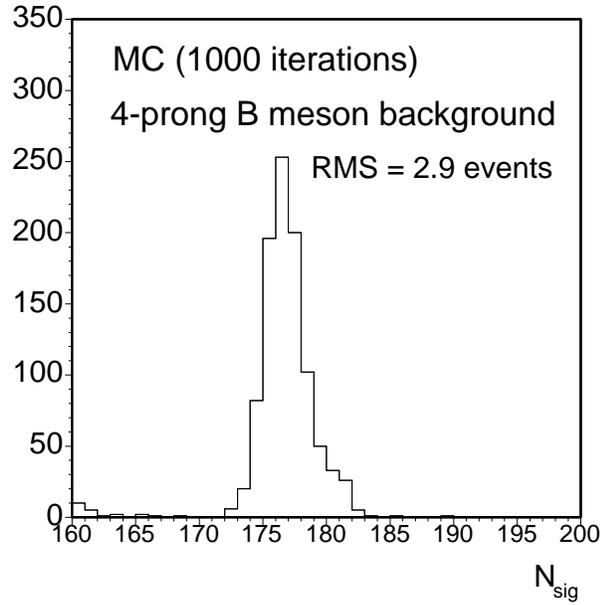} 
     \end{center}
 \caption[\lbhad\ yield from 1000 variations of the 4-prong 
	\B\ meson background \br ]
	{\lbhad\ yield from 1000 variations of the 4-prong 
	\B\ meson background \br. The RMS is recorded as the yield change.}
     \label{fig:b4prongsig}
 \end{figure}

\end{itemize}

\subsubsection{Measured Branching Fractions}
We use the \br\ from the world average in the PDG to estimate the physics 
backgrounds in our semileptonic signals as described in 
Section~\ref{sec-physicsb}. We vary the \br\ of these measured physics 
backgrounds by $\pm$ 1 $\sigma$. We then calculate $\sigma_R$. 
Note that here the variation of the ${\cal B}(\lbhad)$ does not include 
the uncertainty due to the measured \Lb\ \pt\ spectrum. 
The $\sigma_R$ due 
to the measured \br\ is 0.43, 0.75 and ${+0.73 \atop -2.07}$ for 
\(\overline{B}^0 \rightarrow \Dstar\), \(\overline{B}^0 \rightarrow \D\) and 
\(\Lb \rightarrow \Lc\) modes, respectively. For the remainder of this 
section, we quote the systematic uncertainties in the same order.

\subsubsection{Unmeasured Branching Fractions} 
The \br\ of several physics backgrounds in Section~\ref{sec-physicsb} have not 
yet been measured, {\em e.g.}: \lblcfzero, or have just been measured by us 
for this analysis, {\em e.g.}:\lblcstar. For the first case, we use the 
estimated \br\ from the decay file of \evtgen, and 
our own derivation based on HQET. As we have no uncertainty input from the 
estimated \br, we assign a 5$\%$ uncertainty to the \br\ of the excited charm 
meson decays and a 100$\%$ uncertainty to the \br\ of the \B\ hadron decays. 
Because the excited charm hadrons decay via strong interaction and conserve 
isospin symmetry, their \br\ could be inferred from Clebsch-Gordan 
Coefficients. While for the weak decays of \B\ hadrons, allowable decay 
spectrum is wider. For the second case, we add $(20 \oplus 20)\%$ uncertainty 
in quadrature with the uncertainty from the preliminary measurement 
(see Table~\ref{t:physicslc}). The first 20$\%$ arises from the unresolved 
disagreement of measured $\tau_{\Lb}$ with that from the HQET prediction. 
The second 20$\%$ is due to the difference of the soft pion reconstruction 
efficiency between MC and data. 
We vary the \br\ by the uncertainties we assigned and calculate the shift of 
our measurement. The shift due to the unmeasured \br\ is 1.09, 0.91 and 0.50.

\subsubsection{Fake $\mu$ estimate}
As noted in Section~\ref{sec-fakemumethod}, the systematic uncertainties
 from the fake $\mu$ estimate originate from: 
\begin{enumerate}
 \item The uncertainty from the fit to the charm mass spectra.
 \item The uncertainty on the probabilities for the pions, kaons 
	and protons to fake muons.
 \item The uncertainty on the fraction of pion, kaon and proton in the hadron 
tracks.
\end{enumerate}
 For each category, we vary the central value $\pm$ 1 $\sigma$, independently. 
More detailed description can be found in Section~\ref{sec-fakemumethod}. 
The resulting uncertainty on the amount of fake muons together with the 
central value are summarized in Table~\ref{t:fakemuerror}. 
\begin{table}[tbp]
\caption{Summary of fake muon contamination.} 
\label{t:fakemuerror}
\begin{center} 
\begin{tabular}{|c|c|c|}
\hline
 $\overline{B}\rightarrow \Dstar \mu_{fake}$ & 
 $\overline{B}\rightarrow \D \mu_{fake}$ &
 $\overline{B}\rightarrow \Lc \mu_{fake}$ \\
\hline
 45 \(\pm\) 3 & 230 \(\pm\) 19 & 40 \(\pm\) 9 \\
\hline \hline
\end{tabular}
\end{center}
\end{table}
We then vary the number of fake muons $\pm$ 1 $\sigma$ and insert the 
new number into Equation~\ref{eq:bigf2}. 
The total variation on $R$ due to the fake $\mu$ estimate is 0.07, 0.07, 0.17. 

\subsubsection{\bb\ and \cc\ background}
 In Section~\ref{sec-ccbb}, we notice a 10--40$\%$ discrepancy of 
the \Dzero, \Bu, and inclusive $b$ cross-section from \pythia\ with those 
from the data. In addition, we do not possess information about the 
relative \bb\ and \cc\ production rates between flavor creation, flavor 
excitation, and gluon splitting. Therefore, we assign a 100$\%$ uncertainty to 
the amount of \bb\ and \cc\ backgrounds. This changes $R$ by 0.22, 0.22, 0.04. 

\subsubsection{MC sample size}
We have generated large MC samples for calculating the efficiencies of our 
signals and backgrounds, but there is a small statistical uncertainty due to 
the finite MC sample size. We use the uncertainties on the efficiencies to 
calculate $\sigma_R$. $\sigma_R$ is 0.28, 0.18, and 0.32. 

\subsubsection{MC $P_T(B)$ Spectrum}
 We find discrepancies between data and MC in the \pt\ spectrum of \Bd\ and 
\Lb, as described in Section~\ref{sec-mccom}. After reweighting the \pt\ 
spectrum of \Bd\ and \Lb, we have observed good agreement of MC with the data 
as seen in Section~\ref{sec-datamc}. 
However, there is an uncertainty on the exponential slope of data/MC, $p_1$ 
in Figure~\ref{fig:bptbefore}, which is limited by the amount of data used 
for comparison with the MC. We vary $p_1$ $\pm$ 1 $\sigma$ and re-weight the 
MC events after the analysis cuts numerically to calculate the efficiency 
change. In addition, we vary the variables which depend on the \Lb\ \pt\ 
spectrum accordingly, eg: cross-section correction factors and the 
result of \yile\ by Le, \etal~\cite{yile:lblcpi} as described in 
Section~\ref{sec-lbxsec}. The total variation on $R$ due to the MC $\pt$ 
spectrum of \B\ hadrons is 0.38, 0.32, and ${+0.28 \atop -0.50}$.

\subsubsection{Pion Interaction with the Material} 
 One difference between our semileptonic and hadronic final states 
is the muon and the pion. The muon does not interact with the material via the 
hadronic (strong) interaction while the pion does. In order to model the track 
reconstruction efficiency correctly, two things have to be right: 
\begin{enumerate}
\item The type and the amount of material in the detector.
\item The model that describes the hadronic interaction cross-section, 
the final state multiplicities and kinematics.
\end{enumerate}
 We generate MC for the signals as described in Section~\ref{sec-mccom} except 
that we switch off the hadronic interaction in the detector simulation.
We compare the difference in the hadronic to semileptonic signal 
efficiency ratio, between the normal MC and the MC with the hadronic 
interaction off. This difference gives us an idea for the extreme 
situation, when the material is 100$\%$ wrong. For both the \Bd\ and 
\Lb\ modes, the efficiency ratio changes by 4$\%$.
From the study of 
Korn~\cite{cdfnote:6355}, we know that the available CDF detector simulation 
underestimate the amount of material by 15$\%$. 
In addition, a comparison between two programs which model the hadronic 
interactions, {\tt GHEISHA} and {\tt FLUKA}~\cite{Brun:1987ma}, 
has been done by Michael~\cite{numi:1994}. 
The {\tt FLUKA} package is known to better reproduce 
the experimental data but currently it is not available in the CDF detector 
simulation. The effect of the hadronic interaction model estimated by 
Michael is 20$\%$. Adding 15$\%$ and 20$\%$
 in quadrature, we get 25$\%$. We multiply the 100$\%$ efficiency ratio 
difference described earlier, with 0.25, and get 1$\%$.
We apply a scaling factor, 1.01, to all the relative efficiencies, including 
the semileptonic background to hadronic efficiency ratios. 
We then re-calculate $R$ and find $\sigma_R$ is 0.22, 0.17 and 0.22. 

\subsubsection{CMU reconstruction efficiency scaling factor}
The scaling factor to correct the difference of CMU muon reconstruction 
efficiency between MC and data has an uncertainty as described in 
Section~\ref{sec-eff}. We vary the scaling factor $\pm$ 1 $\sigma$ and 
calculate $\sigma_R$= 0.07, 0.05, and 0.07.

\subsubsection{XFT efficiency scaling factor}
\begin{figure}[tbp]
 \begin{center}
 \includegraphics[width=200pt, angle=0]{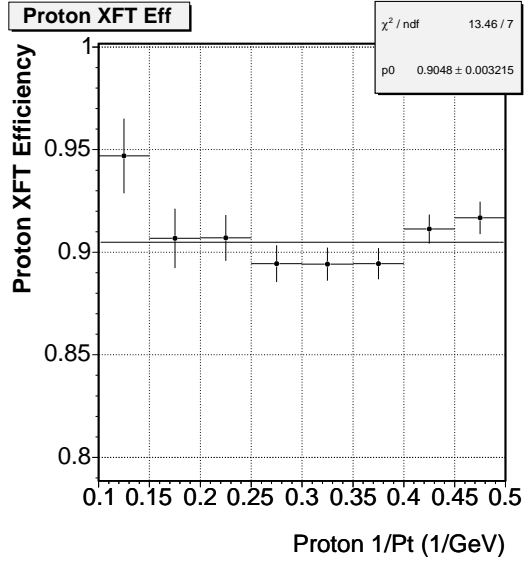}  
 \caption[Proton XFT efficiency scaling factor fit to a constant]
  {The relative proton XFT efficiency between MC and data in bins of 
1/\pt\ fit to a constant by Herndon~\cite{cdfnote:7301}.}
 \label{fig:xftpflat}
 \end{center}
 \end{figure}

We apply the XFT efficiency scaling factors data/MC in bins of inverse \pt\ 
from Herndon~\cite{cdfnote:7301} to correct the signal and background 
efficiencies.
The uncertainty on the kaon and pion scaling factors are varied $\pm$ 1 
$\sigma$ to evaluate $\sigma_R$. For the proton scaling factor, due to the 
limited statistics, we 
evaluate the systematic uncertainty following the suggestion in Herndon's 
analysis: we compare the difference by applying a constant efficiency 
scaling factor as shown in Figure~\ref{fig:xftpflat}, instead of the one 
based on the third order polynomial in Figure~\ref{fig:xftpkpi}. 
For all the three modes, the systematic uncertainties are negligible
as expected, 
since the final states of our semileptonic and hadronic modes are almost 
identical and the difference in the ionization of the pion and muon is 
insignificant. $\sigma_R$ is less than 0.01.

 \begin{figure}[tbp]
 \begin{center}
 \includegraphics[width=220pt, angle=0]
 {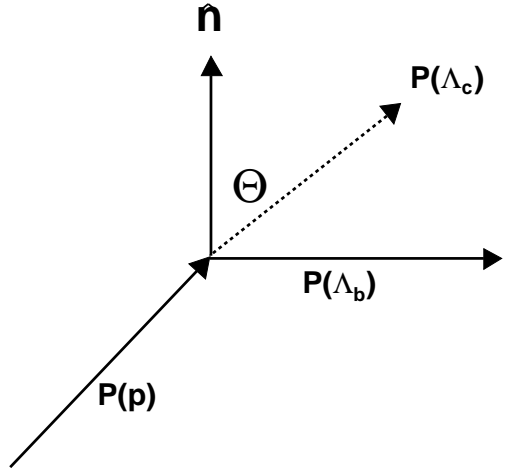}  
 \caption[Angle definition for the \Lb\ production polarization]
  {Angle definition for the \Lb\ production polarization, where the dashed line
   indicates the momentum of \Lamc\ in the rest frame of \Lb, and 
   $\mathbf{\hat{n}}$ is the polarization axis normal to the beam 
   proton-\Lb\ production plane.}
 \label{fig:lbpoldisplay}
 \end{center}
 \end{figure}

\subsubsection{\Lb\ and \Lamc\ polarizations}
 There is not yet a precise measurement of the production 
polarizations of \Lb\ and \Lamc, while the Standard Model predicts both 
particles are produced polarized. The angular distribution of of the \Lb\ 
daughters is parameterized by 
\begin{equation}
 \frac{dN}{d\cos\Theta} \propto 1 + {\cal P_B}\cos\Theta,
\end{equation}
 where ${\cal P_B}$ is the product of the \Lb\ polarization and the asymmetry 
parameter of the weak decay. $\Theta$ is defined as the angle between 
the \Lc\ momentum in the \Lb\ rest frame and the 
axis normal to the beam proton-\Lb\ production plane, $\mathbf{\hat{n}}$. 
Therefore, 
\begin{equation}
 \cos\Theta = \mathbf{\hat{P}}(\Lamc)\cdot \mathbf{\hat{n}},
 \label{eq:lbcost}
\end{equation}
where 
\begin{equation}
 \mathbf{\hat{n}} \equiv 
\frac{\mathbf{\hat{P}}(p)\times\mathbf{\hat{P}}(\Lb)}
{|\mathbf{\hat{P}}(p)\times\mathbf{\hat{P}}(\Lb)|}, 
\end{equation}
Here ``$\times$'' (``$\cdot$'') means vector (scalar) product of two vectors.
See Figure~\ref{fig:lbpoldisplay} for the definition of $\Theta$ and 
$\mathbf{\hat{n}}$.
The angular distribution of \Lamc\ daughters is parameterized in a 
similar way;
\begin{equation}
 \frac{dN}{d\cos\theta} \propto 1 + {\cal P_C}\cos\theta,
\end{equation}
 where $\theta$ is defined as the angle between the proton momentum in 
the rest frame of \Lamc\ and the \Lamc\ momentum in the lab frame, i.e.
\begin{equation}
 \cos\Theta = \mathbf{\hat{P}}(\Lamc)\cdot\mathbf{\hat{p}},
 \label{eq:lccost}
\end{equation}
Note that here $\mathbf{\hat{P}}(\Lamc)$ is in a different frame from that in 
Equation~\ref{eq:lbcost}. The value of ${\cal P_B}$ (${\cal P_C}$) is 
$\pm$ 1 for the polarized state and 0 for the unpolarized state.

The \bgen\ and \evtgen\ do not include the polarization of \Lb\ and \Lamc.
 We use the default settings for the central 
value of $R$. We study the systematics due to a non-null polarization using 
the generator-level signal MC without the detector and trigger simulation. 
We use the ``acceptance-rejection 
(Von Neumann)'' method~\cite{pdg:mctech} and reweight the MC according to:  
\[
   (1 + {\cal P_B}\cos\Theta)\cdot(1 + {\cal P_C}\cos\theta),
\] 
where all combinations of ${\cal P_B}$ and ${\cal P_C}$ for values 
at -1, 0, 1 are used. Each MC starts from a different random seed. 
We apply generator-level analysis-like cuts and 
obtain the efficiency ratio for each combination. We compare these 
efficiency ratios with that from the MC generated with zero ${\cal P_B}$ and 
${\cal P_C}$. We find the efficiency ratios are 
mainly determined by the ${\cal P_C}$ as seen in Table~\ref{t:lbpolarsys}, 
i.e. the efficiency ratios with the same ${\cal P_C}$, but different 
${\cal P_B}$ are consistent with each other. 
 Therefore, we apply scaling factors from the two ${\cal P_C}$ 
values: -1 and 1 ($\pm$ 1.017$\%$) on all the relative efficiencies and 
re-calculate $R$. We find $\sigma_R$ = 0.37.

    	\begin{table}[tbp!]
  	\caption{$\epsilon(\lbhad)/\epsilon(\lbsemi)$ from each 
	combination of  ${\cal P_B}$ and  ${\cal P_C}$.}
        \label{t:lbpolarsys}
	\begin{center}
 	\begin{tabular}{|r|r|r@{\,$\pm$\,}l|r|}
	\hline
	\hline
	 ${\cal P_B}$ &  ${\cal P_C}$ &
	\multicolumn{2}{|c|}{$\frac{\epsilon(\lbhad)}{\epsilon(\lbsemi)}$} &
	Scaling factor \\
	\hline
          0 & 0 & 3.225 & 0.010 & 1\\
          0 & -1& 3.193 & 0.007 & 0.990 $\pm$ 0.004 \\
          0 & 1 & 3.281 & 0.006 & 1.017 $\pm$ 0.004 \\
          1 & 0 & 3.232 & 0.006 & 1.002 $\pm$ 0.004 \\
          1 & -1& 3.193 & 0.009 & 0.990 $\pm$ 0.004 \\
          1 & 1 & 3.275 & 0.009 & 1.016 $\pm$ 0.004 \\
         -1 & 0 & 3.235 & 0.006 & 1.003 $\pm$ 0.004 \\
         -1 & 1 & 3.274 & 0.009 & 1.015 $\pm$ 0.004 \\
         -1 & -1 & 3.175 & 0.011 & 0.985 $\pm$ 0.005 \\
        \hline
        \hline
  	\end{tabular}
       \end{center}
	\end{table}

\subsubsection{\Lamc\ Dalitz structure}
\begin{figure}[tbp]
 \begin{center}
 \includegraphics[width=200pt, angle=0]{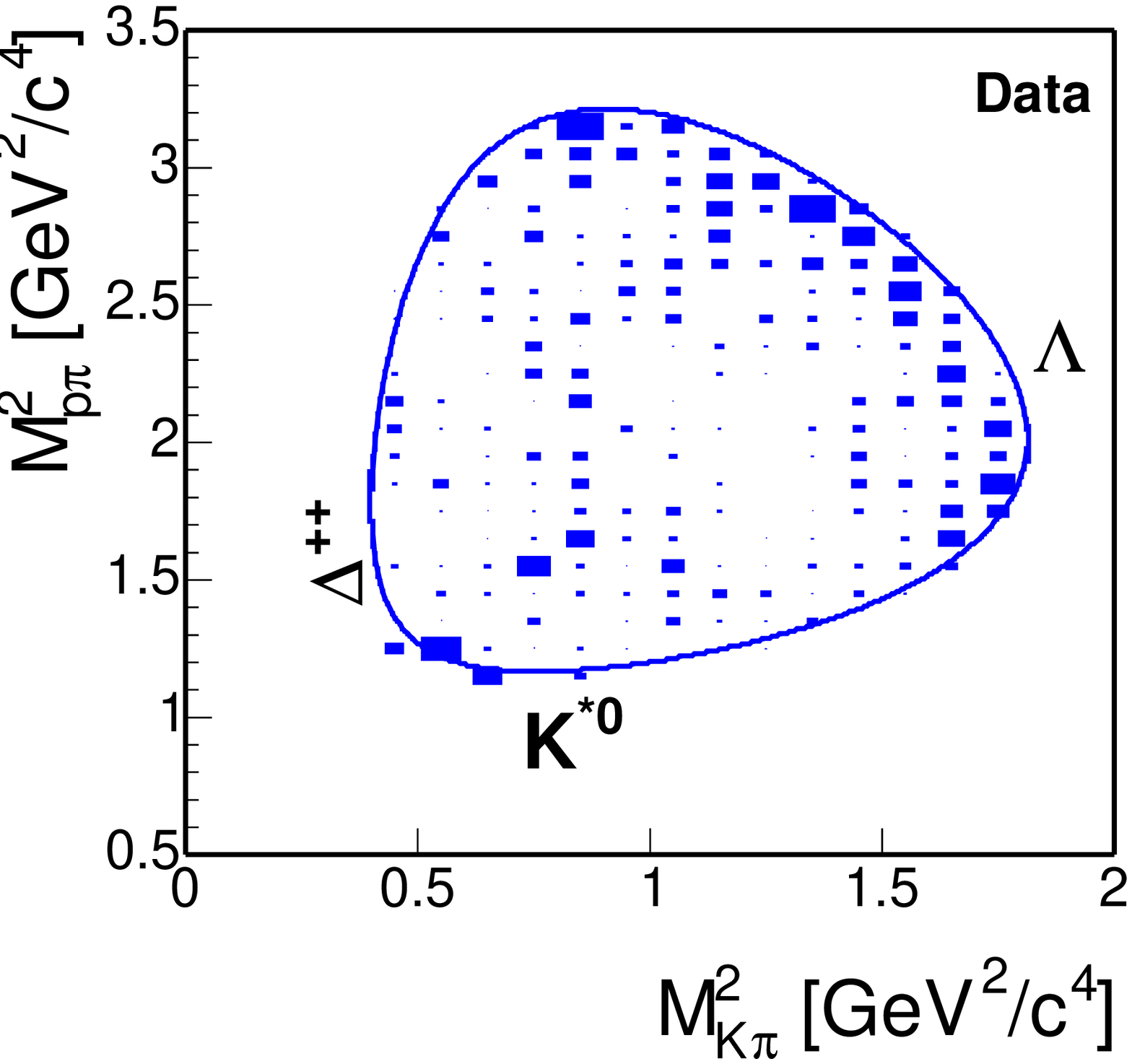}
 \includegraphics[width=200pt, angle=0]{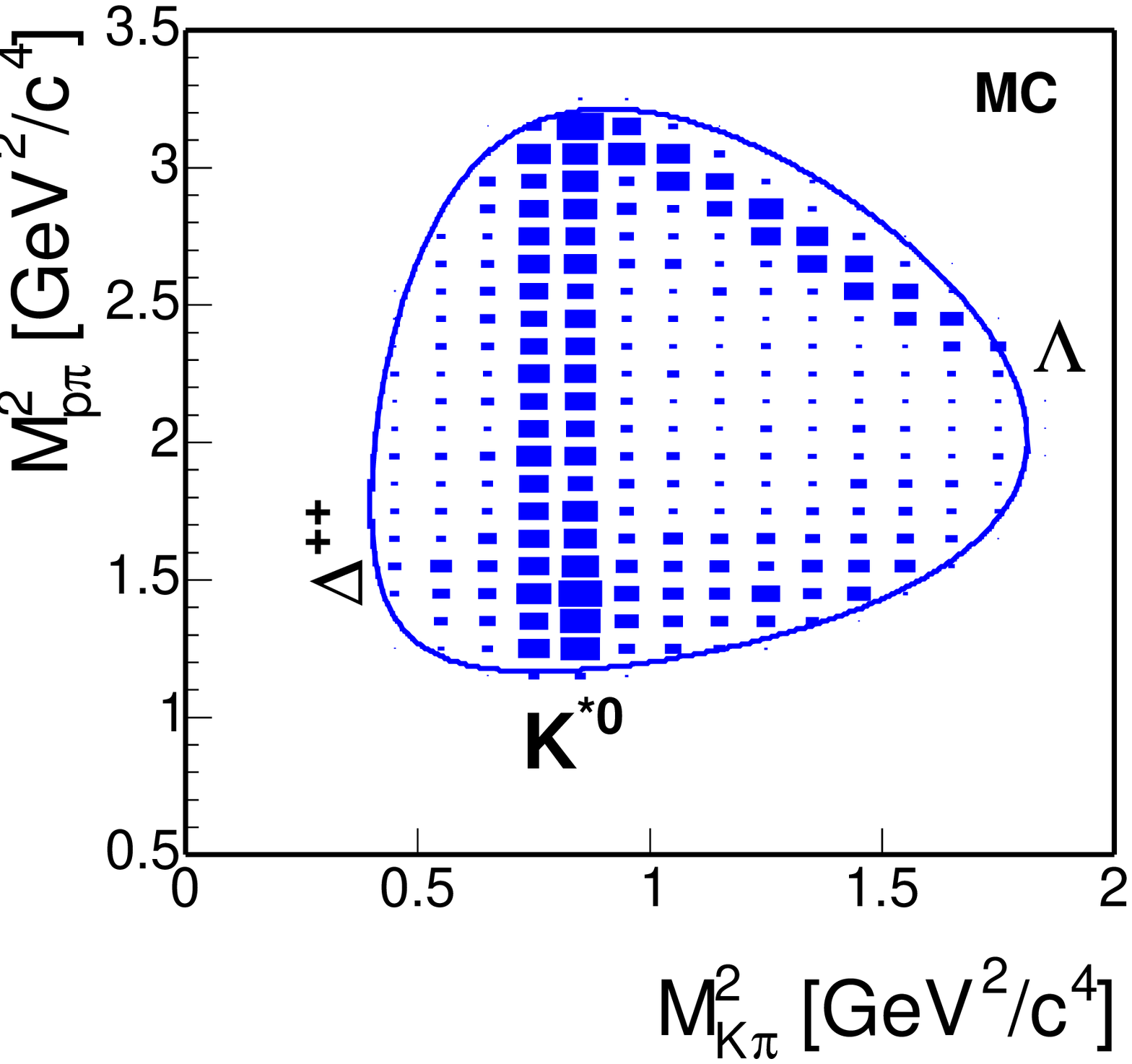}  
 \caption[\Lamc\ Dalitz structure in \inclbsemi\ data and MC]
 {\Lamc\ Dalitz structure in the sideband subtracted \inclbsemi\ data (left) 
  and MC (right). The concentrated areas in the top figure indicate the 
  existence of $K^*(892)^0$ and $\Lambda(1520)$. Clearly, the destructive
 interference between the resonant states are not simulated in the MC.
 \label{fig:lcdalitzmcdata}
}
 
\end{center}
 \end{figure}

The \Lamc\ from our \lbhad\ and \inclbsemi\ signal, decays into $p$, $K$, 
and $\pi$ in the final state. However, any two \Lamc\ daughters 
could form an intermediate resonant state, 
see Table~\ref{t:lcdalitz}. 
The resonant structure is called the ``Dalitz'' structure in the literature, 
and is usually displayed with a Dalitz plot~\cite{dalitz:phil}, where the 
invariant mass square of one pair of daughters is plotted versus 
another pair in the two-dimension. Figure~\ref{fig:lcdalitzmcdata} (left) 
shows the Dalitz plot from the \inclbsemi\ data after sideband subtraction.
If a resonance exists, a concentrated area near the mass of the 
resonant particle will be visible. 
The momenta of $p$, $K$ and $\pi$ are affected by the 
Dalitz structure and \Lb\ decays have different efficiencies for 
various structures. 
However, \evtgen\ does not take into account the interference of each 
resonant state. Each state is considered as an 
independent decay with a \br\ measured by E791~\cite{Aitala:1999uq} and 
listed in Table~\ref{t:lcdalitz}. See Figure~\ref{fig:lcdalitzmcdata} 
(right) for the \Lamc\ Dalitz structure in the MC.
 
Without a better model to describe the \Lamc\ Dalitz structure, we 
study the change in the efficiency ratio of \lbhad\ to \lbsemi\ by varying 
the \br\ in Table~\ref{t:lcdalitz} $\pm$ 1$\sigma$. 
We generate four sets of \lbhad\ and \lbsemi\ MC samples without 
detector and trigger simulation, 
where \Lamc\ decay is forced to one single mode.  The efficiency of \Lb\ decay 
with $\Lc\rightarrow pK^-\pi^+ (\mathrm{total})$, is then the sum of the 
\br\ weighted efficiency of each individual \Lamc\ mode.
\begin{eqnarray}
{\cal E}_c & = & \frac{\sum_i^4 \br_i\cdot\epsilon_i}
	{\sum_i^4\br_i},\\ 
R_c & = &\frac{{\cal E}_c(\lbhad)}{{\cal E}_c(\lbsemi)},
\end{eqnarray}
where ${\cal E}_c$ ($R_c$) is the total weighted efficiency (ratio) 
using the central value of each \Lamc\ \br. 
We re-calculate the absolute and relative efficiency, by varying \br\ of 
each \Lamc\ decay $\pm$ 1 $\sigma$; 
\begin{eqnarray}
{\cal E}_j & = & \frac{\sum_i^3 \br_i\cdot\epsilon_i + 
                           (\br_j+\sigma_j)\cdot\epsilon_j}
	{\sum_i^4\br_i + \sigma_j}, \\
R_j & = & \frac{{\cal E}_j(\lbhad)}{{\cal E}_j(\lbsemi)},
\end{eqnarray}
where ${\cal E}_j$ ($R_j$) is the total weighted efficiency (ratio) with 
\br\ of $j^{th}$ mode varied by $\pm$ 1 $\sigma$ and other \br\ fixed. 

We find a fractional change of $0.3\%$ after adding the difference of 
each $R_j$ from $R_c$ in quadrature. 
We apply this fractional change to the relative efficiencies of 
all the semileptonic backgrounds to the hadronic signal and calculate 
$\sigma_R$= 0.07.   

    	\begin{table}[tbp!]
         \setdec 00.00
  	\caption{\Lamc\ decays with $p$, $K$, $\pi$ in the final state.}
        \label{t:lcdalitz}
	\begin{center}
 	\begin{tabular}{l|r@{\,$\pm$\,}l|r@{\,$\pm$\,}l|r}
	\hline
	 Decay Mode & 
	\multicolumn{2}{c|}{$\br(\%)$} & 
	\multicolumn{2}{|c}{$\frac{\epsilon(\lbhad)}{\epsilon(\lbsemi)}$} &
	\multicolumn{1}{|c}
	{$R_j$}
	\\
	\hline
	\hline
	 $pK^{*}(890)^0$ 
	& \dec 1.6 & \dec 0.5 & \dec 3.17 & \dec 0.01 & 3.22 \\
	 $\Delta(1232)^{++}K^{-}$ 
	& \dec 0.86 & \dec 0.30 & \dec 3.24 & \dec 0.01 & 3.23 \\
         $\Lambda(1520)\pi^+$ 
	& \dec 0.59 & \dec 0.21 & \dec 3.30 & \dec 0.01 & 3.23\\
         non-resonant 
	& \dec 2.8 & \dec 0.8 & \dec 3.24 & \dec 0.01 & 3.23 \\
 	\hline
	\multicolumn{5}{l|}{$R_c$} & 3.23 $\pm$ 0.01 \\
        \hline
        \hline
  	\end{tabular}
       \end{center}
	\end{table}

\subsubsection{\Lb\ Lifetime}
The world average \Lb\ lifetime is lower than the theoretical prediction. 
A smaller \Lb\ lifetime gives a smaller efficiency for reconstructing
 \Lb\ decays. While we cut on the 
\ctau\ of \lbhad, the \lbsemi\
 is not fully reconstructed and we actually cut on the pseudo-\ctau\ of the 
inclusive semileptonic decays. Therefore, systematics due to the 
uncertainty on the \Lb\ lifetime may not cancel in our measurement. 
We study this effect by generating \lbsemi\ and \lbhad\ MC without 
detector and trigger simulation. We vary the lifetime of \Lb\ $\pm$ 15$\%$ 
around the central value: 1.229 $ps$. We compare the difference of the 
relative efficiency ratio from the central value. We then apply a scaling 
factor from the signals, on the efficiency ratios of the semileptonic 
backgrounds to the hadronic mode, and calculate $\sigma_R$=0.22. 

\subsubsection{Semileptonic \Lb\ decay model}
In Section~\ref{sec-eff}, we introduce a scaling factor, $f_c$,
 which accommodates the acceptance difference between the flat phase space MC 
and the form factor weighted MC. We vary the $f_c$ $\pm$ 1 $\sigma$ according 
to its statistical uncertainty and obtain $\sigma_R = \pm 0.57$. 

\subsection{Systematic Uncertainty for Each Mode}
 Tables~\ref{t:sysdstar}--\ref{t:syslc} list the result of systematic 
uncertainties as discussed above. The systematics from the external 
information are separated from the ones from the CDF MC and measurements. 
Table~\ref{t:sysall} summarizes the uncertainties from each category.
The statistical uncertainties on the relative 
branching fractions are also listed for comparison. 
      
      \begin{table}[tbp]
       \caption{Statistical and systematic uncertainties of 
	$\frac{{\cal B}(\dstarsemi)}{{\cal B}(\dstarhad)}$.}
	\label{t:sysdstar}
       \begin{center}
 	\begin{tabular}{lr}  
        \hline
	\hline 
        Source & $\sigma_R$ \\	
        \hline
	\hline
	Statistical & $\pm$ \rdstare \\
        \hline
	\hline
	\multicolumn{2}{c}{Measured \br} \\
	\hline
        \dstarhad & $\pm$ 0.29 \\ 
        \bpdonezeromunu &$\pm$ 0.31 \\ 
        \seqtau & $<$ 0.01 \\
	\hline
  	& $\pm$ 0.43 \\ 
	\hline
	\hline
	\multicolumn{2}{c}{Unmeasured \br}\\
	\hline 
        \seqdonezero & $\pm$ 0.05 \\ 
        \seqdpronezero & $\pm$ 0.04\\ 
        \seqdonep & $\pm$ 0.03\\ 
	\seqdpronep & $\pm$ 0.02\\ 
        \hline
	\bpdpronezeromunu & $\pm$ 0.70\\ 
	\bpdstarpimunu & $\pm$ 0.39\\ 
	\bddstartau & $\pm$ 0.31\\ 
	\bddonemunu & $\pm$ 0.53\\ 
	\bddpronemunu & $\pm$ 0.34\\ 
	\bddstarpizeromunu & $\pm$ 0.19\\ 
	\hline
  	& $\pm$ 1.09 \\ 
	\hline 
	\hline 
	\multicolumn{2}{c}{CDF Internal Systematics}\\
 	\hline
        Fitting of \dstarhad & $<$ 0.01 \\ 
	Fake $\mu$ estimate & $\pm$ 0.07 \\ 
	\bb\ and \cc\ background & $\pm$ 0.22 \\ 
	MC sample size & $\pm$ 0.28 \\ 
	MC $\pt(\Bd)$ & $\pm$ 0.38 \\ 
        $\pi$ interaction with the material & $\pm$ 0.22\\ 
	CMU reconstruction efficiency scaling factor & $\pm$ 0.07 \\
	XFT efficiency scaling factor & $<$ 0.01 \\
	\hline
	& $\pm$ 0.58 \\
        \hline \hline
   	\end{tabular}
       \end{center}
	\end{table}


      \begin{table}[tbp]
       \caption{Statistical and systematic uncertainties of 
	$\frac{{\cal B}(\dsemi)}{{\cal B}(\dhad)}$.}
	\label{t:sysd}
       \begin{center}
 	\begin{tabular}{lr} 
        \hline
	\hline
        Source & $\sigma_R$ \\	
	\hline
        \hline
	Statistical & $\pm$ \rde \\
        \hline
	\hline
	\multicolumn{2}{c}{Measured \br} \\
	\hline 
	 \dhad & $\pm$ 0.70 \\	
	 \dstarsemi & $\pm$ 0.22\\	
	 \bpdonezeromunu & $\pm$  0.08\\
	 $D^{*+} \rightarrow D^{+}\pi^{0}$ & $\pm$  0.11\\	
	 \seqtau & $<$ 0.01 \\
	 $f_s/f_d$ & $\pm$  0.01 \\	
	\hline
	& $\pm$ 0.75 \\
        \hline
	\hline
	\multicolumn{2}{c}{Unmeasured \br}\\ 
	\hline
        \seqdonezero & $\pm$  0.01 \\	
	\seqdpronezero & $\pm$  0.01\\
	\hline 
	\bpdpronezeromunu & $\pm$ 0.17\\
	\bpdpimunu & $\pm$  0.79\\	
	\bddpizeromunu & $\pm$  0.39\\	
	\bddtau & $\pm$  0.10\\	
	\bsdkzero & $\pm$  0.09\\	
	\hline
	& $\pm$ 0.91 \\
	\hline 
	\hline 
	\multicolumn{2}{c}{CDF Internal Systematics}\\
 	\hline
        Fitting of \dhad & $\pm$ 0.38 \\ 
        Fitting of \dsemi & $\pm$ 0.13 \\ 
	Fake $\mu$ estimate & $\pm$ 0.07 \\ 
	\bb\ and \cc\ background & $\pm$ 0.22 \\ 
	MC sample size & $\pm$ 0.18 \\ 
	MC $\pt(\Bd)$ & $\pm$ 0.32 \\ 
        $\pi$ interaction with the material & $\pm$ 0.17\\ 
	CMU reconstruction efficiency scaling factor & $\pm$ 0.05 \\
	XFT efficiency scaling factor & $<$ 0.01 \\
	\hline
	& $\pm$ 0.62 \\
        \hline \hline
   	\end{tabular}
       \end{center}
	\end{table}


      \begin{table}[tbp]
       \caption{Statistical and systematic uncertainties of 
	$\frac{{\cal B}(\lbsemi)}{{\cal B}(\lbhad)}$.}
	\label{t:syslc}
       \begin{center}
	\begin{tabular}{lr} 
        \hline
	\hline
        Source & $\sigma_R$ \\	
	\hline
        \hline
	Statistical & $\pm$ \rlbe \\
        \hline
	\hline
	\multicolumn{2}{c}{Measured \br} \\
        \hline
	\lbhad & ${+0.73 \atop -2.07}$ \\
	\seqtau & $<$ 0.01 \\
	\hline
	Total & ${+0.73 \atop -2.07}$ \\
	\hline \hline
	\multicolumn{2}{c}{Unmeasured \br}\\ 
        \hline
	\hline
 	\lblcstar & $\pm$ 0.21 \\
 	\lblcsstar &  $\pm$ 0.27 \\
 	\lbsigmaczero, 
 	\lbsigmacp,          
 	\lbsigmacpp & $\pm$ 0.24 \\
        \lblcfzero & $\pm$ 0.05 \\
        \lblcpizero , \lblcpim & $\pm$ 0.20 \\
 	\lblctau & $\pm$ 0.10 \\
	\bplcpmunu & $\pm$ 0.11   \\
        \bdlcnmunu & $\pm$ 0.11    \\
	\hline
	Total	& $\pm$ 0.50 \\
	\hline
	\hline 
	\multicolumn{2}{c}{CDF Internal Systematics}\\
 	\hline
        Fitting of \lbhad & $\pm$ 0.63 \\ 
	Fake $\mu$ estimate & $\pm$ 0.17 \\ 
	\bb\ and \cc\ background & $\pm$ 0.04 \\ 
	MC sample size & $\pm$ 0.32 \\ 
	MC $\pt(\Lb)$ & ${+0.28 \atop -0.50}$ \\ 
        $\pi$ interaction with the material & $\pm$ 0.22 \\ 
	CMU reconstruction efficiency scaling factor & $\pm$ 0.07 \\ 
	XFT efficiency scaling factor & $<$ 0.01 \\
	\Lb\ and \Lamc\ polarizations & $\pm$ 0.37 \\
        \Lc\ Dalitz structure & $\pm$ 0.07 \\ 
        \Lb\ lifetime  & $\pm$ 0.22 \\
        Semileptonic \Lb\ decay model & $\pm$ 0.57 \\       
	\hline
	& ${+1.09 \atop -1.15}$ \\
        \hline \hline
   	\end{tabular}
       \end{center}
	\end{table}

      \begin{table}[tbp]
       \caption{Summary of statistical and systematic uncertainties.}
	\label{t:sysall}
       \begin{center}
	\begin{tabular}{l|r|r|r|} 
        \hline\hline
	& \multicolumn{3}{|c|}{$\frac{\sigma_R}{R}$ ($\%$)} \\ \hline
        Source & $\frac{{\cal B}(\dstarsemi)}{{\cal B}(\dstarhad)}$ &
	$\frac{{\cal B}(\dsemi)}{{\cal B}(\dhad)}$ &
        $\frac{{\cal B}(\lbsemi)}{{\cal B}(\lbhad)}$ \\
        \hline 
        Measured \br\ ($\%$)& 2.4 & 7.7 & ${+3.5 \atop -10.5}$\\
        Unmeasured \br\ ($\%$)& 6.2 & 9.3 & 2.5 \\ 
        CDF internal ($\%$)& 3.3 & 6.4 & 6.0\\ 
        Statistical ($\%$) & 13.1 & 10.2 & 15.0\\
        \hline \hline
       \end{tabular}
        \end{center}
        \end{table}

\subsection{Consistency Check of $R$}
\label{sec-check}
\begin{figure}[htbp]
 \begin{center}
 \begin{tabular}{cc}
 \includegraphics[width=200pt, angle=0]{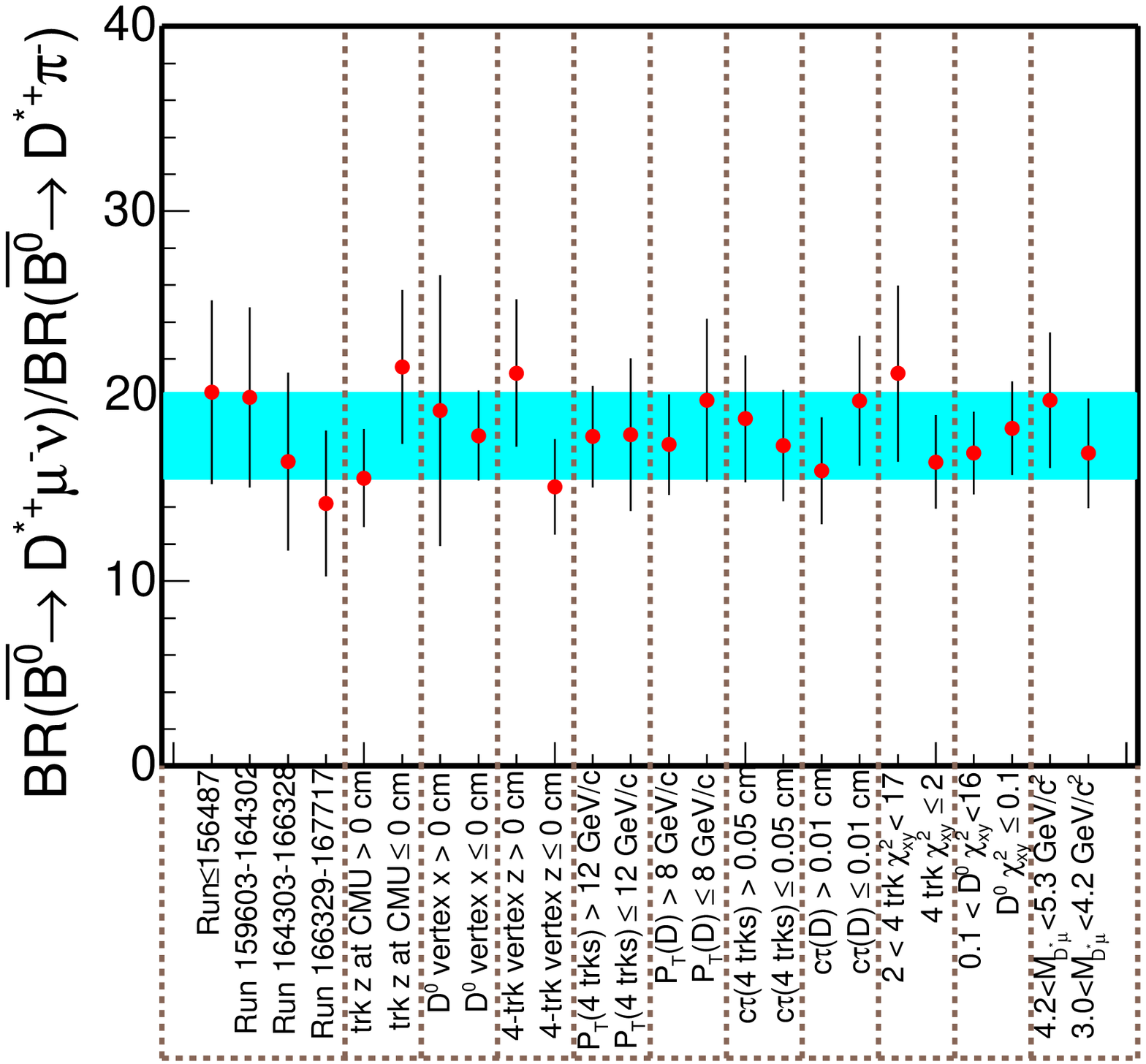}  &
 \includegraphics[width=200pt, angle=0]{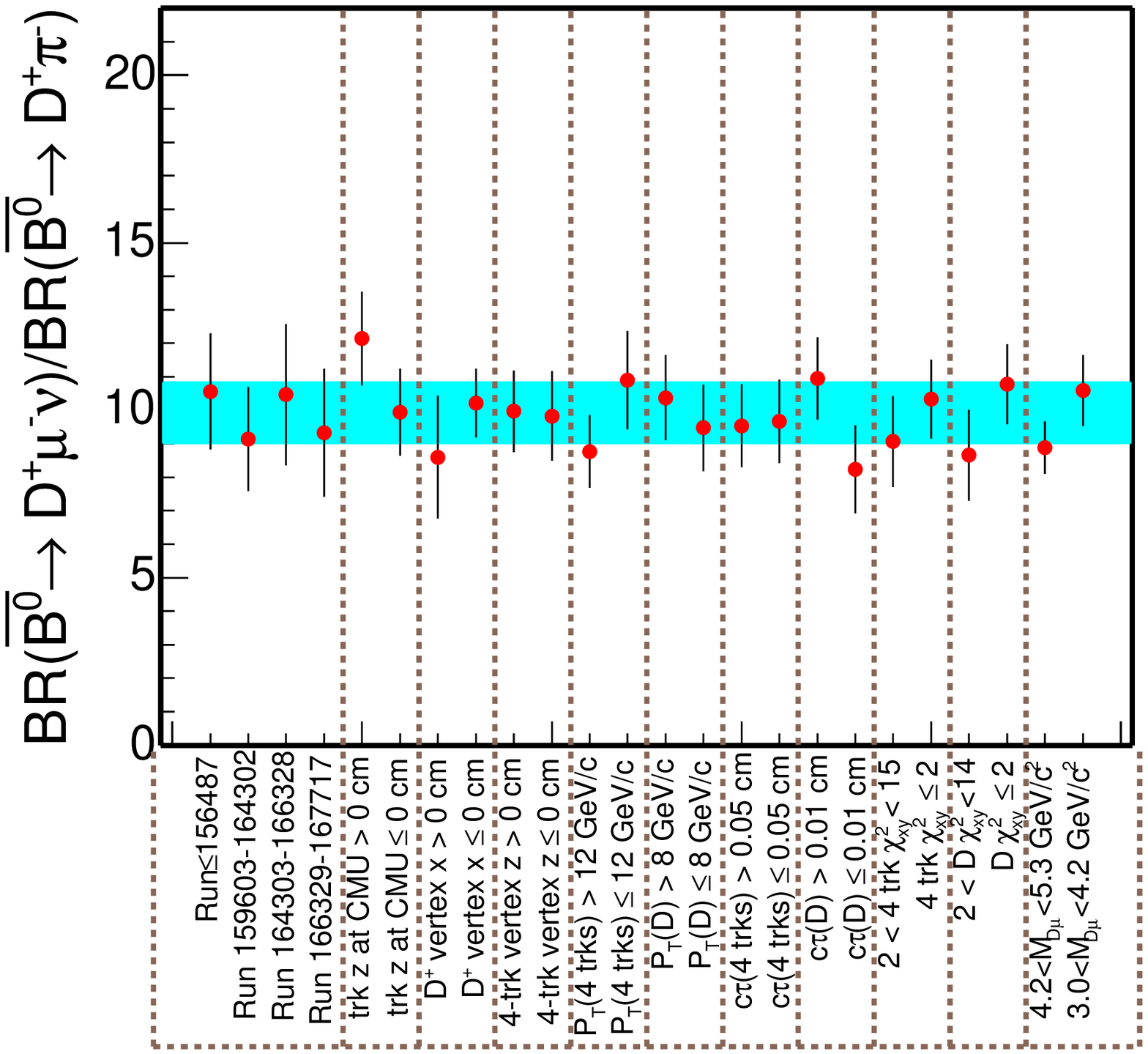} \\ 
 \multicolumn{2}{c}{\includegraphics[width=200pt, angle=0]{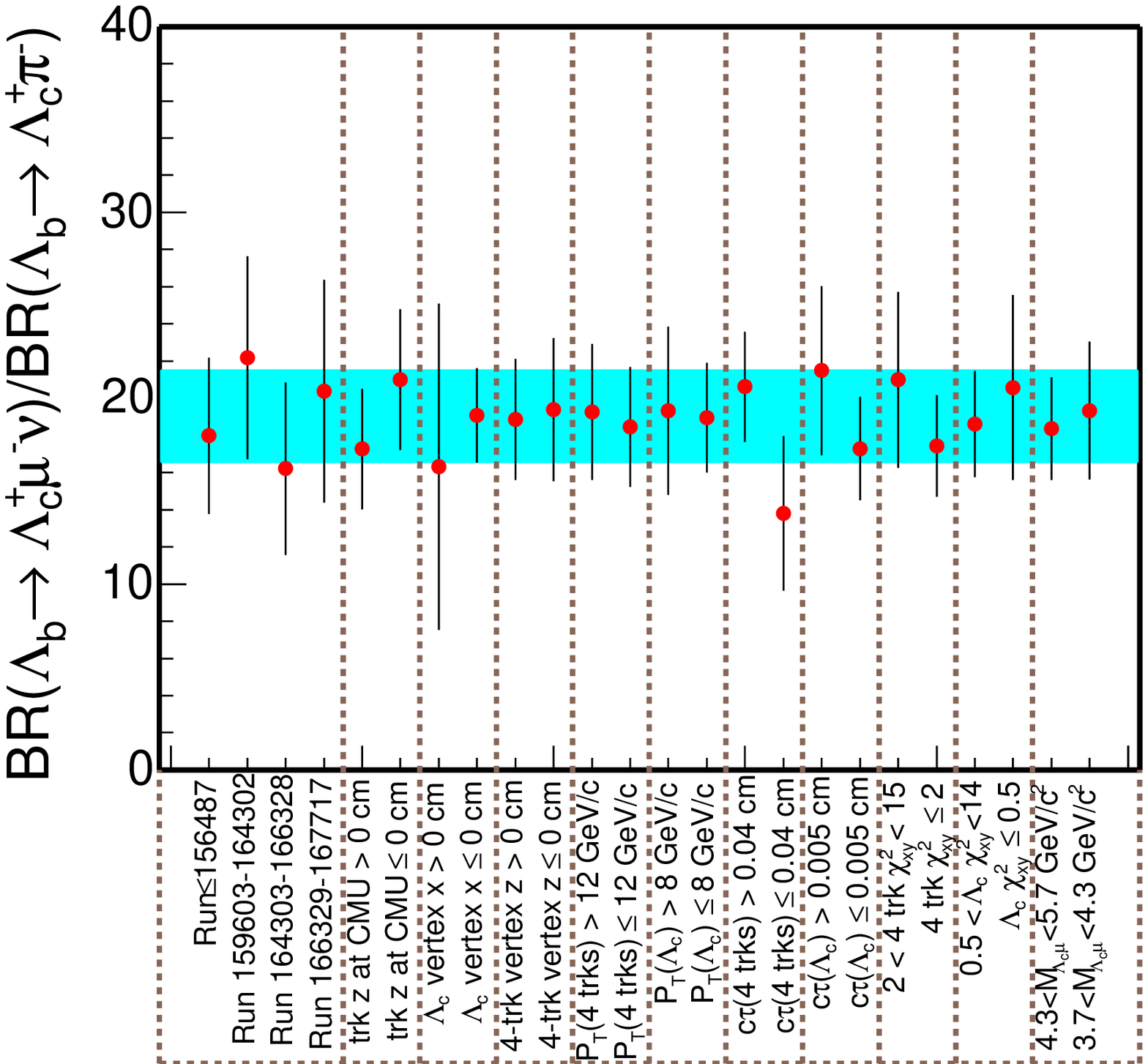}}\\
 \end{tabular}
 \caption[Consistency check of the relative branching fractions]
 {Consistency check of ${\cal B}(\dstarsemi)/{\cal B}(\dstarhad)$ (top left),
 ${\cal B}(\dsemi)/{\cal B}(\dhad)$ (top right) and 
 ${\cal B}(\lbsemi)/{\cal B}(\lbhad)$ (bottom). The uncertainty on each 
 point is statistical only.  
 Each independent group is separated by a vertical dashed line.
\label{fig:consistbr}}
 \end{center}
 \end{figure}

 In order to detect any unexpected systematics in $R$, we separate the data 
and MC into several groups of independent subsets according to the run 
number, vertex position, \ctau\ and \pt\ of the charm and \B\ hadrons, and 
etc. We cross-check the consistency of the $R$ within each group. 
Figure~\ref{fig:consistbr} displays the result of the cross-check, where 
the uncertainties in the figure are statistical only. The $R$ 
from all the subsets are consistent with the other subsets in the same 
group. 

\section{Measurement Result}
\label{sec-br}
 For the control modes, we measure the relative branching fractions to be:
\[ 
\frac{{\cal B}(\dstarsemi)}{{\cal B}(\dstarhad)} = 
	\rdstarc\ \pm \rdstare\ (stat) \pm \rdstarsyste\ (syst)
	         \pm \rdstarbre\ (BR) \pm \rdstarubre\ (UBR),
\]
and 
\[ 
\frac{{\cal B}(\dsemi)}{{\cal B}(\dhad)} = 
	\rdc\ \pm \rde\ (stat) \pm \rdsyste\ (syst)
	         \pm \rdbre\ (BR) \pm \rdubre\ (UBR),
\]
 which are consistent with the ratios obtained by the PDG, \(19.7 \pm 1.7\) 
and \(7.8 \pm 1.0\) at the 0.7 and 1.1 $\sigma$ level. Finally, we measure the 
relative \Lb\ branching fraction to be:
\[ 
\frac{{\cal B}(\lbsemi)}{{\cal B}(\lbhad)} = 
	\rlbc\ \pm \rlbe\ (stat) \pm \rlbsyste\ (syst)
	         \rlbbre\ (BR) \pm \rlbubre\ (UBR).
\]
The uncertainties of the relative branching fractions are from statistics, 
CDF internal systematics, external measured branching ratios and unmeasured 
branching ratios, respectively. 

\section{Estimate of the ${\cal B}(\lbsemi)$}
\label{sec-exptheory}
We have just presented the first measurement of the ratio of 
\Lb\ exclusive semileptonic to hadronic branching fractions. The ratio 
provides important input for the absolute branching fraction of \lbsemi\ or 
\lbhad. 
Leibovich~\etal~\cite{Leibovich:2003tw} predict 
${\cal B}(\lbhad)$ = 0.45$\%$ and ${\cal B}(\lbsemi)$ = 6.6$\%$, which gives a 
relative branching fraction of 14.7. However, the largest theoretical 
uncertainty from the functional form of the Isgur-Wise function is 30$\%$, 
due to the assumption of the large $N_c$ limit (see Section~\ref{sec-heavyq}). 
 Our measurement of the ratio has a $19\%$ 
uncertainty and may stimulate additional theoretical work. 
Multiplying our \Lb\ relative branching fraction, with our derivation of 
${\cal B}(\lbhad)$ from the CDF measurement of \yile~\cite{yile:lblcpi} 
in Section~\ref{sec-lbxsec}:
\[ 
{\cal B}(\lbhad) = \left(\brlbhadc \pm \brlbhade\ (stat \oplus syst)  \brlbhadpte\ ({\it \pt}) \right)\%,
\]
we obtain 
\[
{\cal B}(\lbsemi)=  
      \left(\brlbsemidatac\ \pm \brlbsemidatae\ (stat)	
	\brlbsemidatasyste\ (syst) 
        \pm \brlbsemidatabre\ ({\cal B}(\lbhad)) \right) \%.
\]
which is also consistent with a recent DELPHI 
result derived from the \lbsemi\ form factor measurement~\cite{Abdallah:2003gn}, \[{\cal B}(\lbsemi)^\mathrm{DELPHI}= \left( 
	5.0 {+1.1 \atop -0.9} (stat) {+1.6 \atop -1.2} (syst)\right)\%\]
Combining our and DELPHI's numbers, we 
obtain $\left (5.5 \pm 1.8 (stat \oplus syst)\right)\%$.   
Our relative branching ratios and the derived 
${\cal B}(\lbhad)$, ${\cal B}(\lbsemi)$ are all in agreement with the 
predictions by Leibovich~\etal, within large uncertainties. Note that 
the dominant uncertainties 
of ${\cal B}(\lbhad)$ arise from \rxsec\ and ${\cal B}(\seqlc)$. 
New CDF-II measurements of \rxsec\ are anticipated. However, 
a better measurement of ${\cal B}(\seqlc)$ has only been proposed by 
Dunietz~\cite{Dunietz:1998uz} and Migliozzi~\cite{Migliozzi:1999ca}. 
Improvements in the ${\cal B}(\lbhad)$ will reduce the uncertainties in our 
determination of the exclusive semileptonic branching ratio.

\section{Conclusion}
\label{sec-con}
We analyze 171.5~\pbarn\ of data collected with the CDF-II detector in the 
$p\overline{p}$ collisions at $\sqrt{s} = 1.96 \rm~TeV$. Using a novel
 secondary vertex track trigger, we reconstruct \nlbsemi\ \inclbsemi\ decays 
and \nlbhad\ \lbhad\ decays. This is the largest \Lb\ sample in 
the world, which enables us to measure the relative \Lb\ branching fractions 
and examine the Heavy Quark Effective Theory. 
We have also observed several \Lb\ semileptonic decays which have never been 
seen in the other experiments: \inclcstar, \inclcsstar, \incsigmaczero, and 
\incsigmacpp. 
In addition, we reconstruct the 
\(\overline{B}^0 \rightarrow \Dstar\) and \(\overline{B}^0 \rightarrow \D\) 
decays similar to our \Lb\ decays and use them as the control samples to 
understand the issues associated with the \Lb\ measurement. 
After the estimate and the subtraction of the background in the inclusive 
semileptonic signal, we correct the yield observed in the data with the 
trigger and reconstruction efficiencies obtained from the Monte Carlo. We find 
the relative branching fraction of the control modes in good agreement with 
the values obtained by the PDG. We measure the ratio of \Lb\ 
branching fraction to be:
\[
\frac{{\cal B}(\lbsemi)}{{\cal B}(\lbhad)} = 
	\rlbc \pm \rlbe (stat) \pm \rlbsyste (syst) 
	        \rlbbre (BR) \pm \rlbubre (UBR).
\]
The uncertainty is dominated by the size of the data sample and the 
branching ratio of \lbhad. More data and a more precise measurement 
of ${\cal B}(\lbhad)$ in the future will immediately improve our relative 
branching fraction measurement and our determination of ${\cal B}(\lbsemi)$.

\section{Future}
In addition to the measurement presented in this dissertation, several 
measurements can be performed at CDF and increase our knowledge of the \Lb. 
\paragraph{\Lb\ lifetime}
Previous analyses~\cite{Barate:1997if, Abreu:1999hu,Ackerstaff:1997qi,Abe:1996df} used semileptonic decays to measure the \Lb\ lifetime so 
the \Lb\ was not fully reconstructed. A scaling factor based on 
the MC was applied to convert the total momenta of the observed daughters 
to the \Lb\ parent momentum. On top of that, possible backgrounds from the 
other \Lb\ decays might not have been included. Therefore, a \Lb\ lifetime 
measurement using fully reconstructed decays, such as \lbhad\ or 
\(\Lb\rightarrow J/\Psi \Lambda\), can provide a model-independent comparison 
with the current world average and the theoretical prediction and possibly 
resolve the discrepancy as noted in Chapter~\ref{ch:intro}.
\paragraph{\Lb\ production cross-section}
Instead of measuring the ratio of two branching fractions, we can 
measure an absolute branching ratio directly if the \Lb\ production 
cross-section at the Tevatron is known. An alternative way is to measure 
the ratio of the \Lb\ to \Bd\ production cross-section. As noted in 
Section~\ref{sec-exptheory}, an accurate measurement of the \rxsec\ will 
reduce the uncertainty on the ${\cal B}(\lbhad)$ when normalizing the \Lb\ 
decays to the \B\ meson decays.

\paragraph{\lbsemi\ form factor and $|V_{cb}|$}
As the three-dimensional (3D) vertex reconstruction using the SVX information 
is not fully developed at CDF-II, yet, this analysis made requirements only on 
the two-dimensional vertex (\lxy) when reconstructing \lcmu\ events.
  Once the 3D vertex reconstruction is mature, we can use kinematic 
constraints to obtain the neutrino momentum and calculate the scalar 
product of the \Lb\ and the \Lamc\ four-velocities, $w$. The $w$ distribution 
can be fit to the Isgur-Wise function to obtain the slope parameter, $\rho^2$, 
in Equation~\ref{eq:zetaw}. When an exact form of the Isgur-Wise 
function is established, the \Lb\ semileptonic decays can give a 
$|V_{cb}|$ competitive with that from the \B\ meson decays~\cite{Isgur:1990pm,Georgi:1990cx}.

\paragraph{CP asymmetry in the \(\Lb\rightarrow p\pi\) and \(\Lb\rightarrow pK\) decays} 
Bensalem,~\etal~\cite{Bensalem:2002pz} and Dunietz~\cite{Dunietz:1992ti} 
predict a large T violation (or CP asymmetry) in the charmless \Lb\ decays 
in the Standard Model. The asymmetry can be studied without a need for the 
flavor tagging since the baryon decays are not affected by the mixing. 
However, \Lb\ charmless decays, such as \(\Lb\rightarrow p\pi\) and 
\(\Lb\rightarrow pK\), have not been observed, yet. A recent search for these 
decays at CDF-II by Carosi,~\etal~\cite{cdfnote:7443} set an upper limit on 
the branching fraction \({\cal B}(\Lb\rightarrow ph) \leq 23\cdot 10^{-6}\) 
at 90$\%$ confidence level, and improved the previous upper limit 
by ALEPH~\cite{Buskulic:1996tx}. More data in the future can either improve 
the upper limit or result a first observation of the signal. Once enough
 signal is established, a CP asymmetry measurement can be made.






\begin{thebibliography}{100}
\expandafter\ifx\csname bibnamefont\endcsname\relax
  \def\bibnamefont#1{#1}\fi
\expandafter\ifx\csname bibfnamefont\endcsname\relax
  \def\bibfnamefont#1{#1}\fi
\expandafter\ifx\csname url\endcsname\relax
  \def\url#1{\texttt{#1}}\fi
\expandafter\ifx\csname urlprefix\endcsname\relax\def\urlprefix{URL }\fi
\providecommand{\bibinfo}[2]{#2}
\providecommand{\eprint}[2][]{\url{#2}}

\bibitem{Albajar:1991sq}
\bibinfo{author}{\bibfnamefont{C.}~\bibnamefont{Albajar}} \emph{et~al.}
  (\bibinfo{collaboration}{UA1}), \bibinfo{journal}{Phys. Lett.}
  \textbf{\bibinfo{volume}{B273}}, \bibinfo{pages}{540} (\bibinfo{year}{1991}).

\bibitem{Abreu:1996mi}
\bibinfo{author}{\bibfnamefont{P.}~\bibnamefont{Abreu}} \emph{et~al.}
  (\bibinfo{collaboration}{DELPHI}), \bibinfo{journal}{Phys. Lett.}
  \textbf{\bibinfo{volume}{B374}}, \bibinfo{pages}{351} (\bibinfo{year}{1996}).

\bibitem{Buskulic:1996eq}
\bibinfo{author}{\bibfnamefont{D.}~\bibnamefont{Buskulic}} \emph{et~al.}
  (\bibinfo{collaboration}{ALEPH}), \bibinfo{journal}{Phys. Lett.}
  \textbf{\bibinfo{volume}{B380}}, \bibinfo{pages}{442} (\bibinfo{year}{1996}).

\bibitem{Abe:1996tr}
\bibinfo{author}{\bibfnamefont{F.}~\bibnamefont{Abe}} \emph{et~al.}
  (\bibinfo{collaboration}{CDF}), \bibinfo{journal}{Phys. Rev.}
  \textbf{\bibinfo{volume}{D55}}, \bibinfo{pages}{1142} (\bibinfo{year}{1997}).

\bibitem{cdfnote:6963}
\bibinfo{author}{\bibnamefont{{Andreas Korn}}},
  \emph{\bibinfo{title}{Measurement of the B Hadron Masses in exclusive
  $J/\psi$ Decay Channels}}, \bibinfo{type}{CDF Internal Note}
  \bibinfo{number}{CDF/ANAL/BOTTOM/CDFR/6963},
  \bibinfo{institution}{Massachusetts Institute of Technology}
  (\bibinfo{year}{2004}).

\bibitem{Abreu:1995me}
\bibinfo{author}{\bibfnamefont{P.}~\bibnamefont{Abreu}} \emph{et~al.}
  (\bibinfo{collaboration}{DELPHI}), \bibinfo{journal}{Z. Phys.}
  \textbf{\bibinfo{volume}{C68}}, \bibinfo{pages}{375} (\bibinfo{year}{1995}).

\bibitem{Barate:1997if}
\bibinfo{author}{\bibfnamefont{R.}~\bibnamefont{Barate}} \emph{et~al.}
  (\bibinfo{collaboration}{ALEPH}), \bibinfo{journal}{Eur. Phys. J.}
  \textbf{\bibinfo{volume}{C2}}, \bibinfo{pages}{197} (\bibinfo{year}{1998}).

\bibitem{pdg:2004}
\bibinfo{author}{\bibfnamefont{S.}~\bibnamefont{Eidelman}} \emph{et~al.}
  (\bibinfo{collaboration}{Particle Data Group}), \bibinfo{journal}{Phys.
  Lett.} \textbf{\bibinfo{volume}{B592}}, \bibinfo{pages}{1}
  (\bibinfo{year}{2004}).

\bibitem{Anikeev:2001rk}
\bibinfo{author}{\bibfnamefont{K.}~\bibnamefont{Anikeev}} \emph{et~al.}
  (\bibinfo{year}{2001}), \eprint{hep-ph/0201071}.

\bibitem{Weinberg:1995mt}
\bibinfo{author}{\bibfnamefont{S.}~\bibnamefont{Weinberg}},
  \emph{\bibinfo{title}{The Quantum theory of fields. Vol. 1--3: Foundations}}
  (\bibinfo{publisher}{Cambridge, UK: Univ. Pr.},
  \bibinfo{year}{1995,1996,2000}).

\bibitem{Peskin:1995ev}
\bibinfo{author}{\bibfnamefont{M.~E.} \bibnamefont{Peskin}} \bibnamefont{and}
  \bibinfo{author}{\bibfnamefont{D.~V.} \bibnamefont{Schroeder}},
  \emph{\bibinfo{title}{An Introduction to quantum field theory}}
  (\bibinfo{publisher}{Reading, USA: Addison-Wesley}, \bibinfo{year}{1995}).

\bibitem{Robertson:2005mt}
\bibinfo{author}{\bibfnamefont{R.~G.~H.} \bibnamefont{Robertson}},
  \bibinfo{journal}{Nucl. Phys. Proc. Suppl.} \textbf{\bibinfo{volume}{138}},
  \bibinfo{pages}{243} (\bibinfo{year}{2005}).

\bibitem{wittich:thesis}
\bibinfo{author}{\bibfnamefont{P.}~\bibnamefont{Wittich}},
  \emph{\bibinfo{title}{{First Measurement of the Flux of Solar Neutrinos}}},
  \bibinfo{type}{SNO Thesis}, \bibinfo{institution}{University of Pennsylvania}
  (\bibinfo{year}{2000}).

\bibitem{Ashie:2005ik}
\bibinfo{author}{\bibfnamefont{Y.}~\bibnamefont{Ashie}} \emph{et~al.}
  (\bibinfo{collaboration}{Super-Kamiokande})  (\bibinfo{year}{2005}),
  \eprint{hep-ex/0501064}.

\bibitem{Litchfield:2005np}
\bibinfo{author}{\bibfnamefont{P.~J.} \bibnamefont{Litchfield}}
  (\bibinfo{collaboration}{Soudan 2}), \bibinfo{journal}{Nucl. Phys. Proc.
  Suppl.} \textbf{\bibinfo{volume}{138}}, \bibinfo{pages}{402}
  (\bibinfo{year}{2005}).

\bibitem{Ambrosio:2004ig}
\bibinfo{author}{\bibfnamefont{M.}~\bibnamefont{Ambrosio}} \emph{et~al.}
  (\bibinfo{collaboration}{MACRO}), \bibinfo{journal}{Eur. Phys. J.}
  \textbf{\bibinfo{volume}{C36}}, \bibinfo{pages}{323} (\bibinfo{year}{2004}).

\bibitem{Kane:1987gb}
\bibinfo{author}{\bibfnamefont{G.~L.} \bibnamefont{Kane}},
  \emph{\bibinfo{title}{Modern Elementary Particle Physics}}
  (\bibinfo{publisher}{Reading, USA: Addison-Wesley}, \bibinfo{year}{1987}).

\bibitem{Cabibbo:1963yz}
\bibinfo{author}{\bibfnamefont{N.}~\bibnamefont{Cabibbo}},
  \bibinfo{journal}{Phys. Rev. Lett.} \textbf{\bibinfo{volume}{10}},
  \bibinfo{pages}{531} (\bibinfo{year}{1963}).

\bibitem{Kobayashi:1973fv}
\bibinfo{author}{\bibfnamefont{M.}~\bibnamefont{Kobayashi}} \bibnamefont{and}
  \bibinfo{author}{\bibfnamefont{T.}~\bibnamefont{Maskawa}},
  \bibinfo{journal}{Prog. Theor. Phys.} \textbf{\bibinfo{volume}{49}},
  \bibinfo{pages}{652} (\bibinfo{year}{1973}).

\bibitem{Chau:1984fp}
\bibinfo{author}{\bibfnamefont{L.-L.} \bibnamefont{Chau}} \bibnamefont{and}
  \bibinfo{author}{\bibfnamefont{W.-Y.} \bibnamefont{Keung}},
  \bibinfo{journal}{Phys. Rev. Lett.} \textbf{\bibinfo{volume}{53}},
  \bibinfo{pages}{1802} (\bibinfo{year}{1984}).

\bibitem{Harari:1986xf}
\bibinfo{author}{\bibfnamefont{H.}~\bibnamefont{Harari}} \bibnamefont{and}
  \bibinfo{author}{\bibfnamefont{M.}~\bibnamefont{Leurer}},
  \bibinfo{journal}{Phys. Lett.} \textbf{\bibinfo{volume}{B181}},
  \bibinfo{pages}{123} (\bibinfo{year}{1986}).

\bibitem{Fritzsch:1986gv}
\bibinfo{author}{\bibfnamefont{H.}~\bibnamefont{Fritzsch}} \bibnamefont{and}
  \bibinfo{author}{\bibfnamefont{J.}~\bibnamefont{Plankl}},
  \bibinfo{journal}{Phys. Rev.} \textbf{\bibinfo{volume}{D35}},
  \bibinfo{pages}{1732} (\bibinfo{year}{1987}).

\bibitem{Botella:1985gb}
\bibinfo{author}{\bibfnamefont{F.~J.} \bibnamefont{Botella}} \bibnamefont{and}
  \bibinfo{author}{\bibfnamefont{L.-L.} \bibnamefont{Chau}},
  \bibinfo{journal}{Phys. Lett.} \textbf{\bibinfo{volume}{B168}},
  \bibinfo{pages}{97} (\bibinfo{year}{1986}).

\bibitem{belle:tdr}
\bibinfo{journal}{Nucl. Instrum. Meth.} \textbf{\bibinfo{volume}{A479}},
  \bibinfo{pages}{117} (\bibinfo{year}{2002}).

\bibitem{babar:tdr}
\bibinfo{author}{\bibfnamefont{D.}~\bibnamefont{Boutigny}} \emph{et~al.}
  (\bibinfo{collaboration}{BABAR}) \bibinfo{note}{SLAC-R-0457}.

\bibitem{Arisaka:1992ch}
\bibinfo{author}{\bibfnamefont{K.}~\bibnamefont{Arisaka}} \emph{et~al.}
  \bibinfo{note}{FERMILAB-FN-0580}.

\bibitem{Wilson:1969zs}
\bibinfo{author}{\bibfnamefont{K.~G.} \bibnamefont{Wilson}},
  \bibinfo{journal}{Phys. Rev.} \textbf{\bibinfo{volume}{179}},
  \bibinfo{pages}{1499} (\bibinfo{year}{1969}).

\bibitem{Aglietti:1991rr}
\bibinfo{author}{\bibfnamefont{U.}~\bibnamefont{Aglietti}},
  \bibinfo{journal}{Phys. Lett.} \textbf{\bibinfo{volume}{B281}},
  \bibinfo{pages}{341} (\bibinfo{year}{1992}).

\bibitem{Manohar:2000dt}
\bibinfo{author}{\bibfnamefont{A.~V.} \bibnamefont{Manohar}} \bibnamefont{and}
  \bibinfo{author}{\bibfnamefont{M.~B.} \bibnamefont{Wise}},
  \emph{\bibinfo{title}{Heavy quark physics}} (\bibinfo{publisher}{Camb.
  Monogr. Part. Phys. Nucl. Phys. Cosmol.}, \bibinfo{year}{2000}).

\bibitem{Godfrey:1985xj}
\bibinfo{author}{\bibfnamefont{S.}~\bibnamefont{Godfrey}} \bibnamefont{and}
  \bibinfo{author}{\bibfnamefont{N.}~\bibnamefont{Isgur}},
  \bibinfo{journal}{Phys. Rev.} \textbf{\bibinfo{volume}{D32}},
  \bibinfo{pages}{189} (\bibinfo{year}{1985}).

\bibitem{Isgur:1989ed}
\bibinfo{author}{\bibfnamefont{N.}~\bibnamefont{Isgur}} \bibnamefont{and}
  \bibinfo{author}{\bibfnamefont{M.~B.} \bibnamefont{Wise}},
  \bibinfo{journal}{Phys. Lett.} \textbf{\bibinfo{volume}{B237}},
  \bibinfo{pages}{527} (\bibinfo{year}{1990}).

\bibitem{Georgi:1990ei}
\bibinfo{author}{\bibfnamefont{H.}~\bibnamefont{Georgi}},
  \bibinfo{author}{\bibfnamefont{B.}~\bibnamefont{Grinstein}},
  \bibnamefont{and} \bibinfo{author}{\bibfnamefont{M.~B.} \bibnamefont{Wise}},
  \bibinfo{journal}{Phys. Lett.} \textbf{\bibinfo{volume}{B252}},
  \bibinfo{pages}{456} (\bibinfo{year}{1990}).

\bibitem{Leibovich:2003tw}
\bibinfo{author}{\bibfnamefont{A.~K.} \bibnamefont{Leibovich}},
  \bibinfo{author}{\bibfnamefont{Z.}~\bibnamefont{Ligeti}},
  \bibinfo{author}{\bibfnamefont{I.~W.} \bibnamefont{Stewart}},
  \bibnamefont{and} \bibinfo{author}{\bibfnamefont{M.~B.} \bibnamefont{Wise}},
  \bibinfo{journal}{Phys. Lett.} \textbf{\bibinfo{volume}{B586}},
  \bibinfo{pages}{337} (\bibinfo{year}{2004}), \eprint{hep-ph/0312319}.

\bibitem{Isgur:1989vq}
\bibinfo{author}{\bibfnamefont{N.}~\bibnamefont{Isgur}} \bibnamefont{and}
  \bibinfo{author}{\bibfnamefont{M.~B.} \bibnamefont{Wise}},
  \bibinfo{journal}{Phys. Lett.} \textbf{\bibinfo{volume}{B232}},
  \bibinfo{pages}{113} (\bibinfo{year}{1989}).

\bibitem{Jenkins:1992se}
\bibinfo{author}{\bibfnamefont{E.}~\bibnamefont{Jenkins}},
  \bibinfo{author}{\bibfnamefont{A.~V.} \bibnamefont{Manohar}},
  \bibnamefont{and} \bibinfo{author}{\bibfnamefont{M.~B.} \bibnamefont{Wise}},
  \bibinfo{journal}{Nucl. Phys.} \textbf{\bibinfo{volume}{B396}},
  \bibinfo{pages}{38} (\bibinfo{year}{1993}), \eprint{hep-ph/9208248}.

\bibitem{Huang:2005me}
\bibinfo{author}{\bibfnamefont{M.-Q.} \bibnamefont{Huang}},
  \bibinfo{author}{\bibfnamefont{H.-Y.} \bibnamefont{Jin}},
  \bibinfo{author}{\bibfnamefont{J.~G.} \bibnamefont{Korner}},
  \bibnamefont{and} \bibinfo{author}{\bibfnamefont{C.}~\bibnamefont{Liu}}
  (\bibinfo{year}{2005}), \eprint{hep-ph/0502004}.

\bibitem{Abdallah:2003gn}
\bibinfo{author}{\bibfnamefont{J.}~\bibnamefont{Abdallah}} \emph{et~al.}
  (\bibinfo{collaboration}{DELPHI}), \bibinfo{journal}{Phys. Lett.}
  \textbf{\bibinfo{volume}{B585}}, \bibinfo{pages}{63} (\bibinfo{year}{2004}),
  \eprint{hep-ex/0403040}.

\bibitem{cockcroft:web}
\bibinfo{author}{\bibfnamefont{G.}~\bibnamefont{Aubrecht}} \emph{et~al.},
  \bibinfo{journal}{Contemporary Physics Education Project}
  (\bibinfo{year}{2003}),
  \bibinfo{note}{http://www.lbl.gov/abc/wallchart/teachersguide/pdf/Chap11.pdf%
}.

\bibitem{Schmidt:1993fz}
\bibinfo{author}{\bibfnamefont{C.~W.} \bibnamefont{Schmidt}}
  \bibinfo{note}{Presented at 1993 Particle Accelerator Conference (PAC 93),
  Washington, DC, 17-20 May 1993}.

\bibitem{main_injector}
\bibinfo{author}{\bibnamefont{{Ferrmilab Beam Division}}}
  \bibinfo{note}{{http://www-bd.fnal.gov/runII/index.html}}.

\bibitem{bishai:bxsec}
\bibinfo{author}{\bibfnamefont{D.}~\bibnamefont{Acosta}} \emph{et~al.}
  (\bibinfo{collaboration}{CDF}), \bibinfo{journal}{Phys. Rev.}
  \textbf{\bibinfo{volume}{D71}}, \bibinfo{pages}{032001}
  (\bibinfo{year}{2005}), \eprint{hep-ex/0412071}.

\bibitem{Affolder:2000tj}
\bibinfo{author}{\bibfnamefont{A.}~\bibnamefont{Affolder}} \emph{et~al.}
  (\bibinfo{collaboration}{CDF}), \bibinfo{journal}{Nucl. Instrum. Meth.}
  \textbf{\bibinfo{volume}{A453}}, \bibinfo{pages}{84} (\bibinfo{year}{2000}).

\bibitem{Hill:2004qb}
\bibinfo{author}{\bibfnamefont{C.~S.} \bibnamefont{Hill}}
  (\bibinfo{collaboration}{On behalf of the CDF}), \bibinfo{journal}{Nucl.
  Instrum. Meth.} \textbf{\bibinfo{volume}{A530}}, \bibinfo{pages}{1}
  (\bibinfo{year}{2004}).

\bibitem{Acosta:2004kc}
\bibinfo{author}{\bibfnamefont{D.}~\bibnamefont{Acosta}} \emph{et~al.}
  (\bibinfo{collaboration}{CDF-II}), \bibinfo{journal}{Nucl. Instrum. Meth.}
  \textbf{\bibinfo{volume}{A518}}, \bibinfo{pages}{605} (\bibinfo{year}{2004}).

\bibitem{Balka:1987ty}
\bibinfo{author}{\bibfnamefont{L.}~\bibnamefont{Balka}} \emph{et~al.}
  (\bibinfo{collaboration}{CDF}), \bibinfo{journal}{Nucl. Instrum. Meth.}
  \textbf{\bibinfo{volume}{A267}}, \bibinfo{pages}{272} (\bibinfo{year}{1988}).

\bibitem{Bertolucci:1987zn}
\bibinfo{author}{\bibfnamefont{S.}~\bibnamefont{Bertolucci}} \emph{et~al.}
  (\bibinfo{collaboration}{CDF}), \bibinfo{journal}{Nucl. Instrum. Meth.}
  \textbf{\bibinfo{volume}{A267}}, \bibinfo{pages}{301} (\bibinfo{year}{1988}).

\bibitem{Kuhlmann:2003hx}
\bibinfo{author}{\bibfnamefont{S.}~\bibnamefont{Kuhlmann}} \emph{et~al.},
  \bibinfo{journal}{Nucl. Instrum. Meth.} \textbf{\bibinfo{volume}{A518}},
  \bibinfo{pages}{39} (\bibinfo{year}{2004}), \eprint{physics/0310155}.

\bibitem{Artikov:2004ew}
\bibinfo{author}{\bibfnamefont{A.}~\bibnamefont{Artikov}} \emph{et~al.},
  \bibinfo{journal}{Nucl. Instrum. Meth.} \textbf{\bibinfo{volume}{A538}},
  \bibinfo{pages}{358} (\bibinfo{year}{2005}), \eprint{physics/0403079}.

\bibitem{Acosta:2001zu}
\bibinfo{author}{\bibfnamefont{D.}~\bibnamefont{Acosta}} \emph{et~al.}
  (\bibinfo{collaboration}{CDF}), \bibinfo{journal}{Nucl. Instrum. Meth.}
  \textbf{\bibinfo{volume}{A461}}, \bibinfo{pages}{540} (\bibinfo{year}{2001}).

\bibitem{Acosta:2002hx}
\bibinfo{author}{\bibfnamefont{D.}~\bibnamefont{Acosta}} \emph{et~al.},
  \bibinfo{journal}{Nucl. Instrum. Meth.} \textbf{\bibinfo{volume}{A494}},
  \bibinfo{pages}{57} (\bibinfo{year}{2002}).

\bibitem{knoll:rad}
\bibinfo{author}{\bibfnamefont{G.~F.} \bibnamefont{Knoll}},
  \emph{\bibinfo{title}{Radiation Detection and Measurement}}
  (\bibinfo{publisher}{New York, USA: John Wiley $\&$ Sons (2000) 816 p},
  \bibinfo{year}{2000}).

\bibitem{Veenhof:1998tt}
\bibinfo{author}{\bibfnamefont{R.}~\bibnamefont{Veenhof}},
  \bibinfo{journal}{Nucl. Instrum. Meth.} \textbf{\bibinfo{volume}{A419}},
  \bibinfo{pages}{726} (\bibinfo{year}{1998}).

\bibitem{cdfnote:asdq}
\bibinfo{author}{\bibnamefont{{W. Bokhari, F. M. Newcomer,~\etal}}},
  \emph{\bibinfo{title}{The ASDQ ASIC}}, \bibinfo{type}{CDF Internal Note}
  \bibinfo{number}{CDF/DOC/TRACKING/CDFR/4515},
  \bibinfo{institution}{University of Pennsylvania, Fermilab}
  (\bibinfo{year}{1998}).

\bibitem{yu:dedx}
\bibinfo{author}{\bibnamefont{{Shin-Shan Yu, Joel Heinrich, Nigel
  Lockyer,~\etal}}}, \emph{\bibinfo{title}{COT $dE/dx$ Measurement and
  Corrections}}, \bibinfo{type}{CDF Internal Note}
  \bibinfo{number}{CDF/DOC/BOTTOM/PUBLIC/6361},
  \bibinfo{institution}{University of Pennsylvania} (\bibinfo{year}{2003}).

\bibitem{cdfnote:6932}
\bibinfo{author}{\bibnamefont{{M. Donega, S. Giagu, D. Tonelli,~\etal}}},
  \emph{\bibinfo{title}{Track-based calibration of the COT specific
  ionization}}, \bibinfo{type}{CDF Internal Note}
  \bibinfo{number}{CDF/ANAL/BOTTOM/CDFR/6932}, \bibinfo{institution}{INFN,
  Rome, Pisa, and University of Geneva} (\bibinfo{year}{2004}).

\bibitem{cdfnote:cottracking}
\bibinfo{author}{\bibnamefont{{C. Hays, P. Tamburello, \etal}}},
  \emph{\bibinfo{title}{{The COT Pattern Recognition Algorithm and Offline
  Code}}}, \bibinfo{type}{CDF Internal Note}
  \bibinfo{number}{CDF/DOC/TRACKING/CDFR/6992}, \bibinfo{institution}{Duke
  University, Fermilab and University of Pennsylvania} (\bibinfo{year}{2004}).

\bibitem{Ascoli:1987av}
\bibinfo{author}{\bibfnamefont{G.}~\bibnamefont{Ascoli}} \emph{et~al.},
  \bibinfo{journal}{Nucl. Instrum. Meth.} \textbf{\bibinfo{volume}{A268}},
  \bibinfo{pages}{33} (\bibinfo{year}{1988}).

\bibitem{cdf:tdr}
\bibinfo{author}{\bibfnamefont{P.~T.} \bibnamefont{Lukens}}
  (\bibinfo{collaboration}{CDF IIb}) \bibinfo{note}{FERMILAB-TM-2198}.

\bibitem{Thomson:2002xp}
\bibinfo{author}{\bibfnamefont{E.~J.} \bibnamefont{Thomson}} \emph{et~al.},
  \bibinfo{journal}{IEEE Trans. Nucl. Sci.} \textbf{\bibinfo{volume}{49}},
  \bibinfo{pages}{1063} (\bibinfo{year}{2002}).

\bibitem{Ashmanskas:2003gf}
\bibinfo{author}{\bibfnamefont{B.}~\bibnamefont{Ashmanskas}} \emph{et~al.}
  (\bibinfo{collaboration}{CDF-II}), \bibinfo{journal}{Nucl. Instrum. Meth.}
  \textbf{\bibinfo{volume}{A518}}, \bibinfo{pages}{532} (\bibinfo{year}{2004}),
  \eprint{physics/0306169}.

\bibitem{Gomez-Ceballos:2004jk}
\bibinfo{author}{\bibfnamefont{G.}~\bibnamefont{Gomez-Ceballos}} \emph{et~al.},
  \bibinfo{journal}{Nucl. Instrum. Meth.} \textbf{\bibinfo{volume}{A518}},
  \bibinfo{pages}{522} (\bibinfo{year}{2004}).

\bibitem{lucchesi:bsyield}
\bibinfo{author}{\bibfnamefont{D.}~\bibnamefont{Lucchesi}} \emph{et~al.},
  \emph{\bibinfo{title}{Study of $B_s$ Yields in the Hadronic Trigger}},
  \bibinfo{type}{CDF Internal Note}
  \bibinfo{number}{CDF/PHYS/BOTTOM/CDFR/6345}, \bibinfo{institution}{INFN,
  Padova} (\bibinfo{year}{2003}).

\bibitem{cdfnote:1996}
\bibinfo{author}{\bibfnamefont{J.}~\bibnamefont{Marriner}},
  \emph{\bibinfo{title}{Secondary Vertex Fit with Mass and Pointing Constraints
  (CTVMFT)}}, \bibinfo{type}{Tech. Rep.}
  \bibinfo{number}{CDF/DOC/SEC$\_$VTX/PUBLIC/1996} (\bibinfo{year}{1993}).

\bibitem{cdfnote:strip}
\bibinfo{author}{\bibfnamefont{S.-S.} \bibnamefont{Yu}},
  \bibinfo{author}{\bibfnamefont{R.}~\bibnamefont{Tesarek}},
  \bibinfo{author}{\bibfnamefont{N.}~\bibnamefont{Lockyer}}, \bibnamefont{and}
  \bibinfo{author}{\bibfnamefont{D.}~\bibnamefont{Litvintsev}},
  \emph{\bibinfo{title}{Description of the Datasets for the Ratio of
  Semileptonic to Hadronic Partial Decay Width}}, \bibinfo{type}{CDF Internal
  Note} \bibinfo{number}{CDF/PHYS/BOTTOM/CDFR/6979},
  \bibinfo{institution}{University of Pennsylvania,Fermilab}
  (\bibinfo{year}{2004}).

\bibitem{cdfnote:7559}
\bibinfo{author}{\bibfnamefont{S.-S.} \bibnamefont{Yu}},
  \bibinfo{author}{\bibfnamefont{R.}~\bibnamefont{Tesarek}},
  \bibinfo{author}{\bibfnamefont{D.}~\bibnamefont{Litvintsev}},
  \bibinfo{author}{\bibfnamefont{J.}~\bibnamefont{Heinrich}}, \bibnamefont{and}
  \bibinfo{author}{\bibfnamefont{N.}~\bibnamefont{Lockyer}},
  \emph{\bibinfo{title}{Ratio of \Lb\ Semileptonic to Hadronic Branching
  Fractions in the Two Track Trigger}}, \bibinfo{type}{CDF Internal Note}
  \bibinfo{number}{CDF/PHYS/BOTTOM/CDFR/7559}, \bibinfo{institution}{University
  of Pennsylvania,Fermilab} (\bibinfo{year}{2005}).

\bibitem{James:1975dr}
\bibinfo{author}{\bibfnamefont{F.}~\bibnamefont{James}} \bibnamefont{and}
  \bibinfo{author}{\bibfnamefont{M.}~\bibnamefont{Roos}},
  \bibinfo{journal}{Comput. Phys. Commun.} \textbf{\bibinfo{volume}{10}},
  \bibinfo{pages}{343} (\bibinfo{year}{1975}).

\bibitem{furic:thesis}
\bibinfo{author}{\bibfnamefont{I.}~\bibnamefont{Furic}},
  \emph{\bibinfo{title}{{Measurement of the Ratio of Branching Fractions ${\cal
  B}(B_{s}^0\rightarrow D_s^{-}\pi^+)/{\cal B}(B^0\rightarrow D^-\pi^+)$ at
  CDF-II}}}, \bibinfo{type}{CDF Thesis}
  \bibinfo{number}{CDF/THESIS/BOTTOM/PUBLIC/7352},
  \bibinfo{institution}{Massachusetts Institute of Technology}
  (\bibinfo{year}{2004}).

\bibitem{yile:lblcpi}
\bibinfo{author}{\bibfnamefont{Y.}~\bibnamefont{Le}},
  \bibinfo{author}{\bibfnamefont{M.}~\bibnamefont{Martin}}, \bibnamefont{and}
  \bibinfo{author}{\bibfnamefont{P.}~\bibnamefont{Maksimovi\'{c}}},
  \emph{\bibinfo{title}{{Observation of \lbhad\ and the Measurement of
  $f_{\Lambda_b}{\cal B}(\Lambda_b\rightarrow\Lambda_c^+\pi^-)/f_{B^0}{\cal
  B}(B^0\rightarrow D^-\pi^+)$}}}, \bibinfo{type}{CDF Internal Note}
  \bibinfo{number}{CDF/ANAL/BOTTOM/CDFR/6396}, \bibinfo{institution}{Johns
  Hopkins University} (\bibinfo{year}{2004}).

\bibitem{cdfnote:7388}
\bibinfo{author}{\bibnamefont{{A. Belloni, J. Piedra~\etal}}},
  \emph{\bibinfo{title}{{Unbinned Likelihood Fit for B0 Mixing in Fully
  Reconstructed Decays}}}, \bibinfo{type}{CDF Internal Note}
  \bibinfo{number}{CDF/PHYS/BOTTOM/CDFR/7388} (\bibinfo{year}{2004}),
  \bibinfo{note}{see Chapter 11}.

\bibitem{mit:bgen}
\bibinfo{author}{\bibfnamefont{K.}~\bibnamefont{Anikeev}},
  \bibinfo{author}{\bibfnamefont{P.}~\bibnamefont{Murat}}, \bibnamefont{and}
  \bibinfo{author}{\bibfnamefont{C.}~\bibnamefont{Paus}},
  \emph{\bibinfo{title}{{Description of Bgenerator}}}, \bibinfo{type}{CDF
  Internal Note} \bibinfo{number}{CDF/DOC/BOTTOM/CDFR/5092},
  \bibinfo{institution}{MIT} (\bibinfo{year}{1999}).

\bibitem{pythia:manual}
\bibinfo{author}{\bibfnamefont{T.}~\bibnamefont{Sjostrand}},
  \bibinfo{author}{\bibfnamefont{L.}~\bibnamefont{Lonnblad}}, \bibnamefont{and}
  \bibinfo{author}{\bibfnamefont{S.}~\bibnamefont{Mrenna}}
  (\bibinfo{year}{2001}), \eprint{hep-ph/0108264}.

\bibitem{Nason:1989zy}
\bibinfo{author}{\bibfnamefont{P.}~\bibnamefont{Nason}},
  \bibinfo{author}{\bibfnamefont{S.}~\bibnamefont{Dawson}}, \bibnamefont{and}
  \bibinfo{author}{\bibfnamefont{R.~K.} \bibnamefont{Ellis}},
  \bibinfo{journal}{Nucl. Phys.} \textbf{\bibinfo{volume}{B327}},
  \bibinfo{pages}{49} (\bibinfo{year}{1989}).

\bibitem{Peterson:1982ak}
\bibinfo{author}{\bibfnamefont{C.}~\bibnamefont{Peterson}},
  \bibinfo{author}{\bibfnamefont{D.}~\bibnamefont{Schlatter}},
  \bibinfo{author}{\bibfnamefont{I.}~\bibnamefont{Schmitt}}, \bibnamefont{and}
  \bibinfo{author}{\bibfnamefont{P.~M.} \bibnamefont{Zerwas}},
  \bibinfo{journal}{Phys. Rev.} \textbf{\bibinfo{volume}{D27}},
  \bibinfo{pages}{105} (\bibinfo{year}{1983}).

\bibitem{pdg:mctech}
\bibinfo{author}{\bibfnamefont{C.}~\bibnamefont{Caso}} \emph{et~al.},
  \bibinfo{journal}{Phys. Let. B}  (\bibinfo{year}{2004}), \bibinfo{note}{see
  Monte Carlo techniques, page 289}.

\bibitem{Lange:2001uf}
\bibinfo{author}{\bibfnamefont{D.~J.} \bibnamefont{Lange}},
  \bibinfo{journal}{Nucl. Instrum. Meth.} \textbf{\bibinfo{volume}{A462}},
  \bibinfo{pages}{152} (\bibinfo{year}{2001}).

\bibitem{Brun:1978fy}
\bibinfo{author}{\bibfnamefont{R.}~\bibnamefont{Brun}},
  \bibinfo{author}{\bibfnamefont{R.}~\bibnamefont{Hagelberg}},
  \bibinfo{author}{\bibfnamefont{M.}~\bibnamefont{Hansroul}}, \bibnamefont{and}
  \bibinfo{author}{\bibfnamefont{J.~C.} \bibnamefont{Lassalle}}
  \bibinfo{note}{CERN-DD-78-2-REV}.

\bibitem{heinrich:comm}
\bibinfo{author}{\bibfnamefont{J.}~\bibnamefont{Heinrich}},
  \emph{\bibinfo{title}{{Private communications}}}.

\bibitem{cdfnote:6347}
\bibinfo{author}{\bibnamefont{{Ken Bloom and David Dagenhart}}},
  \emph{\bibinfo{title}{Muon-Reconstruction Efficiency for Winter 2003
  Conferences}}, \bibinfo{type}{CDF Internal Note}
  \bibinfo{number}{CDF/ANAL/MUON/CDFR/6347} (\bibinfo{year}{2003}).

\bibitem{cdfnote:6391}
\bibinfo{author}{\bibnamefont{{S.Giagu, M.Rescigno,~\etal.}}},
  \emph{\bibinfo{title}{BR ratios and direct CP violation in Cabibbo supressed
  decays of \Dzero}}, \bibinfo{type}{CDF Internal Note}
  \bibinfo{number}{CDF/PHYS/BOTTOM/CDFR/6391} (\bibinfo{year}{2003}).

\bibitem{cdfnote:7301}
\bibinfo{author}{\bibnamefont{{M. Herndon,~\etal.}}},
  \emph{\bibinfo{title}{Proton XFT Efficiency Estimate for \lbhad\ Analysis}},
  \bibinfo{type}{CDF Internal Note} \bibinfo{number}{CDF/PHYS/BOTTOM/CDFR/7301}
  (\bibinfo{year}{2004}).

\bibitem{cdfnote:6643}
\bibinfo{author}{\bibfnamefont{T.}~\bibnamefont{Yamashita}} \bibnamefont{and}
  \bibinfo{author}{\bibfnamefont{e.}~\bibnamefont{Ting~Miao}},
  \emph{\bibinfo{title}{Measurement of ${\cal B}(\Lb \rightarrow J/\psi
  \Lambda)$}}, \bibinfo{type}{CDF Internal Note}
  \bibinfo{number}{CDF/PHYS/BOTTOM/CDFR/6643}, \bibinfo{institution}{Okayama
  University, Fermilab and Argonne National Laboratory} (\bibinfo{year}{2003}).

\bibitem{Abbaneo:2001bv}
\bibinfo{author}{\bibfnamefont{D.}~\bibnamefont{Abbaneo}} \emph{et~al.}
  (\bibinfo{collaboration}{ALEPH})  (\bibinfo{year}{2001}),
  \eprint{hep-ex/0112028}.

\bibitem{Affolder:1999iq}
\bibinfo{author}{\bibfnamefont{T.}~\bibnamefont{Affolder}} \emph{et~al.}
  (\bibinfo{collaboration}{CDF}), \bibinfo{journal}{Phys. Rev. Lett.}
  \textbf{\bibinfo{volume}{84}}, \bibinfo{pages}{1663} (\bibinfo{year}{2000}),
  \eprint{hep-ex/9909011}.

\bibitem{taylor:thesis}
\bibinfo{author}{\bibfnamefont{W.}~\bibnamefont{Taylor}},
  \emph{\bibinfo{title}{{A Measurement of b-quark Fragmentation Fractions in
  $p\overline{p}$ Collisions at $\sqrt{s}$ = 1.8 TeV}}}, \bibinfo{type}{CDF
  Thesis} \bibinfo{number}{CDF/THESIS/BOTTOM/PUBLIC/4913},
  \bibinfo{institution}{University of Toronto} (\bibinfo{year}{1999}).

\bibitem{cleo:qq}
\bibinfo{author}{\bibnamefont{{Avery, P. and Read, K. and Trahern, G.}}},
  \emph{\bibinfo{title}{{QQ: A Monte Carlo Generator}}}, \bibinfo{type}{CLEO
  Software Note} \bibinfo{number}{CSN-212} (\bibinfo{year}{1985}).

\bibitem{cdfnote:7546}
\bibinfo{author}{\bibfnamefont{D.}~\bibnamefont{Litvintsev}},
  \bibinfo{author}{\bibfnamefont{S.-S.} \bibnamefont{Yu}}, \bibnamefont{and}
  \bibinfo{author}{\bibfnamefont{R.}~\bibnamefont{Tesarek}},
  \emph{\bibinfo{title}{Observation of $\Lamc^{*}$ and $\Sigma_c$ in
  semileptonic \Lb\ decays}}, \bibinfo{type}{CDF Internal Note}
  \bibinfo{number}{CDF/PHYS/BOTTOM/CDFR/7546}, \bibinfo{institution}{Fermilab,
  University of Pennsylvania} (\bibinfo{year}{2005}).

\bibitem{cdfnote:6599}
\bibinfo{author}{\bibfnamefont{R.}~\bibnamefont{Tesarek}} \bibnamefont{and}
  \bibinfo{author}{\bibfnamefont{S.-S.} \bibnamefont{Yu}},
  \emph{\bibinfo{title}{A Study of Backgrounds to the Decay $\dsemi$}},
  \bibinfo{type}{CDF Internal Note}
  \bibinfo{number}{CDF/PHYS/BOTTOM/CDFR/6599}, \bibinfo{institution}{Fermilab,
  University of Pennsylvania} (\bibinfo{year}{2003}).

\bibitem{cdfnote:7545}
\bibinfo{author}{\bibfnamefont{R.}~\bibnamefont{Tesarek}},
  \bibinfo{author}{\bibfnamefont{S.-S.} \bibnamefont{Yu}}, \bibnamefont{and}
  \bibinfo{author}{\bibfnamefont{D.}~\bibnamefont{Litvintsev}},
  \emph{\bibinfo{title}{A Study of Backgrounds to the decay \lbsemi}},
  \bibinfo{type}{CDF Internal Note}
  \bibinfo{number}{CDF/ANAL/BOTTOM/CDFR/7545}, \bibinfo{institution}{Fermilab,
  University of Pennsylvania} (\bibinfo{year}{2005}).

\bibitem{Edwards:1994ar}
\bibinfo{author}{\bibfnamefont{K.~W.} \bibnamefont{Edwards}} \emph{et~al.}
  (\bibinfo{collaboration}{CLEO}), \bibinfo{journal}{Phys. Rev. Lett.}
  \textbf{\bibinfo{volume}{74}}, \bibinfo{pages}{3331} (\bibinfo{year}{1995}).

\bibitem{Albrecht:1997qa}
\bibinfo{author}{\bibfnamefont{H.}~\bibnamefont{Albrecht}} \emph{et~al.}
  (\bibinfo{collaboration}{ARGUS}), \bibinfo{journal}{Phys. Lett.}
  \textbf{\bibinfo{volume}{B402}}, \bibinfo{pages}{207} (\bibinfo{year}{1997}).

\bibitem{Leibovich:1997az}
\bibinfo{author}{\bibfnamefont{A.~K.} \bibnamefont{Leibovich}}
  \bibnamefont{and} \bibinfo{author}{\bibfnamefont{I.~W.}
  \bibnamefont{Stewart}}, \bibinfo{journal}{Phys. Rev.}
  \textbf{\bibinfo{volume}{D57}}, \bibinfo{pages}{5620} (\bibinfo{year}{1998}),
  \eprint{hep-ph/9711257}.

\bibitem{ash:dmm}
\bibinfo{author}{\bibnamefont{{CDF Collaboration}}}, \bibinfo{journal}{Phys.
  Rev. D} \textbf{\bibinfo{volume}{68}}, \bibinfo{pages}{091101}
  (\bibinfo{year}{2003}).

\bibitem{dmitri:prepare}
\bibinfo{author}{\bibfnamefont{D.}~\bibnamefont{Litvintsev}},
  \emph{\bibinfo{title}{{CDF note in preparation}}}, \bibinfo{type}{Tech. Rep.}

\bibitem{Acosta:2004uq}
\bibinfo{author}{\bibfnamefont{D.}~\bibnamefont{Acosta}} \emph{et~al.}
  (\bibinfo{collaboration}{CDF II})  (\bibinfo{year}{2004}),
  \eprint{hep-ex/0406078}.

\bibitem{lannon:bpythia}
\bibinfo{author}{\bibfnamefont{K.}~\bibnamefont{Lannon}} \bibnamefont{and}
  \bibinfo{author}{\bibfnamefont{K.}~\bibnamefont{Pitts}},
  \emph{\bibinfo{title}{{Bottom Quark Production Using PYTHIA and HERWIG}}},
  \bibinfo{type}{CDF Internal Note}
  \bibinfo{number}{CDF/PHYS/BOTTOM/CDFR/6253}, \bibinfo{institution}{University
  of Illinois at Urbana-Champaign} (\bibinfo{year}{2003}).

\bibitem{rfield:pythia}
\bibinfo{author}{\bibfnamefont{R.}~\bibnamefont{Field}},
  \emph{\bibinfo{title}{{The Sources of b-quarks at the Tevatron}}},
  \bibinfo{type}{CDF Internal Note}
  \bibinfo{number}{CDF/ANAL/BOTTOM/CDFR/5558}, \bibinfo{institution}{University
  of Florida} (\bibinfo{year}{2001}).

\bibitem{bmc:nbot90}
\bibinfo{author}{\bibfnamefont{G.}~\bibnamefont{Gomez-Ceballos}} \emph{et~al.},
  \emph{\bibinfo{title}{B Monte Carlo homepage}}, \bibinfo{type}{Tech. Rep.}
  (\bibinfo{year}{2004}), \bibinfo{note}{see
  http://www-cdf.fnal.gov/internal/physics/bottom/b-montecarlo/db/g020.txt}.

\bibitem{run1:bpxsec}
\bibinfo{author}{\bibnamefont{{CDF Collaboration}}}, \bibinfo{journal}{Phys.
  Rev. D} \textbf{\bibinfo{volume}{65}}, \bibinfo{pages}{052006}
  (\bibinfo{year}{2002}).

\bibitem{cchen:dxec}
\bibinfo{author}{\bibfnamefont{D.}~\bibnamefont{Acosta}} \emph{et~al.}
  (\bibinfo{collaboration}{CDF}), \bibinfo{journal}{Phys. Rev. Lett.}
  \textbf{\bibinfo{volume}{91}}, \bibinfo{pages}{241804}
  (\bibinfo{year}{2003}), \eprint{hep-ex/0307080}.

\bibitem{lannon:bcorel}
\bibinfo{author}{\bibfnamefont{D.}~\bibnamefont{Acosta}} \emph{et~al.}
  (\bibinfo{collaboration}{CDF}), \emph{\bibinfo{title}{Measurements of bottom
  anti-bottom azimuthal production correlations in proton antiproton collisions
  at s**(1/2) = 1.8-TeV}}, \bibinfo{type}{Tech. Rep.} (\bibinfo{year}{2004}),
  \eprint{hep-ex/0412006}.

\bibitem{cdfnote:6953}
\bibinfo{author}{\bibfnamefont{M.}~\bibnamefont{Martin}} \emph{et~al.},
  \emph{\bibinfo{title}{{Evaluation of fit systematics due to background shapes
  in \lbhad}}}, \bibinfo{type}{Tech. Rep.}
  \bibinfo{number}{CDF/DOC/BOTTOM/PUBLIC/6953}, \bibinfo{institution}{Johns
  Hopkins University} (\bibinfo{year}{2004}).

\bibitem{cdfnote:6355}
\bibinfo{author}{\bibnamefont{{A. Korn, G. Bauer and C. Paus}}},
  \emph{\bibinfo{title}{{Update on Calibration of Energy Loss and Magnetic
  Field using $J/\psi$ Events in Run II}}}, \bibinfo{type}{CDF Internal Note}
  \bibinfo{number}{CDF/DOC/BOTTOM/CDFR/6355},
  \bibinfo{institution}{Massachusetts Institute of Technology}
  (\bibinfo{year}{2003}).

\bibitem{Brun:1987ma}
\bibinfo{author}{\bibfnamefont{R.}~\bibnamefont{Brun}},
  \bibinfo{author}{\bibfnamefont{F.}~\bibnamefont{Bruyant}},
  \bibinfo{author}{\bibfnamefont{M.}~\bibnamefont{Maire}},
  \bibinfo{author}{\bibfnamefont{A.~C.} \bibnamefont{McPherson}},
  \bibnamefont{and} \bibinfo{author}{\bibfnamefont{P.}~\bibnamefont{Zanarini}}
  \bibinfo{note}{CERN-DD/EE/84-1}.

\bibitem{numi:1994}
\bibinfo{author}{\bibfnamefont{D.}~\bibnamefont{Michael}}
  (\bibinfo{year}{1994}), \bibinfo{note}{nuMI-NOTE-BEAM-0019}.

\bibitem{dalitz:phil}
\bibinfo{author}{\bibnamefont{{R. H. Dalitz}}}, \bibinfo{journal}{Phil. Mag.}
  \textbf{\bibinfo{volume}{44}}, \bibinfo{pages}{1068} (\bibinfo{year}{1953}).

\bibitem{Aitala:1999uq}
\bibinfo{author}{\bibfnamefont{E.~M.} \bibnamefont{Aitala}} \emph{et~al.}
  (\bibinfo{collaboration}{E791}), \bibinfo{journal}{Phys. Lett.}
  \textbf{\bibinfo{volume}{B471}}, \bibinfo{pages}{449} (\bibinfo{year}{2000}),
  \eprint{hep-ex/9912003}.

\bibitem{Dunietz:1998uz}
\bibinfo{author}{\bibfnamefont{I.}~\bibnamefont{Dunietz}},
  \bibinfo{journal}{Phys. Rev.} \textbf{\bibinfo{volume}{D58}},
  \bibinfo{pages}{094010} (\bibinfo{year}{1998}), \eprint{hep-ph/9805287}.

\bibitem{Migliozzi:1999ca}
\bibinfo{author}{\bibfnamefont{P.}~\bibnamefont{Migliozzi}},
  \bibinfo{author}{\bibfnamefont{G.}~\bibnamefont{D'Ambrosio}},
  \bibinfo{author}{\bibfnamefont{G.}~\bibnamefont{Miele}}, \bibnamefont{and}
  \bibinfo{author}{\bibfnamefont{P.}~\bibnamefont{Santorelli}}
  (\bibinfo{collaboration}{CHORUS}), \bibinfo{journal}{Phys. Lett.}
  \textbf{\bibinfo{volume}{B462}}, \bibinfo{pages}{217} (\bibinfo{year}{1999}),
  \eprint{hep-ph/9906219}.

\bibitem{Abreu:1999hu}
\bibinfo{author}{\bibfnamefont{P.}~\bibnamefont{Abreu}} \emph{et~al.}
  (\bibinfo{collaboration}{DELPHI}), \bibinfo{journal}{Eur. Phys. J.}
  \textbf{\bibinfo{volume}{C10}}, \bibinfo{pages}{185} (\bibinfo{year}{1999}).

\bibitem{Ackerstaff:1997qi}
\bibinfo{author}{\bibfnamefont{K.}~\bibnamefont{Ackerstaff}} \emph{et~al.}
  (\bibinfo{collaboration}{OPAL}), \bibinfo{journal}{Phys. Lett.}
  \textbf{\bibinfo{volume}{B426}}, \bibinfo{pages}{161} (\bibinfo{year}{1998}),
  \eprint{hep-ex/9802002}.

\bibitem{Abe:1996df}
\bibinfo{author}{\bibfnamefont{F.}~\bibnamefont{Abe}} \emph{et~al.}
  (\bibinfo{collaboration}{CDF}), \bibinfo{journal}{Phys. Rev. Lett.}
  \textbf{\bibinfo{volume}{77}}, \bibinfo{pages}{1439} (\bibinfo{year}{1996}).

\bibitem{Isgur:1990pm}
\bibinfo{author}{\bibfnamefont{N.}~\bibnamefont{Isgur}} \bibnamefont{and}
  \bibinfo{author}{\bibfnamefont{M.~B.} \bibnamefont{Wise}},
  \bibinfo{journal}{Nucl. Phys.} \textbf{\bibinfo{volume}{B348}},
  \bibinfo{pages}{276} (\bibinfo{year}{1991}).

\bibitem{Georgi:1990cx}
\bibinfo{author}{\bibfnamefont{H.}~\bibnamefont{Georgi}},
  \bibinfo{journal}{Nucl. Phys.} \textbf{\bibinfo{volume}{B348}},
  \bibinfo{pages}{293} (\bibinfo{year}{1991}).

\bibitem{Bensalem:2002pz}
\bibinfo{author}{\bibfnamefont{W.}~\bibnamefont{Bensalem}},
  \bibinfo{author}{\bibfnamefont{A.}~\bibnamefont{Datta}}, \bibnamefont{and}
  \bibinfo{author}{\bibfnamefont{D.}~\bibnamefont{London}},
  \bibinfo{journal}{Phys. Lett.} \textbf{\bibinfo{volume}{B538}},
  \bibinfo{pages}{309} (\bibinfo{year}{2002}), \eprint{hep-ph/0205009}.

\bibitem{Dunietz:1992ti}
\bibinfo{author}{\bibfnamefont{I.}~\bibnamefont{Dunietz}}, \bibinfo{journal}{Z.
  Phys.} \textbf{\bibinfo{volume}{C56}}, \bibinfo{pages}{129}
  (\bibinfo{year}{1992}).

\bibitem{cdfnote:7443}
\bibinfo{author}{\bibnamefont{{R. Carosi and M. A. Ciocci and S. Torre}}},
  \emph{\bibinfo{title}{{Search for $\Lb\rightarrow p\pi$ and $\Lb\rightarrow
  pK$ decays}}}, \bibinfo{type}{CDF Internal Note}
  \bibinfo{number}{CDF/PUB/BOTTOM/PUBLIC/7443}, \bibinfo{institution}{INFN, Sez
  di Pisa} (\bibinfo{year}{2005}), \bibinfo{note}{{See
  http://fcdfwww.fnal.gov/internal/physics/godparents/lb$\_$ppi$\_$pk/}}.

\bibitem{Buskulic:1996tx}
\bibinfo{author}{\bibfnamefont{D.}~\bibnamefont{Buskulic}} \emph{et~al.}
  (\bibinfo{collaboration}{ALEPH}), \bibinfo{journal}{Phys. Lett.}
  \textbf{\bibinfo{volume}{B384}}, \bibinfo{pages}{471} (\bibinfo{year}{1996}).

\end{thebibliography}

\end{document}